\documentclass[11pt,makeidx,head]{book}
\usepackage{latexsym}                                   %
\makeatletter
\@addtoreset{equation}{section}         
\makeatother
\makeindex

\newcommand{\lo}{\left(}
\newcommand{\ro}{\right)}
\newcommand{\p}{\partial}
\newcommand{\ti}{\tilde}
\newcommand{\g}{\gamma}
\newcommand{\G}{\Gamma}
\newcommand{\s}{\sigma}
\newcommand{\vp}{\varphi}
\newcommand{\ov}{\overline}
\newcommand{\al}{\alpha}
\newcommand{\lbd}{\lambda}

\newcommand{\ve}{\varepsilon}
\newcommand{\e}{\epsilon}
\newcommand{\om}{\omega}
\newcommand{\edi}{\displaystyle}
\newcommand{\ssl}{\scriptscriptstyle}
\font \msbmten=msbm10 scaled\magstephalf

\def \R {\mbox{\msbmten\char'122}}

\def \C {\mbox{\msbmten\char'103}}

\def \N {\mbox{\msbmten\char'116}}
\def \Z {\mbox{\msbmten\char'132}}

\def\stimes{\subset \!\!\!\!\!\! \times}
\def\vark{\mbox{\msbmten\char'173}}

\def\varp{\varphi}
\def\pa{\partial}
\def\ga{\gamma}
\def\al{\alpha}
\def\ov{\overline}
\def\la{\lambda}

\def\wid{\widetilde}
\def\non{\nonumber}
\def\dis{\displaystyle}

\begin{document}  
\begin{titlepage}
\vspace*{25mm}
\begin{center}
{\Large \bf SYMMETRIES AND EXACT SOLUTIONS\\ 
\vspace*{5mm}
OF NONLINEAR DIRAC EQUATIONS}
\end{center}
\vspace*{25mm}

\begin{center}
{\bf Wilhelm Fushchych and Renat Zhdanov}
\end{center}

\vspace*{10mm}
\begin{center}
{\small Institute of Mathematics of the National}\\
{\small Ukrainian Academy of Sciences}
\end{center}

\vfill
\begin{center}
{\sf Ukraine $\quad$ 1997 $\quad$ Kyiv\\
Mathematical Ukraina Publisher}
\end{center}
\end{titlepage}
\pagestyle{myheadings}                
\thispagestyle{empty}
\markboth{CONTENTS}{CONTENTS}
\pagenumbering{roman}
\begin{tableofcontents}
\contentsline{chapter}{PREFACE}{\pageref{preface}}
\contentsline{chapter}{INTRODUCTION}{\pageref{introduction}}
\contentsline{chapter}{1. SYMMETRY OF NONLINEAR SPINOR 
EQUATIONS}{\pageref{ch1}}
\contentsline{section}{1.1. Local and nonlocal symmetry of 
  the Dirac equation}{\pageref{s1.1}}
\contentsline{section}{1.2. Nonlinear spinor equations}{\pageref{s1.2}}
\contentsline{section}{1.3. Systems of nonlinear second-order
  equations for the spinor field}{\pageref{s1.3}}
\contentsline{section}{1.4. Symmetry of systems of nonlinear equations
  for spinor, vector and scalar fields}{\pageref{s1.4}}
\contentsline{section}{1.5. Conditional symmetry and reduction of
  partial differential equations}{\pageref{s1.5}}
\contentsline{section}{1.6. Conservation laws}{\pageref{s1.6}}
\contentsline{chapter}{2. EXACT SOLUTIONS}{\pageref{ch2}}
\contentsline{section}{2.1. On compatibility and general solution of
  the d'Alembert-Ha\-mil\-ton system}{\pageref{s2.1}}
\contentsline{section}{2.2. Ans\"atze for the spinor
  field}{\pageref{s2.2}} 
\contentsline{section}{2.3. Reduction of Poincar\'e-invariant spinor
  equations}{\pageref{s2.3}} 
\contentsline{section}{2.4. Exact solutions of nonlinear spinor
  equations}{\pageref{s2.4}} 
\contentsline{section}{2.5. Nonlinear spinor equations and special
  functions} {\pageref{s2.5}}
\contentsline{section}{2.6. Construction of fields with spins $s= 0,
  1, 3/2$ via the Dirac field}{\pageref{s2.6}}
\contentsline{section}{2.7. Exact solutions of the Dirac-d'Alembert
  equation}{\pageref{s2.7}} 
\contentsline{section}{2.8. Exact solutions of the nonlinear
  electrodynamics equations}{\pageref{s2.8}}
\contentsline{chapter}{3. TWO-DIMENSIONAL SPINOR MODELS}{\pageref{ch3}}
\contentsline{section}{3.1. Two-dimensional spinor equations invariant
  under the infinite-parameter groups}{\pageref{s3.1}}
\contentsline{section}{3.2. Nonlinear two-dimensional Dirac-Heisenberg
  equations}{\pageref{s3.2}} 
\contentsline{section}{3.3. Two-dimensional classical electrodynamics
  equations}{\pageref{s3.3}}
\contentsline{section}{3.4. General solutions of Galilei-invariant
  spinor equations}{\pageref{s3.4}}
\contentsline{chapter}{4. NONLINEAR GALILEI-INVARIANT
  EQUATIONS}{\pageref{ch4}}
\contentsline{section}{4.1. Nonlinear equations for the spinor field
  invariant under the group $G(1,3)$ and its
  extensions}{\pageref{s4.1}} 
\contentsline{section}{4.2. Exact solutions of Galilei-invariant
  spinor equations}{\pageref{s4.2}}
\contentsline{section}{4.3. Galilei-invariant second-order spinor
  equations}{\pageref{s4.3}} 
\contentsline{chapter}{5. SEPARATION OF VARIABLES}{\pageref{ch5}}
\contentsline{section}{5.1. Separation of variables and symmetry of
  systems of partial differential equations}{\pageref{s5.1}}
\contentsline{section}{5.2. Separation of variables in the
  Galilei-invariant spinor equation}{\pageref{s5.2}}
\contentsline{section}{5.3. Separation of variables in the
  Schr\"odinger equation}{\pageref{s5.3}}
\contentsline{chapter}{6. CONDITIONAL SYMMETRY AND REDUCTION OF SPI\-NOR
  EQUATIONS}{\pageref{ch6}} 
\contentsline{section}{6.1. Non-Lie reduction of Poincar\'e-invariant
  spinor equations}{\pageref{s6.1}}
\contentsline{section}{6.2. Non-Lie reduction of Galilei-invariant
  spinor equations}{\pageref{s6.2}}
\contentsline{chapter}{7. REDUCTION AND EXACT SOLUTIONS OF {\boldmath
    $SU$}(2) YANG-MILLS EQUATIONS}{\pageref{ch7}}
\contentsline{section}{7.1. Symmetry reduction and exact solutions of
  the Yang-Mills equations}{\pageref{s7.1}}
\contentsline{section}{7.2. Non-Lie reduction of the Yang-Mills
  equations}{\pageref{s7.2}} 
\contentsline{chapter}{APPENDIX 1. THE POINCAR\'E GROUP AND ITS
  REPRE\-SENTATIONS}{\pageref{ap1}} 
\contentsline{chapter}{APPENDIX 2. THE GALILEI GROUP AND ITS
  REPRE\-SENTATIONS}{\pageref{ap2}}
\contentsline{chapter}{APPENDIX 3. REPRESENTATIONS OF THE POINCAR\'E
  AND GALILEI GROUPS BY LIE VECTOR VIELDS}{\pageref{ap3}}
\contentsline{chapter}{BIBLIOGRAPHY}{\pageref{bibliography}}
\contentsline{chapter}{INDEX}{\pageref{index}}
\end{tableofcontents}

\newpage
\pagestyle{myheadings}
\pagenumbering{arabic}               
\setcounter{page}{1}  
\thispagestyle{empty}

\vspace*{15mm}

\begin{flushright}
{\em Physical law should have\\
 mathematical beauty}
\end{flushright}
\begin{flushright}
P.-M.~Dirac
\end{flushright}
\vspace*{10mm}

\rightline
{\large\bf
P R E F A C E\label{preface}}
\markboth{PREFACE}{PREFACE}              
\vspace{10mm}

{\small

\noindent
The Dirac equation describing motion of an elementary particle with
spin $1/2$ (electron or proton) is an inseparable part of the
modern mathematical and theoretical physics. Together with the Maxwell
and Schr\"odinger equations it forms a basis of the quantum mechanics,
quantum electrodynamics and quantum field theory.

Following Dirac's discovery of the linear equation of an electron
there appeared fundamental papers by D.D.~Ivanenko, W.~Heisenberg,
R.~Finkelstein with collaborators and F.~G\"ursey advocating the idea
of nonlinear description of an elementary particle with spin $1/2$
which made it possible to take into account its self-interaction.
Furthermore, W.~Heisenberg put forward the idea to use a nonlinear
Dirac equation as a possible basis model for a unified field theory.
These ideas have contributed substantially to the modern view of an
elementary particle as a complex dynamical system described (modeled)
by a nonlinear system of partial differential equations. The general
structure of such nonlinear equations is determined by the
Lorentz-Poincar\'e-Einstein or the Galilei relativity principle.

Till now there is no book devoted to a systematic study of nonlinear
generalizations of the classical Dirac equation. So it was our
primary intention to write a book devoted entirely to a comprehensive
and detailed group-theoretical study of first--order nonlinear spinor
partial differential equations satisfying either the
Lorentz--Poincar\'e--Einstein or the Galilei relativity principle.
These equations contain, as particular cases, the nonlinear spinor
models suggested by D.D.~Ivanenko, W.~Heisenberg, R.~Finkelstein and
F.~G\"ursey.

In the course of research we have discovered that the methods and
techniques developed to study nonlinear Dirac equations can be
successfully applied to a wide range of Poincar\'e- and
Galilei-invariant nonlinear multi-dimensional equations of modern
quantum field theory describing interactions of spinor, scalar and
vector fields.

As a result, the book has a \lq two-level\rq\ structure. At the first
level, it may be considered as a self-contained group-theoretical
introduction to the theory of the first-order nonlinear spinor
equations with a particular emphasis on a development of efficient
methods for constructing their exact (classical) solutions. At the
second level, we employ these methods to construct multi-parameter
families of exact solutions of nonlinear wave, Dirac-d'Alembert,
Maxwell-Dirac, d'Alembert-eikonal, $SU(2)$ Yang-Mills, L\'evy-Leblond,
and some other partial differential equations. Furthermore, the
approach used enables us to give a systematic and unified treatment of
the related questions such as conditional symmetry of differential
equations, separation of variables in linear systems of partial
differential equations, and integrability of some nonlinear systems of
differential equations in two-independent variables.

It was our aim to write a book in a form accessible not only for 
\lq pure theoreticians\rq\ but also for those who are interested in
applications of group-theoretical/symmetry methods to concrete
nonlinear systems of partial differential equations. Every opportunity
is taken to illustrate general statements by specific examples and to
reduce to a reasonable minimum the level of abstractness in the
exposition.

The book is based on the authors' results obtained at the Institute of
Mathematics of the National Academy of Sciences of Ukraine in
1984--1996 \cite{90,91,93}, \cite{96}--\cite{107},
\cite{204}--\cite{216}. It also accumulates a rich experience of other
groups working in the related areas of group-theoretical,
algebraic-theoretical analysis of differential equations. The
bibliography is claimed to be the most comprehensive and complete as
far as symmetry and exact solutions of nonlinear spinor equations are
concerned. But it is not our intention to give the full list of
references devoted to application and development of group-theoretical
methods in the mathematical and theoretical physics. Only references
used directly are cited.

When the book was at the last stage of preparation one of the authors
(RZ) was at the Arnold-Sommerfeld Institute for Mathematical Physics
(Clausthal-Zel\-ler\-feld, Germany) as an Alexander von Humboldt
Fellow. He is indebted to Professor \mbox{H.-D.}~Doebner for an
invitation and kind hospitality. His critical remarks as well as
stimulating discussions with participants of the Seminar at the
Institute for Theoretical Physics, V.~Dobrev, J.~Hennig, W.~L\"ucke,
P.~Nattermann and W.~Scherer, are gratefully acknowledged.
Authors would like to thank Soros Foundation for financial
support. 

Our special thanks are addressed to W.M.~Shtelen, I.A.~Yehorchenko
and P.~Ba\-sa\-rab-Hor\-vath for critical reading the manuscript and
and valuable suggestions. 

We express deep gratitude to our colleagues at the Department of
Applied Research of the Institute of Mathematics of the National
Academy of Sciences of Ukraine, A.G.~Nikitin, I.V.~Revenko,
V.I.~Lahno, A.Yu.~Andreitsev, for their fruitful cooperation and also
to G.A.~Zhdanova for her kind help in preparing the manuscript
for publication.}

\begin{flushright} {\em Ukraine, Kyiv} --\phantom {\em Germany,
    Clausthal-Zellerfeld}\\ {\em Germany, Clausthal-Zellerfeld} 
\end{flushright}
\begin{flushright}
{\em 1996, June}
\end{flushright}

\newpage
\thispagestyle{empty}
\vspace*{35mm}

\rightline
{\large\bf
INTRODUCTION\label{introduction}}
\markboth{INTRODUCTION}{INTRODUCTION}
\def\theequation{\arabic{section}.\arabic{equation}}
\setcounter{section}{0}
\setcounter{equation}{0}
\vspace{7mm}

\noindent
In 1913 the outstanding French mathematician Elie Cartan discovered
spinors \cite{31,32}. He made this discovery while investigating 
irreducible representations of the group of rotations in the
$n$-dimensional Euclidean space. He was the first to find and to
describe in full detail spinor representations of the group of
rotations.\index{Rotation group}

The theory of spinors became an inseparable part of mathematical and
theoretical physics after Dirac's discovery of the equation of motion
for an electron (1928) which bears his name \cite{47,49}. The four
complex-valued functions of four arguments contained in the Dirac
equation are the components of a spinor with respect to the Lorentz
group.

It is interesting to note that the methods used by Cartan and Dirac to 
discover spinors are essentially different. These methods lie at the
basis of algebraic-theoretical, group-theoretical investigations in
modern quantum theory.

Today spinors and spinor representations play a basic role in
mathematical and theoretical physics, since all elementary particles,
clas\-si\-cal and quantum fields having half-integer spins $(s =
1/2,3/2,5/2,\ldots)$ are described with their help. Moreover, using de 
Broglie's heuristic idea of "fusion" we can construct particles
(fields) having integer spins $(s = 0,1,2,\ldots)$ from a particle
(field) having the spin $s =1/2$.  That is why the theory of spinors
and spinor analysis as the principal analytical apparatus for
investigation of spinor dynamical systems are useful in solving
problems from other fields of mathematics and quantum physics.

The first paper devoted to a nonlinear generalization of the Dirac
equation was published by Ivanenko in 1938 \cite{129}. Later
Finkelstein with collaborators in 1951 \cite{53,54} and Heisenberg in
1953 \cite{116,117} started analyzing various nonlinear
generalizations of the Dirac equation.

Heisenberg \cite{117}--\cite{121} put forward a vast program on the
construction of a unified field theory of elementary particles. As a
basis of this theory he chose a self-interacting spinor field
described by a nonlinear equation.  According to Heisenberg such a
field is determined by the following Dirac-type nonlinear equation:
\begin{equation}
i\gamma_{\mu}\partial_{\mu}\psi + \lambda\gamma_{\mu}\gamma_{4}
(\bar \psi\gamma^{\mu}
\gamma_{4} \psi)\psi = 0,
\label{0.1}
\end{equation}
where $\psi$ is a four-component Dirac spinor and $\lambda$ is a
parameter.  We will call system (\ref{0.1}) the Dirac-Heisenberg
equation\index{Dirac-Heisenberg equation}.

The present book deals with the following two principal problems: the
first one is to describe systems of nonlinear spinor partial
differential equations of the first and second orders invariant under
the Poincar\'e and the Galilei groups and under their natural
extensions; the second problem is the construction in explicit form of
exact solutions of the classical nonlinear spinor, vector and scalar
differential equations describing interaction of the Dirac, Maxwell
and Yukawa fields.

Unlike the majority of researchers we do not derive nonlinear
equations within the framework of the variational principle. We apply
the symmetry selection principle, namely, from the whole set of
partial differential equations (PDEs) of a given order we select those
on whose sets of solutions some fixed representation of the Poincar\'e
or the Galilei group is realized.\index{Symmetry!selection principle}
Such an approach to the derivation of motion
equations\index{Motion!equation} seems to be more general than the
traditional method based on the Lagrange function \cite{75,77.0,89}.

The major part of the book is devoted to the development of
efficient methods designed to obtain exact solutions of nonlinear
equations. All these methods are based on the idea of reducing
multi-dimensional partial differential equations to equations
having smaller dimensions.

While reducing PDEs a key role is played by substitutions of the
special form \cite{59,60,63,89,103}
\begin {equation}\label{0.2}
\psi(x) = A(x) \varphi(\omega_{1},\omega_{2},\ldots,\omega_{n}),
\end{equation}
where $\varphi(\omega)$ is an unknown function-column and $A(x)$ is a
variable matrix of corresponding dimensions; $\omega_{\alpha} =
\omega_{\alpha} (x)$, $\alpha =$ $1,\ldots,n$ are  real-valued scalar
functions.

Explicit forms of the functions $A(x)$, $\omega_\alpha(x)$ are
obtained by requiring that substitution of the expression (\ref{0.2})
into the PDE under study reduces it to an equation containing only
"new" dependent $(\varphi)$ and independent
$(\omega_{1},\omega_{2},\ldots,\omega_{n})$ variables.
\index{Reduction of PDE} Of course, the availability of an 
effective procedure of
computing the matrix $A(x)$ and the variables $\omega_{\alpha}(x)$
providing the reduction of the initial equation is implied.
Furthermore, the construction described above will be called the
Ansatz\index{Ansatz} for field $\psi(x)$.

Provided the equation under study possesses nontrivial local
symmetry, there exists an effective algorithm for constructing
Ans\"atze (suggested and applied for the first time to some of the
simplest PDEs by Sophus Lie). Ans\"atze obtained in this way will be
called Lie Ans\"atze.

In \cite{62,63} we suggested the generalization of the Lie method.
The idea of this generalization is based on the following observation:
the Lie method of constructing particular solutions, apart from its
group-theoretical foundations, can be considered as addition of
some first-order PDE to a given equation. Within the Lie approach this
additional equation is a linear combination of basis elements of the
invariance algebra of the equation under investigation. In view of
this fact it was suggested to consider the coefficients of that linear
combination as arbitrary functions of $x$, $\psi$, $\psi_{x_{\mu}}$,
$\psi_{x_{\mu}x_{\nu}}$.  In other words the additional constraint on
the set of solutions of the equation under investigation is, generally
speaking, a nonlinear first- or second-order PDE with variable
coefficients. Such a generalization proved to be constructive. In
many cases it provided the possibility of obtaining broad classes of
exact solutions of nonlinear equations which could not be found within
the framework of the classical Lie approach \cite{65.2,65.3},
\cite{68}--\cite{68.2}, \cite{77.1,79,69,81,81.1,81.2,89,95},
\cite{102}--\cite{106.2}, \cite{173.1,211,212}.

With the use of nonlocal and conditional symmetry of linear and
nonlinear spinor equations (the notion of the conditional symmetry of
differential equations was introduced in \cite{62,75,89}) we obtain
wide classes of non-Lie Ans\"atze, which reduce these equations to
systems of ordinary differential equations (ODEs).

Due to large symmetry of equations being considered 
systems of ODEs obtained by reduction via Lie and non-Lie Ans\"atze
are often integrable by quadratures. Their exact solutions, after
being substituted into the corresponding Ans\"atze, give rise to
particular solutions of the nonlinear spinor equations under study.

As shown in Section 2.6, exact solutions of nonlinear spinor
equations make it possible to construct exact solutions of other
Poincar\'e-invariant equations. In particular, we construct a number
of exact solutions of the nonlinear d'Alembert equation via solutions
of the nonlinear Dirac equation.

More detailed information concerning the contents of the book is
provided by chapter and section titles.

For the reader's convenience, we give a brief account of some facts,
terminology and notations from  group theory which are used in
the book (for more details, see 
\cite{5,24,25,30,36,52,52.2,127,148,161,164}).

An $r$-parameter Lie transformation 
group\index{Lie!transformation group} $G_r$ is a set of 
transformations of the space ${\R}^{n}\times {\C}^{m}$
\begin{equation}\label{0.3}
\begin{array}{rcl}
x'_{\alpha}&=&f_{\alpha}(x,u,\theta), \quad  \alpha =
0,\ldots,n-1,\\[2mm] 
u'_{\beta}&=& g_{\beta} (x,u,\theta), \quad   \beta =
0,\ldots,m-1, 
\end{array}
\end{equation}
$\theta \in U, \ U$ is an open sphere in ${\R}^r$,
where  $f_{\alpha}$ and $g_{\beta}$ are real-analytical functions of
$\theta$ satisfying the following relations:
\begin{enumerate}
\item $f_{\alpha}(x,u,0) = x_{\alpha}, \quad g_{\beta}(x,u,0) =
u_{\beta},$
\item $\forall \{\theta_{1}, \theta_{2}\} \subset U, \
\exists \theta_{3} = T(\theta_{1}, \theta_{2}) \in U: \\
f_{\alpha} \Bigl(f(x,u,\theta_{1}),\, g(x,u,\theta_{1}),\,
\theta_{2}\Bigr) = 
f_{\alpha} (x,u,\theta_{3}), \\
g_{\beta} \Bigl(f(x,u,\theta_{1}),\, g(x,u,\theta_{1}),\,
\theta_{2}\Bigr) = 
g_{\beta} (x,u,\theta_{3})$.
\end{enumerate}

Here $T :U \times U \rightarrow U$ is a vector-function whose
components are real-analytical functions satisfying the relations
\begin{enumerate}
\item $T(\theta,0) = T(0,\theta) = \theta, \quad \forall\theta \in U,$ 
\item $\forall\theta \in U, \ \exists \theta^{-1} \in U: \ \
T(\theta,\theta^{-1}) = T(\theta^{-1},\theta) = 0,$
\item $\forall \{\theta_{1}, \theta_{2}, \theta_{3}\} \subset U: \ \
T\Bigl(T(\theta_{1},\theta_{2}),\theta_{3}\Bigr) = T\Bigl(\theta_{1},
T(\theta_{2},\theta_{3})\Bigr)$.
\end{enumerate}

The $r$-parameter Lie transformation group (\ref{0.3}) is related to
the $r$-dimensi\-on\-al vector space $AG_{r}$ whose basis elements are
first-order differential operators
\begin{equation}  
\label{0.4}
Q_{\tau} =\sum\limits_{\alpha = 0}^{n-1} \xi_{\tau \alpha}(x,u)
\frac{\partial}{\partial x_{\alpha}} 
+ \sum\limits_{\beta = 0}^{m-1} \eta_{\tau \beta}(x,u)
\frac{\partial}{\partial u_{\beta}}, 
\end{equation}
the coefficients $\xi_{\tau \alpha},\ \eta_{\tau \beta}$ being
defined by the following formulae:
\begin{equation}
\label{0.5}
\matrix{\xi_{\tau \alpha}(x,u) &=& \left. {\dis\partial
      f_{\alpha}\over\dis\partial 
\theta_{\tau}} \right|_{\dis\theta = 0},\cr\cr
\eta_{\tau \beta}(x,u)  &=& \left. {\dis\partial
    g_\beta\over\dis\partial \theta_{\tau}} \right|_{\dis\theta =
  0}.\cr} 
\end{equation}

The vector space $AG_r$ is closed with respect to the operation
$$
(X,Y) \rightarrow Z = XY -YX \equiv [X,Y]
$$
and, consequently, forms an $r$-dimensional Lie algebra. This algebra
is called the Lie algebra of the group $G_r$.\index{Lie!algebra}

Conversely, given the Lie algebra with basis elements (\ref{0.4}),
where $\xi_{\tau \alpha}$ and $\eta_{\tau \beta}$ are sufficiently
smooth functions, then the $r$-parameter Lie transformation group is
obtained by solving the Lie equations\index{Lie!equations}
\begin{equation}
\label{0.6}
\begin{array}{rcl}
{\dis\partial f_{\alpha}\over\dis\partial 
\theta_{\tau}}&=&\xi_{\tau\alpha}(f,g), \quad  
f_{\alpha}(x,u,0)=x_{\alpha}, \\[2mm]
{\dis\partial g_{\beta}\over\dis\partial 
\theta_{\tau}}&=&\eta_{\tau\beta}(f,g), \quad  
g_{\beta}(x,u,0)=u_{\beta},\quad \tau=1,\ldots,r
\end{array}
\end{equation}
and by constructing the superposition of the resulting 
one-parameter Lie groups.

Thus, there exists a one-to-one correspondence between a Lie 
transformation group $G_r$ and its Lie algebra $AG_r$. To emphasize
this correspondence we say that operators $Q_{\tau}$ generate the
group $G_r$. These operators are called infinitesimal 
operators\index{Infinitesimal operator}
(generators)\index{Generator of transformation group} of the group $G_r$ 
(as a rule, we omit the word "infinitesimal").

We say that the differential equation
\begin{equation}  
\label{0.7}
L\Bigl(x,u(x)\Bigr)=0
\end{equation}
is invariant under the group of transformations $G_r$ (or: admits the
group $G_r$) if the change of variables (\ref{0.3}) transforms the set
of solutions of equation (\ref{0.7}) into itself. The group $G_r$ is
called invariance or symmetry group of equation
(\ref{0.7})\index{Invariance!group}\index{Symmetry!group}. A
corresponding Lie algebra is called invariance or symmetry algebra of
the equation in
question.\index{Invariance!algebra}\index{Symmetry!algebra}

According to Lie \cite{148} the differential equation (\ref{0.7}) is
invariant under the group $G_r$ having generators (\ref{0.4}) if and
only if
\begin{equation}    
\label{0.8}
\left.\matrix{{\widetilde Q}_{\tau} L\cr\cr}\right.
\left|\matrix{&=0,\cr[L]&\cr}\right.
\end{equation}
where $[L]$ means the set of solutions of the equation $L=0$ and
${\widetilde Q}_{\tau}$ is the $N$-th prolongation of the operator
$Q_\tau$ ($N$ is the order of differential equation (\ref{0.7})).

The $N$-th prolongation of the 
operator\index{Prolongation of the operator}
\[
Q=\sum\limits_{\alpha = 0}^{n-1} \xi_{\alpha}(x,u)
\frac{\partial}{\partial x_{\alpha}} + 
\sum\limits_{\beta = 0}^{m-1} \eta_{\beta}(x,u)
\frac{\partial}{\partial u_{\beta}} 
\]
is constructed as follows
\[
{\widetilde Q} = Q + \zeta_{\beta {\alpha}_1}
{\dis\partial\over\dis\partial \left( {\dis\partial u_{\beta}\over
\partial x_{{\alpha}_1}} \right)} + \cdots +
\zeta_{\beta {\alpha}_1 \ldots {\alpha}_N}
{\dis\partial\over\dis \partial \left( {\dis\partial^N u_{\beta}\over
\dis \partial x_{{\alpha}_1} \ldots \partial x_{{\alpha}_N}} \right)},
\]
where
\begin{eqnarray*}
& &\zeta_{{\beta \alpha}_1}=D_{{\alpha}_1} \eta_{\beta} -
\frac{\partial u_{\beta}}{\partial x_{\alpha}} D_{{\alpha}_1}
\xi_{\alpha},\\
& &\zeta_{{\beta \alpha}_1{\alpha}_2}=D_{{\alpha}_2} \zeta_{{\beta
\alpha}_1} - \frac{{\partial}^2 u_{\beta}}{\partial x_{{\alpha}_1}
\partial x_{\alpha}} D_{{\alpha}_2} \xi_{\alpha},\\
& &\zeta_{\beta {\alpha}_1 \ldots {\alpha}_N}=D_{{\alpha}_N}
\zeta_{{{\beta \alpha}_1} \ldots {\alpha}_{N-1}} - \frac{{\partial}^N 
u_{\beta}}{\partial x_{{\alpha}_1} \ldots \partial x_{{\alpha}_{N-1}}
\partial x_{\alpha}} D_{\alpha_N} \xi_{\alpha},\\
& &D_{\alpha}=\frac{\partial}{\partial x_{\alpha}} +
\frac{\partial u^\beta}{\partial x_\alpha}\frac{\partial}{\partial
  u^\beta} +
\sum_{n=1}^{\infty} \frac{{\partial}^{n+1} u_{\beta}}{\partial
  x_{{\alpha}_1} \ldots \partial x_{{\alpha}_n} \partial x_{\alpha}}
\ {\dis\partial\over\dis\partial
\left( {\dis\partial^n u_{\beta}\over\dis\partial x_{{\alpha}_1}
    \ldots \partial x_{{\alpha}_n}} \right)}
\end{eqnarray*}
(summation over repeated indices is implied).

The invariance criterion (\ref{0.8})\index{Invariance!criterion} 
gives rise to a linear system
of PDEs (the determining equations\index{Determining equations}) 
for the functions $\xi_{\alpha},\
\eta_{\beta}$,  whose general solution determines the maximal (in
Lie sense) invariance algebra of the equation considered. The
corresponding Lie group is called the maximal invariance (symmetry)
group of equation (\ref{0.7}).

The procedure described above is just Lie method\index{Lie!method} for
investigating symmetries of differential equations. Application
of this method to equations of mathematical physics requires the
performing of cumbersome computations (this is especially the case for
multi-component systems of PDEs). If we deal with a system of
linear PDEs
\begin{equation}   
\label{0.9}
L(x)u(x)=0, \quad u=(u_0, u_1,\ldots,u_{m-1})^T,
\end{equation}
the computations can be substantially simplified. A symmetry
operator\index{Symmetry!operator} acting in the linear space of
solutions of system (\ref{0.9}) is seeked in the form
\begin{equation}   
\label{0.10}
Q= \sum\limits_{\mu = 0}^{n-1}\xi_{\mu}(x) \frac{\partial}{\partial
  x_{\mu}} + \eta(x), 
\end{equation}
where $\xi_{\mu}(x)$ are smooth real-valued scalar functions,
$\eta(x)$ is some $(m\times m)$-matrix.  Within the
Lie approach operator (\ref{0.10}) is represented in the form
\begin{displaymath}
X = \sum\limits_{\alpha = 0}^{n-1}
\xi_{\alpha} (x) \frac{\partial}{\partial x_{\alpha}} -
\sum\limits_{\beta_1, \beta_2 = 0}^{m-1}\eta_{{\beta}_1 {\beta}_2}
(x)u_{{\beta}_2} \frac{\partial}{\partial u_{{\beta}_1}}.
\end{displaymath}

The invariance criterion for system of PDEs (\ref{0.9}) reads
(see, e.g., \cite{74})
\begin{equation}    
\label{0.11}
\left.\matrix{L Q u(x)\cr\cr}\right.
\left|\matrix{&=0.\cr[Lu]&\cr}\right.
\end{equation}
Condition (\ref{0.11}) means that the operator $Q$ transforms the set
of solutions of (\ref{0.9}) into itself.

Relation (\ref{0.11}) is rewritten in the following equivalent form:
\begin{equation}   
\label{0.12}
[L,Q] = R(x)L,
\end{equation}
where $R(x)$ is some $(m\times m)$-matrix. The above operator equality
is to be understood in the following way: operators on the left- and
right-hand sides of (\ref{0.12}) give the same result when acting on an
arbitrary solution of system (\ref{0.9}).

Let us emphasize that the invariance algebra obtained by solving
relation (\ref{0.11}) or (\ref{0.12}) is not the maximal one because
any system of linear PDEs admits the Lie transformation group
\begin{eqnarray*}
x'_{\mu} &=& x_{\mu}, \quad \mu = 0,\ldots, n-1,\\
u'_\beta &=& u_\beta + \theta u_{0\beta} (x),\quad 
\beta={0,\ldots,m-1},
\end{eqnarray*}
where $\theta$ is a real parameter, $u_{0} (x)$ is an arbitrary
solution of the system considered. But the above Lie group gives no
essential information about the structure of solutions of the equation 
under study and is not considered in the present book.

For many symmetry groups of systems of PDEs of mathematical and
theoretical physics, the matrix $\eta (x)$ possesses very important
algebraic properties which simplify substantially all manipulations
with symmetry operators (\ref{0.10}). Moreover, in most of the
problems considered in this book we use the algebraic relations which
are satisfied by $\eta (x)$, but we do not use their explicit form.
That is why  we will represent the infinitesimal symmetry operators in
the form (\ref{0.10}) (if it is possible and does not lead to
confusion).

In the approach based on the formulae (\ref{0.10}), (\ref{0.11}) the
restrictions of Lie method are quite evident since an operator
transforming the set of solutions of equation (\ref{0.9}) into itself
does not have to be of the form (\ref{0.10}) (a symmetry operator may
belong to the class of differential operators of the order $N \ge 1$
or to the class of integro-differential operators \cite{74,75,77}).

Below we give a list of notations and conventions used throughout
the book.

A scalar product in the Minkowski space\index{Minkowski space} $R(1,3)$ 
with the metric tensor
$$
g_{\mu \nu} = \left\{ \matrix{
\hfill 0, & \mu \not= \nu, \hfill  \cr
\hfill 1, & \mu = \nu  = 0,  \hfill \cr
-1, & \mu = \nu  = {1,2,3}  \hfill } \right.
$$
is denoted by $a \cdot b = g_{\mu \nu} a_{\mu} b_{\nu}, \
\{a, b\} \subset R(1,3)$.

A scalar product in the Euclidean space $R(3)$ with the metric tensor
$\delta_{ab} = -g_{ab}$ is written as follows
$$
\vec n\cdot \vec m = \delta_{ab} \ n_{a} m_{b} =  n_{a}m_{a}.
$$

Summation over  repeated indices is used, indices
being denoted by the Greek  letters $\alpha,\beta,\mu, \nu $
with the values 0, 1, 2, 3 and indices being denoted by the Latin
letters $a, b, c$ with the values 1, 2, 3 (unless otherwise
indicated).

By the symbol $\varepsilon_{abc}$ the antisymmetric tensor of rank
three 
$$
\varepsilon_{abc} = \left\{ \matrix{
\hfill 1,  &   (a, b, c)  =  {\rm cycle}\, (1, 2, 3),  \cr
-1,  &  (a, b, c)  = {\rm cycle}\, (2, 1, 3),  \cr
\hfill 0,  &  \mbox{in other cases} } \right.
$$
is designated.

All the functions considered in the book are supposed to be
differentiable as many times as is necessary. The derivative of a
function of one variable $f = f(z)$ is denoted by a dot over the
symbol $\dot f \equiv df / dz$. To distinguish a partial derivative we
use the symbol $\partial_{z}$, i.e. $\partial f / \partial z
\equiv\partial_{z}f$, and the partial derivative with respect to the
$\mu$-th independent variable is denoted by $\partial_{\mu} f
\equiv\partial_{x_\mu} f$.

Vector and tensor indices are written as subscripts ($x_{\mu},
A_{\mu}, F_{\mu \nu}$, etc.) and spinor indices as superscripts
($\psi^{\alpha}$). Lowering or raising of an index in the Minkowski
space $ R(1,3)$ is carried out by the metric tensor $g_{\mu \nu}$, for
example,
$$
x^{\mu} = g_{\mu \nu} x_{\nu}=\cases{x_0, \ \mu=0,\cr
                                     -x_a, \ \mu=a=1,2,3.}
$$

Complex conjugation is denoted by the asterisk $(x+iy)^*=x-iy$ and the
matrix transposed with respect to a given matrix $A$ is designated by
$A^T$. The symbol $A^{\dagger}$ stands for a complex conjugate of a
transposed matrix, i.e. $(A^T)^*=A^\dagger$. 
\newpage

\pagestyle{myheadings}                
\thispagestyle{empty}
\noindent
{\sl
C H A P T E R \ \  1\label{ch1}}
\vspace{2mm}

\hrule
\vspace{35mm}

\rightline
{\large\bf
SYMMETRY}
\vspace{2mm}

\rightline
{\large\bf
OF NONLINEAR}
\vspace{2mm}

\rightline
{\large\bf
SPINOR EQUATIONS}
\vspace{7mm}

\noindent
The first chapter is of an introductory character. Here we present
well-known facts about different representations of the Dirac equation
\cite{74,75,77}, its local and nonlocal (non-Lorentz) symmetry and
conservation laws for the Dirac field. Detailed group-theoretical
analysis of nonlinear generalizations of the Dirac equation which are
invariant under the Poincar\'e group $P(1,3)$, extended Poincar\'e
group $\wid P(1,3)$ and conformal group $C(1,3)$ is carried out. Some
second-order Poincar\'e- and conformally-invariant spinor equations
are considered.  Wide classes of nonlinear PDEs for spinor, scalar and
vector fields invariant under the groups $P(1,3)$,\ $\wid P(1,3)$,\ 
$C(1,3)$ are described.

We establish correspondence between reducibility of PDEs and their
conditional symmetry (the results obtained play a crucial role when
constructing exact solutions of multi-dimensional partial differential
equations). 
\vspace{10mm}

\noindent
{\large\bf 1.1. Local and nonlocal symmetry of the Dirac
  equation\label{s1.1}} 

\markboth{Chapter 1. SYMMETRY OF NONLINEAR SPINOR EQUATIONS}
{1.1. Local and nonlocal symmetry of the Dirac equation}
\def\theequation{1.\arabic{section}.\arabic{equation}}
\setcounter {section} {1}
\setcounter {equation}{0}
\vspace{7mm}

\noindent
The Dirac equation\index{Dirac!equation} is the system of four 
linear complex partial dif\-fe\-ren\-tial equations
\begin{equation}
\label{1.1.1}
(i \gamma_{\mu} \partial_{\mu} - m) \psi(x) = 0,
\end{equation}
where $ \psi = \psi (x_{0}, x_{1}, x_{2}, x_{3}) $ is the
four-component, complex-valued func\-ti\-on-column, $ m = \mbox{\rm
  const}$, $ \gamma_{\mu}$ are $(4\times 4)$-matrices satisfying the
Clif\-ford-Dirac algebra
\begin{equation}
\gamma_{\mu} \gamma_{\nu} + \gamma_{\nu} \gamma_{\mu} = 2g_{\mu \nu}I,
\quad \mu, \nu = {0,\ldots,3},
\label{1.1.2}
\end{equation}
where $I$ is the unit $(4\times 4)$-matrix.

Under the massless Dirac equation we mean system (\ref{1.1.1})
with $m=0$.\index{Massless Dirac equation}

Since on the set of solutions of the Dirac equation a spinor 
representation of the Lorentz group is realized (see the Appendix 1), 
the function $\psi(x)$ is called the spinor field\index{Spinor!field} 
(or, for brevity, the spinor)\index{Spinor} and equation (\ref{1.1.1}) 
as well as its nonlinear generalizations are called spinor 
equations.\index{Spinor!equation}

If we act with the operator $i \gamma_{\mu} \partial_{\mu} + m $ on
the left-hand side of equality (\ref{1.1.1}) and use relations
(\ref{1.1.2}), then a system of four splitting wave
equations\index{Wave equation} for the spinor $\psi (x)$
\begin{equation}
\label{1.1.3}
(\partial_{\mu} \partial^{\mu} + m^{2}) \psi (x) = 0
\end{equation}
is obtained.

It is worth noting that Dirac derived equation (\ref{1.1.1}) by
factorizing the second-order differential operator $\partial_{\mu}
\partial^{\mu} + m^{2}$, i.e.,\ by representing it in the form of the
product of two first-order operators $Q_{\pm} = i \gamma_{\mu}
\partial_{\mu} \pm m$, whence it followed that $ \gamma_{\mu} $ were
matrices satisfying the algebra (\ref{1.1.2}) \cite{26,47,49}.  
\vspace{2mm}

\noindent
{\bf 1. Algebra of the Dirac matrices\index{Dirac!matrices}.}\ We say
that a 
representation\index{Representation!of the Clifford-Dirac algebra} of 
the Clifford-Dirac algebra\index{Clifford-Dirac algebra}
is given if there are four $(4\times 4)$-matrices satisfying relations
(\ref{1.1.2}). There exist infinitely many representations of the
Clifford-Dirac algebra. But all these representations are equivalent,
namely, for each two sets of matrices $\Bigl\{ \gamma'_{\mu} \Bigr\},
\ \Bigl\{ \gamma_{\mu} \Bigr\} $ satisfying (\ref{1.1.2}) there exists
such a nonsingular $(4\times 4)$-matrix $ V $ that
\begin{equation}
\label{1.1.4}
\gamma'_{\mu} = V \gamma_{\mu} V^{-1},
\quad \mu = {0,\ldots,3}.
\end{equation}

If it is not indicated otherwise, we assume that the matrices
$\gamma_{\mu}$ realize the following representation of the algebra
(\ref{1.1.2}):
\begin{equation}
\label{1.1.5}
\gamma_{0} = \left( \begin{array}{cc}
I & 0 \\
0 & -I
\end{array}
\right),
\qquad
\gamma_{a} = \left( \begin{array}{cc}
0 & \sigma_{a} \\
- \sigma_{a} & 0
\end{array} \right),
\end{equation}
where $I$, $0$ are the unit and zero $(2\times 2)$-matrices,
$\sigma_{a}$ are the Pauli matrices\index{Pauli matrices}
\begin{equation}
\label{1.1.6}
\sigma_{1} = \left( \begin{array} {cc}
0 & 1 \\
1 & 0
\end{array} \right),
\ \ \ \sigma_{2} = \left( \begin{array} {cc}
0 & -i \\
i & 0
\end{array} \right),
\ \ \ \sigma_{3} = \left( \begin{array} {cc}
1 & 0 \\
0 & -1
\end{array} \right).
\end{equation}

In addition, we use the following representations of the
Clifford-Dirac algebra:

\begin{equation}
\label{1.1.8}
\begin{array}{ll}
\gamma_0 = \left( \begin{array} {cc}
0 & iI \\
-iI & 0 \end{array} \right),
&
\gamma_{1} = \left( \begin{array} {cc}
- i \sigma_{3} & 0 \\
0 & i \sigma_{3} \end{array} \right), \\[8mm]
\gamma_{2} = \left( \begin{array} {cc}
iI & 0 \\
0 & -iI \end{array} \right),
&
\gamma_{3} = \left( \begin{array} {cc}
0 & - iI \\
- iI & 0
\end{array} \right);
\end{array}
\end{equation}

\vspace{\baselineskip}
\begin{equation}
\label{1.1.7}
\gamma_{0} = \left( \begin{array} {cc}
0 & I \\
I & 0 \end{array} \right),
\quad
\gamma_{a} = \left( \begin{array} {cc}
0 & \sigma_{a} \\
- \sigma_{a} & 0
\end{array} \right), \ \ a={1,2,3}.
\end{equation}

Straightforward verification shows that the matrix $\gamma_{4} =
\gamma_{0} \gamma_{1} \gamma_{2} \gamma_{3} $ satisfies relations of
the form

$$
\gamma_{4} \gamma_{\mu} + \gamma_{\mu} \gamma_{4} = 0,
\quad
\gamma_{4}^2 = -1, \ \
\mu = {0,\ldots,3}.
$$

Matrices $ \gamma_{0}, \ \gamma_{1}, \ \gamma_{2}, \ \gamma_{3}, \ 
\gamma_{4} $ form the maximal set of generators of the Clifford-Dirac
algebra in the class of $(4\times 4)$-matrices.

The maximal set of generators of the Clifford-Dirac algebra in the
class of $(8\times 8)$-matrices is exhausted up to the equivalence
relation (\ref{1.1.4}) by the following matrices:

\begin{equation}
\label{1.1.9}
\begin{array}{ll}
\tilde \Gamma_{\mu} = \left( \begin{array} {cc}
\gamma_{\mu} & 0 \\
0 & \gamma_{\mu} \end{array} \right),
&
\tilde \Gamma_{4} = \left( \begin{array} {cc}
0 & \gamma_{4} \\
\gamma_{4} & 0 \end{array} \right), \\[8mm]
\tilde \Gamma_{5} = \left( \begin{array} {cc}
0 & i\gamma_{4} \\
- i\gamma_{4} & 0 \end{array} \right),
&
\tilde \Gamma_{6} = \left( \begin{array} {cc}
\gamma_{4} & 0  \\
0 & -\gamma_{4}
\end{array} \right),\ \
\mu = {0,\ldots,3},
\end{array}
\end{equation}
where $0$ is the zero $(4\times 4)$-matrix.

It is known that all possible products of matrices $\gamma_{\mu}$ form
a basis in the linear space of $(4\times 4)$-matrices. The elements of
this basis can be chosen as follows
\begin{equation}
\label{1.1.10}
I, \quad \gamma_{\mu},\quad \gamma_{\mu} \gamma_{\nu},\quad
\gamma_{4} \gamma_{\mu},\quad \gamma_{4},\quad \mu < \nu,
\ \ \mu, \nu = {0,\ldots,3}.
\end{equation}

Sixteen matrices (\ref{1.1.10}) are linearly independent and,
consequently, an arbitrary $(4\times 4)$-matrix is represented as a
linear combination of the basis elements (\ref{1.1.10}).  
\vspace{2mm}

\noindent
{\bf 2. Various formulations of the Dirac equation.} \ The
four-component function-row $\bar{\psi}(x)=\bigl
(\psi(x)\bigr)^{\dagger}\ga_0$ is called a Dirac-conjugate
spinor\index{Conjugate!spinor}. To obtain an equation for
$\bar{\psi}(x)$ we apply a complex conjugation procedure to
(\ref{1.1.1}) with subsequent transposition and multiply the obtained
expression by $\ga_0$ on the right.  Taking into account relations
$\ga_0^{\dagger}=\ga_0$,\ $\ga_a^{\dagger}=-\ga_a$, we have
\begin{equation}
\label{1.1.11}
i\pa_{\mu}\bar{\psi}\ga_{\mu}+m\bar{\psi}=0.
\end{equation}

If we designate
\begin{equation}
\label{1.1.12}
\tilde{\psi}=i\ga_2\psi^*,
\end{equation}
then equation (\ref{1.1.11}) can be rewritten in the form
\begin{displaymath}
(i\ga_{\mu}\pa_{\mu}-m)\tilde{\psi}=0.
\end{displaymath}

Hence it follows that system (\ref{1.1.1}), (\ref{1.1.11}) can be
represented in the form of the eight-component equation
\begin{equation}
\label{1.1.13}
(i\wid{\Gamma}_{\mu}\pa_{\mu}-m)\Psi(x)=0,
\end{equation}
where
$$\Psi(x)=
\pmatrix{
         \psi(x)\cr
         \tilde{\psi}(x)\cr}.
$$

If we choose the matrices \ $\ga_{\mu}$\ in the representation
(\ref{1.1.8}), we can rewrite the Dirac equation (\ref{1.1.1}) as a
system of eight real PDEs 
\begin{equation}
(i \widetilde \Gamma_{\mu} - m)
\widetilde \Psi (x) = 0,
\label{1.1.14}
\end{equation}
where 
$$
\widetilde \Psi (x) = \pmatrix{
{\rm Re}\, \psi (x) \cr
{\rm Im}\, \psi (x)}.
$$

On multiplying equation (\ref{1.1.1}) by the matrix $\ga_{0}$ on the
left we get the Dirac equation\index{Dirac!equation} in the Hamilton
form 
\begin{displaymath} 
i \partial_{0} \psi = H \psi = (-i \ga_{0} \ga_{a}
\partial_{a} + m \ga_{0} ) \psi.  
\end{displaymath}

Choosing the matrices $ \ga_{\mu} $ in the representation
(\ref{1.1.7}) and representing the spinor $ \psi (x) $ in the form
\begin{equation}
\label{1.1.15}
\psi (x) = 
\pmatrix{\varphi_{-} (x) \cr
\varphi_{+} (x)},
\end{equation}
where $\varphi_{\pm} (x) $ are two-component functions, we rewrite the
Dirac equation as follows
\begin{equation}
\begin{array}{l}\label{1.1.16}
(i\partial_{0} + i \sigma_{a} \partial_{a}) \varphi_{+} 
- m \varphi_{-} = 0,\\[2mm]
(i\partial_{0} - i \sigma_{a} \partial_{a}) \varphi_{-} 
- m \varphi_{+} = 0.
\end{array}
\end{equation}

Acting on the first equation of system (\ref{1.1.16}) by the operator
$i \partial_{0} - i \sigma_{a} \partial_{a} $ we have
\begin{displaymath}
(\partial_{\mu} \partial^{\mu} + m^{2}) \varphi_{+} (x) = 0
\end{displaymath}
and what is more $\varphi_{-}(x) = m^{-1} (i \partial_{0} 
+ i \sigma_{a} \partial_{a}) \varphi_{+} (x)$.
Consequently, the system of four first-order differential equations
(\ref{1.1.1}) is equivalent to the system of two splitting wave
equations. 

From (\ref{1.1.16}) it is clear that the massless Dirac 
equation\index{Dirac!equation} 
\begin{equation}
i \ga_{\mu} \partial_{\mu} \psi (x) = 0
\label{1.1.17}
\end{equation}
splits into two Weyl equations\index{Weyl equation} for two-component 
spinors $ \varphi_{\pm} (x) $.

Let us also note that the massless Dirac equation (\ref{1.1.17}) can
be represented in the form of the Maxwell equations with currents.  To 
become convinced of this fact we represent the four-component
function $\psi(x)$ in the following equivalent form:
\begin{equation}
\label{1.1.18}
\psi=\pmatrix{-E_1\cr E_3\cr -H_2\cr F\cr} \ + \
i\pmatrix{E_2\cr G\cr -H_1\cr H_3\cr},
\end{equation}
where $E_a, \ H_a, \ F, \ G$ are some smooth real-valued functions.

Substituting (\ref{1.1.18}) into (\ref{1.1.17}) and splitting with
respect to $i$ we get the Maxwell equations\index{Maxwell equations} 
with currents \cite{89.1}
\begin{equation}
\begin{array}{ll}
\partial_{0} \vec E ={\rm rot}\, \vec H + \vec j,
&{\rm div}\, \vec E = j_{0},\\[2mm]
\partial_{0} \vec H = -{\rm rot}\, \vec E + \vec k,
&{\rm div}\, \vec H = k_{0},
\end{array}
\label{1.1.19}
\end{equation}
where $j_{\mu} = \partial_{\mu} F, \ k_{\mu} = \partial_{\mu} G$.

The above presented formulations of the Dirac equation are, of course,
equivalent but choosing an appropriate one we can substantially
simplify computations when solving the specific problem. In addition,
these formulations enable us to obtain principally different
generalizations of equation (\ref{1.1.1}) for the fields with an
arbitrary spin \cite{74,75}.  
\vspace{2mm}

\noindent
{\bf 3. Lie symmetry of the Dirac equation.} \ We adduce the
assertions describing the maximal (in Lie sense) invariance groups
admitted by the Dirac 
equation.\index{Maximal symmetry!of the Dirac equation}  
\vspace{1.5mm}

\noindent
{\bf Theorem 1.1.1.}\ \ {\em The maximal local invariance group of the
  Dirac equation (\ref{1.1.1}) is the 14-parameter group $ G_{1} =
  P(1,3) \otimes V (4) $,\label{page14}\footnote{Since equation
    (\ref{1.1.1}) is linear, it admits an infinite-parameter group
    $\psi'=\psi+\theta\Psi(x)$, where $\theta$ is a group parameter
    and $\Psi$ is an arbitrary solution of system of PDEs
    (\ref{1.1.1}). Such a symmetry gives no essential information
    about the structure of solutions of the equation under study and
    therefore is neglected.} where $P(1,3)$ is the Poincar\'e
  group\index{Poincar\'e!group} having the generators
\begin{equation}
P_{\mu} = \partial^{\mu},\quad
J_{\mu \nu} = x_{\mu} \partial^{\nu} - x_{\nu} \partial^{\mu} + S_{\mu
  \nu}
\label{1.1.20}
\end{equation}
and $V (4) $ is the 4-parameter group of transformations in the space
$(\psi^{*}, \psi)$ generated by the operators}
\begin{equation}
\begin{array}{l}
Q_{0} = \psi^{\alpha} \partial_{\psi^{\alpha}} +
\psi^{* \alpha} \partial_{\psi^{* \alpha}},\\[2mm]
Q_{1} = i \psi^{\alpha} \partial_{\psi^{\alpha}} -
i \psi^{* \alpha} \partial_{\psi^{* \alpha}},\\[2mm]
Q_{2} = \{ \ga_{2} \psi^{*} \}^{\alpha} \partial_{\psi^{\alpha}} -
\{ \ga_{2} \psi \}^{\alpha} \partial_{\psi^{* \alpha}},\\[2mm]
Q_{3} = \{ i \ga_{2} \psi^{*} \}^{\alpha} \partial_{\psi^{\alpha}} +
\{ i \ga_{2} \psi \}^{\alpha} \partial_{\psi^{* \alpha}}.
\end{array}
\label{1.1.21}
\end{equation}

In formulae (\ref{1.1.20}), (\ref{1.1.21})  $\{ \Psi \}^{\alpha} $ is 
the $\alpha$-th component of the function $\Psi $ and
\begin{eqnarray*}
& &S_{\mu \nu} = \frac{1}{4} [\ga_{\mu}, \ga_{\nu}] = \frac{1}{4}
(\ga_{\mu} \ga_{\nu} - \ga_{\nu} \ga_{\mu} ),\\
& & \partial_{\psi^{\alpha}} = \partial/ \partial  \psi^{\alpha}, \quad 
\partial_{\psi^{* \alpha}} = \partial/ \partial\psi^{* \alpha}.
\end{eqnarray*}
\vspace{1.5mm}

\noindent
{\bf Theorem 1.1.2.}\ {\em The maximal local invariance group of the
  massless Dirac equation (\ref{1.1.17}) is the 23-parameter group
  $G_{2} = C(1,3) \otimes V(8) $,\footnote{See the previous footnote.}
  where $C(1,3)$ is the conformal group\index{Conformal!group} having
  the 
  generators\index{Maximal symmetry!of the massless Dirac equation}
\begin{equation}
\begin{array}{l}
        P_{\mu}=\partial^{\mu}, \quad J_{\mu \nu} = x_{\mu}
        \partial^{\nu} - 
        x_{\nu} \partial^{\mu} + S_{\mu \nu},\\[2mm]
        D = x_{\mu} \partial_{\mu} + 3/2,\\[2mm]
        K_{\mu} = 2x_{\mu} (x_{\nu} \partial_{\nu} + 3/2) - x \cdot
        x \partial^{\mu} + 2S_{\mu \nu} x^{\nu}
\end{array}
\label{1.1.22}
\end{equation}
and $V(8) $ is the 8-parameter group of transformations in the space
$ (\psi^{*}, \psi) $ generated by the operators (\ref{1.1.21}) and}
\begin{equation}
\begin{array}{l}
       Q_{4} = \{ \ga_{4} \psi \}^{\alpha} \partial_{\psi^{\alpha}} -
       \{ \ga_{4} \psi^{*}\}^{\alpha} \partial_{\psi^{*
           \alpha}},\\[2mm] 
       Q_{5} = \{i \ga_{4} \psi \}^{\alpha} \partial_{\psi^{\alpha}} +
       \{i \ga_{4} \psi^{*}\}^{\alpha} \partial_{\psi^{*
           \alpha}},\\[2mm] 
       Q_{6} = \{ \ga_{2} \ga_{4} \psi^{*} \}^{\alpha} 
       \partial_{\psi^{\alpha}} +
       \{\ga_{2} \ga_{4} \psi\}^{\alpha} \partial_{\psi^{*
           \alpha}},\\[2mm] 
       Q_{7} = \{i\ga_{2} \ga_{4} \psi^{*} \}^{\alpha} 
       \partial_{\psi^{\alpha}} -
       \{i\ga_{2} \ga_{4} \psi \}^{\alpha} \partial_{\psi^{* \alpha}}. 
\end{array}
\label{1.1.23}
\end{equation}

The fact that the groups $G_{1}$, \  $G_{2}$ are the maximal
invariance groups admitted by equation (\ref{1.1.1}) is established by
rather cumbersome computations with the help of the Lie method
\cite{43,125}.

Straightforward verification shows that operators $P_{\mu}$,\ 
$J_{\mu \nu}$,\ $ D $ satisfy the following commutation relations:
\begin{eqnarray*}
& &[P_{\mu}, P_{\nu} ] = 0, \quad [P_{\mu}, J_{\alpha \beta } ] =
g_{\mu \alpha} P_{\beta} - g_{\mu \beta} P_{\alpha},\\
& &[P_{\mu}, D] = P_{\mu}, \quad [J_{\mu \nu}, D] = 0,\\
& &[J_{\mu \nu}, J_{\alpha \beta} ] = g_{\mu \beta} J_{\nu \alpha} +
g_{\nu \alpha} J_{\mu \beta} - g_{\mu \alpha} J_{\nu \beta} - g_{\nu
  \beta} J_{\mu \alpha}.
\end{eqnarray*}

Consequently, the operators $ P_{\mu}, \ J_{\mu \nu}, \ D $ form a
basis of the 11-dimensional Lie algebra which is called the 
extended Poincar\'e algebra\index{Extended!Poincar\'e algebra} 
$A\widetilde P(1,3)$. The corresponding Lie group is called the
extended Poincar\'e group\index{Extended!Poincar\'e group} 
$\widetilde P(1,3)$.

Let us adduce explicit forms of transformation groups generated by
operators (\ref{1.1.20})--(\ref{1.1.23}) (corresponding formulae are
obtained by solving the Lie equations (\ref{0.6})).  \vspace{1.5mm}

\noindent
1)\ the group of translations\index{Translation group} 
$(X=\theta^{\mu} P_{\mu})$

\begin{equation}
x^{\prime}_{\mu}=x_{\mu}+\theta_{\mu}, \quad \psi'(x')=\psi(x);
\label{1.1.24a}
\end{equation}
2)\ the Lorentz group $O(1,3)$\index{Lorentz!group}
 
\noindent
a)\ the rotation group\index{Rotation group} $O(3)\, 
(X=\frac{1}{2} \varepsilon_{abc} \theta_{a} J_{bc})$
\begin{eqnarray}
& &x_{0}'=x_{0},\nonumber\\
& &x_{a}'=x_{a} \cos \theta - \theta^{-1}
\sin \theta\, \varepsilon_{abc} \theta_{b}x_{c} +
\theta^{-2} \theta_{a} (1 - \cos \theta)
\theta_{b}x_{b},\label{1.1.24b}\\ 
& &\psi'(x')=\exp \biggl\{-\frac{1}{2} \varepsilon_{abc} 
\theta_{a} S_{bc}\biggr\} \psi (x);\nonumber
\end{eqnarray}
b)\ the Lorentz transformations\index{Lorentz!transformations} 
$(X=J_{0a})$
\begin{eqnarray}
& &x_{0}'=x_{0}\cosh \theta_{0}+x_{a}\sinh\theta_{0},\nonumber\\
& &x_{a}'=x_{a}\cosh\theta_{0}+x_{0} \sinh\theta_{0},\nonumber\\
& &x_{b}' = x_{b}, \ \ b \ne a,\label{1.1.24c}\\
& &\psi^\prime (x')=\exp \biggl\{\frac{\theta_{0}}{2} 
\ga_{0} \ga_{a}\biggr\}\psi(x);\nonumber
\end{eqnarray}
3)\ the group of scale transformations 
$(X=D)$\index{Scale transformation group}
\begin{equation}
x_{\mu}'=e^{\theta_0}x_{\mu}, \quad \psi'(x')=e^{-k\theta_{0}}\psi(x),
\ \ {\rm with} \ k=3/2;
\label{1.1.24d}
\end{equation}
4)\ the group of special conformal 
transformations\index{Conformal!transformations}
$(X=\theta_{\mu}K^{\mu})$ 
\begin{equation}
\begin{array}{rcl}
& &x^{\prime}_{\mu}=(x_{\mu}-\theta_{\mu} x\cdot x)
\sigma^{-1}(x),\\[2mm] 
& &\psi'(x')=\sigma(x)(1-\gamma \cdot \theta \gamma \cdot x)\psi (x);
\end{array}
\label{1.1.24e}
\end{equation}
5)\ the group $V(8)$
\begin{eqnarray}
X= Q_{0}:& &x^{\prime}_{\mu}=x_{\mu},\non\\
& &\psi'(x')=e^{\theta_{0}}\psi(x), \quad
\psi^{*\prime}(x')=e^{\theta_{0}}\psi^{*}(x);\non\\[2mm]
X=Q_{1}:& & x^{\prime}_{\mu}=x_{\mu},\non\\
& &\psi'(x')=e^{i \theta_{0}}\psi(x), \quad 
\psi^{*\prime}=e^{-i \theta_{0}}\psi^{*}(x);\non\\[2mm]
X=Q_{2}:& & x^{\prime}_{\mu}=x_{\mu},\label{1.1.25}\\
& &\psi'(x')=\psi(x)\cosh
\theta_{0}+\ga_{2}\psi^{*}(x)\sinh\theta_{0},\non\\ 
&
&\psi^{*\prime}(x')=\psi^{*}(x)\cosh\theta_{0} - \ga_{2} \psi(x)
\sinh\theta_{0}; \non\\[2mm]
X=Q_{3}:& & x^{\prime}_{\mu}=x_{\mu},\non\\
& &\psi'(x')=\psi(x)\cosh\theta_{0}+i\ga_{2}\psi^{*}(x) \sinh
\theta_{0},\non\\ 
& &\psi^{*\prime}(x')=\psi^{*}(x) \cosh \theta_{0}+ 
i \ga_{2} \psi (x)\sinh\theta_{0};\non\\[4mm]
X=Q_{4}:& & x^{\prime}_{\mu}=x_{\mu},\non\\
& &\psi'(x')= \exp\{\theta_{0} \ga_{4}\}\psi (x),\quad
\psi^{*\prime}(x')= \exp \{-\theta_{0} \ga_{4}\} \psi^{*}
(x);\non\\[2mm] 
X=Q_{5}:& & x^{\prime}_{\mu}=x_{\mu},\non\\
& &\psi'(x')= \exp\{i \theta_{0} \ga_{4}\}\psi (x),\quad
\psi^{*\prime}(x')= \exp \{i \theta_{0} \ga_{4}\} \psi^{*}
(x);\non\\[2mm] 
X=Q_{6}:& & x^{\prime}_{\mu}=x_{\mu},\label{1.1.26}\\
& &\psi'(x')= \psi (x) \cos \theta_{0} + \ga_{2} \ga_{4} \psi^{*}(x)
\sin \theta_{0},\non\\
& &\psi^{*\prime}(x')=\psi^{*} (x) \cos \theta_{0} + \ga_{2} \ga_{4}
\psi (x) 
\sin \theta_{0};\non\\[2mm]
X=Q_{7}:& & x^{\prime}_{\mu}=x_{\mu},\non\\
& &\psi'(x')= \psi (x) \cos \theta_{0} + i \ga_{2} \ga_{4} 
\psi^{*}(x) \sin \theta_{0},\non\\
& & \psi^{*\prime}(x')=\psi^{*} (x) \cos \theta_{0} - i \ga_{2} 
\ga_{4} \psi (x)\sin \theta_{0}.\non
\end{eqnarray}

In the above formulae $\theta_{\mu} \in {\R}^{1}, \ \mu ={0,\ldots,3}
$ are group parameters, $\theta = (\theta_{a} \theta_{a})^{1/2}$,\ 
$\sigma (x) =1-2 \theta \cdot x + (\theta \cdot \theta)(x \cdot x)$,
by the symbol $X$ we designate a generator of the corresponding group.

The direct verification shows that the Dirac equation is invariant
under the Lie transformation groups (\ref{1.1.24a})--(\ref{1.1.26}).
For example, if we make in PDE (\ref{1.1.1}) the change of variables
(\ref{1.1.24b}), then the identity holds

\begin{displaymath}
(i \ga_{\mu} \partial^{\prime}_{\mu} - m) \psi^{\prime}(x')= 
\exp \biggl\{\frac{1}{2}
\varepsilon_{abc}  \theta_{a} S_{bc}\biggr\} 
(i \ga_{\mu} \partial_{\mu} - m) \psi (x),
\end{displaymath}
whence it follows that the set of solutions of equation
(\ref{1.1.1}) is invariant
with respect to the action of the group (\ref{1.1.24b}).

In addition, the Dirac equation admits discrete transformation
groups\index{Discrete group} which cannot be obtained with the help
of the Lie method. We adduce the most important discrete symmetries
of equation (\ref{1.1.1}).

\noindent
1) the spatial inversion\index{Spatial inversion}
\begin{equation}
\begin{array}{l}
x^{\prime}_{0}=x_{0}, \quad x^{\prime}_{a}=-x_{a},\\[2mm]
\psi^{\prime}(x')=\ga_{0}\psi (x), \quad
\psi^{*\prime}(x')=\ga_{0}\psi^{*} (x); 
\end{array}
\label{1.1.27}
\end{equation}
2) the time reversal\index{Time reversal}
\begin{equation}
\begin{array}{l}
x^{\prime}_{0}=-x_{0}, \quad x^{\prime}_{a}=x_{a},\\[2mm]
\psi^{\prime}(x')=\ga_{1}\ga_{3}\psi^{*} (x),\quad
\psi^{*\prime}(x')=\ga_{1}\ga_{3}\psi (x);
\end{array}
\label{1.1.28}
\end{equation}
3) the charge conjugation\index{Charge conjugation}
$$
\begin{array}{l}
x^{\prime}_{\mu}=x_{\mu},\\[2mm]
\psi^{\prime}(x')=i\ga_{2} \psi^{*}(x),\quad
\psi^{*\prime}(x')=i \ga_{2} \psi (x).
\end{array}
$$

Transformation groups (\ref{1.1.27}), (\ref{1.1.28}),
(\ref{1.1.24b}), (\ref{1.1.24c}) form the full Lorentz 
group\index{Lorentz!group} (for more detail, see \cite{77,111}).
\vspace{2mm}

\noindent
{\bf 4. Non-Lie symmetry of the Dirac equation.}\ In the previous
subsection we adduced theorems describing maximal local invariance
groups of equations (\ref{1.1.1}), (\ref{1.1.17}). Such a symmetry can
be defined as invariance with respect to a Lie algebra having basis 
elements of the form
\begin{equation}
X=\xi_{\mu}(x,\psi, \psi^{*})\partial_{\mu}+\eta^{\al}(x, \psi,
\psi^{*}) \partial_{\psi^{\al}}+\eta^{* \al}(x, \psi, \psi^{*})
\partial_{\psi^{* \al}}, 
\label{1.1.29}
\end{equation}
where $\xi_{\mu}$,\ $\eta^{\al}$,\ $\eta^{* \al} $ are some scalar
smooth functions.\index{Lie!symmetry}

As pointed out in the introduction, the above symmetry does not exhaust
all symmetry properties of the Dirac equation because there exist
linear differential and integro-differential symmetry operators which
cannot be represented in the form (\ref{1.1.29}) and, consequently, 
correspond to a non-Lie symmetry of the Dirac 
equation.\index{Non-Lie!symmetry}

Let ${\cal M}_1$ be a class of complex linear first-order
differential operators with variable matrix coefficients acting
on the space of four-component functions, i.e.,\ 
$$
{\cal M}_1=\Bigl\{Q=A_{\mu} \partial_{\mu} + B\Bigr\},
$$
where $A_\mu(x), \ B(x)$ are complex $(4\times
4)$-matrices. Evidently, the class ${\cal M}_1$ contains all Lie
symmetry operators which can be obtained with the help of formula
(\ref{0.12}).

Following \cite{77} we adduce assertions describing all symmetry
operators of the Dirac equation belonging to the class 
${\cal M}_1$.\index{Nonlocal (non-Lie) symmetry!of the Dirac equation} 
\vspace{1.5mm}

\noindent
{\bf Theorem 1.1.3} \cite{77,131,178}.\ {\em Equation (\ref{1.1.1})
has 26 linearly-in\-de\-pen\-dent symmetry operators belonging to the
class  ${\cal M}_1$. The list of these operators is exhausted  
by the generators of the Poincar\'e group (\ref{1.1.20}) and by the
following operators: 
\begin{eqnarray}
& &I, \quad B=\ga_{4} (i x_{\mu} \partial_{\mu}+3i/2 - m\ga \cdot
x),\non\\ & &\omega_{\mu \nu}= (i/2)(\ga_\mu \partial^\nu - 
\ga_{\nu} \partial^{\mu})+m S_{\mu \nu},\label{1.1.30}\\
& &\rho_{\mu}=(1/2)\ga_{4}(i \partial^{\mu} - m \ga_{\mu}),\non\\
& &R_{\mu}=x^{\nu}\omega_{\mu \nu}+\omega_{\mu \nu} x^{\nu},\non
\end{eqnarray}
where $ I $ is the unit $ (4 \times 4)$-matrix.}
\vspace{1.5mm}

\noindent
{\bf Note 1.1.1.}\ Set of operators (\ref{1.1.20}), (\ref{1.1.30}) is
not closed with respect to the algebraic operation 
\begin{displaymath} 
{Q_1, Q_2} \rightarrow Q_3 = [Q_1,\, Q_2].  
\end{displaymath} 
Consequently, it does not form a
Lie algebra. Nevertheless, there exist such subsets of the above set
which are Lie algebras. An important example is provided by the
operators $P_{\mu}, \ J_{\mu \nu} $ satisfying the commutation
relations
\begin{equation}
\begin{array}{l}
[P_{\mu}, P_{\nu}] =  0,\quad
[P_\mu, J_{\al \beta}] = g_{\mu \al} P_{\nu} -
g_{\nu \al} P_\mu,\\[2mm]
[J_{\mu \nu}, J_{\al \beta}] =  g_{\mu \beta}
J_{\nu \al} + g_{\nu \al} J_{\mu \beta} -
 g_{\mu \al} J_{\nu \beta} - g_{\nu \beta} J_{\mu \al}.\\
\end{array}
\label{1.1.31}
\end{equation}

This algebra is called the Lie algebra of the Poincar\'e group 
(or Poincar\'e algebra) and is designated by the symbol 
$AP(1,3)$.\index{Poincar\'e!algebra}

Another interesting example is the eight-dimensional Lie algebra 
\begin{eqnarray*}
& &\Sigma_{\mu \nu}=m S_{\mu \nu} +(i/2)(1-i \ga_{4})( \ga_\mu
\partial^\nu- \ga_\nu \partial^\mu),\\
& &\Sigma_{0} = I, \quad \Sigma_{1}= m\ga_{4} + 
(1-i\ga_{4})\ga_{\mu} \partial_{\mu}
\end{eqnarray*}
obtained in \cite{74,75}.
\vspace{1.5mm}

\noindent
{\bf Note 1.1.2.} As the direct check shows the relations
\begin{eqnarray*}
& &B=-\varepsilon_{\mu \nu \al \beta} J^{\mu \nu}
J^{\al \beta},\quad
[ P_\mu, B ]  =  2\rho_{\mu}, \\
& &[ \rho_{\mu}, B ]  =  (1/2)(P_\mu +
m R_\mu ), \\
&&[ P_\mu, R_\nu ]  =  2\omega_{\mu \nu}
\end{eqnarray*}
hold true. Hence it follows that all symmetry operators of the Dirac
equation $Q \in {\cal M}_1 $ belong to the enveloping
algebra\index{Enveloping algebra} of the Poincar\'e algebra (i.e.,\ to
the algebra whose basis elements are polynomials in $P_\mu, \ 
J_{\mu\nu}$ with constant coefficients). Furthermore, any linear
$N$-th order partial differential operator with matrix coefficients
which is a symmetry operator of the Dirac equation (\ref{1.1.1}) under
$m\ne 0$ belongs to the enveloping algebra of the Poincar\'e algebra
\cite{77}.
\index{Nonlocal (non-Lie) symmetry!of the massless Dirac equation}
\vspace{1.5mm}

\noindent
{\bf Theorem 1.1.4} \cite{77,159}.\ {\em The massless Dirac equation
  (\ref{1.1.17}) has 52 linear\-ly-independent symmetry operators
  belonging to the class ${\cal M}_1 $. A basis of the linear
  vector space of such operators can be chosen as follows}
\begin{eqnarray}
& &P_\mu, \quad J_{\mu \nu}, \quad K_{\mu}, \quad D, \quad I,  
\quad \widetilde P_\mu =i \ga_4 P_\mu,\non\\
& &\widetilde J_{\mu \nu} = i \ga_{4} J_{\mu \nu}, \quad
\widetilde D = i \ga_4 D, \quad \widetilde K_{\mu} = i \ga_{4}
K_{\mu},\non\\ 
& &F= i \ga_4, \quad R_\mu = (D - 1/2) \ga_\mu 
- \ga \cdot x P_\mu,\label{1.1.32}\\
& &\widetilde R_{\mu} = i \ga_{4} R_{\mu}, 
\quad \omega_{\mu \nu} = \ga_{\mu} P_{\nu} - \ga_{\nu} P_{\mu},\non\\
& &Q_{\mu \nu }= i([R_{\mu}, K_{\nu}]-[R_{\nu}, K_{\mu}]).\non
\end{eqnarray}
{\bf Note 1.1.3.} Operators $R_{\mu}$,\ $\widetilde R_{\mu}$,\ 
$\omega_{\mu \nu}$,\ $Q_{\mu \nu}$ are not contained in the enveloping
algebra\index{Enveloping algebra} of the local invariance algebra of
equation (\ref{1.1.17}). Consequently, they are essentially new.

Until now when analyzing non-Lie symmetry of the Dirac equation we
considered only linear transformations of the set of its solutions. To
investigate symmetry of equation (\ref{1.1.1}) in the class of
operators generating both linear and anti-linear transformations
(i.e.,\ transformations of the form
\begin{eqnarray*}
\psi^{\prime}&=&
L_{1} \psi + L_{2}\psi^{*}, \\
\psi^{*\prime}&=&L^{*}_{1} \psi^{*}+L^{*}_{2} \psi,
\end{eqnarray*}
where
$L_{1}, L_{2} $ are some linear differential operators) 
we turn to the eight-component form of the Dirac equation
(\ref{1.1.13}). 

Let ${\cal M}_2$ be a class of complex first-order linear
differential operators with matrix coefficients
$$
X = A_{\mu}(x)\partial_{\mu} + B(x)
$$
acting on the space of eight-component functions $\Psi = \Psi (x)$.
\vspace{1.5mm}

\noindent
{\bf Theorem 1.1.5} \cite{77}. \ {\em The general form of a symmetry
  operator for equation (\ref{1.1.13}) belonging to the class ${\cal
    M}_2$ is given by the formula
\begin{displaymath}
Q= \left( \begin{array} {cc}
Q_{0} &  -\ga_{2}  Q^{*}_{1} \ga_{2} \\
Q_{1} &  -\ga_{2}  Q^{*}_{0} \ga_{2}
\end{array} \right),
\end{displaymath}
where $Q_0, \ Q_1 $ are arbitrary linear combinations of the
generators of the Poin\-ca\-r\'e group and of operators
(\ref{1.1.30}) with complex coefficients.} 
\vspace{1.5mm}

\noindent
{\bf Theorem 1.1.6} \cite{77}.\ {\em The general form of a symmetry
  operator for equation (\ref{1.1.13}) with $ m = 0 $ belonging to the
  class $ {\cal M}_2 $ is given by the formula
\begin{displaymath}
Q= \left(
\begin{array} {cc}
Q_{0} &  -\ga_{2}  Q^{*}_{1} \ga_{2} \\
Q_{1} &  -\ga_{2}  Q^{*}_{0} \ga_{2}
\end{array}
\right),
\end{displaymath}
where $ Q_{0}, \ Q_{1} $ are arbitrary linear combinations of the
operators (\ref{1.1.32}) with complex coefficients. }

A detailed account of symmetry properties of the linear Dirac
equation in the class of high-order differential and
integro-differential operators is given in the monographs
\cite{74,75,77,77.0}.

It is well-known that the maximal (in Lie sense) invariance group of
the Weyl equation\index{Weyl equation}
\begin{equation}
(i \partial_0 + i \sigma_a \partial_a) \varphi(x)=0
\label{1.1.33}
\end{equation}
is the conformal group $C(1,3)$ supplemented by the two-dimensional
transformation group
\begin{eqnarray*}
&&x^{\prime}_\mu = x_\mu,\ \ \mu={0,\ldots,3},\\
&&\varphi'(x')=e^{\theta_1 + i \theta_2} \varphi(x),
\end{eqnarray*}
where $\{\theta_1, \theta_2\} \subset {\R}^1$.

The class ${\cal M}_1$ has no additional symmetry operators. The
class ${\cal M}_2$ contains 52 symmetry operators for the Weyl
equation \cite{159}.
\vspace{2mm}

\noindent
{\bf 5. Absolute time for the Dirac equation.}\ All fundamental
equations of quantum field theory (Maxwell, Dirac, Klein-Gordon-Fock,
d'Alembert etc.) are invariant with respect to the Lorentz
transformations. With these transformations time changes after
transfer from one inertial coordinate system to another. In other
words, the principal motion equations of the quantum field theory are
invariant with respect to the Lorentz group $O(1,3)\in P(1,3)$.

A question arises whether there exist invariance algebras admitted by
the Maxwell, Dirac and Klein-Gordon-Fock equations which generate
transformations for the time variable $x_0\equiv t$ and coordinates
$\vec x\equiv (x_1,x_2,x_3)$ different from the Lorentz and Galilei
transformations. A positive answer to this question was given in the
papers \cite{56}--\cite{57}.  
\vspace{1.5mm}

\noindent
{\bf Theorem 1.1.7} \cite{56}--\cite{57}. \ {\em The Dirac
  equation (\ref{1.1.1}) is invariant under the Poincar\'e algebra
  having the following basis elements:
\begin{equation}
\begin{array}{l}
        P_{0}^{(1)}\equiv H = -\gamma_0\gamma_a\partial_a -
        im\gamma_0,\quad  
        P_a^{(1)}=-\partial_{a},\\[2mm]
        J_{0a}^{(1)} = -x_{0} \partial_{a} -\frac{1}{2}(x_aH +
        Hx_a),\\[2mm] 
        J_{ab}^{(1)} = x_{b} \partial_{a} - x_{a} \partial_{b}
        +\frac{1}{2}\gamma_a\gamma_b,   
\end{array}
\label{f.1}
\end{equation}
  where $a,b=1,2,3,\ a<b$.}

Proof is carried out by direct check.
\vspace{1.5mm}

\noindent
{\bf Note 1.1.4.} The operators $J_{0a}^{(1)}$ generate
{\em non-Lorentz} transformations of the time variable $x_0=t$
and coordinates $x_a$. Time does not change
\index{Absolute time for the Dirac equation}
\begin{equation}
t\to t^\prime = \exp\{v_aJ_{0a}^{(1)}\}t \exp\{-v_aJ_{0a}^{(1)}\}
\equiv t
\label{f.2}
\end{equation}
and the coordinates transform as follows:
\begin{equation}\label{f.3}
\begin{array}{rcl}
x_a\to x_a^\prime &=&
\exp\{v_bJ_{0b}^{(1)}\}x_a \exp\{-v_bJ_{0b}^{(1)}\}\\[2mm] 
&\not = & \underbrace{\exp\{v_bJ_{0b}\}x_a \exp\{-v_bJ_{0b}\}}
_{\mbox {\footnotesize Lorentz transformations}}.
\end{array}
\end{equation}

Here $v_b$ are parameters which are interpreted as components of the
velocity of a moving inertial reference frame with respect to a fixed
one.  \vspace{1.5mm}

\noindent
{\bf Note 1.1.5.} It follows from Theorem 1.1.7 that on the set of
solutions of the Dirac equation two inequivalent representations of
the Poincar\'e algebra are realized.  Operators $P_0^{(1)},
J_{0a}^{(1)}$ from (\ref{f.1}) generate nonlocal transformations of
coordinates $x_a$ leaving the time variable $x_0=t$ invariant. Let us
emphasize that transformations (\ref{f.3}) are different from the
standard Galilei and Lorentz transformations.

As the relations
\begin{eqnarray*}
&& \left(P_0^{(1)}\right)^2 - \left(P_a^{(1)}\right)
\left(P_a^{(1)}\right) = -m^2,\\
&&[P_0^{(1)},\, J_{0a}^{(1)}] = P_a^{(1)},\\ && [P_b^{(1)},\,
J_{0a}^{(1)}] = - g_{ab} P_0^{(1)}
\end{eqnarray*}
hold, the energy $P_0^{(1)}$ and momentum $P_a^{(1)}$ operators
transform according to the standard Lorentz law. But for the time
variable $x_0=t$ and coordinates $x_a$ this is not the case and 
the interval $s^2=x_0^2-x_ax_a$ is not invariant with respect
to the transformations (\ref{f.2}), (\ref{f.3}).

Thus, the Dirac equation as well as the Maxwell and the
Klein-Gordon-Fock equations \cite{56}--\cite{57} have dual symmetry
(Lorentz and non-Lorentz).

The dual symmetry of the Dirac equation is a consequence of the fact
that the spectrum of the operator $H$ has a lacuna in the interval
$(-m,m)$ and the spectrum of the operator $P_0^{(1)}$ is continuous on
the real axis \cite{56}--\cite{57}.

In conclusion we briefly consider symmetry properties of the equation
\begin{equation}
(1 - i \ga_4) \ga_\mu \partial_\mu \psi =0,
\label{1.1.34}
\end{equation} 
which is obtained from the massless Dirac equation (\ref{1.1.17})
by multiplying it by the singular matrix $1- i\ga_4$. This equation is 
distinguished by the fact that two inequivalent representations of the
conformal group\index{Conformal!group} $C(1,3)$ are realized on the
set of its solutions. The first one is given by formulae
(\ref{1.1.24a})--(\ref{1.1.24e}). In addition, equation (\ref{1.1.34})
admits the group $C(1,3)$ with generators $P_\mu, \ J_{\mu \nu}$ of
the form (\ref{1.1.20}) and  
\begin{equation}
\begin{array}{l}
        D=-x_\mu \partial_\mu - 3/2 + \lambda_1 (i \ga_4 -1),\\[2mm]
        K_\mu = 2x_\mu D - x \cdot x \partial^\mu - 2S_{\mu \nu} x^\nu
        + \lambda_2 (i \ga_4 -1)\ga_\mu,
\end{array}
\label{1.1.35}
\end{equation}
where $\lambda_1,\ \lambda_2 $ are non-zero constants.

From \cite{151} it follows that formulae (\ref{1.1.35}) determine the
most general form of generators of groups of scale and special
conformal transformations from the group $C(1,3)$ if the generators of
the group $P(1,3) $ are given in covariant form (\ref{1.1.20}).

It will be shown in Section 2.2 that the representation (\ref{1.1.35})
plays an important role when constructing conformally-invariant
solutions of spinor equations.  
\vspace{10mm}

\noindent
{\large\bf 1.2. Nonlinear spinor equations\label{s1.2} }

\markboth{Chapter 1. SYMMETRY OF NONLINEAR SPINOR EQUATIONS}
{1.2. Nonlinear spinor equations }
\def\theequation{1.\arabic{section}.\arabic{equation}}
\setcounter {section} {2}
\setcounter {equation}{0}
\vspace{7mm}

\noindent
This section is devoted to symmetry analysis of quasi-linear systems
of PDEs for the spinor field of the form 
\begin{equation}
i \ga_\mu \partial_\mu
\psi - F(\bar{\psi}, \psi)=0,
\label{1.2.1}
\end{equation}
where $ F=(F^0, F^1, F^2, F^3 )^T$, \ $F^\mu \in C^1({\C}^8,
{\C}^1)$.

It is clear that an arbitrary equation of the type (\ref{1.2.1})
cannot be taken as a true nonlinear generalization of the Dirac
equation\index{Nonlinear!Dirac equation}.  A natural restriction on
the choice of functions $F^\mu$ is the condition of invariance under
the Poincar\'e group. This condition provides independence of the
choice of inertial reference frame for physical processes described by
equation (\ref{1.2.1}) (i.e.,\ nonlinear PDE (\ref{1.2.1}) has to
satisfy the Lorentz--Poincar\'e--Einstein relativity
principle)\index{Lorentz-Poincar\'e-Einstein relativity principle}.
Mathematical expression of the above principle is a condition of
invariance under the group $ P(1,3) $ with generators (\ref{1.1.20}).
In addition, it is of interest to select subclasses of
Poincar\'e-invariant equations of the form (\ref{1.2.1}) admitting
wider symmetry groups -- the extended Poincar\'e group and the
conformal group.  
\vspace{1.5mm}

\noindent
{\bf Theorem 1.2.1 } \cite{100,103}.\ {\em  System of nonlinear PDEs
(\ref{1.2.1}) is invariant under the Poincar\'e group $P(1,3) $ iff }
\begin{equation}
F(\bar{\psi},\psi)=\{f_1 (\bar{\psi}\psi, \bar{\psi}\gamma_4\psi) 
+ f_2(\bar{\psi}\psi, \bar{\psi}\gamma_4\psi) \ga_4 \} \psi,
\label{1.2.4}
\end{equation}
where $\{f_1,\ f_2\} \subset C^1(\R^2, \C^1)$ are arbitrary functions.
\vspace{1.5mm}

\noindent
{\em Proof.}{$\quad$}  Without loss of generality equation
(\ref{1.2.1}) can be  rewritten in the following form:
\begin{equation}
\{i \ga_\mu \partial_\mu + \Phi (\bar{\psi}, \psi )\}\psi = 0,
\label{1.2.5}
\end{equation}
where $ \Phi (\bar{\psi}, \psi) $ is a $(4 \times 4)$-matrix.

It is evident that equation (\ref{1.2.5}) with an arbitrary matrix
function $ \Phi $ is invariant under the 4-parameter group of
translations (\ref{1.1.24a}). Consequently, to prove the theorem it is
enough to describe all $ \Phi $ such that PDE (\ref{1.2.5}) admits the
Lorentz transformations (\ref{1.1.24c}), whence due to the commutation
relations of the algebra $AO(1,3) $ it follows that PDE in question is
invariant under the Poincar\'e group.

Acting with the first prolongation of the operator $ J_{0a} $ on
equation (\ref{1.2.5}) and passing to the set of its solutions
we obtain a system of PDEs for an unknown matrix function $ \Phi
(\bar{\psi},\psi ) $ 
\begin{equation}
Q_{0a} \Phi + (1/2) (\Phi \ga_0 \ga_a - \ga_0 \ga_a \Phi ) = 0.
\label{1.2.6}
\end{equation}

Let us expand the matrix $\Phi$ in the complete system of the Dirac
matrices $I$,\ $\ga_{\mu}$,\ $S_{\mu \nu }$,\ $\ga_4\ga_\mu$,\ $\ga_4$
\begin{equation}
\begin{array}{rcl}
\Phi& =& A(\bar{\psi}, \psi ) + B^\mu (\bar{\psi}, \psi ) \ga_\mu +
C^{\mu \nu } (\bar{\psi}, \psi ) S_{\mu \nu }\\[2mm]
& & + D^\mu (\bar \psi, \psi ) \ga_4 \ga_\mu +
E(\bar{\psi}, \psi ) \ga_4.
\label{1.2.7}
\end{array}
\end{equation}

Substituting expression (\ref{1.2.7}) into (\ref{1.2.6}) and 
taking into account the identities
\begin{eqnarray*}
& &[ \ga_4, \ga_0 \ga_a ] = 0,\quad
[ \ga_\mu, \ga_0 \ga_a ] =2(g_{\mu 0} \ga_a - g_{\mu a } \ga_0),\\
& & [\ga_\mu \ga_\nu, \ga_0 \ga_a ] =2(g_{\mu 0 } \ga_a \ga_\nu -
g_{\mu a} \ga_0 \ga_\nu +
 g_{\nu 0 } \ga_\mu \ga_a - g_{\nu a } \ga_\mu \ga_0),
\end{eqnarray*}
where $ g_{\mu \nu } $ is the metric tensor of the Minkowski space
$R(1,3) $, with a subsequent equating to zero of coefficients of
linearly independent matrices $I$, \ $\ga_\mu$, $\ldots$, $\ga_4 $
one gets an over-determined system of PDEs for functions $ A, \ B^\mu,
\ldots, E$
\begin{eqnarray}
& &Q_{0a} A= 0, \quad Q_{0a} E = 0, \label{1.2.8a}\\
& &Q_{0a} B_\mu + B^\al (g_{\al 0} g_{\mu a } -
g_{\al a} g_{\mu 0}) = 0,\label{1.2.8b}\\
& &Q_{0a} D_{\mu} + D^\al (g_{\al 0} g_{\mu a} -
g_{\al a } \g_{\mu 0}) = 0,\label{1.2.8c}\\
& &Q_{0a} C^{\mu \nu } + (1/2) C^{\al \beta } (g_{\al a}
\delta^{\mu \nu}_{\beta 0} + g_{\beta 0} \delta^{\mu \nu}_{\al a} 
\non\\ 
& &\quad -g_{\al 0} \delta^{\mu \nu}_{\beta a} -
g_{\beta a} \delta^{\mu \nu }_{\al 0} )=0.\label{1.2.8d}
\end{eqnarray}

In formulae (\ref{1.2.8a})--(\ref{1.2.8d}) we use the following
notations: 
\begin{eqnarray*}
& &Q_{\mu \nu} = (1/2)\{\ga_\mu \ga_\nu \psi \}^\al
\partial_{\psi^\al} - (1/2) 
\{\bar{\psi} \ga_\mu \ga_\nu\}^\al \partial_{\bar{\psi}^\al},
\ \ \mu <\nu,\\
& &\delta^{\mu \nu}_{\al \beta} = \delta_{\mu\alpha} \delta_{\nu\beta} 
-\delta_{\mu\beta} \delta_{\nu\al},
\ \
a={1,2,3},
\ \
\mu, \nu, \al, \beta = 0, 1, 2, 3.
\end{eqnarray*}

Since $[Q_{0a},Q_{0b}]=Q_{ab}$, functions $ A(\bar{\psi}, \psi) $, \ 
$B(\bar{\psi}, \psi) $ satisfy the system of PDEs
\begin{equation}
Q_{\mu\nu}f(\bar\psi,\psi)=0, \ \ \mu <\nu.
\label{1.2.3}
\end{equation}

According to \cite{41}, the general solution of this system is an
arbitrary smooth function of a complete set of its first integrals
$\omega$.

If we denote by $ r $ the rank of the $(6\times 8)$-matrix of
coefficients of the operators $Q_{\mu \nu} $
\begin{displaymath}
\left[ \begin{array} {rrrrrrrr}
-\frac{i}{2} \psi^3 & -\frac{i}{2} \psi^2 &
 -\frac{i}{2} \psi^1 & -\frac{i}{2} \psi^0 &
\frac{i}{2} \bar{\psi}^3 & \frac{i}{2} \bar{\psi}^2 &
\frac{i}{2} \bar{\psi}^1 & \frac{i}{2} \bar{\psi}^0 \\ \ \\

-\frac{1}{2} \psi^3 & \frac{1}{2} \psi^2 &
 -\frac{1}{2} \psi^1 & \frac{1}{2} \psi^0 &
-\frac{1}{2} \bar{\psi}^3 & \frac{1}{2} \bar{\psi}^2 &
-\frac{1}{2} \bar{\psi}^1 & \frac{1}{2} \bar{\psi}^0 \\  \ \\

-\frac{i}{2} \psi^2 & \frac{i}{2} \psi^3 &
 -\frac{i}{2} \psi^0 & \frac{i}{2} \psi^1 &
\frac{i}{2} \bar{\psi}^2 & -\frac{i}{2} \bar{\psi}^3 &
\frac{i}{2} \bar{\psi}^0 & -\frac{i}{2} \bar{\psi}^1 \\ \ \\

-\frac{1}{2} \psi^0 & \frac{1}{2} \psi^1 &
-\frac{1}{2} \psi^2 & \frac{1}{2} \psi^3 &
\frac{1}{2} \bar{\psi}^0 & -\frac{1}{2} \bar{\psi}^1 &
\frac{1}{2} \bar{\psi}^2 & -\frac{1}{2} \bar{\psi}^3 \\  \ \\

-\frac{1}{2} \psi^1 & -\frac{1}{2} \psi^0 &
 -\frac{1}{2} \psi^3 & -\frac{1}{2} \psi^2 &
\frac{1}{2} \bar{\psi}^1 & \frac{1}{2} \bar{\psi}^0 &
\frac{1}{2} \bar{\psi}^3 & \frac{1}{2} \bar{\psi}^2 \\ \ \\

\frac{i}{2} \psi^1 & -\frac{i}{2} \psi^0 &
\frac{i}{2} \psi^3 & -\frac{i}{2} \psi^2 &
\frac{i}{2} \bar{\psi}^1 & -\frac{i}{2} \bar{\psi}^0 &
\frac{i}{2} \bar{\psi}^3 & -\frac{i}{2} \bar{\psi}^2
\end{array} \right]
\end{displaymath}
(the representation of $ \ga $-matrices is given by formulae
(\ref{1.1.5})), then a maximal set of functionally-independent first
integrals of system (\ref{1.2.3}) consists of $ 8-r $ integrals
\cite{41}. In the case considered $ r = 6 $, whence it follows that
the general solution is represented as an arbitrary smooth function of
two functionally-independent first integrals. As a rule, they are
chosen in the form $ \bar{\psi} \psi $, $\bar{\psi} \ga_4 \psi $.
Thus, the general solution of system (\ref{1.2.8a}) is given by the
formulae
\begin{equation}
A= \widetilde A (\bar{\psi} \psi,\, \bar{\psi} \ga_4 \psi ),
\quad
E =\widetilde E (\bar{\psi} \psi,\, \bar{\psi} \ga_4 \psi ),
\label{1.2.9}
\end{equation}
where $\{\widetilde A, \widetilde E\} \subset C^1 ({\R}^2, {\C}^1 )$ 
are arbitrary functions.

We expand the four-component function with components $ B_\mu $ in the
system of four linearly independent vectors $e_1, \ e_2, \ e_3, \ e_4$
having the  components 
$ \bar{\psi} \ga_\mu \psi $,\ $\bar{\psi} \ga_4 \ga_\mu
\psi$,\ $\psi^T \ga_0 \ga_2 \ga_\mu \psi $,\ 
$\psi^T \ga_0 \ga_2 \ga_4 \ga_\mu \psi $
\begin{eqnarray*}
B_\mu& =& R_1 (\bar{\psi}, \psi ) \bar{\psi} \ga_\mu \psi +
R_2 (\bar{\psi}, \psi ) \bar{\psi} \ga_4 \ga_\mu \psi \\
& &+ R_3 (\bar{\psi}, \psi ) \psi^T \ga_0 \ga_2 \ga_\mu \psi
+ R_4 (\bar{\psi}, \psi ) \psi^T \ga_0 \ga_2 \ga_4 \ga_\mu \psi.
\end{eqnarray*}

Let us prove that the functions $ B_\mu = B_\mu (\bar{\psi}, \psi ),
\ \mu = {0,\ldots,3} $ satisfy system of PDEs (\ref{1.2.8b}) iff the
conditions 
\begin{displaymath}
R_i = \widetilde B_i (\bar{\psi} \psi, \bar{\psi} \ga_4 \psi ),
\quad
\widetilde B_i \in C^1 ({\R}^2,\ {\C}^1 ),\ \  i = {1,\ldots,4}
\end{displaymath}
hold.

Indeed, if we designate by $V_\mu (\bar{\psi}, \psi )$ the components
of one of the vectors $e_i$, then $V_\mu $ satisfy the equalities 
of the form
\begin{eqnarray}
Q_{0a} V_0 = V_a, \ \ a= {1,2,3}
\label{1.2.10}
\end{eqnarray}
(the above fact is established by straightforward
computation). Consequently, we have
\begin{eqnarray}
Q_{0a} B_0 &=& (Q_{0a} R_1) \bar{\psi}\ga_0 \psi + (Q_{0a} R_2)
\bar \psi \ga_4 \ga_0 \psi + (Q_{0a} R_3) \non\\
& &\times\psi^T \ga_0 \ga_2 \ga_0 \psi +
(Q_{0a} R_4) \psi^T \ga_0 \ga_2 \ga_4 \ga_0 \psi 
+ R_1 \bar{\psi} \ga_a \psi \label{1.2.11}\\
& &+R_2 \bar{\psi} \ga_4 \ga_a \psi +
R_3 \psi^T \ga_0 \ga_2 \ga_a \psi 
+ R_4 \psi^T \ga_0 \ga_2 \ga_4 \ga_a \psi.\non
\end{eqnarray}

Setting $ \mu = 0 $ in (\ref{1.2.8b}) we find
\begin{equation}
Q_{0a} B_0 = B_a.
\label{1.2.12}
\end{equation}

Comparing (\ref{1.2.11}) and (\ref{1.2.12}) yields the
following equality: 
\begin{equation}
\begin{array}{l}
(Q_{0a} R_1) \bar{\psi} \ga_0 \psi + (Q_{0a} R_2)
\bar{\psi} \ga_4 \ga_0 \psi +(Q_{0a} R_3)\psi^T \ga_0 \ga_2 \ga_0 \psi 
\\[2mm]
\quad
+(Q_{0a} R_4) \psi^T \ga_0 \ga_2 \ga_4 \ga_0 \psi = 0. 
\end{array}
\label{1.2.13}
\end{equation}

In the same way we obtain equalities of the form
\begin{equation}
\begin{array}{l}
(Q_{0a} R_1) \bar \psi \ga_b \psi + (Q_{0a} R_2 ) \bar{\psi} \ga_4
\ga_b \psi + 
Q_{0a} R_3) \psi^T \ga_0 \ga_2 \ga_b \psi \\[2mm]
\quad +(Q_{0a} R_4)\psi^T \ga_0 \ga_2 \ga_4 \ga_b \psi = 0,
\label{1.2.14}
\end{array}
\end{equation}
where $ a, b = {1,2,3}$.

Since four-vectors with components $ \bar{\psi} \ga_\mu \psi, 
\ldots, \psi^T \ga_0 \ga_2 \ga_4
\ga_\mu \psi$ are linearly-in\-de\-pen\-dent, from
(\ref{1.2.13}), (\ref{1.2.14})
it follows that $Q_{0a} R_i = 0$,\ $a= {1,2,3}$,\  
$i={1,\ldots,4}$ or $R_i = \widetilde B_i
(\bar{\psi} \psi, \bar \psi \ga_4 \psi )$,\ $i = {1,\ldots,4}$.

Taking into account that system of PDEs (\ref{1.2.8c}) coincides with
system (\ref{1.2.8b}) it is easy to write down its general solution
\begin{eqnarray*}
D_\mu (\bar \psi, \psi )&=&\bar \psi \ga_\mu \psi 
\widetilde D_1 (\bar \psi \psi,
\bar \psi \ga_4 \psi ) +
\bar \psi \ga_4 \ga_\mu \psi \widetilde D_2 (\bar \psi \psi, 
\bar \psi \ga_4 \psi ) \\
& &+\psi^T \ga_0 \ga_2 \ga_\mu \psi D_3(\bar \psi \psi, \bar \psi
\ga_4 \psi ) + 
\psi^T \ga_0 \ga_2 \ga_4 \ga_\mu \psi \widetilde D_4 (\bar \psi \psi,
\bar \psi \ga_4 \psi ),
\end{eqnarray*}
where $\widetilde D_i \in C^1 ({\R}^1, {\C}^1), \ i = {1,\ldots,4} $
are arbitrary functions.

Integration of equations (\ref{1.2.8d}) is carried out in the same
way, as a result we have
\begin{eqnarray*}
C_{\mu \nu } (\bar \psi, \psi)& =& \bar \psi \ga_\mu \ga_\nu \psi 
\widetilde C_1 +\bar \psi \ga_4 \ga_\mu \ga_\nu \psi \widetilde C_2 \\ 
& & + \psi^T \ga_0 \ga_2 \ga_\mu \ga_\nu \psi \widetilde C_3 
+ \psi^T \ga_0 \ga_2 \ga_4
\ga_\mu \ga_\nu \psi \widetilde C_4,
\end{eqnarray*}
where $ \widetilde C_i = \widetilde C_i (\bar \psi \psi, 
\bar \psi \ga_4 \psi ),
\  i = {1,\ldots,4} $
are arbitrary smooth functions.

Thus, we have proved that equation (\ref{1.2.1}) is invariant
under the Poincar\'e group iff
\begin{eqnarray}
F(\bar \psi, \psi ) & = & \Phi (\bar \psi, \psi ) \psi \non\\
&\equiv &\biggl\{ \widetilde A I + \widetilde B_1 \ga_\mu 
 (\bar \psi \ga^\mu \psi ) +
 \widetilde B_2 \ga_\mu (\bar \psi \ga_4 \ga^\mu \psi ) \non\\
&&+\widetilde B_3 \ga_\mu (\psi^T \ga_0 \ga_2 \ga^\mu \psi ) +
\widetilde B_4 \ga_\mu (\psi^T \ga_0 \ga_2 \ga_4 \ga^\mu \psi )\non\\
&&+\widetilde C_1 S_{\mu \nu } (\bar \psi S^{\mu \nu } \psi ) +
\widetilde C_2 S_{\mu \nu } (\bar \psi \ga_4 S^{\mu \nu } \psi )
\label{1.2.15}\\
&&+\widetilde C_3
S_{\mu \nu } ( \psi^T \ga_0 \ga_2 S^{\mu \nu } \psi ) +
\widetilde C_4 S_{\mu \nu } (\psi^T \ga_0 \ga_2 \ga_4 
S^{\mu \nu } \psi )\non\\
&&+\widetilde D_1 \ga_4 \ga_\mu (\bar \psi \ga^\mu \psi ) +
 \widetilde D_2 \ga_4 \ga_\mu (\bar \psi \ga_4 \ga^\mu \psi )\non\\
&&+\widetilde D_3 \ga_4 \ga_\mu
(\psi^T \ga_0 \ga_2 \ga^\mu \psi ) +
\widetilde D_4 \ga_4 \ga_\mu (\psi^T \ga_0 \ga_2 \ga_4 \ga^\mu \psi ) +
\widetilde E \ga_4 \biggr\} \psi.\non
\end{eqnarray}

Here $\widetilde A, \ \widetilde B_1, \ldots, \widetilde E $ are
arbitrary smooth functions of $\bar \psi \psi, \ \bar \psi \ga_4 \psi
$. 

Let us show that formula (\ref{1.2.15}) without loss of generality can
be rewritten in the form (\ref{1.2.4}). To this end, we need the
following identity:
\begin{equation}
(\bar{\psi_1} \ga_\mu \psi_2) \ga^\mu \psi_2 = (\bar{\psi_1} \psi_2 )
\psi_2 + (\bar{\psi_1} \ga_4 \psi_2) \ga_4 \psi_2,
\label{1.2.16}
\end{equation}
where $\psi_1, \ \psi_2 $ are arbitrary four-component functions.

The validity of (\ref{1.2.16}) is checked by direct computation.
Choosing $\ga $-matrices in the representation (\ref{1.1.5}) we have
\begin{eqnarray*}
& &\ga_0 \psi_2 = (\psi^0_2,\, \psi^1_2,\, -\psi^2_2,\,
-\psi^3_2 )^T,\\
& &\ga_1 \psi_2  = (\psi^3_2,\, \psi^2_2,\, -\psi^1_2,\, -\psi^0_2
)^T, \\
& &\ga_2 \psi_2 = (i \psi^{3}_{2},\, -i \psi ^2_2,\, -i \psi^1_2,\,
i \psi^0_2 )^T,\\
& &\ga_3 \psi_2  =  (\psi^{2}_{2},\, -\psi ^3_2,\, 
-\psi^0_2,\, \psi^1_2 )^T,
\\
& &\bar{\psi_1} \ga_0 \psi_2 = \bar{\psi}^0_1 \psi^0_2 +
\bar{\psi}^1_1 \psi^1_2 - 
\bar{\psi}^2_1 \psi^2_2 - \bar{\psi}^3_1 \psi^3_2, 
\\
& &\bar{\psi_1} \ga_1 \psi_2 = \bar{\psi}^0_1 \psi^3_2 +
\bar{\psi}^1_1 \psi^2_2 - 
\bar{\psi}^2_1 \psi^1_2 - \bar{\psi}^3_1 \psi^0_2,
\\
& &\bar{\psi}_1 \ga_2 \psi_2 = i( \bar{\psi}^0_1 \psi^3_2 -
\bar{\psi}^1_1 \psi^2_2 - 
\bar{\psi}^2_1 \psi^1_2 + \bar{\psi}^3_1 \psi^0_2 ),
\\
& &\bar{\psi}_1 \ga_3 \psi_2 = \bar{\psi}^0_1 \psi^2_2 -
\bar{\psi}^1_1 \psi^3_2 - 
\bar{\psi}^2_1 \psi^0_2 + \bar{\psi}^3_1 \psi^1_2,
\\
& &\bar{\psi}_1 \psi_2 = \bar{\psi}^0_1 \psi^0_2 + \bar{\psi}^1_1
\psi^1_2 + \bar{\psi}^2_1 \psi^2_2 + \bar{\psi}^3_1 \psi^3_2,
\\
& &\bar{\psi}_1 \ga_4 \psi =-( \bar{\psi}^0_1 \psi^2_2 +
\bar{\psi}^1_1 \psi^3_2 + 
\bar{\psi}^2_1 \psi^0_2 + \bar{\psi}^3_1 \psi^1_2),
\end{eqnarray*}
whence it follows
\begin{displaymath}
(\bar{\psi}_1 \ga_\mu \psi_2 ) \ga^\mu \psi_2 = (\bar{\psi}^0_1
\psi^0_2 + \bar{\psi}^1_1 \psi^1_2 + \bar{\psi}^2_1 \psi^2_2 +
\bar{\psi}^3_1 \psi^3_2) 
\end{displaymath}
\begin{displaymath}
\qquad
\times
\left( \begin{array} {r}
\psi^0_2 \\
\psi^1_2 \\
\psi^2_2 \\
\psi^3_2
\end{array} \right)
- (\bar \psi^0_1 \psi^2_2 + \bar \psi^1_1 \psi^3_2 + \bar \psi^2_1
\psi^0_2 + \bar \psi^3_1 \psi^1_2) 
\left( \begin{array} {r}
\psi^2_2 \\
\psi^3_2 \\
\psi^0_2 \\
\psi^1_2
\end{array} \right) 
\end{displaymath}
\begin{displaymath}
\qquad
=\{\bar \psi_1 \psi_2 + (\bar \psi_1 \ga_4 \psi_2) \ga_4\} \psi_2.
\end{displaymath}

On making in (\ref{1.2.16}) the change of variables $\bar \psi_1
\rightarrow \bar \psi_1 \ga_4 $ we arrive at the identity
\begin{equation}
(\bar \psi_1 \ga_4 \ga_\mu \psi_2 ) \ga^\mu \psi_2  =
\{\bar \psi_1 \ga_4 \psi_2 - (\bar \psi_1 \psi_2 ) \ga_4 \} \psi_2.
\label{1.2.17a}
\end{equation}

Similarly, we obtain from (\ref{1.2.16}) two other identities
\begin{eqnarray}
& &(\bar \psi_1 \ga_4 \ga_\mu \psi_2 ) \ga_4 \ga^\mu \psi_2 =
\{(\bar \psi_1 \ga_4 \psi_2) \ga_4 + \bar \psi_1 \psi_2 \} \psi_2,
\label{1.2.17b}\\
& &(\bar \psi_1 S_{\mu \nu} \psi_2) S^{\mu \nu} \psi_2 
= (1/2)\{\bar \psi_1 \psi_2 -
(\bar \psi_1 \ga_4 \psi_2 ) \ga_4 \}\psi_2.
\label{1.2.17c}
\end{eqnarray}

In (\ref{1.2.17a})--(\ref{1.2.17c}) \  $\psi_1, \psi_2 $ are
arbitrary four-component functions.

Choosing in (\ref{1.2.16}), (\ref{1.2.17a})--(\ref{1.2.17c}) 
functions\ $ \psi_1, \ \psi_2 $
in an appropriate way we arrive at the following relations:
\begin{eqnarray*}
& &(\bar \psi \ga_\mu \psi ) \ga^\mu \psi = \{\bar \psi \psi +
(\bar \psi \ga_4 \psi ) \ga_4 \} \psi,\\
& &(\bar \psi \ga_4 \ga_\mu \psi) \ga^\mu \psi= \{\bar \psi \ga_4 \psi
- (\bar \psi \psi ) \ga_4 \} \psi,\\
& &\ldots \\
& &(\psi^T \ga_0 \ga_2 \ga_4 \ga_\mu \psi ) \ga^\mu \psi 
= \{\psi^T \ga_0 \ga_2 \psi 
+ (\psi^T \ga_0 \ga_2 \ga_4 \psi) \ga_4 \} \psi = 0,
\end{eqnarray*}
whence the existence of such smooth functions
\ $f_1(\bar \psi \psi,
\bar \psi \ga_4 \psi )$, \ $f_2(\bar \psi \psi, \linebreak
\bar \psi \ga_4 \psi ) $ that
$\Phi (\bar \psi, \psi ) \psi = (f_1 + f_2 \ga_4) \psi $ follows.
The theorem is proved. $\rhd$
\vspace{1.5mm}

\noindent
{\bf Note 1.2.1.}\ If we choose in (\ref{1.2.15})
$\widetilde D_2 = \lambda = \mbox{\rm const},\ \
\widetilde A = \widetilde B_1 = \ldots = \widetilde D_1 = \widetilde
D_3 = \widetilde D_4 = \widetilde E = 0 $, then equation 
(\ref{1.2.1}) coincides with the nonlinear spinor equation (\ref{0.1})
suggested by Heisenberg.
\vspace{1.5mm}

\noindent
{\bf Note 1.2.2.} From formulae (\ref{1.2.16})--(\ref{1.2.17c})
the well-known Pauli--Fierz identities follow 
\cite{42,192,193}\index{Pauli-Fierz identities}
\begin{displaymath}
v_\mu v^\mu = s^2 + p^2,\quad
w_\mu w^\mu = s^2 + p^2,\quad
\sigma_{\mu \nu } \sigma^{\mu \nu } = (1/2) (s^2 - p^2),
\end{displaymath}
where  
\begin{eqnarray*}
& & s = \bar \psi \psi, \quad  p = \bar \psi \ga_4 \psi,
 \quad v_\mu = \bar \psi \ga_\mu \psi, \\ 
& &w_\mu = \bar \psi \ga_4 \ga_\mu \psi, \quad
\sigma_{\mu \nu } = \bar \psi S_{\mu \nu } \psi,\ \
\mu, \nu = {0,\ldots,3}.
\end{eqnarray*}

Further we will select subclasses of equations of the form
(\ref{1.2.1}) which in addition to the group $P(1,3) $ admit the
one-parameter group of scale transformations (\ref{1.1.24d}) with
arbitrary non-zero $k \in {\R}^1 $ and the 4-parameter group of
special conformal transformations.
\vspace{1.5mm}

\noindent
{\bf Theorem 1.2.2} \cite{100,103}.\ {\em Equation (\ref{1.2.1})
is invariant under the extended Poincar\'e 
group, iff the function $F(\bar \psi, \psi )$  has the form
(\ref{1.2.4}),  
where }
\begin{equation}
f_i = (\bar \psi \psi )^{1/2k} \tilde f_i\Bigl(\bar \psi 
\psi (\bar \psi \psi \ga_4\psi )^{-1}\Bigr),
\ i = 1,2.
\label{1.2.19}
\end{equation}
{\em Proof.}$\quad$ The necessity. Since PDE (\ref{1.2.1}) is invariant
under the group $\widetilde P(1,3)$, it admits the group $P(1,3)
\subset \widetilde P (1,3)$. Applying Theorem 1.2.1 we conclude
that it is necessary to describe all functions $f_1 (\bar \psi \psi,\,
\bar \psi \ga_4 \psi ), \ f_2 (\bar \psi \psi,\, \bar \psi
\ga_4 \psi )$ such that equation (\ref{1.2.1}) with $ F $ of the form
(\ref{1.2.4}) is invariant under the group of transformations
(\ref{1.1.24d}). Acting by the first prolongation of the
infinitesimal generator of the group (\ref{1.1.24d})

\begin{displaymath}
D=x_\mu \partial_\mu - k \psi^\al \partial \psi^\al - k \bar \psi_\al
\partial_{\bar \psi^\al} 
\end{displaymath}
on equation (\ref{1.2.1}) with $F$ of the form
(\ref{1.2.4}) and passing to the set of its solutions yield
determining equations for $ f_1, \ f_2 $   
\begin{equation}
\Bigl(\omega_1 \partial_{\omega_1} + \omega_2 \partial_{\omega_2} 
- (2k)^{-1}\Bigr) f_i = 0, \ i = 1,2,
\label{1.2.20}
\end{equation}
where $\omega_1 = \bar \psi \psi, \ \omega_2 = \bar \psi \ga_4 \psi $.

The general solutions of the above equations are given by formulae
(\ref{1.2.19}). The necessity is proved.

The sufficiency. Let us introduce a notation
\begin{equation}
G(\bar \psi, \psi ) = i \ga_\mu \partial_\mu \psi - (\tilde f_1 +
\tilde f_2 \ga_4 ) 
(\bar \psi \psi )^{1/2k} \psi.
\label{1.2.21}
\end{equation}

The direct computation yields the following identity:
\begin{displaymath}
G(\bar \psi', \psi' ) = e^{(k+1)\theta} G(\bar \psi, \psi ), \
\ \theta \in {\R}^1,
\end{displaymath}
where $ \psi' $ is given by formulae (\ref{1.1.24d}).

In other words, the group of scale transformations leaves the set of
solutions of equation $ G = 0 $ invariant. Hence it follows that
equation (\ref{1.2.1}), where the function $ F(\bar \psi, \psi ) $ is
determined by (\ref{1.2.4}), (\ref{1.2.19}), admits the extended
Poincar\'e group.  Theorem is proved. $\rhd$ 
\vspace{1.5mm}

\noindent
{\bf Theorem 1.2.3} \cite{100,103}. \ {\em Equation (\ref{1.2.1}) is
  invariant under the conformal group\index{Conformal!group} $C(1,3)$
  iff 
\begin{equation}
F(\bar \psi, \psi ) = (\bar \psi \psi )^{1/3} 
(\tilde f_1 + \tilde f_2 \ga_4) \psi,
\label{1.2.22}
\end{equation}
where $ f_1, \ f_2 $ are arbitrary smooth functions of 
$\bar \psi \psi (\bar \psi\ga_4 \psi)^{-1} $. }
\vspace{1.5mm}

\noindent
{\em Proof.}{$\quad$} The necessity. Since the group $C(1,3) $ contains
the extended Poincar\'e group, the function $F(\bar \psi, \psi ) $ has
the form (\ref{1.2.4}), (\ref{1.2.19}), the conformal 
degree $k$ being equal to $3/2$.

The sufficiency is established by direct verification. Making
the change of variables (\ref{1.1.24e}) in equation $G = 0$, where
$G$ is given by (\ref{1.2.21}) under $k=3/2$, we get the identity

\begin{displaymath}
G(\bar \psi', \psi' ) = \sigma^2 (x) (1- \ga \cdot\, \theta\ga \cdot
x) G(\bar \psi, \psi ),
\end{displaymath}
whence it follows that equation $G=0$ admits the 4-parameter group of
special conformal transformations. The theorem is proved. $\rhd$
\vspace{1.5mm}

\noindent
{\bf Note 1.2.3.}\ If we choose in (\ref{1.2.22}) $\tilde f_1 =
\lambda = \mbox{\rm const}$, \ $\tilde f_2 = 0 $, then the
conformally-invariant spinor equation suggested by G\"ursey \cite{113} 

\begin{equation}
\{i \ga_\mu \partial_\mu - \lambda (\bar \psi \psi)^{1/3}\} \psi = 0
\label{1.2.23}
\end{equation}
is obtained. In addition, by using formulae
(\ref{1.2.16})--(\ref{1.2.17c}) it is 
not difficult to become convinced of that the 
conformally-invariant spinor equation
\begin{displaymath}
i\{ \ga_\mu \partial_\mu - \lambda [(\bar \psi \ga_4 \ga_\mu \psi )
(\bar \psi \ga_4 \ga^\mu \psi )]^{-1/3}
(\bar \psi \ga_4 \ga_\mu \psi ) \ga_4 \ga^\mu \}\psi = 0
\end{displaymath}
suggested in \cite{90,91} is also included into the class of nonlinear
PDEs (\ref{1.2.1}), (\ref{1.2.22}). 
\vspace{1.5mm}

\noindent
{\bf Note 1.2.4.}\ Applying the Lie method we can establish that
Po\-in\-ca\-r\'e-invariant equations (\ref{1.2.1}), (\ref{1.2.4})
admit the three-parameter Pauli-G\"ursey 
group\index{Pauli-G\"ursey group} having generators 
$Q_1, \ Q_2, \ Q_3$ (\ref{1.1.21}) iff
the functions $f_1, \ f_2$ are real-valued ones.

It should be noted that there exist nonlinear spinor equations
which admit infinite-parameter symmetry 
groups\index{Infinite-parameter Lie group}. As an example, we
give the following $P(1,3)$-invariant spinor equation:
\begin{equation}
(\bar \psi \ga_\mu \psi ) \pa_\mu \psi = 0
\end{equation}
which is obtained from (\ref{1.1.17}) by a formal change $\ga_\mu
\rightarrow \bar\psi \ga_\mu \psi$. The maximal symmetry group of the
above equation is generated by the infinitesimal operator
\cite{103,106.2}
\begin{displaymath}
X= \xi_\mu (x, \bar \psi, \psi ) \partial_\mu 
+ \eta^\al (x, \bar \psi, \psi ) \partial_{\psi \al} +
 \bar \eta^\al (x, \bar \psi, \psi) \partial_{\bar \psi^\al},
\end{displaymath}
where
\begin{eqnarray*}
\xi_\mu &=& f_\mu (w,\bar \psi, \psi) + \bar \psi \ga_\mu \psi
f(x, \bar \psi, \psi ) + \bar \psi \ga \cdot x\psi \\
& &\times(\bar R \ga_\mu \psi + \bar \psi \ga_\mu R) \{ (\bar \psi
\ga_\nu \psi ) (\bar \psi \ga^\nu \psi ) \}^{-1},\\
\eta^\al &=& R^\al (w, \bar \psi, \psi ),\\
w &=& \{x_\mu (\bar \psi \ga_\nu \psi)
(\bar \psi \ga^\nu \psi) - (\bar \psi \ga_\mu \psi )
 (\bar \psi \ga \cdot x \psi ) \},
\end{eqnarray*}
$f, \ f_\mu, \ R^\al $ are arbitrary smooth functions and
$ \mu, \nu, \al = {0,\ldots,3} $. 
\vspace{10mm}

\noindent
{\large\bf 1.3. Systems of nonlinear second-order
  equations\label{s1.3} 
\vspace{1.5mm}

\noindent
\phantom{\large\bf 1.3. }for the spinor field}
\markboth{Chapter 1. SYMMETRY OF NONLINEAR SPINOR EQUATIONS}
{1.3. Systems of nonlinear second-order equations}
\def\theequation{1.\arabic{section}.\arabic{equation}}
\setcounter {section} {3}
\setcounter {equation}{0}
\vspace{7mm}

\noindent
As a rule, the spinor field is described by the first-order system of
PDEs. Such description is considered to be the most adequate to
the nature of the spinor field. But there exists another approach
based on the second-order equations \cite{60,62,169,170}.

Each component of the Dirac spinor satisfies the second-order wave
equation (see Section 1.1)
\begin{equation}
(\partial_\mu \partial^\mu + m^2) \psi (x) = 0.
\label{1.3.1}
\end{equation}

The above equations form a system of splitting wave equations for four
functions \ $\psi^0,\ \psi^1,\ \psi^2,\ \psi^3$. That is why they can
be used to describe particles with different spins $s= 0, 1/2, 1, 3/2,
\ldots$. For system (\ref{1.3.1}) to describe a field (particle) with
the spin $s=1/2$ it is necessary to impose an additional constraint
(equation) on the function $\psi (x)$. Possible Poincar\'e-invariant
additional conditions
\begin{equation}
\partial_\mu (\bar \psi \ga_\mu \psi) = \lambda_1 \bar \psi \psi +
\lambda_2 \bar \psi \ga_4 \psi + \lambda_3
\label{1.3.2}
\end{equation}
and
\begin{equation}
\bar \psi (i \ga_\mu \partial_\mu - m) \psi = \lambda_1 \bar \psi \psi
+ \lambda_2 \bar \psi\ga_4 \psi + \lambda_3,
\label{1.3.3}
\end{equation}
where  $\lambda_1,\ \lambda_2,\ \lambda_3$ are constants, have been
suggested in \cite{62}.

Nonlinear conditions (\ref{1.3.2}), (\ref{1.3.3}) select from the
set of solutions of equation (\ref{1.3.1}) the ones which correspond
to a particle with the spin $s=1/2$. On the set of solutions of the system
of PDEs (\ref{1.3.1}), (\ref{1.3.2}) the spinor representation of the
Poincar\'e group having the generators (\ref{1.1.20}) is realized.

It is interesting to note that the system of nonlinear equations
(\ref{1.3.1}), (\ref{1.3.2}) with $\lambda_1 = \lambda_2 = \lambda_3 = 
0 $ admits the group of nonlocal 
transformations\index{Nonlocal!transformation group}
\begin{displaymath}
x'_\mu = x_\mu,\quad
\psi'(x')=\psi(x) + \theta \ga_4 (i \ga_\mu \partial_\mu - m) \psi (x),
\end{displaymath}
where $\theta \in {\R}^1 $ is a group parameter.

Another possibility of describing fields with spin $s=1/2$
by the use of second-order equations is to consider a nonlinear
equation of the form 
\begin{equation}
(\partial_\mu \partial^\mu + m^2)\psi = 
R(\bar \psi,\, \psi,\, \mathop{\bar\psi}\limits_{\scriptscriptstyle
  1},\, \mathop{\psi}\limits_{\scriptscriptstyle 1}),
\label{1.3.4}
\end{equation}
where $\mathop{\psi}\limits_{\scriptscriptstyle 1}=
\Bigl\{\partial \psi^\al / \partial x_\mu, \  \al, 
\mu = {0,\ldots,3}\Bigr\}, \ \  R $
is a four-component func\-ti\-on. 

The complete group-theoretical analysis of the above system can be
carried out in the same way as it is done in Section 1.2.  We will
investigate symmetry properties of the important subclass of equations
of the form 
(\ref{1.3.4})\index{Poincar\'e-invariant second-order spinor equation}
\begin{equation}
\begin{array}{rcl}
\partial_\mu \partial^\mu + m^2 \psi &=& 
\Bigl\{F_1\Bigl(\partial_\mu(\bar \psi \psi ),\, \partial_\mu (\bar \psi
\ga_4 \psi),\, \bar \psi \psi,\, \bar \psi \ga_4 \psi \Bigr)\\[2mm]
& &\times \ga_\mu \partial_\mu + F_2 (\bar \psi,
\psi ) \Bigr\} \psi.
\end{array}
\label{1.3.5}
\end{equation}

In (\ref{1.3.5}) $F_1,\ F_2 $ are variable
$(4\times 4)$-matrices, $m= \mbox{\rm const} $.
\vspace{1.5mm}

\noindent
{\bf Theorem 1.3.1.}\ {\em  System of PDE (\ref{1.3.5}) is 
invariant under the Poincar\'e group with the generators 
(\ref{1.1.20}) iff
\begin{eqnarray}
F_1&=&g_1 + g_2 \ga_4 + (g_3 + g_4 \ga_4 )\ga \cdot v  \non\\
& &+ (g_5 + g_6 \ga_4 )\ga \cdot w  + g_7\ga \cdot v \ga w,
\label{1.3.6a}\\
F_2&=&f_1 + f_2 \ga_4,\label{1.3.6b}
\end{eqnarray}
where}
\begin{eqnarray*}
& &g_l = g_l (\bar \psi \psi,\, \bar \psi \ga_4 \psi,\, v \cdot v,\, 
v \cdot w,\, w \cdot w), \ \ l = {1,\ldots,7},\\
& &v_\mu = \partial_\mu(\bar \psi \psi),\quad
w_\mu = \partial_\mu (\bar \psi \ga_4 \psi ),\ \ \mu = {0,\ldots,3},\\
& &f_i = f_i (\bar \psi \psi,\, \bar \psi \ga_4 \psi),\ i=1,2
\end{eqnarray*}
and $g_l, \ f_i$ are arbitrary smooth functions.

The proof is carried out with the help of the Lie method. First of all
we note that system (\ref{1.3.5}) admits the 4-parameter group of
translations (\ref{1.1.24a}). To obtain constraints on $F_1,\ F_2 $
providing invariance of system (\ref{1.3.5}) under the Lorentz group
$O(1,3) \subset P(1,3) $ we act with the first prolongation of the
operator $J_{0a} $ given in (\ref{1.1.20}) on the equation in question
and pass to the set of its solutions. This procedure yields a system
of determining equations for the matrix functions $ F_1,\ F_2 $.  The
system of PDEs for $ F_2 $ coincides with system (\ref{1.2.6}) whose
general solution is represented in the form (\ref{1.3.6b}).

On introducing the notations
\begin{displaymath}
Q_{0a} = v_0 \partial_{v_a} + v_a \partial_{v_0} 
+ w_0 \partial_{w_a} + w_a \partial_{w_0},\quad
v_\mu =\partial_\mu (\bar \psi \psi ), \quad
w_\mu = \partial_\mu (\bar \psi \ga_4 \psi )
\end{displaymath}
we rewrite the system of determining equations for $F_1$ in the
form (\ref{1.2.6}).

Expanding the $(4\times 4)$-matrix $F_1$ in the complete system
of the Dirac matrices\index{Dirac!matrices}
\begin{equation}
F_1 = A +B_\mu \ga^\mu + C_{\mu \nu } S^{\mu \nu} +
 D_\mu \ga_4 \ga^\mu + E \ga_4
\label{1.3.7}
\end{equation}
and substituting the expression obtained into (\ref{1.2.6}) we arrive
at the system of PDEs for the functions $ A,\ B_\mu,\ \ldots, E $ of
the type (\ref{1.2.8a})--(\ref{1.2.8d}). Its general solution is given
by the following formulae: 
\begin{equation}
\begin{array}{l}
A=g_1,\quad E=g_2,\quad
B_\mu = g_3 v_\mu  + g_5 w_\mu,\\[2mm]
D_\mu = g_4 v_\mu  + g_6 w_\mu,\quad
C_{\mu \nu }=g_7(v_\mu w_\nu - v_\nu w_\mu ),
\end{array}
\label{1.3.8}
\end{equation}
where $ g_1, \ g_2, \ldots, g_7 $ are arbitrary smooth functions of
the invariants of the group $O(1,3)$\ \
$ \bar \psi \psi,\  \bar \psi \ga_4 \psi,\  v \cdot
v, \ v \cdot w, \ w \cdot w $.

Substitution of (\ref{1.3.8}) into (\ref{1.3.7}) gives rise to formula
(\ref{1.3.6a}). The theorem is proved. $\rhd$
\vspace{1.5mm}

\noindent
{\bf Theorem 1.3.2. }\  {\em System of PDEs
(\ref{1.3.5}) is invariant under the
conformal group $C(1,3)$ with generators (\ref{1.1.22}) iff}
\begin{eqnarray}
F_1 &=& (1/3)\ga \cdot v (\bar \psi \psi )^{-1} + (h_1 + h_2 \ga_4)
       \Bigl\{\ga \cdot v (\bar \psi \psi )^{-1}\non\\ 
& &-\ga \cdot w (\bar \psi \ga_4\psi )^{-1}\Bigr\} +
\ga_4 (\bar \psi \psi )^{1/3} (h_3 + h_4 \ga_4 ),\label{1.3.9}\\
F_2 &=& (\bar \psi \psi )^{2/3} (\tilde f_1 + \tilde f_2 \ga_4 ),
\quad  m=0.\non
\end{eqnarray}

In (\ref{1.3.9})\ $h_1,\ldots, h_4 $ are arbitrary smooth
complex-valued functions of the invariants of the group $C(1,3)\
(\bar \psi \psi ) (\bar \psi\ga_4 \psi)^{-1}, \  
\{(\bar \psi \psi )^2 w \cdot w - 2v \cdot w (\bar \psi\psi)
(\bar \psi \ga_4 \psi)$
$+(\bar \psi \gamma_4\psi )^2 v \cdot v\} (\bar \psi \psi)^{-14/3} $ 
and $\tilde f_1$,\ $\tilde f_2$ are arbitrary smooth functions of $(\bar
\psi \psi)$ $\times (\bar \psi \ga_4 \psi)^{-1}$.
\vspace{1.5mm}

\noindent
{\em Proof.}$\quad$ According to Theorem 1.3.1, the necessary and
sufficient conditions for equation (\ref{1.3.5}) to be invariant under
the group $P(1,3) \subset C(1,3) $ are given by equalities
(\ref{1.3.6a}), (\ref{1.3.6b}). Acting by the first prolongation of
the generator of the group of special conformal transformations
$\theta_\mu K^\mu$,\ $\theta_\mu =\mbox{\rm const}$ on system of PDEs
(\ref{1.3.5}) with $F_1,\ F_2$ of the form (\ref{1.3.6a}),
(\ref{1.3.6b}) and passing to the set of its solutions we obtain the
system of PDEs for $A, \ B_1, \ldots, E, \ f_1, \ f_2$
\begin{eqnarray}
& & L_1 g_1 = 2g_1,\quad L_2 g_1 =L_3 g_1 = 0,\non\\
& &L_1 g_2 = 2g_2, \quad L_2 g_2 = L_3 g_2 = 0,\non\\
& &L_1 g_j= -6g_j,\quad L_2 g_j = L_3 g_j = 0, \ j = {3,\ldots,6},
\label{1.3.10}\\
& &z_1 g_3 +z_2 g_5 = 1/3, \quad z_1 g_4 + z_2 g_6 = 0,
\quad g_7=0,\non\\
& & (z_1 \partial_{z_1} +z_2 \partial_{z_2} 
- 2/3) f_i =0,\ i= {1,2}.\non
\end{eqnarray}

Here
\begin{eqnarray*}
& &L_1 = 6(z_1 \partial_{z_1} +z_2 \partial_{z_2}) +16(z_3
\partial_{z_3} + z_4 \partial_{z_4} + z_5 \partial_{z_5}),\\
& &L_2 = z_1 \partial_{z_5} + 2z_2 \partial_{z_4},
\quad
L_3 = z_2 \partial_{z_5} + 2z_1 \partial_{z_3},\\
& &z_1 = \bar \psi \psi,\quad z_2 = \bar \psi \ga_4 \psi,
\quad z_3 = v \cdot v,\quad z_4 = v \cdot w,
\quad z_5 = w \cdot w.
\end{eqnarray*}

System of the first-order PDEs (\ref{1.3.10}) is integrated in a
standard way, its general solution having the form 
\begin{eqnarray*}
& &g_1=z^{1/3}_1 h_3,\quad g_2 = z^{1/3}_1 h_4,
\quad
g_3 = (1/3)z^{-1}_1+h_1 z^{-1}_1,\\
& &g_5 = -z^{-1}_2 h_1,
\quad
g_4 = z^{-1}_1 h_2,
\quad
g_6 = -z^{-1}_2 h_2,\\
& &f_1 =z^{2/3}_1 \tilde f_1 (z_1 / z_2),
\quad
f_2 = z^{2/3}_1 \tilde f_2 (z_1 / z_2),
\end{eqnarray*}
where $h_1,\ h_2 $ are arbitrary smooth complex-valued functions of
$z_1 z^{-1}_2$,
\ $(z_1^2 z_5$  $+ z^2_2 z_3 - 2z_1 z_2 z_4 )z_1^{-14/3}$; 
\ $\tilde f_i \in C^1({\R}^1,\  {\C}^1 )$.

Substitution of the above results into (\ref{1.3.6a}), (\ref{1.3.6b})
yields (\ref{1.3.9}). The theorem is proved. $\rhd$
\vspace{1.5mm}

\noindent
{\bf Consequence 1.3.1.}\ {\em System of PDEs
\begin{equation}
\{\partial_\mu \partial^\mu - F(\bar \psi, \psi )\} \psi = 0,
\label{1.3.11}
\end{equation}
where $F$ is a variable $(4\times 4)$-matrix, 
is not invariant with respect
to the group $C(1,3)$.}

The proof follows from the fact that the class of
conformally-invariant equations (\ref{1.3.5}), (\ref{1.3.9}) does not
contain equations of the form (\ref{1.3.11}).

If we put in (\ref{1.3.9}) $ h_1 = h_2 = h_3 = h_4 = 0$, $ \tilde
f_1 = -\lambda^2 = \mbox{\rm const}$, $\tilde f_2 = 0 $, then the
conformally-invariant second-order PDE
\index{Conformally-invariant second-order spinor equation}
\begin{equation}
\Bigl\{\partial_\mu \partial^\mu - 
(1/3)(\bar \psi \psi)^{-1}\Bigl(\ga_\mu \partial_\mu
(\bar \psi \psi )\Bigr) \ga_\nu \partial_\nu +
\lambda^2 (\bar \psi \psi)^{2/3}\Bigr\}\psi = 0
\label{1.3.12}
\end{equation} 
suggested in \cite{103} is obtained. The direct verification shows
that any solution of the Dirac-G\"ursey equation satisfies PDE
(\ref{1.3.12}). That is why  equation (\ref{1.3.12}) as well as the
Dirac-G\"ursey equation can be used in conformally-invariant quantum
field theories to describe a massless particle with the spin $s =
1/2$.  \vspace{10mm}

\noindent
{\large\bf 1.4. Symmetry of systems of nonlinear equations\label{s1.4} 
\vspace{1.5mm}

\noindent
\phantom{\large\bf 1.4. }for spinor, vector and scalar fields}
\markboth{Chapter 1. SYMMETRY OF NONLINEAR SPINOR EQUATIONS}
{1.4. Symmetry of systems of nonlinear equations}
\def\theequation{1.\arabic{section}.\arabic{equation}}
\setcounter {section} {4}
\setcounter {equation}{0}
\vspace{7mm}

\noindent
It is well-known (see, for example, \cite{94}) that the classical
electrodynamics equations\index{Classical electrodynamics equati\-ons}
\begin{equation}
\begin{array}{l}
       (i \g_\mu \p_\mu - e \g_\mu A^\mu ) \psi = 0,\\[2mm]
      \p_\mu \p^\mu A_\nu - \p^\nu \p_\mu A_\mu = -e \bar\psi \g_\nu
      \psi, 
\end{array}
\label{1.4.1}
\end{equation}
where $A_\mu (x) $ is the vector-potential of electro-magnetic field,
$e = \mbox{\rm const}$,\ $\mu, \nu = {0,\ldots,3}$, are invariant
under the conformal group $C(1,3)$ having the following generators:
\begin{eqnarray}
&&\begin{array}{l}
P_\mu = \p^\mu, \quad
J_{\mu \nu } = x_\mu P_\nu - x_\nu P_\mu +
A_\mu \partial_{A^\nu} - A_\nu \partial_{A^\mu} \\[2mm]
\phantom{P_\mu =}-(1/2)\{\g_\mu \g_\nu \psi \}^\al  \p_{\psi^\al} +
(1/2)\{ \bar{\psi} \g_\mu \g_\nu\}^\al \p_{\bar\psi^\al},
 \ \mu \ne \nu,
\end{array}
\label{1.4.2}\\[2mm]
&&\ D = x_\mu \p_\mu - (3/2) (\psi^\al \p_{\psi^\al} + \bar\psi^\al
\p_{\bar\psi^\al }) - A_\mu \p_{A_\mu },\label{1.4.3}\\[2mm]
&&\begin{array}{l}
K_\mu = 2x_\mu D - (x \cdot x)\p^\mu - x_\mu
(A_\nu \p_{A_\nu} -\psi^\al \p_{\psi^\al} - \bar\psi^\al
\p_{\bar\psi^\al})\\[2mm] 
\phantom{K_\mu =} -\{\g_\mu \g \cdot x \psi \}^\al \p_{\psi^\al} 
-\{\bar\psi \g \cdot x \g_\mu \}^\al\p_{\bar\psi^\al} +
2A_\mu x_\nu \p_{A_\nu}\\[2mm]
\phantom{K_\mu =}  - 2A \cdot x \p_{A^\mu}.
\end{array}
\label{1.4.4}
\end{eqnarray}

In formulae (\ref{1.4.2})--(\ref{1.4.4}) $\p_{A_\mu} = \p /
\p A_\mu, \ \p_{\psi^\al} = \p / \p \psi^\al,\ \p_{\bar\psi^\al } 
= \p / \p \bar\psi^\al$; \
$\{\Psi\}^\al $ means the $\al$-th component of the spinor $\Psi$; \
$\mu, \nu, \al = {0,\ldots,3} $.

Let us note that the operators $K_\mu $ (\ref{1.4.4}) generate a
4-parameter group of special conformal 
transformations\index{Conformal!transformations}
\begin{equation}
\begin{array}{l}
x'_\mu= (x_\mu - \theta_\mu x \cdot x) \s^{-1} (x),\\[2mm]
\psi'(x')= \s(x)(1-\g \cdot \theta\, \g \cdot x) \psi(x),\\[2mm]
 A_\mu ' (x') = \{ \s (x) g_{\mu \nu } +
 2(x_\mu \theta_\nu - x_\nu \theta_\mu  + 2 \theta \cdot x \theta_\mu
 x_\nu \\[2mm] 
 \phantom{A_\mu'(x') =} - x \cdot x \theta_\mu \theta_\nu - \theta
 \cdot \theta x_\mu x_\nu )\} A^\nu (x), 
\end{array}
\label{1.4.5}
\end{equation}
where $\s(x) = 1 - 2 \theta \cdot x + (\theta \cdot \theta) (x \cdot
x)$. 

In \cite{86,94} another conformally-invariant system of PDEs for
spinor and vector fields
\begin{equation}
\begin{array}{l}
(i \g_\mu \p_\mu - e \g_\mu A^\mu ) \psi (x) = 0,\\[2mm]
\p_\mu \p^\mu A_\nu - \p^\nu \p_\mu A_\mu =
\lbd A_\nu (A \cdot A)
\end{array}
\label{1.4.6}
\end{equation}
was suggested. A conjecture arises that there exist more
general systems of nonlinear equations
\begin{equation}
\begin{array}{l}
i \g_\mu \p_\mu\psi - F(\bar\psi,\, \psi,\, A) = 0,\\[2mm]
\p_\mu \p^\mu A_\nu - \p^\nu \p_\mu A_\mu = R_\nu
(\bar\psi,\, \psi,\,  A)
\end{array}
\label{1.4.7}
\end{equation}
invariant under the conformal group. 

In the present section we solve the problem of group-theoretical
classification of systems of PDEs (\ref{1.4.7}). Namely, we describe
all functions $ F = (F^0,\, F^1,\, F^2,\, F^3 )^T$, \ $ R_\mu $ such
that system (\ref{1.4.7}) is invariant with respect to the groups
$P(1,3), \ \wid P(1,3)$, \ $C(1,3)$.

In addition, symmetry analysis of systems of nonlinear equations for
spinor and scalar fields  
\begin{equation}
\begin{array}{l}
i \g_\mu \p_\mu \psi - F(u^{*},\, u,\, \bar\psi,\, \psi) = 0,\\[2mm]
\p_\mu \p^\mu u - H(u^{*},\, u,\, \bar\psi,\, \psi) = 0;
\end{array}
\label{1.4.8}
\end{equation}
vector and scalar fields
\begin{equation}
\begin{array}{l}
\p_\mu \p^\mu u - H(u^{*},\, u,\, A) = 0,\\[2mm]
\p_\mu \p^\mu A_\nu - \p^\nu \p_\mu A_\mu = R_\nu (u^{*},\, u,\, A)
\end{array}
\label{1.4.9}
\end{equation}
is carried out.

In (\ref{1.4.8}), (\ref{1.4.9}) $F = (F^0,\, F^1,\, F^2,\, F^3 )^T$;
\ $F^\mu, \ H, \ R_\mu $ are some
smooth functions; $u(x) \in C^2 ({\R}^4, {\C}^1)$.
\vspace{1.5mm}

\noindent
{\bf Theorem 1.4.1.}\ {\em System (\ref{1.4.8}) is invariant under
\vspace{1.5mm}

\noindent
1) the Poincar\'e group iff
\begin{equation}
F = (f_1 + f_2 \g_4)\psi,
\quad
H = h(u^{*},\,  u,\, 
\bar\psi \psi,\,  \bar\psi \g_4 \psi),
\label{1.4.11}
\end{equation}
where $f_1, \ f_2 $ are arbitrary smooth complex-valued functions
of $u^{*}$,\ $u$,\ $\bar\psi\psi$,\ $\bar\psi\g_4\psi $;
\vspace{1.5mm}

\noindent
2) the extended Poincar\'e group $\wid P(1,3) = P(1,3)
\mbox{$\stimes$} D(1)$, where $D(1)$ is the one-parameter group of
scale transformations\index{Scale transformation group}
\begin{equation}
x'_\mu = x_\mu e^\theta,\quad
u' = u e^{-k_2 \theta},\quad
\psi' = \psi e^{-k_1 \theta},\quad
\theta, \ k_1, \ k_2 = \mbox{\rm const},
\label{1.4.12}
\end{equation}
iff $F, \ H $ are given by (\ref{1.4.11}) with
\begin{eqnarray}
& &f_i = (\bar\psi \psi)^{1/2 k_1} \ti f_i (w_1,\,  w_2,\,  w_3),
\quad
h = (u^{*} u)^{1/k_2} u \ti h (w_1,\,  w_2,\,  w_3),\non\\
& &w_1 = u/u^{*}, \quad
w_2 = u^{2k_1} (\bar\psi \psi)^{-k_2},
\quad
w_3 = u^{2k_1} (\bar\psi\gamma_4 \psi )^{-k_2},\label{1.4.13}\\
& &\{\ti f_i, \ \ti h\} \subset
C^1 ({\C}^3, {\C}^1),\ \ i= 1,2;\non
\end{eqnarray}
3) the conformal group $ C(1,3) = P(1,3) \mbox{$\stimes$} D(1)
\mbox{$\stimes$} K(1,3) $, where $D(1) $ is given by (\ref{1.4.12})
with $k_1 = 3/2, \ k_2 = 1$ and the 4-parameter group of special
conformal transformations $K(1,3) $ 
has the form
\begin{eqnarray}
& &x'_{\mu} = (x_\mu - \theta_\mu x \cdot x) \s^{-1} (x),\non\\
& &\psi' (x') = \s(x) (1 - \g \cdot \theta \g \cdot x ) \psi (x),
\label{1.4.14}\\
& &u'(x') = \s (x) u(x),\non
\end{eqnarray}
iff $ F, \ H $ are given by formulae (\ref{1.4.11}), (\ref{1.4.13})
with $k_1 = 3/2, \ k_2 = 1 $;
\vspace{1.5mm}

\noindent
4) the group $C(1,3) \otimes U(1)$, where $U(1)$ is the
one-parameter 
group of gauge transformations
\begin{displaymath}
x'_\mu = x_\mu,\quad
\psi'(x) = e^{i \theta} \psi(x),
\quad
u'(x) = e^{i \theta} u(x),
\ \ \theta \in {\R}^1,
\end{displaymath}
iff}
\begin{eqnarray}
& &F = (\bar\psi \psi)^{1/3} \{\ti f_1 (z_1,\,  z_2) + \g_4 \ti f_2
(z_1,\,  z_2)\} \psi,\non\\
& &H = |u|^2 u \ti h (z_1,\,  z_2),
\quad
\{\ti f_1, \ \ti f_2, \ \ti h\} \subset C^1 ({\R}^2, {\C}^1),
\label{1.4.15}\\
& &z_1 = \bar\psi \psi |u|^{-3},\quad
z_2 = \bar\psi \g_4 \psi |u|^{-3}.\non
\end{eqnarray}

The proof is carried out with the help of the Lie method.
Acting on system of PDEs (\ref{1.4.8}) by the first prolongation of
the operator $J_{0a}$ (\ref{1.1.20}) and passing to the set of
its solutions we get necessary and sufficient conditions of
Lorentz invariance of system (\ref{1.4.8}) in the form
\begin{equation}
Q_{0a} \Phi - (1/2) [\Phi,\,  \g_0 \g_a] = 0,
\quad
Q_{0a} H = 0, \ \
a =  {1,2,3},
\label{1.4.16}
\end{equation}
where
$
Q_{0a} = -(1/2) \{\g_0 \g_a \psi \}^\al\partial_{\psi^\al} +
(1/2)\{\bar\psi \g_0 \g_a \}^\al \partial_{\bar\psi^\al},
$
$\Phi = \Phi (u^{*},\,
u,\,  \bar\psi,\, \psi) $ is a $(4\times 4)$-matrix (we have
represented the four-component function $F$ in  the form $ \Phi
\psi$). 

Since the first equation of system (\ref{1.4.16}) coincides with
(\ref{1.2.6}) and the second one with (\ref{1.2.8a}), we can write 
down their general solutions using the results obtained in Section
1.2.  According to (\ref{1.2.4}), (\ref{1.2.9}) the general solution
of system of PDEs (\ref{1.4.16}) has the form (\ref{1.4.11}). Taking
into account the fact that system (\ref{1.4.8}) is invariant under the 
4-parameter group of translations (\ref{1.1.24a}) we arrive at the
assertion 1 of Theorem 1.4.1.

Acting on system of PDEs (\ref{1.4.8}), (\ref{1.4.11}) by the first 
prolongation of the generator of the group of scale transformations 
\begin{displaymath}
D = x_\mu \p_\mu
- k_1 \psi^\al \p_{\psi^\al} - k_1 \bar\psi^\al \p_{\bar\psi^\al} -
k_2 u \p_u - k_2 u^{*} \p_{u^*}
\end{displaymath}
and passing to the set of its
solutions we get the following system of PDEs for $f_1, \ f_2, \ h $:
\begin{eqnarray*}
& &k_2(\rho_1 f_{i \rho_1} + \rho_2 f_{i \rho_2}) +
 2k_1(\rho_3 f_{i \rho_3} + \rho_4 f_{i \rho_4}) = 1,
\ \ i = {1,2},\\
& &k_2 (\rho_1 h_{\rho_1} + \rho_2 h_{\rho_2}) +
 2k_1 (\rho_3 h_{\rho_3} + \rho_4 h_{\rho_4}) = 2,\\
& &f_{i \rho_n} = \p f_i / \p \rho_n,
\quad
h_{\rho_n} =
\p h / \p \rho_n, \ \
n = {1,\ldots,4},
\end{eqnarray*}
where $\rho_1 = u^{*},\ \rho_2 = u, \ \rho_3 = \bar\psi \psi, \ 
\rho_4 = \bar\psi \g_4 \psi $ is a complete system of
functionally-independent invariants of the group $P(1,3)$. General
solution of the above system is given by the formulae (\ref{1.4.13}),
$w_1, \ w_2, \ w_3 $ being a complete system of
functionally-independent invariants of the extended Poincar\'e group.
Since the conformal group contains the group $\wid P(1,3)$, the
requirement of\ $C(1,3)$-invariance of system of PDEs (\ref{1.4.8})
leads to formulae (\ref{1.4.11}), (\ref{1.4.13}) under $k_1 = 3/2,\ 
k_2 = 1 $.  The sufficiency of assertion 3 is established by
direct verification.

To select from the class of conformally-invariant equations of the
form (\ref{1.4.8}) the equations which admit the group $U(1)$ we act
with the first prolongation of the generator of this group on system
(\ref{1.4.8}) and pass to the set of its solutions. As a result, we
have

\begin{eqnarray*}
& &2w_1 \ti f_{i w_1} + 3w_2 \ti f_{i w_2}+3w_3 \ti f_{i w_3} = 0,
\ \
i = {1,2},\\
& &2w_1 \ti h_{w_1} + 3w_2 \ti h_{w_2}+3w_3 \ti h_{w_3} = 0.
\end{eqnarray*}

General solution of the above equations is represented in the form
\begin{displaymath}
\ti f_i = \ti f_i (w^{3/2}_1 w^{-1}_2,\, w^{3/2}_1 w^{-1}_3 ),\quad
\ti h = \ti h (w^{3/2}_1 w^{-1}_2,\, w^{3/2}_1 w^{-1}_3 ).
\end{displaymath}

Putting
\begin{displaymath}
w_1 = u(u^*)^{-1},
\quad
w_2 = u^3 (\bar\psi \psi)^{-1},
\quad
w_3 = u^3 (\bar\psi \g_4 \psi)^{-1}
\end{displaymath}
yields formulae (\ref{1.4.15}). The theorem is proved. $\rhd$
\vspace{1.5mm}

\noindent
{\bf Note 1.4.1.}\  In \cite{61} a model for description of
interaction of spinor and real-valued scalar fields based on
the relativistic Hamilton equation
\begin{equation}
\begin{array}{l}
i \g_\mu \p_\mu \psi - F(u,\, \bar\psi,\, \psi ) = 0,\\[2mm]
(\p_\mu u) (\p^\mu u) = H(u,\, \bar\psi,\, \psi)
\end{array}
\label{1.4.17}
\end{equation}
was suggested. Using the Lie method we can prove that system of
PDEs (\ref{1.4.17}) admits the Poincar\'e group iff
\begin{equation}
\begin{array}{l}
F = \{f_1 (u,\,  \bar\psi \psi,\,  \bar\psi \g_4 \psi ) + \g_4 f_2
(u,\,  \bar\psi \psi,\,  \bar\psi \g_4 \psi) \}\psi,\\[2mm]
H =  h (u,\,  \bar\psi \psi,\,  \bar\psi \g_4 \psi).
\end{array}
\label{1.4.18}
\end{equation}
Provided
\begin{displaymath}
f_i = (\bar\psi \psi)^{1/2k_1} \ti f_i (w_1,\,  w_2),
\quad h = u^{2(k_2 +1) /k_2 } \ti h (w_1,\,  w_2),\ \ i = 1,2,
\end{displaymath}
where $w_1 = u^{2k_1}
(\bar\psi \psi)^{-k_2}, \  w_2 = u^{2k_1} (\bar\psi \g_4 \psi)^{-k_2}
$, system of PDEs (\ref{1.4.17}) is invariant with respect to the
extended Poincar\'e group.

The next two theorems are given without proof.
\vspace{1.5mm}

\noindent
{\bf Theorem 1.4.2.} \ {\em System of nonlinear equations (\ref{1.4.7})
is invariant under
\vspace{1.5mm}

\noindent
1) the Poincar\'e group with generators (\ref{1.4.2}) iff
\begin{equation}
\begin{array}{l}
F(\bar\psi,\,  \psi,\,  A) = \{\g \cdot A f_1 + \g_4
\g \cdot A f_2 + f_3 + \g_4 f_4 \} \psi,\\[2mm]
R_\mu (\bar\psi,\,  \psi,\,  A) = A_\mu g_1 + \bar\psi \g_\mu \psi g_2  
+\bar\psi \g_4 \g_\mu \psi g_3 + \psi^T \g_0 \g_2 \g_\mu \psi g_4,
\end{array}
\label{1.4.19}
\end{equation}
where $f_i $ are arbitrary complex-valued functions and $h_i$
are arbitrary real-valued functions of
\begin{displaymath}
\bar\psi \psi, \quad \bar\psi \g_4 \psi, \quad
\bar\psi \g \cdot A \psi, \quad \bar\psi \g_4 \g \cdot A \psi,
\quad
\psi^T \g_0 \g_2 \g \cdot A \psi,\quad A \cdot A;
\end{displaymath}
2) the extended Poincar\'e group $\wid P(1,3)$ with generators
(\ref{1.4.2}) and 
\begin{displaymath}
D=x_\mu\p_\mu - k_1\psi_\al\p_{\psi_\al} -
k_1\psi_\al\p_{\psi_\al} - k_2A_\mu\p_{a_\mu},\ \ \{k_1, k_2\} \subset\R^1
\end{displaymath}
iff functions $F, \ R_\mu $ are given by formulae
(\ref{1.4.19}), where 
\begin{eqnarray}
& &f_i = (A \cdot A)^{(1-k_2)/2k_2}\ti f_i,\ \ i = 1,2,\non\\
& &f_j = (\bar\psi \psi)^{1/2k_1} \ti f_j,
\ \
j ={3,4},\quad
g_1 = (A \cdot A)^{1/k_2} \ti g_1,\label{1.4.20}\\
& &g_i = (\bar\psi \psi)^{(k_2 - 2k_1 + 2)/2k_1} \ti g_i,
\ \ i= {2,3,4},\non
\end{eqnarray}
$\ti f_1,\ldots, \ti f_4,\ \ti g_1,\ldots, \ti g_4 $ being
arbitrary smooth functions of
\begin{eqnarray*}
& &(\bar\psi \psi)(\bar\psi \g_4 \psi)^{-1},
\quad
(\bar\psi \psi)^{k_2}
(A \cdot A)^{-k_1},
\quad
(\bar\psi \psi)^{2k_1 + k_2} (\bar\psi \g \cdot A \psi )^{-2k_1},\\
& &(\bar\psi \psi)^{2k_1 + k_2} (\bar\psi \g_4 \g \cdot A
\psi)^{-2k_1}, 
\quad
(\bar\psi \psi )^{2k_1 + k_2} (\psi^T \g_0 \g_2 \g \cdot A \psi )
^{-2k_1};
\end{eqnarray*}
3) the group $C(1,3)$ with generators (\ref{1.4.2})--(\ref{1.4.4}) iff
the functions $F,\ R_\mu $ are given by (\ref{1.4.19}), (\ref{1.4.20})
under $k_1 = 3/2,\   k_2 = 1$;
\vspace{1.5mm}

\noindent
4) the group $C(1,3) \otimes U(1)$, where $U(1)$ is the group of gauge 
transformations
\begin{equation}
\begin{array}{l}
       x'_\mu = x_\mu, \quad
       \psi' = \psi e^{i \lbd \theta(x)},\\[2mm]
       A'_\mu = A_\mu + \p^\mu \theta (x),\quad
       \theta (x) \in C^3 ({\R}^4, {\R}^1),
\end{array}
\label{1.4.21}
\end{equation}
iff
\begin{equation}
\begin{array}{l}
F (\bar\psi,\, \psi,\, A) = \{ \lbd \g \cdot A + f_1 (\bar\psi \psi
(\bar\psi \g_4 \psi)^{-1}) + \g_4 f_2 (\bar\psi \psi (\bar\psi \g_4
\psi)^{-1}) \} \psi,\\[2mm] 
R_\mu(\bar\psi,\,  \psi,\,  A) = \bar\psi \g_\mu \psi g_1 (\bar\psi
\psi (\bar\psi \g_4 \psi)^{-1}) + \bar\psi \g_4 \g_\mu \psi g_2
(\bar\psi \psi (\bar\psi \g_4 \psi)^{-1}), 
\end{array}
\label{1.4.22}
\end{equation}
where} $f_i \in C^1 ({\R}^1, {\C}^1),\ g_i \in C^1 ({\R}^1, {\R}^1), \ 
 i = 1,2, \ \lbd = \mbox{\rm const}$.
\vspace{1.5mm}

\noindent
{\bf Consequence 1.4.1.}\  {\em On the set of solutions of system of 
PDEs (\ref{1.4.7}), (\ref{1.4.22}) an infinite-dimensional
representation of the Lie algebra $AC(1,3)$ is realized, basis
elements of the algebra having the form
\begin{equation}
\begin{array}{l}
\wid  P_\mu =P_\mu,
\quad
\wid J_{\mu \nu} = J_{\mu \nu },
\quad
\wid D = D +  i \lbd (\psi^\al \p_{\psi^\al} -
\bar\psi^\al \p_{\bar\psi^\al}),\\[2mm]
\wid K_\mu = K_\mu + \p_{A_\mu} +
i \lbd x_\mu (\psi^\al \p_{\psi^\al} - \bar\psi_\al
\p_{\bar\psi^\al}),
\label{1.4.23}
\end{array}
\end{equation}
where the operators $P_\mu,\ J_{\mu \nu}, \ D,\ K_{\mu}$
are given by} \/(\ref{1.4.4}).

The proof is reduced to verification of the commutation
relations of the algebra $AC(1,3)$ if we note that
the operators
\begin{equation}
\begin{array}{l}
Q_1 = i \lbd (\psi^\al \p_{\psi^\al} 
- \bar\psi^\al \p_{\bar\psi^{\al}}),\\[2mm]
Q_{2 \mu} = i \lbd x_\mu (\psi^\al \p_{\psi^\al} -
\bar\psi^\al \p_{\bar\psi^\al}) + \p_{A_\mu},\ \
\mu = {0,\ldots,3}
\label{1.4.24}
\end{array}
\end{equation}
generate transformation groups of the form (\ref{1.4.21}). $\rhd$

Thus, system (\ref{1.4.7}), (\ref{1.4.22}) possesses a dual conformal
symmetry. To fix a definite representation of $AC(1,3)$ it is
necessary to impose an additional constraint on the vector field
$A_\mu (x)$.  In \cite{55} the nonlinear equation
\begin{equation}
\p_\mu (A_\mu A \cdot A) = 0
\label{1.4.25a}
\end{equation}
invariant under the algebra (\ref{1.4.4}) was suggested. Since PDE
(\ref{1.4.25a}) is not invariant under transformation groups generated
by the operators $ Q_{2 \mu} $ from (\ref{1.4.24}), it does not admit
the 4-parameter group with generators $\wid K_\mu $ from (\ref{1.4.23}).
Consequently, system of PDEs (\ref{1.4.7}), (\ref{1.4.22}),
(\ref{1.4.25a}) is invariant under the conformal 
algebra\index{Conformal!algebra} (\ref{1.4.23}). 

Analogously, using results obtained in the paper \cite{94} we
conclude that system (\ref{1.4.7}), (\ref{1.4.22}) supplemented by the 
additional condition
\begin{equation}
\p_\mu A_\mu - 2 A \cdot A = 0
\label{1.4.25b}
\end{equation}
is invariant under the conformal algebra (\ref{1.4.23}) and is
not invariant under the algebra (\ref{1.4.4}).
\vspace{1.5mm}

\noindent
{\bf Theorem 1.4.3.}\ {\em System of nonlinear PDEs (\ref{1.4.9}) is
  invariant under
\vspace{1.5mm}

\noindent
1) the Poincar\'e group iff
\begin{displaymath}
H = h(u^{*},\, u,\, A\cdot A),
\quad
R_\mu = A_\mu g(u^*,\, u,\, A\cdot A),
\end{displaymath}
where $ h \in C^1 ({\C}^2 \times {\R}^1, {\C}^1),
\ \ g \in C^1({\C}^2 \times {\R}^1, {\R}^1);$
\vspace{1.5mm}

\noindent
2) the extended Poincar\'e group $ \wid P(1,3) = P(1,3)
\mbox{$\stimes$} D(1)$, where $D(1)$ is a one-parameter group of scale  
transformations\index{Scale transformation group}
\begin{equation}
\begin{array}{l}
x'_\mu = x_\mu e^\theta,
\quad
A'_\mu = A_\mu e^{-k_1 \theta},\\[2mm]
u' = u e^{-k_2 \theta},
\quad
u^{*\prime} = u^* e^{-k_2 \theta},\ \ \theta \in {\R}^1,
\label{1.4.26}
\end{array}
\end{equation}
iff
\begin{equation}
\begin{array}{l}
H = |u|^{2/k_2} u h\Bigl(u^* u^{-1},\, |u|^{-2k_1} (A \cdot
A)^{k_2}\Bigr),\\[2mm] 
R_\mu = (A \cdot A)^{1/k_1} g\Bigl(u^*  u^{-1},\, |u|^{-2k_1}
(A \cdot A)^{k_2}\Bigr) A_\mu;
\label{1.4.27}
\end{array}
\end{equation}
3) the conformal group $C(1,3) = P(1,3) \mbox{$\stimes$} D(1)
\mbox{$\stimes$} K(1,3) $, 
where $D(1)$ is the group (\ref{1.4.26}) with $k_1=1$,\ $k_2=1$
and $K(1,3)$ is the 4-parameter group of special conformal  
transformations\index{Conformal!transformations}
\begin{eqnarray}
& &x'_\mu = (x_\mu - \theta_\mu x \cdot x) \s^{-1}(x),
\quad u' = \s(x)u,\non\\
& &A'_\mu = \{\s (x) g_{\mu \nu} + 2(x_\mu \theta_{\nu} 
- x_\nu \theta_\mu + 2 \theta \cdot x  \theta_\mu x_\nu\label{1.4.28}\\
& &\quad - x \cdot x \theta_\mu \theta_\nu -
\theta \cdot \theta x_\mu x_\nu )\} A^\nu,\non
\end{eqnarray}
where $s (x) = 1 - 2 \theta \cdot x + \theta \cdot \theta x \cdot x,
\ \theta_\mu = \mbox{\rm const},\ \mu = {0,\ldots,3}$,
iff $H, \ R_\mu $ are of the form (\ref{1.4.27}) under $k_1 = k_2 = 1$;
\vspace{1.5mm}

\noindent
4) the group $C(1,3) \otimes U(1)$, where $U(1)$ is the group of gauge
transformations
\begin{displaymath}
x'_\mu = x_\mu,
\quad
A'_\mu =A_\mu,
\quad
\ u' = u e^{i \theta},
\quad
\theta \in {\R}^1,
\end{displaymath}
iff }
\begin{eqnarray*}
& &H = |u|^2 u h (|u|^{-2} A \cdot A),\quad
R_\mu = A_\mu A \cdot A g (|u|^{-2} A \cdot A),\\
& &h \in C^1 ({\R}^1, {\C}^1),
\quad
g \in C^1 ({\R}^1, {\R}^1).
\end{eqnarray*}

Thus, using the symmetry selection 
principle\index{Symmetry!selection principle} we narrow substantially 
clas\-ses of physically admissible nonlinear generalizations of the
Maxwell-Dirac, Dirac-d'Alembert and Maxwell-d'Alembert equations.
\vspace{10mm}

\noindent
{\large\bf 1.5. Conditional symmetry and reduction of partial
\vspace{1.5mm}

\noindent
\phantom{\large\bf 1.5. }differential equations\label{s1.5}} 

\markboth{Chapter 1. SYMMETRY OF NONLINEAR SPINOR EQUATIONS}
{1.5. Conditional symmetry and reduction of PDEs}
\setcounter {section} {5}
\setcounter {equation}{0}
\vspace{7mm}

\noindent
Analyzing already known methods of construction of exact solutions of
nonlinear partial differential equations we come to conclusion that a
majority of them is based on the idea of narrowing the set of
solutions, i.e.,\ selecting from the whole set of solutions specific
subsets which admit analytic description. To implement this idea we
have to impose some additional constraints (equations) on the set of
solutions of the equation under consideration selecting such subsets.
Clearly, additional equations are supposed to be simpler than the
initial one. Supplementing the initial equation with additional
conditions we come, as a rule, to an over-determined system of PDEs.
So there arises a problem of investigating the matter of its
compatibility.

To clarify the above points we will consider an instructive
example. Let  
\begin{equation}
U(x_1,\, u,\, \mathop{u}\limits_{\scriptscriptstyle 1},\,
\mathop{u}\limits_{\scriptscriptstyle 2})=0   
\label{1.5.0z}
\end{equation}
be a second-order PDE with two independent variables $x_0,\ x_1$ which 
does not depend explicitly on $x_0$.

Since coefficients of PDE (\ref{1.5.0z}) do not contain the variable
$x_0$, substitution of the expression 
\begin{equation}
u=\vp(x_1)
\label{1.5.1z}
\end{equation}
into (\ref{1.5.0z}) results in a differential equation containing
$x_1,\ \vp,\ \dot\vp,\ \ddot\vp$ only, i.e.,\ 
\begin{equation}
\wid U(x_1,\, \vp,\, \dot\vp,\, \ddot\vp)=0.
\label{1.5.2z}
\end{equation}

Consequently, using the fact that PDE (\ref{1.5.0z}) does not contain
the variable $x_0$ we {\em reduce} \/it to an ODE assuming that a
particular solution also does not depend on $x_0$.

But from the group-theoretical point of view the independence of PDE
(\ref{1.5.0z}) of $x_0$ means that it is invariant under the
one-parameter translation group with respect to the variable $x_0$
\begin{equation}
x_0^\prime=x_0+\theta,\quad x_1^\prime=x_1,\quad u^\prime=u,\ \ 
\theta\in \R^1
\label{1.5.3z}
\end{equation}
having the generator $X=\p_{x_0}$. And what is more, formula
(\ref{1.5.1z}) defines the most general manifold in the
three-dimensional space of variables $x_0,\ x_1,\ u$ which is
invariant with respect to the above group. Expression (\ref{1.5.1z}) 
is called a solution (an Ansatz\index{Ansatz}) invariant under the 
one-parameter group (\ref{1.5.3z}).  
 
The above said can be summarized in the form of the following
statement: a solution invariant under the group of translations
(\ref{1.5.3z}) reduces PDE admitting the same group to ODE. When
generalized to the case of an arbitrary admissible one-parameter
group, this statement plays a key role in applications of Lie
transformation groups to construction of exact solutions of
mathematical physics equations.

The way for obtaining an invariant solution is entirely algorithmic.
Since we are looking for a manifold $u=f(x_0,x_1)$ which does not
contain explicitly the variable $x_0$ (is invariant with respect to
the group (\ref{1.5.3z})) we should require that $\p f/\p x_0 = 0$.
Consequently, to find a solution of PDE (\ref{1.5.0z}) invariant under
the group (\ref{1.5.3z}) it is necessary to solve an over-determined
system of PDEs
\begin{displaymath}
U(x_1,\, u,\, \mathop{u}\limits_{\scriptscriptstyle 1},\,
\mathop{u}\limits_{\scriptscriptstyle 2})=0,\quad u_{x_0}=0. 
\end{displaymath}

We have paid so much attention to a very simple example,
since it gives an adequate illustration to ideas of the symmetry
reduction method\index{Symmetry!reduction} pioneered by Sophus Lie. 
Moreover, a general case of PDE
\begin{equation}
U(x_0,\, x_1,\, u,\, \mathop{u}\limits_{\scriptscriptstyle 1},\,
\mathop{u}\limits_{\scriptscriptstyle 2})=0 
\label{1.5.4z}
\end{equation}
invariant under a one-parameter transformation group having a
generator $X=
\linebreak \xi_0(x,u)\p_{x_0} + \xi_1(x,u)\p_{x_1} + \eta(x,u)\p_u$
is reduced to the particular case considered above. Indeed, it is
known from the general theory of PDEs that there is a change of
variables 
\begin{displaymath}
\tilde x_0=F_0(x,u),\quad \tilde
x_1=F_1(x,u),\quad \tilde u=G(x,u)
\end{displaymath}
transforming the operator $X$ to the form $X^\prime=\p_{\tilde x_0}$.
Consequently, PDE (\ref{1.5.4z}) after being rewritten in the
variables $\tilde x,\ \tilde u$ is invariant under the one-parameter
transformation group with the generator $X^\prime =\p_{\tilde x_0}$,
i.e.,\ under the group (\ref{1.5.3z}). According to the above proved a 
substitution $\tilde u=\vp (\tilde x_1)$ reduces the equation
transformed to ODE for a function $\vp$. Hence, we conclude that the
substitution $F_0(x,u)$ $=$ $\varphi\Bigl(F_1(x,u)\Bigr)$ reduces the
initial equation to ODE.

Thus, given a one-parameter transformation group admitted by partial
differential equation (\ref{1.5.4z}), we can reduce it to an ODE by means
of a substitution of a special form (invariant solution or Ansatz) 
\begin{equation}
u=f\Bigl(x,\, \vp(\omega(x,u))\Bigr),
\label{1.5.5z}
\end{equation}
where $f,\ \omega$ are some functions determined by the form of the
generator of the group. A natural question arises: do invariant
solutions exhaust the set of substitutions (\ref{1.5.5z}) reducing
given PDE to an ODE? A negative answer to this question has led us to
the notion of {\em conditional symmetry}\index{Conditional!symmetry} 
of partial differential equations. 

The notion and terminology of conditional symmetry of PDEs was
introduced for the first time in \cite{62,63,75} and developed in a
series of papers and monographs \cite{65.2,65.3},
\cite{68}--\cite{68.2}, \cite{77.1,79,69,81,81.1,81.2,89,95},
\cite{102}--\cite{106.2}, \cite{173.1,211,212} (see also
\cite{24.0,37.1,142.1,162}). The principal idea of conditional
symmetry of PDE is illustrated by the following example. The equation
\begin{displaymath}
U(x_1,\, u,\, \mathop{u}\limits_{\scriptscriptstyle 1},\,
\mathop{u}\limits_{\scriptscriptstyle 2}) 
+ V(x_0,\, x_1,\, u,\, \mathop{u}\limits_{\scriptscriptstyle 1},\,
\mathop{u}\limits_{\scriptscriptstyle 2})u_{x_0}=0,\quad \p
V/\p_{x_0}\ne 0 
\end{displaymath}
is not invariant under the translations with respect to
$x_0$. Nevertheless, Ansatz (\ref{1.5.1z}) invariant under the
translation group (\ref{1.5.3z}) reduces it to an ODE. An explanation for 
this phenomenon is quite simple. The matter is that the second
``non-invariant'' term of the equation in question vanishes on the
manifold (\ref{1.5.1z}). Saying it in another way, the system of two
PDEs 
\begin{equation}
U(x_1,\, u,\, \mathop{u}\limits_{\scriptscriptstyle 1},\,
\mathop{u}\limits_{\scriptscriptstyle 2}) 
+V(x_0,\, x_1,\, u,\, \mathop{u}\limits_{\scriptscriptstyle 1},\,
\mathop{u}\limits_{\scriptscriptstyle 2})u_{x_0}=0,\quad u_{x_0}=0
\end{equation}
is invariant under the group (\ref{1.5.3z}).

Consequently, from the point of view of reducibility of PDE
(\ref{1.5.4z}) by means of the Ansatz invariant under the
one-parameter transformation group with the generator
$X=\xi_0(x,u)\p_{x_0} + \xi_1(x,u)\p_{x_1} + \eta(x,u)\p_u$ it is
enough to require the invariance of a constrained system of PDEs
\begin{displaymath}
U(x_0,\, x_1,\, u,\, \mathop{u}\limits_{\scriptscriptstyle 1},\,
\mathop{u}\limits_{\scriptscriptstyle 2})=0,\quad  
\xi_0(x,u)u_{x_0} + \xi_1(x,u)u_{x_1} - \eta(x,u)=0.
\end{displaymath}

This is a source of the term conditional
symmetry\index{Conditional!symmetry}. Equation (\ref{1.5.4z}) is
non-invariant with respect to the group having the generator $X$ but
being taken together with a condition $Xu=0$ it admits the mentioned
group. Consequently, it is {\em conditionally-invariant} \/under the
Lie group with the generator $X$.  \vspace{2mm}

\noindent
{\bf 1. Reduction of PDEs.}\ Consider an over-determined system of
PDEs of the form
\begin{eqnarray}
&&U_{\ssl A}(x, u, \mathop{u}\limits_{\scriptscriptstyle 1}, \ldots,
\mathop{u} \limits_{\scriptscriptstyle r}) = 0,\ \ A
={1,\ldots,M},\label{1.5.1}\\ 
&&\xi_{a \mu}(x, u)u^\al_{x_\mu} - \eta_a^\al (x, u) = 0,\ \ a=
{1,\ldots,N},\label{1.5.7} 
\end{eqnarray}
where $x = (x_0, x_1, \ldots, x_{n-1}),\ u = (u^0, u^1, \ldots,
u^{m-1})$, 
\begin{displaymath}
\mathop{u}\limits_{\scriptscriptstyle s}=\{\p^su^\al/\p x_{\mu_1}
\ldots \p x_{\mu_s},\ 0\le\al\le m-1, \ 0\le \mu_i \le n-1 \}, 
\end{displaymath}
$U_{\ssl A},\ \xi_{a \mu},\ \eta^\al $ are smooth enough functions, $
N \le n-1 $. In the following, we suppose that the condition
\begin{equation}
{\rm rank}\, \| \xi_{a\mu} (x,u) \|^{N \ \ n-1}_{a=1 \mu =0} =N
\label{1.5.2}
\end{equation}
holds.
\vspace{1.5mm}

\noindent
{\bf Definition 1.5.1.}\ Set of the first-order differential operators   
\begin{equation}
Q_a = \xi_{a \mu}(x,u) \p_{x_\mu} + \eta_a^{\al}(x,u) \p_{u^\al},
\label{1.5.5}
\end{equation}
where $\xi_{a \mu},\ \eta^\al_{a}$ are smooth functions, is called 
involutive\index{Involutive set} if there exist such smooth functions 
$f^c_{ab}(x,u)$ that 
\begin{equation}
[Q_a,\, Q_b]=f^c_{ab} Q_c, \ \ a,b,c = {1,\ldots,N}.
\label{1.5.6}
\end{equation}

The simplest example of an involutive set of operators is given by
first-order differential operators forming a Lie algebra. In such a
case $f^c_{ab}=\mbox{\rm const}$,\ $a,b,c = {1,\ldots,N}$ are called
structure constants\index{Structure constants} of the Lie algebra.

It is common knowledge that conditions (\ref{1.5.6}) are 
necessary and sufficient for the system of PDEs (\ref{1.5.7}) to
be compatible (the Frobenius theorem \cite{176}). Its general solution
can be represented in the form
\begin{equation}
F^\al (\om_1,\, \om_2,\, \ldots, \om_{n+m-N}) = 0,
\ \ \al = {0,\ldots, m-1},
\label{1.5.8}
\end{equation}
where $F^\al \in C^1 ({\C}^{n+m-N}, {\C}^1) $ are arbitrary functions,
$\om_i = \om_i(x, u) $ are func\-ti\-onally-independent first
integrals of system of PDEs (\ref{1.5.7}).

Due to condition (\ref{1.5.2}) we can choose $m$ first
integrals $\om_{j_1}, \ldots, \om_{j_m} $ satisfying the condition
${\rm det}\, \| \p \om_{j_i} / \p u^\al \|^{m \ \ m-1}_{i=1
\beta =0} \ne 0$, since otherwise integrals $\om_1$,\ $\om_2$,
$\ldots$, $\om_{n+m-N}$ would be functionally-dependent.

Changing, if necessary, numeration we can put $j_i=i$ and thus get $m$  
first integrals $\om_1, \ldots, \om_m$ of the system of PDEs
(\ref{1.5.7}) satisfying the following condition:
\begin{equation}
{\rm det} \| \p \om_{i} / \p u^\al \|^{m \ \ m-1}_{i=1
\beta =0} \ne 0.
\label{1.5.9}
\end{equation}

Resolving relations (\ref{1.5.8}) with respect to $\om_1, \ldots,
\om_m $ we have 
\begin{equation}
\om_i = \varphi_i (\om_{m+1}, \ldots, \om_{n+m-N}),
\label{1.5.10}
\end{equation}
where $\varphi_i \in C^1 ({\C}^{n-N}, {\C}^1), \ i = {1,\ldots,m}$ are
arbitrary functions.
\vspace{1.5mm}

\noindent 
{\bf Definition 1.5.2.}\ Expression (\ref{1.5.10}) is called an
Ansatz\index{Ansatz} for the field $u^\al = u^\al (x) $ invariant
under the set of operators (\ref{1.5.5}) provided (\ref{1.5.9}) holds.  

Formulae (\ref{1.5.10}) take an especially simple and clear form
provided  
\begin{equation}
\xi_{a \mu}=\xi_{a \mu}(x),\quad
\eta^\al_a=A^{\al\beta}_a(x) u^\beta,
\ \ a = {1,\ldots,N},\ \ \al = {0,\ldots,m-1}.
\label{1.5.11}
\end{equation}

Given the condition (\ref{1.5.11}) operators (\ref{1.5.5}) are
rewritten in a non-Lie form
\begin{equation}
Q_a = \xi_{a\mu} (x) \p_{x_\mu} + \eta_a (x),
\ \ a={1,\ldots,N},
\label{1.5.12}
\end{equation}
where $ \eta_a = \| -A^{\al\beta}_a(x)\|^{m-1}_{\al,\beta =0} $
are $(m\times m)$-matrices and system (\ref{1.5.7}) is rewritten as a
system of linear PDEs
\begin{equation}
\xi_{a \mu} (x) u_{x_\mu} + \eta_a (x) u = 0,
\ \ a={1,\ldots,N}.
\label{1.5.13}
\end{equation}

Here $u =(u^0, \ldots, u^{m-1})^T$.
\vspace{1.5mm}

\noindent
{\bf Lemma 1.5.1.}\ {\em Assume that conditions (\ref{1.5.2}),
  (\ref{1.5.11}) hold. Then, a set of functionally-independent first
  integrals of system of PDEs (\ref{1.5.7}) can be chosen as follows
\begin{eqnarray*}
&&\om_i = b_i^\al(x) u^\al,\ \ i= {1,\ldots,m},\\
&&\om_{m+j} = \om_{m+j} (x), \ \ j = {1,\ldots,n-N}
\end{eqnarray*}
and besides} ${\rm det}\, \| b_{i}^{\al} (x) \|_{i=1 \al =0}^{m \ \ m-1}
\ne 0$.
\vspace{1.5mm}

\noindent
{\em Proof.}$\quad$ Consider the following system of matrix PDEs:
\begin{equation}
\xi_{a \mu}(x)F_{x_\mu}=F\eta_a(x),
\label{1.5.13z}
\end{equation}
where $F=\|f^{\al\beta}(x)\|_{\al,\beta=0}^{m-1}$ is an $(m\times
m)$-matrix and $\xi_a,\ \eta_a=\|
-A^{\al\beta}_a(x)\|^{m-1}_{\al,\beta =0} $ are coefficients of the
operators $Q_a$. Since the operators $Q_a$ form an involutive set, the
above system is compatible and its general solution has the form
\begin{displaymath}
F(x)=\Theta B(x), 
\end{displaymath}
where $\Theta$ is an $(m\times m)$-matrix whose elements are arbitrary
functions of a complete set of functionally-independent first
integrals of the system 
\begin{equation}
\xi_{a\mu}\p_\mu \om=0,\ \ a={1,\ldots,N}
\label{1.5.16}
\end{equation}
and $B(x)=\|b_i^\al(x)\|_{i,\al=0}^{m-1}$ is a particular solution of
(\ref{1.5.13z}) with ${\rm det}\, B(x)\ne 0$. 

It is straightforward to check that from the involutivity of the set
of operators $Q_a$ it follows that the operators $Q_a^\prime
=\xi_{a\mu}\p_\mu$ form an involutive set.  Consequently, system
(\ref{1.5.16}) is compatible and what is more due to the condition
(\ref{1.5.2}) the number of its functionally-independent first
integrals is equal to $n-N$. We denote these as: $\om_{m+1}(x)$,\ 
$\om_{m+2}(x)$,$\ldots$,$\om_{n+m-N}(x)$.

As ${\rm det}\, \|b_i^\al(x)\|_{i,\al=0}^{m-1}\ne 0$, the
expressions $b_1^\al(x) u^\al$, $\ldots$, $b_m^\al(x)u^\al$,
\ $\om_{m+1}(x)$, $\ldots$, $\om_{m+n-N}(x)$ are
functionally-independent. If we prove that the functions $b_i^\al(x)
u^\al,\ i={1,\ldots,m}$ are first integrals, the proof of the lemma
will be completed.

Acting by the operators $Q_a$ on the functions $b_i^\al(x) u^\al$ one
has  
\begin{displaymath}
\Bigl(\xi_{a\mu}\p_\mu+A^{\g\beta}_a(x) u^\beta \p_{u^\g}\Bigr)  
\Bigl(b_i^\al(x) u^\al\Bigr)=\Bigl(\xi_{a\mu}\p_\mu b_i^\beta(x) +
b_i^\al(x)A^{\al\beta}_a(x)\Bigr)u^\beta=0
\end{displaymath}
(we have taken into account that the matrix
$B(x)=\|b_i^\al(x)\|_{i,\al=0}^{m-1}$ satisfies (\ref{1.5.13z})) the
same which is required. The lemma is proved. $\rhd$ 

Due to Lemma 1.5.1 we can resolve formulae (\ref{1.5.10})
with respect to $u^\alpha$ and thus transform an Ansatz invariant
under operators (\ref{1.5.12}) to the form
\begin{displaymath}
u^\al = a^{\al \beta} (x) \varphi^\beta (\om_{m+1}, \ldots,
\om_{m+n-N}) 
\end{displaymath}
or (in the matrix notation)
\begin{equation}
u=A(x)\varphi(\om_{m+1}, \ldots, \om_{m+n-N}),
\label{1.5.14}
\end{equation}
where $A(x)=\|a_{\al \beta}(x) \|^{m-1}_{\al,\beta =0}$ is the
inverse of the matrix $B$.

Since the matrix function $B(x)$ satisfies the system of PDEs
(\ref{1.5.13z}), the following equalities hold
\begin{eqnarray*}
&&\xi_{a\mu}\p_\mu A(x)=\xi_{a\mu}\p_\mu
B^{-1}(x)=-B^{-1}(x)\Bigl(\xi_{a\mu}\p_\mu B(x)\Bigr)B^{-1}(x)\\
&&\quad=-B^{-1}(x)B(x)\eta_a B^{-1}(x)=-\eta_a A(x).
\end{eqnarray*}

Consequently, we have established that the Ansatz invariant under the
involutive set of operators (\ref{1.5.12}) satisfying condition
(\ref{1.5.2}) is represented in the form (\ref{1.5.14}), where $A(x)$
is a nonsingular $(m\times m)$-matrix satisfying the system of PDEs
\begin{equation}
\xi_{a\mu}\p_\mu A(x)+\eta_a A(x)=0,\ \ a={1,\ldots,N}
\label{1.5.15} 
\end{equation}
and functions $\om_{m+1}(x), \ldots,\om_{m+n-N} (x) $ form a
complete set of functionally-in\-de\-pen\-dent first integrals of the
system of PDEs (\ref{1.5.16}).

We say that Ansatz (\ref{1.5.10}) reduces system of PDEs (\ref{1.5.1}) 
if the substitution of formulae (\ref{1.5.10}) into (\ref{1.5.1}) gives
rise to a system of PDEs which is equivalent to one containing
"new" independent $ \om_{m+1}$,\ $\om_{m+2}$, $\ldots$,
$\om_{m+n-N}$ and dependent $\varphi^0$,\ $\varphi^1$, $\ldots$,
$\varphi^{m-1}$ variables only.

Let us recall the classical theorem about reduction of PDEs by means
of group-invariant solutions: {\em a solution invariant under the
  $N$-dimensional Lie algebra with basis elements (\ref{1.5.5})
  satisfying the condition (\ref{1.5.2}), which is a subalgebra of the
  symmetry algebra of PDE under study, reduces it to an
  $(n-N)$-dimensional PDE} \cite{25,127,162,163}.

We will prove that for a given PDE to be reducible by means of the
Ansatz (\ref{1.5.10}) it is enough to require conditional invariance
with respect to the corresponding involutive set of differential
operators.  Such a condition is essentially weaker than a requirement
of invariance in the Lie sense and makes it possible to obtain
principally new reductions of PDEs as compared with those obtainable
within the framework of the classical Lie approach.  
\vspace{1.5mm}

\noindent
{\bf Definition 1.5.3.}\ We say that the system of PDEs (\ref{1.5.1}) is
conditionally-invari\-ant under the involutive set of differential
operators (\ref{1.5.5}) if the system of PDEs
\begin{equation}
\cases{U_{\ssl A}(x, u, \mathop{u}\limits_{\scriptscriptstyle 1},
  \ldots, \mathop{u}\limits_ 
{\scriptscriptstyle r}) = 0,\ \ A ={1,\ldots,M},\cr
\xi_{a \mu}(x, u)u^\al_{x\mu} - \eta_a^\al (x, u) = 0,\ \ a=
{1,\ldots,N},\cr D\Bigl(\xi_{a \mu}(x, u)u^\al_{x\mu} - \eta_a^\al (x,
u) 
= 0\Bigr), \ \ a= {1,\ldots,N},\cr
\ldots \cr
D^{r-1}\Bigl(\xi_{a \mu}(x, u)u^\al_{x\mu} - \eta_a^\al (x, u) =
0\Bigr), \ \ a= {1,\ldots,N},\cr}
\label{1.5.17}
\end{equation}
where the symbol $D^s(L=0)$ denotes a set of all differential
consequences of the equation $L=0$ of order $s$,
is invariant in the Lie sense under the one-parameter transformation
groups having the generators $Q_a,\ a={1,\ldots,N}$.

Before formulating the reduction theorem we will prove two
auxiliary assertions.
\vspace{1.5mm}

\noindent
{\bf Lemma 1.5.2.}\ {\em Let us suppose that operators (\ref{1.5.5})
  form an   involutive set. Then the set of differential operators
\begin{equation}
Q'_a = \lbd_{a b}(x) Q_b, \quad {\rm det}\, \|\lbd_{ab}(x)
\|^N_{a,b=1}\ne 0 
\label{1.5.18}
\end{equation}
is also involutive.}
\vspace{1.5mm}

\noindent
{\em Proof.}$\quad$ The lemma is proved by direct computation. Indeed,
\begin{eqnarray*}
[Q'_a,\, Q'_b]&=&[\lbd_{ac} Q_c,\, \lbd_{bd} Q_d] =
\lbd_{ac}(Q_c \lbd_{bd}) Q_d - \lbd_{bd}(Q_d \lbd_{ac}) Q_c \\
&&+ \lbd_{ac} \lbd_{bd} f^{d_1}_{cd} Q_{d_1} = \ti f^c_{ab} Q_c = \ti
f^c_{ab} 
\lbd^{-1}_{cd} Q'_d.
\end{eqnarray*}
where $\lbd^{-1}_{cd} $ are elements of the matrix inverse to
the matrix $\| \lbd_{ab}(x) \|^N_{a,b=1} $. $\rhd$
\vspace{1.5mm}

\noindent
{\bf Lemma 1.5.3.} \ {\em Let system of PDEs (\ref{1.5.1}) be
  conditionally-invariant under the involutive set of differential
  operators (\ref{1.5.5}). Then, it is conditionally-invariant under
  the involutive set (\ref{1.5.18}) with arbitrary smooth functions
  $\lbd_{a b}$.}
\vspace{1.5mm}

\noindent
{\em Proof.}$\quad$ To prove the lemma we need the following identity
for coefficients of the $s$-th prolongation of the operator
$\xi_{\mu}\p_\mu +\eta^\al\p_{u^\al}$:
\begin{equation}
\zeta_{\mu_1\ldots\mu_i}^\al=D_{\mu_1}\ldots D_{\mu_i}
(\eta^\al-\xi_{\mu}u^\al_{x_\mu})-\xi_\mu u^\al_{x_\mu x_{\mu_1}\ldots
x_{\mu_i}},\ \ i=1,2,\ldots,s,
\label{1.5.18z}
\end{equation}
where
\begin{eqnarray*}
D_{\alpha}=\partial_{x_\alpha} + 
u_{x_\alpha}^\beta {\partial\over \partial u^\beta} +
\sum_{n=1}^{\infty}
u^\beta_{x_{\alpha_1} \ldots x_{\alpha_n} x_\alpha}
\ {\dis\partial\over\dis\partial
(u^{\beta}_{x_{\alpha_1} \ldots x_{\alpha_n}})}
\end{eqnarray*}
is a total differentiation operator with respect to the variable
$x_\al$. The above identity is proved by the method of mathematical
induction. First, we will prove it under $i=1$. From the prolongation
formulae given in the introduction we have
\begin{displaymath}
\zeta_{\mu}^{\alpha}=D_{\mu} \eta^{\al} -
u^{\al}_{x_\beta} D_{\mu}\xi_{\beta}
=D_{\mu}(\eta^\al-\xi_{\beta}u^\al_{x_\beta})
-\xi_\beta u^\alpha_{x_\beta x_{\mu}},
\end{displaymath}
whence it follows that the identity (\ref{1.5.18z}) holds for
$i=1$. Consequently, the base of induction is established.

Let us suppose now that the identity (\ref{1.5.18z}) holds for all
$i\le k-1$. We will prove that hence its validity for $i=k$ follows.

Indeed, 
\begin{eqnarray*}
&&\zeta_{\mu_1 \ldots \mu_k}^\al=D_{\mu_k}\zeta_{\mu_1\ldots
  \mu_{k-1}}^\al -u^{\al}_{x_{\mu_1} \ldots x_{\mu_{k-1}}x_\beta} 
D_{\mu_k} \xi_{\beta}\\
&&\quad=D_{\mu_k}\Bigl(D_{\mu_1}\ldots D_{\mu_{k-1}}
(\eta^\al-\xi_{\mu}u^\al_{x_\mu})-\xi_\mu u_{x_\mu x_{\mu_1}\ldots
x_{\mu_{k-1}}}\Bigr)\\
&&\quad-u^{\al}_{x_{\mu_1} \ldots x_{\mu_{k-1}} 
\partial x_{\beta}} D_{\mu_k} \xi_{\beta}
=D_{\mu_1}\ldots D_{\mu_k}(\eta^\al-\xi_{\mu}u^\al_{x_\mu})
-\xi_\mu u^\al_{x_\mu x_{\mu_1}\ldots x_{\mu_k}},
\end{eqnarray*}
the same which is required.

Due to the identity proved above the $r$-th prolongation of the
operator $Q_a^\prime$ being restricted to the set of solutions of
system of PDEs (\ref{1.5.17}) takes the form
\begin{eqnarray*}
\wid Q_a^\prime &=& Q_a^\prime+ D_{\mu_1}\ldots D_{\mu_r}
(\eta_a^{\prime\al}-\xi_{a\mu}^\prime
u^\al_{x_\mu}){\edi\partial\over\edi \partial 
(u^{\al}_{x_{\mu_1} \ldots \partial x_{\mu_r}})}\\
&&-\xi_{a\mu}^\prime u^\al_{x_\mu x_{\mu_1}\ldots x_{\mu_r}}
{\edi\partial\over\edi \partial (u^{\al} 
_{x_{\mu_1} \ldots x_{\mu_r}})}.
\end{eqnarray*}
Substituting the formulae $\eta_a^{\prime\al}= \lbd_{a b}\eta_b^\al, \ 
\xi_{a\mu}^\prime=\lbd_{a b}\xi_{b\mu}$ into the above equality and
taking into account that the relations
\begin{displaymath}
D_{\mu_1}\ldots D_{\mu_i}(\eta_a^\al-\xi_{a\mu} u^\al)=0,\ \
i=1,2,\ldots, r-1
\end{displaymath}
hold on the set of solutions of system (\ref{1.5.17}) we get $\wid
Q_a^\prime=\lbd_{a b}(x,u)\wid Q_b$.

If we denote by the symbol $L^i$ one of the equations of system
(\ref{1.5.17}) and by the symbol $[L]$ the set of its solutions, then 
the following equalities hold
\begin{displaymath}
\left.\matrix{\wid Q_a^\prime L^i\cr\cr}\right.
\left|\matrix{&=\cr[L]&\cr}\right.
\left.\matrix{\lbd_{a b}(x,u)\wid Q_b L^i\cr\cr}\right.
\left|\matrix{&=\cr[L]&\cr}\right.
\left.\matrix{\lbd_{a b}(x,u)
(\wid Q_b L^i\cr\cr}\right.
\left|\matrix{)&=0,\cr[L]&\cr}\right.
\end{displaymath}
whence it follows that the system of PDEs (\ref{1.5.1}) is
conditionally-invariant under the involutive set of operators
(\ref{1.5.18}). The lemma is proved. $\rhd$
\vspace{1.5mm}

\noindent
{\bf Theorem1.5.1.}\ {\em Let the system of PDEs (\ref{1.5.1}) be
  conditionally-invariant under the involutive set of differential
  operators (\ref{1.5.5}) satisfying condition (\ref{1.5.2}). Then,
  the Ansatz (\ref{1.5.10}) invariant under the involutive set
  (\ref{1.5.5}) reduces system of PDEs (\ref{1.5.1}).}  
\vspace{1.5mm}

\noindent
{\em Proof.}$\quad$ Due to condition (\ref{1.5.2}) there exists
such a nonsingular $(N\times N)$-matrix $ \| \lbd_{ab} (x, u)
\|^N_{a,b=1} $ that 
\begin{displaymath}
Q_a^\prime=\lbd_{ab}(\xi_{b\mu}u^\al_\mu -\eta^\al_b)=u^\al_{x_{a-1}} 
+ \sum^{n-1}_{\mu =N}\xi_{a \mu}^\prime u^\al_{x_\mu} -
\eta_a^{\prime\al}, \ \ a= {1,\ldots,N}
\end{displaymath}
and what is more the operators $Q_a^\prime$ form the involutive set
(Lemma 1.5.1) such that system of PDEs (\ref{1.5.1}) is
conditionally-invariant with respect to it (Lemma 1.5.2).

Since the set of operators $Q_a^\prime,\ a={1,\ldots,N}$ is involutive,
there exist such functions $f^c_{ab}(x, u) $ that
\begin{equation}
[Q_a^\prime,\, Q_b^\prime]= f^c_{ab} Q'_c,\ \ a,b = {1,\ldots,N}.
\label{1.5.23}
\end{equation}

Computing commutators on the left-hand sides of the above equalities
and equating coefficients of the linearly independent differential
operators $\p_{x_0}$,\ $\p_{x_1}$, $\ldots$, $\p_{x_{N-1}}$ we have
$f^c_{ab} = 0,\ a,b,c = {1,\ldots,N} $. Consequently, operators
$Q_a^\prime$ form a commutative Lie algebra.

Furthermore, systems of PDEs $Q_a\om(x,u)=0,\ a={1,\ldots,N}$ and
$\lbd_{a b}(x,u) \linebreak \times Q_b\om(x,u)=0,\ a={1,\ldots,N}$
with ${\rm det}\, \|\lbd_{a b}(x,u)\|_{a,b=1}^{N}\ne 0$ have the same
set of functionally-in\-de\-pen\-dent first integrals. Hence we
conclude that the involutive sets of operators $Q_a$ and $Q_a^\prime$
give rise to the same Ansatz (\ref{1.5.10}).

From the definition of the conditional invariance it follows that
the system of PDEs 
\begin{equation}
\cases{U_{\ssl A}(x, u, \mathop{u}\limits_{\scriptscriptstyle 1}, \ldots,
\mathop{u}\limits_{\scriptscriptstyle r}) = 0,\ \ A ={1,\ldots,M},\cr
u^\al_{x_{a-1}} 
+ \sum^{n-1}_{\mu =N}\xi_{a \mu}^\prime u^\al_{x_\mu} -
\eta_a^{\prime\al}=0,\ \ a= {1,\ldots,N},\cr
D\Bigl(u^\al_{x_{a-1}} 
+ \sum^{n-1}_{\mu =N}\xi_{a \mu}^\prime u^\al_{x_\mu} -
\eta_a^{\prime\al}=0\Bigr),\ \ a= {1,\ldots,N},\cr
\ldots \cr
D^{r-1}\Bigl(u^\al_{x_{a-1}} 
+ \sum^{n-1}_{\mu =N}\xi_{a \mu}^\prime u^\al_{x_\mu} -
\eta_a^{\prime\al}=0\Bigr),\ \ a= {1,\ldots,N}\cr}
\label{1.5.23z}
\end{equation}
is invariant in Lie sense under the one-parameter groups generated
by the mutually commuting operators $Q_a^\prime$. Consequently, the
above system is invariant in Lie sense under the commutative Lie
algebra $\langle Q_1^\prime,\, Q_2^\prime,\, \ldots
,Q_N^\prime\rangle$.

Now we can apply the classical theorem about sym\-met\-ry
(gro\-up-theo\-re\-ti\-cal) reduction of PDEs and conclude that the
Ansatz invariant under the involutive set of operators (\ref{1.5.5})
(or, which is the same, under the commutative Lie algebra $\langle
Q_1^\prime,\, Q_2^\prime,\, \ldots,Q_m^\prime\rangle$) reduces system
of PDEs (\ref{1.5.23z}). But by construction all equations from the
system (\ref{1.5.23z}) with the exception of the first $m$ equations
(which form the initial system of PDEs (\ref{1.5.1})) vanish
identically on the manifold (\ref{1.5.5}). Consequently, the Ansatz
(\ref{1.5.5}) reduces system (\ref{1.5.1}). The theorem is proved.  
$\rhd$ 
\vspace{1.5mm}

\noindent
{\bf Note 1.5.1.}\ There exists a deep relation between reducibility
of PDE (\ref{1.5.1}) condi\-ti\-on\-al\-ly-invariant under the
involutive set of operators (\ref{1.5.5}) and compatibility of the
over-determined system of PDEs (\ref{1.5.1}), (\ref{1.5.7}). But as is 
shown below from conditional invariance of PDE (\ref{1.5.1}) with respect
to the involutive set of operators (\ref{1.5.5}) it does not
follow a compatibility of system (\ref{1.5.1}), (\ref{1.5.7}) and {\it
  vise versa}.

The equation
\begin{displaymath}
(x_a x_a)(u_{x_b}u_{x_b})-(x_a u_{x_a})^2=m^2,\ \ m\ne 0,
\end{displaymath}
where $a,b=1,2,3$, is invariant with respect to the rotation group
$O(3)$ having the generators $J_{a b}=x_a\p_{x_b}-x_b\p_{x_a},\ a<b,\
a,b=1,2,3$. However, the system of PDEs
\begin{displaymath}
\cases{(x_a x_a)(u_{x_b}u_{x_b})-(x_a u_{x_a})^2=m^2,\ \ m\ne 0,\cr
J_{a b}u=0,\ \ a,b=1,2,3\cr}
\end{displaymath}
is incompatible, because substitution of the general solution of the
last three equations $u=\vp\,(x_a x_a)$ into the first one yields an
inconsistent equality $0=m^2$. 

On the other hand, system of PDEs
\begin{displaymath}
\cases{u_{xx}+u_{yy}-u+y(u_x-u)=0,\cr
u_y=0\cr}
\end{displaymath}
is compatible (it has a solution $u=C{\rm e}^x,\ C=\mbox{\rm const}$)
but the equation $u_{xx}+u_{yy}-u+y(u_x-u)=0$ is not
conditionally-invariant under the operator $Q=\p_y$.
\vspace{1.5mm}

\noindent
{\bf Note 1.5.2.}\ We have proved Theorem 1.5.1 under assumption
that the condition (\ref{1.5.2}) holds. It is not difficult to prove
that Theorem 1.5.1 is still valid, provided 
\begin{equation}
{\rm rank}\, \|\xi_{a\mu}\|^{N \ \ n-1}_{a=1 \mu =0}
={\rm rank}\, \|\xi_{a\mu}\, \eta_a^1,\ldots \eta_a^{m-1} \|^{N \ \
  n-1}_{a=1 \mu =0} =N^\prime < N.
\label{1.5.25z}
\end{equation}

Indeed, using transformation (\ref{1.5.18}) we can reduce involutive
set of operators (\ref{1.5.5}) satisfying (\ref{1.5.25z}) to the form
$Q_1^\prime$, $\ldots$, $Q_{N^\prime}$,\ $Q_{N^\prime+1}=0$, $\ldots$,
$Q_{N}=0$. Now, we can apply Theorem 1.5.1 with $N=N^\prime$.
Consequently, if the system of PDEs (\ref{1.5.1}) is
conditionally-invariant under the involutive set of operators
(\ref{1.5.5}) satisfying (\ref{1.5.25z}), then the Ansatz
(\ref{1.5.10}) invariant under the involutive set (\ref{1.5.5})
reduces it to $(n-N^\prime)$-dimensional PDE.

In the case when the condition (\ref{1.5.25z}) is not satisfied,
so-called partially-invariant 
solutions\index{Partially-invariant solution} (the term was introduced
by Ovsjannikov \cite{164}) are obtained. Reduction of PDEs
conditionally-invariant under the involutive set of differential
operators (\ref{1.5.5}) not obeying the condition (\ref{1.5.25z}) is
studied in detail in our paper \cite{106.1}.  
\vspace{2mm}

\noindent
{\bf 2.$\!$ Symmetry and compatibility of over-determined systems of
  linear PDEs.}\ This subsection is devoted to the investigation of the
following systems of PDEs:
\begin{equation}
{B_{\mu \nu}(x)\p_{x_\nu}+B_{\mu}(x)}u(x)=0,\ \ \mu={0,\ldots,n-1},
\label{1.5.3}
\end{equation}
where $ x= (x_0,\, x_1, \ldots , x_{n-1}),
\ u(x)=\Bigl(u^0(x),\, u^1(x),\,  \ldots, u^{m-1}(x)\Bigr)^T $,
\ $B_{\mu \nu},\ B_\mu $ are variable $(m\times m)$-matrices
satisfying the condition 
\begin{equation}
{\rm rank}\, \| B_{\mu \nu} (x) \|^{n-1}_{\mu,\nu =0} = n \times m.
\label{1.5.4}
\end{equation}
The problem of investigating compatibility of an over-determined
system of the form (\ref{1.5.4}) is closely connected with the
problem of separation of variables in systems of linear
PDEs\index{Separation of variables} (see \cite{97,156,179} 
and Chapter 5).
\vspace{1.5mm}

\noindent
{\bf Theorem 1.5.2.}\ {\em System of PDEs (\ref{1.5.3}) is compatible
  iff 
\begin{equation}
[B_{\mu \nu} \p_\nu + B_\mu,\, B_{\al \beta} \p_\beta + B_\al ] =
R_{\mu \al \beta } (B_{\beta\nu} \p_\nu + B_\beta),
\label{1.5.31}
\end{equation}
where $R_{\mu \al \beta} $ are some linear first-order differential
operators with matrix coefficients, $\mu, \al ={0,\ldots,n-1}$.}
\vspace{1.5mm}

\noindent
{\em Proof.}$\quad$ The necessity. Let system (\ref{1.5.4}) be
compatible. We will show that hence it follows that (\ref{1.5.31})
holds. Due to (\ref{1.5.4}) the block 
$(nm\times nm)$-matrix $\| B_{\mu \nu} \|^{n-1}_{\mu,\nu=0} $ is
invertible. That is why  there exists such a block $(nm\times
nm)$-matrix $ \| C_{\mu \nu} \|^{n-1}_{\mu,\nu = 0}$ that
\begin{equation}
C_{\mu \nu} (x) B_{\nu \al} (x) = B_{\mu \nu}(x) C_{\nu \al} (x) =
\delta_{\mu \al} I,
\label{1.5.32}
\end{equation}
where $I$ is the unit $(m\times m)$-matrix.

Let us  rewrite (\ref{1.5.3}) in the equivalent form
\begin{equation}
\p_{\mu} u = F_\mu (x) u,
\label{1.5.33}
\end{equation}
where $F_\mu = -C_{\mu \al} B_\al $.

It is well-known (see, for example, \cite{33,41,183}) that the
necessary and sufficient compatibility conditions of system of PDEs
(\ref{1.5.33}) read
\begin{equation}
\p_\mu F_\nu - \p_\nu F_\mu + [F_\mu, F_\nu ] = 0,
\ \ \mu, \nu = {0,\ldots,n-1}.
\label{1.5.34}
\end{equation}

Introducing notations $Q_\mu = \p_\mu - F_\mu(x) $ we rewrite
(\ref{1.5.34}) in the form
\begin{displaymath}
[Q_\mu,\, Q_\nu ] = 0.
\end{displaymath}

Representing the operators $B_{\mu \nu } \p_\nu + B_\mu $ in the form
$B_{\mu \nu} \p_\nu + B_\mu = B_{\mu \nu } Q_\nu $ we compute the
commutator 
\begin{eqnarray*}
[B_{\mu \nu} Q_\nu,\, B_{\al \beta} Q_\beta ] &=& [B_{\mu \nu},B_{\al
  \beta}] 
Q_\nu Q_\beta + B_{\mu \nu} [Q_\nu,\, B_{\al \beta}] Q_\beta\\ 
&&- B_{\al \beta} [Q_\nu,\, B_{\mu \beta}] Q_\beta.
\end{eqnarray*}

Finally, substituting formulae $Q_\mu = C_{\mu \al} (B_{\al \nu}
\p_\nu + B_\al )$ into the equality obtained we arrive at
(\ref{1.5.31}) and besides
\begin{displaymath}
R_{\mu \al \beta} = \{[B_{\mu \nu},\, B_{\al \beta_1}] Q_\nu +
B_{\mu \nu} [Q_\nu,\, B_{\al \beta_1}] - B_{\al \nu}[Q_\nu,\, B_{\mu
\beta_1}]\} C_{\beta_1 \beta}.
\end{displaymath}

The sufficiency. Given (\ref{1.5.4}), we will prove
that there exist such linear first-order operators $ \wid R_{\mu \nu
  \beta} $ that the equality 
\begin{equation}
[Q_\mu,\, Q_\nu ] = \wid R_{\mu \nu \beta } Q_\beta
\label{1.5.35}
\end{equation}
holds.

 Indeed,
\begin{eqnarray*}
[Q_\mu,\, Q_\al ] &=& [C_{\mu \al_1} (B_{\al_1 \nu} \p_\nu +
B_{\al_1} ),\, 
  C_{\al \beta} (B_{\beta \nu_1} \p_{\nu_1} + B_\beta)] \\
&&=C_{\mu \al_1}[B_{\al_1 \nu} \p_\nu +B_{\al_1},\, C_{\al \beta}]
  (B_{\beta \nu_1} \p_{\nu_1} + B_\beta) \\
&&+C_{\mu \al_1}C_{\al \beta}[B_{\al_1 \nu} \p_\nu +B_{\al_1},\,
  B_{\beta \nu_1} \p_{\nu_1} + B_\beta] \\
&&+[C_{\mu \al_1},\, C_{\al \beta}](B_{\beta_1 \nu} \p_\nu +
B_{\beta_1}) 
  (B_{\al \nu_1} \p_{\nu_1} +B_\al) \\
&&+ C_{\al \beta} [C_{\mu \al_1},\, B_{\beta \nu_1} \p_{\nu_1} +
B_\beta] 
  (B_{\al_1 \nu} \p_\nu + B_{\al_1}) \\
&&= P_{\mu \al \beta} (B_{\beta \nu} \p_\nu +B_\beta) =
  P_{\mu \al \beta} B_{\beta \nu} Q_\nu.
\end{eqnarray*}

Choosing $P_{\mu \al \beta} B_{\beta \nu} = \wid R_{\mu \al \nu} $ we
arrive at (\ref{1.5.35}).

Computing the commutator on the left-hand side of (\ref{1.5.35}) and
equating coefficients of linearly independent operators $\p_\mu $ we
get the equalities
\begin{displaymath}
\wid R_{\mu \al \beta} = 0.
\end{displaymath}

Consequently, operators $Q_\mu$ commute, i.e.,\  conditions
(\ref{1.5.34}) hold identically. Hence it follows that system
(\ref{1.5.33}) is compatible. Since equations (\ref{1.5.33}) are
equivalent to the initial system of PDEs (\ref{1.5.3}), the
sufficiency is proved. $\rhd$ \vspace{1.5mm}

\noindent
{\bf Note 1.5.3.}\ In the theory of non-Abelian gauge fields
(Yang-Mills fields\index{Yang-Mills!field}) conditions (\ref{1.5.34})
are called the zero curvature equations. The general solution of the
system of matrix PDEs (\ref{1.5.34}) has the form
\begin{equation}
F_\mu = V_{x_\mu} V^{-1},\ \ \mu = {0,\ldots,3},
\label{1.5.36}
\end{equation}
where $V(x)$ is an arbitrary nonsingular $(m\times m)$-matrix which
elements are smooth functions of $x$. Formula (\ref{1.5.36})
establishes an one-to-one correspondence between over-determined
systems (\ref{1.5.3}) and solutions of the equation 
\begin{equation}
{B_{0 \mu}(x)\p_{x_\mu} + B_0(x)}u(x) = 0
\label{1.5.37}
\end{equation}
of the form
\begin{displaymath}
u(x) =V(x)\chi,
\end{displaymath}
where $V(x)$ is a nonsingular $(m\times m)$-matrix, $\chi = (\chi^0,\, 
\chi^1,\, \chi^2,\, \chi^3)^T$.

Thus, to construct particular solutions of the system of PDEs
(\ref{1.5.37}) it is necessary to classify algebraic objects of the type
(\ref{1.5.3}), (\ref{1.5.31}). Up to now this problem is solved for a
number of the Lie algebras and some simplest superalgebras 
\cite{8,9}, \cite{13}--\cite{16}, \cite{66,165,166}.

The most simple is the case where in (\ref{1.5.31}) $R_{\mu \al
  \beta}=0 $ i.e.,\ the operators $\Sigma_\mu = B_{\mu \nu}
\p_\nu + B_\mu $ commute. For many fundamental mathematical and
theoretical physics equations (in particular, for the Dirac
equation\index{Dirac!equation} \cite{133,179}) it is possible to
obtain complete description of commuting operators $\Sigma_\mu,\ 
\mu = {0,\ldots,n-1} $, where $\Sigma_0 \psi = 0 $ is the equation under
investigation, and to construct solutions with separated variables. In
this respect, we will consider the following particular case of system
(\ref{1.5.3}):
\begin{equation}
\Sigma_\mu u \equiv \Bigl(B_{\mu \nu}(x)\p_\nu + B_\mu(x)\Bigr) u =
\lbd_\mu u, \ \ \mu = {0,\ldots,n-1},
\label{1.5.38}
\end{equation}
where $(\lbd_0, \lbd_1, \ldots, \lbd_{n-1}) \in \Lambda \subset
{\R}^n $, matrices $B_{\mu \nu} (x),\ B_\mu (x) $ being independent of 
$\lbd_\al $.

When proving the principal assertion we will essentially use the
following lemma.
\vspace{1.5mm}

\noindent
{\bf Lemma 1.5.4.}\ {\em If one of the systems of algebraic
  equations 
\begin{eqnarray}
&&P_{\mu \al} B_{\mu \beta } + P_{\mu \beta } B_{\mu \al} = 0,
\label{1.5.39}\\
&&P_{\al \mu} C_{\mu \beta} + P_{\beta \mu} C_{\mu \al} =
0,\label{1.5.40} 
\end{eqnarray}
where $\|B_{\mu \nu} (x) \|^{n-1}_{\mu, \nu = 0}$ is a nonsingular
block $(nm\times nm)$-matrix, $\| C_{\mu \nu}(x) \|^{n-1}_{\mu,
  \nu = 0} $ is its inverse and $P_{\mu \al} $ are some variable
$(m\times m)$-matrices, holds true, then $P_{\mu \al} = 0$.}
\vspace{1.5mm}
  
\noindent
{\em Proof.}$\quad$ We prove the lemma under assumption that
(\ref{1.5.39}) holds. Let us rewrite equalities (\ref{1.5.39}) in the
equivalent form
\begin{equation}
P_{\mu \mu_1} C_{\mu_1 \nu_1} T_{\nu_1 \mu \al \beta} = 0.
\label{1.5.41}
\end{equation}
Here $T_{\nu_1 \mu \al \beta} = B_{\nu_1 \al} B_{\mu \beta} +
B_{\nu_1 \beta} B_{\mu \al} $.

Since the identities
\begin{displaymath}
C_{\mu \nu} C_{\mu_1 \nu_1} T_{\nu_1 \nu \al \beta} = (\delta_{\mu_1
  \al} 
\delta_{\mu \beta} + \delta_{\mu_1 \beta} \delta_{\mu \al} ) I
\end{displaymath}
hold, the block matrix $\| T_{\nu_1 \mu \al \beta} \|$ is invertible.
Consequently, equation (\ref{1.5.41}) is equivalent to the relation
\begin{equation}
P_{\mu \al} C_{\al \nu} = 0.
\label{1.5.42}
\end{equation}

Multiplying (\ref{1.5.42}) by $B_{\nu \beta} $ and summing over $\nu $
we have 
\begin{displaymath}
P_{\mu \beta} = 0,\ \ \mu, \beta = {0,\ldots, n-1}.
\end{displaymath}

In the case, where (\ref{1.5.40}) holds, the proof is
analogous. $\rhd$ 
\vspace{1.5mm}

\noindent
{\bf Theorem 1.5.3.}\ {\em Provided (\ref{1.5.4}) holds the system of PDEs
  (\ref{1.5.38}) is compatible iff}
\begin{equation}
[B_{\mu \nu} \p_\nu + B_\mu,\, B_{\al \beta} \p_\beta +B_\al]=0,
\ \ \mu,\al={0,\ldots,n-1}.
\label{1.5.43}
\end{equation}
{\em Proof.}$\quad$ According to Theorem 1.5.2, the compatibility
criterion for the system of PDEs (\ref{1.5.38}) reads 
\begin{equation}
\begin{array}{l}
[B_{\mu \nu} \p_\nu + B_\mu - \lbd_\mu,\, B_{\al \beta} \p_\beta
+B_\al - \lbd_\al ] \\[2mm]
\quad=(R_{\mu \al \beta \nu} \p_\nu + R_{\mu \al \beta})(B_{\beta
  \nu_1 }\p_{\nu_1} + B_\beta - \lbd_\beta).
\label{1.5.44}
\end{array}
\end{equation}

Computing the commutator in the left-hand side and equating
coefficients of the linearly independent operators $\p_\nu \p_\beta,\
\p_\beta,\ I $ we get the system of PDEs for matrix functions $B_{\mu
  \nu},\ B_\mu,\ R_{\mu \al \beta \nu},\ R_{\mu \al \beta}$ 
\begin{eqnarray}
&&\ [B_{\mu \nu },\, B_{\al \beta} ] +
[B_{\mu \beta},\, B_{\al \nu} ]
= R_{\mu \al \mu_1 \nu } B_{\mu_1 \beta } + R_{\mu \al \mu_1
  \beta} B_{\mu_1 \nu },
\label{1.5.45}\\[2mm]
&&\begin{array}{l}
B_{\mu \nu} \p_\nu B_{\al \beta } - B_{\al \nu } \p_\nu B_{\mu \beta}
+ [B_{\mu \beta},\, B_\al ] - [B_{\al \beta },\, B_\mu]\\[2mm]
\quad = R_{\mu \al \mu_1 \nu } \p_\nu B_{\mu_1 \beta} 
+ R_{\mu \al \mu_1 \beta } (B_{\mu_1} - \lbd_{\mu_1}) +
R_{\mu \al \mu_1} B_{\mu_1 \beta },
\end{array}\label{1.5.46}\\[2mm]
&&\begin{array}{l}
B_{\mu \nu} \p_\nu B_\al - B_{\al \nu} \p_\nu B_\mu +
[B_\mu,\, B_\nu ]= R_{\mu \al \mu_1 \nu } \p_\nu B_{\mu_1}\\[2mm]
\quad +R_{\mu \al \mu_1}
(B_{\mu_1} - \lbd_{\mu_1}).
\end{array}\label{1.5.47}
\end{eqnarray}

Differentiating (\ref{1.5.45}) with respect to $\lbd_{\al_1} $ we
arrive at the relations
\begin{displaymath}
 \frac{\p R_{\mu \al \mu_1 \nu }}{\p \lbd_{\al_1}} B_{\mu_1 \beta} +
\frac{\p R_{\mu \al \mu_1 \beta}}{\p \lbd_{\al_1}} B_{\mu_1 \nu} = 0,
\end{displaymath}
whence due to Lemma 1.5.4 it follows that
\begin{displaymath}
  \frac{\p R_{\mu \al \beta \nu }}{\p \lbd_{\mu_1}}=0,\ \ \mu, \al,
  \beta, \nu, \mu_1 = {0,\ldots, n-1}.
\end{displaymath}

Differentiation of equality (\ref{1.5.46}) with respect to
$\lbd_{\al_1}$ yields 
\begin{displaymath}
\frac{\p R_{\mu \al \mu_1 }}{\p \lbd_{\al_1}} B_{\mu_1 \beta} -
R_{\mu \al \al_1 \beta } = 0.
\end{displaymath}

Multiplying the above equality by $C_{\beta \beta_1} $ and summing
over $\beta $ we have
\begin{displaymath}
\frac{\p R_{\mu \al \beta_1 }}{\p \lbd_{\al_1}} = R_{\mu \al \al_1
  \beta} C_{\beta \beta_1}
\end{displaymath}
or
\begin{equation}
 R_{\mu \al \beta_1} = \lbd_{\al_1} R_{\mu \al \al_1 \beta}
C_{\beta \beta_1} + \wid R_{\mu \al \beta_1},
\label{1.5.48}
\end{equation}
and besides $ \p \wid R_{\mu \al \beta_1} / \p \lbd_{\al_1} = 0,
\ \al_1 = {0,\ldots,n-1}$.

Substituting (\ref{1.5.48}) into (\ref{1.5.47}) and equating
coefficients of $\lbd_{\al_1}$,\ $\lbd_{\mu_1} \lbd_{\al_1}$ we
come to the following relations: 
\begin{displaymath}
R_{\mu \al \al_1 \beta} C_{\beta \beta_1} + R_{\mu \al \beta_1 \beta} 
C_{\beta \al_1} = 0,
\end{displaymath}
\begin{equation}
R_{\mu \al \al_1 \beta} C_{\beta \beta_1} B_{\beta_1} -
\wid R_{\mu \al \al_1} = 0.
\end{equation}

According to Lemma 1.5.4 $R_{\mu \al \al_1 \beta} = 0 $, whence it
follows that $\wid R_{\mu \al \al_1 } = 0$.

Thus, the necessary and sufficient compatibility conditions for system
(\ref{1.5.38}) are given by relations (\ref{1.5.44}) with $ R_{\mu \al
  \beta \nu} = R_{\mu \al \beta} = 0$ or, which is the same, by
relations (\ref{1.5.43}). The theorem is proved. $\rhd$

Results obtained in the present section are applied in a sequel to
reduce multi-dimensional nonlinear partial differential equations to
ODEs and to construct their exact solutions in explicit form. 
In addition, Theorems 1.5.2, 1.5.3 form a basis of our approach 
to separation of variables in systems of linear PDEs (see Chapter 5).  
\vspace{10mm}

\noindent
{\large\bf 1.6. Conservation laws\label{s1.6}}

\markboth{Chapter 1. SYMMETRY OF NONLINEAR SPINOR EQUATIONS}
{1.6. Conservation laws}
\def\theequation{1.\arabic{section}.\arabic{equation}}
\setcounter {section} {6}
\setcounter {equation}{0}
\vspace{7mm}

\noindent
One of the important properties of equations admitting a
nontrivial symmetry group is the existence of constants of
motion (by a constant of motion we mean some combination
of solutions of the equation considered which preserves its
value in time).\index{Motion!constant} The well-known examples of 
constants of motion
are the energy, the momentum\index{Momentum} and the angular 
momentum\index{Angular momentum}.

Within the framework of the traditional approach to the problem of
construction of constants of motion, going back to Noether's works, we
have to investigate symmetry of the Lagrangian of the equation in
question and to construct conservation laws with the help of the
Noether theorem \cite{26,127}.\index{Noether theorem} This theorem
establishes correspondence between one-parameter subgroups of the
symmetry group of the Lagrangian and conservation laws. However the above
approach has restricted applicability since there exist mathematical
physics equations which can not be derived via the Lagrange function.
In addition, there are examples of conservation laws which cannot be
obtained with the help of the Noether theorem even for equations
derived in the framework of the variational principle
\cite{74,75,77,127,197}.

That is why  we apply a method of construction of constants
of motion for the Dirac equation based on the direct calculation
of a conserved quantity as a zero component of the four-vector
of current with components $j_\mu = j_\mu (x, \bar\psi, \psi,
\mathop{\bar\psi}\limits_{\scriptscriptstyle 1},
\mathop\psi\limits_{\scriptscriptstyle 1}, \ldots),$ 
\ $\mu = {0,\ldots,3} $ which satisfies the continuity condition
\begin{equation}
\p_\mu j_\mu = 0
\label{1.6.1}
\end{equation}
for each solution $ \psi = \psi(x) $ of the Dirac equation.
\vspace{1.5mm}

\noindent
{\bf Lemma 1.6.1.}\ {\em Let us suppose that there exists the
  four-vector of current satisfying the relation (\ref{1.6.1}) and
  besides conditions 
\begin{displaymath} 
j_a \rightarrow 0, \quad a = {1,2,3}\ \ \mbox{\it
under }\ \ |\vec x | \rightarrow + \infty 
\end{displaymath} 
hold true. Then, the quantity
\begin{equation}
I = \int\limits_{{\R}^3} j_0 \, d^3 x
\label{1.6.2}
\end{equation}
is conserved in time, i.e.,\  $\p I / \p x_0 = 0 $.
}
\vspace{1.5mm}

\noindent
{\em Proof.}$\quad$  Differentiating (\ref{1.6.2}) with respect to
$x_0$ yields
\begin{displaymath}
\p I / \p x_0 = \int\limits_{{\R}^3} (\p j_0 / \p x_0 )\, d^3 x,
\end{displaymath}
whence it follows
\begin{displaymath}
\p I / \p x_0 = - \int\limits_{{\R}^3} (\p_a j_a )\, d^3 x.
\end{displaymath}

Applying the Gauss-Ostrogradski theorem we get $ \p I / \p x_0 = 0$.
The lemma is proved. $\rhd$

For brevity we will call the four-vector of current satisfying
relation (\ref{1.6.1}) on the set of solutions of the Dirac equation 
the conservation law\index{Conservation law}.

Up to now there is no effective algorithm making it possible
to obtain all conservation laws admitted by an arbitrary PDE.
We will construct conservation laws for the Dirac equation
following an approach suggested in \cite{74,75} which utilizes
its Lie and non-Lie symmetry.
\vspace{1.5mm}

\noindent
{\bf Lemma 1.6.2.}\ {\em Let $ Q $ be a symmetry operator of the Dirac
equation (\ref{1.1.1}). Then, the four-vector with components
\begin{equation}
j_\mu = \bar\psi \gamma_\mu Q \psi,
\label{1.6.3}
\end{equation}
where $\psi = \psi(x) $ is an arbitrary solution of
PDE (1.1.1) vanishing under $ |\vec x | \rightarrow + \infty $,
is a conservation law. }

The proof is carried out by direct verification
\begin{eqnarray*}
\p_\mu j_\mu &=& \p_\mu (\bar\psi \gamma_\mu Q \psi ) = (\p_\mu
\bar\psi \gamma_\mu ) Q \psi + \bar\psi \gamma_\mu \p_\mu Q \psi = im
\bar\psi Q \psi\\  
& & -im \bar\psi Q \psi = 0
\end{eqnarray*}
(we use the fact that any symmetry operator $ Q $ transforms the
set of solutions of PDE (\ref{1.1.1}) into itself, i.e.,\ 
$i \gamma_\mu \p_\mu Q \psi = m Q \psi $). $\rhd$

Let us find explicit expressions of conservation laws corresponding
to the symmetry operators of the Dirac equation which
belong to the class ${\cal M}_1 $ (see Section 1.1). Substituting
the basis elements of the Poincar\'e algebra
$AP(1,3)$\ \  $P_{\mu},\ J_{\mu \nu} $
into (\ref{1.6.3}) we get the well-known expressions of the
energy-momentum and angular-momentum tensors
\begin{equation}
T_{\mu \nu} = \bar\psi \gamma_\mu \p_\mu \psi,\quad
\Omega_{\mu \al \beta} = \bar\psi \gamma_\mu J_{\al \beta} \psi
\label{1.6.4}
\end{equation}
satisfying continuity equation (\ref{1.6.1}) on the index $\mu $.

A trivial identity operator $ I $ gives rise to the current
of a probability density
\begin{equation}
T_\mu = \bar\psi \gamma_\mu \psi.
\label{1.6.5}
\end{equation}

Substitution of zero components of currents (\ref{1.6.4}),
(\ref{1.6.5}) into formula (\ref{1.6.2}) yields the following
conserved quantities: 
\vspace{1.5mm}

\noindent
a) the energy\index{Energy}
\begin{displaymath}
E = \int\limits_{{\R}^3} \psi^\dagger
\p_0 \psi\, d^3 x ;
\end{displaymath}
b) the momentum\index{Momentum}
\begin{displaymath}
P_a =\int\limits_{{\R}^3} \psi^\dagger \p_a \psi\, d^3 x;
\end{displaymath}
c) the angular  momentum\index{Angular momentum}
\begin{displaymath}
\omega_{\al \beta} = \int\limits_{{\R}^3} \psi^\dagger
\Bigl(x_\al\p^\beta -x_\beta \p^\al -(1/2) \gamma_\al
\gamma_\beta\Bigr) \psi\, d^3 x, \ \ \al \ne \beta;
\end{displaymath}
d) the probability
\begin{displaymath}
p = \int\limits_{{\R}^3} \psi^\dagger \psi\, d^3 x.
\end{displaymath}

Constants of motion corresponding to the non-Lie
symmetry\index{Non-Lie!symmetry} operators of the Dirac equation
(\ref{1.1.30}) are obtained in the same way.

In the case of the massless Dirac equation (\ref{1.1.17}) there arise
additional conserved quantities (for more detail see \cite{77}). We
restrict ourselves to adducing constants of motion which correspond to
symmetry operators of equation (\ref{1.1.17}) not belonging to an
enveloping algebra of the conformal algebra\index{Conformal!algebra}
$AC(1,3)$
\begin{eqnarray*}
& &I^1_{\mu \nu} = \int\limits_{{\R}^3} \psi^\dagger (\gamma_\mu
\p^\nu - 
\gamma_\nu \p^\mu ) \psi\, d^3 x,\quad
I^1_\mu = \int\limits_{{\R}^3} \psi^\dagger A_\mu \psi d^3 x,\\
& &I^2_{\mu \nu} = \int\limits_{{\R}^3} \psi^\dagger \Bigl([K_\mu,\,
A_\nu ] - [K_\nu,\, A_\mu ]\Bigr)\psi d^3 x,\quad
I^2_\mu = \int\limits_{{\R}^3} \psi^\dagger \gamma_4 A_\mu \psi d^3 x, 
\end{eqnarray*}
where
$
A_\mu = \gamma_\mu x_\nu \p_\nu -x^\nu \gamma_\nu \p^\mu 
- 2\gamma_\mu,\ \mu,\nu = {0,\ldots,3}$.

\newpage
\thispagestyle{empty}
\noindent
{\sl
C H A P T E R \ \  2\label{ch2}}
\vspace{2mm}

\hrule
\vspace{35mm}

\rightline
{\large\bf
EXACT}
\vspace{2mm}

\rightline
{\large\bf
SOLUTIONS}
\vspace{7mm}

\noindent

The present chapter is devoted to exact solutions of
Poincar\'e-invariant systems of nonlinear PDEs for spinor, 
vector and scalar fields. We establish the necessary and
sufficient compatibility conditions and construct the general
solution of the system of nonlinear PDEs which consists of the
nonlinear d'Alembert and Hamilton equations.  With the use of 
subgroup structure of the groups $P(1,3), \wid P(1,3), C(1,3) $ we
construct Ans\"atze reducing multi-dimensional spinor and vector
equations to PDEs of lower dimension.  These Ans\"atze enable us to
obtain multi-parameter families of exact solutions of the nonlinear
Dirac, Maxwell-Dirac and Dirac-d'Alembert equations, some of the
families containing arbitrary functions.  In particular, the exact
solutions of the nonlinear Dirac equation expressed via the Bessel,
Weierstrass, Gauss and Chebyshev-Hermite functions are constructed.
In addition, a method of constructing exact solutions of PDEs for 
scalar, vector and tensor fields via solutions of a nonlinear
spinor equation is suggested.

\vspace{10mm}

\noindent
{\large\bf 2.1. On compatibility and general solution
\vspace{1.5mm}

\noindent
\phantom{\large\bf 2.1. }of the d'Alembert--Hamilton
system\label{s2.1}} 
\markboth{Chapter 2. EXACT SOLUTIONS}
{2.1. On compatibility and general solution}
\def\theequation{2.\arabic{section}.\arabic{equation}}
\setcounter{section}{1}
\setcounter{equation}{0}
\vspace{7mm}

\noindent
As shown in \cite{78,104,108} the 
substitution\index{Ansatz!for scalar field}
\begin{equation}
w(x) = \vp\Bigl(u(x)\Bigr),
\quad
\vp \in C^2 ({\R}^1, {\R}^1)
\label{2.1.1}
\end{equation}
reduces the $n$-dimensional nonlinear d'Alembert 
equation\index{d'Alembert equation}
\begin{equation}
\Box_n w \equiv {\p^2 w\over \p x^2_0} - \triangle_{n-1} w = F_0(w)
\label{2.1.2}
\end{equation}
to ODE for a function  $ \vp(u) $ iff the scalar function $u =
u(x_0, x_1, \ldots, x_{n-1}) $ satisfies the nonlinear d'Alembert
and Hamilton equations 
\begin{eqnarray}
&&\Box_n u = F_1 (u),\label{2.1.3}\\
&&(\p_{\ssl A} u)(\p^{\ssl A} u) = F_2 (u),\label{2.1.4}
\end{eqnarray}
simultaneously.
\index{Non-Lie reduction!of the d'Alembert equation}

In the above formulae $F_1,\ F_2 $ are arbitrary smooth functions
depending on $u$ only. Hereafter in the present section we suppose
that indices denoted by $A,\ B,\ C $ take the values $0, \ldots, n-1
$ and besides the summation convention in the pseudo-Euclidean space
$M(1, n-1) $ with the metric tensor $g_{\ssl AB} = {\rm diag}\, (1, -1,
\ldots, -1) $ is implied.

Thus, to obtain all Ans\"atze of the form (\ref{2.1.1}) reducing
equation (\ref{2.1.2}) to an ODE one has to construct the general
solution of system (\ref{2.1.3}), (\ref{2.1.4}). Let us emphasize that
such an approach to the problem of reduction of equation (\ref{2.1.2})
does not require the knowledge of a subgroup structure of the
invariance group.

Following \cite{102,104} we call the system of PDEs (\ref{2.1.3}),
(\ref{2.1.4}) the d'Alem\-bert-Ha\-milton 
system\index{d'Alembert-Hamilton system}. 

The d'Alembert-Hamilton system plays an important role in the theory
of Poincar\'e-invariant equations for the scalar
\cite{89,102,104,107}, spinor \cite{99,103} and vector fields. In
particular, any second-order $P(1, n-1)$-invariant scalar equation can
be reduced to ODE with the use of solutions of system (\ref{2.1.3}),
(\ref{2.1.4}) (without applying the symmetry reduction technique).

The three-dimensional elliptic analogue of system of PDEs
(\ref{2.1.3}), (\ref{2.1.4})
\begin{displaymath}
u_{x_1x_1}+u_{x_2x_2}+u_{x_3x_3}=0,\quad
u_{x_1}^2+u_{x_2}^2+u_{x_3}^2=0
\end{displaymath}
with a complex-valued function $u(\vec x)$ was studied by Jacobi
\cite{20.2}, who constructed the following class of its exact
solutions 
\begin{equation}
C_0(u)+C_1(u)x_1+C_2(u)x_2+C_3(u)x_3=0,
\label{2.1.4z}
\end{equation}
where $C_0(u),\ldots,C_3(u)$ are arbitrary smooth complex-valued
functions satisfying the equality
\begin{equation}
C_1(u)^2+C_2(u)^2+C_3(u)^2=0.
\label{2.1.5z}
\end{equation}
 
Later on, Smirnov and Sobolev \cite{184.1,184.2} proved that the
formulae (\ref{2.1.4z}), (\ref{2.1.5z}) give the general solution of
the above over-determined system of PDEs. Some classes of exact
solutions of the system of PDEs (\ref{2.1.3}), (\ref{2.1.4}) were obtained
by Bateman \cite{20.1}, Cartan \cite{31.1} and Erugin \cite{52.0}.

Recently, Collins \cite{39} has obtained the general solution of the
three-dimen\-si\-on\-al d'Alembert-Hamilton system using the methods
of differential geometry. How\-ev\-er approach cannot be applied to systems of
PDEs (\ref{2.1.3}), (\ref{2.1.4}) having $n>3$ independent variables.

In the present section we will establish the necessary compatibility
conditions of system (\ref{2.1.3}), (\ref{2.1.4}) for arbitrary $n \in
{\N} $ and obtain its compatibility criterion in the case $n =
4$. Next, we will construct the general solution of the
four-dimensional d'Alembert-Hamilton system.
\vspace{2mm}

\noindent
{\bf 1. Compatibility of over-determined system of PDEs (\ref{2.1.3}),
  (\ref{2.1.4}).}\
We study the matter of compatibility of the d'Alembert-Hamilton
system under assumption that $u(x) $ is a complex-valued
function of $n$ complex variables $x_0,\ x_1, \ldots, x_{n-1} $.
Provided $F_2 (u) \ne 0 $, we can transform  system (\ref{2.1.3}),
(\ref{2.1.4}) by means of changing the dependent variable
\begin{equation}
u \rightarrow u' = \int\limits^{\edi u} 
\Bigl(F_2(\tau)\Bigr)^{-1/2} d \tau
\label{2.1.5}
\end{equation}
as follows
\begin{displaymath}
\Box_n u^\prime = F(u^\prime), \quad 
(\p_{\ssl A}u^\prime)(\p^{\ssl A}u^\prime)=1.
\end{displaymath}

Consequently, the problem of investigating compatibility
of the d'Alem\-bert-Ha\-mil\-ton system is reduced to 
studying compatibility of the system of PDEs
\begin{equation}
\Box_n u = F(u),
\quad
(\p_{\ssl A} u) (\p^{\ssl A} u) = \lbd,
\label{2.1.6}
\end{equation}
where $\lbd $ is a discrete parameter taking the values $0,\ 1$.

To solve the above problem we will need the following auxilliary
results.
\vspace{1.5mm}

\noindent
{\bf Lemma 2.1.1}\cite{107}.\ {\em Solutions of the system
  (\ref{2.1.6}) satisfy the identities
\begin{equation}
\begin{array}{l}
u_{\ssl AB} u^{\ssl AB} = -\lbd \dot{F}(u),\\[2mm]
u_{\ssl AB_1} u^{\ssl B_1 B_2} u^{\ssl A}_{\ssl B_2} = 
{\edi\lbd^2\over \edi 2!}\ddot{F} (u),\\
\ldots,\\
u_{\ssl AB_1} u^{\ssl B_1B_2}\cdot\cdots\cdot u^{\ssl B_mA} = 
{\edi(-\lbd)^m\over\edi m!}F^{(m)}(u), 
\ \ m \ge 1,
\end{array}
\label{2.1.7}
\end{equation}
where} $u_{\ssl AB} = \p_{\ssl A} \p_{\ssl B} u,\ u_{\ssl A}^{\ssl B}
=g_{\ssl BC}u_{\ssl CA},\ A,B,C = {0,\ldots, n-1}$,\ $F^{(m)}=d^mF/du^m$.
\vspace{1.5mm}

\noindent
{\em Proof.}$\quad$ We prove the assertion by means of the mathematical
induction method by m. Differentiating the second equation of
system (\ref{2.1.6}) with respect to $x_{\ssl B},\ x_{\ssl C} $ we have
\begin{equation}
u_{\ssl ABC} u^{\ssl A} + u_{\ssl AB} u^{\ssl A}_{\ssl C} = 0.
\label{2.1.8}
\end{equation}

Convoluting (\ref{2.1.8}) with the metric tensor $g_{\ssl AB} $ we
arrive at the equality
\begin{displaymath}
u_{\ssl AB} u^{\ssl AB} + u^{\ssl A} \Box_n u_{\ssl A} = 0.
\end{displaymath}

Since $\Box_n u_{\ssl A} = \p_{\ssl A} F(u) = u_{\ssl A} \dot F(u)$, 
the above expression is rewritten in the form
\begin{displaymath}
u_{\ssl AB} u^{\ssl AB} + \lbd \dot F(u) = 0.
\end{displaymath}

Consequently, the base of induction is ensured. Let us assume that
the assertion holds for $m = k \in {\N}$. We will prove that it
holds for $ m = k+1$ as well.

Convoluting (\ref{2.1.8}) with the tensor
\begin{displaymath}
u^{\ssl BB_2} u_{\ssl B_2B_3}\cdot \cdots\cdot u^{\ssl B_k C},
\end{displaymath}
gives
\begin{equation}
\begin{array}{l}
u_{\ssl AB} u^{\ssl BB_2} u_{\ssl B_2B_3}\cdot \cdots\cdot 
u^{\ssl B_k C} u^{\ssl A}_{\ssl C}\\[2mm]
\quad +u^{\ssl A} u_{\ssl ABC} u^{\ssl BB_2} u_{\ssl B_2B_3}\cdot 
\cdots\cdot u^{\ssl B_k C} = 0.
\end{array}
\label{2.1.9}
\end{equation}

Since, according to the assumption of the induction, the equalities
\begin{eqnarray*}
&& u^{\ssl A} u_{\ssl ABC} u^{\ssl BB_2} u_{\ssl B_2B_3}\cdot 
\cdots\cdot u^{\ssl B_k C} = (k+1)^{-1} u^{\ssl A} \p_{\ssl A}\\ 
&&\quad \times \Bigl(u_{\ssl BC} u^{\ssl BB_2} u_{\ssl B_2B_3}
\cdot\cdots\cdot u^{\ssl B_kC}\Bigr)= (k+1)^{-1} u^{\ssl A} \p_{\ssl
  A}\\  
&&\quad \times {(k!)^{-1}(-\lbd)^k F^{(k)} (u)} =
-\Bigl((k+1)!\Bigr)^{-1}(-\lbd)^{k+1} F^{(k+1)} (u)
\end{eqnarray*}
hold, from (\ref{2.1.9}) it follows that
\begin{displaymath}
u_{\ssl AB_1}u^{\ssl B_1B_2}\cdot\cdots\cdot u^{\ssl B_{k+1}A}
=\Bigl((k+1)!\Bigr)^{-1}(-\lbd)^{k+1}F^{(k+1)} (u).
\end{displaymath}

The lemma is proved. $\rhd$
\vspace{1.5mm}

\noindent
{\bf Lemma 2.1.2}\cite{107}.\ {\em Solutions of the system of PDEs
  (\ref{2.1.6}) satisfy the $n$-di\-men\-si\-on\-al Monge-Amp\`ere
  equation}\index{Monge-Amp\`ere equation}
\begin{equation}
{\rm det}\, \| u_{x_{\ssl A} x_{\ssl B}} \|^{n-1}_{{\ssl A,B}=0} = 0.
\label{2.1.10}
\end{equation}
{\em Proof.}$\quad$ The assertion follows from the fact that
(\ref{2.1.10}) is a criterion of functional dependence of functions
$u_{x_0},\ u_{x_1},\ldots, u_{x_{n-1}}$. $\rhd$
\vspace{1.5mm}

\noindent
{\bf Theorem 2.1.1.}\ {\em Let the d'Alembert-Hamilton system
  (\ref{2.1.6})  be compatible. Then
\begin{equation}
F(u) = \lbd \dot f(u) f^{-1} (u),
\label{2.1.11}
\end{equation}
and what is more}
\begin{equation}
{d^n f(u)\over du^n} = 0.
\label{2.1.12}
\end{equation}
{\em Proof.}$\quad$ The cases $\lbd =1$ and $\lbd =0$ have to be
considered separately. 

\noindent
{\bf \/The case $\lbd =1$.}\ Due to the Hamilton-Cayley
theorem\index{Hamilton-Cayley theorem} \cite{110} an arbitrary
$(n\times n)$-matrix $W = \|W_{\ssl AB} \|^{n-1}_{{\ssl A,B}=0} $
satisfies the following identity:
\begin{equation}
\sum\limits^{n-1}_{k=0} (-1)^k \Sigma(M_k)\ {\rm tr}\,(W^{n-k}) +
(-1)^n n\ {\rm det}\, W = 0,
\label{2.1.13}
\end{equation}
where ${\rm tr}\, \|W_{\ssl AB}\|_{{\ssl A,B}=0}^{n-1}= 
\sum\limits^{n-1}_{{\ssl C}=0}W_{\ssl CC}$ is the trace of a 
matrix $W$.

In (\ref{2.1.13}) we designate the sum of $k$-th order principal
minors of the matrix $M$ by the symbol $\Sigma(M_k)$. This sum is
determined by the recurrent formula
\begin{equation}
\begin{array}{l}
\Sigma(M_k) = k^{-1} (-1)^{k-1}
\left\{ \sum\limits^{k-1}_{l=0}(-1)^{l} \Sigma(M_l)
\ {\rm tr}\, (W^{k-l}) \right\}, \ \ k \ge 1,\\[2mm]
\Sigma(M_0)\stackrel{\rm def}{=} 1.
\end{array}
\label{2.1.14}
\end{equation}

We choose the matrix elements $W_{\ssl AB}$ as follows
\begin{displaymath}
W_{\ssl AB} = \p_{\ssl A} \p^{\ssl B} u(x), \ \ A, B = {0,\dots, n-1},
\end{displaymath}
whence due to Lemmas 2.1.1, 2.1.2 we conclude that
\begin{equation}
{\rm tr}\, (W^k) = {1\over (k-1)!} F^{(k-1)},\quad
{\rm det}\, W = 0.
\label{2.1.15}
\end{equation}

Substitution of the above formulae into (\ref{2.1.13}) gives rise to
an ODE for $F(u)$. Let us prove that this ODE is transformed to the
form (\ref{2.1.12}) by means of a nonlocal change of the dependent
variable (\ref{2.1.11}).

Introducing the notation
\begin{displaymath}
Y_N = \sum\limits^N_{k=0} (-1)^k \Sigma(M_k) {\rm tr}\, (W^{N-k+1}),
\end{displaymath}
we rewrite formula (\ref{2.1.14}) as follows
\begin{displaymath}
\Sigma(M_k) = {(-1)^{k-1}\over k} Y_{k-1},\ \ k\ge 1,
\end{displaymath}
whence
\begin{eqnarray}
&&Y_N = {\rm tr}\, (W^{N+1})- \sum^N_{k=1} {1\over k}Y_{k-1}{\rm tr}\,
(W^{N-k+1}) ={(-1)^N\over N!}\left ({\dot f\over
  f}\right)^{(N)}\non\\
&&\quad +\sum^N_{k=1}{(-1)^{N-k-1}\over k(n-k)!} 
Y_{k-1} \left({\dot f\over f}\right)^{(N-k)}, \ \ N\ge
1,\label{2.1.16}\\  
&&Y_0 ={\dot f\over f}.\non
\end{eqnarray}

Using the mathematical induction method we will prove the
equalities
\begin{equation}
Y_N = {(-1)^N\over N!}{f^{(N+1)}\over f},\ \ N \ge 1.
\label{2.1.17}
\end{equation}
Let us prove that (\ref{2.1.17}) holds under $N=1$. Due to
(\ref{2.1.15}) an expression for $Y_1$ can be rewritten in the
following way:
\begin{displaymath}
Y_1={\rm tr}\, (W^2)-\Sigma(M_1){\rm tr}\, W=-\dot F-F^2.
\end{displaymath}
 
Substitution of $F=\dot f/f$ into the above equality yields
$Y_1=-\ddot f/f$. The base of induction is established.

Let us assume that (\ref{2.1.17}) holds for all $m \le N-1$. We will
prove that (\ref{2.1.17}) holds for $m=N$ as well.

Indeed,
\begin{eqnarray*}
Y_N &=&{(-1)^N\over N!}\left({\dot f\over f}\right)^{(N)}+
 \sum^N_{k=1}{(-1)^{N-k-1}\over k(N-k)!}\left({\dot f\over
     f}\right)^{(N-k)}
{(-1)^{k-1}\over (k-1)!}\left({f^{(k)}\over f}\right)\\
&&= {(-1)^N\over N!}\left({\dot f\over f}\right)^{(N)}+
\sum^N_{k=1}{(-1)^N\over k! (N-k)!}\left({\dot f\over
     f}\right)^{(N-k)}\left({f^{(k)}\over f}\right)\\
&&={(-1)^N\over N! f}\sum_{k=0}^N{\rm C}_N^k \left({\dot f\over
     f}\right)^{(N-k)}f^{(k)}
={(-1)^N\over N! f}{f^{(N+1)}\over f}.
\end{eqnarray*}

Consequently, relation (\ref{2.1.17}) holds for all $N \in
{\N}$. Putting $N = n-1 $ yields 
\begin{displaymath}
Y_{n-1} = {(-1)^{n-1}\over (n-1)!}{f^{(n)}\over f}.
\end{displaymath}

On the other hand, using (\ref{2.1.13}), (\ref{2.1.15}) we come to
the following relation:
\begin{displaymath}
Y_{n-1} = (-1)^{n+1} n\, {\rm det}\, W = 0,
\end{displaymath}
whence $f^{(n)}(u) = 0$.
\vspace{1.5mm}

\noindent
{\bf The case $\lbd = 0$.}\ Taking into account Lemmas 2.1.1,
2.1.2 yields
\begin{displaymath}
{\rm det}\ W = 0,\quad {\rm tr}\,(W^k) = 0, \ \ k= {2,\ldots, n-1}.
\end{displaymath}

Due to these equalities formulae (\ref{2.1.13}), (\ref{2.1.14})
take the form 
\begin{eqnarray}
&&(-1)^{n-1}F \Sigma(M_{n-1}) = 0,\label{2.1.18a}\\
&&\Sigma(M_0)=1, \quad
\Sigma(M_k) = {F\over k}\Sigma(M_{k-1}).\label{2.1.18b}
\end{eqnarray}

Resolving the recurrent relations (\ref{2.1.18b}) with respect to
$\Sigma(M_k)$ we get
\begin{displaymath}
\Sigma(M_k) = {F^k\over k!},\ \ k \ge 1.
\end{displaymath}

Inserting $\Sigma(M_{n-1})=\Bigl((n-1)!\Bigr)^{-1}F^{n-1}$ into
(\ref{2.1.18a}) we have
\begin{displaymath}
{(-1)^{n-1}\over (n-1)!}F^n = 0,
\end{displaymath}
whence $F = 0 $. The theorem is proved. $\rhd$
\vspace{1.5mm}

\noindent
{\bf Consequence 2.1.1.}\ {\em The over-determined system of PDEs
\begin{equation}
\Box_n u = F(u), \quad (\p_{\ssl A} u)(\p^{\ssl A} u) = 0
\label{2.1.19}
\end{equation}
is compatible iff} $F(u) \equiv 0$.
\vspace{1.5mm}

\noindent
{\em Proof.}$\quad$ The necessity is a direct consequence of 
Theorem 2.1.1.  To prove the sufficiency we will show that
system (2.1.19) with $F(u) = 0 $ possesses nontrivial solutions. It
is straightforward to check that the function $u(x) = C_1(x_0 + x_3) + 
C_2$, where $C_1,\ C_2 $ are constants, satisfies equations (2.1.19)
under $F(u)=0$, the same as what was to be proved. $\rhd$

Let us note that the compatibility criterion for the system of
PDEs (2.1.19) with a real-valued function $u=u(x)$ has been
established in \cite{37}.

Let us say a few words about geometrical interpretation of the
d'Alembert-Hamilton system. If we designate by $P_k(u)$ a $k$-th
order polynomial, then the necessary compatibility conditions
(\ref{2.1.11}), (\ref{2.1.12}) can be represented in the form
\begin{displaymath}
F(u) = \lbd{d \over du}\ln P_k (u), \ \  0 \le k \le n-1.
\end{displaymath}

Let $\al_1,\ \al_2, \ldots, \al_k $ be the roots of the polynomial
$P_k(u) $. Then, the above relations read
\begin{equation}
F(u) = \lbd \sum\limits^k_{i=1} {1\over u + \al_i},
\ \ 1 \le k \le n-1
\label{2.1.20}
\end{equation}
or
\begin{equation}
F(u) = 0,\ \  k=0.
\label{2.1.21}
\end{equation}

According to \cite{37,108,188.1} the parameters
\begin{eqnarray*}
\vark_i &=& (\al_i)^{-1},\ \ i = {1,\ldots,k},\\
\wid{\vark}_j &=& 0,\ \ j = {k+1,\ldots, n-1}
\end{eqnarray*}
can be interpreted as the principal curvatures of the level surface
of the solution of system (\ref{2.1.6}), (\ref{2.1.20}) under $\lbd = 1$.
Consequently, solutions of the d'Alembert-Hamilton system have
the remarkable geometrical property: their level surfaces have
all principal curvatures constant (for the first time this fact was
established by Cartan \cite{31.1}).

Now we adduce an assertion giving the compatibility criterion
of the nonlinear d'Alembert-Hamilton system (\ref{2.1.3}),
(\ref{2.1.4}) in the case $n=4$
\begin{equation}
\Box u = F_1(u), \quad
(\p_\mu u) (\p^\mu u) = F_2(u).
\label{2.1.22}
\end{equation}

Here $u = u(x_0, x_1, x_2, x_3) \in C^2 ({\C}^4,\, {\C}^1)$,
\ $\{F_1, F_2\} \subset C ({\C}^1,\, {\C}^1)$.
\vspace{1.5mm}

\noindent
{\bf Theorem 2.1.2.}\ {\em System of PDEs (\ref{2.1.22}) is
  compatible iff the functions $F_1,\ F_2 $ have the form
\index{Compatibility of the d'Alembert-Ha\-mil\-ton system}
\begin{eqnarray}
&1)& F_1(u) = F_2(u) = 0,\ \mbox{\it or}\non\\
&2)& F_1(u) = N(\dot f f)^{-1} - \ddot{f} (\dot f)^{-3},\quad
F_2(u) = (\dot f)^{-2},\label{2.1.23}
\end{eqnarray}
where $f = f(u) \in C^2 ({\C}^1,\ {\C}^1) $ is an arbitrary function
satisfying the condition $\dot f \not\equiv 0$,\ $ N $ is a discrete
parameter taking the values $0, 1, 2, 3$.}

The proof can be found in \cite{108}.
\vspace{1.5mm}

\noindent
{\bf Note 2.1.1.}\ System of PDEs (\ref{2.1.22}) with $F_1,
\ F_2$ given by formulae (\ref{2.1.23}) is transformed to the form
\begin{equation}
\Box u = Nu^{-1}, \quad
 (\p_\mu u) (\p^\mu u) = 1
\label{2.1.24}
\end{equation}
by means of the change of the dependent variable
\begin{equation}
u \rightarrow u^\prime = f(u).
\label{2.1.25}
\end{equation} 
{\bf Theorem 2.1.3.}\ {\em Let $u = u(x) $ be a real-valued
  function of four real variables $x_0,\ x_1$,\ $x_2$,\ $x_3 $. Then,
  system (\ref{2.1.22}) is compatible iff the functions $F_1,\ F_2 $
  have the form
\index{Compatibility of the d'Alembert-Ha\-mil\-ton system}
\begin{eqnarray}
&1)& F_1(u) = F_2(u) = 0,\ \mbox{\it or}\non\\
&2)& F_1(u) = \ve N(\dot ff)^{-1} - \ve \ddot{f} (\dot f)^{-3},\quad
F_2(u) = \ve (\dot f)^{-2},\label{2.1.26}
\end{eqnarray}
where $f = f(u) \in C^2 ({\R}^1, {\R}^1) $ is an arbitrary
function satisfying the condition $ \dot f \not\equiv 0 ;\ N $ is a
discrete parameter taking the values $0, 1, 2, 3;\ \ve = \pm 1$.}
\vspace{1.5mm}

\noindent
{\bf Note 2.1.2.}\ System of PDEs (\ref{2.1.22}) with $F_1,\ F_2 $
given by formulae (\ref{2.1.26}) is transformed to the form
\begin{equation}
\Box u = \ve N u^{-1}, \quad (\p_\mu u) (\p^\mu u) = \ve
\label{2.1.27}
\end{equation}
by means of the change of the dependent variable (\ref{2.1.25}).
\vspace{1.5mm}

\noindent
{\bf Note 2.1.3.}\ It follows from Theorem 2.1.3 that the nonlinear
differential operator $\ve u^2 \Box $ defined on the set of
solutions of the PDE $(\p_\mu u)(\p^\mu u) = \ve$ has a discrete
spectrum, i.e.,\  
\begin{equation}
\ve u^2 \Box u = Nu,\ \  N = 0,1,2,3
\label{2.1.28}
\end{equation}
and what is more, the spectrum is determined by the dimension
of the space of independent variables only. Consequently, the
nonlinear additional constraint $(\p_\mu u)(\p^\mu u) = \ve $
plays the same role as the boundary conditions in the Sturm-Liouville
problem \cite{41}.

It is natural to expect that an additional constraint
changes the symmetry properties of the d'Alembert equation.
This conjecture is confirmed by comparison of results
given in the Tables 2.1.1, 2.1.2.
\vspace{2mm}

\noindent
{\em Table 2.1.1.}\ {\bf Symmetry of the nonlinear d'Alembert}\\
\phantom{{\em Table 2.1.1.}\ }{\bf equation (\ref{2.1.1}) with
  {\boldmath $n$ = 4}}  
\vspace{1.5mm}

\noindent
\begin{tabular}{|l|l|l|}
\hline
\ N & \ \ \ Invariance group &\qquad\quad $F(u)$ \\
\hline
1. & the Poincar\'e group $P(1,3)$ & arbitrary smooth function \\
2. & the extended Poincar\'e group & $C_1 (u + C_2)^k $, \\
& $\wid P(1,3)$ \cite{78,89} & $C_1 \exp\{ku\} $ \\
3. & the conformal group & $C_1 (u + C_2)^3 $ \\
& $C(1,3) $ \cite{78,126} & \\
\hline
\end{tabular}
\vspace{1.5mm}

\noindent
Here $C_1,\ C_2,\ k $ are arbitrary constants.
\vspace{2mm}

\noindent
{\em Table 2.1.2.}\ \ {\bf Symmetry of the system}\\
{\phantom{{\em Table 2.1.2.}\ }{\boldmath $\Box u = F(u),\ \
    (\p_\mu u)(\p^\mu u) = \lbd$} 
\vspace{1.5mm}

\noindent
\begin{tabular}{|l|l|l|l|}
\hline
\ N &\ \ \  Invariance group &\qquad\qquad $F(u)$ &\quad $\lbd $ \\
\hline
1. & the Poincar\'e group & arbitrary smooth function & $ \lbd \in
{\R}^1 $\\ 
& $P(1,3) $ & & \\
2. & the extended Poincar\'e & $C_1 (u + C_2)^{-1} $ & $ \lbd \in
{\R}^1 $\\ 
& group $ \wid P(1,3) $ \cite{89} & & \\
3. & the conformal group & $3 \lbd (u + C_1)^{-1} $ & $  \lbd \in
{\R}^1 $\\ 
& $C(1,3) $ \cite{102,104} & & \\
4. & the generalized Poincar\'e & 0 & $\lbd > 0 $ \\
& group $P(1,4)$ & & \\
5. & the generalized Poincar\'e & 0 & $\lbd < 0 $ \\
& group $P(2,3) $ & & \\
6. & infinite-dimensional & 0 & $ 0 $ \\
& group & & \\
\hline
\end{tabular}
\vspace{1.5mm}

\noindent
Here $C_1,\ C_2$ are arbitrary constants.
\vspace{2mm}

Comparing Tables 2.1.1, 2.1.2 we come to the conclusion that
the conformally non-invariant nonlinear d'Alembert equation $\Box u = 3
u^{-1}$ after being restricted to the set of solutions of the Hamilton
equation $(\p_\mu u)(\p^\mu u) = 1$ admits the conformal group
$C(1,3)$. Consequently, an additional constraint $(\p_\mu u)(\p^\mu u)
= 1$ ``selects'' a subset of solutions which is invariant under the
group $C(1,3)$. In other words, the nonlinear d'Alembert equation
$\Box u = 3 u^{-1}$ is {\em conditionally-invariant} \/with respect to
the conformal group.

Such a definition of conditional
invariance\index{Conditional!invariance} is much more general than
that introduced in Chapter 1. Indeed, when defining in Section 1.5 a
conditional invariance of a given PDE we restricted ourselves to
considering additional constraints which were first-order quasi-linear
PDEs. It is straightforward to verify that the nonlinear d'Alembert
equation mentioned in the previous paragraph is not
conditionally-invariant with respect to conformal group in the sense
of Definition 1.5.3. Nevertheless, its generalized conditional
invariance can be used effectively to construct exact solutions. The
peculiarity is that Ans\"atze invariant under three-dimensional
subalgebras of the conformal algebra not belonging to the Lie algebra
of the extended Poincar\'e group reduce the equation $\Box u = 3
u^{-1}$ to two ODEs.

But we are not going to apply the symmetry reduction procedure to
constructing solutions of the d'Alembert-Hamilton system, since we
have developed a method enabling us to construct its general solution.
\vspace{1.5mm}

\noindent
{\bf 2. Integration of the d'Alembert-Hamilton system.}\ It follows
from Theorem 2.1.2 that the compatible system of PDEs
(\ref{2.1.3}), (\ref{2.1.4}) is equivalent either to (\ref{2.1.24}) or
to the following system: 
\begin{equation} 
\Box u=0,\quad (\p_\mu u)(\p^\mu u)=0.
\label{2.1.29z}
\end{equation}

General solutions of systems of PDEs (\ref{2.1.24}), (\ref{2.1.29z})
are given by the following assertions.
\index{General solution!of the d'Alembert-Hamilton system}
\vspace{1.5mm}

\noindent
{\bf Theorem 2.1.4.}\ {\em The general solution of system of PDEs
(\ref{2.1.24}) is given by one of the following formulae: 

\noindent
1)\ $N = 0$,
\begin{displaymath}
u = A_\mu(\tau) x^\mu + R_1 (\tau),
\end{displaymath}
where $\tau = \tau(x)$ is determined in implicit way
\begin{displaymath}
B_\mu (\tau) x^\mu + R_2(\tau) = 0
\end{displaymath}
and $A_\mu(\tau),\ B_\mu(\tau),\ R_1(\tau),\ R_2(\tau)$ are
arbitrary smooth complex-valued functions satisfying the conditions
\begin{displaymath}
A_\mu A^\mu = 1, \quad A_\mu B^\mu = 0,
\quad \dot A_\mu B^\mu = 0,\quad B_\mu B^\mu = 0;
\end{displaymath}
2)\ $N = 1$,
\begin{displaymath}
u^2 = (a_{\mu} x^{\mu} + G_1)^2 - (b_{\mu} x^{\mu} + G_2)^2,
\end{displaymath}
where $G_i = G_i(\theta_\mu x^{\mu}) \in C^2({\C}^1, {\C}^1) $ are
arbitrary functions, $a_{\mu},\ b_{\mu},\ \theta_{\mu}$ are arbitrary
complex parameters satisfying the conditions
\begin{displaymath}
a_{\mu} a^{\mu} =- b_{\mu} b^{\mu} = 1,
\quad
a_{\mu} b^{\mu} = a_{\mu} \theta^{\mu} =
b_{\mu} \theta^{\mu} = \theta_{\mu} \theta^{\mu} = 0;
\end{displaymath}
3)\ $N = 2$,
\vspace{1.5mm}

\noindent
a)\ $u^2 = \Bigl(x_{\mu} + A_{\mu} (\tau)\Bigr)\Bigl(x^{\mu} +A^{\mu}
(\tau)\Bigr) + \Bigl\{B_{\mu} (\tau)\Bigl(x^{\mu} + A^{\mu}
(\tau)\Bigr)\Bigr\}^2,$ 
where $\tau = \tau(x) $ is determined in implicit way
\begin{displaymath}
\Bigl(x_{\mu} +A_{\mu} (\tau)\Bigr)\dot B^{\mu}(\tau) = 0,
\end{displaymath}
$A_{\mu} (\tau),\ B_{\mu} (\tau)$ are arbitrary smooth complex-valued
functions satisfying the conditions
\begin{displaymath}
B_{\mu} B^{\mu} = -1,
\quad
\dot {B}_{\mu} \dot {B}^{\mu} =0,
\quad
\dot A_{\mu} = R(\tau) \dot B_{\mu}
\end{displaymath}
with arbitrary $R(\tau) \in C^1 ({\C}^1,\ {\C}^1)$;
\vspace{1.5mm}

\noindent
b)\ $u^2 = \Bigl(x_\mu + A_\mu (\tau)\Bigr)\Bigl(x^\mu + A^\mu
(\tau)\Bigr)+ 
\Bigl\{b_\mu\Bigl(x^\mu + A^\mu (\tau)\Bigr)\Bigr\}^2$,
where $\tau = \tau(x)$ is determined in implicit way
\begin{displaymath}
\Bigl(x_\mu + A_\mu (\tau)\Bigr)\dot A^\mu(\tau) + \Bigl(x_\mu + A_\mu
(\tau)\Bigr) b^\mu b_\nu \dot A^\nu(\tau) = 0,
\end{displaymath}
$A_\mu (\tau)$ are arbitrary smooth complex-valued functions
satisfying the condition
\begin{displaymath}
\dot A_\mu \dot A^\mu + (b_\mu \dot A^\mu)^2 = 0,
\end{displaymath}
$b_\mu $ are arbitrary complex constants satisfying the condition
$b_\mu b^{\mu} = -1$;
\vspace{1.5mm}

\noindent
4)\ $N = 3$,
\begin{equation}
u^2 = \Bigl(x_\mu + A_\mu (\tau)\Bigr)\Bigl(x^\mu + A^\mu
(\tau)\Bigr), 
\label{2.1.30z}
\end{equation}
where $\tau = \tau(x)$ is determined in implicit way
\begin{equation}
\Bigl(x_\mu + A_\mu (\tau)\Bigr)B^\mu (\tau) = 0,
\label{2.1.31a}
\end{equation}
$A_\mu (\tau),\ B_\mu(\tau) $ are arbitrary smooth complex-valued
functions satisfying the conditions }
\begin{equation}
\dot A_\mu B^\mu = 0, \quad B_\mu B^\mu = 0.
\label{2.1.31b}
\end{equation}
{\em Proof.}$\quad$ We will give a detailed proof of the theorem for the 
case $N=3$. In the remaining cases only the schemes of the proofs will
be outlined.

Our approach for integration of the d'Alembert-Hamilton system is based
on the generalization of the nonlocal transformation method
\cite{92,93} to a case of multi-dimensional PDEs suggested in
\cite{108}--\cite{108.2}. 

By a nonlocal transformation\index{Nonlocal!transformation} of the
order $r$ we mean the transformation 
\begin{equation}
\begin{array}{l}
x_\mu^\prime=f_\mu(x,\, u,\, \mathop u \limits_{\scriptscriptstyle
  1},\ldots, \mathop u\limits_{\scriptscriptstyle k}),\\[2mm]
u^\prime=f(x,\, u,\, \mathop u \limits_{\scriptscriptstyle 1},\ldots,
\mathop u\limits_{\scriptscriptstyle k}),
\end{array}
\label{2.1.32z}
\end{equation}
where $\{f_\mu, f\} \subset C^r(\C^n,\C^1)$, the symbol $\mathop
u\limits_{\scriptscriptstyle s}$ denotes the set of second-order
derivatives of the function $u=u(x)$. 

A principal idea of the mentioned method is to linearize a PDE under
study by means of the proper nonlocal transformation (\ref{2.1.32z}).
If we succeed in constructing a solution of the linear equation
(general or particular), then a solution of the initial equation is
obtained by inverting transformation (\ref{2.1.32z}).

Especially important are the contact 
transformations\index{Contact transformation} (first-order
nonlocal transformations)
\begin{equation}
\label{2.1.33z}
x_\mu^\prime=f_\mu(x,\, u,\, \mathop{u}\limits_{\scriptscriptstyle
  1}),\quad u^\prime=f(x,\, u,\, \mathop{u}
\limits_{\scriptscriptstyle 1}),\quad 
u^\prime_{x_\mu}=g_\mu(x,\, u,\, \mathop{u}
\limits_{\scriptscriptstyle 1}), 
\end{equation}
which preserve the first-order tangency condition
\begin{displaymath}
du-u_{x_\mu}dx_\mu=0\Rightarrow
du^\prime-u^\prime_{x_\mu^\prime}dx_\mu^\prime=0.
\end{displaymath}

This fact is explained by that any two first-order PDEs can be
transformed one into another by means of a proper contact
transformation \cite{127,148}.

According to Lemma 2.1.2 ${\rm det}\, \|u_{x_\mu
  x_\nu}\|_{\mu,\nu=0}^3=0$. Consequently, the rank of the matrix
$U=\|u_{x_\mu x_\nu}\|_{\mu,\nu=0}^3=0$ takes the values 1,2,3. Each
case listed has to be considered separately.
\vspace{1.5mm}

\noindent
{\bf Case 1.} ${\rm rank}\, U=3$. With such a condition there is a
non-vanishing third-order minor of the matrix $U$. Making, if
necessary, changes $x_0\to ix_a,\ x_a\to ix_0$ or $x_a\to x_b,\ x_b\to
x_a$ which leave system (\ref{2.1.24}) invariant we can choose 
\begin{equation} 
{\rm det}\, \|u_{x_a x_b}\|_{a,b=1}^3\ne 0.
\label{2.1.34z}
\end{equation}

Performing the generalized Euler-Amp\`ere transformation
\cite{108}: 
\begin{eqnarray}
&&y_0=x_0,\quad y_a=u_{x_a},\quad
H(y)=x_au_{x_a}-u,\non\\
&&H_{y_0}=-u_{x_0},\quad H_{y_a}=x_a,\ \ a=1,2,3,\non\\
&&H_{12}=-\left|\matrix{u_{12}&u_{23}\cr
u_{13}&u_{33}\cr}\right|\Delta^{-1},\quad
H_{11}=\left|\matrix{u_{22}&u_{23}\cr
u_{23}&u_{33}\cr}\right|\Delta^{-1},\non\\
&&H_{31}=-\left|\matrix{u_{12}&u_{22}\cr
u_{13}&u_{23}\cr}\right|\Delta^{-1},\quad
H_{22}=\left|\matrix{u_{11}&u_{13}\cr
u_{13}&u_{33}\cr}\right|\Delta^{-1},\non\\
&&H_{23}=-\left|\matrix{u_{11}&u_{12}\cr
u_{13}&u_{23}\cr}\right|\Delta^{-1},\quad
H_{33}=\left|\matrix{u_{11}&u_{12}\cr
u_{12}&u_{22}\cr}\right|\Delta^{-1},\label{2.1.35z}\\
&&H_{01}=-\left|\matrix{u_{01}&u_{02}&u_{03}\cr
u_{12}&u_{22}&u_{23}\cr
u_{13}&u_{23}&u_{33}\cr}\right|\Delta^{-1},\non\\
&&H_{02}=-\left|\matrix{u_{11}&u_{12}&u_{13}\cr
u_{01}&u_{02}&u_{03}\cr
u_{13}&u_{23}&u_{33}\cr}\right|\Delta^{-1},\non\\
&&H_{03}=-\left|\matrix{u_{11}&u_{12}&u_{13}\cr
u_{12}&u_{22}&u_{23}\cr
u_{01}&u_{02}&u_{03}\cr}\right|\Delta^{-1},\non\\
&&H_{00}=-\Delta^{-1}{\rm det}\,\|u_{\mu\nu}\|_{\mu,\nu=0}^3,\non
\end{eqnarray}
where $u_{\mu\nu}=u_{x_\mu x_\nu},\ H_{\mu\nu}=H_{y_\mu y_\nu}$,\
$|W|={\rm det}\, \|W\|$,\ $\Delta={\rm det}\,\|u_{a b}\|_{a,b=1}^3$,
in (\ref{2.1.24}) we get 
\begin{equation}
\begin{array}{l}
{\rm det}\,\|H_{y_\mu y_\nu}\|_{\mu,\nu=0}^3 +\Sigma_2(H)+
3[T(H)]^{-1}{\rm det}\,\|H_{y_a y_b}\|_{a,b=1}^3=0,\\[2mm]
H_{y_0}=-(1+y_ay_a)^{1/2}.
\end{array}
\label{2.1.36z}
\end{equation}

Hereafter $T(H)=y_a H_{y_a} - H$,\ $\Sigma_2(H)$ is the sum of the
second-order principal minors of the matrix $\|H_{y_\mu
  y_\nu}\|_{\mu,\nu=0}^3$.

Thus, instead of the nonlinear Hamilton equation, we have a 
simple linear PDE which is easily integrated
\begin{equation}
H=-y_0(1+y_ay_a)^{1/2}-B(y_1,y_2,y_3),
\label{2.1.37}
\end{equation}
where $B\in C^2(\C^3,\C^1)$ is an arbitrary function.

Inserting (\ref{2.1.37}) into the first equation from (\ref{2.1.36z})
and multiplying by $T(H)$ we note that the equation obtained is
rewritten in the following way:
\begin{equation}
a_1y_0^2+a_2y_0+a_3=0,
\label{2.1.38}
\end{equation}
where
\begin{eqnarray*}
a_1&=&\triangle_3B+y_a y_b B_{y_ay_b}+3T(B),\\
a_2&=&\Sigma_2(B)+ y_a y_b B_{y_ay_b}\triangle_3B
-y_a y_b B_{y_ay_c}B_{y_cy_b}-3[T(B)]^2,\\
a_3&=&(1+y_ay_a){\rm det}\,\|B_{y_a y_b}\|_{a,b=1}^3
+[T(B)]^3.
\end{eqnarray*}

Since $a_1,\ a_2,\ a_3$ are independent of $y_0$, from (\ref{2.1.38})
it follows that $a_1=a_2=a_3=0$.

Thus, we reduce d'Alembert-Hamilton system (\ref{2.1.24}) with ${\rm
  rank}\, U=3$ to the system of three nonlinear PDEs with three
independent variables

\begin{eqnarray}
&1)&\triangle_3B+y_a y_b B_{y_ay_b}=-3T(B),\non\\
&2)&\Sigma_2(B)+ y_a y_b B_{y_ay_b}\triangle_3B
-y_a y_b B_{y_ay_c}B_{y_cy_b}=3[T(B)]^2,\label{2.1.39}\\
&3)&{\rm det}\,\|B_{y_a y_b}\|_{a,b=1}^3=-[T(B)]^3(1+y_ay_a)^{-1}.\non
\end{eqnarray}

The above system is simplified substantially by means of the following
change of variables: 
\begin{equation}
\begin{array}{l}
z_a=y_a(1+y_by_b)^{-1/2},\\[2mm]
P(z_1,z_2,z_3)=(1+y_ay_a)^{-1/2}B(y_1,y_2,y_3).
\end{array}
\label{2.1.40}
\end{equation}

In the new variables $z,\ P(z)$ system (\ref{2.1.39}) reads
\begin{eqnarray}
&1)&\triangle_3P-z_a z_b P_{z_az_b}=0,\non\\
&2)&\Sigma_2(P)-z_a z_b P_{z_az_b}\triangle_3P
+z_a z_b P_{z_az_c}P_{z_cz_b}=0,\label{2.1.40z}\\
&3)&{\rm det}\,\|P_{z_a z_b}\|_{a,b=1}^3=0.\non
\end{eqnarray}

Since ${\rm   det}\, \wid P = {\rm det}\, \|P_{z_az_b}\|_{a,b=1}^3=0$,
the rank of the $(3\times 3)$-matrix $\wid P$ is equal either to $1$ or
to $2$.    
\vspace{1.5mm}

\noindent
{\bf Subcase 1.1.}\ ${\rm rank}\,\wid P=1$. Hence, according to the
theorem about an implicit function, it follows that there are such
functions $\{R_1, R_2\} \subset C^2(\C^1,\C^1)$ that
\begin{equation}
P_{z_k}=R_k(P_{z_3}),\ \ k=1,2.
\label{2.1.41}
\end{equation}

Substitution of (\ref{2.1.41}) into the second equation of system
(\ref{2.1.40z}) shows that its left-hand side vanishes under arbitrary 
$R_1,\ R_2$. The first equation takes the form
\begin{displaymath}
\Bigl(1+\dot R_k\dot R_k-(z_k\dot R_k +z_3)^2\Bigr)P_{z_3z_3}=0,
\end{displaymath}
whence 
\begin{equation}
P_{z_3z_3}=0
\label{2.1.42}
\end{equation}
or
\begin{equation}
1 + \dot R_k\dot R_k - (z_k\dot R_k +z_3) = 0,
\label{2.1.43}
\end{equation}
where $\dot R_k=dR_k/dP_{z_3},\ k=1,2$. Hereafter in this section, the
summation over the repeated indices denoted by the letters $k,\ l,\ n$
from $1$ to $2$ is understood.

Let the equality (\ref{2.1.42}) hold true. Then, differentiating 
(\ref{2.1.41}) with respect to $z_3$ we have $P_{z_1z_3} = P_{z_2z_3}
= 0$. Next, differentiating (\ref{2.1.41}) with respect to $z_1,\ z_2$
we conclude that $P_{z_az_b}=0,\ a,b=1,2,3$, whence
\begin{equation}
P=C_az_a+C_0,\ C_\mu\in\C^1.
\label{2.1.44}
\end{equation}

Now we turn to the case $P_{z_3z_3}\ne 0$. Hence it follows that the
equality (\ref{2.1.43}) holds. To integrate system of the first-order
PDEs (\ref{2.1.41}), (\ref{2.1.43}) we make the contact transformation
\begin{eqnarray*}
&&t_k=z_k,\quad t_3=P_{z_3},\quad G(t_1,t_2,t_3)=z_3P_{z_3}-P,\\
&&G_{t_k}=-P_{z_k},\quad G_{t_3}=z_3,\ \ k=1,2.
\end{eqnarray*}

As a result, we get
\begin{equation}
\begin{array}{l}
G_{t_k}=-R_k(t_3),\ \ k=1,2,\\[2mm]
1+\dot R_k(t_3)\dot R_k(t_3)-\Bigl(t_k\dot R_k(t_3)
+G_{t_3}\Bigr)^2=0. 
\end{array}
\label{2.1.45}
\end{equation}

Integration of the first two equations of system (\ref{2.1.45}) yields     
\begin{equation}
G=-t_k R_k(t_3)+Q(t_3),
\label{2.1.46}
\end{equation}
where $Q\in C^2(\C^1,\C^1)$ is an arbitrary function.

Substituting the result obtained into the third equation of system
(\ref{2.1.45}) we have
\begin{equation}
1+\dot R_k\dot R_k-\Bigl(t_k\dot R_k 
-t_k\dot R_k +\dot Q\Bigr)^2\equiv
1+\dot R_k\dot R_k-\dot Q^2=0.
\label{2.1.47}
\end{equation}

Thus, formulae (\ref{2.1.46}), (\ref{2.1.47}) determine the general
solution of system of PDEs (\ref{2.1.45}). Returning to the initial
variables $z,\ P(z)$ we obtain the general solution of system
(\ref{2.1.41}), (\ref{2.1.43})
\begin{equation}
P=z_k R_k(t_3)+t_3z_3-Q(t_3),\quad 
1+\dot R_k\dot R_k-\dot Q^2=0,
\label{2.1.48}
\end{equation}
where $t_3=t_3(z)$ is determined by the relation $G_{t_3}=z_3$, whence
\begin{equation}
z_k\dot R_k(t_3)+z_3-\dot Q(t_3)=0.
\label{2.1.49}
\end{equation}

To represent formulae (\ref{2.1.48}), (\ref{2.1.49}) in a manifestly
$O(3)$-invariant form we re-determine the parametric function to be
$t_3(z)=\wid R_3\Bigl(\tau(z)\Bigr)$ and designate
\begin{displaymath}
\wid R_k(\tau)=R_k\Bigl(\wid R_3(\tau)\Bigr),\quad
\wid Q(\tau)=-Q\Bigl(\wid R_3(\tau)\Bigr),\ \ k=1,2.
\end{displaymath}

With such notations formulae (\ref{2.1.48}), (\ref{2.1.49}) read
\begin{equation}
P=z_a \wid R_a(\tau)+\wid Q(\tau),\quad 
\dot {\wid R_a}\dot {\wid R_a}-\dot {\wid Q^2}=0,
\label{2.1.50}
\end{equation}
where $\tau=\tau(z)$ is a smooth function defined by the relation
\begin{equation}
z_a\dot {\wid R}_a(\tau)+\dot {\wid Q}(\tau)=0.
\label{2.1.51}
\end{equation}

Thus, the general solution of system of PDEs (\ref{2.1.40z}) is given
by one of formulae (\ref{2.1.44}) or (\ref{2.1.50}), (\ref{2.1.51}).
Making the change of variables (\ref{2.1.40}) we obtain the general
solution of the system of nonlinear PDEs (\ref{2.1.39})
\begin{eqnarray}
B(y)&=&C_ay_a+C_0(1+y_ay_a)^{1/2},\label{2.1.52}\\
B(y)&=&y_a\wid R_a(\tau)+\wid Q(\tau) (1+y_ay_a)^{1/2},\quad
\dot {\wid R}_a\dot {\wid R}_a-\dot {\wid Q^2}=0,\label{2.1.53}
\end{eqnarray}
where $\tau=\tau(y)$ is a smooth function determined in implicit way
\begin{equation}
y_a\dot {\wid R}_a(\tau)+\dot {\wid Q}(\tau)(1+y_ay_a)^{1/2}=0.
\label{2.1.54}
\end{equation}

Evidently, the solution (\ref{2.1.52}) is contained in the class
(\ref{2.1.53}), (\ref{2.1.54}). Inserting the expression for the
function $B(y)$ from (\ref{2.1.53}) into (\ref{2.1.37}) we have
\begin{displaymath}
H(y)=- (1+y_ay_a)^{1/2}\Bigl(y_0+\wid Q(\tau)\Bigr)-y_a\wid R_a(\tau),
\end{displaymath}
where the function $\tau=\tau(y)$ is determined by (\ref{2.1.54}).

At last, rewriting the expression obtained in the initial variables
$x,\ u(x)$ we arrive at the following class of solutions of the
d'Alembert-Hamilton system (\ref{2.1.24}):
\begin{equation}
u(x)=x_ay_a-H=\Bigl(x_a+\wid R_a(\tau)\Bigr)y_a
+(1+y_ay_a)^{1/2}\Bigl(x_0+\wid Q(\tau)\Bigr),
\label{2.1.55}
\end{equation}
where $y_a=y_a(x)$ are determined by the equalities
\begin{displaymath}
x_a=H_{y_a}=-\wid R_a(\tau)-y_a(1+y_by_b)^{-1/2}
\Bigl(x_0+\wid Q(\tau)\Bigr),\ \ a=1,2,3.
\end{displaymath}
Resolving the above equalities with respect to $y_a$ we get
\begin{displaymath}
y_a=-(x_a+\wid R_a)\Bigl((x_0+\wid Q)^2-(x_b+\wid R_b)(x_b+\wid
R_b)\Bigr)^{-1/2}.
\end{displaymath}

Substitution of the expressions obtained into (\ref{2.1.55}) yields
\begin{displaymath}
u(x)=\Bigl((x_0+\wid Q)^2-(x_b+\wid R_b)(x_b+\wid
R_b)\Bigr)^{1/2},
\end{displaymath}
where $\tau=\tau(x)$ is a smooth function determined by the equality
\begin{displaymath}
y_a\dot {\wid R}_a(\tau)+\dot {\wid Q}(\tau)(1+y_ay_a)^{1/2}
\equiv \Bigl(x_0+\wid Q(\tau)\Bigr)\dot {\wid Q}(\tau)-
\Bigl(x_a+\wid R_a(\tau)\Bigr)\dot{\wid R}_a(\tau)=0
\end{displaymath}
and $\wid Q,\ \wid R_a$ are arbitrary smooth functions satisfying the
relation $\dot {\wid R}_a\dot {\wid R}_a-\dot {\wid Q}^2=0$.
Introducing the notations $A_0=\wid Q,\ A_a=\wid R_a$ we obtain
formulae (\ref{2.1.30z})--(\ref{2.1.31b}) under $B_\mu\equiv {\dot
  A}_\mu,\ \mu={0,\ldots,3}$. 
\vspace{1.5mm}

\noindent
{\bf Subcase 1.2.}\ ${\rm rank}\, \wid P=2$. Without loss of
generality, we can assume that
\begin{displaymath}
{\rm det}\, \left\|\matrix{P_{z_1z_1}&P_{z_1z_2}\cr
P_{z_1z_2}&P_{z_2z_2}\cr}\right\|\ne 0.
\end{displaymath}

Consequently, there is such a function $R\in C^3(\C^2,\C^1)$ that
the relation $P_{z_3} = R(P_{z_1},\, P_{z_2})$ holds. With account of
this fact system (\ref{2.1.40z}) is rewritten in the following way:
\begin{eqnarray}
&&P_{z_kz_k} + (z_k+z_3R_k)(z_n+z_3R_n)
P_{z_kz_n}=0,\non\\
&&(1-z_kz_k-z_3^2)(1+R_kR_k)+(z_3-z_kR_k)^2=0,\label{2.1.56}\\
&&P_{z_3}=R(P_{z_1},\, P_{z_2}).\non
\end{eqnarray}
 
Here $R_k=\p R/\p(P_{z_k}),\ k=1,2$.

Let us perform in (\ref{2.1.56}) the following contact transformation:
\begin{eqnarray*}
&&t_k=P_{z_k},\quad t_3=z_3,\quad G(t_1,t_2,t_3)=z_kP_{z_k}-P,\\
&&G_{t_k}=z_k,\quad G_{t_3}=-P_{z_3},\ \ k=1,2,\\
&&G_{t_1t_1}=\delta^{-1}P_{z_2z_2},\quad
G_{t_1t_2}=-\delta^{-1}P_{z_1z_2},\\ 
&&G_{t_2t_2}=\delta^{-1}P_{z_1z_1}, \quad
G_{t_3t_3}=-\delta^{-1}{\rm det}\, \|P_{z_az_b}\|_{a,b=1}^3,\\
&&G_{t_1t_3}=\delta^{-1}(P_{z_1z_2}P_{z_2z_3}-P_{z_2z_2}P_{z_1z_3}),\\
&&G_{t_2t_3}=\delta^{-1}(P_{z_1z_2}P_{z_1z_3}-P_{z_1z_1}P_{z_2z_3}),
\end{eqnarray*}
where $\delta=P_{z_1z_1}P_{z_2z_2}-P_{z_1z_2}^2\ne 0$.

Being rewritten in the new variables $t,\ G(t)$ system (\ref{2.1.56})
takes the form
\begin{eqnarray}
&1)& \Bigl(1+R_{t_2}^2-(G_{t_2}+t_3R_{t_2})^2\Bigr)G_{t_1t_1}
-2\Bigl(R_{t_1}R_{t_2}-(G_{t_1}+t_3R_{t_1})\non\\
&&\times(G_{t_2}+t_3R_{t_2})\Bigr)G_{t_1t_2}
+\Bigl(1+R_{t_1}^2-(G_{t_1}+t_3R_{t_1})^2\Bigr)G_{t_2t_2}=0,\non\\
&2)&(1-t_3^2-G_{t_k}G_{t_k})(1+R_{t_k}R_{t_k})
+(t_3-R_{t_k}G_{t_k})^2=0,\label{2.1.57}\\
&3)& G_{t_3}=R(t_1,t_2).\non
\end{eqnarray}

Integrating equation 3 from (\ref{2.1.57}) we have
\begin{equation}
G=-t_3R(t_1,t_2)+iQ(t_1,t_2),
\label{2.1.58}
\end{equation}
where $Q\in C^3(\C^2,\C^1)$ is an arbitrary function.

Substituting the expression (\ref{2.1.58}) into the equations 1,2 from
(\ref{2.1.57}) and splitting with respect to the variable $t_3$ we
arrive at the two-dimensional system of PDEs for the functions
$R(t_1,t_2),\ Q(t_1,t_2)$:
\begin{eqnarray}
&1)& (1+Q_{t_k}Q_{t_k})(1+R_{t_n}R_{t_n})-(Q_{t_k}R_{t_k})^2=0,
\label{2.1.59}\\
&2)& (1+Q_{t_k}Q_{t_k}+R_{t_k}R_{t_k})\triangle_2 Q
-(Q_{t_k}Q_{t_n}+R_{t_k}R_{t_n})Q_{t_kt_n}=0,\non \\
&3)& (1+Q_{t_k}Q_{t_k}+R_{t_k}R_{t_k})\triangle_2 R
-(Q_{t_k}Q_{t_n}+R_{t_k}R_{t_n})R_{t_kt_n}=0,\non
\end{eqnarray}
where $\triangle_2=\p_{t_1}^2+\p_{t_2}^2$.

We have succeeded in integrating the over-determined system
(\ref{2.1.59}). Making use of formulae (\ref{2.1.37}),
(\ref{2.1.40}), (\ref{2.1.58}), we rewrite its general solution
in the initial variables $x,\ u(x)$. After representing the result
obtained in a manifestly covariant form we arrive at formulae 
(\ref{2.1.30z})--(\ref{2.1.31b}).
\vspace{1.5mm}

\noindent
{\bf Case 2.}\ ${\rm rank}\, U<3$. When studying the
compatibility of the d'Alembert-Hamilton system, we have 
established that system of PDEs (\ref{2.1.24}) with $N=3$ is
incompatible provided ${\rm rank}\, \|u_{x_\mu
  x_\nu}\|_{\mu,\nu=0}^3<3$ \cite{108}.

Consequently, any solution of system (\ref{2.1.24}) can be reduced by
means of one of the transformations $x_0\to ix_a,\ x_a\to ix_0$ or
$x_a\to x_b,\ x_b\to x_a$ to the form
(\ref{2.1.30z})--(\ref{2.1.31b}).  Since the class of functions $u(x)$
determined by formulae (\ref{2.1.30z})--(\ref{2.1.31b}) is invariant
with respect to the above transformations, hence it follows that any
solution of the d'Alembert-Hamilton system (\ref{2.1.24}) with $N=3$
is contained in it. To complete the proof for the case $N=3$ it
suffices to check that any function $u(x)$ determined by
(\ref{2.1.30z})--(\ref{2.1.31b}) satisfies the d'Alembert-Hamilton
system (\ref{2.1.24}). The check is performed by direct computation.
Differentiating the equalities (\ref{2.1.30z}), (\ref{2.1.31a}) with
respect to $x_\mu$ and excluding from the equalities obtained
$\tau_{x_\mu}$ we get
\begin{equation}
u_{x_\mu}=\Bigl((x_\nu+A_\nu)(x^\nu+A^\nu)\Bigr)^{-1/2}
\Bigl(x^\mu+A^\mu-\rho(\dot A\cdot x+\dot A\cdot A)B^\mu\Bigr),
\label{2.1.60}
\end{equation}
where $A\cdot x=A_\mu x^\mu$,\ $\rho =(\dot B\cdot x+\dot B\cdot
A)^{-1}$. 

Since 
\begin{eqnarray*}
g_{\mu\nu}u_{x_\mu}u_{x_\nu}&=&
\Bigl((x_\nu+A_\nu)(x^\nu+A^\nu)\Bigr)^{-1}
\Bigl(x_\mu+A_\mu-\rho(\dot A\cdot x+\dot A\cdot A)B_\mu\Bigr)\\
&&\times\Bigl(x^\mu+A^\mu-\rho(\dot A\cdot x+\dot
A\cdot A)B^\mu\Bigr)=1 
\end{eqnarray*}
(we have used the equalities (\ref{2.1.31b})), the Hamilton equation
is identically satisfied.

Next, differentiating (\ref{2.1.60}) with respect to $x_\mu$ and
excluding $\tau_{x_\mu}$ we get
\begin{eqnarray*}
u_{x_\mu x_\nu}&=&-\Bigl((x_\al+A_\al)(x^\al+A^\al)\Bigr)^{-3/2}
\Bigl(x^\mu+A^\mu-\rho(\dot A\cdot x+\dot A\cdot A)B^\mu\Bigr)\\
&&\times\Bigl(x^\nu+A^\nu-\rho(\dot A\cdot x+\dot A\cdot A)B^\nu\Bigr)
\Bigl((x_\al+A_\al)(x^\al+A^\al)\Bigr)^{-1/2}\\
&&\times\Bigl(g_{\mu\nu}-\rho(\dot A^\mu B^\nu+\dot A^\nu B^\mu)
+\rho^2[\ddot A\cdot x+\ddot A\cdot A+\dot A\cdot \dot A\\
&&+(\dot A\cdot x+\dot A\cdot A)(\ddot B\cdot x+\ddot B\cdot A+\dot
B\cdot \dot A)]B^{\mu}B^{\nu}\\
&&+\rho^2(\dot A\cdot x+\dot A\cdot A)(\dot B^\mu B^\nu +\dot B^\nu
B^\mu) \Bigr).
\end{eqnarray*}

Convoluting $u_{x_\mu x_\nu}$ with the metric tensor $g_{\mu\nu}$ and
taking into account the equalities (\ref{2.1.31a}) we come to the
following relation:
\begin{displaymath}
\Box u=\Bigl((x_\al+A_\al)(x^\al+A^\al)\Bigr)^{-1/2}
(g_{\mu\nu}g_{\mu\nu}-1)=3u^{-1},
\end{displaymath}
the same as what was to be proved.

Further, we will outline the scheme of the proof of the theorem
provided $N=0,1,2$ in (\ref{2.1.24}).

According to \cite{108} system of PDEs (\ref{2.1.24}) with $N=2$ is
compatible if and only if ${\rm rank}\,
\|u_{x_\mu x_\nu}\|_{\mu,\nu=0}^3=2$. Consequently, without loss
of generality, we can suppose that the condition
\begin{displaymath}
\delta=\left|\matrix{u_{x_1x_1}&u_{x_1x_2}\cr
u_{x_1x_2}&u_{x_2x_2}\cr}\right|\ne 0
\end{displaymath}
holds.

Since ${\rm rank}\, \|u_{x_\mu x_\nu}\|_{\mu,\nu=0}^3=2$ and
$\delta\ne 0$, there exists such a function $S\in C^2(\C^4,\C^1)$ that
solutions of the d'Alembert-Hamilton equation (\ref{2.1.24}) with
$N=2$ satisfy an additional constraint $S(u_{x_0},\, u_{x_1},\,
u_{x_2},\, u_{x_3})=0$.  Consequently, in the case involved we have to
solve the following over-determined system of PDEs: 
\begin{displaymath} 
\Box u = 2u^{-1}, \quad (\p_\mu u) (\p^\mu u) = 1,\quad S(u_{x_0},\,
u_{x_1},\, u_{x_2},\, u_{x_3})=0.  
\end{displaymath}

Due to the condition $\delta\ne 0$ we can resolve the last two
equations with respect to $u_{x_0},\ u_{x_3}$ and rewrite the above
system as follows
\begin{equation}
\begin{array}{l}
u_{x_0}=\Bigl(1+u_{x_k}u_{x_k}+W^2(u_{x_1},\,
u_{x_2})\Bigr)^{1/2},\\[2mm] 
u_{x_3}=W(u_{x_1},\, u_{x_2}),\quad \Box u = 2u^{-1}.
\end{array}
\label{2.1.61}
\end{equation}

Let us apply to the system of PDEs (\ref{2.1.61}) the contact
transformation 
\begin{eqnarray}
&&x_0=y_0,\quad x_k=H_{y_k},\quad x_3=y_3,\non\\
&&u=y_kH_{y_k}-H,\non\\
&&u_{x_0}=-H_{y_0},\quad u_{x_k}=y_k,\quad u_{x_3}=-H_{y_3},\non\\
&&H_{11}=u_{22}\delta^{-1},\quad H_{12}=-u_{12}\delta^{-1},\quad
H_{22}=u_{11}\delta^{-1},\non\\
&&H_{01}=-\left|\matrix{
u_{01}&u_{12}\cr
u_{02}&u_{22}\cr}\right|\delta^{-1},\quad
H_{23}=-\left|\matrix{
u_{11}&u_{13}\cr
u_{12}&u_{23}\cr}\right|\delta^{-1},\non\\
&&H_{13}=-\left|\matrix{
u_{13}&u_{12}\cr
u_{23}&u_{22}\cr}\right|\delta^{-1},\quad
H_{02}=-\left|\matrix{
u_{11}&u_{01}\cr
u_{12}&u_{02}\cr}\right|\delta^{-1},\label{2.1.62}\\
&&H_{00}=-\left|\matrix{
u_{00}&u_{01}&u_{02}\cr
u_{01}&u_{11}&u_{12}\cr
u_{02}&u_{12}&u_{22}\cr}\right|\delta^{-1},\quad
H_{03}=-\left|\matrix{
u_{01}&u_{02}&u_{03}\cr
u_{11}&u_{12}&u_{13}\cr
u_{12}&u_{22}&u_{23}\cr}\right|\delta^{-1},\non\\
&&H_{33}=-\left|\matrix{
u_{11}&u_{12}&u_{13}\cr
u_{12}&u_{22}&u_{23}\cr
u_{13}&u_{23}&u_{33}\cr}\right|\delta^{-1}.\non
\end{eqnarray}

Here $H_{\mu\nu}=\p^2 H/\p y_\mu \p y_\nu,\ u_{\mu\nu}=\p^2 u/\p x_\mu
\p x_\nu,\ \mu,\nu={0,\ldots,3}$.

The first two equations of system (\ref{2.1.62}) are linearized by the
transformation 
(\ref{2.1.62}) 
\begin{eqnarray*}
H_{y_0}&=&-\Bigl(1+y_ky_k+W^2(y_1, y_2)\Bigr)^{1/2},\\
H_{y_3}&=& - W(y_1, y_2).
\end{eqnarray*}

Integrating the above system, inserting the obtained expression for
$H(y)$
\begin{equation}
H=-y_0\Bigl(1+y_ky_k+W^2(y_1, y_2)\Bigr)^{1/2}-y_3W(y_1, y_2)-
B(y_1,y_2),
\label{2.1.63}
\end{equation}
where $B\in C^2(\C^2,\C^1)$ is an arbitrary function, into the last
equation of system (\ref{2.1.61}) and splitting with respect to $y_0,\
y_3$ we come to the over-determined system of five PDEs for two
functions $B(y_1,y_2),\ W(y_1,y_2)$
\begin{eqnarray*}
&1)& (\triangle_2 W+ y_ky_nW_{y_ky_n})\Bigl(1+W_{y_k}W_{y_k}
+T^2(W)\Bigr)\\
&&-\Bigl(T(W)y_k+W_{y_k}\Bigr)\Bigl(T(W)y_n+W_{y_n}\Bigr)W_{y_ky_n}\\
&&=-2T(W)\Bigl(1+W_{y_k}W_{y_k}+T^2(W)\Bigr),\\
&2)& {\rm det}\,
\|W_{y_ky_n}\|_{k,n=1}^2=T^2(W)\Bigl(1+W_{y_k}W_{y_k}+T^2(W)\Bigr)^{-1}\\
&&\times(1+y_ky_k+W^2)^{-1},\\
&3)& (\triangle_2 B+ y_ky_nB_{y_ky_n})\Bigl(1+W_{y_k}W_{y_k}
+T^2(W)\Bigr)\\
&&-\Bigl(T(W)y_k+W_{y_k}\Bigr)\Bigl(T(W)y_n+W_{y_n}\Bigr)B_{y_ky_n}\\
&&=-2T(B)\Bigl(1+W_{y_k}W_{y_k}+T^2(W)\Bigr),\\
&4)& {\rm det}\,
\|B_{y_ky_n}\|_{k,n=1}^2=T^2(B)\Bigl(1+W_{y_k}W_{y_k}+T^2(W)\Bigr)\\
&&\times(1+y_ky_k+W^2)^{-1},\\
&5)& (\triangle_2 W)(\triangle_2 B)-B_{y_ky_n}W_{y_ky_n}=T(W)T(B)\\
&&\times\Bigl(1+W_{y_k}W_{y_k}+T^2(W)\Bigr)
(1+y_ky_k+W^2)^{-1}.
\end{eqnarray*}
 
Here the notations $T(F)=y_kF_{y_k}-F,\ \triangle_2F=F_{y_ky_k}$
are used. 

Integrating the above system and returning to the initial variables
$x,\ u(x)$ according to the formulae (\ref{2.1.62}) we get the general
solution of the d'Alem\-bert-Hamilton system (\ref{2.1.24}) with $N=2$
which is contained in the class of functions $u(x)$ determined by the
formulae $3$ from the statement of Theorem 2.1.4.

According to \cite{108} system of PDEs (\ref{2.1.24}) with $N=1$ is
compatible only in the following cases
\vspace{1.5mm}

\noindent
$a$)\ ${\rm rank}\, \|u_{x_\mu x_\nu}\|_{\mu,\nu=0}^3=2$;
\vspace{1.5mm}

\noindent
$b$)\ ${\rm rank}\, \|u_{x_\mu x_\nu}\|_{\mu,\nu=0}^3=1$.

In the case $a$, we apply to the system under study the contact
transformation (\ref{2.1.62}). The general solution of the Hamilton
equation being written in the variables $y,\ H(y)$ takes the form
(\ref{2.1.63}) with arbitrary smooth functions $B(y_1,y_2),\
W(y_1,y_2)$. Inserting (\ref{2.1.63}) into the d'Alembert equation
written in the variables $y,\ H(y)$ and splitting the equality
obtained with respect to $y_0,\ y_3$ we arrive at the following system  
of four PDEs:
\begin{eqnarray*}
&1)&1+W_{y_k}W_{y_k}+T^2(W)=0,\\
&2)&{\rm det}\, \|W_{y_ky_n}\|_{k,n=1}^2=0,\\
&3)&(1+y_ky_k+W^2){\rm det}\,
\|B_{y_ky_n}\|_{k,n=1}^2=T(B)\\
&&\times\Bigl(T(W)y_k+W_{y_k}\Bigr)\Bigl(T(W)y_n
+W_{y_n}\Bigr)B_{y_ky_n},\\
&4)&(1+y_ky_k+W^2)\Bigl((\triangle_2 W)(\triangle_2
B)-B_{y_ky_n}W_{y_ky_n}\Bigr)\\ 
&&=T(W)\Bigl(T(W)y_k+W_{y_k}\Bigr)\Bigl(T(W)y_n
+W_{y_n}\Bigr)B_{y_ky_n}.
\end{eqnarray*}

Integrating these equations and returning to the initial variables
$x,\ u(x)$ yield the general solution of system (\ref{2.1.24}) with
$N=1$ provided the condition $a$ holds.

Let us turn now to the case $b$. Since ${\rm rank}\, \|u_{x_\mu
  x_\nu}\|_{\mu,\nu=0}^3=1$, there exist such functions
$W_a=W_a(u_{x_0}) \in C^1(\C^1,\C^1)$ that
\begin{displaymath}
u_{x_a}=W_a(u_{x_0}),\ \  a=1,2,3.
\end{displaymath}

With this remark system (\ref{2.1.24}) is rewritten in the form
\begin{equation}
\Box u=u^{-1},\quad u_{x_a}=W_a(u_{x_0}),
\label{2.1.64}
\end{equation}
where $W_a(\tau)$ are arbitrary smooth functions satisfying the
equality $\tau^2-W_a(\tau)
\linebreak\times W_a(\tau)=1$.

Make in (\ref{2.1.64}) the following contact transformation:
\begin{equation}
\begin{array}{l}
y_0=u_{x_0},\quad y_a=x_a,\quad H=x_0u_{x_0}-u,\\[2mm]
H_{y_0}=x_0,\quad H_{y_a}=-u_{x_a},\\[2mm]
H_{00}=u_{00}^{-1},\quad H_{0a}=-u_{0a}u_{00}^{-1},\\[2mm]
H_{ab}=(u_{0a}u_{0b}-u_{00}u_{ab})u_{00}^{-1}.
\end{array}
\label{2.1.65}
\end{equation}

Here $u_{\mu\nu}=u_{x_\mu x_\nu},\ H_{\mu\nu}=H_{y_\mu y_\nu},\
a,b=1,2,3$. 

The last three equations from (\ref{2.1.64}) are linearized
\begin{displaymath}
H_{y_a}=-W_a(y_0),\ \  a=1,2,3.
\end{displaymath}

Inserting the general solution of the above system
\begin{equation}
H=y_aW_a(y_0)-B(y_0),
\label{2.1.66}
\end{equation}
where $B\in C^2(\C^1,\C^1)$ is an arbitrary function,
into the first equation of the system of PDEs (\ref{2.1.64}) and
splitting with respect to $y_0$ we come to the system of ODEs for the
functions $W_a,\ B$
\begin{eqnarray*}
&1)&\ddot W_a=(1-\dot W_b\dot W_b)(y_0\dot W_a-W_a),\ \ a=1,2,3,\\
&2)&\ddot B=(1-\dot W_b\dot W_b)(y_0\dot B-B)
\end{eqnarray*}
and what is more $W_aW_a=y_0^2-1$.

Integrating the system of ODEs obtained and returning to the initial
variables $x,\ u(x)$ we obtain a particular case of the formulae $2$
from the statement of Theorem 2.1.4.

Provided $N=0$, the general solution of system of PDEs (\ref{2.1.24}) 
splits into two classes satisfying one of the conditions:\
${\rm rank}\, \|u_{x_\mu x_\nu}\|_{\mu,\nu=0}^3=1,2$.

If ${\rm rank}\, \|u_{x_\mu x_\nu}\|_{\mu,\nu=0}^3=2$, then we can
apply the contact transformation (\ref{2.1.62}). The general solution
of the d'Alembert-Hamilton system is given by the formula
(\ref{2.1.63}), where $B(y_1,y_2),\ W(y_1,y_2)$ are solutions of the
system of two PDEs
\begin{eqnarray*}
&1)&1+W_{y_k}W_{y_k}+(y_kW_{y_k}-W)^2=0,\\
&2)&\Bigl(y_k(y_nW_{y_n}-W)+W_{y_k}\Bigr)
\Bigl(y_l(y_nW_{y_n}-W)+W_{y_l}\Bigr)B_{y_ky_l}=0.
\end{eqnarray*}

Integrating it and returning to the initial variables $x,\ u(x)$ we
arrive at the formulae $1$ from the statement of Theorem 2.1.4.

Provided solutions of the d'Alembert-Hamilton system (\ref{2.1.24})
with $N=0$ satisfy the condition ${\rm rank}\, \|u_{x_\mu
  x_\nu}\|_{\mu,\nu=0}^3=1$, we can perform the contact transformation 
(\ref{2.1.65}). The general solution of the system obtained is of the
form (\ref{2.1.66}), where $W_a(y_0),\ B(y_0)$ are solutions of the
system of two ODEs $\dot W_a\dot W_a=1,\ W_aW_a=y_0^2-1$.
Rewriting (\ref{2.1.66}) in the initial variables $x,\ u(x)$ according
to the formulae (\ref{2.1.65}) yields the formulae $1$ from the
statement of Theorem 2.1.4 with $B_\mu\equiv \dot A_\mu,\ R_2=\dot
R_1$.

Thus, we have established that the general solutions of the system of
PDEs (\ref{2.1.24}) with $N=0,1,2$ are contained in the classes of
functions given by the formulae $1$--$3$ from the statement of 
Theorem 2.1.4. To complete the proof we have to check that the
function $u(x)$ determined by these formulae satisfies the
d'Alembert-Hamilton system. This check is carried out by direct
computation. The theorem is proved. $\rhd$
\vspace{1.5mm}

\noindent
{\bf Theorem 2.1.5.}\ {\em The general solution of system of PDEs
(\ref{2.1.29z}) has the form
\begin{equation}
A_\mu (u,\, \tau) x^\mu + A(u,\, \tau)=0,
\label{2.1.67}
\end{equation}
where $\tau=\tau(x,u)$ is determined in implicit way
\begin{equation}
B_\mu(u,\tau)x^\mu + B(u,\, \tau)=0
\label{2.1.68}
\end{equation}
and $ A_\mu(u,\, \tau),\ B_\mu(u,\, \tau),\ A(u,\, \tau),\ B(u,\,
\tau)$ are arbitrary complex-valued functions satisfying the
conditions} 
\begin{equation}
A_{\mu} A^{\mu} = A_\mu B^\mu = B_\mu B^\mu = 0,\quad 
B_\mu {\partial A^\mu\over \partial\tau}=0.
\label{2.1.69}
\end{equation}
\vspace{1.5mm}

\noindent
{\em Proof.}$\quad$ If $u(x)\ne \mbox{\rm const}$, then making, when
necessary, the change of independent variables 
\begin{equation}
\begin{array}{l}
x_0\to ix_a,\ x_a\to ix_0,\\[2mm]
x_b\to x_c,\quad x_c\to x_b
\end{array}
\label{2.1.69z}
\end{equation}
with some fixed $a,b,c=1,2,3$ we can without loss of generality
suppose that $u_{x_3}\ne 0$. With this condition we can make in
(\ref{2.1.29z}) the hodograph transformation 
\begin{equation}
y_\al=x_\al,\ \ \al={0,1,2},\quad
y_3=u,\quad U=x_3,
\label{2.1.70}
\end{equation}
where $y_0,\ldots,y_3$ are new independent variables and $U=U(y)$ is a 
new dependent variable. As a result, the following system of
PDEs
\begin{equation}
U_{y_0y_0}-U_{y_1y_1}-U_{y_2y_2}=0,\quad
U_{y_0}^2-U_{y_1}^2-U_{y_2}^2=1
\label{2.1.71}
\end{equation}
is obtained.

Therefore, the four-dimensional system of PDEs (\ref{2.1.29z}) is
transformed to the system with three independent variables (the fourth
variable $y_3$ is contained in (\ref{2.1.71}) as a parameter).

Equations (\ref{2.1.71}) are obtained from the d'Alembert-Hamilton
system (\ref{2.1.24}) with $N=0$ by assuming that its solutions do not
depend on $x_3$ and by identifying $x_\al$ with $y_\al,\ \al=0,1,2$
and $u$ with $U$. Consequently, the general solution of (\ref{2.1.71})
is given by the formulae $1$ from the statement of Theorem 2.1.4
provided the indices take the values $0,1,2$. And what is more, all
arbitrary functions included into the general solution contain $y_3$
as an argument.

Thus, the general solution of system of PDEs (\ref{2.1.71}) is
determined by the formulae 
\begin{eqnarray*}
&&U=a_0(\tau,\, y_3)y_0-a_1(\tau,\, y_3)y_1-a_2(\tau,\, y_3)y_2
+R_1(\tau,\, y_3),\\
&&b_0(\tau,\, y_3)y_0-b_1(\tau,\, y_3)y_1-b_2(\tau,\, y_3)y_2
+R_2(\tau,\, y_3)=0,
\end{eqnarray*} 
where $a_\al(\tau,\, y_3),\ b_\al(\tau,\, y_3),\ \al=0,1,2$ are
arbitrary complex-valued functions satisfying the equalities
\begin{equation}
\begin{array}{l}
a_0^2-a_1^2-a_2^2=1,\quad b_0^2-b_1^2-b_2^2=0,\\[2mm]
a_0b_0-a_1b_1-a_2b_2=0,\quad {\edi\p a_0\over \edi \p\tau}b_0-{\edi\p
  a_1\over \edi \p\tau}b_1-{\edi\p a_2\over \edi \p\tau}b_2=0.
\end{array}
\label{2.1.72}
\end{equation}

Rewriting the result obtained in the initial variables $x,\ u(x)$
according to (\ref{2.1.70}) we arrive at the following representation
of the general solution of system (\ref{2.1.29z}):
\begin{displaymath}
x_3=a_0(\tau,\, u)x_0-a_1(\tau,\, u)x_1-a_2(\tau,\, u)x_2
+R_1(\tau,\, u), 
\end{displaymath}
where $\tau=\tau(x,u)$ is a complex-valued function defined implicitly 
\begin{displaymath}
b_0(\tau,\, u)x_0-b_1(\tau,\, u)x_1-b_2(\tau,\, u)x_2
+R_2(\tau,\, u)=0
\end{displaymath}
and $a_\al(\tau,\, u),\ b_\al(\tau,\, u),\ \al=0,1,2$ are
arbitrary complex-valued functions satisfying (\ref{2.1.72}).
 
It is readily seen that the above formulae are obtained from 
(\ref{2.1.67})--(\ref{2.1.69}) under
\begin{eqnarray*}
&&A_\al=a_\al,\quad A_3=1,\quad A=R_1,\\ 
&&B_\al=b_\al,\quad B_3=0,\quad B=R_2,
\end{eqnarray*}
where $\al=0,1,2$. 

We have proved that any solution of system (\ref{2.1.29z}) satisfying
the relation $u(x)\ne \mbox{\rm const}$ can be reduced to the form
(\ref{2.1.67})--(\ref{2.1.69}) by the change of the independent
variables (\ref{2.1.69z}). Since the class of functions
$F$ determined by the relations (\ref{2.1.67})--(\ref{2.1.69}) is
invariant with respect to the transformations (\ref{2.1.69z}) and
contains the solution $u(x)=\mbox{\rm const}$, hence it follows that
$G \subset F$, where $G$ is the class of functions $u(x)$ determining
the general solution of system of PDEs (\ref{2.1.29z}). Let us prove
the inverse inclusion $G \subset F$. This assertion will be established
if we show that any function $u(x)$ determined by the formulae
(\ref{2.1.67})--(\ref{2.1.69}) satisfies equations (\ref{2.1.29z}).

Differentiating equalities (\ref{2.1.67}), (\ref{2.1.68}) with respect
to $x_\mu$ we find $u_{x_\mu}$ and $\tau_{x_\mu}$ as
\begin{equation}
\begin{array}{rcl}
u_{x_\mu}&=&{\edi 1\over \edi\Delta}\Bigl((x\cdot B_\tau
+R_{2\tau})A^\mu 
-(x\cdot A_\tau +R_{1\tau})B^\mu\Bigr),\\[2mm]
\tau_{x_\mu}&=&{\edi 1\over \edi\Delta}\Bigl((x\cdot A_u +R_{1u})B^\mu
-(x\cdot B_u +R_{2u})A^\mu\Bigr),
\end{array}
\label{2.1.73}
\end{equation}
where $\Delta=(x\cdot A_\tau + R_{1\tau}) (x\cdot B_u + R_{2u})-
(x\cdot A_u + R_{1u}) (x\cdot B_\tau + R_{2\tau}),\ x\cdot A=x_\mu
A^\mu$. 

Since
\begin{eqnarray*}
u_{x_\mu}u_{x^\mu}&=&\Delta^{-2}\Bigl((x\cdot B_\tau +R_{2\tau})^2
A\cdot A-2(x\cdot B_\tau +R_{2\tau}) \\
&&\times(x\cdot A_\tau+R_{1\tau})A\cdot B+ (x\cdot A_\tau
+R_{1\tau})^2B\cdot B\Bigr)=0 
\end{eqnarray*}
(we have used the identities (\ref{2.1.69})), the Hamilton equation is
satisfied. 

Differentiating the first equation from (\ref{2.1.73}) with respect to
$x_\nu $ we get
\begin{eqnarray*}
u_{x_\mu x_\nu}&=&-{1\over \Delta^2}\Bigl((x\cdot B_\tau
+R_{2\tau})A^\mu -(x\cdot A_\tau +R_{1\tau})B^\mu\Bigr)\\
&&\times\left({\p\Delta\over\p\tau}\tau_{x_\nu}
+ {\p\Delta\over\p u}u_{x_\nu}\right)
+ {1\over \Delta}(A^\mu B_\tau^\nu - A^\nu B_\tau^\mu)\\
&&+ {1\over \Delta}\biggl\{\tau_{x_\nu}{\p \over\p\tau}\Bigl((x\cdot
B_\tau +R_{2\tau})A^\mu -(x\cdot A_\tau +R_{1\tau})B^\mu\Bigr)\\
&&+u_{x_\nu}{\p \over\p u}\Bigl((x\cdot
B_\tau +R_{2\tau})A^\mu -(x\cdot A_\tau
+R_{1\tau})B^\mu\Bigr)\biggr\}.
\end{eqnarray*} 
 
Convoluting $u_{x_\mu x_\nu}$ with the metric tensor $g_{\mu\nu}$ and
taking into account identities (\ref{2.1.69}) we arrive at the
equality $\Box u = 0$.

Thus, we have established that the relations $F \subset G,\ G \subset F$
hold, whence it follows that $F=G$. In other words, formulae
(\ref{2.1.67})--(\ref{2.1.69}) (the class $F$) give the general
solution of the d'Alembert-Hamilton equation (\ref{2.1.29z}) (the
class $G$). The theorem is proved. $\rhd$
\vspace{1.5mm}

\noindent
{\bf Note 2.1.4.}\ Assuming that the functions $A_\mu,\ B_\mu$ do not
depend on $\tau$ and excluding $\tau$ from the relations
(\ref{2.1.67}), (\ref{2.1.68}) we get the following class of the exact
solutions of system (\ref{2.1.29z}):
\begin{equation}
g\Bigl(A_\mu(u) x^\mu,\, B_\mu(u)x^\mu,\, u\Bigr)=0,
\label{2.1.74}
\end{equation}  
where $g\in C^2(\C^3,\C^1)$ is an arbitrary function.

Provided $A_\mu,\ B_\mu$ are constants, formula (\ref{2.1.74})
gives the class of exact solutions of the d'Alembert-Hamilton system
obtained by Erugin \cite{52.0}.

Furthermore, if the function $g$ does not depend on
$B_\mu(u)x^\mu$, we can resolve (\ref{2.1.29z}) with respect to $A_\mu
(u) x^\mu$ and thus get the generalization of the
Jacobi-Smirnov-Sobolev formula (\ref{2.1.4z}) 
\begin{equation}
A_\mu(u) x^\mu +A(u)=0,\quad A_{\mu} A^{\mu}=0.
\label{2.1.75}
\end{equation}

It has been proved in \cite{108.3,213.6} that formulae (\ref{2.1.75}), 
where indices take the values $0,1, \ldots$, $n-1$, give the general
solution of the d'Alembert-Hamilton system $\Box_n u=0$,\ $(\p_{\ssl A}
u)(\p^{\ssl A} u)=0$, provided $u$ is a real-valued function of $n$ real
variables $x_0$,\ $x_1$, $\ldots$, $x_{n-1}$.  
\vspace{1.5mm}

\noindent
{\bf Note 2.1.5.}\ If we choose in (\ref{2.1.67})--(\ref{2.1.69})
\begin{displaymath}
A_\mu=C_\mu(\tau),\quad B_\mu=\dot C_\mu(\tau),\quad A=C(\tau),\quad
B=\dot C(\tau),
\end{displaymath}
then we get the class of exact solutions
\begin{displaymath}
u=C_\mu(\tau)x^\mu+C(\tau),\quad \dot C_\mu(\tau)x^\mu+\dot
C(\tau)=0,\quad \dot C_\mu \dot C^\mu=0
\end{displaymath}
which was constructed by Bateman \cite{20.1}.
\vspace{1.5mm}

\noindent 
{\bf 3. Explicit solutions of the d'Alembert-Hamilton system.}\ 
Theorems 2.1.4, 2.1.5 give a description of the general solution of
systems of nonlinear PDEs (\ref{2.1.24}), (\ref{2.1.29z}) in the
parametric form. But for some special choices of the arbitrary
functions it is possible to obtain particular solutions in explicit
form which is very important for applications of the above results.
Below we will construct some real solutions of system (\ref{2.1.27})
using Theorem 2.1.4.

Take, for example, system (\ref{2.1.27}) with $N=3,\ \ve=-1$
\begin{equation}
\Box u=-3u^{-1},\quad (\p_\mu u)(\p^\mu u)=-1.
\label{2.1.76}
\end{equation}

To obtain the general solution of (\ref{2.1.76}) it is necessary 
to make in (\ref{2.1.30z}), (\ref{2.1.31a}), (\ref{2.1.31b}) the 
change $u\to iu$. As a result, we get the following formulae:
\begin{eqnarray}
&&u^2=-\Bigl(x_\mu+A_\mu(\tau)\Bigr)\Bigl(x^\mu
+A^\mu(\tau)\Bigr),\non\\
&&\Bigl(x_\mu+A_\mu(\tau)\Bigr)B^\mu(\tau)=0,\label{2.1.77}\\
&&B_\mu{\dot A}^\mu=0,\quad B_\mu B^\mu=0.\non
\end{eqnarray}

Putting in the above formulae $A_\mu=0,\ B_\mu=0$ we get the
well-known $O(1,3)$-invariant solution of system (\ref{2.1.24}) with
$N=3$:\ $u(x)=(x_\mu x^\mu)^{1/2}$. This solution can be obtained by
means of the symmetry reduction of PDE (\ref{2.1.24}) with the use of
the $O(1,3)$-invariant Ansatz $u(x)=\vp(x_\mu x^\mu)$.

A more interesting solution is obtained by putting
\begin{eqnarray*}
&&A_0=\tau,\quad A_1=C\sin (\tau/C),\quad A_2=C\cos (\tau/C),\quad
A_3=0,\\
&&B_\mu={\dot A}_\mu,\ \ \mu={0,\ldots,3},
\end{eqnarray*}
where $C\in \R^1,\ C\ne 0$.

With the chosen $A_\mu,\ B_\mu$ formulae (\ref{2.1.77}) take the form
\begin{eqnarray*}
&&u^2=[x_1+C\sin (\tau/C)]^2+[x_2+C\cos (\tau/C)]^2
+x_3^2-(x_0+\tau)^2,\\
&&x_0+\tau-x_1\cos(\tau/C)+x_2\sin(\tau/C)=0.
\end{eqnarray*}

After making some simple algebraic manipulations we find an explicit
form of the parametric function $\tau$
\begin{displaymath}
\tau(x,u)=\pm\{2C(u^2-x_3^2)^{1/2}+x_ax_a-u^2-C^2\}^{1/2},
\end{displaymath}
whence we conclude that the function $u(x)$ is determined by the
formula
\begin{displaymath}
x_0+\tau(x,u)=x_1\cos\Bigl(\tau(x,u)/C\Bigr) -
x_2\sin\Bigl(\tau(x,u)/C\Bigr)=0. 
\end{displaymath}

This solution is new and cannot be in principle obtained within the
framework of the Lie approach.

In a similar way we have constructed other particular solutions of
the d'Alembert-Hamilton system (\ref{2.1.27}) with different $N,\ \ve$
which are listed below
\index{Exact solutions!of the d'Alembert-Hamilton system}
\vspace{1.5mm}

\noindent
1)\ \ $N=0, \ \ \ve = 1$
\begin{equation}
u(x)=x_0;
\label{2.1.29}
\end{equation}
2)\ \ $N=1,\ \   \ve =1$
\begin{equation}
u(x) = \pm (x^2_0 - x^2_3)^{1/2} ;
\label{2.1.30}
\end{equation}
3)\ \  $N=2, \ \ \ve = 1$
\begin{equation}
u(x) = \pm (x^2_0 - x^2_1 - x^2_3)^{1/2} ;
\label{2.1.31}
\end{equation}
4)\ \  $N=3, \ \ \ve = 1$
\begin{equation}
u(x) = \pm (x^2_0 - x^2_1 - x^2_2 - x^2_3)^{1/2} ;
\label{2.1.32}
\end{equation}
5)\ \  $N = 0,\ \  \ve = -1 $
\begin{eqnarray}
&&u(x) = x_1 \cos W_1(x_0 + x_3) + x_2 \sin W_1 (x_0 + x_3) +
 W_2 (x_0 + x_3 ),\non\\
&&x_0 + x_1 \sin W_1 \Bigl(u(x) + x_3\Bigr) + x_2 \cos W_1 
\Bigl(u(x) + x_3\Bigr)\label{2.1.33}\\
&&\phantom{x_0} + W_2\Bigl(u(x) + x_3\Bigr) = 0;\non
\end{eqnarray}
6) \ \ $N = 1, \ \ \ve = -1 $
\begin{equation}
u(x) = \pm \Bigl\{\Bigl(x_1 + W_1(x_0 + x_3)\Bigr)^2 +
 \Bigl(x_2 + W_2(x_0 + x_3)\Bigr)^2\Bigr\}^{1/2};
\label{2.1.34}
\end{equation}
7)\ \  $N = 2,\ \  \ve = -1$
\begin{eqnarray}
&&\pm u(x)+C = x_0\sinh (\tau /C) - x_1 \cosh (\tau /C),\non\\
&&\quad\tau = - x_2 \pm\Bigl\{x^2_0 - x^2_1 + \Bigl(C \pm
u(x)\Bigr)^2\Bigr\}^{1/2} ;\non\\ 
&&\pm u(x)-C = x_1 \sin (\tau /C) + x_2 \cos (\tau /C),\non\\
&&\quad\tau = - x_0 \pm \Bigl\{x^2_1 + x^2_2 -\Bigl(-C \pm
u(x)\Bigr)^2\Bigr\}^{1/2};\label{2.1.35}\\ 
&&x_0\sinh\tau - x_3\cosh\tau = 2^{-1/2} \{\pm (-u^2(x) - x_\mu x^\mu 
)^{1/2}\pm u(x)\},\non\\
&&\quad\tau = \arcsin\Bigl\{\Bigl(\sqrt 2 (x^2_1 +
x^2_2)^{1/2}\Bigr)^{-1} 
\Bigl(\pm u(x) \mp (-u^2(x) - x_\mu x^\mu)\Bigr)^{1/2}\Bigr\}\non\\ 
&&\phantom{\quad\tau =} - \arcsin \Bigl\{x_2 (x^2_1 +
x^2_2)^{-1/2}\Bigr\},\non\\ 
&&u(x) = \pm (x^2_1 + x^2_2 + x^2_3)^{1/2};\non
\end{eqnarray}
8)\ \ $N = 3, \ \ \ve = -1 $
\begin{eqnarray}
&&\pm \Bigl(u^2(x) - x^2_3\Bigr)^{1/2}+C = x_0 \sinh (\tau /C) -
x_1 \cosh (\tau /C),\non\\
&&\quad\tau = - x_2 \pm \Bigl\{x^2_0 - x^2_1 + \Bigr(C \pm [u^2 (x) -
x^2_3]^{1/2}\Bigl)^2\Bigr\}^{1/2};\non\\
&&\pm \Bigl(u^2(x) - x^2_3\Bigr)^{1/2}-C = x_1 \sin (\tau /C) + x_2
\cos (\tau /C), 
\label{2.1.36}\\
&&\quad\tau = - x_0 \pm \Bigl\{x^2_1 + x^2_2 - \Bigl(C \mp [u^2 (x) -
x^2_3]^{1/2}\Bigl)^2\Bigr\}^{1/2}.\non 
\end{eqnarray}

Here $\{W_1, W_2\} \subset C^2 ({\R}^1, {\R}^1) $ are arbitrary functions,
$C$ is a real non-zero constant.
\vspace{1.5mm}

\noindent
{\bf 4. Conditional symmetry of the nonlinear d'Alembert equation.}\
According to the remark made in the very beginning of the section
substitution of the Ansatz (\ref{2.1.1}), where $u(x)$ is an arbitrary
solution of the d'Alembert-Hamilton system (\ref{2.1.27}), into the
nonlinear d'Alembert equation\index{d'Alembert equation}
\begin{equation}
\Box w = F_0(w),
\label{2.1.78}
\end{equation}
reduces it to an ODE for a function $\vp$.

It occurs that the class of Ans\"atze obtained in this way is
substantially wider that the one obtainable by means of the symmetry
reduction.

Indeed, within the framework of the symmetry reduction approach to
reduce the nonlinear d'Alembert equation (\ref{2.1.78}) to an ODE one
has to construct Ans\"atze invariant under the three-parameter
subgroups of its symmetry group.  It is well-known that, provided
$F_0$ is an arbitrary function, the maximal symmetry group admitted by
PDE (\ref{2.1.78}) is the ten-parameter Poincar\'e group $P(1,3)$
having the generators 
\begin{equation} 
P_\mu=\p^\mu,\quad J_{\mu\nu}=x_\mu P_\nu -
x_\nu P_\mu.  
\label{2.1.79}
\end{equation}

Furthermore, the general form of mentioned Ans\"atze is given by the
formula (\ref{2.1.1}), where $u(x)$ is an invariant of some
three-parameter subgroup of the group $P(1,3)$. An exhaustive
description of the invariants of the Poincar\'e group having the
generators (\ref{2.1.79}) is obtained in \cite{167}. In particular, it
is established that any invariant of a three-parameter subgroup of the
group $P(1,3)$ can be reduced by an appropriate transformation from
the Poincar\'e group either to the forms
(\ref{2.1.29})--(\ref{2.1.32}) or to the forms
\begin{displaymath} 
x_0+x_3,\quad x_1+\theta\ln(x_0+x_3),\quad x_1+\theta
(x_0+x_3)^2,\quad x_1^2+x_2^2,\quad x_1^2+x_2^2+x_3^2, 
\end{displaymath} 
where $\theta$ is a constant.

But the invariants listed above are very special cases of the formulae
(\ref{2.1.33})--(\ref{2.1.35}) which in its turn determine only
particular solutions of the d'Alembert-Hamilton system.

Such substantial extension of the class of the Ans\"atze reducing the
nonlinear d'Alembert equation is achieved at the expense of its
conditional symmetry. 

Consider, as an illustration, the 
Ansatz\index{Conditional symmetry!of the d'Alembert equation}
\begin{equation}
w(x)=\vp\Bigl(x_1+\rho(x_0 + x_3 )\Bigr),
\label{2.1.80}
\end{equation}
where $\rho$ is an arbitrary smooth function, obtained by substitution
of the first formula from (\ref{2.1.33}) with $W_1=0,\ W_2=\rho$ into
(\ref{2.1.1}).

In spite of the fact that the Ansatz (\ref{2.1.80}) is not
Poincar\'e-invariant, it reduces PDE (\ref{2.1.78}) to the ODE $\ 
-\ddot \vp=F_0(\vp)$. This phenomenon cannot be in principle
understood within the framework of the classical Lie approach because the
existence of such Ans\"atze is a consequence of conditional invariance
of the nonlinear d'Alembert equation.
 
Indeed, the manifold (\ref{2.1.80}) is invariant under the
three-parameter Abe\-li\-an Lie group with the generators
\begin{displaymath}
Q_1=\p_0-\p_3,\quad Q_2=\p_0+\p_3- 2\dot\rho\p_1,\quad
Q_3=\p_2
\end{displaymath}
(this fact is established by direct computation). Obviously,
the operator $Q_2$ cannot be represented as a linear combination of
the operators $P_\mu,\ J_{\mu\nu}$ with constant coefficients which
means that equation (\ref{2.1.78}) is not invariant under the Lie
algebra $A=\langle Q_1,\, Q_2,\, Q_2\rangle$. 

We will prove that PDE (\ref{2.1.78}) is conditionally-invariant under
the algebra $A$. Acting by the second prolongations of the operators
$Q_a$ on (\ref{2.1.78}) we have
\begin{displaymath}
\wid Q_1L=0,\quad \wid Q_2L=4\ddot\rho\p_1Q_1u,\quad \wid Q_3L=0,
\end{displaymath}
where $L=\Box u-F_0(u)$.

Hence it follows that the system of PDEs
\begin{displaymath}
\Box u=F_0(u),\quad Q_au=0,\ \ a=1,2,3
\end{displaymath}
is invariant under the Lie algebra $A$, the same as what was to be
proved. 

All Ans\"atze obtained by substitution of the formulae for $u(x)$
listed in (\ref{2.1.33})--(\ref{2.1.36}) (with the only exception of
the last formula from (\ref{2.1.35})) into (\ref{2.1.1}) correspond to  
the conditional invariance of the
nonlinear d'Alem\-bert equation and give rise to the new (non-Lie)
reductions\index{Non-Lie!reduction} of PDE (\ref{2.1.78}). Hence it
follows, in particular, that the nonlinear d'Alembert equation admits
an {\em infinite} \/conditional symmetry. It will be shown that the
nonlinear Dirac and Yang-Mills equations have the same property (see
Chapters 6,7).  
\vspace{10mm}

\noindent
{\large\bf 2.2. Ans\"atze for the spinor field\label{s2.2}}

\markboth{Chapter 2. EXACT SOLUTIONS}
{2.2. Ans\"atze for the spinor field}
\def\theequation{2.\arabic{section}.\arabic{equation}}
\setcounter {section} {2}
\setcounter {equation}{0}
\vspace{7mm}

\noindent
We will apply the results given in Chapter 1 to construct
Ans\"atze (\ref{1.5.10}) reducing Poincar\'e-invariant
multi-dimensional PDEs for the spinor field to equations having a
lower dimension.

According to Theorem 1.5.1 to construct an Ansatz (\ref{1.5.10})
reducing a given equation to PDE with the less number of independent
variables we have (see also \cite{66,103,161,164})
\begin{itemize}
\item {to obtain operators $Q_1,\ Q_2, \ldots, Q_N $ of the form
    (\ref{1.5.5}) satisfying conditions of Theorem 1.5.1;}
\item{to integrate the corresponding system of PDEs (\ref{1.5.7}).}
\end{itemize}

In the present section we consider the case when operators
$Q_a $ form a basis of the $N$-dimensional real Lie algebra which is a
subalgebra of the Lie algebra of the invariance group $G$ of the
equation under study. 

Let $\Sigma_1, \,\Sigma_2, \ldots, \Sigma_M,\  M \ge N $ be the
basis elements of the Lie algebra $AG$.
\vspace{1.5mm}

\noindent
{\bf Definition 2.2.1.}\ Two sets of operators $\{Q_1, \ Q_2, \ldots,
Q_N\}$ and $\{Q_1',$ $\ Q_2', \ldots$, 
\linebreak
$Q_N'\} $ are called $G$-conjugate
if there exist such real parameters $\theta_1, \ldots$,
$\theta_M $
that
\begin{equation}
\exp  \left\{\theta_i \Sigma_i\right\} Q_j\ \exp
\left\{- \theta_i \Sigma_i\right\} = Q_j',
\quad j = {1,\ldots,M},
\label{2.2.1}
\end{equation}
summation over repeated indices being implied.

In other words, sets of operators $\{Q_1, \ Q_2, \ldots, Q_N\}$ and
$\{Q_1', \ Q_2', \ldots, Q_N'\} $ are $G$-conjugate if there exists
a group transformation from the Lie group $G$ having generators
$\Sigma_1, \ \Sigma_2, \ldots, \Sigma_M $ which transforms $Q_j $
into $Q_j', \ j = {1,\ldots,N}$. Two Lie algebras with basis elements
$Q_i$,\ $i={1,\ldots,N}$ and $Q_i^\prime$,\ $i={1,\ldots,N}$ are called
$G$-conjugate\index{Conjugate!Lie algebras} if the sets of the 
first-order differential operators
$\{Q_1$, $\ldots$, $Q_N\}$ and $\{Q_1^\prime$, $\ldots$,
$Q_N^\prime\}$ are $G$-conjugate. Two Lie transformation groups are
called $G$-conjugate if their Lie algebras are 
$G$-conjugate\index{Conjugate!Lie groups}.

It is evident that Ans\"atze invariant under $G$-conjugate subgroups
of the Lie group $G$ are equivalent in a sense that they can be
transformed one into another by a suitable group transformation from
the group $G$.  That is why we will consider non-conjugate subgroups
(subalgebras).

Since the group generated by operators $Q_1, \ldots, Q_N$ is
transformed by (\ref{2.2.1}) into the group having generators $Q_1',
\ldots, Q_N' $, Definition 2.2.1 introduces some relation on the set
of subgroups of the Lie group $G$. It is not difficult to become
convinced of the fact that this relation is the equivalence relation
on the set of subgroups of the group $G$ and, consequently, it
separates this set into mutually disjoint classes. The problem of
complete description of such classes (called the problem of a subgroup
classification of the group $G$)\index{Subgroup classification} has
been solved for many important invariance groups of mathematical and
theoretical physics equations \cite{8,9}, \cite{13}--\cite{16},
\cite{66,142,165,166,186}.  In particular, a complete description of
non-conjugate subgroups of the Poincar\'e group $P(1,3)$ 
\cite{8,9,142,165}, extended Poincar\'e group $\ti P(1,3)$ 
\cite{13,66,166} and conformal group $C(1,3)$ \cite{14,66} is
obtained.

We will construct Ans\"atze invariant under one- and
three-parameter subgroups of the groups $P(1,3),\ \wid P(1,3),\ 
C(1,3)$.
\vspace{1.5mm}

\noindent
{\bf 1. {\boldmath $P$}(1,3)-invariant Ans\"atze} \cite{98,100}.\  The
Lie algebra of the Poincar\'e group has thirteen $P(1,3)$
non-conjugate one-dimensional 
subalgebras\index{Subalgebras!of the Poincar\'e algebra}
\begin{equation}
\begin{array}{l}
A_1 = \langle J_{03}\rangle,
\quad
A_2 = \langle J_{12}\rangle,
\quad
A_3 = \langle J_{03} + \al J_{12}\rangle,\\[2mm]
A_4 = \langle J_{01} - J_{03}\rangle,
\quad
\ A_5 = \langle P_0\rangle,
\quad
\ A_6 = \langle P_3\rangle,\\[2mm]
A_7 = \langle P_0 + P_3\rangle,
\quad
\ A_8 = \langle J_{03} + \al P_1\rangle,\\[2mm]
A_9 = \langle J_{12}+ \al P_3\rangle,
\quad
A_{10} = \langle J_{12} + \al P_0\rangle,\\[2mm]
A_{11} = \langle J_{12} + \al (P_0 + P_3)\rangle,
\quad
A_{12} = \langle J_{01} - J_{13} + \al P_3\rangle,\\[2mm]
A_{13} = \langle J_{01} - J_{13} + \al P_2 \rangle,
\end{array}
\label{2.2.2}
\end{equation}
where $ \al \in {\R}^1, \ \al \not= 0 $.

Thus, to construct all inequivalent Ans\"atze invariant under 
one-para\-meter subgroups of the group $P(1,3)$ it suffices to
integrate system (\ref{1.5.7}) for each of the operators listed in
(\ref{2.2.2}). The problem of integrating equations (\ref{1.5.7}) 
is substantially simplified by the fact that operators (1.1.22) 
realize a linear representation of the algebra $AP(1,3)$.

At first, we adduce the Ans\"atze constructed and then consider an
example of integration of equations (\ref{1.5.7}).

A general form of the Ansatz invariant under the group with 
generators (\ref{2.2.2}) is as follows\index{Ansatz!for spinor field}
\begin{equation}
\psi (x) = A(x) \vp (\omega_1, \, \omega_2, \, \omega_3),
\label{2.2.3}
\end{equation} 
where $\vp = \vp (\vec {\omega}) $ is a new unknown four-component
function. A $(4\times 4)$-matrix $A(x)$ and scalar functions $\omega_a 
=\omega_a (x)$ are determined by the choice of a subalgebra from
$A_1,\ A_2, \ldots$, $A_{13}$ and are given 
below\index{Ansatz!$P(1,3)$-invariant}
\begin{eqnarray*}
1)\ \psi(x)&=&\exp\{(1/2) \g_0 \g_3 \ln (x_0 + x_3)\}
  \vp\, (x_0^2 - x_3^2,\, x_1,\, x_2),\\
2)\ \psi(x)&=&\exp\{-(1/2)\g_1 \g_2 \arctan (x_1/x_2)\}
  \vp\, (x_0,\, x_1^2 + x_2^2,\, x_3),\\
3)\ \psi(x)&=&\exp \{(1/2) \g_0 \g_3 \ln (x_0 + x_3) -(1/2)\ga_1\ga_2
  \arctan (x_1/x_2)\}\\
  & &\times\vp\, \Bigl(x_0^2 - x_3^2,\,  x_1^2 + x_2^2,\,
  \al \ln (x_0 + x_3) + \arctan (x_1/x_2)\Bigr),\\
4)\ \psi(x)&=&\exp\Bigl\{ x_1 \Bigl(2(x_0 + x_3)\Bigr)^{-1} 
  (\g_0 + \g_3) \g_1\Bigr\}\vp\, (x_0 + x_3,\, x_0^2 - x_1^2 \\
  & &- x_3^2,\, x_2),\\
5)\ \psi(x)&=&\vp\, (x_1,\, x_2,\, x_3), \\
6)\ \psi(x)&=&\vp\, (x_0,\, x_1,\, x_2), \\
7)\ \psi(x)&=&\vp\, (x_0+x_3,\, x_1,\, x_2), \\
8)\ \psi(x)&=&\exp\{(1/2) \g_0 \g_3 \ln (x_0 + x_3)\}
  \vp\, \Bigl(x_0^2 - x_3^2,\, x_2,\, \al \ln (x_0 + x_3)\\ & &-
  x_1\Bigr),\\ 
9)\ \psi(x)&=&\exp\{-(1/2) \g_1 \g_2 \arctan (x_1/x_2)\} 
  \vp\, \Bigl(x_0,\, x_1^2 + x_2^2,\, x_3 \\
  & &+ \al \arctan (x_1/x_2)\Bigr), \\
10)\ \psi(x)&=&\exp\{-(1/2) \g_1 \g_2 \arctan (x_1/x_2)\}
  \vp\, \Bigl(x_3,\, x_1^2 + x_2^2,\, x_0 \\
  & &- \al \arctan (x_1/x_2)\Bigr),\\
11)\ \psi(x)&=&\exp\{-(1/2) \g_1 \g_2 \arctan (x_1/x_2)\}
  \vp\, \Bigl(x_0 +x_3,\, x_1^2 + x_2^2,\, x_0 - x_3\\ 
  & &- 2 \al \arctan (x_1/x_2)\Bigr), \\
12)\ \psi(x)&=&\exp\{(1/2\al)(x_0+x_3)(\g_0+\g_3)\g_1\}
  \vp\, \Bigl((x_0 +x_3)^2-2\al x_1,\, x_2,\\ 
  & &(x_0+x_3)^3-3\al x_1(x_0+x_3)+3\al^2x_0\Bigr), \\
13)\ \psi(x)&=&\exp \{x_2(2\al)^{-1} (\g_0 + \g_3) \g_1 \} 
  \vp\, \Bigl(x_0 + x_3,\, x_0^2 - x_1^2 - x_3^2,\, \al x_1 \\
  & &- (x_0 + x_3)x_2\Bigr).
\end{eqnarray*}

In the above formulae $\exp \{R\} = \sum\limits^\infty_{n=1} (n!)^{-1} 
R^n +I,$ \ $I$ is the unit $(4\times 4)$-matrix.

We will construct the Ansatz invariant under the algebra $A_1$.  Since
the operator $Q_1 = J_{03} = -x_0 \p_3 - x_3 \p_0 + (1/2) \g_0 \g_3$
satisfies conditions (\ref{1.5.11}), the above Ansatz can be looked
for in the form (\ref{1.5.14}) with $n =4, \ m = 4, \ N =1,$ \ a
$(4\times 4)$-matrix $A(x) $ and functions $\om_1(x), \ \om_2(x), \ 
\om_3(x) $ satisfying equations (\ref{1.5.16}), (\ref{1.5.15}). Thus,
to construct the Ansatz for the field $\psi (x) $ we have to find a
particular solution of the matrix PDE
\begin{equation}
\Bigl(x_0 \p_3 + x_3 \p_0 - (1/2) \g_0 \g_3\Bigr) A(x) = 0
\label{2.2.4}
\end{equation}
and to obtain a complete system of functionally-independent
first integrals of the PDE

\begin{equation}
(x_0 \p_3 + x_3 \p_0) \omega (x) = 0.
\label{2.2.5}
\end{equation}

Hereafter, when integrating a matrix PDEs of the type (\ref{2.2.4}) we
use the following identity:
\begin{equation}
\p_\mu \exp \{Tf(x)\} = \Bigl(\p_\mu f(x)\Bigr)T \exp \{Tf(x)\},
\label{2.2.6}
\end{equation}
which holds true for an arbitrary constant $(4\times 4)$-matrix $T$ and
a smooth scalar function $f(x)$.

We look for a solution of (\ref{2.2.4}) in the form
\begin{displaymath}
A(x) = \exp \{\g_0 \g_3 f(x)\}.
\end{displaymath}

Substituting the above expression into (\ref{2.2.4}) and applying
(\ref{2.2.6}) we arrive at the equality
\begin{displaymath}
\{(x_0 \p_3 + x_3 \p_0)f - 1/2\}\g_0 \g_3  
 \exp \{\g_0 \g_3 f\} = 0
\end{displaymath}
or
\begin{displaymath}
(x_0 \p_3 + x_3 \p_0) f = 1/2.
\end{displaymath}

A particular solution of the above PDE is of the form
$f(x) = (1/2)
\ln (x_0 + x_3)$, whence it follows that
$A(x) = \exp \{(1/2) \ln (x_0 + x_3) \g_0 \g_3\}$.

PDE (\ref{2.2.5}) is equivalent to the Euler-Lagrange 
system\index{Euler-Lagrange equations}
\begin{displaymath}
{dx_0 \over x_3} = {dx_1 \over 0} = {dx_2 \over 0} =
{dx_3 \over x_0},
\end{displaymath}
whose first integrals can be chosen in the form
$\omega_1 = x^2_0 - x^2_3, \ \omega_2 = x_1,
\ \omega_3 = x_2 $.

Substituting the results obtained into the formula (\ref{2.2.3}) we
obtain an Ansatz invariant under the one-dimensional Lie algebra
$A_1$.  The remaining algebras $A_2, \ldots, A_{13} $ are treated in
a similar way.

Now we give a complete list of $P(1,3)$ non-conjugate
three-dimensional subalgebras of the Lie algebra $AP(1,3)$ following 
\cite{66,165}:\index{Subalgebras!of the Poincar\'e algebra}
\begin{eqnarray}
& &A_1 = \langle P_0,\,  P_1,\,  P_2\rangle,\quad
A_2 = \langle P_1,\,  P_2,\,  P_3\rangle,\non\\
& &A_3 = \langle P_0 + P_3,\,  P_1,\,  P_2\rangle,\quad
A_4 = \langle J_{03},\,  P_1,\,  P_2\rangle,\non\\
& &A_5 = \langle J_{03},\,  P_0 + P_3,\,  P_1\rangle,\quad
A_6 = \langle J_{03} + \al P_2,\,  P_0,\,  P_3\rangle ,
\non\\
& &A_7 = \langle J_{03} + \al P_2,\,  P_0 + P_3,\,  P_1\rangle, \quad
A_8 = \langle J_{12},\,  P_0,\,  P_3\rangle,\non\\
& &A_9 = \langle J_{12} + \al P_0,\,  P_1,\,  P_2 \rangle, \quad
A_{10} = \langle J_{12} + \al P_3,\,  P_1,\,  P_2 \rangle,\non\\
& &A_{11} = \langle J_{12} +P_0 + P_3,\,  P_1,\,  P_2\rangle,\quad
A_{12} = \langle G_1,\,  P_0 + P_3,\,  P_2\rangle,\non\\
& &A_{13} = \langle G_1,\,  P_0 + P_3,\,  P_1 + \al P_2\rangle,\non\\ 
& &A_{14} = \langle G_1 + P_2,\,  P_0 + P_3,\,  P_1\rangle,\non\\
& &A_{15} = \langle G_1+ P_0,\,  P_0+ P_3,\,  P_2\rangle
,\label{2.2.7}\\ 
& &A_{16} = \langle G_1 + P_0,\,  P_1 + \al P_2,\,  P_0 + P_3\rangle
,\non\\ 
& &A_{17} = \langle J_{03} + \al J_{12},\, P_0,\, P_3\rangle,\quad
A_{18} = \langle J_{03}+ \al J_{12},\,  P_1,\,  P_2\rangle,\non\\ 
& &A_{19} = \langle J_{12},\,  J_{03},\, P_0 + P_3\rangle,\quad
A_{20} = \langle G_1,\,  G_2,\,  P_0+ P_3\rangle,\non\\ 
& &A_{21} = \langle G_1 + P_2,\,  G_2 + \al P_1 +
\beta P_2,\,  P_0 + P_3 \rangle ,\non\\
& &A_{22} = \langle G_1,\,  G_2 + P_1 +\beta P_2,\,  P_0 + P_3 \rangle
, \non\\
& &A_{23} = \langle G_1,\,  G_2 + P_2,\,  P_0 + P_3 \rangle,\quad
A_{24} = \langle G_1,\,  J_{03},\,  P_2\rangle,\non\\
& &A_{25} = \langle J_{03} + \al P_1 + \beta P_2,\,  G_1,\,  P_0 +
P_3\rangle, \non\\
& &A_{26} = \langle J_{12} + P_0 + P_3,\,  G_1,\,  G_2\rangle,\quad
A_{27} = \langle J_{03} + \al
J_{12},\,  G_1,\,  G_2\rangle,\non\\
& &A_{28} = \langle G_1,\,  G_2,\,  J_{12}\rangle,\quad
A_{29} = \langle J_{01},\,  J_{02},\,  J_{12}\rangle,\quad
A_{30} = \langle J_{12},\,  J_{23},\,  J_{31}\rangle.\non
\end{eqnarray}

In (\ref{2.2.7}) $G_i = J_{0i} - J_{i3},\ i = {1,2}$ and $\langle
Q_1,\, Q_2,\, Q_3\rangle$ designates the linear span of operators
$Q_a$.

Ans\"atze invariant under the algebras (\ref{2.2.7}) were constructed
in \cite{100,103}. They can be represented in the form
\begin{equation}
\psi(x) = A(x)\vp(\omega),
\label{2.2.8}
\end{equation}
where $\vp(\omega) $ is a new unknown four-component function, a
$(4\times 4)$-matrix $A(x)$ and scalar function $\omega(x)$ being
given below.\index{Ansatz!$P(1,3)$-invariant}

\begin{eqnarray*}
1)\ \psi(x)&=&\vp\, (x_3),\\
2)\ \psi(x)&=&\vp\, (x_0),\\
3)\ \psi(x)&=&\vp\, (x_0+x_3),\\
4)\ \psi(x)&=&\exp \{(1/2) \g_0 \g_3 \ln (x_0 + x_3)\}\vp\, (x_0^2 -
  x_3^2),\\
5)\ \psi(x)&=&\exp\{(1/2)\g_0 \g_3 \ln (x_0 + x_3)\}\vp\, (x_2),\\
6)\ \psi(x)&=&\exp \{(x_2/2 \al)\g_0 \g_3\}\vp\, (x_1),\\
7)\ \psi(x)&=&\exp \{(x_2/2 \al)\g_0 \g_3\}\vp\, \Bigl(\al 
  \ln (x_0 + x_3) - x_2\Bigr),\\
8)\ \psi(x)&=&\exp \{-(1/2) \g_1 \g_2 \arctan (x_1/x_2)\}\vp\, (x_1^2
+ x_2^2),\\
9)\ \psi(x)&=&\exp \{-(x_0/2 \al)\g_1 \g_2\}\vp\, (x_3),\\
10)\ \psi(x)&=&\exp \{(x_3/2 \al)\g_1 \g_2\}\vp\, (x_0),\\
11)\ \psi(x)&=&\exp \{(1/4)(x_3 - x_0) \g_1 \g_2\}\vp\, (x_0 + x_3),\\ 
12)\  \psi(x)&=&\exp \Bigl\{\Bigl(x_1/2(x_0 + x_3)\Bigr)(\g_0 
  + \g_3)\g_1\Bigl\}\vp\, (x_0 + x_3   ),\\
13)\  \psi(x)&=&\exp \Bigl\{\Bigl((\al x_1 - x_2)/2(x_0 + x_3)\Bigr)
  (\g_0 + \g_3)\g_1\Bigl\}\vp\, (x_0 + x_3),\\
14)\  \psi(x)&=&\exp \{(x_2/2)(\g_0 + \g_3)\g_1\}\vp\, (x_0 + x_3),\\
15)\  \psi(x)&=&\exp \Bigl\{-\Bigl((x_0 + x_3)/2\Bigr) 
  (\g_0 + \g_3)\g_1\Bigr\}
  \vp\, \Bigl(2x_1+(x_0 + x_3)^2\Bigr),\\
16)\ \psi(x)&=&\exp \Bigl\{-\Bigl((x_0 + x_3)/2\Bigr)(\g_0 +
\g_3)\g_1\Bigr\} 
  \vp\, \Bigl(2(x_2 - \al x_1) \\
  & &- \al (x_0 + x_3)^2\Bigr),\\
17)\ \psi(x)&=&\exp \{-(1/2 \al)(\g_0 \g_3 + \al \g_1 \g_2)
\arctan (x_1/x_2)\}\vp\, (x_1^2 + x_2^2 ),\\
18)\ \psi(x)&=&\exp \{(1/2)(\g_0 \g_3 + \al \g_1 \g_2)
\ln (x_0+x_3)\}\vp\, (x_0^2 - x_3^2 ),\\
19)\ \psi(x)&=&\exp \{(1/2)\g_0 \g_3 \ln (x_0 + x_3) - (1/2) \g_1 \g_2
  \arctan (x_1/x_2)\}\\
  & &\times\vp\, (x_1^2 + x_2^2),\\
20)\ \psi(x)&=&\exp \Bigl\{\Bigl(1/2(x_0 + x_3)\Bigr)(\g_0 + \g_3)
(\g_1 x_1 + \g_2 x_2)\Bigr\}\vp\, (x_0 + x_3),\\
21)\ \psi(x)&=&\exp \Bigl\{\Bigl[2\Bigl((x_0 + x_3)(x_0 + x_3 
  + \beta) - \al\Bigl)\Bigl]^{-1}(\g_0 + \g_3)\\
  & &\times \Bigl[\g_1 \Bigl((x_0 + x_3 
  + \beta) x_1 - \al x_2\Bigr) +
  \g_2 \Bigl((x_0 + x_3) x_2- x_1\Bigr)\Bigr] \Bigr\} \\
  & &\times\vp\, (x_0+x_3),\\
22)\ \psi(x)&=&\exp \Bigl\{\Bigl(2(x_0 + x_3)(x_0 + x_3 
  + \beta) \Bigl)^{-1}(\g_0 + \g_3) \\
  & &\times\Bigl[\g_1 \Bigl((x_0 + x_3 
  + \beta) x_1 - x_2\Bigr) +
  \g_2 x_2 (x_0 + x_3)\Bigr] \Bigr\}
  \vp\, (x_0+x_3),\\
23)\ \psi(x)&=&\exp \Bigl\{\Bigl(2(x_0 + x_3)(x_0 + x_3 + 1)
\Bigl)^{-1} 
  (\g_0 + \g_3) \\
  & &\times\Bigl(\g_1 x_1(x_0 + x_3 +1) +
  \g_2 x_2 (x_0 + x_3)\Bigr) \Bigr\}
  \vp\, (x_0+x_3),\\
24)\ \psi(x)&=&\exp \Bigl\{\Bigl(x_1/2(x_0 + x_3)\Bigr)
  (\g_0 + \g_3) \g_1\Bigr\}
  \exp \{(1/2)\g_0 \g_3 \\
  & &\times\ln (x_0 + x_3)\}
  \vp\, (x_0^2 - x_1^2 - x_3^2),\\
25)\ \psi(x)&=& \exp \Bigl\{ \Bigl(1/2(x_0 + x_3)\Bigr)\Bigl(x_1 
  - \al \ln (x_0 + x_3)\Bigr)   (\g_0 + \g_3) \g_1\Bigr\}\\
  & &\times\exp \{(1/2)\g_0 \g_3 \ln (x_0 + x_3)\} 
 \vp\, \Bigl(x_2 - \beta \ln (x_0 + x_3)\Bigr),\\
26)\ \psi(x)&=&\exp \Bigl\{\Bigl(1/2(x_0 + x_3)\Bigr)(\g_0 + \g_3) 
  (\g_1 x_1 + \g_2 x_2)\Bigr\}\\
  & &\times\exp \Bigl(-\Bigl(1/4(x_0 + x_3)\Bigr) 
  (x \cdot x) \g_1 \g_2\Bigr\} 
  \vp\, (x_0 + x_3),\\
27)\ \psi(x)&=&\exp \Bigl\{\Bigl(1/2(x_0 + x_3)\Bigr)(\g_0 +
\g_3)(\g_1 x_1 + \g_2 x_2)\Bigr\}\\ 
  & &\times\exp \{(1/2)(\g_0 \g_3 + \al \g_1 \g_2) \ln (x_0 + x_3)\} 
  \vp\, (x \cdot x).\\
\end{eqnarray*}

Let us note that triplets of operators $Q_a$ which are basis elements
of the algebras $A_{28}$--$A_{30} $ do not satisfy condition
(\ref{1.5.2}). Consequently, they lead to partially-invariant
solutions which are not considered here.

As an example, we will carry out integration of equations
(\ref{1.5.16}), (\ref{1.5.15}) for the algebra $A_4$ from
(\ref{2.2.7}). Choosing in (\ref{1.5.16}), (\ref{1.5.15}) $n= 4, \ m =
4, \ N= 3,$\ $Q_1 = -x_0\p_3 - x_3 \p_0 + (1/2) \g_0 \g_3,$ \ $Q_2 =
\p_1$,\ $ Q_3 = \p_2 $ yields the following system of PDEs for $A(x),
\ \omega(x)$:
\begin{eqnarray}
& &(x_0 \p_3 + x_3 \p_0 - (1/2) \g_0 \g_3) A = 0,
\quad
\p_1 A = \p_2 A = 0,\label{2.2.9}\\
& &(x_0 \p_3 + x_3 \p_0) \omega = 0,
\quad
\p_1 \omega = \p_2 \omega = 0.\label{2.2.10}
\end{eqnarray}

From the last two equations of system (\ref{2.2.9}) it follows
that $A = A(x_0, x_3)$. Substituting this expression into the
first equation we get
\begin{displaymath}
(x_3 \p_0 + x_0 \p_3 - (1/2) \g_0 \g_3) A(x_0, x_3) = 0,
\end{displaymath}
whence
\begin{displaymath}
A(x) = \exp \{(1/2) \g_0 \g_3 \ln (x_0 + x_3)\}.
\end{displaymath}

It is easy to see that a complete set of functionally-independent
first integrals of system (\ref{2.2.10}) consists of one integral
which can be chosen in the form $\omega (x) = x_0^2 - x_3^2 $.
Thus, we obtain the Ansatz numbered by 4.
\vspace{2mm}

\noindent
{\bf 2.\ {\boldmath $\wid P$}(1,3)-invariant Ans\"atze }
\cite{96,98,100,103}.\ 
Subgroup classification of the extended Poincar\'e group was carried
out in \cite{13,66,166}. One-dimensional subalgebras of the algebra
$ A\wid P(1,3)$ which are $\wid P(1,3)$ non-conjugate to subalgebras
of the algebra $AP(1,3)$ are equivalent to the following 
ones:\index{Subalgebras!of the extended Poincar\'e algebra}
\begin{equation}
\begin{array}{l}
\langle J_{01} - J_{13} + \al D\rangle,
\quad
\langle J_{12} + \al D\rangle,\\
\langle J_{03} + \beta J_{12} + \al D\rangle,
\quad
\langle J_{03} + \beta J_{12} -
D + \al P_0\rangle,
\end{array}
\label{2.2.11}
\end{equation}
where $\{\al, \beta\} \subset {\R}^1, \  \al \not= 0,\ 
D = x_\mu \p_\mu + k, \ k \in {\R}^1$ is the
infinitesimal operator of the group of scale transformations
(\ref{1.1.24d}).

Ans\"atze invariant under operators (\ref{2.2.11}) are given by the
formulae\index{Ansatz!$\wid P(1,3)$-invariant} 
\begin{eqnarray}
\psi(x)&=&(x_0 - x_3)^{-k} \ \exp \{(1/2 \al)(\g_0 + \g_3)
  \g_1 \ln (x_0 + x_3)\} \vp\, (\vec \omega),\non\\
  & &\omega_1 = (x_0^2 - x_1^2 - x_3^2) x_2^{-2},
  \quad
  \omega_2 = (x_0 + x_3) x_2^{-1},\non\\
  & &\omega_3 = \al x_1(x_0 + x_3)^{-1} + \ln (x_0 + x_3);\non\\
\psi(x)&=&(x_1^2 + x_2^2)^{-k/2}
  \exp \{-(1/2) \g_1 \g_2 \arctan (x_1/x_2)\}
  \vp\, (\vec \omega),\non\\
  & &\omega_1 = x_0 x_3^{-1},\quad
  \omega_2 = 2 \al\arctan (x_1/x_2) -
  \ln (x_1^2 + x_2^2),\non\\
  & &\omega_3 = (x_0^2 - x_3^2)(x_1^2 + x_2^2)^{-1};\non\\
\psi(x)&=&(x_0^2 - x_3^2)^{-k/2}
  \exp \Bigl\{(1/4)(\g_0 \g_3+ \beta \g_1 \g_2)\non\\ 
  & &\times \ln \Bigl((x_0 + x_3)/
  (x_0 - x_3)\Bigr)\Bigr\} \vp\,  (\vec \omega),\label{2.2.12}\\
  & &\omega_1 = (x_0^2 - x_3^2)^{\al - 1} (x_0 + x_3)^{2 \al},
  \quad
  \omega_2 = (x_0^2 - x_3^2)\non\\
  & &\times (x_1^2 + x_2^2)^{-1},\quad
  \omega_3 = \beta\ln (x_1^2 + x_2^2) \non\\
  & &- 2 \al \arctan (x_1/x_2);\non\\
\psi(x)&=&(2x_0 + 2x_3 - \al)^{-k/2}
  \exp \Bigl\{(1/4)(\g_0 \g_3+ \beta \g_1 \g_2)\non\\
  & &\times\ln \Bigl((x_0 + x_3)/
  (x_0 - x_3)\Bigr)\Bigr\} \vp\,  (\vec \omega),\non\\
  & &\omega_1 = (2x_0 + 2x_3 - \al) \exp \{(2/ \al) (x_0 -
  x_3)\},\non\\ 
  & &\omega_2 = (2x_0 + 2x_3 - \al) (x_1^2 + x_2^2)^{-1},\non\\
  & &\omega_3 = \beta\ln (x_1^2 + x_2^2) + 2\arctan (x_1/x_2).\non
\end{eqnarray}

Here $ \vp = \vp(\omega_1, \omega_2, \omega_3) $ is an arbitrary
four-component function.

Three-dimensional subalgebras of the algebra $A \wid P(1,3)$ which
are $\wid P(1,3)$ non-conjugate to subalgebras of the Poincar\'e 
algebra are as 
follows\index{Subalgebras!of the extended Poincar\'e algebra}
\begin{eqnarray}
& &A_1 = \langle -J_{03} + D + P_0 + P_3,\,  P_1,\,  P_2\rangle,\non\\
& &A_2 = \langle -J_{03} + D + P_0 + P_3,\,  P_0 - P_3,\,  
P_1\rangle,\non\\
& &A_3 = \langle J_{12}+ \al (-J_{03} + D + P_0 + P_3),\,  
  P_1,\,  P_2\rangle,\non\\
& &A_4 = \langle -J_{03} + D + P_0 + P_3,\,  J_{12} + \al(P_0 +
  P_3),\, P_0 - P_3\rangle,\non\\
& &A_5 = \langle -J_{03} + D,\, J_{12} + P_0 + P_3,\,  P_0 -
P_3\rangle,\non\\ 
& &A_6 = \langle -J_{03} + 2D,\,  \wid G_1 + P_0 + P_3,\,  P_0 -
P_3\rangle,\non\\ 
& &A_7 = \langle -J_{03} + 2D,\,  \wid G_1 + P_0 + P_3,\,  P_2\rangle
,\non\\ 
& &A_8 = \langle -J_{03} + D,\,  \wid G_1 - P_2,\,  P_0 - P_3 \rangle
,\non\\ 
& &A_9 = \langle -J_{03} - D + P_0 - P_3,\,  \wid G_1,\,  P_2\rangle
,\non\\ 
& &A_{10} = \langle -J_{03} + D,\,  \wid G_1,\,  \wid G_2 - P_2\rangle
,\non\\ 
& &A_{11} = \langle -J_{03} - D + P_0 - P_3,\,  \wid G_1,\,
  \wid G_2\rangle,\non\\
& &A_{12} = \langle J_{12} - \al(J_{03} + D - P_0 + P_3),\,  
  \wid G_1,\,  \wid G_2\rangle,\non\\
& &A_{13} = \langle J_{12} - \al D,\,  P_0,\,  P_3\rangle,
  \quad
A_{14} = \langle J_{12} - \al D,\,  P_1,\,  P_2\rangle,\non\\
& &A_{15} = \langle J_{03} + \al D,\,  P_0,\,  P_3\rangle,
  \quad
A_{16} = \langle J_{03} + \al D,\,  P_0 - P_3,\,  P_1\rangle,\non\\
& &A_{17} = \langle J_{03} + \al D,\,  P_1,\,  P_2\rangle,
  \quad
A_{18} = \langle J_{12} - \al J_{03} - \beta D,\,  P_0,\,  P_3\rangle
,\non\\ 
& &A_{19} = \langle J_{12} - \al J_{03} - \beta D,\,  P_1,\,  
  P_2\rangle,\non\\
& &A_{20} = \langle \wid G_1 - \al D,\,  P_0 - P_3,\,  P_2 \rangle,
  \label{2.2.13}\\
& &A_{21} = \langle \wid G_1 - \al D,\,  P_0 - P_3,\,  P_1 \rangle
,\non\\ 
& &A_{22} = \langle \wid G_1 - \al D,\,  P_0 - P_3,\,  P_1 + \beta
  P_2\rangle,  \non\\
& &A_{23} = \langle \wid G_1 - \al D,\,  \wid G_2 - \beta D,\,  P_0 -
  P_3\rangle,\non\\
& &A_{24} = \langle \wid G_1,\,  J_{03} + \al D,\,  P_0 - P_3 \rangle
,\non\\ \quad
& &A_{25} = \langle \wid G_1,\,  J_{03} + \al D,\,  P_2 \rangle,\non\\
& &A_{26} = \langle J_{12} - \al D,\,  J_{03} + \beta D,\, P_0 -
P_3\rangle, \non\\
& &A_{27} = \langle J_{03},\,  J_{12},\,  D \rangle,\quad
A_{28}=\langle P_0, \,  P_3, \,  D\rangle, \nonumber\quad  \\
& &A_{29}=\langle P_1, \,  P_2, \,  D\rangle, \quad 
A_{30}=\langle P_0+P_3, \,  P_1, \,  D\rangle, \quad \nonumber \\
& &A_{31}=\langle P_0, \,  J_{12}, \,  D\rangle, \quad 
A_{32}=\langle P_3, \,  J_{12}, \,  D\rangle, \quad \nonumber\\
& &A_{33}=\langle P_0+P_3, \,  J_{12}, \,  D\rangle, \quad 
A_{34}=\langle P_1, \,  J_{03}, \,  D\rangle, \quad \nonumber\\
& &A_{35}=\langle P_0+P_3, \,  J_{03}, \,  D\rangle, \quad 
A_{36}=\langle P_0+P_3, \,  J_{12}+\alpha J_{03}, \,  
  D\rangle, \quad \nonumber\\
& &A_{37}=\langle P_0+P_3, \,  \wid G_1, \,  D\rangle, \quad 
A_{38}=\langle P_2, \,  \wid G_1, \,  D\rangle, \quad \nonumber\\
& &A_{39}=\langle \wid G_1, \,  \wid G_2, \,  D\rangle, \quad 
A_{40}=\langle \wid G_1, \,  J_{03}, \,  D\rangle, \quad \nonumber\\
& &A_{41}=\langle P_0+P_3, \, \wid G_1, \,  \wid G_2+D\rangle, \quad 
A_{42}=\langle \wid G_1, \,  \wid G_2, \,  J_{03}+\alpha D\rangle,
\quad \nonumber\\ 
& &A_{43}=\langle \wid G_1, \, \wid G_2, \,  J_{12}+\alpha D\rangle,
\quad A_{44}=\langle \wid G_1, \,  \wid G_2, \, J_{12}+\alpha J_{03}
+\beta D\rangle, \quad \nonumber
\end{eqnarray}
where $\wid G_i = - J_{0i} - J_{i3},  \ i = 1,2;  \ \{\al, \beta\}
\subset {\R}^1$.

Without going into details of integration of equations (\ref{1.5.15}),
(\ref{1.5.16}) we list the Ans\"atze for the spinor field $\psi(x)$
invariant under the three-dimensional subalgebras (\ref{2.2.13}).
\begin{eqnarray*}
1)\ \psi(x)&=&(x_0+x_3)^{-k/2}\exp \{(1/4)\g_0 \g_3 \ln (x_0 + x_3)\}
   \vp\, \Bigl(\ln (x_0 + x_3) \\
   & &- x_0 + x_3\Bigr),\\
2)\ \psi(x)&=&x_2^{-k}\exp \{(1/2)\g_0 \g_3\ln x_2\}\vp\, 
   (x_0 - x_3 - 2 \ln x_2),\\
3)\ \psi(x)&=&(x_0+x_3)^{-k/2}\exp \{(1/4\alpha) (\alpha\g_0 \g_3 -
\g_1 \g_2) \ln (x_0 + x_3)\}\\
   & &\times\vp\, \Bigl(\ln(x_0 + x_3) - x_0 + x_3\Bigr),\\
4)\ \psi(x)&=&(x_1^2 + x_2^2)^{-k/2}\exp \{(1/2) \g_1 \g_2 \arctan
{(x_2/x_1)} + (1/4)\g_0 \g_3  \\
   & &\times\ln (x_1^2+x_2^2)\}\vp\, \Bigl(
   \al \arctan {(x_2/x_1)} + (1/2)(x_0 - x_3) \\
   & & - (1/2) \ln (x_1^2 + x_2^2)\Bigr),\\
5)\ \psi(x)&=&(x_1^2 + x_2^2)^{-k/2}\exp \{(1/2)\g_1 \g_2 \arctan
(x_2/x_1)  
   + (1/4)\g_0 \g_3 \\
   & &\times\ln (x_1^2 + x_2^2)\}
   \vp\, \Bigl(\arctan (x_2/x_1) + (x_0 - x_3)/2\Bigr),\\
6)\ \psi(x)&=&x_2^{-k}\exp \{-(1/4) \g_1 (\g_0 - \g_3) (x_0 - x_3)\}
   \exp \{(1/4) 
   \g_0 \g_3 \ln x_2\}\\
   & &\times\vp\, \Bigl(x_2[(x_0 - x_3)^2 - 4x_1]^{-1}\Bigr),\\
7)\ \psi(x)&=&\Bigl((x_0 - x_3)^2 - 4x_1\Bigr)^{-k}\exp \{-(1/4) 
   \g_1 (\g_0 - \g_3) (x_0 - x_3)\}\\
   & &\times\exp \Bigl\{ (1/4)\g_0 \g_3 \ln 
   \Bigl((x_0 - x_3)^2 - 4x_1\Bigr)\Bigr\}
   \vp\, \Bigl([(x_0 - x_3)^2 - 4x_1]^3\\ 
   & &\times[(x_0 - x_3)^3 - 6(x_0 - x_3) x_1 
   + 6(x_0 + x_3)]^{-2}\Bigr),\\
8)\ \psi(x)&=&\Bigl(x_1(x_0 - x_3)^{-1} - x_2\Bigr)^{-k}
   \exp \{-(1/2)\g_1 (\g_0 - \g_3)x_1(x_0 - x_3)^{-1}\}\\
   & &\times\exp \Bigl\{(1/2) \g_0 \g_3 \ln \Bigl(x_1(x_0 - x_3)^{-1}
   - x_2\Bigr)\Bigr\}\vp\, (x_0-x_3),\\
9)\ \psi(x)&=&(x_0-x_3)^{-k/2} \exp\{-(1/2) \g_1 (\g_0 - \g_3)
   x_1(x_0 - x_3)^{-1}\}\\ 
   & &\times\exp \{-(1/4) \g_0 \g_3 \ln (x_0 - x_3)\}
   \vp\, \Bigl((x_0^2 - x_1^2 - x_3^2)(x_0 - x_3)^{-1}\\
   & &+ \ln (x_0 - x_3)\Bigr),\\
10)\ \psi(x)&=&\Bigl(x \cdot x + (x_0^2 - x_1^2 - x_3^2)(x_0 -
   x_3)^{-1}\Bigr)^{-k/2} 
   \exp \{-(1/2)x_1\\
   & &\times (x_0 - x_3)^{-1}\g_1 (\g_0 - \g_3)\}
   \exp \{-(1/2)x_2(x_0 - x_3+1)^{-1}\\
   & &\times\g_2 (\g_0 - \g_3)\} 
   \exp \Bigl\{(1/4)\g_0 \g_3 \ln \Bigl(x \cdot x + (x_0^2 - x_1^2 -
   x_3^2)\\ 
   & &\times(x_0 - x_3)^{-1}\Bigr)\Bigr\}\vp\, (x_0 - x_3),\\
11)\ \psi(x)&=&(x_0-x_3)^{-k/2}\exp\{(1/2)(x_0 - x_3)^{-1} 
   (\g_0 - \g_3) (\g_1 x_1 + \g_2 x_2)\}\\
   & &\times\exp \{-(1/4)\g_0 \g_3\ln (x_0 - x_3)\}
   \vp\, \Bigl((x \cdot x)(x_0 -x_3)^{-1} \\
   & &+ \ln (x_0 - x_3)\Bigr),\\
12)\ \psi(x)&=&(x_0-x_3)^{-k/2}\exp\{(1/2)(x_0 - x_3)^{-1} 
   (\g_0 - \g_3) (\g_1 x_1 + \g_2 x_2)\}\\
   & &\times\exp \{(1/4\alpha)(\g_1\g_2-\alpha\g_0 \g_3)\ln (x_0 -
   x_3)\} 
   \vp\, \Bigl((x \cdot x)\\ 
   & &\times(x_0 -x_3)^{-1}+ \ln (x_0 - x_3)\Bigr),\\
13)\ \psi(x)&=&(x_1^2 + x_2^2)^{-k/2}\exp \{(1/2)\g_1 \g_2 \arctan
(x_2/x_1)\} 
   \vp\, \Bigl(\alpha \arctan (x_2/x_1) \\
   & &-(1/2)\ln (x_1^2 + x_2^2)\Bigr),\\
14)\ \psi(x)&=&x_3^{-k}\exp \{(1/2\alpha) \g_1 \g_2 \ln x_3\}
   \vp\, (x_0/x_3),\\
15)\ \psi(x)&=&x_2^{-k}\exp \{-(1/2\alpha)\g_0 \g_3\ln x_2\}\vp\,
(x_1/x_2),\\ 
16)\ \psi(x)&=&x_2^{-k}\exp \{-(1/2\alpha)\g_0 \g_3\ln x_2\}
   \vp\, \Bigl((x_0 - x_3)x_2^{-(\alpha + 1)/\alpha}\Bigr),\\
17)\ \psi(x)&=&(x_0+x_3)^{k\alpha/2}(x_0-x_3)^{-k\alpha/2}
   \exp \Bigl\{(1/4)\g_0 \g_3\ln \Bigl((x_0 + x_3) \\
   & &\times(x_0 - x_3)^{-1}\Bigr)\Bigr\} 
   \vp\, \Bigl((x_0 + x_3)^{(1 + \alpha)/2}
   (x_0 - x_3)^{(1- \alpha)/2}\Bigr),\\
18)\ \psi(x)&=&(x_1^2 + x_2^2)^{-k/2}\exp \{(1/2)(\g_1 \g_2 - \al \g_0
\g_3)  
   \arctan (x_2/x_1)\}\\
   & &\times\vp\, \Bigl(\beta \arctan (x_2/x_1)
   - (1/2) \ln (x_1^2 + x_2^2)\Bigr),\\
19)\ \psi(x)&=&(x_0+x_3)^{k\beta/2\alpha}(x_0-x_3)^{-k\beta/2\alpha} 
   \exp \Bigl\{(1/4\al)(\al\g_0 \g_3 \\
   & &- \g_1 \g_2)
   \ln \Bigl((x_0 + x_3)(x_0 - x_3)^{-1}\Bigr)\Bigr\}
   \vp\, \Bigl((x_0 + x_3)^{(\al + \beta)/2}\\
   & &\times(x_0 - x_3)^{(\al - \beta)/2}\Bigr),\\
20)\ \psi(x)&=&(x_0-x_3)^{-k} \exp\{-(1/2)x_1(x_0 
   - x_3)^{-1} \g_1 (\g_0 - \g_3) \}\\ & &\times
   \vp\, \Bigl(\ln (x_0 - x_3) + \alpha x_1(x_0 - x_3)^{-1}\Bigr),\\
21)\ \psi(x)&=&x_2^{-k}\exp \{(1/2\alpha) \g_1 (\g_0 - \g_3) \ln (x_0
- x_3)\} 
   \vp\, \Bigl((x_0 - x_3)/x_2\Bigr),\\
22)\ \psi(x)&=&(x_0-x_3)^{-k}\exp \{(1/2\beta) \g_1 (\g_0 - \g_3) 
   (x_2 - \beta x_1)(x_0 - x_3)^{-1}\}\\
   & &\times\vp\, \Bigl((x_2 - \beta x_1)(x_0 - x_3)^{-1}-
   (\beta / \alpha) \ln (x_0 - x_3)\Bigr),\\
23)\ \psi(x)&=&\exp \Bigl\{ (1/2) \Bigl(2 \al k - \g_1 (\g_0 -
\g_3)\Bigr) 
   x_1(x_0 - x_3)^{-1}\Bigr\}
   \exp \Bigl\{ (1/2) \\
   & &\times\Bigl(2 \beta k - \g_2 (\g_0 - \g_3)\Bigr)
   x_2(x_0 - x_3)^{-1}\Bigr\} \vp\, \Bigl(\exp \{(\al x_1 \\
   & &+ \beta x_2)(x_0 - x_3)^{-1}\}
   (x_0 - x_3)\Bigr),\\
24)\ \psi(x)&=&x_2^{-k}\exp \{-(1/2) x_1(x_0 
   - x_3)^{-1}\g_1 (\g_0 - \g_3) \}
   \exp \{-(1/2\alpha) \\
   & &\times\g_0 \g_3\ln x_2\}
   \vp\, \Bigl((x_0 - x_3) x_2^{-(\alpha + 1)/\alpha}\Bigr),\\
25)\ \psi(x)&=&(x_0^2-x_1^2-x_3^2)^{-k/2}
   \exp\{-(1/2)x_1(x_0 - x_3)^{-1} \g_1 (\g_0 - \g_3)\}\\
   & &\times\exp \{-(1/4\alpha) \g_0 \g_3\ln (x_0^2-x_1^2-x_3^2)\}
   \vp\, \Bigl((x_0 - x_3)\\
   & &\times(x_0^2 - x_1^2 - x_3^2)^
   {- (\alpha + 1)/2\alpha}\Bigr),\\
26)\ \psi(x)&=&(x_1^2 + x_2^2)^{-k/2}\exp \Bigl\{(1/2) 
   \g_1 \g_2 \arctan (x_2/x_1)
   +(1/4) \g_0 \g_3 \\
   & &\times\Bigl(\ln (x_1^2 + x_2^2) 
   -2 \ln (x_0 - x_3)\Bigr)\Bigr\}
   \vp\, \Bigl((1/2)(\beta + 1) \ln (x_1^2 + x_2^2)\\ 
   & &- \beta \ln (x_0 - x_3) - \al \arctan (x _2/x_1)\Bigr),\\
27)\ \psi(x)&=&(x\cdot x)^{-k/2}\exp \Bigl\{(1/4)\g_0\g_3\ln
   \Bigl((x_0+x_3)(x_0-x_3)^{-1}\Bigr)\Bigr\}\\
   & &\times\exp\{-(1/2)\g_1\g_2\arctan (x_1/x_2)\}
   \varphi \Bigl((x_0^2-x_3^2)(x_1^2+x_2^2)^{-1}\Bigr),\\
28)\ \psi(x)&=&(x_1^2+x_2^2)^{-k/2}\vp\, (x_1/x_2), \\
29)\ \psi(x)&=&(x_0^2-x_3^2)^{-k/2}\vp\, (x_0/x_3), \\
30)\ \psi(x)&=&(x_0-x_3)^{-k}\vp\, \Bigl(x_2(x_0-x_3)^{-1}\Bigr), \\
31)\ \psi(x)&=&(x_1^2+x_2^2)^{-k/2}\exp \{(1/2)
   \g_1\g_2\arctan (x_2/x_1)\}\\
   & &\times \vp\,  \Bigl((x_1^2+x_2^2)^{1/2}x_3^{-1}\Bigr),\\
32)\ \psi(x)&=&(x_1^2+x_2^2)^{-k/2}\exp \{(1/2)
   \g_1\g_2\arctan (x_2/x_1)\}\\
   & &\times\vp\,  \Bigl((x_1^2+x_2^2)^{1/2}x_0^{-1}\Bigr),\\
33)\ \psi(x)&=&(x_1^2+x_2^2)^{-k/2}\exp \Bigl\{(1/2)
   \g_1\g_2\arctan (x_2/x_1)\}\\
   & &\times\vp\,  \Bigl((x_1^2+x_2^2)^{1/2}(x_0-x_3)^{-1}\Bigr),\\
34)\ \psi(x)&=&(x_0^2-x_3^2)^{-k/2}\exp \Bigl\{(1/4)
   \g_0\g_3\ln \Bigl((x_0+x_3)/(x_0-x_3)\Bigr)\Bigr\}\\
   & &\times\vp\,  \Bigl((x_0^2-x_3^2)^{1/2} x_2^{-1}\Bigr),\\
35)\ \psi(x)&=&x_1^{-k}\exp \Bigl\{(1/2)
   \g_0\g_3\ln \Bigl(x_1(x_0-x_3)^{-1}\Bigr)\Bigr\}
   \vp\,  (x_2/x_1),\\
36)\ \psi(x)&=&(x_1^2+x_2^2)^{-k/2}\exp \Bigl\{(1/2)\g_0\g_3\ln
   \Bigl((x_1^2+x_2^2)^{1/2}(x_0-x_3)^{-1}\Bigr)\\
   & &+(1/2)\g_1\g_2\arctan (x_2/x_1)\Bigr\}
   \vp\,  \Bigl(\ln (x_1^2+x_2^2)^{1/2}
   -\ln (x_0-x_3)\\
   & &+\alpha\arctan (x_2/x_1)\Bigr),\\
37)\ \psi(x)&=&(x_0-x_3)^{-k}\exp \{-(1/2)
   \g_1(\g_0-\g_3)x_1(x_0-x_3)^{-1}\}\\
   & &\times\vp\,  \Bigl(x_2 (x_0-x_3)^{-1}\Bigr),\\
38)\ \psi(x)&=&(x_0-x_3)^{-k}\exp \{-(1/2)
   \g_1(\g_0-\g_3)x_1(x_0-x_3)^{-1}\}\\
   & &\times\vp\,  \Bigl((x_0^2-x_1^2-x_3^2) (x_0-x_3)^{-2}\Bigr),\\
39)\ \psi(x)&=&(x_0-x_3)^{-k}\exp \{(1/2)(x_0-x_3)^{-1}
   (\g_0-\g_3)(\g_1x_1+\g_2x_2)\}\\ 
   & &\times\vp\,  \Bigl(x\cdot x (x_0-x_3)^{-2}\Bigr),\\
40)\ \psi(x)&=&x_2^{-k}\exp \{-(1/2)
   \g_1(\g_0-\g_3)x_1(x_0-x_3)^{-1}\}
   \exp \{(1/2)\g_0\g_3\\
   & &\times\ln \Bigl(x_2(x_0-x_3)^{-1}\Bigr)\Bigr\} 
   \vp\,  \Bigl((x_0^2-x_1^2-x_3^2) x_2^{-2}\Bigr),\\
41)\ \psi(x)&=&(x_0-x_3)^{-k}\exp \Bigl\{(1/2)
   (\g_0-\g_3)\Bigl(\g_1x_1(x_0-x_3)^{-1}\\
   & &-\g_2\ln (x_0-x_3)\Bigr)\Bigr\} 
   \vp\, \Bigl(\ln (x_0-x_3)+x_2(x_0-x_3)^{-1}\Bigr), \\
42)\ \psi(x)&=&(x\cdot x)^{-k/2}\exp \{(1/2)(x_0-x_3)^{-1}
   (\g_0-\g_3)(\g_1x_1+\g_2x_2)\} \\ 
   & &\times\exp \Bigl\{-(1/ 4\alpha)
   \g_0\g_3 \, x\cdot x\Bigr\} 
   \vp\,  \Bigl((x\cdot x)^{\alpha+1}
   (x_0-x_3)^{-2\alpha}\Bigr), \\
43)\ \psi(x)&=&(x_0-x_3)^{-k}\exp \{(1/2)(x_0-x_3)^{-1}
   (\g_0-\g_3)(\g_1x_1+\g_2x_2)\} \\ 
   & &\times\exp \{(1/ 2\alpha )
   \g_0\g_3 \ln (x_0-x_3)\} 
   \vp\,  \Bigl(x\cdot x (x_0-x_3)^{-2}\Bigr), \\
44)\ \psi(x)&=&(x\cdot x)^{-k/2}\exp \{(1/2)(x_0-x_3)^{-1}
   (\g_0-\g_3)(\g_1x_1+\g_2x_2)\} \\
   & &\times\exp \{(1/ 4\beta )(\g_1\g_2 - \alpha\g_0\g_3)\ln (x\cdot
   x) \} \vp\, \Bigl((x\cdot x)^{\alpha+\beta}\\
   & &\times(x_0-x_3)^{-2\beta}\Bigr).
\end{eqnarray*}
\vspace{2mm}

\noindent
{\bf 3. Conformally-invariant Ans\"atze.}\ Complete classification of
$C(1,3)$ non-con\-ju\-gate subgroups of the conformal
group\index{Conformal!group} was obtained quite recently \cite{14,66}.
We use this classification to construct Ans\"atze for the spinor
field $ \psi (x) $ invariant under one- and three-parameter
subgroups of the group $C(1,3)$.

$C(1,3)$ non-conjugate one-parameter subgroups of the conformal group
which are not $C(1,3)$-conjugate to subgroups of the group $\wid
P(1,3)$ are generated by the following 
operators:\index{Subalgebras!of the conformal algebra}
\begin{eqnarray}
&&Q_1 = Q,\quad
Q_2 = Q + \ve (P_0 - P_3),
\non\\
&&Q_3 = J_{12} + \al Q,\quad
Q_4 = Q + \al (D - J_{03}),
\non\\
&&Q_5 = \beta J_{12} + \al Q + \ve (P_0 - P_3),
\non\\
&&Q_6 = \al J_{12} + Q - J_{01} - J_{13} - P_2,
\non\\
&&Q_7 = \delta J_{12} + \al Q + \beta(D - J_{03}),
\label{2.2.14}\\
&&Q_8 = P_0 + K_0,
\quad
Q_9 = \al (P_0 + K_0) + J_{12},
\non\\
&&Q_{10} = \al (P_0 + K_0) + J_{12} + \beta (P_3 - K_3),
\non\\
&&Q_{11} = J_{12} + \beta (P_3 - K_3).\non
\end{eqnarray}

Here $Q=(1/2)(K_0 - K_3 + P_0 + P_3), \ \{\al, \beta\} \subset {\R}^1, \
\ve = \pm 1$. 

Operators (\ref{2.2.14}) unlike generators of the extended Poincar\'e
group $\wid P(1,3)$ have quadratic dependence on $x_\mu$.  That is
why the corresponding system (\ref{1.5.16}), (\ref{1.5.15}) is 
nonlinear with respect to the independent variables $x_\mu$ (in
particular, equations (\ref{1.5.16}) with $Q$ of the form
(\ref{2.2.14}) lead to a Riccati-type system of ODEs). To avoid a
necessity to integrate a nonlinear Riccati-type system of ODEs we will
apply the trick used by Dirac when investigating conformal invariance
of equation (\ref{1.1.17}) \cite{48}. Relying on the well-known fact
of isomorphism of Lie algebras of the groups $C(1,3)$ and $O(2,4)$ he
obtained a change of variables connecting the transformation group
$C(1,3)$ of the form (\ref{1.1.24a})--(\ref{1.1.24e}) with the group
of homogeneous linear transformations of some six-dimensional
projective space preserving the quadratic form $z_1^2 + z_2^2 - z_3^2
- z_4^2 - z_5^2 - z_6^2$.  And what is more, generators of the group
$O(2,4)$ were linear in the variables $z_{\ssl A}, \ A = {1,\ldots,6}$.

Consider the following representation of the Lie algebra $AO(2,4)$: 
\begin{eqnarray}
& &\Omega_{1\, 2} = z_1 \p_{z_2} - z_2 \p_{z_1} + (i/2) \g_4 \g_0,
\non
\\
& &\Omega_{1 \, 2+a} = -z_1 \p_{z_{2+a}} - z_{2+a} \p_{z_1} +
(i/2) \g_4 \g_a,
\non
\\
& &\Omega_{2 \, 2+a} = -z_2 \p_{z_{2+a}} - z_{2+a} \p_{z_2} + (1/2)
\g_0 \g_a,
\non
\\
& &\Omega_{2+a \, 2+b} = -z_{2+a} \p_{z_{2+b}} + z_{2+b} \p_{z_{2+a}}
+ (1/2) \g_a \g_b,
\label{2.2.15}
\\
& &\Omega_{1\, 6} = z_1 \p_{z_6} - z_6 \p_{z_1} + (i/2) \g_4,
\non
\\
& &\Omega_{2\, 6} = z_2 \p_{z_6} - z_6 \p_{z_2} + (1/2) \g_0,
\non
\\
& &\Omega_{2+a \, 6} = - z_{2+a} \p_{z_6} + z_6 \p_{z_{2+a}} +
(1/2) \g_a, \ \ a,b = {1,2,3}\non
\end{eqnarray}
(the remaining elements of $ AO(2,4) $ are obtained by the rule
$\Omega_{\ssl AB} = - \Omega_{\ssl BA}, \linebreak  
A,B = {1,\ldots,6}, \ A \not= B$).

It is straightforward to verify that operators (\ref{2.2.15}) do
satisfy commutation relations of the algebra $AO(2,4)$
\begin{displaymath}
\left [ \Omega_{\ssl AB},\, \Omega_{\ssl CD} \right ] = 
(\rho_{\ssl AD} \Omega_{\ssl BC}
+ \rho_{\ssl BC} \Omega_{\ssl AD} - \rho_{\ssl AC} \Omega_{\ssl BD} 
- \rho_{\ssl BD} \Omega_{\ssl AC}), 
\end{displaymath}
where $ \rho_{\ssl AB} = \mbox{\rm diag}(1, \ 1, -1, -1, -1, -1) $ is
the metric tensor of the pseudo-Eucli\-de\-an space $R(2,4)$. Next,
the isomorphism of the algebras $AO(2,4)$ and $AC(1,3)$ is established
by the formulae
\begin{equation}
\begin{array}{l}
P_0 = - \Omega_{1\, 2} - \Omega_{2\, 6},
\quad
P_a = - \Omega_{1 \, 2+a} -
\Omega_{2+a\, 6},\\[2mm]
J_{0a} = \Omega_{2 \, 2+a},
\quad
J_{ab} = \Omega_{2+a \, 2+b},\\[2mm]
D = - \Omega_{16},
\quad
K_0 = - \Omega_{1\, 2} + \Omega_{2\, 6},\\[2mm]
K_a = - \Omega_{1 \, 2+a} + \Omega_{2 + a\, 6},
\ \
a,b = {1,2,3},
\ \ a \not= b.\non
\end{array}
\label{2.2.16}
\end{equation}

The transformation relating the groups $O(2,4)$ and $C(1,3)$ can be
represented in the form
\begin{eqnarray}
x_\mu &=& z_{\mu + 2}\, (z_6 - z_1)^{-1},\non\\
\psi(x) &=& (z_6 - z_1)^2 \Bigl\{1 - (1/2)(z_6 - z_1)^{-1}(1+ i \g_4)
\label{2.2.17}\\
& &\times (\g_0 z_2 - \g_a z_{2+a})\Bigr\} \Psi(z),\non
\end{eqnarray}
coordinates $z_1, \ldots, z_6 $ satisfying an additional constraint
\begin{displaymath}
z_{\ssl A} z^{\ssl A} \equiv z_1^2 + z_2^2 - z_3^2 - z_4^2 
- z_5^2 - z_6^2=0.
\end{displaymath}

It is important to note that the Lie groups $O(2,4)$ and
$C(1,3)$ are not isomorphic. Formulae (\ref{2.2.17}) determine a 
projection of the group $O(2,4)$ on the group $C(1,3)$.

On rewriting operators (\ref{2.2.16}) in the variables $x, \ \psi(x)$ 
according to (\ref{2.2.17}) we get the following expressions
for the generators of the conformal group $C(1,3)$:
\begin{eqnarray}
& &P_\mu = \p^\mu, \quad J_{\mu\nu} = x_\mu P_\nu - x_\nu P_\mu 
+ S_{\mu\nu},\non\\
& &D = x_\mu \p_\mu + 3/2 + (1/2) (1 - i \g_4),\label{2.2.18}\\
& &K_\mu = 2x_\mu D - (x \cdot x) \p^\mu + 2S_{\mu \nu} x^\nu +
(1/2)(1 - i \g_4) \g_\mu,\non
\end{eqnarray}
where $S_{\mu\nu}=(1/4)(\g_\mu\g_\nu-\g_\nu\g_\mu), \
\mu,\nu={0,\ldots,3},\ \mu < \nu$.

Hence we conclude that an Ansatz invariant under a subgroup of the
group $O(2,4)$ with generators (\ref{2.2.15}) is transformed by
(\ref{2.2.17}) into an Ansatz invariant under a subgroup of the group
$C(1,3)$ with generators (\ref{2.2.18}). But the above arguments
cannot be immediately applied to construct conformally-invariant
Ans\"atze reducing the massless Dirac equation (\ref{1.1.17}). The
matter is that on the set of solutions of equation (\ref{1.1.17}) a
representation of the algebra $AC(1,3)$ inequivalent to the
representation (\ref{2.2.18}) is realized (see Section 1.1). To avoid
this difficulty we will modify the change of variables (\ref{2.2.17}).
Let us consider the group $O(2,4)$ acting on the space of
eight-component spinors $ \wid \Psi $ which depend on six variables
$z_1, \ldots, z_6$. Its generators are chosen as follows
\begin{eqnarray}
& &\Omega_{1\, 2} = z_1 \p_{z_2} - z_2 \p_{z_1} + (1/2) \s \G_0,
\non
\\
& &\Omega_{1 \, 2+a} = -z_1 \p_{z_{2+a}} - z_{2+a} \p_{z_1}
+ (1/2) \s \G_a,
\non
\\
& &\Omega_{2 \, 2+a} = -z_2 \p_{z_{2+a}} - z_{2+a} \p_{z_2} + (1/2)
\G_0 \G_a,
\non
\\
& &\Omega_{2+a \, 2+b} = -z_{2+a} \p_{z_{2+b}} + z_{2+b} \p_{z_{2+a}}
+ (1/2) \G_a \G_b,
\non
\\
& &\Omega_{1\, 6} = z_1 \p_{z_6} - z_6 \p_{z_1} + (1/2) \s,
\label{2.2.19}
\\
& &\Omega_{2\, 6} = z_2 \p_{z_6} - z_6 \p_{z_2} + (1/2) \G_0,
\non
\\
& &\Omega_{2+a \, 6} = - z_{2+a} \p_{z_6} + z_6 \p_{z_{2+a}} +
(1/2) \G_a,
\non
\\
& &\Omega_{A\, B} = - \Omega_{BA}, \ \  A \not= B.\non
\end{eqnarray}

Here $\G_\mu, \ \s $ are $(8\times 8)$-matrices of the form
\begin{displaymath}
 \G_\mu = \lo \begin{array} {cc}
0 & \g_\mu \\
\g_\mu & 0
\end{array} \ro,
\qquad
\s = \lo \begin{array} {cc}
I & 0 \\
0 & -I
\end{array} \ro.
\end{displaymath}

Using the relations
\begin{displaymath}
\G_\mu \G_\nu + \G_\nu \G_\mu = 2g_{\mu \nu}I,
\quad
\G_\mu \s = - \s \G_\mu,
\end{displaymath}
we can become convinced of the fact that operators (\ref{2.2.19}) do form a
basis of the Lie algebra $AO(2,4)$.

The change of variables
\begin{eqnarray}
x_\mu &=& z_{\mu + 2}\, (z_6 - z_1)^{-1},\non\\
\ti \psi (x) &=& (z_6 - z_1)^2\Bigl\{1 - (1/2)(1 + \s)\label{2.2.20}\\  
& &\times (\G_0 z_2 - \G_a z_{2+a})(z_6 - z_1)^{-1}\Bigr\}
\wid \Psi (z),\non
\end{eqnarray}
where $ \ti \psi (x) $ is an eight-component spinor, establishes 
a correspondence bet\-we\-en the group $O(2,4)$ having the generators
(\ref{2.2.19}) and the group $C(1,3) $ having the generators
\begin{equation}
\begin{array}{l}
\wid P_\mu = \p^\mu, \quad \wid J_{\mu \nu} = x_\mu \p^\nu - x_\nu
\p^\mu + \wid S_{\mu \nu},\\[2mm]
\wid D = x_\mu \p_\mu + 3/2 + (1/2) (1 - \s),\\[2mm]
\wid K_\mu = 2x_\mu \wid D - (x \cdot x) \p^\mu + 2 \wid S_{\mu
  \nu}x^\nu + (1/2)(1 - \s) \G_\mu,
\end{array}
\label{2.2.21}
\end{equation}
where $\wid S_{\mu \nu} = (1/4)[ \G_\mu,\, \G_\nu]$.
\vspace{1.5mm}

\noindent
{\bf Lemma 2.2.1.}\ {\em Let $\ti \psi (x)$ satisfy the equation
\begin{equation}
\wid Q \ti \psi (x) = (\al_\mu \wid P_\mu + \beta_{\mu \nu} \wid
J_{\mu \nu} + \delta\wid D + \theta_\mu \wid K_\mu) \ti \psi (x) = 0, 
\label{2.2.22}
\end{equation}
where $ \al_\mu,\ \beta_{\mu \nu}, \ \delta,
\ \theta_\mu $ are some real parameters.
Then, the four-component spinor $ \psi = (\ti \psi^0, \ti \psi^1, \ti
\psi^2, \ti \psi^3)^T $ satisfies the following equation:
\begin{equation}
Q \psi (x) = (\al_\mu P_\mu + \beta_{\mu \nu} J_{\mu \nu} + \delta D +
\theta_\mu K_\mu ) \psi (x) = 0,
\label{2.2.23}
\end{equation}
the operators $P_\mu, \ldots, K_\mu $ being of the form
(\ref{1.1.22}).} 
\vspace{1.5mm}

\noindent
{\em Proof.}$\quad$ We represent the eight-component function $ \ti \psi
(x) $ as follows
\begin{displaymath}
\ti \psi (x) = (1/2) (1 + \s) \ti \psi_1 (x) + (1/2) (1 - \s)
\ti \psi_2 (x).
\end{displaymath}

Substitution of the above expression into (\ref{2.2.22}) yields
\begin{displaymath}
(1/2) (1 + \s) \wid Q \ti \psi_1 + (1/2)(1 - \s) \wid Q \ti \psi_2 =
0, 
\end{displaymath}
whence it follows that
\begin{equation}
(1/2)(1 + \s) \wid Q \ti \psi_1 = 0.
\label{2.2.24}
\end{equation}

Since
\begin{eqnarray*}
& &(1/2)(1 + \s ) \wid D \ti \psi_1 (x) = (x_\mu \p_\mu + 3/2)(1/2)(1
+ \s) 
\ti \psi_1 (x),\\
& &(1/2) (1 + \s) \wid K_\mu \ti \psi_1 (x) =
\{2x_\mu (x_\nu \p_\nu + 3/2)-(x \cdot x) \p^\mu \\
& &\quad + 2 \wid S_{\mu \nu}x^\nu\} (1/2)(1 + \s) \ti \psi_1,\\
& &\wid S_{\mu \nu} = \lo \begin{array} {cc}
S_{\mu \nu} & 0 \\
0 & S_{\mu \nu} \end{array} \ro,
\end{eqnarray*}
we conclude that due to (\ref{2.2.24}) equality (\ref{2.2.23}) holds
true. $\rhd$ 

The above arguments can be summarized in the form of the following
algorithm of constructing conformally-invariant Ans\"atze for the
spinor field $ \psi (x) $:
\begin{itemize}
\item{using the isomorphism (\ref{2.2.16}) we establish the
    correspondence between $C(1,3)$ non-conjugate subalgebras of the
    algebra $AC(1,3)$ and $O(2,4)$ non-conjugate subalgebras of the
    algebra $AO(2,4) $;}
\item{integrating the systems of PDEs (\ref{1.5.16}), (\ref{1.5.15}) we
    construct Ans\"atze invariant under non-conjugate subalgebras of
    the algebra $AO(2,4)$ having the basis elements (\ref{2.2.19});}
\item{using the change of variables (\ref{2.2.20}) we rewrite the
    obtained Ans\"atze in variables $x, \ \psi (x) $;}
\item{acting on the eight-component spinor $ \ti \psi (x) $ by the
    projector $ P = (1/2) \linebreak \times (1 + \s) $ we arrive at
    the conformally-invariant Ans\"atze for the spinor field $ \psi
    (x) $.} 
\end{itemize}

We will realize the above algorithm for the operator $Q_2$ from
(\ref{2.2.14}), the remaining operators being treated in an analogous
way.

Due to (\ref{2.2.16}), (\ref{2.2.19}) the operator $Q_2$ 
takes the form
\begin{eqnarray*}
& &Q_2 = z_2 \p_{z_1} - z_1 \p_{z_2} + z_5 \p_{z_6} - z_6 \p_{z_5} 
+ \ve (z_6 - z_1)(\p_{z_2} + \p_{z_5})\\
& &+ \ve(z_2 - z_5)(\p_{z_1} + \p_{z_6})
- (1/2) \{\G_3 + \s \G_0 + \ve (1 + \s) (\G_0 - \G_3)\}.
\end{eqnarray*}

Consequently, to determine matrix function $A(z)$ and scalar
functions $\omega_1(z), \ldots$, $\omega_5 (z)$ it is necessary 
to integrate the equations 
\begin{eqnarray}
& &Q_2 A(z) = 0,\non\\
& &\Bigl\{z_2 \p_{z_1}-z_1 \p_{z_2} + z_5 \p_{z_6} - z_6 \p_{z_5} +
\ve (z_6 -z_1) 
\label{2.2.25}\\
& &\quad\times (\p_{z_2} + \p_{z_5}) +\ve (z_2 - z_5)(\p_{z_1} +
\p_{z_6})\Bigr\}\omega_i (z) = 0, \non
\end{eqnarray}
where $i = {1,\dots,5}$.

It is convenient to rewrite (\ref{2.2.25}) by introducing new
independent variables
\begin{eqnarray*}
& &u_1 = (z_1 - z_6)^2 + (z_2 - z_5)^2,\\
& &u_2 = 2(z_1 - z_6)(z_2 - z_5),
\quad
 u_3 = z_3,
\quad
\ u_4 = z_4,\\
& &u_5 = (z_1 - z_6)(z_1 + z_6) + (z_2 - z_5)(z_2 + z_5),\\
& &u_6 = (z_1 - z_6)(z_2 + z_5) - (z_2 - z_5)(z_1 + z_6).\\
\end{eqnarray*}
As a result, equations (\ref{2.2.25}) read
\begin{eqnarray}
& &\Bigl\{-2 (u_1^2 - u_2^2)^{1/2} \p_{u_2} - 2 \ve u_1
\p_{u_6}\Bigr\} \omega_i (u) = 0, 
\non\\
& &\Bigl\{-2(u_1^2 - u_2^2)^{1/2} \p_{u_2} - 2 \ve u_1 \p_{u_6}
-(1/2)\Bigl(\s \G_0 + \G_3 \label{2.2.26}\\
& &\quad +\ve (1+ \s)(\G_0 - \G_3)\Bigr)\Bigr\} A(u) = 0.\non
\end{eqnarray}

The first equation of system (\ref{2.2.26}) implies that $\omega_1
(u), \ldots, \omega_5(u) $ are the first integrals of the
following Euler-Lagrange system\index{Euler-Lagrange equations}: 
\begin{displaymath}
{du_1 \over 0} = {du_2 \over -2(u_1^2 - u_2^2)^{1/2} } = {du_3 \over
  0} = {du_4 \over 0} = {du_5 \over 0 } = {du_6 \over -2 \ve u_1}.
\end{displaymath}

A complete set of functionally-independent first integrals
of the above system can be chosen in the form
\begin{eqnarray*}
& &\omega_1 = \arcsin (u_2/u_1) - \ve (u_6/u_1),\quad
\omega_2=u_1,\\
& &\omega_3 = u_2, \quad \omega_4 = u_3,\quad \omega_5 = u_4.
\end{eqnarray*}

Using the identity (\ref{2.2.6}) we get the following particular
solution of the second equation of system (\ref{2.2.26}):
\begin{displaymath}
A(u) = \exp \{- (\ve u_6/4u_1) (\s \G_0 + \G_3 + \ve (1 + \s)
(\G_0 - \G_3) \}.
\end{displaymath}

Rewriting the above expressions in the variables $z_{\ssl A} $ and
substituting these into Ansatz (\ref{1.5.14}) we have
\begin{eqnarray}
& &\wid \Psi (z) = \exp \Bigl\{-(\ve /4) [(z_1 - z_6) (z_2 + z_5) -
(z_2 - z_5)\non\\
& &\quad\times (z_1 + z_6)][(z_1 - z_6)^2 + (z_2 - z_5)^2]^{-1}
\label{2.2.27}\\
& &\quad\times [ \G_3 + \s \G_0 + (1+ \s) 
(\G_0 - \G_3)]\Bigr\} \ti \vp (\omega),\non
\end{eqnarray}
where $ \ti \vp $ is an arbitrary eight-component function of
$\omega_1, \ldots, \omega_5 $ and
\begin{eqnarray*}
& &\omega_1 = \arcsin \Bigl\{2(z_1 - z_6)(z_2 - z_5)\Bigl((z_1 -
z_6)^2 + (z_2 - z_5)^2\Bigr)^{-1}\Bigr\}\\
& &\quad- \ve\{(z_1 - z_6)(z_2 + z_5) - (z_2 - z_5)(z_1 + z_6)\}\\
& &\quad\times \Bigl((z_1 - z_6)^2 + (z_2 - z_5)^2\Bigr)^{-1},\\
& &\omega_2 = (z_1 - z_6)^2 + (z_2 - z_5)^2,
\quad
\omega_3 = z_3,
\quad
\omega_4 = z_4,\\
& &\omega_5 = z_1^2 + z_2^2 - z_3^2 -z_4^2 - z_5^2 - z_6^2.
\end{eqnarray*}

In the initial variables $x, \ \ti \psi (x)$ Ansatz (\ref{2.2.27})
reads 
\begin{eqnarray*}
& &\ti\psi (x) = x_1^{-2} \Bigl\{1 - (1/2)(1+ \s) \G \cdot x\}\\
& &\quad
\times \exp \Bigl\{\tau (x) \Bigl(\G_3 + \s \G_0 + (1 + \s) 
(\G_0 - \G_3)\Bigr)\Bigr\}
\ti \vp\, (\omega_1, \omega_2, \omega_3),\\
& &\omega_1 = - \arcsin \Bigl\{2(x_0 - x_3) 
\Bigl(1 + (x_0 - x_3)^2\Bigr)^{-1}\Bigr\} \\
& &\quad+\ve\Bigl(x_0 + x_3 + (x_0 - x_3) x \cdot x\Bigr) \Bigl(1 +
(x_0 - x_3)^2\Bigr)^{-1},\\
& &\omega_2 = \Bigl(1 + (x_0 - x_3)^2\Bigr) x_1^{-2},
\quad
\omega_3 = (1 + (x_0 - x_3)^2) x_2^{-2},\\
& &\tau(x) = (\ve /4)\Bigl(x_0 + x_3 + (x_0 - x_3) x \cdot x\Bigr)
\Bigl(1 + (x_0 - x_3)^2\Bigr)^{-1}
\end{eqnarray*}
(when deriving the above formulae we use the identities $z_1(z_6 -
z_1)^{-1} = (1/2)(x \cdot x - 1)$,\ $z_6(z_6 - z_1)^{-1} = (1/2)
(x \cdot x + 1) $ which follow directly from (\ref{2.2.20})).

Acting on the Ansatz obtained by the projector $P = (1/2) (1 + \s)$
we get\index{Ansatz!$C(1,3)$-invariant} 
\begin{displaymath}
\psi (x) = x_1^{-2}\{\cos^2 \tau + \g_0 \g_3 \sin^2 \tau 
+\g \cdot x (\g_0 - \g_3)\cos \tau \sin \tau \}
\vp (\omega_1, \omega_2, \omega_3),
\end{displaymath}
where $ \vp(\omega) $ is a new four-component function, scalar
functions $ \tau (x), \ \omega_a(x) $ are determined above.

Below we adduce Ans\"atze invariant under operators
$Q_1, \ Q_3, \ldots, Q_{11}$.
\vspace{1.5mm}

\noindent
\underline{the operator $Q_1$}
\begin{eqnarray*}
& &\psi (x) = R(x, \tau, 1) x_1^{-2} \vp\,  ( \vec \omega),\\
& &\omega_1 = [(x \cdot x -1)^2 + 4x_0^2] x_1^{-2},\quad
   \omega_2 = [(x \cdot x -1)^2 + 4x_0^2] x_2^{-2},\\
& &\omega_3 = \arctan \{(x \cdot x -1)(2x_0)^{-1}\}- \arctan
\{(x \cdot x +1)(2x_3)^{-1}\},\\
& &\tau = (1/2) \arctan \{(x \cdot x -1)(2x_0)^{-1}\} + \pi/2;
\end{eqnarray*}

\noindent
\underline{the operator $Q_3$}
\begin{eqnarray*}
& &\psi(x) = R(x, \tau, \al) (x_1^2 + x_2^2)^{-1} \exp \{-\tau \g_1
\g_2\} \vp\,  ( \vec \omega),
\\
& &
\omega_1 = [(x \cdot x -1)^2 + 4x_0^2] (x_1^2 + x_2^2)^{-1},
\\
& &
\omega_2 = \arctan \{(x \cdot x -1)(2x_0)^{-1}\}- \al \arctan
(x_1/x_2),
\\
& &
\omega_3 = \arctan \{(x \cdot x -1)(2x_0)^{-1}\}- \arctan
\{(x \cdot x +1)(2x_3)^{-1}\},
\\
& &
\tau = (1/2) \arctan (x_1/x_2);
\end{eqnarray*}
\underline{the operator $Q_4$}
\begin{eqnarray*}
& &\psi (x) = R(x, \tau, 1) x_1^{-2} \exp \{\tau \al (1 + \g_0
\g_3)\} \vp\,  (\vec \omega),
\\
& &
\omega_1 = \ln \{x_1^2 [1 + (x_0 - x_3)^2]^{-1}\} 
\\
& &
\quad
- \al \arcsin \{2(x_0 - x_3) [1 + (x_0 - x_3)^2]^{-1}\},
\\
& &
\omega_2 = x_1^{-2} [x_0 + x_3 + (x_0 - x_3) x \cdot x],
\\
& &
\omega_3 = x_2^{-2} [x_0 + x_3 + (x_0 - x_3) x \cdot x],
\\
& &
\tau = (1/4\al) \ln \{x_1^2 [1 + (x_0 - x_3)^2]^{-1}\}; 
\end{eqnarray*}
\underline{the operator $Q_5$}
\begin{eqnarray*}
& &
\psi(x) = R(x, \tau, \al) (x_1^2 + x_2^2)^{-1} \exp
\{- \beta \tau \g_1 \g_2\}\vp\,  ( \vec \omega),
\\
& &
\omega_1 = [1 + (x_0 - x_3)]^2(x_1^2 + x_2^2)^{-1},
\\
& &
\omega_2 = \ve \arcsin \{2(x_0 - x_3) [1 + (x_0 - x_3)^2]^{-1}\}  
\\
& &
\quad
+ \al  [x_0 + x_3 + (x_0 - x_3) x \cdot x] [1 + (x_0 -
x_3)^2]^{-1}, 
\\
& &
\omega_3 = -2 \ve \arctan (x_1/x_2) + \beta [x_0 + x_3 + (x_0 -
x_3) x \cdot x]
\\
& &
\quad
\times [1 + (x_0 - x_3)^2]^{-1},
\\
& &
 \tau = (\ve/2) [x_0 + x_3 + (x_0 - x_3) x \cdot x] [1 +
(x_0 - x_3)^2]^{-1};
\end{eqnarray*}
\underline{the operator $Q_6$}
\begin{eqnarray*}
& &
\psi(x) = [1 + (x_0 - x_3)^2]^{-1}\{ \cos^2 (\tau /2) + \g_0 \g_3
\sin^2(\tau /2) 
\\
& &
\quad + \sum_{j=1}^2(f_j \cos \tau - g_j \sin \tau) \g_j (\g_0 - \g_3) 
+ \g \cdot x(\g_0 - \g_3)  
\\
& &
\quad
\times \cos ( \tau /2) \sin (\tau /2)\} 
\exp \{ - ( \al \tau /2) \g_1 \g_2\} \vp\,  ( \vec \omega), \ \ j =
1,2, 
\end{eqnarray*}
a)\ under $\al = 1$
\begin{eqnarray*}
& &
f_1 = g_2 = - \tau /2, \quad f_2 = - g_1 = 1,
\\
& &
\omega_1 = [x_2(x_0 - x_3) - x_1][1 + (x_0 - x_3)^2]^{-1}, 
\\
& &
\omega_2 = -\arcsin \{2(x_0 - x_3) [1 + (x_0 - x_3)^2]^{-1}\} 
\\
& &
\quad
+2 [x_1(x_0 - x_3) + x_2][1 + (x_0 - x_3)^2]^{-1}, 
\\ 
& &
\omega_3 = \arcsin \{2(x_0 - x_3)[1 + (x_0 - x_3)^2
]^{-1}\}  
\\
& &
\quad
+ \Bigl(x_0 + x_3 + (x_0 - x_3)x \cdot x\Bigr) \Bigl((x_0 - x_3)x_2 -
x_1\Bigr)^{-1}, 
\\
& &
\tau = \arctan (x_0 - x_3);
\end{eqnarray*}
b)\ under $ \al \not= 1 $
\begin{eqnarray*}
& &
f_1 = g_2 = \sin (1 - \al) \tau,
\\
& &
f_2 = - g_1 = [2(\al - 1)]^{-1} 
[2 (\al - 1) \cos (\al - 1) \tau +1],
\\
& &
\omega_1 = [2(x_0 - x_3) x_2 - 2x_1 - (1 - \al)(x_1^2 +
x_2^2)][1 + (x_0 - x_3)^2]^{-1},
\\
& &
\omega_2 = ( \al - 1) \Bigl\{\Bigl((x_0 -x_3) x_2 - x_1\Bigr)^2 +
\Bigl((x_0 - x_3) x_1 + x_2\Bigr)^2\Bigr\} 
\\
& &
\quad
\times \Bigl(1 + (x_0 - x_3)^2\Bigr)^{-2} + 2\Bigl((x_0 - x_3) x_2 -
x_1\Bigr) \Bigl(1 + (x_0 - x_3)^2\Bigr)^{-1},
\\
& &
\omega_3 = 2 \arcsin \biggl\{ \Bigl[(\al - 1) \Bigl((x_0 - x_3) x_2 
- x_1\Bigr) + 1 + (x_0 - x_3)^2\Bigr]
\\
& &
\quad
\times  \Bigl\{\Bigl[(\al - 1) \Bigl((x_0 - x_3) x_2 
- x_1\Bigr) + 1 + (x_0 - x_3)^2\Bigr]^2 \\
& &\quad + (\al - 1)^2 [(x_0 - x_3) x_1 + x_2]^2\Bigr\}^{-1/2}
\biggr\} \\
& &
\quad
+ (\al - 1) \arcsin \Bigl\{2(x_0 - x_3) \Bigl(1 + (x_0 -
x_3)^2\Bigr)^{-1}\Bigr\}, \\
& &
\tau = \arctan (x_0 - x_3);
\end{eqnarray*}
\underline{the operator $Q_7$}
\begin{eqnarray*}
& &
\psi(x) = R(x, \tau, \al) (x_1^2 + x_2^2)^{-1} \exp \{\beta \tau(1
 + \g_0 \g_3) - \delta \tau \g_1 \g_2\} \vp\,  ( \vec \omega),
\\
& &
\omega_1 = (x_1^2 + x_2^2)^{-1} [x_0 + x_3 + (x_0 - x_3) x \cdot x],
\\
& &
\omega_2 = \delta \ln \{(x_1^2 + x_2^2) [1 + (x_0 - x_3)^2]^{-1}\} 
- 2 \beta \arcsin\, \{ x_1 (x_1^2 + x_2^2)^{-1}\},
\\
& &
\omega_3 = \al\ln \{(x_1^2 + x_2^2) [1 + (x_0 - x_3)^2]^{-1}\} 
\\
& &
\quad
-  \beta \arcsin \{2 (x_0 - x_3) [1 + (x_0 - x_3)^2]^{-1}\},
\\
& &
\tau = (1/2\delta)\arctan (x_1/x_2);
\end{eqnarray*}
\underline{the operator $Q_8$}
\begin{eqnarray*}
& &
\psi (x) = x_1^{-2} (\cos \tau - \g \cdot x \g_0 \sin \tau)
\vp\,  ( \vec w),
\\
& &
\omega_1 = x_2/x_1, \quad \omega_2 = x_3/x_1,\quad
\omega_3 = (1 + x \cdot x)^2 (x_1^2 + x_2^2 + x_3^2)^{-1},
\\
& &
\tau = (1/2) \arctan \{(x \cdot x -1 ) (2x_0)^{-1}\}+ \pi/2;
\end{eqnarray*}
\underline{the operator $Q_9$}
\begin{eqnarray*}
& &
\psi (x) = x_3^{-2} (\cos \al \tau - \g \cdot x \g_0 \sin \al \tau)
\exp  \{- \tau \g_1 \g_2\} \vp\,  ( \vec \omega),
\\
& &
\omega_1 = (x_1^2 + x_2^2) x_3^{-2}, \quad \omega_2 = (x \cdot x + 1)
x_3^{-1}, \\
& &
\omega_3 = \arctan \{(x \cdot x - 1) (2x_0)^{-1}\}
- \al \arctan (x_1/x_2),
\\
& &
\tau = (1/2) \arctan \{(x \cdot x -1) (2x_0)^{-1}\} + \pi/2;
\end{eqnarray*}
\underline{the operator $Q_{10}$ }
\begin{eqnarray*}
& &
\psi (x) = (x^2_1 + x^2_2)^{-1} \Bigl(\cos \al \tau  \cos \beta \tau  
+ \g_0 \g_3 \sin \al \tau  \sin \beta \tau 
\\
& &
+ \g \cdot x (\g_0 \sin\al \tau \cos \al \tau 
- \g_3 \cos \al \tau \sin \al \tau)\Bigr) \exp \{- \tau \g_1 \g_2 \}
\vp\,  ( \vec \omega),
\\
& &
\omega_1 = \al \arctan (x_1/x_2) - \arctan \{(x \cdot x -
1)(2x_0)^{-1}\}, \\
& &
\omega_2 = \beta \arctan (x_1/x_2) - \arctan \{(x \cdot x +
1)(2x_0)^{-1}\}, \\
& &
\omega_3 = [(x \cdot x -1)^2 + 4x_0^2](x_1^2 + x_2^2)^{-1},
\\
& &
\tau = (1/2) \arctan (x_1/x_2);
\end{eqnarray*}
\underline{the operator $Q_{11}$}
\begin{eqnarray*}
& &
\psi (x) = (x^2_1 + x^2_2)^{-1} (\cos \beta \tau  - \g \cdot x \g_3
\sin \beta \tau ) \exp \{- \tau \g_1 \g_2 \} \vp\,  ( \vec \omega),
\\
& &
\omega_1 = (x_1^2 + x_2^2) x_0^{-2},
\\
& &
\omega_2 = - \beta \arctan (x_1/x_2) + \arctan \{(x \cdot x + 1)
(2x_3)^{-1}\},
\\
& &
\omega_3 = (x \cdot x -1) (x_1^2 + x_2^2)^{-1/2},
\quad \tau = (1/2) \arctan (x_1/x_2).
\end{eqnarray*}

In the above formulae we use the following notation:
\begin{displaymath}
R(x, \tau, \al ) = \cos^2 \al \tau  + \g_0 \g_3 \sin^2 \al \tau +
 \g \cdot x (\g_0 - \g_3) \cos \al \tau  \sin \al \tau,
\end{displaymath}
$\vp = \vp( \vec \omega) $ is an arbitrary four-component function of
$\omega_1, \ \omega_2, \ \omega_3$.

Three-dimensional $C(1,3)$ non-conjugate subalgebras of the conformal
al\-geb\-ra which are $C(1,3)$ inequivalent to subalgebras of the
algebra $A\wid P(1,3)$ are as 
follows\index{Subalgebras!of the conformal algebra}
\begin{eqnarray}
& &
A_1 = \langle Q +J_{12},\,   \ -J_{01} - J_{12} -
P_2,\,  D - J_{03}\rangle,
\non\\
& &
A_2 = \langle Q + \al J_{12},\,  \
D - J_{03},\,  P_0 - P_3\rangle,
\non\\
& &
A_3 = \langle Q +J_{12} - J_{01} - J_{13} - P_2,\, 
 - J_{02} - J_{23} + P_1,\,  P_0 - P_3\rangle,
\non\\
& &
A_4 = \langle Q +J_{12} + \al (D - J_{03}),\, 
-J_{01} - J_{13} - P_2,\,  P_0 - P_3\rangle,
\non\\
& &
A_5 = \langle J_{12} + \al (D - J_{03}),\,  Q
+ \beta (D - J_{03}),\,  P_0 -P_3\rangle, 
\non\\
& &
A_6 = \langle J_{03} + D,\,  (1/2)(K_0 - K_3),\,  (1/2)(P_0 +
P_3)\rangle, \non\\
& &
A_7 = \langle Q,\,  D - J_{03},\,  J_{12}\rangle,
\quad
A_8 = \langle J_{12},\,  Q,\,  P_0 - P_3\rangle,
\label{2.2.28}\\
& &
A_9 = \langle Q + J_{12},\,  - J_{01} - J_{13} - P_2,
P_0 - P_3\rangle,\, 
\non\\
& &
A_{10} = \langle J_{12},\,  (1/2) (K_0 + P_0),\,  (1/2)(P_3 -
K_3)\rangle, \non\\
& &
A_{11} = \langle J_{23} + (1/2)(P_1 - K_1),\,  J_{31} 
+ (1/2)(P_2 - K_2),\, J_{12}\non\\ 
& &\phantom{A_{11} = }+ (1/2)(P_3 - K_3)\rangle,
\non\\
& &
A_{12} = \langle J_{12} + (1/2)(P_3 - K_3),\,  - J_{03} - (1/2)(P_1 + 
K_1),\, \non\\
& &\phantom{A_{12} = }(1/2)(P_0 -K_0) + (1/2)(P_2 + K_2)\rangle,
\non\\
& &
A_{13} = \langle  \sqrt{3} J_{01} - J_{02} - D,\,  \ P_0 + K_0 + 2(K_2 
- P_2),\, K_0 - P_0 \non\\
& &\phantom{A_{13} = } - K_2 - P_2 - \sqrt{3}(K_1 + P_1)\rangle,
\quad
A_{14} = \langle K_0,\,  P_0,\,  D\rangle.\non
\end{eqnarray}

Here $Q=(1/2)(K_0 - K_3 +P_0 + P_3)$, \ $\{\al, \beta\} \subset {\R}^1$.

The algorithm of constructing conformally-invariant Ans\"atze
formulated above proves to be very efficient when obtaining Ans\"atze
invariant under three-dimensional subalgebras of the algebra $C(1,3)$
listed in (\ref{2.2.28}) but computations are much more cumbersome.
That is why we omit intermediate computations and write down the final
result:\ the Ans\"atze for the eight-component spinor field $ \Psi (x)
$ invariant under three-parameter subgroups of the group $C(1,3)$ with
generators (\ref{2.2.28}).

\begin{eqnarray*}
&1)& \Psi(x) = \Bigl(1 + (x_0 - x_3)^2\Bigr)\Bigl(x_2(x_0 - x_3) -
x_1\Bigr)^{-2}  
\\
& &
\quad\times \Bigl(1 - (1/2) (1 + \s)\G \cdot x\Bigr)
\exp \Bigl\{(1/2) (\s \G_0 +\G_3  
\\
& &
\quad- \G_1 \G_2)\tau_1 + (1/2)
\Bigl(\G_1(\G_0 - \G_3) + (1 + \s) \G_2\Bigr) \tau_2\Bigr\}
\\
& &
\quad\times \exp \{(1/2)(\s + \G_0 \G_3)\tau_3\} \vp\, (\omega),
\\
& &
\omega = 2 \Bigl(x_1(x_0 - x_3) + x_2\Bigr) \Bigl(x_2 (x_0 - x_3) -
x_1\Bigr)^{-1} + \Bigl(1 + (x_0 - x_3)^2\Bigr) 
\\
& &
\quad\times \Bigl(x_0 + x_3 + (x_0 - x_3) x \cdot x\Bigr) 
\Bigl(x_2(x_0 - x_3) -x_1\Bigr)^{-2},
\\
& &
\tau_1 = \arctan (x_0 - x_3),
\\
& &
\tau_2 = (1/2)\Bigl(x_0 + x_3 + (x_0 - x_3) x \cdot x\Bigr)
\Bigl(x_2(x_0 - x_3 )- x_1)\Bigr)^{-1},
\\
& &
\tau_3 = - \ln \Bigl\{\Bigl(1 + (x_0 - x_3)^2\Bigr)
\Bigl(x_2(x_0 - x_3) - x_1\Bigr)^{-1}\Bigr\};
\\
&2)& \Psi(x) = (x_1^2 + x_2^2)^{-1} \Bigl(1 - (1/2) (1 + \s)\G \cdot
x\Bigr) \exp \{(1/2) (\s \G_0 +\G_3\\
& &
\quad
- \al \G_1 \G_2)\tau_1\}
\exp  \{(1/2)(1 + \s)(\G_0 - \G_3) \tau_3\} 
\\
& &
\quad
\times \exp \{(1/2)(\s + \G_0 \G_3)\tau_2\} \vp\, (\omega),
\\
& &
\omega = \arctan (x_1/x_2) - \al \arctan (x_0 - x_3),\quad
\tau_1 = \arctan (x_0 - x_3),
\\
& & 
\tau_2 = -(1/2)\ln \Bigl\{\Bigl(1 +(x_0 -
x_3)^2\Bigr) (x_1^2 + x_2^2)^{-1}\Bigr\},
\\
& &
\tau_3 = (1/2)\Bigl(x_0 + x_3 + (x_0 - x_3) x \cdot x\Bigr)
\Bigl(1 + (x_0 - x_3)^2\Bigr)^{-1};
\\
&3)& \Psi(x) = \Bigl(1 + (x_0 - x_3)^2\Bigr)^{-1} \Bigl(1 - (1/2)
(1 + \s)\G \cdot x\Bigr) 
\\
& &
\quad
\times \exp \Bigl\{(1/2) \Bigl(\s \G_0 +\G_3 - \G_1 \G_2 -
\G_1(\G_0 - \G_3) - (1 + \s) \G_2\Bigr) \tau_1\Bigr\} 
\\
& &
\quad
\times \exp \{(1/2)(1 + \s)(\G_0 - \G_3)\tau_2\}
\exp \Bigl\{(1/2)\Bigl((1 + \s) \G_1 
\\
& &
\quad
- \G_2(\G_0 - \G_3)\Bigr) \tau_3 -
(1 + \s) (\G_0 - \G_3) \omega \tau_3\Bigr\} \vp\, (\omega),
\\
& &
\omega = - \arctan (x_0 - x_3) + \Bigl(x_1(x_0 - x_3) +
x_2\Bigr)\Bigl(1 + (x_0 - x_3)^2\Bigr)^{-1},
\\
& &
\tau_1 = \arctan (x_0 - x_3),
\\
& &
\tau_2 = \Bigl(x_1 (x_0 - x_3) + x_2\Bigr)\Bigl(x_2(x_0 - x_3 ) 
- x_1\Bigl)\Bigl(1 +(x_0 - x_3)^2\Bigr)^{-2} 
\\
& &
\quad
+ (1/2)\Bigl(x_0 + x_3 + (x_0 - x_3) x \cdot x\Bigr) 
\Bigl(1 + (x_0 - x_3)^2\Bigr)^{-1},
\\
& &
\tau_3 = \Bigl(x_2(x_0 - x_3)- x_1\Bigr)\Bigl(1 +(x_0 -
x_3)^2\Bigr)^{-1}; 
\\
&4)&\Psi(x) = \Bigl(1 + (x_0 - x_3)^2\Bigr)\Bigl(x_2(x_0 - x_3) -
x_1\Bigr)^{-2} \Bigl(1 - (1/2) (1 + \s)\G \cdot x\Bigr) 
\\
& &
\quad
\times \exp \Bigl\{(1/2) \Bigl(\G_3 + \s \G_0 -  \G_1 \G_2 + \al
(\s + \G_0 \G_3)\Bigr)\tau_1 \Bigr\} 
\\
& &
\quad
\times \exp \Bigl\{-(1/2)\Bigl(\G_1(\G_0 - \G_3) + (1 + \s) \G_2\Bigr) 
\tau_2\Bigr\}  
\\
& &
\quad
\times \exp \{(1/2)(1 +\s)(\G_0 - \G_3)\tau_3\} \vp\, (\omega),
\\
& &
\omega = \ln \Bigl\{ \Bigl(1 + (x_0 - x_3)^2\Bigr) 
\Bigl(x_2 (x_0 - x_3) - x_1\Bigr)^{-1}\Bigr\} 
\\
& &
\quad + 2 \al \arctan (x_0 - x_3),\quad \tau_1 = \arctan (x_0 - x_3),
\\
& &
\tau_2 = \Bigl(x_1 (x_0 - x_3) + x_2\Bigr)\Bigl(1 + (x_0 - x_3
)^2\Bigr)^{-1} \\
& &
\quad
\times \exp \{- \al \arctan (x_0 - x_3)\},
\\
& &
\tau_3 = \Bigl\{\Bigl(x_1 (x_0 - x_3) + x_2\Bigr)\
\Bigl(x_2(x_0 - x_3) - x_1\Bigr)\Bigl(1 +
(x_0 - x_3)^2\Bigr)^{-2} \\
& &+ (1/2) \Bigl(x_0 + x_3 + (x_0 -x_3) x \cdot x\Bigr) 
\Bigl(1 + (x_0 - x_3)^2\Bigr)^{-1}\Bigr\}\\
& &
\quad
\times\exp \{- \al \arctan (x_0 - x_3)\};
\\
&5)& \Psi (x) = (x_1^2 + x_2^2)^{-1} \Bigl(1 - (1/2) (1 +
\s) \G \cdot x\Bigr)\exp \Bigl\{(1/2) \Bigl(\s \G_0 + \G_3 
\\
& &
\quad
+ \beta (\s + \G_0 \G_3)\Bigr)
\tau_2\Bigr\} \exp \Bigl\{(1/2) \Bigl(\G_1 \G_2 
- \al (\s + \G_0 \G_3)\Bigr)\tau_1\Bigr\}
\\
& &
\quad
\times \exp \{(1/2) (1 + \s) (\G_0 - \G_3) \tau_3\} \vp\,  (\omega),
\\
& &
\omega = 2 \al \arctan (x_2/x_1)- 2 \beta \arctan (x_0 - x_3) 
\\
& &
\quad
 + \ln \Bigl \{(x_1^2 + x_2^2)\Bigl(1 + (x_0 -
 x_3)^2\Bigr)^{-1}\Bigr\}, \\
& &
\tau_1 = \arctan (x_2/x_1), \quad \tau_2 = \arctan (x_0 - x_3),
\\
& &
\tau_3 = (1/2) \exp \{2 \al \arctan (x_2/x_1) - 2 \beta \arctan
(x_0 - x_3)\} 
\\
& &
\quad
 \times \Bigl(x_0 + x_3 + (x_0 - x_3) x \cdot x\Bigr) 
\Bigl(1 + (x_0 - x_3)^2\Bigr)^{-1};
\\
&6)& \Psi (x) = x_1^{-2} \Bigl(1 - (1/2) (1 + \s) \G \cdot x\Bigr)
\exp \{(1/2) (1 - \s)(\G_0 - \G_3) \tau_1\} 
\\
& &
\quad
\times \exp \{(1/2)(\s - \G_0 \G_3) \tau_2\} 
\exp \{(1/2)(1 + \s)(\G_0 + \G_3) \tau_3\} \vp\,  (\omega),
\\
& &
\omega = x_1/x_2, \quad \tau_1 = (1/2)(x_0 + x_3) (x \cdot
x)^{-1},\quad \tau_2 = \ln \Bigl((x \cdot x)/x_1\Bigr),
\\
& &
\tau_3 =(1/2)x_1^2(x_3- x_0) \Bigl((x_1^2 + x_2^2) x \cdot
x\Bigr)^{-1}; \\
&7)& \Psi (x) = (x_1^2 + x_2^2)^{-1} \Bigl(1 - (1/2)(1 + \s)
\G \cdot x\Bigr)\exp \{(1/2)( \s \G_0 + \G_3) \tau_1
\\
& &
\quad
+ (1/2)(\s + \G_0 \G_3) \tau_2 -
(1/2) \G_1 \G_2 \tau_3\} \vp\,  (\omega),
\\
& &
\omega = \Bigl(x_0 + x_3 + (x_0 - x_3) x \cdot x\Bigr) (x_1^2 +
x_2^2)^{-1}, \\
& &
\tau_1 = \arctan (x_0 - x_3),\quad
\tau_2 = (1/4) \ln \Bigl\{\Bigl((x \cdot x)^2 + (x_0 + x_3)^2\Bigr) 
\\
& &
\times\Bigl(1 + (x_0 - x_3)^2\Bigr)^{-1}\Bigr\},\quad \tau_3 = \arctan
(x_1/x_2); 
\\
&8)& \Psi (x) = (x_1^2 + x_2^2)^{-1}\Bigl(1 - (1/2) (1
+ \s) \G \cdot x\Bigr) \exp \{-(1/2) \G_1 \G_2 \tau_1 
\\
& &
\quad
+ (1/2) (\s \G_0 + \G_3) \tau_2 +
(1/2)(1 + \s) (\G_0 - \G_3) \tau_3\} \vp\, (\omega),
\\
& &
\omega = (x_1^2 + x_2^2) \Bigl(1 + (x_0 - x_3)^2\Bigr)^{-1},
\\
& &
\tau_1 = \arctan (x_1/x_2), \quad \tau_2 = \arctan (x_0 - x_3),
\\
& &
\tau_3 = (1/2)\Bigl(x_0 + x_3 + (x_0 - x_3) x \cdot x\Bigr) \Bigl(1
+ (x_0 - x_3)^2\Bigr)^{-1};
\\
&9)& \Psi (x) = \Bigl(1 + (x_0 - x_3)^2\Bigr)^{-1} \Bigl(1 - (1/2) (1
+ \s) \G \cdot x\Bigr) 
\\
& &
\quad
\times \exp \Bigl\{ (1/2) (\s \G_0 + \G_3 - \G_1 \G_2) \tau_1 -
(1/2)\Bigl((1 + \s) \G_2 
\\
& &
\quad
+ \G_1 (\G_0 - \G_3)\Bigr) \tau_2 
 + (1/2) (1 + \s) (\G_0 - \G_3) \tau_3 \Bigr\} \vp\,  (\omega),
\\
& &
\omega = \Bigl(x_2 (x_0 - x_3) - x_1\Bigr)\Bigl(1 + (x_0 -
x_3)^2\Bigr)^{-1}, 
\\
& &
\tau_1 = \arctan (x_0 - x_3), \quad \tau_2 = 
\Bigl(x_1(x_0 - x_3) + x_2\Bigr)\Bigl(1 +
(x_0 - x_3)^2\Bigr)^{-1},
\\
& &
\tau_3 = (1/2) (x_0 + x_3) - (1/2)x_1^2(x_0 - x_3)^{-1} +
(1/2) \Bigl(1 - (x_0 - x_3)^2\Bigr) \times
\\
& &
\quad
\times \Bigl(x_2 (x_0 - x_3)- x_1\Bigr)^2 (x_0 - x_3)^{-1} 
\Bigl(1 + (x_0 - x_3)^2\Bigr)^{-2};
\\
&10)& \Psi (x) = (x_1^2 + x_2^2)^{-1} \Bigl(1 - (1/2)(1
+ \s) \G \cdot x\Bigr) 
\\
& &
\quad
\times \exp \{-(1/2) \G_1 \G_2 \tau_1 + (1/2) \s \G_0 \tau_2 +
(1/2) \G_3 \tau_3\} \vp\,  (\omega),
\\
& &
\omega = \Bigl(4 x_0^2 + (x \cdot x - 1)^2\Bigr) (x_1^2 + x_2^2)^{-1}, 
\\
& &
\tau_1 = \arctan (x_1/x_2),\quad
\tau_2 = \arctan \Bigl((x \cdot x - 1)(2 x_0)^{-1}\Bigr) + \pi /2,
\\
& &
\tau_3 = \arctan \Bigl((x \cdot x + 1) (2x_3)^{-1}\Bigr);
\\
&11)& \Psi (x) = x_0^{-2}\Bigl(1 - (1/2)(1 + \s) \G \cdot x\Bigr)
\\
& &
\quad
\times (R_1 B_1 C_1 - R_1 B_2 C_2 - R_2 B_1 C_2 - R_2 B_2 C_1) \vp\,
(\omega), 
\\
& &
R_1 = 1 + (\tau_1/2) (\G_2 \G_3 - \G_1) - (\tau_2 /2)(\G_3\G_1 -
\G_2), 
\\
& &
R_2 = (\tau_1/2) (\G_3 \G_1 - \G_2) - (\tau_2 /2)(\G_2\G_3 - \G_1),
\\
& &
B_1 = (1/2)\Bigl\{ 1 - \G_1 \G_2 \G_3 + (1/2)
\Bigl(x_1^2 + x_2^2 + (x \cdot x - 1)^2\Bigr)
\\
& &
\quad
\times  (x \cdot x -1)^{-1}(x_1^2 + x_2^2)^{-1} \Bigl(x_1(1 + \G_1
\G_2 \G_3) + x_2(\G_1 \G_2 - \G_3)\Bigr)\Bigr\}, 
\\
& &
B_2 = (1/4)\Bigl(x_1^2 + x_2^2 - (x \cdot x - 1)^2\Bigr)
(x \cdot x - 1)^{-1} (x_1^2 + x_2^2)^{-1} 
\\
& &
\quad
\times \Bigl(x_2 (1 + \G_1 \G_2 \G_3) + x_1 (\G_3 - \G_1 \G_2)\Bigr), 
\\
& &
C_1 = 1 + (\rho_1 /2)(\G_2 \G_3 - \G_1) + (\rho_2 /2)(\G_3 \G_1 -
\G_2), \\
& &
C_2 = (\rho_1 /2)(\G_2  - \G_3 \G_1) + (\rho_2 /2)(\G_2 \G_3 - \G_1),
\\
& &
\omega = (x \cdot x -1) x_0^{-1},
\\
& &
\tau_1 = (1/4)\Bigl(2x_2 x_3 - x_1 (x \cdot x +1)\Bigr) (x_1^2 +
x_2^2)^{-1}, 
\\
& &
\tau_2 = (1/4)\Bigl(2x_1 x_3 + x_2 (x \cdot x +1)\Bigr) (x_1^2 +
x_2^2)^{-1},
\\
& &
\rho_1 = - (x \cdot x - 1)^2 \Bigl(2x_2 x_3 + x_1 (x \cdot x
+1)\Bigr)(x_1^2 + x_2^2)^{-1} 
\\
& &
\quad
\times \Bigl((x \cdot x -1)^2 + 4x_0^2\Bigr)^{-1},
\\
& &
\rho_2 = - (x \cdot x - 1)^2 \Bigl(2x_1 x_3 - x_2 (x \cdot x
+1)\Bigr)(x_1^2 + x_2^2)^{-1} \times
\\
& &
\quad
\times \Bigl((x \cdot x -1)^2 + 4x_0^2\Bigr)^{-1};
\\
&12)& \Psi(x) = \Bigl((x \cdot x - 1)^{2} - 4x^2_1 - 4x^2_2\Bigr)^{-1} 
\Bigl(1 - (1/2)(1 + \s)\G \cdot x\Bigr)
\\
& &
\quad
\times \exp \{(\G_1 \G_2 - \G_3 + \s \G_1 - \G_0 \G_3)\tau_1\}
\exp \{(\G_0 + \s \G_2)\tau_2\} 
\\
& &
\quad
\times \exp \Bigl\{(1/2)(\G_1 \G_2 - \G_3 + \s \G_1 + \G_0 \G_3) 
\arcsin \Bigl((2\tau_4 - \omega)\\
& &\quad\times(\omega^2 + 4)^{-1/2}\Bigr)\Bigr\} \vp\,  (\omega),
\\
& &
\omega = \tau_4^{-1} (\tau^2_3 + \tau_4^2 -1),\quad
\tau_1 = x_1 (x \cdot x -1 - 2x_2)^{-1},
\\
& &
\tau_2 = -(1/2) \ln \Bigl\{(1/2)(x \cdot x -1 - 2x_2)
\Bigl((x \cdot x -1)^2 - 4x_1^2 -
4x_2^2\Bigr)^{-1/2}\Bigr\},
\\
& &
\tau_3 = 2\Bigl(x_3 (x \cdot x - 1 - 2x_2) - x_1(x \cdot x + 1 -
2x_0)\Bigr)  
\\
& &
\quad
\times (x \cdot x - 1 - 2x_2)^{-1} \Bigl((x \cdot x -1)^2 - 4x_1^2
- 4x_2^2\Bigr)^{-1/2},
\\
& &
\tau_4 = (2x_0 - x \cdot x - 1)(x \cdot x - 1 - 2x_2)^{-1};
\\
&13)& \Psi(x) = x_3^{-2}\bigl(1 - (1/2)(1 + \s) \G \cdot x\Bigr)
\exp\{\tau_1 q_{+}\} \exp\{\tau_2 q\} 
\\
& &
\quad
\times \exp\{\tau_3 q_{+}\} \exp\{\tau_4 q\} \exp\{\tau_5 q_{-}\}
\vp\, (\omega), 
\\
& &
q_{\pm} = (1/2) (\s - \G_0 \G_2 + \sqrt{3} \G_0 \G_1) \pm (1/2)
(2 \G_2 - \s \G_0),
\\
& &
q = (1/2) (\g_0 + \s \G_2 + \sqrt{3} \s \G_1),
\\
& &
\tau_1 = (1/2)(x \cdot x + 1 + 2x_0)(x \cdot x - 1 + x_2 -
\sqrt{3} x_1)^{-1},
\\
& &
\tau_2 = (1/2) \ln \{2(x \cdot x - 1 + x_2 + \sqrt{3} x_1)
(x \cdot x - 1)^{-1}\},
\end{eqnarray*}
functions $\omega(y_1, y_2), \ \tau_k (y_1, y_2), \ k = {3,4,5} $
being determined by the following relations:
\begin{eqnarray*}
& &Q \tau_3 - \tau_3^2 + 3y_1 = 0,
\quad Q \tau_4 + 2 \tau_3 = 0,\\
& &
Q \tau_5 + \exp\{- \tau_4\} = 0, \quad \omega = y_2^2 - 4y_1 (y_1^2 +
1), 
\end{eqnarray*}
where
\begin{eqnarray*}
& &Q = 2y_2 \p_{y_1} + 4 (3y_1^2 + 1) \p_{y_2},\\
& &
y_1 = (\sqrt{3} x_2 - x_1)(\sqrt{3} x_3)^{-1} + (1/4)(2x_0 + x
\cdot x + 1)^2 \\
& &
\quad
\times x_3^{-1}(x \cdot x - 1 + x_2 + \sqrt{3} x_1)^{-1},\\
& &
y_2 = (1/4) x_3^{-3/2} (x \cdot x - 1
+ x_2 + \sqrt{3} x_1)^{-3/2} \\
& &
\quad
\times \{ (2x_0 + x \cdot x + 1)^3 + 2 \sqrt{3} (2x_0 + x \cdot x +1)
(\sqrt{3} x_2 - x_1) \\
& &
\quad
\times  (x \cdot x - 1 + x_2 +\!\sqrt{3} x_1) + 2 (2x_0 - x \cdot x -
1)(x \cdot x - 1 + x_2 +\!\sqrt{3} x_1)^2 \}.
\end{eqnarray*}

Basis operators of the algebra $A_{14} $ do not satisfy condition
(\ref{1.5.2}). Consequently, they give rise to a partially-invariant
solution which is not considered here.

In the above formulae $\vp\, (\omega) $ is an arbitrary
eight-component complex-valued function.

To obtain conformally-invariant Ans\"atze for the
four-component Dirac field we act with the projector $P = (1/2)(1 + \s)
$ on expressions 1--13. As a result, we 
have\index{Ansatz!$C(1,3)$-invariant}
\begin{eqnarray}
1)\ \psi (x) &=& [1 + (x_0 - x_3)^2][x_2(x_0 - x_3) -
x_1]^{-2} R(\tau_1) 
\non\\
& &
\times
\exp \{-(\tau_1/2) \g_1 \g_2 \} \exp\{-(\tau_2/2) \g_1 (\g_0 - 
\g_3)\} \non\\
& &\times\exp\{(\tau_3/2)(1 + \g_0 \g_3 )\} \vp(\om),
\non\\
2)\ \psi(x) &=& (x_1^2 + x_2^2)^{-1} R(\tau_1) \exp\{-(\al \tau_1/2)\g_1
\g_2 \} 
\non\\
& &
\times \exp\{(\tau_2/2)(1 + \g_0 \g_3)\} \vp(\om),
\non\\
3)\ \psi(x) &=& [1 + (x_0 - x_3)^2]^{-1} R(\tau_1)
\exp\{-(\tau_1/2) \g_1 \g_2 \} 
\non\\
& &
\times \exp\{-(1/2)(\g_1\tau_1 + \g_2\tau_3)(\g_0 - \g_3) \}\vp(\om), 
\non\\
4)\ \psi(x) &=& [1 + (x_0 - x_3)^2][x_2(x_0 - x_3)
- x_1]^{-2}R(\tau_1) 
\non\\
& &
\times \exp\{-(\tau_1/2) \g_1 \g_2 + (\al\tau_1 /2) (1 + \g_0 \g_3)\}
\non\\
& &
\times\exp\{-(\tau_2 / 2) \g_1 (\g_0 - \g_3)\} \vp (\om),
\non\\
5)\ \psi(x) &=& (x_1^2 + x_2^2)^{-1} R(\tau_2) \exp\{(\beta \tau_2/2)
(1 + \g_0 \g_3) \} 
\label{2.2.30}\\
& &
\times \exp \{(\tau_1/2)[\g_1 \g_2- \al (1 + \g_0 \g_3)] \} \vp
(\om), \non\\
6)\ \psi(x) &=& (x \cdot x)^{-2} (\g \cdot x) 
\exp \{(1/2)(3 - \g_0 \g_3)\ln [(x \cdot x)/x_1]\} \vp(\om),
\non\\
7)\ \psi(x) &=& (x_1^2 + x_2^2)^{-1} R(\tau_1)
\exp\{(\tau_2 /2)(1 +\g_0 \g_3) - (\tau_3 /2) \g_1 \g_2\} \vp(\om), 
\non\\
8)\ \psi(x) &=& (x_1^2 + x_2^2)^{-1} R(\tau_2) \exp\{-(\tau_1 /2)
\g_1 \g_2\} \vp(\om),
\non\\
9)\ \psi (x) &=& [1 + (x_0 - x_3)^2]^{-1} R(\tau_1)
\exp\{- (\tau_1 /2) \g_1 \g_2\} 
\non\\
& &
\times \exp \{- (\tau_2 /2) \g_1 (\g_0 - \g_3)\} \vp(\om),
\non\\
10)\ \psi(x) &=& \{x_1^2 + x_2^2)^{-1}(\cos (\tau_2 /2)
\cos (\tau_3 /2) + \g_0 \g_3 \sin (\tau_2 / 2) \sin (\tau_3 /2)
\non\\
& &
 + \g \cdot x [\g_0 \sin (\tau_2 /2) 
\cos (\tau_3 /2) - \g_3 \cos (\tau_2 /2) \sin (\tau_3/2)]\}
\non\\
& &
\times \exp\{-(\tau_1 /2) \g_1 \g_2\} \vp(\om).\non
\end{eqnarray}

Ans\"atze invariant under the algebras $A_{11}, \ A_{12}, \ A_{13} $
are given by very cumbersome formulae. Therefore they are not adduced
here.

In (\ref{2.2.30}) $\vp (\omega) $ is an arbitrary four-component
function; $\omega, \ \tau_1, \ \tau_2, \ \tau_3$ are real-valued
functions defined above in the formulae 1--13;
\begin{displaymath}
R(\tau) = \cos^2 (\tau /2) +
\g_0 \g_3 \sin^2 (\tau /2) + (1/2) \g \cdot x (\g_0 - \g_3) \sin \tau. 
\end{displaymath}

Thus, the problem of construction of Ans\"atze for the spinor
field invariant under the $C(1,3)$ non-conjugate one- and
three-parameter subgroups of the conformal group is
completely solved. It is important to note that these
Ans\"atze can be applied to reduction of any spinor equation
invariant under the groups $P(1,3), \ \widetilde P(1,3),\ C(1,3) $
in representation (\ref{1.1.24a})--(\ref{1.1.24e}).

Now we will say a few words about Ans\"atze reducing
Poincar\'e-invariant equations for particles with arbitrary spins. 
Suppose that on the set of solutions of the PDE under
study a covariant representation\index{Covariant representation} 
of the Poincar\'e algebra\index{Poincar\'e!algebra}
\begin{equation}
P_\mu = \p^\mu,\quad
J_{\mu \nu} = x_\mu \p^\nu - x_\nu \p^\mu + S_{\mu \nu},
\label{2.2.31}
\end{equation}
where $S_{\mu \nu} $ are constant matrices fulfilling the
commutation relations of the Lie algebra of the Lorentz group 
$O(1,3)$, is realized. Then Ans\"atze invariant under the $P(1,3)$ 
non-conjugate one- and three-dimensional subalgebras of the 
algebra with basis elements (\ref{2.2.31}) are obtained by 
making in the $P(1,3)$-invariant Ans\"atze for the spinor 
field the following replacement:  
\begin{displaymath}
\g_0 \g_a \to  2 S_{0a}, \quad \g_a \g_0 \to - 2 S_{a0},\quad
\g_a \g_b \to 2 S_{ab}, \ \ a \not= b.
\end{displaymath}

On applying the same trick to the $\wid P(1,3)$ Ans\"atze for the
spinor field we get the Ans\"atze invariant under the $\wid P(1,3) $
non-conjugate subalgebras of the algebra $A \wid P(1,3) $ having
generators (\ref{2.2.31}) and $D = x_\mu \p_\mu + k$ ($ k $ may be a
constant matrix commuting with $S_{\mu \nu}$).

Another method of constructing Poincar\'e and conformally-invariant
An\-s\"atze for fields with spins $s = 0,1, 3/2 $ via
Ans\"atze for the Dirac field is suggested in Section 2.6.

In conclusion we mention nonlocal Ans\"atze for the Dirac
equation. As established in \cite{101} the real
eight-component Dirac equation (\ref{1.1.14}) admits the Poincar\'e
algebra having the following basis elements:
\begin{equation}
\begin{array}{l}
P_\mu = \p^\mu + \theta (\wid \G_4 + \wid \G_5)(\p^\mu + im \wid
\G_\mu),\\ 
J_{\mu \nu} = x_\mu \p^\nu - x_\nu \p^\mu + (1/4)(\wid \G_\mu \wid
\G_\nu - \wid \G_\nu \wid \G_\mu).
\end{array}
\label{2.2.32}
\end{equation}
Here $\wid \G_\mu $ are $(8\times 8)$-matrices defined in Section 1.1,
$\nu = {0,\ldots,3}, \ \theta = \mbox{\rm const} $.

Let us emphasize that the operators $P_\mu $ are non-Lie operators
because the coefficients of $\p^\mu$ are matrices not proportional
to the unit matrix. 

We have succeeded in solving systems of PDEs (\ref{1.5.16}),
(\ref{1.5.15}) for each inequivalent subalgebra of the algebra
$AP(1,3)$ listed in (\ref{2.2.7}). As a result we get
$P(1,3)$-inequivalent Ans\"atze for the spinor field 
\begin{displaymath} 
\Psi (x) = A(x) \vp (\omega), 
\end{displaymath} 
where $A(x) $ is an $(8\times 8)$-matrix and $\omega = \omega(x) $ is
a scalar function, reducing (\ref{1.1.14}) to systems of ODEs for
$\vp = \vp(\omega) $. These Ans\"atze cannot be, in principle, 
obtained within the framework of the traditional Lie approach (for
more details, see \cite{101}).
\vspace{10mm}

\noindent
{\large\bf 2.3. Reduction of Poincar\'e-invariant spinor
  equations\label{s2.3}} 

\markboth{Chapter 2. EXACT SOLUTIONS}
{2.3. Reduction of Poincar\'e-invariant spinor equations}
\def\theequation{2.\arabic{section}.\arabic{equation}}
\setcounter {section} {3}
\setcounter {equation}{0}
\vspace{7mm}

\noindent
According to Consequence 1.5.1, substitution of $P(1,3)$-invariant
Ans\"atze (2.2.3) obtained in the previous section into the
Poincar\'e-invariant equation
\begin{equation}
\{i \g_\mu \p_\mu - f_1 - f_2 \g_4\} \psi (x) = 0,
\label{2.3.1}
\end{equation}
where $ f_i = f_i (\bar\psi \psi,\, \bar\psi \g_4 \psi) $, 
yields a three-dimensional system of PDEs for a four-component
function $\vp = \vp (\omega_1, \omega_2, \omega_3)$. As a direct
computation shows these Ans\"atze  satisfy the relations
\begin{eqnarray}
& &\bar\psi \psi = \bar\vp \vp, \quad \bar\psi \g_4 \psi
= \bar\vp \g_4 \vp,\non\\
& &A^{-1} (x) \{i \g_\mu \p_\mu - f_1 - f_2 \g_4\} A(x) \vp (\omega) 
\label{2.3.2}\\
\quad
& &= \{\g_\mu f_{\mu a} \p_{\omega_a} + (g_\mu + h_\mu g_4) \g_\mu
+ f_1 + f_2 \g_4\} \vp (\omega),\non
\end{eqnarray}
where $ f_i = f_i (\bar\vp \vp, \bar\vp \g_4 \vp) ;\
f_{\mu a}, \ g_\mu, \ h_\mu $
are rational functions of $\omega_1, \ \omega_2, \ \omega_3$.

Omitting intermediate computations we adduce a final
result:\ reduced equations\index{Reduced equation} 
for four-component functions $\vp (\vec
\omega)$ 
\begin{eqnarray}
&1)& (1/2)(\g_0 + \g_3)\vp + \Bigl( \omega_1(\g_0 + \g_3)
 + \g_0 - \g_3\Bigr) \vp_{\omega_1}+ \g_1 \vp_{\omega_2}  
 \non\\
 & &
 + \g_2 \vp_{\omega_3} = R,
 \non\\
&2)& (1/2)  \omega_2^{-1/2} \g_2\vp + \g_0 \vp_{\omega_1} +
 2 \omega_2^{1/2} \g_2 \vp_{\omega_2} + \g_3 \vp_{\omega_3} = R,
 \non\\
&3)& (1/2)\Bigl(\g_0 + \g_3 +  \omega_2^{-1/2}\g_2\Bigr) \vp 
 + \Bigl( \omega_1(\g_0 + \g_3) + \g_0 - \g_3\Bigr)
 \vp_{\omega_1}\non\\ 
& &
 +2 \omega_2^{1/2} \g_2\vp_{\omega_2} 
 + \Bigl(\al (\g_0 + \g_3) +  \omega_2^{-1/2}\g_1\Bigr) \vp_{\omega_3}
 = R, \non\\
&4)& (1/2\omega_1) (\g_0 + \g_3)\vp +
 (\g_0 + \g_3) \vp_{\omega_1}+ \Bigl(\omega_1(\g_0 - \g_3)   
 \non\\
 & &
 +  (\omega_2/\omega_1)(\g_0 + \g_3)\Bigr)
 \vp_{\omega_2} + \g_2 \vp_{\omega_3} = R,
 \non\\
&5)& \g_1 \vp_{\omega_1} + \g_2 \vp_{\omega_2} +
 \g_3 \vp_{\omega_3} = R,
 \non\\
&6)& \g_0 \vp_{\omega_1} + \g_1 \vp_{\omega_2} + \g_2
 \vp_{\omega_3} = R,
\non\\
&7)& (\g_0 + \g_3)\vp_{\omega_1} + \g_1 \vp_{\omega_2}
 + \g_2 \vp_{\omega_3} = R,
 \label{2.3.3}\\
&8)& (1/2)(\g_0 + \g_3) \vp + \Bigl(\omega_1(\g_0 + \g_3)
 + \g_0 - \g_3\Bigr) \vp_{\omega_1} + \g_2 \vp_{\omega_2} 
 \non\\
 & &
 + \Bigl(\al (\g_0 + \g_3) - \g_1\Bigr) \vp_{\omega_3} = R,
 \non\\
&9)& (1/2)\omega_2^{-1/2} \g_2 \vp + \g_0 \vp_{\omega_1} +
 2 \omega_2^{1/2} \g_2 \vp_{\omega_2} 
 + (\g_3 + \al  \omega_2^{-1/2} \g_1)\vp_{\omega_3} = R,
 \non\\
&10)& (1/2)\omega_2^{-1/2} \g_2 \vp + \g_3 \vp_{\omega_1} +
 2 \omega_2^{1/2} \g_2 \vp_{\omega_2} 
 + (\g_0 - \al \omega_2^{-1/2}\g_1) \vp_{\omega_3} = R,
 \non\\
&11)& (1/2)\omega_2^{-1/2} \g_2 \vp + (\g_0 + \g_3) \vp_{\omega_1}
 + 2  \omega_2^{1/2} \g_2\vp_{\omega_2} +
 \non\\
 & &
 + (\g_0 - \g_3 - 2\al \omega_2^{-1/2}\g_1) \vp_{\omega_3} = R,
 \non\\
&12)& - 2\al \g_1 \vp_{\omega_1} + \g_2 \vp_{\omega_2} 
 + (3/2)\Bigl(2 \al^2 \g_0 + \omega_1(\g_0 + \g_3) \Bigr)
 \vp_{\omega_3} = R, \non\\
&13)& (2\al)^{-1} \g_4 (\g_0 + \g_3) \vp + (\g_0 + \g_3)
 \vp_{\omega_1}+ \Bigl(\omega_1(\g_0 + \g_3)  - 2 \al^{-1}\omega_3
 \g_1 \non\\
 & &
 + (\omega_2 + \al^{-2} \omega_3^2)\omega_1^{-1}(\g_0 + \g_3) \Bigr) 
 \vp_{\omega_2} + (\al \g_1 - \omega_1\g_2 ) \vp_{\omega_3} = R.
 \non
\end{eqnarray}

In (\ref{2.3.3}) $\vp_{\omega_a} = \p_{\omega_a}\vp$,\ \
$R = -if_1 (\bar\vp \vp, \bar\vp \g_4 \vp)\vp -
if_2 (\bar\vp \vp, \bar\vp \g_4 \vp) \g_4 \vp $.

$P(1,3)$-invariant Ans\"atze (2.2.8) also satisfy conditions
of the form (\ref{2.3.2})
\begin{eqnarray}
& &\bar\psi \psi = \bar\vp \vp, \quad 
\bar\psi \g_4 \psi=\bar\vp \g_4 \vp,\non\\
& &A^{-1} (x) (i \g_\mu \p_\mu - f_1 - f_2 \g_4) A(x) \vp (\omega)
\label{2.3.4} \\
& &\quad = \Bigl(\rho_\mu \g_\mu \p_\omega 
+ (g_\mu + h_\mu \g_4) \g_\mu - f_1
- f_2 \g_4\Bigr) \vp(\omega),\non
\end{eqnarray}
where $f_i = f_i (\bar\vp \vp, \bar\vp \g_4 \vp); \ \rho_\mu, \ 
\g_\mu, \ h_\mu $ are rational functions of $ \omega$. Using this
result we get the following set of reduced 
equations\index{Reduced equation} for four-component functions $ \vp $:
\begin{eqnarray}
&1)& \g_3 \dot{\vp} = R,
\non\\
&
2)& \g_0 \dot{\vp} = R,
\non\\
&
3)& (\g_0 + \g_3) \dot{\vp} = R,
\non\\
&
4)& (1/2)(\g_0 + \g_3) \vp + \Bigl(\omega(\g_0 + \g_3)  +
\g_0 - \g_3\Bigr) \dot{\vp} = R,
\non\\
&
5)& (1/2)(\g_0 + \g_3) \vp + \g_2 \dot{\vp} = R,
\non\\
&
6)&  - (1/2 \al) \g_1 \g_4 \vp + \g_1 \dot{\vp} = R,
\non\\
&
7)&  - (1/2 \al)\g_1 \g_4 \vp + \Bigl(\al 
\exp \{ - \omega / \al\} (\g_0 + \g_3)
- \g_2\Bigr)\dot{\vp} = R,
\non\\
&
8)&  (1/2) \omega^{-1/2} \g_2 \vp + 2 \omega^{1/2} \g_2 \dot{\vp} = R,  
\non\\
&
9)&  -(1/2 \al)\g_3 \g_4 \vp + \g_3 \dot{\vp} = R,
\non\\
&
10)&  (1/2 \al) \g_0 \g_4 \vp + \g_0 \dot{\vp} = R,
\non\\
&
11)& (1/4)(\g_0 - \g_3) \g_4 \vp + (\g_0 + \g_3) \dot{\vp} = R,
\non\\
&
12) & (1/2\omega)(\g_0 + \g_3) \vp + (\g_0 + \g_3) \dot{\vp} = R,
\non\\
&
13)& (1/2 \al\omega)(\al + \g_4) (\g_0 + \g_3) \vp +
(\g_0 + \g_3) \dot{\vp} = R,
\non\\
&
14)& (1/2)(\g_0 + \g_3) \g_4 \vp + (\g_0 + \g_3) \dot{\vp} = R,
\label{2.3.5}\\
&
15)&  2 \g_1 \dot{\vp}  = R,
\non\\
&
16)&  2 (\g_2 - \al \g_1) \dot{\vp} = R,
\non\\
&
17)& (1/2 \al) \omega^{-1/2} \g_2 (\al - \g_4) \vp +
 2 \omega^{1/2} \g_2 \dot{\vp} = R,
\non\\
&
18)&  (1/2) (\g_0 + \g_3)(1 + \al \g_4) \vp +
\Bigl(\omega (\g_0 + \g_3) + \g_0 - \g_3\Bigr) \dot{\vp} = R,
\non\\
&
19)& (1/2) (\g_0 + \g_3 + \omega^{-1/2}\g_2 ) \vp +
2\omega^{1/2} \g_2 \dot{\vp} = R,
\non\\
&
20)&  \omega^{-1} (\g_0 + \g_3) \vp + (\g_0 + \g_3) \dot{\vp} = R,
\non\\
&
21)& (1/2)\Bigl(\omega(\omega + \beta) - \al\Bigr)^{-1} 
\Bigl((1 - \al) \g_4 +2 \omega + \beta\Bigr) (\g_0 + \g_3)\vp
\non\\
& & + (\g_0 + \g_3) \dot{\vp} = R,
\non\\
&
22)&  (1/2)\Bigl(\omega(\omega + \beta)\Bigr)^{-1} (2\omega
+ \beta - \g_4)(\g_0 + \g_3) \vp  + (\g_0 + \g_3) \dot{\vp} = R,
\non\\
&
23)& (1/2)\Bigl(\omega(\omega + 1)\Bigr)^{-1} (2\omega + 1)(\g_0 +
\g_3) \vp + (\g_0 + \g_3) \dot{\vp} = R,
\non\\
&
24)&  (\g_0 + \g_3) \vp + \Bigl(\omega(\g_0 + \g_3) + \g_0 -
\g_3\Bigr) \dot{\vp} = R,
\non\\
&
25)&  (\g_0 + \g_3) \vp + \Bigl( \g_2 - \beta (\g_0 + \g_3)\Bigr)
\dot{\vp} = R,
\non\\
&
26)&  \Bigl(\omega^{-1}(\g_0 + \g_3) + (1/4)(\g_0 - \g_3) \g_4\Bigr)
\vp + (\g_0 + \g_3) \dot{\vp} = R,
\non\\
&
27)& (1/2) (\g_0 + \g_3) (3 + \al \g_4) \vp +
\Bigl(\omega (\g_0 + \g_3) + \g_0 - \g_3\Bigr) \dot{\vp} = R.\non
\end{eqnarray}

Here $ \dot{\vp} = d \vp / d \omega,\ 
\ R = -i f_1 (\bar\vp \vp, \bar\vp \g_4 \vp) \vp -
i f_2 (\bar\vp\vp, \bar\vp \g_4 \vp) \g_4 \vp $.

Formulae (\ref{2.3.2}), (\ref{2.3.4}) can be applied to reduce the
equation 
\begin{equation}
\p_\mu \p^\mu \psi (x) = 0
\label{2.3.6}
\end{equation}
by means of $P(1,3)$-invariant Ans\"atze for the spinor field
$\psi(x)$. To this end we make use of the identity
\begin{equation}
\p_\mu \p^\mu = \g_\mu \p_\mu A(x) A^{-1} (x) \g_\mu \p_\mu
\label{2.3.7}
\end{equation}
which holds for each invertible $(4\times 4)$-matrix $A(x)$. By
force of (\ref{2.3.7}), we have
\begin{eqnarray*}
& &A^{-1} (x) \p_\mu \p^\mu A(x) \vp (\vec\omega) =
 A^{-1} (x) \g_\mu \p_\mu A(x) A^{-1} (x)
 \g_\mu \p_\mu A(x) \vp (\vec\omega) \\
& &\quad=A^{-1} (x) \g_\mu \p_\mu A(x)
 \Bigl\{\g_\mu f_{\mu a} (\vec\omega) \vp_{\omega_a} 
 +\g_\mu \Bigl(g_\mu (\vec\omega) + h_\mu
 (\vec\omega) \g_4\Bigr) \vp\Bigr\} \\
& &\quad=A^{-1}(x)\Bigl\{\g_\mu f_{\mu a} (\vec\omega) \pa_{\omega_a}
 + \g_\mu \Bigl(g_\mu (\vec\omega) +
 h_\mu (\vec\omega) \g_4\Bigr)\Bigr\}^2\vp,
\end{eqnarray*}
the forms of functions $f_{\mu a}, \ g_\mu, \ h_\mu $ 
being determined by (\ref{2.3.3}).

In the same way we establish that $P(1,3)$-invariant Ans\"atze 
(\ref{2.2.8}) reduce equation (\ref{2.3.6}) to ODE
\begin{displaymath}
\Bigl\{\rho_\mu (\omega) \g_\mu \p_\omega + \Bigl(g_\mu (\omega) +
h_\mu (\omega) \g_4\Bigr) \g_\mu\Bigr\}^2 \vp = 0,
\end{displaymath}
where functions $ \rho_\mu (\omega), \ g_\mu (\omega), \ h_\mu
(\omega) $ are determined by (\ref{2.3.5}).

Provided reduced equations (\ref{2.3.3}), (\ref{2.3.5}) possess
nontrivial symmetry, their dimension can also be decreased with the
use of Theorem 1.5.1. But direct application of the infinitesimal
Lie method to investigation of the symmetry of systems of PDEs with
variable coefficients (\ref{2.3.3}), (\ref{2.3.5}) is, in many cases,
impossible without applying symbolic computation packages
\cite{69,70,137,177} (for multi-component systems of PDEs with $n>2$
independent variables these packages are also of little help).

In the papers \cite{100,103} we suggested a purely algebraic
method of investigation of invariance properties of reduced
equations. It is based on the following assertion.
\vspace{1.5mm}

\noindent
{\bf Theorem 2.3.1.}\ {\em Let $G$ be a Lie invariance
  group\index{Invariance!group} of some PDE and $H$ be a normal
  divisor\index{Normal divisor} in $G$.  Then an equation obtained
  via reduction with the help of an $H$-invariant Ansatz admits the
  group $G /H $ (here the symbol $/$ means factorization). }

Proof can be found in \cite{164}. $\rhd$

We use an equivalent formulation of the above theorem: {\em if there
  is a PDE admitting a Lie algebra $AG$ whose subalgebra $Q$ is an
  ideal\index{Ideal} in $AG$, then an equation obtained by reduction
  with the help of a $Q$-invariant Ansatz is invariant under the Lie
  algebra $AG/Q$.}

To apply Theorem 2.3.1 to algebras (2.2.2), (2.2.7) we have have to
select the maximal subalgebras of the algebra $AP(1,3)$ such that
algebras (2.2.2), (2.2.7) are ideals in these.

From the general theory of Lie algebras (see, e.g.
\cite{18,129.1,164}) it follows that the algebra $A\wid G=\langle Q_1,
\ldots, Q_N\rangle$ is an ideal in the Lie algebra $AG
=\langle\Sigma_1$,\, $\Sigma_2$, $\ldots$, $\Sigma_M\rangle$, $M \ge N
$ iff
\begin{equation}
[Q_i,\, \Sigma_j] = \lbd_{ij}^k Q_k,
\ \ \lbd_{ij}^k = \mbox{\rm const},
\label{2.3.8}
\end{equation}
the summation over repeated indices being implied.

Given an explicit form of the elements $Q_i$, we compute with the aid
of (\ref{2.3.8}) the maximal subalgebra of the algebra $AG$ such that
the algebra $A\wid G$ is an ideal in it. Next, we compute a
factor-algebra whose basis elements according to Theorem 2.5.1
generate an invariance group of the corresponding reduced equation.

The above scheme will be realized for the algebra $A_5$
from (\ref{2.2.2}). Substituting $Q = P_0$ into
(\ref{2.3.8}) and putting $N = 1$ we arrive at the following relations 
for $ \Sigma_j = \theta_j^{\mu \nu} J_{\mu \nu} + \theta_j^\mu P_\mu $:
\begin{equation}
[P_0,\, \theta_j^{\mu \nu} J_{\mu \nu} +
\theta_j^\mu P_\mu] = \lbd_j P_0, \quad j= {1,\ldots,M}.
\label{2.3.9}
\end{equation}

Computing the commutators and equating coefficients of
the linearly-in\-de\-pen\-dent operators $P_\mu, \ J_{\mu \nu} $  
yield the system of linear algebraic equations for constants
$\theta_j^{\mu \nu}, \ \theta_j^\mu $ 
\begin{displaymath}
\theta_j^{0a} = \theta_j^{a0} = 0, \ \ a = {1,2,3},
\end{displaymath}
$\theta_j^{ab}, \ \theta_j^\mu $ are arbitrary real constants.

Consequently, the basis of a maximal subalgebra of the algebra
$AP(1,3)$ containing the algebra $A_5 = \langle P_0\rangle$ as an
ideal consists of the operators
\begin{equation}
P_\mu, \quad J_{12}, \quad J_{23}, \quad J_{31}.
\label{2.3.10a}
\end{equation}

The basis of the factor-algebra $\langle\,P_\mu, J_{ab}\rangle\, /
\langle\,P_0\rangle\, $ is formed by those operators from
(\ref{2.3.10a}) which are linearly independent of $P_0$. As a result,
we come to the Lie algebra 
\begin{equation}
\wid A_5 = \langle P_1,\, P_2,\, P_3,\, J_{12},\, J_{23},\,
J_{31}\rangle
\label{2.3.10b}
\end{equation}
which, according to Theorem 2.3.1, is the invariance algebra of the
system 5 from (\ref{2.3.3}). The explicit form of symmetry operators
is obtained by passing from the "old " variables $x, \ \psi (x) $ to
the "new" ones $\omega, \ \vp(\omega) $ according to formula
(\ref{2.2.3}).

Below we write down the invariance algebras of equations (\ref{2.3.3})
\begin{eqnarray}
& &\wid A_1 = \langle -\omega_2 \p_{\omega_3} +
\omega_3 \p_{\omega_2} + (1/2) \g_1 \g_2,\, 
\p_{\omega_2},\,  \p_{\omega_3}\rangle,
\non\\
& &\wid A_2 = \langle \omega_3 \p_{\omega_1} +
\omega_1 \p_{\omega_3} - (1/2) \g_0 \g_3,\, 
\p_{\omega_1},\,  \p_{\omega_3} \rangle,
\quad
\wid A_3 = \langle \p_{\omega_3}\rangle,
\non\\
& &\wid A_4 = \langle 2 \omega_1 \omega_3 \p_{\omega_2} +
\omega_1 \p_{\omega_3} - (1/2)(\g_0 + \g_3)\g_1,
- \omega_1 \p_{\omega_1} + (1/2) \g_0 \g_3,\non\\
& &
\phantom{\wid A_4 = }\omega_1 \p_{\omega_2},\,  \p_{\omega_3} \rangle,
\non\\
& &\wid A_5 = \langle -\omega_1 \p_{\omega_2} +
\omega_2 \p_{\omega_1} + (1/2) \g_1 \g_2,\,  - \omega_2 \p_{\omega_3}
+ \omega_3 \p_{\omega_2} +(1/2) \g_2 \g_3,
\non\\
& &
\phantom{\wid A_5 = } -\omega_3 \p_{\omega_1}
+ \omega_1 \p_{\omega_3} + (1/2) \g_3 \g_1,\, 
\p_{\omega_1},\,  \p_{\omega_2},\,  \p_{\omega_3} \rangle,
\non\\
& &\wid A_6 = \langle \omega_1 \p_{\omega_2} +
\omega_2 \p_{\omega_1} - (1/2) \g_0 \g_1,\,  \omega_1 \p_{\omega_3} +
\omega_3 \p_{\omega_1} - (1/2) \g_0 \g_2,\, 
\non\\
& &
\phantom{\wid A_6 = } -\omega_2 \p_{\omega_3} + \omega_3 \p_{\omega_2} +
(1/2) \g_1 \g_2,\,  \p_{\omega_1},\,  
\p_{\omega_2},\,  \p_{\omega_3} \rangle,
\label{2.3.11}\\
& &\wid A_7 = \langle \omega_1 \p_{\omega_2} 
- (1/2) (\g_0 + \g_3) \g_1,\,  \omega_1 \p_{\omega_3}
- (1/2) (\g_0 + \g_3) \g_2,\, - \omega_1 \p_{\omega_1}  
\non\\
& &
\phantom{\wid A_7 = } + (1/2) \g_0 \g_3,\,  -\omega_2 \p_{\omega_3}
+ \omega_3 \p_{\omega_2} 
+ (1/2) \g_1 \g_2,\,  \p_{\omega_1},\,  \p_{\omega_2},\, 
\p_{\omega_3} \rangle,
\non\\
& &\wid A_8 = \langle  \p_{\omega_2},\,  \p_{\omega_3} \rangle,
\quad
\wid A_9 = \langle  \p_{\omega_1},\,  \p_{\omega_3} \rangle,
\quad
\wid A_{10} = \langle  \p_{\omega_1},\,  \p_{\omega_3} \rangle,
\non\\
& &\wid A_{11} = \langle  \p_{\omega_1},\,  \p_{\omega_3} \rangle,
\quad
\wid A_{12} = \langle  \p_{\omega_2},\,  \p_{\omega_3} \rangle,
\non\\
& &\wid A_{13} = \langle 2 \al \omega_1 \p_{\omega_3} -
(\g_0 + \g_3) \g_1,\,  2\omega_3 \p_{\omega_2} +
(\omega_1^2 - \al^2) \p_{\omega_3} 
\non\\
& &
\phantom{\wid A_{13} = }+ (1/2 \al)(\g_0 + \g_3) (\al \g_2 -
\g_1 \omega_1),\,  \omega_1 \p_{\omega_2} \rangle.\non
\end{eqnarray}

The invariance algebras of systems of ODEs listed in (\ref{2.3.5})
are obtained in a similar way
\begin{eqnarray}
& &
\wid A_1 = \langle \p_\omega,\,  \g_0 \g_1,\,  
\g_0 \g_2,\,  \g_1 \g_2\rangle,\quad
\wid A_2 = \langle \p_\omega,\,  \g_1 \g_2,\,  \g_2 \g_3,\,  \g_3
\g_1\rangle, \non\\
& &
\wid A_3 = \langle \g_1(\g_0 + \g_3),\,  \g_2(\g_0 + \g_3),\,  \g_1
\g_2,\, \omega \p_\omega - (1/2) \g_0 \g_3,\,  \p_\omega \rangle,
\non\\
& &
\wid A_4 = \langle \g_1 \g_2\rangle,\quad
\wid A_5 = \langle  \p_\omega\rangle,\quad
\wid A_6 = \langle \p_\omega,\,  \g_0 \g_3\rangle,\quad
\wid A_7 = \langle 2 \al \p_\omega - \g_0 \g_3\rangle,
\non\\
& &
\wid A_8 = \langle \g_0 \g_3\rangle,\quad
\wid A_9 = \langle \p_\omega,\,  \g_1 \g_2\rangle,\quad
\wid A_{10} = \langle \p_\omega,\,  \g_1 \g_2\rangle,\quad
\wid A_{11} = \langle \p_\omega,\,  \g_1 \g_2\rangle,
\non\\
& &
\wid A_{12} = \langle \g_2(\g_0 + \g_3),\,  \omega^{-1}
\g_1 (\g_0 + \g_3),\, 
\omega \p_\omega - (1/2) \g_0 \g_3\rangle,
\non\\
& &
\wid A_{13} = \langle (\g_1 + \al \g_2) (\g_0 + \g_3),\, 
\omega^{-1} \g_1 (\g_0 + \g_3),\, 
\omega \p_\omega - (1/2) \g_0 \g_3 \rangle,
\non\\
& &
\wid A_{14} = \langle \g_1 (\g_0 + \g_3 ),\,  \g_2 (\g_0 + \g_3),\, 
\p_\omega \rangle,\quad
\wid A_{15} = \langle \p_\omega,\,  \g_2 (\g_0 + \g_3) \rangle,
\non\\
& &
\wid A_{16} = \langle \p_\omega\rangle,\quad
\wid A_{17} = \langle \g_0 \g_3\rangle,\quad
\wid A_{18} = \langle \g_1 \g_2\rangle,\quad
\wid A_{19} = \oslash,
\label{2.3.12}\\
& &
\wid A_{20} = \langle \omega \p_\omega - (1/2) \g_0 \g_3,\, 
\omega^{-1} \g_1 (\g_0 + \g_3),\, 
\omega^{-1} \g_2 (\g_0 + \g_3),\,  \g_1 \g_2 \rangle,
\non\\
& &
\wid A_{21} = \langle [\omega (\omega + \beta) - \al]^{-1} (\g_0
+ \g_3)[(\omega + \beta)
\g_1 - \g_2],\, 
[\omega (\omega + \beta) 
\non\\
& &
\phantom{\wid A_{21} = }
 - \al]^{-1} (\g_0 + \g_3)
(\omega \g_2 - \al \g_1)\rangle,
\non\\
& &
\wid A_{22} = \langle \omega^{-1} \g_1 (\g_0 + \g_3),\, 
[\omega (\omega + \beta) - \al]^{-1} (\g_0 + \g_3)
(\omega \g_2 - \g_1)\rangle,
\non\\
& &
\wid A_{23} = \langle \omega^{-1} \g_1 (\g_0 + \g_3),\, 
(\omega + 1)^{-1} \g_2 (\g_0 + \g_3)\rangle,\quad
\wid A_{24} = \langle \g_1 \g_2\rangle,
\non\\
& &
\wid A_{25} = \langle \p_\omega,\,   \g_1 (\g_0 + \g_3)\rangle,\quad
\wid A_{26} = \langle \g_1 \g_2\rangle,\quad
\wid A_{27} = \langle \g_1 \g_2\rangle.\non
\end{eqnarray}

It is worth noting that any Poincar\'e-invariant spinor PDE after
being reduced by means of the $P(1,3)$-invariant Ans\"atze
(\ref{2.2.3}), (\ref{2.2.2}) is invariant under Lie algebras
(\ref{2.3.11}), (\ref{2.3.12}).  But for the specific reduced
equations these algebras are not, generally speaking, the maximal
ones. We will consider in more detail symmetry properties of the
systems 5--7 from (\ref{2.3.3}).

By the Lie method we can prove the following assertions.
\vspace{1.5mm}

\noindent
{\bf Theorem 2.3.2.}\ {\em Equation 5 from (\ref{2.3.3}) is invariant
under the conformal group\index{Conformal!group} $C(3)$ iff }
\begin{equation}
f_j = (\bar\psi \psi)^{1/2} \ti f_j \Bigl(\bar\psi \psi (\bar\psi
\g_4 \psi)^{-1}\Bigr), \ \ j = 1,2.
\label{2.3.13}
\end{equation}
{\bf Theorem 2.3.3.}\ {\em Equation 6 from (\ref{2.3.3}) is invariant 
under the conformal group $C(1,2)$ iff (\ref{2.3.13}) holds.}
\vspace{1.5mm}

\noindent
{\bf Theorem 2.3.4.}\ {\em Equation 7 from (\ref{2.3.3}) admits an
infinite-parameter invariance group with the following generators:}
\vspace{1.5mm}

\noindent
a)\ \ {\em with arbitrary $f_1,\ f_2$ }
\begin{eqnarray}
& &Q_1 = \p_{\omega_1},
\quad
Q_2 = -\omega_2 \p_{\omega_3} +
\omega_3 \p_{\omega_2} + (1/2) \g_1 \g_2,\non\\
& &Q_3 = w_1 \p_{\omega_2} + w_2
\p_{\omega_3} + (1/2) (\dot w_1 \g_1 
+\dot w_2 \g_2) (\g_0 + \g_3),\label{2.3.14}\\
& &Q_4 = \omega_1 \p_{\omega_1} - (1/2) \g_0 \g_3;\non
\end{eqnarray}
b)\ {\em  with $ f_1 = f_1(\bar\psi \psi ), \ f_2 = 0$}
\begin{eqnarray*}
& &Q_1 = \p_{\omega_1},
\quad
Q_2 = -\omega_2 \p_{\omega_3} + \omega_3 
\p_{\omega_2} + (1/2) \g_1 \g_2,\\
& &Q_3 = w_1 \p_{\omega_2} + w_2 \p_{\omega_3} +
(1/2) (\dot w_1 \g_1 +\dot w_2 \g_2) (\g_0 + \g_3),\\
& &Q_4 = w_3 \g_4 (\g_0 + \g_3),
\quad
Q_5 = \omega_1 \p_{\omega_1} - (1/2) \g_0 \g_3;
\end{eqnarray*}
c)\ {\em with $f_i = (\bar\psi \psi)^{1/2k} \ti f_i \Bigl(\bar\psi
  \psi (\bar\psi \g_4 \psi)^{-1}\Bigr), \ i = 1,2$}
\begin{eqnarray*}
& &Q_1 = \p_{\omega_1},
\quad
Q_2 = -\omega_2 \p_{\omega_3} + \omega_3 \p_{\omega_2}
+ (1/2) \g_1 \g_2,\\
& &Q_3 = w_1 \p_{\omega_2} + w_2 \p_{\omega_3} +
(1/2) (\dot w_1 \g_1 +\dot w_2 \g_2) (\g_0 + \g_3),\\
& &Q_4 = \omega_a \p_{\omega_a} + k,
\quad
Q_5 = \omega_1 \p_{\omega_1} - (1/2) \g_0 \g_3;
\end{eqnarray*}
d)\ {\em with $f_i = (\bar\psi \psi)^{1/2} \ti f_i \Bigl(\bar\psi
  \psi (\bar\psi \g_4 \psi)^{-1}\Bigr), \ i = {1,2}$ }
\begin{eqnarray*}
& &Q_1 = w_1 \p_{\omega_2} + w_2 \p_{\omega_3} 
+(1/2) (\dot w_1 \g_1 + \dot w_2 \g_2) (\g_0 + \g_3),\\
& &Q_2 = -\omega_2 \p_{\omega_3} + \omega_3 \p_{\omega_2} 
+ (1/2) \g_1 \g_2,\\
& &Q_3 = w_0 \p_{\omega_1} + \dot w_0(\omega_2 \p_{\omega_2} 
+ \omega_3 \p_{\omega_3}) + \dot w_0+ (1/2) \ddot w_0\\ 
& &\quad\times(\g_1 \omega_2 + \g_2 \omega_3) (\g_0 + \g_3),\quad
Q_4 = \omega_1 \p_{\omega_1} - (1/2) \g_0 \g_3.
\end{eqnarray*}

Here $w_\mu=w_\mu (\omega_1)$ are arbitrary smooth real-valued
functions, an overdot means differentiation with respect to
$\omega_1$. 

Consequently, the invariance algebras of PDEs 5--7 from (\ref{2.3.3})
are substantially wider than the algebras $ \wid A_5 - \wid A_7 $
adduced in (\ref{2.3.11}).

Using the above results we have constructed the Ans\"atze for field $
\psi (x) $ reducing PDE (\ref{2.3.1}) with $f_1 = f_1 (\bar\psi
\psi)$,\ $f_2 = 0 $ which cannot be obtained within the framework of
the Lie approach: 
\vspace{1.5mm}

\noindent
1)\ $ f_1 \in C({\R}^1, {\R}^1) $ is an arbitrary function
\begin{eqnarray*}
\psi (x) &=& \exp \{w_3 \g_4 (\g_0 + \g_3) -
(1/2)(\dot w_1 \g_1 + \dot w_2 \g_2 ) (\g_0 + \g_3)\} \\
& &\times \left \{
\begin{array} {l}
\vp (x_1 + w_1), \\
\exp \{-(1/2) \g_1 \g_2 \arctan [(x_1 + w_1)(x_2 + w_2)^{-1}] \} \\
\quad\times \vp\, [(x_1 + w_1)^2 + (x_2 + w_2)^2];
\end{array} \right.
\end{eqnarray*}
2)\ $f_1 = \lbd (\bar\psi \psi)^{1/2},
\ \lbd \in {\R}^1 $
\begin{eqnarray}
\psi (x) &=& w_0^{-1} \exp \left \{ w_3 \g_4 (\g_0 + \g_3) -
(1/2)(\dot w_1 \g_1 + \dot w_2 \g_2 ) (\g_0 + \g_3)  \right.\non\\
& &\left. - (\dot w_0 / 2w_0) [\g_1 (x_1 + w_1) + \g_2 (x_2 + w_2)]
(\g_0 + \g_3) \right \}\label{2.3.15} \\
& &\times  \left \{
\begin{array} {l}
\vp\, [w_0^{-1}(x_1 + w_1)], \non\\
\exp \{-(1/2) \arctan [(x_1 + w_1)(x_2 + w_2)^{-1}] \} \non\\
\quad\times \vp\, [(x_1 + w_1)^2 w_0^{-2} + (x_2 + w_2)^2 w_0^{-2}] ;
\end{array} \right.\non\\[1mm]
\psi(x) &=& (\g_0 x_0 - \g_1 x_1 - \g_2 x_2)(x_0^2 - x_1^2 -x_2^2)
^{-3/2} \non\\
& &\times \left \{
\begin{array} {l}
\vp\, [ x_0 (x_0^2 - x_1^2 - x_2^2)^{-1}], \non\\
\vp\, [x_1 (x_0^2 - x_1^2 - x_2^2)^{-1}], \non\\
\exp \{-(1/2) \arctan (x_1/x_2)\}  \non\\
\quad\times\vp\, [(x_1^2 + x_2^2)(x_0^2 - x_1^2 - x_2^2)^{-2}];
\end{array} \right.\non\\[1mm]
\psi(x) &=& (\g_1 x_1 + \g_2 x_2 + \g_3 x_3)(x_1^2 +
x_2^2 +x_3^2)^{-3/2} \non\\
& &\times \left \{
\begin{array} {l}
\vp\, [x_1 (x_1^2 + x_2^2 + x_3^2)^{-1}], \non\\
\exp \{-(1/2)\g_1 \g_2 \arctan (x_1/x_2)\} \non\\
\quad\times \vp\, [(x_1^2 + x_2^2)(x_1^2 +x_2^2 + x_3^2)^{-2}].
\end{array} \right.\non
\end{eqnarray}

In (\ref{2.3.15}) $w_0, \ldots, w_3$ are arbitrary smooth functions
of $ x_0 + x_3; \ \vp = \vp\, (\omega) $ are unknown four-component
functions.

Substitution of Ans\"atze (\ref{2.3.15}) into PDE (\ref{2.3.1}) with
corresponding $f_1,\ f_2$ gives rise to the following systems of ODEs:
\begin{eqnarray} & &i \g_1 \dot \vp = f_1 (\bar\vp \vp) \vp, \non\\ & & (i/2)
\omega^{-1/2} \g_2 \vp + 2i\omega^{1/2} \g_2 \dot \vp = f_1 (\bar\vp
\vp) \vp, \non\\ & & i \g_1 \dot \vp = \lbd (\bar\vp \vp)^{1/2} \vp,
\non\\ & & (i/2) \omega^{-1/2} \g_2 \vp + 2i\omega^{1/2} \g_2 \dot \vp
= \lbd (\bar\vp \vp)^{1/2} \vp, \non\\ 
& &
i \g_0 \dot \vp = - \lbd (\bar\vp \vp)^{1/2} \vp,
\label{2.3.16}\\
& &
i \g_1 \dot \vp = - \lbd (\bar\vp \vp)^{1/2} \vp,
\non\\
& &
(i/2) \omega^{-1/2} \g_2 \vp + 2i\omega^{1/2} \g_2 \dot \vp =
- \lbd (\bar\vp \vp)^{1/2} \vp,
\non\\
& &
i \g_1 \dot \vp = - \lbd (\bar\vp \vp)^{1/2} \vp,
\non\\
& &
(i/2) \omega^{-1/2} \g_2 \vp + 2i\omega^{1/2} \g_2 \dot \vp =
- \lbd (\bar\vp \vp)^{1/2} \vp.\non
\end{eqnarray}

From Theorem 2.3.4 it follows that the Dirac equation (\ref{1.1.1}) is
conditi\-onally-invariant under an infinite-parameter Lie
group\index{Infinite-parameter Lie group}.  As established in
\cite{100,103} a broad class of Poincar\'e-invariant equations (the
Bhabha-type equations)\index{Bhabha equation}
\begin{equation}
(i \beta_s \p_s - m) \Psi (x) = 0,
\quad m = \mbox{\rm const}
\label{2.3.17}
\end{equation}
possess such a property.

In (\ref{2.3.17}) $ \Psi = (\Psi^1, \Psi^2, \ldots, \Psi^n)^T;$\ 
$x = (x_0, x_1, \ldots, x_N)$,\  $ N \ge 2;$\ $\beta_0, 
\ \beta_1, \ldots, \linebreak 
\beta_N $ are $(n\times n)$-matrices satisfying the conditions
\begin{equation}
[\beta_s,\, S_{\tau \rho}] = (g_{s \tau} \beta_\rho -
g_{s \rho} \beta_\tau),
\label{2.3.18}
\end{equation}
where $S_{\tau \rho} = (\beta_\tau \beta_\rho - \beta_\rho
\beta_\tau), \ g_{s \tau} = {\rm diag}\, (1, -1, -1, \ldots, -1)$.

It is well-known that the Bhabha equation is invariant
under the Poincar\'e group $P(1,N)$ having the generators \cite{22}
\begin{displaymath}
P_\tau = g_{\tau \rho} \p_{x_\rho},\quad
J_{\tau \rho} = x_\tau P_\rho - x_\rho P_\tau + S_{\tau \rho}.
\end{displaymath}

Imposing an additional condition $(\p_{x_0} - \p_{x_N}) \Psi(x) = 0$ 
on $ \Psi(x) $ we get the following system of PDEs for $\Psi(\omega) =
\Psi (x_0 + x_N,\, x_1, \ldots, x_{N-1})$:
\begin{equation}
\Biggl\{i (\beta_0 + \beta_N)\p_{\omega_0} +
\sum\limits^{N-1}_{j=1} \beta_j \p_{\omega_j} - m\Biggr\} \Psi(\omega)
= 0. 
\label{2.3.19}
\end{equation}
{\bf Theorem 2.3.5.}\ {\em Equation (\ref{2.3.19}) is invariant under
  the infinite-parameter Lie group having the generators
\begin{equation}
\begin{array}{l}
Q_1 = \p_{\omega_0}, \quad Q_{jk} = - \omega_j 
\p_{\omega_k} + \omega_k \p_{\omega_j} + S_{jk},\\[2mm]
Q_2 = \sum\limits^{N-1}_{k=1}\{W_k (\omega_0) \p_{\omega_k} -
\dot W_k (\omega_0) (S_{0k} - S_{kN})\},
\end{array}
\label{2.3.20}
\end{equation}
where $W_1, \ W_2, \ldots, W_{N - 1} $ are arbitrary smooth
functions}, 
$ \dot W_k = dW_k/d\omega_0$,\ $j$, $k = {1,\ldots, N-1}$.
\vspace{1.5mm}

\noindent
{\em Proof.}$\quad$ It is evident that the operators $Q_1, \ Q_{jk}$
belong to the invariance algebra of equation (\ref{2.3.19}). Let us
prove that the operator $Q_2$ commutes with the operator
of equation (\ref{2.3.20})
\begin{displaymath}
L = i(\beta_0 + \beta_N) \p_{\omega_0} +
i \sum\limits^{N-1}_{j=1} \beta_j \p_{\omega_j} - m.
\end{displaymath}

Computing the commutator $[L, Q ] $ we have
\begin{eqnarray*}
[L,Q] &=& i \sum\limits^{N-1}_{k=1} \Bigl\{- \ddot W_k (\beta_0
+ \beta_N) (S_{0k} - S_{kN})\\ 
& &- \dot W_k [\beta_0 + \beta_N,\, S_{0k} - S_{kN}]
\p_{\omega_0} \Bigr\}.
\end{eqnarray*}

Resulting from relations (\ref{2.3.18}), the equalities
\begin{eqnarray*}
& &[\beta_0 + \beta_N,\, S_{0k} - S_{kN}] = 0,\\
& &(\beta_0 + \beta_N) (S_{0k} - S_{kN}) =
(\beta_0 + \beta_N)
(\beta_0 \beta_k  - \beta_k \beta_0 - \beta_k \beta_N \\
& &\quad
+\beta_N \beta_k) =
(\beta_0 \beta_0 \beta_k - \beta_0 \beta_k \beta_0) +
(\beta_N \beta_N \beta_k -
\beta_N \beta_k \beta_N) \\
& &\quad
+(\beta_0 \beta_N \beta_k -
\beta_N \beta_k \beta_0) 
+(\beta_N \beta_0 \beta_k -
\beta_0 \beta_k \beta_N)\\
& &\quad = \beta_k -  \beta_k = 0, \ \ k = {1,\ldots, N-1}
\end{eqnarray*}
hold, whence it follows that $[L, Q] = 0$. The theorem is proved. $\rhd$

$\wid P(1,3)$-invariant Ans\"atze for the spinor field $ \psi =
\psi(x) $ (\ref{2.2.8}) obtained in the previous section reduce a
$ \wid P(1,3)$-invariant spinor equation
\begin{displaymath}
i \g_\mu \p_\mu \psi - (\bar\psi \psi)^{1/2k}
\Bigl \{ \ti f_1\Bigl(\bar\psi \psi
(\bar\psi \g_4 \psi)^{-1}\Bigr) + \ti f_2 \Bigl(\bar\psi \psi 
(\bar\psi \g_4 \psi)^{-1}\Bigr) \g_4 \Bigr\}\psi = 0
\end{displaymath}
to systems of ODEs of the form
\begin{eqnarray}
&1)& 2i \g_3 \dot \vp + (i/4)(\g_0 + \g_3) (\g_0 \g_3 - 2k) \vp = R,
\non\\
&
2)& i (\g_0 - 2 \g_2 - \g_3) \dot \vp + (i /2)\g_2(\g_0
\g_3 - 2k) \vp = R,
\non\\
&
3)& 2 i \g_3 \dot \vp + (i /4\alpha) (\g_0 + \g_3)
(\alpha\g_0 \g_3 - \g_1 \g_2 - 2k\alpha) \vp = R,
\non\\
&
4)& (i/2) (\g_0 - \g_3 - 2 \g_1 + 2 \al \g_2) \dot \vp +
(i/2) (1 - 2k + \g_0 \g_3) \vp = R,
\non\\
&
5)& (i/2)(\g_0 - \g_3 + 2 \g_2) \dot \vp + (i/2) \g_1 (1 -
2k + \g_0 \g_3) \vp = R,
\non\\
&
6)& i\omega (4 \omega \g_1 + \g_2) \dot \vp + (1/4) \g_2 (\g_0 \g_3 -
4k) \vp = R,
\non\\
&
7)& - i\omega \Bigl(12 \g_1 + \omega^{1/2} (15 \g_0 + 9 \g_3)\Bigl) 
\dot \vp - i \g_1 (\g_0 \g_3 - 4k ) \vp = R,
\non\\
&
8)&  i (\g_0 - \g_3) \dot \vp + (i /2\omega)\Bigl(\g_0 - \g_3
+ (\g_1 - \omega\g_2) 
\non\\
& &
\times (\g_0 \g_3 - 2k )\Bigr) \vp = R,
\non\\
&
9)& 2i \g_0 \dot \vp + (i /4)(\g_0 - \g_3) (2 - 2k -
\g_0 \g_3) \vp = R,
\non\\
&
10)& i (\g_0 - \g_3) \dot \vp + (i/2) (\g_0 - \g_3) \Bigl(\omega^{-1}
+ (\omega + 1)^{-1}\Bigr) \vp + (i/4)
\non\\
& &
\times \Bigl(( \g_0 + \g_3) (1 + \omega) + (\g_0 - \g_3) (1 +
\omega)^{-1}\Bigr) (\g_0 \g_3 - 2k) \vp = R,
\non\\
&
11)& 2i \g_0 \dot \vp + (i/4) (\g_0 - \g_3) (4 -
2k - \g_0 \g_3) \vp = R,
\non\\
&
12)&  2i \g_0 \dot \vp + (i/4\alpha) (\g_0 - \g_3)
\Bigl((4 - 2k)\alpha -\alpha\g_0 \g_3 + \g_1 \g_2\Bigr) \vp = R,
\non\\
&
13)& i (\alpha \g_2 - \g_1) \dot \vp + (i/2)(1 - 2k) \g_1  \vp = R,
\non\\
&
14)& i (\g_0 - \g_3 \omega) \dot \vp + (i/2\alpha) \g_3 (\g_1 \g_2
- 2k\alpha) \vp = R,
\non\\
&
15)&  i (\g_1 - \g_2 \omega) \dot \vp - (i/2\alpha) \g_2 (2k\alpha +
\g_0 \g_3) \vp = R,
\non\\
&
16)& (i/\alpha) \Bigl(\alpha(\g_0 - \g_3) - (\alpha + 1)\omega
\g_2\Bigr) \dot \vp - (i/2\alpha) \g_2 (2k\alpha + \g_0 \g_3) \vp = R,
\non\\
&
17)&  i\omega^2 (\g_0 + \alpha \g_3) \dot \vp + (i/2\alpha)\omega 
(\g_0 + 2k \alpha\g_3 ) \vp = R,
\non\\
&
18)& i (\beta \g_2 - \g_1) \dot \vp + (i/2\beta) \g_1 \Bigl((1 -
2k)\beta - \al \g_0 \g_3\Bigr) \vp = R,
\non\\
&
19)& i\omega^{(\al + 1)/ \al} (\al \g_0 + \beta \g_3)
\dot \vp + (i /2 \al) \omega^{1/ \al} \g_3
( \al \g_0 \g_3 - \g_1 \g_2 
\non\\
& & + 2 \beta k ) \vp = R,
\non\\
&
20)& i (\g_0 - \g_3 + \alpha \g_1) \dot \vp + (i/2) (1
- 2k) (\g_0 - \g_3) \vp = R,
\non\\
&
21)& i (\g_0 - \g_3 - \omega \g_2 ) \dot \vp - ik \g_2 \vp = R,
\non\\
&
22)& (i/\alpha) \Bigl(\alpha(\g_2 - \beta \g_1) 
+ \beta (\g_3 - \g_0) \Bigr) \dot \vp 
-(i/2\beta) (\g_0 - \g_3) (2k\beta + \g_1 \g_2) \vp \non\\
& &= R,
\non\\
&
23)& i (\al \g_1 + \beta \g_2 + \g_0 - \g_3) \dot \vp +
(i/\omega) (\g_0 - \g_3) \vp = R,
\non\\
&
24)& (i/\al) \Bigl(\al(\g_0 - \g_3) - (\al + 1)\omega
\g_2\Bigr) \dot \vp + (i/2\omega)( \g_0 - \g_3 ) \vp
\non\\
& &
- (i/2 \al) \g_2 (\g_0 \g_3 +
2 k\al ) \vp = R,
\non\\
&
25)& (i/2\al)\Bigl((\al - 1)(\g_0 -
\g_3) -(\alpha + 1) \omega^2(\g_0 + \g_3)\Bigr)\dot\vp +
(i/2\omega)(\g_0 - \g_3) \vp \non\\
& &
- (i/4\al\omega)\Bigl((\g_0 - \g_3) 
+ (\g_0 + \g_3)\omega^2\Bigr) (\g_0 \g_3 + 2k\al) \vp = R,
\non\\
&
26)& i\Bigl((\beta + 1) \g_1 - \beta (\g_0 - \g_3) -
\al \g_2\Bigr) \dot \vp + (i/2) \g_1 (1 - 2k + \g_0 \g_3) \vp
\non\\
& &
+ (i/2) (\g_0 - \g_3) \vp = R,
\label{2.3.21}\\
&27)& (i/2)(1-\omega)^{-1}(\omega^{-1/2}\g_0-\g_2)\dot\varphi
+i\Bigl((1/2)(\omega^{-1/2}\g_0+\g_2) \non\\
&
&+k(\omega-1)^{-1}(\omega^{1/2}\g_0-\g_1)\Bigr)\varphi
=R(\omega-1)^{-1/2}, \non\\ 
&28)&i (\g_1-\omega\g_2)\dot\varphi-ik(\omega^2+1)^{-1}
(\omega\g_1+\g_2)\varphi =R(\omega^2+1)^{-1/2}, \non\\
&29)& (i\g_0-\omega\g_3)\dot\varphi-ik(\omega^2-1)^{-1}
(\omega\g_0-\g_3)=R(\omega^2-1)^{-1/2},\non\\
&30)& i \Bigl( \g_2 - \omega(\g_0-\g_3)\Bigr )\dot \varphi
-ik(\g_0-\g_3)\varphi=R, \non\\
&31)& i(\omega\g_1-\omega^2\g_3)\dot \varphi+(i/2)(1-2k)\g_1\varphi=R,
\non\\ 
&32)& i(\omega\g_1-\omega^2\g_0)\dot \varphi+(i/2)(1-2k)\g_1\varphi=R,
\non\\ 
&33)& i\Bigl(\omega\g_1-\omega^2(\g_0-\g_3)\Bigr)\dot \varphi
+(i/2)(1-2k)\g_1\varphi=R, \non\\
&34)& i(\omega\g_0-\omega^2\g_2\Bigr)\dot \varphi
+(i/2)(1-2k)\g_1\varphi=R, \non\\
&35)& i(\gamma_2-\omega\gamma_1)\dot \varphi+(i/2)(\gamma_0-\gamma_3
+\gamma_2\gamma_4-2k\gamma_1)\varphi=R, \non\\
&36)& i(\gamma_1+\alpha \gamma_2-\gamma_0+\gamma_3)\dot \varphi
+(i/2)\Bigl(\gamma_0-\gamma_3+\gamma_2\gamma_4 
+(1-2k)\gamma_1\Bigr)\varphi=R, \non\\
&37)& i\Bigl(\gamma_2-\omega(\gamma_0-\gamma_3)\Bigr)\dot \varphi
+(i/2)(1-2k)(\gamma_0-\gamma_3)\varphi=R, \non\\
&38)& i\Bigl(\gamma_0+\gamma_3-\omega(\gamma_0-\gamma_3)\Bigr)\dot
\varphi 
+(i/2)(1-2k)(\gamma_0-\gamma_3)\varphi=R, \non\\
&39)& i\Bigl(\gamma_0+\gamma_3-\omega(\gamma_0-\gamma_3)\Bigr)\dot
\varphi 
+i(1-k)(\gamma_0-\gamma_3)\varphi=R, \non\\
&40)& i\Bigl(\omega(\gamma_0-2\gamma_2-\gamma_3)+\gamma_0
+\gamma_3\Bigr)\dot \varphi+(i/2)\Bigl(2(\gamma_0
-\gamma_3)-\gamma_1\gamma_4-2k\gamma_2\Bigr)\varphi \non\\
& &=R, \non\\
&41)& i\Bigl((1-\omega)(\gamma_0-\gamma_3)+\gamma_2\Bigr)\dot
\varphi+(i/2)(1-2k)(\gamma_0-\gamma_3)\varphi=R, \non\\
&42)& i\omega\Bigl((1-\alpha)\omega^{1/2\alpha}(\gamma_0-\gamma_3)+
(1+\alpha)
\omega^{-1/2\alpha}(\gamma_0+\gamma_3)\Bigr)\dot \varphi \non\\
& &+(i/4\alpha)\Bigl[\Bigl(1+2\alpha(2-k)\Bigr)
\omega^{1/2\alpha}(\gamma_0-\gamma_3)-(1+2k\alpha)\omega^{-1/2\alpha}
\non\\ 
& &\times(\gamma_0+\gamma_3)\Bigr]\varphi=R, \non\\
&43)& i\Bigl(\gamma_0+\gamma_3-\omega(\gamma_0-\gamma_3)\Bigr)\dot
\varphi +(i/2\alpha)\Bigl(2\alpha(1-k)+\gamma_4\Bigr)
(\gamma_0-\gamma_3)\varphi =R, \non\\
&44)& i\omega\Bigl((\alpha-\beta)\omega^{1/2\beta}(\gamma_0 -
\gamma_3)+(\alpha+\beta)\omega^{-1/2\beta}(\gamma_0 +
\gamma_3)\Bigr)\dot \varphi \non\\ 
& &+(i/4\beta)\Bigl[\Bigl(\alpha+4\beta(1-k)-\gamma_4\Bigr)
\omega^{1/2\beta}(\gamma_0-\gamma_3) \non\\
&
&-(\alpha+4k\beta-\gamma_4)\omega^{-1/2\beta}(\gamma_0 +
\gamma_3)\Bigr]\varphi=R,\non 
\end{eqnarray}
where $ R = (\bar\vp \vp)^{1/2k} \{\ti f_1 (\bar\vp \vp
(\bar\vp \g_4 \vp)^{-1}) +
\ti f_2 (\bar\vp \vp (\bar\vp \g_4 \vp )^{-1}) \g_4 \} \vp$.

At last, Ans\"atze (\ref{2.2.30}) invariant under $C(1,3)$
non-conjugate three-dimen\-sional subalgebras of the algebra
$AC(1,3)$ listed in (2.2.29) after being substituted into
a conformally-invariant spinor equation
\begin{displaymath}
i \g_\mu \p_\mu \psi - (\bar\psi \psi)^{1/3}
\Bigl\{ \ti f_1 \Bigl(\bar\psi \psi (\bar\psi \g_4 \psi)^{-1}\Bigr) +
\ti f_2 \Bigl(\bar\psi \psi (\bar\psi \g_4 \psi)^{-1}\Bigr) 
\g_4 \Bigr\} \psi = 0
\end{displaymath}
give rise to the following systems of ODEs for
$ \vp = \vp (\omega)$ :
\begin{eqnarray}
&1)& i \Bigl(-(3/4) (\omega^2  + 4)(\g_0 - \g_3) + \g_0 + \g_3 +
\omega \g_1 + 2 \g_2\Bigr)\dot \vp  
\non\\
&
&
+ i\Bigl(\g_1 - \omega (\g_0 - \g_3) + (1/2) \g_1 \g_2 (\g_0 -
\g_3)\Bigr) \vp = R, 
\non\\
&
2)& i\Bigl(\g_1 \cos \omega - \g_2 \sin \omega - \al (\g_0 -
\g_3)\Bigr) \dot \vp - (i/2) (3 - \g_0 \g_3) 
\non\\
&
&
\times(\g_1 \sin \omega + \g_2 \cos \omega) \vp 
- (i \al /2) (\g_0 - \g_3) \g_1 \g_2 \vp = R,
\non\\
&
3)& i (\g_2 - \g_0 + \g_3) \dot \vp = R,
\non\\
&
4)& i \Bigl( \g_1 + \al (\g_0 - \g_3) e^{-\omega}\Bigr) \dot \vp + 2i
\g_1 \vp = R e^{\omega/3},
\non\\
&
5)& i \Bigl( \g_1 + \al \g_2 - \beta (\g_0 - \g_3) e^\omega\Bigr) \dot
\vp - (i/2) \Bigl(3 \g_1 
\label{2.3.22}\\
&
&
+ 2 \al \g_2 (1 + \g_0 \g_3)\Bigr) \vp = R e^{-\omega/3},
\non\\
&
6)& i (\omega^2 \g_2 - \omega\g_1) \dot \vp 
- i  \omega (1 + \omega^2)^{-1} \g_2 (1 +\g_0 \g_3)\vp 
+ (i/2) (\omega^2 + 3) 
\non\\
&
&
\times(\omega^2 + 1)^{-1}\g_1  \vp - (i/2) 
(3\omega^2 + 1)(\omega^2 + 1)^{-1} \g_2 \g_4 \vp = R,
\non\\
&
7)& i \Bigl( (\omega^2 + 1)^{1/4}(\g_0 + \g_3) + 
(\omega^2 + 1)^{-1/4}(\g_0 - \g_3) - 2  \omega\g_2\Bigr) \dot \vp 
\non\\
&
&
+  i \Bigl( \omega (\omega^2 + 1)^{-3/4}(\g_0 + \g_3) 
- (\omega^2 + 1)^{-1}(1 + \g_0 \g_3)\g_2   
\non\\
&
&
-(3/2) \g_2\Bigr) \vp = R (\omega^2 + 1)^{1/4},
\non\\
&
8)& 2i  \omega\g_2 \dot \vp - (3i/2) \g_2 \vp = R \omega^{-1/6},
\non\\
&
9)& - i \g_1 \dot \vp = R,
\non\\
&
10)& - 2i \omega (\omega - 4) \g_2 \dot \vp - i \Bigl( (1/2)
\omega^{1/2} (\omega - 4)^{1/2} +
\omega\ - 2 \Bigr) \g_2 \vp 
\non\\
&
&
= R \omega^{1/2} (\omega - 4)^{1/2} \Bigl((1/2) \Bigl[\omega^{1/2} 
+ (\omega - 4)^{1/2}\Bigr]\Bigr)^{1/3},\non
\end{eqnarray}
where $ R = (\bar\vp \vp)^{1/3} \Bigl\{ \ti f_1 \Bigl(\bar\vp \vp
(\bar\vp \g_4 \vp)^{-1}\Bigr) +
\ti f_2 \Bigl(\bar\vp \vp (\ov \g_4 \vp)^{-1}\Bigr) \g_4 \Bigr\} \vp$.
\vspace{10mm}

\noindent
{\large\bf 2.4. Exact solutions of nonlinear spinor
  equations\label{s2.4}} 

\markboth{Chapter 2. EXACT SOLUTIONS}
{2.4. Exact solutions of nonlinear spinor equations}
\def\theequation{2.\arabic{section}.\arabic{equation}}
\setcounter {section} {4}
\setcounter {equation}{0}
\vspace{7mm}

\noindent
Using the results obtained in Sections 2.2, 2.3 we will
construct in explicit form multi-parameter families of
exact solutions of the following systems of nonlinear PDEs:
\begin{eqnarray}
& &\{ i \g_\mu \p_\mu - \lbd (\bar\psi \psi)^{1/2k}\} \psi = 0,
\label{2.4.1}\\[2mm]
& &\{i \g_\mu \p_\mu - m - \lbd (\bar\psi \psi)^k \} \psi = 0,
\label{2.4.2}
\end{eqnarray}
which are obtained from (\ref{2.3.1}) by putting
$f_1 = \lbd (\bar\psi \psi)^{1/2k}$,
\ $ f_2 = 0$ and $ f_1 = m + \lbd (\bar\psi \psi)^k$,
\ $f_2 = 0$, respectively.

In (\ref{2.4.1}), (\ref{2.4.2}) $m,\ \lbd,\ k $ are real constants,
$ m \not= 0, \ \ k \not= 0$.

Equation (\ref{2.4.1}) with $k = 1/2$ was considered by Heisenberg
\cite{116}--\cite{120} (see also \cite{121}) and equation
(\ref{2.4.2}) with $k = 1 $ was suggested by Ivanenko
\index{Dirac-Ivanenko equation} as a possible
basic model for the unified field theory \cite{129}.

According to Theorem 1.2.1 equations (\ref{2.4.1}), (\ref{2.4.2})
are invariant under the Poincar\'e group. In addition, system of PDEs
(\ref{2.4.1}) admits the one-parameter group of scale transformations
(\ref{1.1.24e}). 

To reduce equations (\ref{2.4.1}),  (\ref{2.4.2}) we apply $P(1,3)$-, \
$\wid P(1,3)$- and $C(1,3)$-
\linebreak invariant Ans\"atze constructed in Section 2.2.
\vspace{2mm}

\noindent
{\bf 1. Poincar\'e-invariant solutions of system of PDEs
  (\ref{2.4.1}).} \\ {\bf 1.1.}\ {\em Integration of reduced ODEs}.\ 
Substitution of the $P(1,3)$-invariant Ans\"atze (\ref{2.2.8}) into
(\ref{2.4.1}) gives rise to systems of ODEs (\ref{2.3.5}) with $R=-i
\lbd (\bar\vp \vp)^{1/2k}$ $\times\vp $. When integrating these we
will use essentially the following assertions.  
\vspace{1.5mm}

\noindent
{\bf Lemma 2.4.1.}\ {\em  Solutions of equations
3, 12--14, 20--23 from (\ref{2.3.5})
satisfy the relation $ \bar\vp \vp = 0$. }
\vspace{1.5mm}

\noindent
{\em Proof.}$\quad$ Multiplication of the ODE 3 from (\ref{2.3.5}) by
the matrix $ \g_0 + \g_3 $ on the left yields the following
consistency condition:
\begin{displaymath}
- i \lbd (\bar\vp \vp)^{1/2k} (\g_0 + \g_3) \vp = 0,
\end{displaymath}
whence $ \bar\vp \vp = 0$ or $ (\g_0 + \g_3) \vp = 0$. The general
solution of the algebraic equation $ (\g_0 + \g_3) \vp = 0$ is
represented in the form 
\begin{displaymath}
\vp = (\g_0 + \g_3) \vp_1,
\end{displaymath}
where $ \vp_1 $ is an arbitrary four-component function-column.

Since $ \bar\vp = \{ \vp_1 (\g_0 + \g_3) \}^{\dagger} \g_0
= \bar\vp_1 (\g_0 + \g_4)$,
an identity $ \bar\vp \vp = \bar\vp_1 (\g_0 + \g_4)^2 \vp_1 = 0$
holds. Other equations are treated in the same way. $\rhd$
\vspace{1.5mm}

\noindent
{\bf Lemma 2.4.2.}\ {\em The quantity $ \bar\vp \vp $ 
is the first integral of
systems of ODEs 1, 2, 5, 15, 16, 25 from (\ref{2.3.5}).}

We prove the assertion for the system 1. Multiplying it by 
$ - \g_3 $ yields
\begin{equation}
\dot {\vp} = i \lbd (\bar\vp \vp)^{1/2k} \g_3 \vp.
\label{2.4.3}
\end{equation}

The conjugate spinor\index{Conjugate!spinor} satisfies the 
following equation:
\begin{equation}
\dot{\bar{\vp}} = - i \lbd (\bar\vp \vp)^{1/2k} \bar\vp \g_3.
\label{2.4.4}
\end{equation}

Multiplying (\ref{2.4.3}) by $ \bar\vp$ on the left, (\ref{2.4.4}) by
$ \vp $ on the right and summing the expressions obtained we
arrive at the relation
\begin{displaymath}
\dot{\bar\vp} \vp + \bar \vp \dot \vp = 0,
\end{displaymath}
whence $ d(\bar \vp \vp) / d\om =
\dot{\bar\vp} \vp + \bar \vp \dot \vp = 0$.
The lemma is proved. $\rhd$

Due to Lemma 2.4.1 we conclude that the Ans\"atze
numbered by 3, 12--14, 20--23 give rise to the solutions of
equation (\ref{2.4.1}) which satisfy the condition
$ \bar\psi \psi = \bar\vp \vp = 0$. Consequently, a factor
$ \lbd (\bar\psi \psi)^{1/2k} $ determining the nonlinear
self-coupling of the spinor field $ \psi (x) $ vanishes. Such
solutions are of low interest and are not considered here.

According to Lemma 2.4.2 the system of ODEs 1 from
(\ref{2.3.5}) is equivalent to the linear equation
\begin{equation}
\dot \vp = i \lbd C^{1/2k} \g_3 \vp
\label{2.4.5}
\end{equation}
with a nonlinear additional constraint
\begin{equation}
\bar\vp \vp = C = \mbox{\rm const}.
\label{2.4.6}
\end{equation}

Integrating ODE (\ref{2.4.5}) we get
\begin{equation}
\vp (\om) = \exp \{i \lbd C^{1/2k} \g_3 \om\}\chi,
\quad
\bar\vp (\om) = \bar \chi\, \exp \{ - i \lbd C^{1/2k} \g_3 \om \}.
\label{2.4.7}
\end{equation}

Hereafter $ \chi $ is an arbitrary constant four-component column.

Substitution of expressions (\ref{2.4.7}) into (\ref{2.4.6}) yields
\begin{displaymath}
\bar\chi \exp \{- i \lbd C^{1/2k} \g_3 \om\}
\exp \{i \lbd C^{1/2k} \g_3 \om \} \chi = C,
\end{displaymath}
whence $ \bar\chi \chi = C$. Thus, the general solution of the system 
of nonlinear ODEs 1 from (\ref{2.3.5}) is given by the formula
\begin{displaymath}
\vp (\om) = \exp \{ i \lbd (\bar\chi \chi)^{1/2k} \g_3 \om\} \chi.
\end{displaymath}

The general solutions of equations 2, 5, 15, 16, 25 are constructed
in the same way. As a result, we have
\begin{eqnarray}
\vp (\om) &=& \exp \{- i \lbd (\bar\chi \chi)^{1/2k} \g_0 \om\} \chi,
\non\\
\vp (\om) &=& \exp \{ i \g_2 \Bigl((\bar\chi \chi)^{1/2k} -
(i/2) (\g_0 + \g_3)\Bigr)\om\} \chi,
\non\\
\vp (\om) &=& \exp \{(i\lbd /2) (\bar\chi \chi)^{1/2k}
\g_1 \om\} \chi,
\label{2.4.8}\\
\vp (\om) &=& \exp \{(i\lbd /2) (1 + \al^2)^{-1} (\bar\chi \chi)^{1/2k}
(\g_2 - \al \g_1) \om \} \chi,
\non\\
\vp (\om) &=& \exp \Bigl\{\Bigl[ \g_2 (\g_0 + \g_3) + i \lbd (\bar\chi
\chi)^{1/2k} \Bigl(\g_2 - \beta (\g_0 + \g_3)\Bigr)\Bigr]\om\Bigr\}
\chi.\non 
\end{eqnarray}

To integrate systems of ODEs 6, 9--11 from (\ref{2.3.5}) we
will use their symmetry properties. As established in 
Section 2.3 the equation 6 is invariant under the Lie algebra
with the basis elements $ \p_\om, \ \g_0 \g_3$. We seek for a
solution which is invariant under the one-dimensional subalgebra
of this algebra $\langle \p_\om - \theta \g_0 \g_3\rangle,
\ \theta \in {\R}^1$.

In other words, a four-component function $ \vp = \vp (\om) $
has to satisfy the additional constraint
\begin{displaymath}
Q \vp = (\p_\om - \theta \g_0 \g_3) \vp = 0.
\end{displaymath}

The general solution of the above equation reads
\begin{equation}
\vp (\om) = \exp \{ \theta \g_0 \g_3 \om \} \chi^\prime,
\label{2.4.9}
\end{equation}
where $ \chi^\prime $ is an arbitrary constant four-component column.
Substituting (\ref{2.4.9}) into the system of ODEs 6 from
(\ref{2.3.5}) we have
\begin{displaymath}
\Bigl(\theta \g_1 \g_0 \g_3 - (1/2\al)\g_1 \g_4\Bigr) \exp
\{ \theta \g_0 \g_3 \om \} \chi^\prime 
= - i \lbd \tau \exp \{ \theta \g_0 \g_3 \om \}\chi^\prime,
\end{displaymath}
where $ \tau = (\bar{\chi^\prime} \chi^\prime)^{1/2k}$.

Multiplying both parts of the above equality by $\exp \{- \theta \g_0
\g_3 \om\} $ on the left we arrive at the system of algebraic
equations for $ \chi^\prime $
\begin{equation}
\Bigl\{\Bigl(\theta \g_2 - (1/2 \al)\g_1\Bigr) \g_4 + i \lbd
\tau\Bigr\} \chi^\prime = 0. 
\label{2.4.10}
\end{equation}

Consequently, substitution (\ref{2.4.9}) reduces the system 6 to
algebraic equations (\ref{2.4.10}). Making in (\ref{2.4.10}) the
transformation
\begin{displaymath}
\chi^\prime = \Bigl([\theta \g_2 - (1/2\al) \g_1] \g_4 - i \lbd
\tau\Bigr) \chi
\end{displaymath}
yields
\begin{displaymath}
[\lbd^2 \tau^2 - \theta^2 - (2 \al)^{-2}] \chi = 0.
\end{displaymath}

As $\chi \not= 0$, the equality
\begin{equation}
\theta = (\ve/2\al) (4 \lbd^2 \tau^2 \al^2 -1)^{1/2},
\quad
\ve = \pm 1
\label{2.4.11}
\end{equation}
has to be satisfied.

The condition $ \tau = (\bar{\chi^\prime} \chi^\prime)^{1/2k} $ gives
rise to the nonlinear algebraic equation for $ \tau $
\begin{equation}
\tau^{2k} = 2 \lbd^2 \tau^2 (\bar\chi \chi) +
2 i \lbd \tau \theta ( \bar\chi \g_2 \g_4 \chi) 
- i \lbd \tau \al^{-1} (\bar\chi \g_1 \g_4 \chi).
\label{2.4.12}
\end{equation}

Thus, we have constructed a particular solution of the system of ODEs
6 from (\ref{2.3.5})
\begin{displaymath}
\vp (\om) = \exp \{\theta \g_0 \g_3 \om \} \Bigl([\theta \g_2
- (1/2\al)\g_1] \g_4 - i \lbd \tau\Bigr) \chi,
\end{displaymath}
where $ \theta,\ \tau $ are  determined by (\ref{2.4.11}),
(\ref{2.4.12}). 

Particular solutions of systems of ODEs 9--11 from (\ref{2.3.5}) are 
obtained in an analogous way
\begin{equation}
\vp (\om) = \exp \{ \theta \g_1 \g_2 \om\}
\Bigl([\theta \g_0 - (1/2\al) \g_3 ] \g_4 - i \lbd \tau\Bigr) \chi,
\label{2.4.13}
\end{equation}
parameters $ \theta, \ \tau $ being defined by the formulae
\begin{equation}
\begin{array}{l}
\tau^{2k} = 2 \lbd^2 \tau^2 (\bar\chi \chi) + 2i \lbd
\tau \theta (\bar\chi \g_0 \g_4 \chi ) - i \lbd
\tau \al^{-1} (\bar\chi \g_3 \g_4 \chi ),\\[1mm]
\theta = (\ve/2\al) (1 - 4 \al^2 \lbd^2 \tau^2)^{1/2};
\end{array}
\label{2.4.14}
\end{equation}
\begin{equation}
\vp (\om) = \exp \{ \theta \g_1 \g_2 \om \}
\Bigl([ \theta \g_3 + (1/2 \al)\g_0] \g_4 -
i \lbd \tau\Bigr) \chi,
\label{2.4.15}
\end{equation}
parameters $ \theta, \ \tau$ being defined by the formulae
\begin{equation}
\begin{array}{l}
\tau^{2k} = 2 \lbd^2 \tau^2 (\bar\chi \chi) +
2 i \lbd \tau \theta ( \bar\chi \g_3 \g_4 \chi)
+i \lbd \tau \al^{-1} (\bar\chi \g_0 \g_4 \chi),\\[1mm]
\theta = (\ve/2\al) (1 + 4 \al^2 \lbd^2 \theta^2)^{1/2} ;
\end{array}
\label{2.4.16}
\end{equation}
\begin{equation}
\vp (\om) = \exp \{ \theta \g_1 \g_2 \om\} \Bigl(4 \theta
( \g_0 + \g_3) \g_4 
+( \g_0 - \g_3) \g_4 - 4 i \lbd \tau\Bigr) \chi,
\label{2.4.17}
\end{equation}
parameters $ \theta, \ \tau$ being defined by the formulae
\begin{equation}
\begin{array}{l}
\tau^{2k} = 32 \lbd^2 \tau^2 (\bar\chi \chi) - 8i
\lbd \tau [ \bar\chi ( \g_0 - \g_3) \chi]\\[1mm]
\quad 
- 32 i \lbd^3 \tau^3[ \bar\chi ( \g_0 + \g_3) \g_4 \chi],\quad
\theta = - \lbd^2 \tau^2.
\label{2.4.18}
\end{array}
\end{equation}

Equation 8 from (\ref{2.3.5}) by virtue of the change of variables
\begin{displaymath}
 \vp (\om) = \om^{-1/4} \phi (\om),
\end{displaymath}
where $ \phi (\om) $ is a new unknown four-component function, is
reduced to the following system of ODEs:
\begin{displaymath}
2 i \om^{1/2} \g_2 \dot \phi = \lbd \om^{-1/4k}
( \bar\phi \phi)^{1/2k} \phi.
\end{displaymath}

Multiplying both parts of the above equality by $ (i/2) \g_2
\om^{-1/2} $ we come to the equation
\begin{equation}
\dot \phi = (i \lbd /2) \om^{-(1+2k)/4k} \g_2 ( \ov
\phi \phi)^{1/2k} \phi,
\label{2.4.19a}
\end{equation}
the conjugate spinor satisfying the following equation:
\begin{equation}
\dot {\bar\phi} = -(i \lbd /2) \om^{-(1 + 2k)/4k} \g_2
( \bar\phi \phi)^{1/2k} \phi.
\label{2.4.19b}
\end{equation}

Multiplying (\ref{2.4.19a}) by $ \bar\phi$ on the left,
(\ref{2.4.19b}) by $ \phi $ on the right and summing the equalities
obtained we get 
\begin{displaymath}
\dot{\bar\phi} \phi + \bar\phi \dot\phi = 0,
\end{displaymath}
whence $ \bar\phi \phi = C =\mbox{\rm const}$. Consequently, equation
(\ref{2.4.19a}) is equivalent to the linear ODE 
\begin{equation}
\dot \phi = (i \lbd/2)\om^{-(2k+1)/4k} C^{1/2k} \g_2\phi
\label{2.4.19c}
\end{equation}
which is supplemented by the additional constraint $ \bar\phi \phi =
C$. 

Integration of (\ref{2.4.19c}) yields
\begin{eqnarray*}
k \not= 1/2,\quad
\phi(\om) &=& \exp \{2i \lbd k (1 - 2k)^{-1} C^{1/2k} \g_2
\om^{(2k-1)/4k} \} \chi,\\
k = 1/2,\quad
\phi (\om) &=& \exp \{(i \lbd /2) C \g_2 \ln \om\}.
\end{eqnarray*}

Since $ C = \bar\phi \phi = \bar\chi \chi $, 
the general solution of the initial
equation 8 is given by the formulae
\begin{eqnarray}
k \not= 1/2,
\quad
\vp(\om) &=& \om^{-1/4}
\exp \{ 2i \lbd k (1 - 2k)^{-1}(\bar\chi \chi)^{1/2k}\non\\
& &\times \g_2 \om^{(2k-1)/4k} \} \chi,\label{2.4.19d}\\
k = 1/2,
\quad
\vp(\om) &=& \om^{-1/4} \exp \{(i \lbd/2) ( \bar\chi \chi)
\g_2 \ln \om \} \chi.\non
\end{eqnarray}

To integrate the system of ODEs 19 from (\ref{2.3.5}) we make the
change of variables $ \vp (\om) = \om^{-1/4} \phi (\om) $ 
transforming it to the form
\begin{equation}
2 \om^{1/2} \g_2 \dot \phi + (1/2) ( \g_0 + \g_3) \phi =
- i \lbd \om^{-1/4k}( \bar\phi \phi)^{1/2k} \phi.
\label{2.4.20a}
\end{equation}

Solutions of the above system of ODEs satisfy the condition
$ \bar\phi \phi = C =\mbox{\rm const} $, whence it follows
that equation (\ref{2.4.20a}) is linearized
\begin{equation}
2\om^{1/2} \g_2 \dot \phi + (1/2) ( \g_0 + \g_3) \phi =
-i \lbd C^{1/2k} \om^{-1/4k} \phi.
\label{2.4.20b}
\end{equation}

A general solution of (\ref{2.4.20b}) is looked for in the form
\begin{equation}
\begin{array}{rcl}
\phi (\om) &=& \bigl\{ f_1(\om) + \g_2 f_2 (\om) 
+ ( \g_0 + \g_3)f_3(\om)\\[2mm]
& &+\g_2 (\g_0 + \g_3) f_4 (\om) \bigr\} \chi,
\label{2.4.21}
\end{array}
\end{equation}
where $ f_i (\om) $ are some real-valued scalar functions.

Substituting (\ref{2.4.21}) into (\ref{2.4.20b}) we arrive at the
following system of four linear ODEs:
\begin{eqnarray*}
& &2\om^{1/2} \dot f_1 = - i \lbd C^{1/2k} \om^{-1/4k} f_2,\\
& &2\om^{1/2} \dot f_2 = i \lbd C^{1/2k} \om^{-1/4k} f_1,\\
& &2\om^{1/2} \dot f_3 = (1/2) f_2 - i \lbd C^{1/2k} \om^{-1/4k}
f_4,\\ 
& &2\om^{1/2} \dot f_4 = (1/2) f_1 + i \lbd C^{1/2k} \om^{-1/4k} f_3.
\end{eqnarray*}

Integration of the above system is carried out by standard
methods. As a result, we have
\begin{eqnarray}
f_1 &=& \cosh f(\om), \quad f_2 \ \; =\ \; i\sinh f(\om),\non\\
f_3 &=& (i/4)\Biggl\{\cosh f(\om) \int\limits^{\edi\om} z^{-1/2}
\sinh [2f(z)] dz \non\\
& &-\sinh f(\om) \int\limits^{\edi\om} z^{-1/2}
\cosh [2f(z)] dz\Biggr\},
\label{2.4.22}\\
f_4 &=& (1/4) \Biggl\{\cosh f(z) \int\limits^{\edi\om} z^{-1/2}
\cosh [2f(z)] dz \non\\
& &- \sinh f(\om) \int\limits^{\edi\om} z^{-1/2}
\sinh [2f(z)] dz \Biggr\},\non
\end{eqnarray}
where
\begin{equation}
f(\om) = \left\{
\begin{array}{ll}(\lbd C/2) \ln \om, & k = 1/2, \\[2mm]
2 \lbd k (1 - 2k)^{-1}C^{1/2k} \om^{(2k-1)/4k}, & k \not= 1/2.
\end{array}\right.
\label{2.4.23}
\end{equation}

From (\ref{2.4.22}), (\ref{2.4.23}) it follows that $ \bar\phi \phi =
\bar\chi \chi $, whence we conclude that $C = \bar\chi \chi $. Thus,
the general solution of the system of ODEs 19 from (\ref{2.3.5}) is
given by the formula 
\begin{displaymath}
\vp (\om) = \om^{-1/4}
\{f_1 + \g_2 f_2 + (\g_0 + \g_3) f_3 + \g_2 (\g_0 + \g_3) f_4 \}\chi,
\end{displaymath}
functions $ f_1 (\om), \ldots, f_2 (\om) $ 
being defined by (\ref{2.4.22}), (\ref{2.4.23})
with $C = \bar\chi \chi $.

In addition, we have succeeded in integrating the systems of ODEs
num\-be\-red by 4, 24, 27 (with $\al = 0$). These systems can be
written as follows:
\begin{displaymath}
(N/2)(\g_0 + \g_3) \vp + \Bigl( \om(\g_0 + \g_3) 
+ \g_0  - \g_3\Bigr) \dot \vp =
- i \lbd ( \bar\vp \vp)^{1/2k} \vp,
\end{displaymath}
where cases $ N = 1, 2, 3$ correspond to equations $4, 24, 27$ (with
$\alpha=0$) from (\ref{2.3.5}).

Multiplying both parts of the above equality by the matrix
$\om(\g_0 + \g_3) + \g_0 - \g_3 $ on the left yields
\begin{equation}
4\om \dot \vp = - \Bigl\{ N(1 + \g_0 \g_3) + i \lbd
( \bar\vp \vp)^{1/2k}
\Bigl(\om (\g_0 + \g_3) + \g_0 - \g_3\Bigr) \Bigr\} \vp,
\label{2.4.24a}
\end{equation}
the equation for the conjugate spinor taking the form
\begin{equation}
4 \om\dot{\bar\vp} = - \bar\vp \Bigl\{ N ( 1
- \g_0 \g_3) - i \lbd ( \bar\vp \vp)^{1/2k} 
\Bigl(\om ( \g_0 + \g_3 ) + \g_0 - \g_3 \Bigr) \Bigr\}.
\label{2.4.24b}
\end{equation}

Multiplying (\ref{2.4.24a}) by $ \bar\vp $, 
(\ref{2.4.24b}) by $ \vp $ and summing we arrive at the relation
\begin{displaymath}
\dot \vp \vp + \bar\vp \dot \vp = - 2N \bar\vp \vp,
\end{displaymath}
whence it follows that $ \bar\vp \vp = C \om^{-N/2},
\ C= \mbox{\rm const}$.

Substitution of the result obtained into (\ref{2.4.24a}) gives rise
to a linear equation for $ \vp (\om) $
\begin{displaymath}
4\om \dot \vp = \{ - N (1 + \g_0 \g_3) + i \tau \om^{\theta}
(\om(\g_0 + \g_3) + \g_0 - \g_3)\} \vp,
\end{displaymath}
where $\tau = - \lbd C^{1/2k},\ \theta = - N/4k$.

Writing this equation component-wise (we assume that
$ \g$-matrices are of the form
(\ref{1.1.7})) we get a system of four ODEs 
\begin{equation}
\begin{array}{l}
2\om \dot \vp^0 = i \tau \om^{\theta +1} \vp^2,\quad
2\om \dot \vp^1 = - N \vp^1 + i \tau \om^{\theta } \vp^3,\\[2mm]
2\om \dot \vp^3 = i \tau \om^{\theta +1} \vp^2,\quad
2\om \dot \vp^2 = - N \vp^2 + i \tau \om^{\theta } \vp^0,
\label{2.4.25}
\end{array}
\end{equation}
which is equivalent to the following second-order system of ODEs:
\begin{eqnarray*}
& &\om^2 \ddot \vp^0 + (1/2) (N - 2 \theta)\om \dot \vp^0 + (\tau^2
/4) \om^{2 \theta +1} \vp^0 = 0,\\
& &\om^2 \ddot \vp^3 + (1/2) (N-2 \theta) \om \dot \vp^3 + (\tau^2 /4)
\om^{2 \theta + 1} \vp^3 = 0,\\
& &\vp^1 =-(2i / \tau) \om^{- \theta} \dot \vp^3, 
\quad \vp^2 = -(2 i / \tau) \om^{- \theta} \dot \vp^0.
\end{eqnarray*}

The first two equations of the above system are the 
Bessel-type\index{Bessel!equation} equations
\cite{20.3,132,199}. Provided $ \theta \not= -1/2 $, their 
general solutions are given by the formulae
\begin{equation}
\begin{array}{rcl}
\vp^0 &=& \om^{(2 + 2 \theta - N)/4} \Bigl(\chi^0 J_\nu (z)
+ \chi^2 Y_\nu (z)\Bigr),\\[2mm]
\vp^3 &=& \om^{(2 + 2 \theta - N)/4 }\Bigl( \chi^1 Y_\nu (z)
+ \chi^3 J_\nu (z)\Bigr),
\end{array}
\label{2.4.26}
\end{equation}
where $J_\nu, \ Y_\nu $ are the Bessel
functions,\index{Bessel!function}  
$ z = \tau (2 \theta + 1)^{-1} \om^{(2 \theta + 1)/2}$,
\ \ $\nu = (\theta + 1 
\linebreak - N/2)(1 + 2 \theta)^{-1}, \ \chi^0,
\ldots, \chi^3 $ are arbitrary complex
constants. Consequently, the general solution of 
system of ODEs (\ref{2.4.25}) is given by
(\ref{2.4.26}) and by the following formulae:
\begin{equation}
\begin{array}{rcl}
\vp^2 &=& \om^{(2 + 2 \theta - N)/4} \Bigl\{ (i / 2 \tau)
(N - 2 \theta - 2) \om^{- \theta - 1} \\[2mm]
& &\times \Bigl(\chi^0 J_\nu (z) + \chi^2 Y_\nu (z)\Bigr) -
i\om^{-1/2}\Bigl( \chi^0 \dot J_\nu (z)
+\chi^2 \dot Y_\nu (z)\Bigr) \Bigr\},\\[2mm]
\vp^1 &=& \om^{(2 + 2 \theta - N)/4} \Bigl\{ (i / 2 \tau)
(N- 2 \theta - 2) \om^{- \theta - 1} \\[2mm]
& &\times \Bigl(\chi^3 J_\nu (z) + \chi^1 Y_\nu (z)\Bigr) -
i\om^{-1/2} 
\Bigl( \chi^3 \dot J_\nu (z)
+ \chi^1 \dot Y_\nu (z)\Bigr)\Bigr\}.
\end{array}
\label{2.4.27}
\end{equation}

Formulae (\ref{2.4.26}), (\ref{2.4.27}) determine the general solution
of nonlinear equations (\ref{2.4.24a}), (\ref{2.4.24b}) provided
\begin{displaymath}
\bar\vp \vp = \vp^{0*} \vp^2 + \vp^0 \vp^{2*} + \vp^{3*} \vp^1 +
\vp^3 \vp^{1*} = C\om^{-N/2}.
\end{displaymath}

Substitution of expressions (\ref{2.4.26}), (\ref{2.4.27}) into this
formula gives rise to the following equality:
\begin{displaymath}
2 i\, (2 \theta + 1)(\tau \pi)^{-1} ( \chi^0 \chi^{2*} -
\chi^2 \chi^{0*} + \chi^3 \chi^{1*}
\end{displaymath}
\begin{displaymath}
\qquad
-\chi^1 \chi^{3*}) \om^{-N/2} = C\om^{-N/2}
\end{displaymath}
(we have used a well-known identity for the Bessel functions
$J_\nu (z) \dot Y_\nu (z) - Y_\nu (z) \dot J_\nu (z) =
2(\pi z)^{-1} $ \cite{199}).

Comparing the both parts of the above equality yields
\begin{displaymath}
C = 2i\, (2 \theta + 1)\,(\tau \pi)^{-1} (\chi^0 \chi^{2*} -
\chi^2 \chi^{0*} + \chi^3
\chi^{1*} - \chi^1 \chi^{3*}),
\end{displaymath}
whence
\begin{displaymath}
C = \bigl\{i\,(2k-N)\,(\pi k \lbd)^{-1} ( \chi^{0*} \chi^2 -
\chi^0 \chi^{2*} +\chi^{3*} \chi^1 -
\chi^3 \chi^{1*})\bigr\}^{2k/(2k+1)}.
\end{displaymath}

System (\ref{2.4.25}) with $ \theta = - 1/2\ (\Leftrightarrow \ k =
-N/2) $ is integrated in elementary functions. Omitting intermediate
calculations we present the final result
\vspace{1.5mm}

\noindent
1)\ $ \tau^2 \not= N-1, \quad N = 2,3 $
\begin{equation}
\begin{array}{rcl}
\vp^0 &=& \chi^0 \om^{\theta_+} + \chi^2 \om^{\theta_{-}},\\[2mm]
\vp^1 &=& -(2i/\tau)\om^{-1/2} (\theta_{+} \chi^3 \om^{\theta_{+}}
+ \theta_{-} \chi^1 \om^{\theta_{-}}),\\[2mm]
\vp^2 &=& -(2i/\tau)\om^{-1/2} (\theta_{+} \chi^0 \om^{\theta_{+}}
+ \theta_{-} \chi^2 \om^{\theta_{-}}),\\[2mm]
\vp^3 &=& \chi^3 \om^{\theta_{+}} + \chi^1 \om^{\theta_{-}},
\label{2.4.28}
\end{array}
\end{equation}
where $ \theta_{\pm} = (1/4) \Bigl(1 - N \pm [(N - 1)^2 -
4 \tau^2]^{1/2}\Bigr)$,
\ $ \chi^0, \ldots, \chi^3 $ are arbitrary constants; $ \tau $
satisfies the equality
\begin{eqnarray*}
& &i ( \chi^{0*} \chi^2 - \chi^0 \chi^{2*} + \chi^{3*} \chi^1
- \chi^3 \chi^{1*})\Bigl((N - 1)^2 - 4 \tau^2\Bigr)^{1/2}  \\
& &\quad
= (-1)^{N+1} \tau^{N+1} \lbd^{-N};
\end{eqnarray*}
2)\ $ \tau \not= 0, \ N = 1 $
\begin{equation}
\begin{array}{rcl}
\vp^0 &=& \chi^0 \cos [(\tau  /2)\ln \om] + \chi^2
\sin [(\tau /2) \ln \om],\\[2mm]
\vp^1 &=& - i\om^{-1/2} \Bigl( \chi^1 
\cos [(\tau  /2) \ln \om] - \chi^3 \sin
[(\tau /2) \ln \om]\Bigr),\\[2mm]
\vp^2 &=& - i\om^{-1/2} \Bigl( \chi^2 
\cos [(\tau  /2) \ln \om] - \chi^0 \sin
[(\tau /2) \ln \om] \Bigr),\\[2mm]
\vp^3 &=& \chi^3 \cos [(\tau  /2) \ln \om] + \chi^1
\sin [(\tau /2) \ln \om],
\end{array}
\label{2.4.29}
\end{equation}
where $ \chi^0, \ldots, \chi^3 $ are arbitrary complex constants;
\ $ \tau $ is determined by the equality
\begin{displaymath}
\tau = - i \lbd (\chi^0 \chi^{2*} - \chi^{0*} \chi^2 + \chi^3
\chi^{1*} - \chi^1 \chi^{3*});
\end{displaymath}
3)\ $ \tau = \ve (N - 1)/2, \quad \ve = \pm 1 $
\begin{equation}
\begin{array}{rcl}\vp^0 &=& \om^{(1-N)/4} (\chi^0 + \chi^2 \ln \om),
\\[2mm]
\vp^1 &=& (i/2 \tau)(N-1)\om^{-1/2} \vp^3 + 4i
\ve (1 - N)^{-1} \om^{-(N+1)/4 } \chi^1,
\\[2mm]
\vp^2 &=& (i/2 \tau)(N-1)\om^{-1/2} \vp^0 + 4i
\ve (1 - N)^{-1} \om^{-(N+1)/4 } \chi^2,
\\[2mm]
\vp^3 &=& \om^{(1-N)/4} (\chi^3 + \chi^1 \ln \om),
\end{array}
\label{2.4.30}
\end{equation}
where $ \chi^0, \ldots, \chi^3 $ are complex constants satisfying
the equality
\begin{displaymath}
2i (\chi^0 \chi^{2*} - \chi^{0*} - \chi^{0*} \chi^2 +
\chi^3 \chi^{1*} - \chi^{3*} \chi^1) 
=(-1)^N \Bigl((N-1)/2 \ve \lbd\Bigr)^{N+1}.
\end{displaymath}

Thus, the general solution of system (\ref{2.4.24a}) is given 
by formulae (\ref{2.4.26}),
(\ref{2.4.27}) under $ k \not= N/2 $ and by formulae
(\ref{2.4.28})--(\ref{2.4.30}) under $k = N/2$.

Now we turn to Ans\"atze (\ref{2.3.15}) which were obtained by
reducing the nonlinear Dirac equation (\ref{2.2.1}) by means of the
one-parameter subgroups of the group $P(1,3)$ and then by means of
symmetry groups of the reduced equations 5--7 from (\ref{2.3.3}).  As
established in Section 2.3 Ans\"atze (\ref{2.3.15}) reduce system of
PDEs (\ref{2.4.1}) to systems of ODEs (\ref{2.3.16}) with $ f_1 = \lbd
(\bar\psi \psi)^{1/2k} $. Up to the sign at the nonlinear term $ \lbd
(\bar\vp \vp)^{1/2k} \vp $, they coincide with systems of ODEs 1, 2, 8
from (\ref{2.3.5}).  Using this fact it is not difficult to
construct their general solutions 
\begin{eqnarray}
\vp(\om) &=& \exp \{i \lbd \g_1 ( \bar\chi \chi)^{1/2k}\om\} \chi,
\non\\
\vp(\om) &=& \om^{-1/4} \left\{ \begin{array} {l}
\exp \left\{ 2i \lbd k (1-2k)^{-1} (\bar\chi \chi)^{1/2k} \g_2
\right. \non\\ 
\quad\times \left. \om^{(2k-1)/4k} \right\} \chi, \ \ k \not= 1/2,
\non\\ 
\exp \left\{(i \lbd / 2) ( \bar\chi \chi) \g_2 \ln \om \right\} 
\chi,\ \ k=1/2,
\end{array} \right.
\non\\
\vp(\om) &=& \exp \{i \lbd ( \bar\chi \chi )^{1/2} \g_1 \om \} \chi,
\non\\
\vp(\om) &=& \om^{-1/4} \exp \{-2 i \lbd ( \bar\chi
\chi )^{1/2} \g_2 \om^{1/4}\} \chi,
\label{2.4.31}\\
\vp(\om) &=& \exp \{i \lbd ( \bar\chi \chi )^{1/2} \g_0 \om\} \chi,
\non\\
\vp(\om) &=& \exp \{-i \lbd ( \bar\chi \chi )^{1/2} \g_1 \om\} \chi,
\non\\
\vp(\om) &=& \om^{-1/4} \exp \{2 i \lbd ( \bar\chi \chi )^{1/2}
\g_2 \om^{1/4} \} \chi,
\non\\
\vp(\om) &=& \exp \{-i \lbd ( \bar\chi \chi )^{1/2} \g_1 \om\} \chi,
\non\\
\vp(\om) &=& \om^{-1/4} \exp \{2 i \lbd ( \bar\chi \chi )^{1/2}
\g_2 \om^{1/4} \} \chi.\non
\end{eqnarray}

Here $ \chi $ is an arbitrary constant four-component column.

The fact that many of nonlinear systems (\ref{2.3.5}) are integrable
in quadratures is closely connected with their nontrivial symmetry.
The last property, in its turn, is the consequence of the broad
symmetry admitted by the initial PDE (Theorem 2.3.1).  Therefore, the
wider the symmetry group of the equation under study the more
effective is the application of the group-theoretical methods for
construction of its exact solutions.  
\vspace{2mm}

\noindent
{\bf 1.2.}\ {\em Exact solutions of equation (\ref{2.4.1}).} \ 
Substitution of formulae (\ref{2.4.8}), (\ref{2.4.9}), (\ref{2.4.13}),
(\ref{2.4.15}), (\ref{2.4.17}), (\ref{2.4.19d}),
(\ref{2.4.21})--(\ref{2.4.23}), (\ref{2.4.26})--(\ref{2.4.31}) into
the corresponding $P(1,3)$-invariant Ans\"atze (\ref{2.2.8}) and
(\ref{2.3.15}) yields the following classes of exact solutions of
nonlinear spinor equation (\ref{2.4.1}): 
\index{Exact solutions!of the nonlinear Dirac equation}
\vspace{1.5mm}

\noindent
\underline{the case $ k \in { \R}^1 $ }
\begin{eqnarray*}
\psi_1 (x) &=& \exp \{- i \lbd ( \bar\chi \chi )^{1/2k}
\g_0 x_0 \} \chi,
\\
\psi_2 (x) &=& \exp \{ i \lbd ( \bar\chi \chi )^{1/2k}
\g_3 x_3 \} \chi,
\\
\psi_3 (x) &=& \exp \{ (1/2) \g_0 \g_3 \ln (x_0 + x_3)\} 
\\
& &\times \exp \{ i \g_2 [ ( \bar\chi \chi )^{1/2k} - (i/2)
( \g_0 + \g_3)] x_2\} \chi,
\\
\psi_4 (x) &=& \exp \{- (1/2) (\g_0 + \g_3) \g_1 (x_0 + x_3)\} 
\\
& &\times \exp \{ (i \lbd /2) ( \bar\chi \chi)^{1/2k} \g_1
[ 2x_1 + (x_0 + x_3)^2]\} \chi,
\\
\psi_5 (x) &=& \exp \{- (1/2) (\g_0 + \g_3) \g_1 (x_0 + x_3)\} 
\exp \{ (i \lbd /2) (1 + \al^2)^{-1}\\
& &
\times(\bar\chi \chi)^{1/2k}(\g_2 - \al \g_1) 
[2 (x_2 - \al x_1)- \al (x_0 + x_3)^2] \} \chi,
\\
\psi_6 (x) &=& \exp \{ (1/2) [x_1 - \al \ln (x_0 +x_3)]
(x_0 + x_3)^{-1} (\g_0 + \g_3) \g_1 \} 
\\
& &
\times \exp \{(1/2) \g_0 \g_3 \ln (x_0 + x_3) \} 
\exp \Bigl\{\Bigl(\g_2( \g_0 + \g_3) + i \lbd ( \bar\chi \chi)^{1/2k}
\quad\\
& &
\times [\g_2 - \beta (\g_0 + \g_3)]\Bigr)
[x_2 - \beta \ln (x_0 + x_3)] \Bigr\} \chi,\\
\psi_7(x) &=& \exp \{(2 \al)^{-1} (x_2 + 2 \al \theta x_1)
\g_0 \g_3 \} 
[( \theta \g_2 - (2 \al)^{-1} \g_1) \g_4 - i \lbd \tau ] \chi,\\
& &\al \in { \R}^1,\ \theta,\ \tau \ {\rm \ are\ determined\ by}
\ (\ref{2.4.11}),\  (\ref{2.4.12});\\
\psi_8 (x) &=& \exp \{ (2 \al)^{-1} (2 \al \theta x_3 -
x_0) \g_1 \g_2 \} 
[( \theta \g_0 - (2 \al)^{-1} \g_3) \g_4 - i \lbd \tau ] \chi,\\
& &\al \in { \R}^1 ;\ \theta,\ \tau \ {\rm are\ determined\ by}\
(\ref{2.4.14});\\
\psi_9 (x) &=& \exp \{[(1/4) (x_3 - x_0) + \theta (x_0 +
x_3)] \g_1 \g_2 \} \\
& &\times [ 4 \theta (\g_0 + \g_3) \g_4 + (\g_0 - \g_3) \g_4 -
4 i \lbd \tau ] \chi,\\
& & \theta,\ \tau  \ {\rm are\ determined\ by}\ (\ref{2.4.18});\\
\psi_{10} (x) &=& \exp \{ [ - (1/2) (\dot w_1 \g_1 + \dot w_2 \g_2 )
+ w_3 \g_4 ] (\g_0 + \g_3 ) \} \\
& &\times \exp \{ i \lbd ( \bar\chi \chi ) ^{1/2k} 
\g_1 (x_1 + w_1 )\} \chi ;
\end{eqnarray*}
\vspace{1.5mm}

\noindent
\underline{the case $k \in {\R}^1,\ k \not= 1/2$ }
\begin{eqnarray*}
\psi_{11} (x) &=& \exp \{ (1/2) \g_0 \g_3
\ln (x_0 + x_3) \} \vp (x_0^2 - x_3^2),\\
& & \vp (\om) \ {\rm is\ determined\ by}\ (\ref{2.4.26}),\ 
(\ref{2.4.27})\ {\rm under}\ N = 1 ;\\
\psi_{12} (x) &=& [(x_1 + w_1)^2 + (x_2 + w_2)^2 ]^{-1/4} 
\exp \{[- (1/2) ( \dot w_1 \g_1\\
& &+ \dot w_2 \g_2)
+ w_3 \g_4 ] ( \g_0 + \g_3)\}
\exp \{-(1/2)\g_1\g_2 \\
& &\times\arctan [(x_1+w_1)/(x_2+w_2)]\}
\exp \{ 2i \lbd k (1 - 2k)^{-1}  \\
& &\times ( \bar\chi \chi )^{1/2k} \g_2
[(x_1 + w_1)^2 +
(x_2 + w_2)^2]^{(2k-1)/4k} \} \chi,\\
\psi_{13}(x) &=& (x_1^2 + x_2^2)^{-1/4} \exp \{ (1/2) \g_0
\g_3 \ln (x_0 + x_3) - (1/2) \g_1 \g_2 \\
& &\times \arctan (x_1/x_2)\}
\{ f_1 + \g_2 f_2 + (\g_0 + \g_3 ) f_3
+ \g_2 (\g_0 + \g_3) f_4 \} \chi,\\
& & f_i = f_i (x_1^2 + x_2^2) \ {\rm are\ determined\ by}\
(\ref{2.4.22}),\ (\ref{2.4.23})\\ 
& &{\rm under}\ k \not= 1/2 ;
\end{eqnarray*}
\vspace{1.5mm}

\noindent
\underline{the case $ k \in {\R}^1,\ k \not= 1 $}
\begin{eqnarray*}
\psi_{14} (x) &=& \exp \{(1/2) x_1 (x_0 + x_3)^{-1}
(\g_0 + \g_3) \g_1 \} \\
& &\times \exp \{(1/2) \g_0 \g_3 \ln (x_0 + x_3) \}
\vp (x_0^2 - x_1^2 - x_3^2),\\
& & \vp (\om) \ {\rm is\ determined\ by}\ (\ref{2.4.26}),\
(\ref{2.4.27})\  
{\rm under}\ N = 2;\qquad \;
\end{eqnarray*}
\vspace{1.5mm}

\noindent
\underline{the case $k \in {\R}^1,\ k \not= 3/2$}
\begin{eqnarray*}
\psi_{15} (x) &=& \exp \{(1/2) (x_0 + x_3)^{-1}
(\g_0 + \g_3) (\g_1 x_1 + \g_2 x_2) \}\\
& &\times \exp \{(1/2) \g_0 \g_3 \ln (x_0 + x_3) \}
\vp (x \cdot x),\\
& & \vp (\om) \ {\rm is\ determined\ by}\ (\ref{2.4.26}),\
(\ref{2.4.27})\ 
{\rm under}\ N = 3;\qquad \;
\end{eqnarray*}
\vspace{1.5mm}

\noindent
\underline{the case $k = 1/2$}
\begin{eqnarray*}
\psi_{16} (x) &=& \exp \{(1/2) \g_0 \g_3 \ln (x_0 + x_3) \}
\vp (x_0^2 - x_3^2),\\
& & \vp (\om) \ {\rm is\ determined\ by}\ (\ref{2.4.29});\\
\psi_{17}(x) &=& (x_1^2 + x_2^2)^{-1/4} \exp \{ (1/2) \g_0 \g_3
\ln (x_0 + x_3) - (1/2) \g_1 \g_2 \\
& &\times\arctan (x_1/x_2)\} [ f_1 + \g_2 f_2  
+ (\g_0 + \g_3) f_3 + \g_2 (\g_0 + \g_3) f_4 ] \chi,\\
& &f_i = f_i (x_1^2 + x_2^2) \ {\rm are\ 
determined\ by}\ (\ref{2.4.22}),\ (\ref{2.4.23})\\ 
& &{\rm under}\ k = 1/2 ;\\
\psi_{18} (x) &=& [(x_1 + w_1)^2 + (x_2 + w_2)^2]^{-1/4} \\ 
& &\times \exp \{[-(1/2) (\dot w_1 \g_1 + \dot w_2 \g_2) + w_3 \g_4 ]
(\g_0 + \g_3) \} \\
& &\times \exp \{-(1/2) \g_1 \g_2 \arctan [(x_1 + w_1)/(x_2
+ w_2)] \} \\
& &\times \exp \{( i \lbd /2) ( \bar\chi \chi ) \g_2  \ln [(x_1 +
w_1)^2 + (x_2 + w_2)^2] \} \chi ;
\end{eqnarray*}
\underline{the case $k = 1 $ }
\begin{eqnarray*}
\psi_{19} (x) &=& w_0^{-1} \exp \Bigl\{ \Bigl(-(1/2)
(\dot w_1 \g_1 + \dot w_2 \g_2 ) + w_3 \g_4
\\
&
&
 - (1/2) \dot w_0 w_0^{-1} [\g_1 (x_1 + w_1) + \g_2
(x_2 + w_2)]\Bigr) (\g_0 + \g_3) \Bigr\} 
\\
&
&
\times \exp \{ i \lbd w_0^{-1} (\bar\chi \chi)^{1/2}
\g_1 (x_1 + w_1) \} \chi ;
\\
\psi_{20} (x) &=& w_0^{-1/2} [(x_1 + w_1)^2 + (x_2 +
w_2)^2]^{-1/4} 
\\
&
&
\times \exp \Bigl\{ \Bigl(-(1/2)(\dot w_1 \g_1 + \dot w_2 \g_2 )
+ w_3 \g_4 - (1/2) \dot w_0 w_0^{-1} 
\\
&
&
\times [\g_1 (x_1 + w_1) + \g_2
(x_2 + w_2)] (\g_0 + \g_3) \Bigr\} 
\\
&
&
\times \exp \{- (1/2) \g_1 \g_2 \arctan [(x_1 + w_1)/(x_2
+ w_2] \} 
\\
&
&
\times \exp \{-2 i \lbd (\bar\chi \chi)^{1/2} \g_2
[(x_1 + w_1)^2 + (x_2 + w_2)^2]^{1/4} w_0^{-1/2} \} \chi,
\\
\psi_{21} (x) &=& (\g_0 x_0 - \g_1 x_1 - \g_2 x_2)(x_0^2 -
x_1^2 - x_2^2)^{-3/2} 
\\
&
&
\times \exp \{ i \lbd (\bar\chi \chi)^{1/2} \g_0 x_0
(x_0^2 - x_1^2 - x_2^2)^{-1} \} \chi;
\\
\psi_{22} (x) &=& (\g_0 x_0 - \g_1 x_1 - \g_2 x_2)(x_0^2 -
x_1^2 - x_2^2)^{-3/2} 
\\
&
&
 \times \exp \{- i \lbd (\bar\chi \chi)^{1/2} \g_1 x_1
(x_0^2 - x_1^2 - x_2^2)^{-1} \} \chi;
\\
\psi_{23} (x) &=& (\g_0 x_0 - \g_1 x_1 - \g_2 x_2)(x_0^2 -
x_1^2 - x_2^2)^{-1}(x_1^2 + x_2^2)^{-1/4} 
\\
&
&
\times \exp \{- (1/2) \g_1 \g_2 \arctan (x_1/x_2) \} 
\exp \{2i \lbd (\bar\chi \chi)^{1/2}\\
&
&
\times  \g_2 (x_1^2 + x_2^2)^{1/4}
(x_0^2 - x_1^2 - x_2^2)^{-1/2} \} \chi ;
\\
\psi_{24}(x) &=& \g_a x_a (x_b x_b)^{-3/2} 
\exp \{ - i \lbd ( \bar\chi \chi)^{1/2}
\g_1 x_1 (x_a x_a)^{-1} \} \chi ;
\\
\psi_{25} (x) &=& \g_a x_a (x_a x_a)^{-1} (x_1^2 + x_2^2)^{-1/4} 
\exp \{- (1/2) \g_1 \g_2  
\\
&
&
\times \arctan (x_1/x_2) \}\exp \{ 2i \lbd ( \bar\chi \chi)^{1/2}
(x_1^2 + x_2^2)^{1/4} (x_a x_a)^{-1/2} \} \chi;
\\
\psi_{26} (x) &=& \exp \{(1/2) x_1 (x_0 + x_3)^{-1}
(\g_0 + \g_3) \g_1 \} 
\\
&
&
\times \exp \{ (1/2) \g_0 \g_3 \ln (x_0 + x_3) \}
\vp (x_0^2 - x_1^2 - x_3^2),\\
& & \vp(\om) \ {\rm is\ determined\ by}\ (\ref{2.4.28})\ {\rm or}\
(\ref{2.4.30})\\ 
& &{\rm under}\  N = 2 ;
\end{eqnarray*}
\underline{the case $k = 3/2 $}
\begin{eqnarray*}
\psi_{27}(x) &=& \exp \{(1/2)(\g_0 + \g_3) (\g_1 x_1
+ \g_2 x_2)(x_0 + x_3)^{-1} \}\qquad \qquad \qquad
\\
&
&
\times \exp \{(1/2) \g_0\g_3 \ln (x_0+x_3)\}\vp(x\cdot x),\\
& & \vp (\om) \ {\rm is\ determined\ by}\ (\ref{2.4.28})\ {\rm or}\
(\ref{2.4.30})\\
& &{\rm under}\ N = 3.
\end{eqnarray*}

In the above formulae $ w_0,\ w_1,\ w_2,\ w_3 $ are arbitrary smooth
functions of $x_0+x_3$, an overdot denotes differentiation with
respect to $x_0+x_3$.

In addition, in \cite{100,103} we have constructed two other classes of
exact solutions of system of PDEs (\ref{2.4.1})
\vspace{1.5mm}

\noindent
\underline{the case $k = 1/2$ }
\begin{eqnarray*}
\psi_{28} (x) &=&  \om^{-1}\exp \{ (1/2) \g_1 (\g_0 + \g_3)(x_0 + x_3) 
\}  
\Bigl\{[\g_2 + \beta (\g_0 + \g_3)]\\
& &\times [x_2 + \beta (x_0 + x_3)]
+ (1/2) \g_1 [2x_1 + (x_0 + x_3)^2]\Bigl\} \\
& &\times \exp \Bigl\{i \lbd (\bar\chi \chi) (\beta_1^2 +
\beta_2^2)^{-1} \om^{-1} \Bigl( \beta_1 [\g_2 
+ \beta(\g_0 + \g_3)] + \beta_2 \g_1\Bigr)\\
& &\Bigl(\beta_1[x_2 + \beta (x_0 + x_3)] + (\beta_2 /2) 
[2x_1 + (x_0 + x_3)^2] \Bigr) \Bigr\} \chi;\\
\end{eqnarray*}
\underline{the case $ k \in {\R}^1,\ k < 0 $}
\begin{eqnarray*}
\psi_{29} (x) &=& \exp \{ (1/2) \g_1 (\g_0 + \g_3) (x_0 + x_3) \} 
\Bigl\{\Bigl([\g_2 + \beta (\g_0 + \g_3)][x_2 + \beta\\
& &\times (x_0 + x_3)] + (1/2) \g_1 [2x_1 + (x_0 + x_3)^2]\Bigr)f(\om) 
+ i g(\om)\Bigr\} \chi. 
\end{eqnarray*}

Here $\al,\ \beta,\ \beta_1,\ \beta_2 $ are arbitrary constants,
\begin{eqnarray*}
& &\om = [x_2 + \beta (x_0 + x_3)]^2 + (1/4) [2x_1 
+ (x_0 + x_3)^2]^2,\\
& &f(\om) = |k|^{1/2}\Bigl(\ve (1 - k)^{1/2} \lbd^{-1}
(\bar\chi \chi)^{-1/2k} \Bigr)^k \om^{-(k+1)/2},\\
& &g(\om) = -\ve (1-k)^{1/2}\Bigl(\ve (1 - k)^{1/2} \lbd^{-1}
(\bar\chi \chi)^{-1/2k} \Bigr)^k \om^{-k/2},
\quad \ve = \pm 1.
\end{eqnarray*}

Thus, we have constructed wide classes of exact solutions of
the nonlinear Dirac equation (2.41), some of them containing
arbitrary functions. By a special choice of these functions
we can select subclasses of exact solutions possessing
important additional properties.

For example, if we put
\begin{displaymath}
w_0 = \exp \{ \theta^2 (x_0 + x_3)^2 \}, \quad \theta \in {\R}^1,
\quad
w_1 = w_2 = w_3 = 0
\end{displaymath}
in the solution $ \psi_{19} (x) $, then it takes the form
\begin{eqnarray}
\psi (x) &=& \exp \{- \theta^2 (x_0 + x_3)^2 \} 
\Bigl(1 + \theta^2 (x_0 + x_3)(\g_1 x_1 + \g_2 x_2)\label{2.4.32}\\
& &\times(\g_0 + \g_3)\Bigr) 
\exp \Bigl\{i\lbd(\bar\chi\chi)^{1/2k}\g_1 x_1 \exp \{- \theta^2 (x_0
+ x_3)^2 \}\Bigr\} \chi. \non
\end{eqnarray}

This solution is localized inside the infinite cylinder
having the generatix parallel to the coordinate axis $ Ox_3 $.
In addition, it decreases exponentially as $ x_0 \to  + \infty$.

It is worth noting that (\ref{2.4.32}) under $ \theta = 0 $ becomes
the plane-wave solution
\begin{equation}
\psi (x) = \exp \{ i \lbd ( \bar\chi \chi )^{1/2}
\g_1 x_1\} \chi.
\label{2.4.33}
\end{equation}
Consequently, (\ref{2.4.32}) can be considered as a perturbation of
the stationary state (\ref{2.4.33}).
\vspace{2mm}

\noindent
{\bf 1.3.}\ {\em Generation of solutions.} \ Solutions $ \psi_1 (x) -
\psi_{29} (x) $ depend on the variables $ x_\mu $ in asymmetrical way,
while in equation (\ref{2.4.1}) all independent variables are enjoying
equal rights. Using the language of physics we can say that system of
PDEs (\ref{2.4.1}) is solved in some fixed reference frame. To obtain
solutions (more precisely, families of solutions) not depending on the
choice of a reference frame it is necessary to apply the procedure
of generating solutions by transformations from the Poincar\'e group
\cite{89,90,91,103}.

Let the equation under study be invariant under the Lie group of
transformations of the form
\begin{equation}
x'_\mu = f_\mu (x, \theta), \quad \psi'(x') = A(x, \theta) \psi (x),
\label{2.4.34a}
\end{equation}
where $ A(x, \theta) $ is an invertible $(m \times m)$-matrix, $
\theta = (\theta_1, \ \theta_2, \ldots, \theta_r ) $ are group
parameters. In addition, there is some particular solution $ \psi_I(x) $
of the equation considered.  
\vspace{1.5mm}

\noindent
{\bf Theorem 2.4.1.}\ {\em The $m$-component function
$ \psi_{II} (x) $ determined by the equality
\begin{equation}
\psi_{II} = A^{-1} (x, \theta) \psi_I \Bigl(f(x, \theta)\Bigr)
\label{2.4.34b}
\end{equation}
is a solution of PDE admitting the Lie group (\ref{2.4.34a}).}
\vspace{1.5mm}

\noindent
{\em Proof.}$\quad$ According to the definition of the invariance
group, the Lie group (\ref{2.4.34a}) transforms the set of solutions
of the equation under study into itself. In other words, provided $
\psi = \psi(x) $ is a solution of the equation written in coordinates
$ x, \ \psi (x) $ the function constructed by means of formulae
(\ref{2.4.34a}) is a solution of the same equation written in
coordinates $ x', \ \psi'(x') $. Resolving (\ref{2.4.34a}) with
respect to $ \psi(x) $ we have
\begin{displaymath}
\psi (x) = A^{-1} (x, \theta) \psi ' (x'),
\end{displaymath}
whence due to (\ref{2.4.34a}) we get
\begin{displaymath}
\psi (x) = A^{-1} (x, \theta) \psi ' \Bigl(f (x, \theta)\Bigr).
\end{displaymath}

Denoting $ \psi = \psi_{II}, \ \psi ' = \psi_I $ yields 
(\ref{2.4.34b}). $\rhd$

Using Theorem 2.4.1 it is possible to obtain a $r$-parameter
family of exact solutions starting from a single solution.
\vspace{1.5mm}

\noindent
{\bf Definition 2.4.1.}\ Formula (\ref{2.4.34b}) is called the formula 
for generating solutions\index{Solution generation} by 
transformations from the group (\ref{2.4.34a}).
\vspace{1.5mm}

\noindent
{\bf Definition 2.4.2.}\ A family of solutions of the form
\begin{eqnarray*}
& &\psi(x) = \psi_0 (x, \tau), \quad \tau = (\tau_1, \tau_2,
\ldots, \tau_s) \in {\R}^s,\\
& &R_i (\tau) = 0, \ \ i = {1,\ldots, s - n + 1}, \ \ 1 \le n \le s
\end{eqnarray*}
is called $G $-ungenerable (or ungenerable)\index{Ungenerable family} 
provided the equality
\begin{displaymath}
A^{-1}(x, \theta) \psi_0 \Bigl(f(x, \theta), \tau\Bigr) =
\psi_0 \Bigl(x, \tau' (\tau, \theta)\Bigr)
\end{displaymath}
holds and what is more $ R_i (\tau ' (\tau, \theta)) = 0,
\ i = {1,\ldots,s - n + 1} $.

Using the final transformations from the group
$C(1,3)$ (\ref{1.1.24a})--(\ref{1.1.24e}) and Theorem 2.4.1
we obtain formulae of generating solutions by transformations from the
conformal group $C(1,3)$.
\vspace{1.5mm}

\noindent
1) the group of translations
\index{Solution generation!with translations}
\begin{equation}
\psi_{II} (x) = \psi_I (x'), \quad x^{\prime}_\mu = x_\mu
+ \theta_\mu ;
\label{2.4.35}
\end{equation}
2) the Lorentz group $O(1,3)$
\vspace{1.5mm}

\noindent
a) the group of rotations $O(3)$
\index{Solution generation!with rotation group}
\begin{eqnarray}
& &\psi_{II} (x) = \exp \{(1/2) \ve_{abc}
\theta_a S_{bc} \} \psi_I (x'),\non\\
& &x_0' = x_0, \quad x_a' = x_a \cos \theta - \theta^{-1} \ve_{abc}
\theta_b x_c \sin \theta \label{2.4.36}\\
& &\quad + \theta^{-2} \theta_a (\theta_b x_b) (1 - \cos \theta) ;\non 
\end{eqnarray}
b) the Lorentz transformations
\index{Solution generation!with Lorentz transformations}
\begin{eqnarray}
& &\psi_{II} (x) = \exp \{-(\theta_0/2) \g_0 \g_a \} \psi_I
(x'),\non\\ 
& &x_0' = x_0 \cosh \theta_0 + x_a \sinh \theta_0,
\label{2.4.37}\\
& &x_a' = x_a \cosh \theta_0 + x_0 \sinh \theta_0,
\quad
x_b' = x_b, \ b \not= a;\non
\end{eqnarray}
3) the group of scale transformations
\index{Solution generation!with scale transformations}
\begin{equation}
\psi_{II} (x) = e^{k \theta_0} \psi_I(x'), \quad x_\mu' = x_\mu
e^{\theta_0};
\label{2.4.38}
\end{equation}
4) the group of special conformal 
transformations
\index{Solution generation!with special conformal transformations}
\begin{equation}
\begin{array}{l}
\psi_{II}(x) = \s^{-2} (x) (1 - \g \cdot x \g
\cdot \theta) \psi_{I}(x'),\\[2mm]
x_\mu' = (x_\mu - \theta_\mu x \cdot x) \s^{-1} (x).
\end{array}
\label{2.4.39}
\end{equation}

Here $ \theta_0,\ldots,\theta_3 $ are real constants, $ \theta =
(\theta_a \theta_a)^{1/2},
\ \s (x) = 1 - 2 \theta\cdot x + \theta\cdot\theta
x\cdot x $.

As an example, we will consider the procedure of generating the
solution $\psi_1(x) $. Let us apply formula (\ref{2.4.37}) with $a = 3
$ to $ \psi_1 (x) $ 
\begin{displaymath}
\psi_{II} (x) = \exp \{ - (\theta_0 / 2) \g_0 \g_3 \}  
\exp \{ - i \lbd ( \bar\chi \chi)^{1/2k} (x_0 \cosh \theta_0
+ x_3 \sinh \theta_0) \g_0 \} \chi.
\end{displaymath}

We rewrite this expression as follows
\begin{eqnarray*}
& &\psi_{II}(x) = \exp \{-(\theta_0/2) \g_0 \g_3 \}
\Bigl\{ \cos \Bigl(\lbd ( \bar\chi \chi)^{1/2k}
(x_0 \cosh \theta_0 \\
& &\quad + x_3 \sinh \theta_0)\Bigr) -
i \g_0 \sin \Bigl(\lbd (\bar\chi \chi)^{1/2k} (x_0 \cosh \theta_0
+ x_3 \sinh \theta_0)\Bigr) \Bigr\} \\
& &\quad
\times \exp \{-(\theta_0 /2) \g_0 \g_3 \} \chi.
\end{eqnarray*}

Taking into account the identities
\begin{displaymath}
V\g_\mu V^{-1}= \left\{ \begin{array} {lll}
\g_0 \cosh \theta_0 +\g_3\sinh \theta_0, & \mu = 0, \\
\g_3 \cosh \theta_0 + \g_0\sinh \theta_0, & \mu = 3, \\
\g_\mu, & \mu = 1,2,  \\
\end{array} \right.
\end{displaymath}
where $V=\exp \{-(\theta_0/2)\g_0 \g_3 \} $,
which are proved with the help of the Campbell-Hausdorff
formula \cite{30,115} we have
\index{Campbell-Hausdorff formula}
\begin{eqnarray*}
& &\psi_{II} (x) = \Bigl\{ \cos \Bigl(\lbd (\bar\chi' \chi')^{1/2k}
(x_0 \cosh \theta_0 + x_3 \sinh \theta_0 )\Bigr) 
- i (\g_0 \cosh \theta_0 \\
& &\quad + \g_3 \sinh \theta_0) \sin \Bigl( \lbd 
(\bar\chi' \chi')^{1/2k} 
(x_0 \cosh \theta_0 + x_3 \sinh \theta_0)\Bigr)\Bigr\} \chi',
\end{eqnarray*}
where $\chi' = \exp \{-(\theta_0/2) \g_0 \g_3 \}$.

Using formula (\ref{2.4.36}) yields the following family of
exact solutions:
\begin{equation}
\begin{array}{l}
\psi_{II}(x) = \Bigl\{ \cos \Bigl( \lbd ( \bar\chi \chi)^{1/2k} 
a \cdot x\Bigr) - i \g \cdot a 
\sin \Bigl( \lbd ( \bar\chi \chi)^{1/2k} a \cdot
x\Bigr)\Bigr\}\chi\\[2mm] 
\quad =\exp \{- i \lbd ( \bar\chi \chi)^{1/2k} 
(\g \cdot a)(a \cdot x) \} \chi,
\end{array}
\label{2.4.40}
\end{equation}
where $ a_\mu $ are arbitrary real parameters satisfying the
condition $a \cdot a = 1$.

It is not difficult to verify that family (\ref{2.4.40}) is invariant
with respect to transformations (\ref{2.4.35}), (\ref{2.4.38}).

The family of solutions (\ref{2.4.40}) depends on the variables $x_\mu
$ in symmetrical way.  Let us show that it is invariant under the
Lorentz group $O(1,3)$. Applying, for example, formula (\ref{2.4.36})
to (\ref{2.4.40}) and grouping terms in a proper way we arrive at the
following family of solutions of PDE (\ref{2.4.1}):
\begin{displaymath}
\psi_{II} (x) = \exp \{- i \lbd ( \bar\chi' \chi')^{1/2k}
( \g \cdot a') ( a' \cdot x) \} \chi',
\end{displaymath}
where
\begin{eqnarray*}
& &a'_0 = a_0,
\quad a_b' = a_b \cos \theta - \theta^{-1} \ve_{bcd} a_c \theta_d 
+ \theta^{-2} \theta_b ( \theta_c a_c) (1 - \cos \theta),\\
& &\chi' = \exp \{(1/2) \ve_{abc} \theta_a S_{bc} \} \chi.
\end{eqnarray*}

Since $a' \cdot a' = 1 $, the obtained family coincides with
(\ref{2.4.40}). Thus, we have constructed the $\wid
P(1,3)$-ungenerable family of exact solutions of the nonlinear Dirac
equation. The transition from $ \psi_1 (x) $ to (\ref{2.4.40}) seems
to be of principal importance because we obtain the class of exact
solutions having the same invariance group as the initial equation
(\ref{2.4.1}). In other words, the family of solutions (\ref{2.4.40})
contains complete information about the Lie symmetry of the nonlinear
Dirac equation (\ref{2.4.1}).

Generating in the same way solutions $ \psi_2 (x) - \psi_6(x) $
we obtain the following $ \wid P(1,3)$-ungenerable families of exact
solutions of system of nonlinear PDEs (\ref{2.4.1}):
\begin{eqnarray*}
\psi_2 (x)&=& \exp \{ i \lbd ( \bar\chi \chi)^{1/2k} 
(\g \cdot b)(b \cdot x) \} \chi,
\\
\psi_3 (x) &=& \exp \{ (1/2) (\g \cdot a) (\g \cdot d)
\ln [\theta (a \cdot z + d \cdot z)]\}
\\
& &
\times \exp \{ i \g \cdot c [( \bar\chi \chi)^{1/2k} -
(i/2)(\g \cdot a + \g \cdot d)] c \cdot z \} \chi,
\\
\psi_4 (x) &=& \exp \{-(\theta/2) (\g \cdot a + \g \cdot d) (\g \cdot
b) (a \cdot z + d \cdot z) \} 
\\
& &
\times \exp \{ ( i \lbd/2) ( \bar\chi \chi)^{1/2k} (\g \cdot b)
[2b \cdot z + \theta (a \cdot z + d \cdot z)^2] \} \chi,
\\
\psi_5 (x) &=& \exp \{ -(\theta/2) (\g \cdot a + \g \cdot d) (\g \cdot
b) (a \cdot z + d \cdot z) \} 
\\
& &
\times \exp \{ ( i \lbd/2)(1 + \al^2)^{-1} ( \bar\chi \chi)^{1/2k}
(\g \cdot c - \al \g \cdot b) \\
& &\times [2 (c \cdot z - \al b \cdot z) 
- \al \theta ( a \cdot z + d \cdot z)^2] \} \chi,
\\
\psi_6 (x) &=& \exp \Bigl\{ (2 \theta)^{-1} \Bigl( \theta b \cdot z
- \al \ln [\theta (a \cdot z + d \cdot z )] \Bigr)
(a \cdot z + d \cdot z)^{-1} 
\\
& &
\times ( \g \cdot a
+ \g \cdot d) \g \cdot b \Bigr\} 
\exp \{ (1/2) (\g \cdot a) (\g \cdot d)
\ln [ \theta (a \cdot z + d \cdot z )] \Bigr\} 
\\
& &
\times \exp \Bigl\{ \Bigl(\g \cdot c (\g \cdot a +  \g \cdot d) + i
\lbd ( \bar\chi \chi)^{1/2k} [\g \cdot c - \beta (\g \cdot a
+ \g \cdot d)]\Bigr)
\\
& &
\times\Bigl(c \cdot z - \beta \theta^{-1} \ln [\theta
(a \cdot z + d \cdot z)]\Bigr) \Bigr\} \chi,
\end{eqnarray*}
where $z_\mu = x_\mu + \theta_\mu$;\ $\al,\ \beta,
\ \theta,\ \theta_\mu $ are arbitrary constants.

Hereafter we denote by $a_\mu,\ b_\mu,\ c_\mu,\ d_\mu,\
\mu={0,\ldots,3}$ arbitrary real constants satisfying the following
conditions:  
\begin{equation}
\begin{array}{l}
a \cdot a = - b \cdot b = - c \cdot c = -d \cdot d = 1, \\
a \cdot b = a \cdot c = a \cdot d = b \cdot c
= b \cdot d = c \cdot d = 0.
\end{array}
\label{2.4.40z}
\end{equation}

Evidently, the four-vectors with components $a_\mu,\ b_\mu,\ c_\mu,\ 
d_\mu $ form a basis in the Minkowski space 
$R(1,3)$\index{Minkowski space} with the scalar product 
$x\cdot y=x_\mu y^\mu $.

Provided the parameter $k $ in (\ref{2.4.1}) is equal to 3/2, this
equation admits the conformal group $C(1,3)$. Consequently, we can
generate solutions by transformations (\ref{2.4.39}). Let us give an
example of the $C(1,3)$-ungenerable family of exact solutions of the
conformally-invariant Dirac-G\"ursey 
equation\index{Dirac-G\"ursey equation}  
\begin{eqnarray*}
\psi (x) &=& \s^{-2} (x) (1 - \g \cdot x \g \cdot \theta) \exp \{ - i
\lbd (\bar\chi \chi)^{1/2k} (\g \cdot a) \\
& &\times (a \cdot x - a \cdot \theta
x \cdot x) \s^{-1} (x) \} \chi.
\end{eqnarray*}
{\bf 2. {\boldmath $\wid P $}(1,3)-invariant solutions of the nonlinear
Dirac equation (\ref{2.4.1}).}\ Now we turn to reduced equations
(\ref{2.3.21}) putting
$R = \lbd (\bar\vp \vp)^{1/2k}\vp $. To integrate these we need some
well-known facts from the general theory of systems of linear ODEs.
\vspace{1.5mm}

\noindent
{\bf Definition 2.4.3.}\ By a normalized solution of the system of
linear ODEs
\begin{equation}
\dot \vp (\om) = B (\om) \vp (\om)
\label{2.4.41}
\end{equation}
we mean the $(4\times 4)$-matrix $ \Omega^\om_{\om_0} (B) $
satisfying the following
conditions:
\begin{displaymath}
{d \Omega_{\om_0}^\om \over d\om} = B(\om) \Omega^\om_{\om_0},
\quad \Omega_{\om_0}^{\om_0} = I,
\end{displaymath}
where $\om_0=\mbox{\rm const}$,\ $I$ is the unit $(4\times 4)$-matrix.

The normalized solution of system (\ref{2.4.41}) is given by 
the following infinite series \cite{132}:
\begin{equation}
\Omega_{\om_0}^\om = I + \int\limits^{\edi\om}_{\edi\om_0} B(\tau) d\tau
+ \int\limits^{\edi\om}_{\edi\om_0}B(\tau)\int
\limits^{\edi\tau}_{\edi\om_0} B(\tau_1) 
d \tau_1 d \tau + \ldots
\label{2.4.42}
\end{equation}

If we succeed in constructing the normalized solution of
system of ODEs (\ref{2.4.41}) in explicit form, then its 
general solution is given by the formulae
\begin{displaymath}
\vp (\om) = \Omega_{\om_0}^\om (B) \chi,
\quad \vp (\om_0) = \chi,
\end{displaymath}
where $ \chi $ is an arbitrary  constant four-component column.

We will consider in detail a procedure of integration of
system of ODEs 1 from (\ref{2.3.21}). On multiplying it by the matrix
$(i/2) \g_3 $ on the left we get
\begin{equation}
\dot \vp = (1/8) (2k - 1) (1 + \g_0 \g_3) \vp + (i \lbd/2)
\g_3 (\bar\vp \vp)^{1/2k} \vp,
\label{2.4.43}
\end{equation}
while the conjugate spinor $ \bar\vp (\om) $ satisfies system of ODEs
of the form
\begin{equation}
\dot{\bar\vp} = (1/8)(2k - 1) \bar\vp
(1 - \g_0 \g_3)  - (i \lbd/2) \bar\vp \g_3
(\bar\vp \vp)^{1/2k}.
\label{2.4.44}
\end{equation}
Multiplying (\ref{2.4.43}) by $ \bar\vp $ on the left, (\ref{2.4.44})
by $ \vp $ on the right and summing we come to the linear ODE for
$\bar\vp \vp$  
\begin{displaymath}
\bar\vp \dot \vp + \dot{\bar\vp} \vp = ( \bar\vp \vp)^{\edi{.}} =
(1/4) (2k - 1) \bar\vp \vp,
\end{displaymath}
which general solution reads
\begin{equation}
\bar\vp \vp = C \exp \{ (1/4)(2k - 1) \om \},
\quad C \in {\R}^1.
\label{2.4.45}
\end{equation}

Substitution of (\ref{2.4.45}) into (\ref{2.4.43}) gives rise to the
system of linear ODEs 
\begin{equation}
\begin{array}{rcl}
\dot \vp &=& \Bigl\{ (1/8)\,(2k - 1)\,(1 + \g_0 \g_3) +
(i \lbd/2) C^{1/2k} \\
& &\times \exp \{ (2k - 1) (8k)^{-1}\om \} \g_3
\Bigr\} \vp.
\end{array}
\label{2.4.46}
\end{equation}

Let $ \Omega_0^\om $ be a normalized solution of system
(\ref{2.4.45}). Then the general solution of (\ref{2.4.45}) is given
by the formula
\begin{equation}
\vp (\om) = \Omega_0^\om \chi.
\label{2.4.47}
\end{equation}

Substituting (\ref{2.4.47}) into (\ref{2.4.45}) we have
\begin{displaymath}
\bar\chi {\bar \Omega}_0^\om \Omega_0^\om \chi =
C \exp \{ (1/4) (2k - 1) \om \},
\end{displaymath}
where $ \ov \Omega_0^\om = \g_0 (\Omega_0^\om)^{\dagger} 
\g_0 $. As $ \Omega_0^\om\vert_{\om = 0} = I $ and 
$ {\bar\Omega}_0^\om\vert_{\om = 0} = I $, from the above 
relation it follows that $ \bar\chi \chi = C $.
Inserting $ C = \bar\chi \chi $ into (\ref{2.4.47}) we obtain the 
general solution of the initial system of nonlinear ODEs
(\ref{2.4.43}). 

Substitution of the result obtained into the corresponding 
$\wid P(1,3)$-invariant Ansatz gives rise to the exact solution 
of the nonlinear spinor equation (\ref{2.4.1})
\begin{equation}
\psi (x) = \exp \{ (1/4) (\g_0 \g_3 - 2k) \ln (x_0 + x_3) \}
\Omega_0^\om \chi,
\label{2.4.48}
\end{equation}
where $\om = \ln (x_0 + x_3) - x_0 + x_3 $.

In a similar way we can integrate systems of ODEs 2, 4, 7, 9, 11, 13,
16, 17, 20, 23--25 from (\ref{2.3.21}) (a detailed analysis of these
equations has been performed in \cite{7}). Here only the cases, when
infinite series (\ref{2.4.42}) can be summed up, are considered.

If we put in (\ref{2.4.46}) $k = 1/2 $, then a system of linear ODEs
with constant coefficients is obtained. Its general solution
has the form (\ref{2.4.47}), where
\begin{eqnarray}
\Omega_0^\om &=& I + \int\limits_{\edi 0}^{\edi\om} B d \tau +
\int\limits_{\edi 0}^{\edi\om} B \int\limits_{\edi 0}^{\edi\tau} B d
\tau_1 d \tau + \ldots \non\\
&=& I + \om B + (2!)^{-1}\om^2B^2 + (3!)^{-1} \om^3 B^3 + \ldots
\label{2.4.49}\\
&=& \exp \{ B\om \}.\non
\end{eqnarray}

In (\ref{2.4.49}) $ B = (i \lbd /2) (\bar\chi \chi) \g_3 $.

Substitution  of (\ref{2.4.49}) into (\ref{2.4.48}) gives rise to the
exact solution of system of nonlinear PDEs (\ref{2.4.1}) with $ k =
1/2 $ 
\begin{equation}
\begin{array}{rcl}
 \psi (x) &=& \exp \{ (1/4) (\g_0 \g_3 - 1) \ln (x_0 + x_3) \} \\[2mm] 
& &\times \exp \{ (i \lbd / 2) (\bar\chi \chi ) \g_3 [\ln (x_0 + x_3)
- x_0 + x_3 ] \} \chi.
\end{array}
\end{equation}

Similarly, computing the normalized solutions of systems of ODEs 4
(under $\al = 0, k = 1/2$), 9 (under $k = 3/2$), 11 (under
$k = 5/2$), 13 (under $k = 1/2$), 20, 22 from (\ref{2.3.21}) we
get their general solutions in the form (\ref{2.4.47})
\begin{eqnarray*}
\vp (\om) &=& \exp \{(i \lbd /2) (\bar\chi \chi ) (\g_3 - \g_0
+ 2 \g_1) \om \} \chi,
\\
\vp (\om) &=& \exp \{-(i \lbd /2) (\bar\chi \chi )^{1/3} \g_0 \om \}
\chi, 
\\
\vp (\om) &=& \exp \{-(i \lbd /2) (\bar\chi \chi )^{1/5} \g_0 \om \}
\chi, 
\\
\vp (\om) &=& \exp \{ i \lbd (1 + \alpha^2)^{-1} (\alpha \g_2 - \g_1)
 (\bar\chi \chi ) \om \} \chi,
\\
\vp (\om) &=& \exp \{ [ (1/2 \alpha)(2k - 1)(\g_0 - \g_3) \g_1  
\\
& &
 - i \lbd \alpha^{-2} (\g_0 - \g_3 + \alpha \g_1) (\bar\chi \chi)^
{1/2k}] \om \} \chi,
\\
\vp (\om) &=& \exp \Bigl\{ \Bigl( (1/2\beta) (1 
+ \beta^2)^{-1} [(2k \beta^2 - 1) \g_1 
\\
& &
- \beta (2k + 1) \g_2] (\g_0 - \g_3 ) - i \lbd (1 + \beta^2)^{-1} 
\\
& &
\times [ \g_2 - \beta \g_1 - (\beta/ \alpha)(\g_0 - \g_3)] ( \bar\chi
\chi)^{1/2k}] \om \Bigr\} \chi.
\end{eqnarray*}

We have also succeeded in integrating systems of ODEs 30, 37, 41. They 
can be represented in the following unified form: 
\begin{equation}
i\Bigl(\g_2-(\g_0-\g_3)z\Bigr){d\varphi\over dz}
=\Bigl(i\theta(\g_0-\g_3)+\lambda(\bar\varphi
\varphi)^{1/2k}\Bigr)\varphi,
\label{2.4.49z}
\end{equation}
where the case $\theta=k,\ z=\om $ corresponds to the system 30, 
the case $\theta=(1/2)(1-2k),\ z=\om $ to the system 37 and the
case $\theta=k,\ z=\om - 1$ to the system 41.

Rewrite equation (\ref{2.4.49z}) in the equivalent form
\begin{displaymath}
\dot\varphi=\Bigl\{\theta\g_2(\g_0-\g_3)-i\lambda(\bar\varphi
\varphi)^{1/2k}\Bigl(\g_2-(\g_0-\g_3)z\Bigr)\Bigr\}
\varphi.
\end{displaymath}

Since $\bar\varphi\varphi\equiv\tau=\mbox{\rm const}$, the above
equation is linearized 
\begin{equation}
\dot\varphi=\Bigl\{\theta\g_2(\g_0-\g_3)-i\lambda\tau^{1/2k}
\Bigl(\g_2-(\g_0-\g_3)z\Bigr)\Bigr\}
\varphi.
\label{2.4.50z}
\end{equation}

The general solution of the system of ODEs (\ref{2.4.50z}) can be
represented in the form
\begin{equation}
\varphi=\Bigl(f_1(z)+f_2(z)\g_2+f_3(z)(\g_0-\g_3)
+f_4(z)\g_2(\g_0-\g_3)\Bigr)\chi,
\label{2.4.51z}
\end{equation}
where $\chi $ is an arbitrary four-component constant column and 
functions $f_1,\ldots,
\linebreak f_4$ satisfy the following system of ODEs:
\begin{eqnarray}
& &\dot f_1=i\lambda\tau^{1/2k} f_2,\quad 
\dot f_2=-i\lambda\tau^{1/2k} f_1,\non\\
& &\dot f_3=i\lambda\tau^{1/2k} f_4+\theta f_2+i\lambda\tau^{1/2k}z
f_1,\non\\ 
& &\dot f_4=-i\lambda\tau^{1/2k} f_3+\theta f_1-i\lambda\tau^{1/2k}z
f_2.\non 
\end{eqnarray}
 
The above system is integrated by the standard methods, its particular 
solution reads
\begin{eqnarray}
&&f_1=\cosh(\lambda\tau^{1/2k} z), \quad 
f_2=-i\sinh(\lambda\tau^{1/2k} z),\label{2.4.52z} \\
&&f_3=-\Bigl ((2\theta+1)i/4\lambda\Bigr
)\tau^{-1/2k}\cosh(\lambda\tau^{1/2k} z) + (i/2)(1+z)
\sinh(\lambda\tau^{1/2k} z),\non \\
&&f_4=\Bigl ((2\theta+1)/4\lambda\Bigr
)\tau^{-1/2k}\sinh(\lambda\tau^{1/2k} z) + (1/2)(1-z)
\cosh(\lambda\tau^{1/2k} z). \non 
\end{eqnarray}

As a direct computation shows, the function (\ref{2.4.51z}) satisfies
an identity
\begin{eqnarray*}
\bar\varphi\varphi&=&\bar\chi\Bigl(|f_1|^2-|f_2|^2
+(f_1f_2^*+f_1^*f_2)\g_2 +(f_1^*f_3+f_1f_3^*
-f_2^*f_4\\
& &-f_2f_4^*)(\g_0-\g_3) 
+(f_1^*f_4-f_1f_4^*+f_2^*f_3-f_2f_3^*)\g_2(\g_0-\g_3)\Bigr)\chi.
\end{eqnarray*}

Substituting into its right-hand side formulae (\ref{2.4.52z}) we
get 
\begin{equation}
\tau =\bar\varphi \varphi =\bar\chi \chi.
\label{2.4.53z}
\end{equation}

Consequently, we have established that the general solution of the
system of nonlinear ODEs (\ref{2.4.49z}) is given by the formulae
(\ref{2.4.51z})--(\ref{2.4.53z}).

Substitution of the expressions obtained above into the corresponding
$\wid P(1,3)$-invariant Ans\"atze (\ref{2.2.8}) yields the following
exact solutions of the nonlinear Dirac equation (\ref{2.4.1}):
\index{Exact solutions!of the nonlinear Dirac equation}
\vspace{1.5mm}

\noindent
\underline {the case $k = 1/2$} 
\begin{eqnarray*}
\psi (x) &=& \exp \{ (1/2) \g_1 \g_2 \arctan (x_2/x_1)
+ (1/4) (\g_0 \g_3 - 1) \ln (x_1^2 + x_2^2) \} \ \ \\
& &\times \exp \{ ( i \lbd /4) (\g_3 - \g_0 + 2 \g_1) (\bar\chi \chi )
[ x_0 - x_3 - \ln (x_1^2 + x_2^2)] \} \chi ;\\
\psi (x) &=& \exp \{ (1/2) \g_1 \g_2 \arctan (x_2/x_1)- (1/4) \ln
(x_1^2 + x_2^2 ) \}\\ 
& &\times \exp\{ i \lbd (1 + \alpha^2)^{-1} (\bar\chi \chi )(\alpha \g_2 -
\g_1) [ \alpha \arctan (x_2/x_1) \\
& & - (1/2) \ln ( x_1^2  + x_2^2 )] \} \chi ;
\end{eqnarray*}
\underline {the case $k = 3/2$}
\begin{eqnarray*}
\psi (x) &=& \exp \{ -(1/2) \g_1 (\g_0 - \g_3) x_1 (x_0
- x_3)^{-1} \} \\
& &\times \exp \{ - (1/4) (\g_0 \g_3 + 3) \ln (x_0 - x_3) \} 
\exp \{ - (i \lbd /2) ( \bar\chi \chi)^{1/3}\quad \ \, \\ 
& &\times \g_0 [(x_0^2 - x_1^2 - x_3^2) (x_0 - x_3)^{-1} +
\ln (x_0 - x_3)] \} \chi,
\end{eqnarray*}
\underline{the case $k = 5/2$}
\begin{eqnarray*}
\psi (x) &=& \exp \{-(1/2) \g_1 (\g_0 - \g_3)
x_1 (x_0 - x_3)^{-1} - (1/2) \g_2 (\g_0 - \g_3)\\ 
& &\times x_2 (x_0 - x_3)^{-1} \} 
\exp \{ - (1/4) ( \g_0 \g_3 + 5) \ln (x_0 - x_3) \}\\ 
& &\times\exp \{ - (i \lbd /2) ( \bar\chi \chi )^{1/5} \g_0 [ x \cdot
x (x_0 - x_3 )^{-1 } + \ln (x_0 - x_3 )] \} \chi ;\ \ \;
\end{eqnarray*}
\underline {the case of arbitrary $k$ }
\begin{eqnarray*}
\psi (x) &=& \exp \{ - (1/2) \g_1 ( \g_0 - \g_3)
x_1 (x_0 - x_3)^{-1} - k \ln (x_0 - x_3) \}\\ 
& &\times \exp \{ [ (1/2 \alpha) \g_1 (\g_0 - \g_3) (1 - 2k ) 
 - i \lbd \alpha^{-2} ( \g_0 - \g_3 \\
& &+ \alpha \g_1 )( \bar\chi \chi)^{1/2k} ][\ln (x_0 - x_3) 
+ \alpha x_1 (x_0 - x_3)^{-1}] ) \} \chi ;\\
\psi (x) &=& \exp \{ (1/2 \beta )
\g_1 (\g_0 - \g_3) (x_2 - \beta x_1 ) (x_0 - x_3)^{-1} \\
& &- k \ln (x_0 - x_3) \} \exp \Bigl\{ \Bigl((1/2 \beta)
(1 + \beta^2)^{-1} [(2 \beta^2 k + 1 ) \g_1\\ 
& &- \beta (2k + 1) \g_2 ] (\g_0 - \g_3) - i \lbd (1 + \beta^2)^{-1}
[\g_2 - \beta \g_1 \\
& &- (\beta/\al)(\g_0 - \g_3)]( \bar\chi \chi)^{1/2k} \Bigr) 
[( x_2 - \beta x_1) (x_0 - x_3)^{-1} \\
& &- (\beta/\al)\ln (x_0 - x_3)] \} \chi ;\\
\psi (x)&=&(x_0-x_3)^{-k}\Bigl(f_1+f_2\g_2+f_3(\g_0-\g_3)
+f_4\g_2(\g_0-\g_3)\Bigr)\chi,\\
& & {\rm where}\ f_i=f_i[x_2(x_0-x_3)^{-1}]\ {\rm are\ 
determined\ by}\ (\ref{2.4.52z}),\\
& & (\ref{2.4.53z})\ {\rm with}\ \theta=k;\\
\psi (x)&=&(x_0-x_3)^{-k}\exp \{-(1/2)
\g_1(\g_0-\g_3)x_1(x_0-x_3)^{-1}\}\\
& &\times \Bigl(f_1+f_2\g_2+f_3(\g_0-\g_3)
+f_4\g_2(\g_0-\g_3)\Bigr)\chi,\\
& & {\rm where}\ f_i=f_i[x_2(x_0-x_3)^{-1}]\ {\rm are\ determined\
  by}\ (\ref{2.4.52z}),\\
& &(\ref{2.4.53z})\ {\rm with}\ \theta=(2k-1)/2;\\
\psi (x)&=&(x_0-x_3)^{-k}\exp \{(1/2)
(\g_0-\g_3)[\g_1x_1(x_0-x_3)-\g_2\\
& &\times\ln (x_0-x_3)]\}\Bigl(f_1+f_2\g_2+f_3(\g_0-\g_3)
+f_4\g_2(\g_0-\g_3)\Bigr)\chi,\\
& & {\rm where}\ f_i=f_i[\ln (x_0-x_3)+x_2(x_0-x_3)^{-1}-1]\ 
{\rm are\ determined}\\ & &{\rm by}\
(\ref{2.4.52z}),\ (\ref{2.4.53z})\ {\rm with}\ \theta=(2k-1)/2.
\end{eqnarray*}

{\bf 3. Conformally-invariant solutions of the nonlinear
  Dirac-G\"ursey equation.} \ Substitution of the $C(1,3)$-invariant
Ans\"atze for spinor field listed in (\ref{2.2.30}) into the
Dirac-G\"ursey equation yields systems of ODEs (\ref{2.3.22}) with $ R
= \lbd ( \bar\vp \vp)^{1/3} $.

In spite of the extremely complicated structure of equations
(\ref{2.3.22}) some of them can be integrated in quadratures
within the framework of the above described approach.
\vspace{1.5mm}

\noindent
{\bf Lemma 2.4.3.}\ {\em The quantities
\begin{eqnarray*}
& &I_3 = \bar\vp \vp, \quad I_4 = \bar\vp \vp \exp \{ 4 \om \},\\
& &I_8 = \bar\vp \vp\, \om^{-3/2},\quad 
I_9 = \bar\vp \vp,\\
& &I_{10} = \bar\vp \vp\, \om^{1/2} (\om - 4)^{1/2} [\om^{1/2}
+ (\om - 4)^{1/2} ]
\end{eqnarray*}
are the first integrals of the systems of ODEs 
3, 4, 8--10 from (\ref{2.3.22}).}

We will prove the lemma for the system of ODEs 8, other systems
are  treated in the same way.

Multiplying the mentioned system by the matrix $ - ( i /2\om) \g_2 $ on
the left yields 
\begin{equation}
\dot \vp = (3/4) \om^{-1} \vp - (i \lbd /2) \om^{-7/6}
(\bar\vp \vp)^{1/3} \g_2 \vp,
\label{2.4.50}
\end{equation}
the conjugate spinor satisfying the equation
\begin{equation}
\dot{\bar{\vp}} = (3/4) \om^{-1}
\bar\vp + (i \lbd /2) \om^{-7/6}
(\bar\vp \vp)^{1/3} \bar\vp \g_2.
\label{2.4.51}
\end{equation}

Multiplying (\ref{2.4.50}) by $ \bar\vp $ on the left, (\ref{2.4.51})
by $ \vp $ on the right and summing we come to the ODE for $ \bar\vp
\vp $ 
\begin{displaymath}
( \bar\vp \vp )^{\edi{.}} = (3/2\om)\bar\vp \vp,
\end{displaymath}
whence $ \bar\vp \vp= C\om^{3/2} $ or $ \bar\vp \vp\, \om^{-3/2} =
C = \mbox{\rm const}$. The assertion is proved. $\rhd$

Applying the above lemma we can construct general solutions
of nonlinear systems of ODEs 3, 4, 8--10 from (\ref{2.3.22}) with
the help of normalized solutions of their linearized
versions. And what is more, normalized solutions of the linearized
systems of ODEs 3, 8--10 can be obtained in explicit form.
This fact enables us to integrate in quadratures the systems
of nonlinear ODEs 3, 8--10 from (\ref{2.3.22}).
\begin{eqnarray*}
\vp (\om) &=& \exp \{ i \lbd (\bar\chi \chi )^{1/3}
(\g_2 + \g_3 - \g_0 ) \om \} \chi,\\
\vp (\om) &=& \om^{3/4} \exp \{ (3 i \lbd /2)
( \bar\chi \chi )^{1/3} \g_2 \om^{1/3} \} \chi,\\
\vp (\om) &=& \exp \{ - i \lbd ( \bar\chi \chi)^{1/3} \g_1 \om \} \chi
,\\ 
\vp (\om) &=& \om^{-1/4} (\om - 4)^{-1/4} \Bigl(\om^{1/2} + (\om -
4)^{1/2} \Bigr)^{-1/2} \\
& &\times \exp \biggl\{ - i 2^{-4/3} \lbd (\bar\chi \chi )^{1/3} 
\g_2 \int^\om z^{-2/3}(z - 4)^{-2/3} dz \biggr\} \chi,
\end{eqnarray*}
where $ \chi $ is an arbitrary constant four-component column.

Substitution of the above expressions into the corresponding Ans\"atze
for the spinor field $\psi(x)$ listed in (\ref{2.2.30}) yields four
classes of exact solutions of the confor\-mal\-ly-invariant nonlinear
Dirac-G\"ursey equation (\ref{1.2.23})
\index{Exact solutions!of the Dirac-G\"ursey equation}
\begin{eqnarray*}
\psi (x) &=& [1 + (x_0 - x_3)^2]^{-1} R [\arctan (x_0 - x_3)] 
\exp \{ - (1/2) \g_1 \g_2
\\
&
&
\times  \arctan (x_0 - x_3) \} \exp \{ - (1/2)
\g_1 (\g_0 - \g_3) \arctan (x_0 - x_3) \}
\\
&
&
\times   \exp \{ -
(1/2) \g_2 (\g_0 - \g_3) [x_2 (x_0 - x_3) - x_1]
\\
&
&
\times [1 + (x_0 - x_3)^2]^{-1} \} \exp \Bigl\{ i \lbd (\bar
\chi \chi)^{1/3} ( \g_2 + \g_3 - \g_0) 
\\
&
&
\times \Bigl(- \arctan (x_0 - x_3) + [x_1 (x_0 - x_3) + x_2]
[1 + (x_0 - x_3)^2 ]^{-1} \Bigr) \Bigr\} \chi,
\\
\psi (x) &=& (x_1^2 + x_2^2)^{-1/4} [1 + (x_0 - x_3)^2 ]^{-3/4}
R[\arctan (x_0 - x_3)]  
\\
&
&
\times \exp \{ - (1/2)\g_1 \g_2 \arctan (x_1/x_2)\}
\exp \{ (3 i \lbd /2) (\bar\chi \chi )^{1/3} \g_2 
\\
&
&
\times (x_1^2 + x_2^2)^{1/3}[1 + (x_0 - x_3)^2]^{-1/3} \} \chi, 
\\
\psi(x) &=& [1 + (x_0 - x_3)^2]^{-1} R[\arctan (x_0 - x_3)] 
\exp \{ -(1/2) \g_1 \g_2 
\\
&
&
\times \arctan (x_0 - x_3) \} 
\exp \{-(1/2) \g_1 (\g_0 - \g_3) [x_1 (x_0 - x_3) + x_2 ]
\\
&
&
\times [1 + (x_0 - x_3)^2 ]^{-1} \} 
\exp \{ - i \lbd (\bar\chi \chi )^{1/3} \g_1 [x_2 (x_0 - x_3) - x_1]
\\
&
&
\times[1 + (x_0 - x_3)^2]^{-1} \} \chi,
\\
\psi(x) &=& (x_1^2 + x_2^2)^{-1} \{\cos (\tau_2 /2) \cos (\tau_3 /2) 
+ \g_0 \g_3 \sin (\tau_2 /2) \sin (\tau_3 /2)
\\
&
&
 + \g \cdot x [\g_0 \sin (\tau_2 /2) 
\cos (\tau_3 /2) - \g_3 \cos (\tau_2 /2 ) \sin (\tau_3 /2)]\} 
\\
&
&
\times \exp \{-(1/2) \g_1 \g_2 \arctan (x_1/x_2) \} \om^{-1/4}
(\om - 4)^{-1/4} 
\\
&
&
\times [\om^{1/2} + (\om - 4)^{1/2}]^{-1/2} \exp \biggl\{-i 2^{-4/3}
\lbd (\bar\chi \chi)^{1/3}\g_2 \int^\om z^{-2/3}
\\
&
& 
\times(z - 4)^{-2/3} dz \biggr\} \chi,
\end{eqnarray*}
where the following notations are used
\begin{eqnarray*}
& &R(\tau) = \cos^2(\tau/2) + \g_0 \g_3 \sin^2 (\tau/2) 
+ (1/2) \g \cdot x
(\g_0 - \g_3) \sin \tau,\\
& &\tau_2 = \arctan [(x \cdot x - 1) (2x_0)^{-1}] + \pi /2 ,\\
& &\tau_3 = \arctan [(x \cdot x + 1) (2x_3)^{-1}] - \pi /2,\\
& &\om = [4 x_0^2 + (x \cdot x - 1)^2 ] (x_1^2 + x_2^2)^{-1}.
\end{eqnarray*}
\vspace{2mm}

\noindent
{\bf 4. Exact solutions of equation (\ref{2.4.2}).}\ To construct
exact solutions of system of nonlinear PDEs (\ref{2.4.2}) we use
$P(1,3)$-invariant Ans\"atze for the spinor field (\ref{2.2.8}) and
Ans\"atze (2.3.16).  Omitting intermediate computations we give the
$P(1,3)$-ungenerable families of exact solutions of the nonlinear
spinor equation (\ref{2.4.2}) (see, also, \cite{87,89}):
\begin{eqnarray*}
\psi_1 (x) &=& \exp \{ - i \theta (\g \cdot a) (a \cdot x) \} \chi,
\\
\psi_2 (x) &=& \exp \{ i \theta (\g \cdot b) (b \cdot x) \} \chi,
\\
\psi_3 (x) &=& \exp \{ (1/2) (\g \cdot a) (\g \cdot d) 
\ln (a \cdot z + d \cdot z) \}
\\
&
&
\times \exp \{ i \g \cdot c [ \theta - (i /2) (\g \cdot a +
\g \cdot d)] c \cdot z \} \chi,
\\
\psi_4 (x) &=& \exp \{ -(1/2) (\g \cdot a + \g \cdot d)(\g \cdot b)
 (a \cdot z + d \cdot z) \} 
\\
&
&
\times \exp \{ (i \theta /2)(\g \cdot b) [2b \cdot z +
(a \cdot z + d \cdot z)^2 ] \} \chi,
\\
\psi_5 (x) &=& \exp \{ -(1/2) (\g \cdot a + \g \cdot d)(\g \cdot b)
 (a \cdot z + d \cdot z) \} 
\\
&
&
\times \exp \{ (i \theta /2) (1 + \al^2)^{-1} (\g \cdot c
- \al \g \cdot b) 
\\
&
&
\times [2 (c \cdot z - \al b \cdot z)- \al (a \cdot z +
d \cdot z)^2 ]\} \chi,
\\
\psi_6 (x) &=& \exp \{ (1/2) [b \cdot z - \ln (a \cdot z +
d \cdot z)](a \cdot z + d \cdot z)^{-1} 
\\
&
&
\times  (\g \cdot a + \g \cdot d) \g \cdot b \} 
\exp \{ (1/2) (\g \cdot a) (\g \cdot d) \ln (a \cdot z +
d \cdot z) \} 
\\
&
&
\times \exp \Bigl\{ \Bigl( \g \cdot c (\g \cdot a + \g \cdot d)
+ i \theta [\g \cdot c - \beta (\g \cdot a + \g \cdot d)]\Bigr)
\\
&
&
\times ( c \cdot z - \beta \ln [a \cdot z +
d \cdot z ]) \Bigr\} \chi,
\\
\psi_7 (x) &=& \exp \{ [ -(1/2) (\dot w_1 \g \cdot b +
\dot w_2 \g \cdot c) + w_3 \g_4](\g \cdot a + \g \cdot d) \}
\\
&
&
\times  \exp
\{ i \theta \g \cdot b ( b \cdot z + w_1) \} \chi,
\\
\psi_8 (x) &=& [(b \cdot z + w_1)^2 + (c \cdot z + w_2)^2]^{-1/4} 
\\
&
&
\times\exp \{ (-(1/2) [\dot w_1 \g \cdot b+ \dot w_2 \g \cdot c)
+ w_3 \g_4] (\g \cdot a + \g \cdot d) \} 
\\
&
&
\times \exp \{ - (1/2)(\g \cdot b ) (\g \cdot c) \arctan [(b \cdot z +
w_1)/ (c \cdot z + w_2)] \} 
\\
&
&
\times \exp \{ i \g \cdot c f [(b \cdot z + w_1)^2 +
(c \cdot z + w_2)^2 ] \} \chi.
\end{eqnarray*}

Here we use the following notations:
\begin{displaymath}
f(\om) = \cases{m \om^{1/2} + \lbd (1 - k)^{-1}
(\bar\chi \chi )^k \om^{(1-k)/2},\quad k \not= 1, \cr
m \om^{1/2} + (\lbd /2) (\bar\chi \chi )\ln \om,\quad
k = 1; \cr}
\end{displaymath}
$z_\mu = x_\mu + \theta_\mu ;\ \theta = m + \lbd (\bar\chi \chi)^k$;
$w_a = w_a (d \cdot z + d \cdot z) $ are arbitrary smooth real-valued 
functions; $\al, \ \beta, \ \theta_\mu $ are real
constants.

As earlier, we denote by $a_\mu, \ b_\mu, \ c_\mu, \ d_\mu $
arbitrary real parameters satisfying (\ref{2.4.40z}).
\vspace{10mm}

\noindent
{\large\bf 2.5. Nonlinear spinor equations and special
  functions\label{s2.5}} 

\markboth{Chapter 2. EXACT SOLUTIONS}
{2.5. Nonlinear spinor equations and special functions}
\def\theequation{2.\arabic{section}.\arabic{equation}}
\setcounter {section} {5}
\setcounter {equation}{0}
\vspace{7mm}

\noindent
Here we will establish a rather unexpected fact: there exists a
correspondence between exact solutions of the nonlinear Dirac 
equation\index{Nonlinear!Dirac equation}
\begin{equation}
\{i\ga_{\mu}\,\pa_{\mu}-F(\bar\psi\psi)\}\psi (x)=0,
\label{2.5.1}
\end{equation}
where $F\in C^1\,(\R^1,\, \R^1)$, and special functions satisfying a
second-order ODE of the form
\begin{equation}
\ddot U+a_1(\omega)\dot U+ a_2(\omega) U = 0.
\label{2.5.2}
\end{equation}

The above facts enable us to construct exact solutions of
equation (\ref{2.5.1}) in terms of the Weierstrass, Gauss and 
Chebyshev-Hermite functions.

To obtain exact solutions of PDE (\ref{2.5.1}) we use the following
Ans\"atze:
\begin{equation}
\begin{array}{rcl}
\psi(x)&=&\exp \{(1/2)x_1 (x_0+x_3)^{-1} (\ga_0+\ga_3)\ga_1\}\\
& &\times \exp \{(1/2) \ga_0\ga_3 \ln (x_0+x_3)
\} \varp\, (x_0^2-x_1^2-x_3^2),
\end{array}
\label{2.5.3}
\end{equation}
\begin{equation}
\begin{array}{rcl}
\psi (x)&=&\exp  \{(1/2)\,(x_0+x_3)^{-1}(\ga_0+\ga_3)
(\ga_1x_1+\ga_2x_2)\}\\
& &\times \exp \{(1/2) \ga_0\ga_3 \ln (x_0+x_3)
\}\varp\,(x\cdot x).
\end{array}
\label{2.5.4}
\end{equation}

Substituting (\ref{2.5.3}), (\ref{2.5.4}) into the initial equation
(\ref{2.5.1}) we get systems of ODEs for the four-component functions  
$\varp=\varp\,(\omega)$
\begin{equation}
4\omega\dot\varp=-\Bigl\{n (1+\ga_0\ga_3)+
iF (\bar{\varp}\varp)\Bigl (\omega (\ga_0+\ga_3)+
\ga_0-\ga_3\Bigr )\Bigr\}\varp,
\label{2.5.5}
\end{equation}
where the cases $n=2$ and $n=3$ correspond to Ans\"atze (\ref{2.5.3})
and (\ref{2.5.4}) accordingly.

The equation for the conjugate spinor
$\bar{\varp}$ has the form
\begin{equation}
4\omega\dot{\bar{\varp}}=-\bar{\varp}\Bigl\{n(1-\ga_0\ga_3)-
iF(\bar{\varp}\varp)\Bigl (\omega (\ga_0+\ga_3)+
\ga_0-\ga_3\Bigr )\Bigr\}.
\label{2.5.6}
\end{equation}

Multiplying equation (\ref{2.5.5}) by
$\bar{\varp}$, equation (\ref{2.5.6}) by $\varp$ and summing yield
the ODE for a scalar function $\bar{\varp}\varp$
\begin{displaymath}
4\om(\bar{\varp}\varp)^{\edi{.}} = -2n\bar{\varp}\varp,
\end{displaymath}
which general solution reads
\begin{equation}
\bar{\varp}\varp=C\om^{-n/2},
\ \
C= \mbox{\rm const}.
\label{2.5.7}
\end{equation}

Thus, equation (\ref{2.5.5}) is reduced to the linear ODE
\begin{equation}
4\omega\dot{\varp}=-\Bigl\{n(1+\ga_0\ga_3)+
iF(C\om^{-n/2})\Bigl (\omega (\ga_0+\ga_3)+
\ga_0-\ga_3\Bigr )\Bigr\}\varp
\label{2.5.8}
\end{equation}
with the nonlinear additional condition (\ref{2.5.7}).

If we choose $\ga$-matrices in the representation (\ref{1.1.7}),
then equation (\ref{2.5.8}) in component-wise notation takes the form
\begin{eqnarray*}
& &2\dot\varp^0=-iF(C\om^{-n/2})\varp^2,\quad
2\om\dot\varp^1=-iF(C\om^{-n/2})\varp^3-n\varp^1,\\
& &2\dot\varp^3=-iF(C\om^{-n/2})\varp^1,\quad
2\om\dot\varp^2=-iF(C\om^{-n/2})\varp^0-n\varp^2.
\end{eqnarray*}

On making the change of the independent variable
\begin{equation}
t=C \om^{-n/2},
\quad
\om=(t/C)^{-2/n},
\label{2.5.9}
\end{equation}
we get
\begin{equation}
\begin{array}{l}
n C^{-2/n} t^{(n+2)/n} \varp_t^0=iF(t)\varp^2,\quad
nt\varp_t^1=iF (t) \varp^3+n\varp^1,\\[2mm]
n C^{-2/n} t^{(n+2)/n} \varp_t^3=i F (t) \varp^1,\quad
nt\varp_t^2=iF (t) \varp^0+n\varp^2.
\end{array}
\label{2.5.10}
\end{equation}

System of ODEs (\ref{2.5.10}) by means of the change of the
independent variable 
\begin{equation}
\xi=\int\limits_{\edi a}^{\edi t} F (\tau) \tau^{-n/2} d\tau
\label{2.5.11}
\end{equation}
is reduced to the form
\begin{equation}
\begin{array}{l}
nC^{-2/n}\varp_{\xi}^0=it^{-1}\varp^2,\quad
nt^{(n-2)/n}\varp_{\xi}^1=i\varp^3+nF^{-1}(t)\varp^1,\\[2mm]
nC^{-2/n}\varp_{\xi}^3=it^{-1} \varp^1,\quad
nt^{(n-2)/n}\varp_{\xi}^2=i\varp^0+nF^{-1}(t)\varp^2.
\end{array}
\label{2.5.12}
\end{equation}

Differentiating the first equation with respect to
$\xi $ we get a second-order ODE of the form
\begin{equation}
R_{\xi\xi}+C^{2/n}n^{-2}t^{2(1-n)/n}R=0,
\label{2.5.13}
\end{equation}
where the function
$t=t (\xi)$ is determined by (\ref{2.5.11}).

Consequently, system (\ref{2.5.12}) is equivalent to the 
following second-order system of ODEs:
\begin{equation}
\begin{array}{l}
\varp^0_{\xi\xi}+C^{2/n}n^{-2} \Bigl (t (\xi)
\Bigr )^{2(1-n)/n} \varp^0=0,\quad
\varp^1=-i n\, t(\xi) C^{-2/n} \varp^3_{\xi},\\[2mm]
\varp^3_{\xi\xi}+C^{2/n}n^{-2} \Bigl (t (\xi)\Bigr )^{2(1-n)/n} 
\varp^3=0,\quad
\varp^2=-i n\, t(\xi) C^{-2/n} \varp^0_{\xi}.
\end{array}
\label{2.5.14}
\end{equation}

Let $u(\xi),\ v(\xi)$\  be a fundamental system of
solutions of equation (\ref{2.5.13}). Then, the general
solution of system (\ref{2.5.14}) is represented in the form
\begin{equation}
\begin{array}{rcl}
\varp^0&=&\chi^0u(\xi)+\chi^2v(\xi),\\[2mm]
\varp^1&=&-i n\, t(\xi) C^{-2/n} \Bigl (\chi^3
\dot u(\xi)+\chi^1\dot v(\xi)\Bigr ),\\[2mm]
\varp^2&=&-i n\, t(\xi) C^{-2/n} \Bigl (\chi^0
\dot u(\xi)+\chi^2\dot v(\xi)\Bigr ),\\[2mm]
\varp^3&=&\chi^3u(\xi)+\chi^1v(\xi),
\end{array}
\label{2.5.15}
\end{equation}
where $\chi^0,\ \chi^1,\ \chi^2,\ \chi^3$ are arbitrary complex
constants. 

Formulae (\ref{2.5.15}) give the general solution of system of
nonlinear ODEs (\ref{2.5.5}) if equality (\ref{2.5.7}) holds.
Substitution of (\ref{2.5.15}) into (\ref{2.5.7}) gives rise to the
following relation for $C,\ \chi^{\mu}$:
\begin{eqnarray*}
\bar{\psi}\psi&=&\psi^{0*}\psi^2+\psi^{2*}\psi^0+\psi^{1*}\psi^3+
\psi^{3*}\psi^1=i n t C^{-2/n} \\
& &\times(\chi^0\chi^{2*}- \chi^2\chi^{0*}+ \chi^3\chi^{1*}-
\chi^1\chi^{3*}) w(u,v)=t=C\om^{-n/2}.
\end{eqnarray*}

Here $w(u,v)=u\dot v-\dot u v$ is the Wronskian of the fundamental
system of solutions of equation (\ref{2.5.13}) which is constant for
any $u,\ v$ satisfying (\ref{2.5.13}).

The above relation is rewritten in the form
\begin{equation}
C=\{in (\chi^0\chi^{2*}- \chi^2\chi^{0*}+
\chi^3\chi^{1*}- \chi^1\chi^{3*}) w(u,v)\}^{n/2}.
\label{2.5.16}
\end{equation}

It is well-known that each ODE of the form (\ref{2.5.2}) is
transformed to equation (\ref{2.5.13}) by an appropriate change of
variables (one has to take into account that the function $t=t(\xi)$
depends on arbitrary function $F$). Consequently, choosing the
function $F (\bar{\varp}\varp)$ in an appropriate way we can obtain
exact solutions of the nonlinear Dirac equation in terms of any
special function described by equation (\ref{2.5.2}).

We will consider several particular cases of equation (\ref{2.5.13}).
First of all, we recall that solutions of equation (\ref{2.5.13})
(and, consequently, solutions of the nonlinear Dirac equation
(\ref{2.5.1}) of the form (\ref{2.5.3}), (\ref{2.5.4})) under $F=\la
(\bar{\varp}\varp) ^{1/2k}$ are expressed in terms of the Bessel
functions\index{Bessel!function} (see Section 2.4).  
\vspace{1.5mm}

\noindent
{\bf 1.}\  Choosing 
\begin{equation}
n^{-2}C^{2/n}t^{2(1-n)/n}=
2N+1-\xi^2,\ \ N\in \N,
\label{2.5.17}
\end{equation}
in (\ref{2.5.13}) yields the Weber equation\index{Weber equation}
\begin{displaymath}
\ddot R+(2N+1-\xi^2) R =0.
\end{displaymath}

The fundamental system of solutions of the above equation reads
\cite{132}
\begin{equation}
\begin{array}{rcl}
u (\xi)&=&\exp \Bigl\{-(1/2)\xi^2\Bigr\} H_N (\xi),\\[2mm]
v (\xi)&=& u (\xi)\int\limits_{\edi a}^{\edi \xi} \Bigl (u (\tau)\Bigr
)^{-2} d\tau,\quad a= \mbox{\rm const},
\end{array}
\label{2.5.18}
\end{equation}
where
\begin{displaymath}
H_N (\xi)=(-1)^N \exp \{\xi^2\} {d^N\over d\xi^N} \exp \{-\xi^2\}
\end{displaymath}
is the Chebyshev-Hermite 
polynomial\index{Chebyshev-Hermite po\-ly\-no\-mi\-al}. 

It is not difficult to verify that functions (\ref{2.5.18})
satisfy the identity 
\begin{displaymath}
w (u,v)=1.
\end{displaymath}

Thus, substitution of formulae (\ref{2.5.15}), (\ref{2.5.18}) into
(\ref{2.5.3}), (\ref{2.5.4}) with account of (\ref{2.5.16}) under
$w (u,v)=1$ gives rise to a class of the exact solutions of the
nonlinear Dirac equation in terms of the Chebyshev-Hermite polynomials  
and what is more
\begin{equation}
\xi^2=2N+1-n^{-2}C^{(4-2n)/n} \om^{n-1}.
\label{2.5.19}
\end{equation}

To obtain an explicit form of
$F=F (t)$ we differentiate equality (\ref{2.5.17}) with respect
to $t$
\begin{displaymath}
2 (1-n) n^{-3} C^{2/n}t^{(2-3n)/n}=-2\xi
 {d\xi\over dt},
\end{displaymath}
whence it follows that
\begin{displaymath}
F (t)=(n-1) n^{-3}C^{2/n}t^{(4-3n)/n}\Bigl(
2N-n^{-2}C^{2/n}t^{2(1-n)/n}\Bigr)^{-1/2}.
\end{displaymath}

Let us note that under $n=3$,\ $i (\chi^0\chi^{2*}-\chi^2\chi^{0*}+
\chi^3\chi^{1*}-\chi^1\chi^{3*})<0$ the solution obtained is localized
in the Minkowski space with exception of the hyperplane $x_3=- x_0$,
where it has a non-integrable singularity.  \vspace{1.5mm}

\noindent
{\bf 2.}\  If we choose 
\begin{equation}
n^{-2}C^{2/n}t^{2(1-n)/n}=-(3/4) {\rm We} (\xi),
\label{2.5.20}
\end{equation}
where ${\rm We} (\xi)$ is the Weierstrass 
function\index{Weierstrass function} having the invariants 
$\om_1,\ \om_2$, in (\ref{2.5.13}), 
then the Lam\'e equation\index{Lam\'e equation} is obtained
\begin{equation}
\ddot R-(3/4) {\rm We} (\xi) R=0.
\label{2.5.21}
\end{equation}

The fundamental system of solutions of ODE (\ref{2.5.21}) is as
follows \cite{132} 
\begin{equation}
\begin{array}{rcl}
u (\xi)&=&\{{\rm\dot We} (\xi/2)\}^{-1/2},\\[2mm]
v (\xi)&=&
{\rm We} (\xi/2) \{{\rm\dot We} (\xi/2)\}^{-1/2}
\end{array}
\end{equation}
and what is more $w (u,v)=1/2$. Hence, using formulae (\ref{2.5.3}),
(\ref{2.5.4}), (\ref{2.5.15}), (\ref{2.5.16}) (under $w (u,v)=1/2$) we
obtain the exact solutions of the initial PDE (\ref{2.5.1}) in terms of
the Weierstrass function, the equalities 
\begin{eqnarray*}
& &\xi=\int\limits_{-\edi\infty}^{\displaystyle
  (4/3)n^{-2}C^{(4-2n)/n}\om^{n-1}} 
(-4\tau^3+\om_1\tau-\om_2)^{-1/2}d\tau,\\
& &F (t)=(4/3) (n-1) n^{-2}C^{2/n}t^{(2-3n)/n}
\Bigl\{-4\Bigl ((4/3) n^{-2}C^{2/n}\\
& &\quad
\times t^{2(1-n)/n}\Bigr )^3 +(4/3) \om_1n^{-2}C^{2/n}
t^{2(1-n)/n}-\om_2\Bigr\}^{-1/2}
\end{eqnarray*}
holding.
\vspace{1.5mm}

\noindent
{\bf 3.}\  Choosing in (\ref{2.5.13})
\begin{displaymath}
n^{-2}C^{2/n}t^{2(1-n)/n}=(1/4) \xi^{-2} 
\{2 (a+b+1)-(a+b+1)^2\}-ab
\end{displaymath}
we get the hypergeometric equation\index{Hypergeometric!equation}
\begin{displaymath}
\ddot R+\Bigl ((1/4) \xi^{-2}[1-(a+b)^2]-ab\Bigr ) R
=0.
\end{displaymath}

The fundamental system of solutions of this equation is as follows
\cite{132} 
\begin{equation}
\begin{array}{rcl}
u (\xi)&=&\xi^{(1+a+b)/2}F (a,\, b,\, a+b+1,\, \xi),\\
v (\xi)&=&\xi^{(1-a-b)/2}F (-b,\, -a\,,1-a-b,\, \xi),
\end{array}
\label{2.5.23}
\end{equation}
where $F=F (a,\, b\,,c,\, \xi)$ is the hypergeometric Gauss 
function\index{Hypergeometric!function} and besides
\begin{equation}
\begin{array}{l}
w (u,v)=(a+b) \Gamma (1+a+b) \Gamma (1-a-b)\\
\quad\times
\Bigl\{\Gamma (1+a) \Gamma (1+b) \Gamma (1-a) \Gamma 
(1-b)\Bigr\}^{-1}.
\end{array}
\label{2.5.24}
\end{equation}

Here $\Gamma=\Gamma (a)$ is the Euler $\ga$-function.

Substitution of formulae (\ref{2.5.15}), (\ref{2.5.16}),
(\ref{2.5.23}), (\ref{2.5.24}) into the An\-s\"atze (\ref{2.5.3}),
(\ref{2.5.4}) yields the exact solutions of the nonlinear Dirac
equation (\ref{2.5.1}), the relations  
\begin{eqnarray*}
& &\xi=(1/2) \Bigl (1-(a+b)^2\Bigr )^{1/2}
\Bigl (n^{-2}C^{(4-2n)/n}\om^{n-1}+ab\Bigr )^{-1/2},\\
& &F (t)= (1/2) n^{-3}(n-1) C^{2/n} \Bigl (1-
(a+b)^2\Bigr )^{1/2}t^{(4-3n)/n}\\
& &\phantom{F(t)=}\times
\Bigl(n^{-2}C^{2/n}t^{2(1-n)/n}+ab\Bigr)^{-3/2}
\end{eqnarray*}
holding.

Let us note that solutions of equation (\ref{2.5.1}) of the form
(\ref{2.5.3}), (\ref{2.5.4}) can be treated as solutions of the linear
Dirac equation 
\begin{equation}
\bigl\{i\ga_{\mu} \pa_{\mu}-U (x)\bigr\} \psi (x)=0
\label{2.5.25}
\end{equation}
with potentials $U (x)=F [C (x_0^2-x_1^2-x_3^2)^{-1}]$ and $U (x)=F [C
(x\cdot x)^{-3/2}]$.  That is why there exists an analogy between
equations (\ref{2.5.1}) and (\ref{2.5.25}). The principal difference
is that in the case of a linear equation the potential characterizes
interaction of the spinor field with some external field (for example,
with the scalar field $u (x)=U (x)$), while in the nonlinear case the
"potential" is determined by self-interaction of the spinor field
$\psi (x)$.  
\vspace{10mm}

\noindent
{\large\bf 2.6. Construction of fields with spins {\boldmath $s=$}\ 0,
  1, 3/2 
\vspace{1.5mm}

\noindent
\phantom{\large\bf 2.6. }via the Dirac field\label{s2.6}}
\markboth{Chapter 2. EXACT SOLUTIONS}
{2.6. Construction of fields with spins $s=0,1,3/2$}
\def\theequation{2.\arabic{section}.\arabic{equation}}
\setcounter {section} {6}
\setcounter {equation}{0}
\vspace{7mm}

\noindent
In \cite{100} we have suggested a purely algebraic method of
construction of Ans\-\"atze for scalar\index{Scalar field},  
vector\index{Vector field} and tensor
fields\index{Tensor field} by the use of Ans\"atze for the spinor 
field $\psi (x)$. The method is based on the following well-known fact: 
provided the spinor field $\psi (x)$ transforms according to
formulae (\ref{1.1.24a})--(\ref{1.1.24c}), then the quantities
\begin{eqnarray}
& &u (x)=\bar{\psi}\psi,\label{2.6.1a}\\
& &A_{\mu} (x)=\bar{\psi} \ga_{\mu} \psi,\label{2.6.1b}\\
& &F_{\mu\nu} (x)=i\bar{\psi} \ga_{\mu}\ga_{\nu} \psi,\label{2.6.1c}
\end{eqnarray}
transform with respect to the Poincar\'e group as the scalar, 
vector and second-rank tensor correspondingly.  Consequently,
substitution of the $P(1,3)$-invari\-ant Ans\"a\-tze for $\psi (x)$
obtained in Section 2.2 into formulae
(\ref{2.6.1a})--(\ref{2.6.1c}) with subsequent replacement 
$\bar{\varp}\varp\to B (\om)$, \ $\bar{\varp}\ga_{\mu}\varp\to B_{\mu}
(\om)$, \ $i\bar{\varp}\ga_{\mu}\ga_{\nu}\varp\to B_{\mu\nu} (\om)$
yields the Ans\"atze for the scalar, vector and tensor
fields invariant under the one- and three-dimensional
subalgebras of the algebra $AP(1,3)$.

It is worth noting that the above described procedure of 
construction of invariant Ans\"atze is much simpler than 
integration of system of PDEs (\ref{1.5.15}), (\ref{1.5.16}).

Furthermore, if we substitute Ans\"atze for $\psi (x)$ invariant under
one- and three-dimensional subalgebras of the Lie algebra of the 
extended Poincar\'e group $A\wid P (1,3)$ into formulae 
(\ref{2.6.1a})--(\ref{2.6.1c}),
then $\wid P (1,3)$-invariant Ans\"atze for fields $u (x)$,\ 
$A_{\mu} (x)$,\ $F_{\mu\nu} (x)$ are obtained.

To construct conformally-invariant Ans\"atze for the scalar, 
vector and ten\-sor fields we introduce into formulae
(\ref{2.6.1a})--(\ref{2.6.1c}) the normalizing factors of the form
$(\psi \bar{\psi})^\alpha$
\begin{eqnarray}
& &u (x)=(\bar{\psi}\psi)^{1/3},\label{2.6.2a}\\
& &A_{\mu} (x)=\bar{\psi} \ga_{\mu} \psi 
(\bar{\psi}\psi)^{-2/3},\label{2.6.2b}\\
& &F_{\mu\nu}(x)=i\bar{\psi} \ga_{\mu}\ga_{\nu} \psi (\bar{\psi}
\psi)^{-1/3}\label{2.6.2c}
\end{eqnarray}
(it is not difficult to ascertain that the fields
$u (x)$,\ $A_{\mu} (x)$ transform according
to formulae (\ref{1.4.5}), (\ref{1.4.14}) provided the spinor field
$\psi (x)$ transforms according to (\ref{1.1.24e})).

We apply the procedure described to obtain Poincar\'e-invariant
Ans\"atze for the vector field $A_{\mu} (x)$ which reduce
the corresponding $P(1,3)$-invariant system of PDEs to ODEs. Before 
substituting Ans\"atze for $\psi (x)$ into formula (\ref{2.6.1b}) 
we generate them by transformations from the Poincar\'e group 
(formulae (\ref{2.4.35})--(\ref{2.4.37})). Substitution of 
$P(1,3)$-ungenerable Ans\"atze for the spinor field $\psi (x)$ 
into (\ref{2.6.1b}) yields $P(1,3)$-ungenerable Ans\"atze for the 
vector field $A_{\mu} (x) $ that can be represented in the 
following unified 
form:\index{Ansatz!for vector field}\index{Ansatz!$P(1,3)$-invariant}
\begin{eqnarray}
A_{\mu}(x) &=& \Bigl\{(a_{\mu}a_{\nu} - d_{\mu}d_{\nu})\cosh\theta_{0} 
+ (d_{\mu}a_{\nu} - d_{\nu}a_{\mu})\sinh\theta_0 \non\\
& &+2(a_{\mu} + d_{\mu})[(
\theta_{1}\cos\theta_{3} + \theta_{2}\sin\theta_{3})b_{\nu} +
( \theta_{2}\cos\theta_{3} \non\\
& &-\theta_{1}\sin\theta_{3})c_{\nu}+(\theta_{1}^{2} 
+ \theta_{2}^{2})e^{-\theta_{0}}(a_{\nu} +
d_{\nu})] + (b_{\mu}c_{\nu} \label{2.6.3}\\
& &- b_{\nu}c_{\mu})\sin\theta_{3} -
(c_{\mu}c_{\nu} + b_{\mu}b_{\nu}) 
\cos\theta_{3} - 2 e^{-\theta_{0}}\non\\
& &\times(\theta_{1}b_{\mu} + \theta_{2}c_{\mu})(a_{\nu}
+ d_{\nu})\Bigr\} B^{\nu}(\omega),\non
\end{eqnarray}
where $a_\mu,\ b_\mu,\ c_\mu,\ d_\mu $ are arbitrary real constants
satisfying equalities (\ref{2.4.40z}) and $B^\nu$ are arbitrary smooth 
functions. Explicit forms of the functions $\theta_\mu,\ \omega$
depend on the choice of a three-dimensional subalgebra of
the Poincar\'e algebra (\ref{2.2.7}) and are given below
\begin{eqnarray}
&1)& \theta_{\mu} = 0, \quad \omega = d\cdot z ;\non\\
&2)& \theta_{\mu} = 0, \quad \omega = a\cdot z ;\non\\
&3)&  \theta_{\mu} = 0, \quad\omega = k\cdot z ;\non\\
&4)&  \theta_{0} = -\ln (k\cdot z),
\quad\theta_{1} = \theta_{2} = \theta_{3} = 0,
\quad\omega = (a\cdot z)^{2} - (d\cdot z)^{2} ;\non\\
&5)&  \theta_{0} = -\ln(k\cdot z),
\quad\theta_{1} = \theta_{2} = \theta_{3} = 0,
\quad\omega = b\cdot z ;\non\\
&6)&  \theta_{0} = -\alpha^{-1}(c\cdot z),
\quad\theta_{1} = \theta_{2} = \theta_{3} = 0,
\quad\omega = b\cdot z, \ \ \alpha \not={0} ;\non\\
&7)&  \theta_{0} = -\alpha^{-1}(c\cdot z),
\quad\theta_{1} = \theta_{2} = \theta_{3} = 0,
\quad\omega = \alpha \ln(k\cdot z) - c\cdot z, \ \ \alpha \not={0}
;\non\\ 
&8)&  \theta_{0} = \theta_{1} = \theta_{2} = 0,
\quad\theta_{3} = -\arctan(b\cdot z/c\cdot z),
\quad\omega = (b\cdot z)^{2} + (c\cdot z)^{2} ;\non\\
&9)&  \theta_{0} = \theta_{1} = \theta_{2} = 0,
\quad\theta_{3} = -\alpha^{-1}(a\cdot z),
\quad\omega = d\cdot z, \ \ \alpha \not={0} ;\non\\
&10)& \theta_{0} = \theta_{1} = \theta_{2} = 0,
\quad\theta_{3} = \alpha^{-1}(d\cdot z),
\quad\omega = a\cdot z, \ \ \alpha \not={0} ;\non\\
&11)& \theta_{0} = \theta_{1} = \theta_{2} = 0,
\quad\theta_{3} = (d\cdot z - a\cdot z)/2, \quad\omega = k\cdot z
;\non\\ 
&12)& \theta_{0} = \theta_{2} = \theta_{3} = 0,
\quad\theta_{1} = b\cdot z/2k\cdot z, \quad\omega = k\cdot z ;\non\\
&13)& \theta_{0} = \theta_{2} = \theta_{3} = 0,
\quad\theta_{1} = (\alpha b\cdot z - c\cdot z)
(2 \alpha k\cdot z)^{-1}, \quad\omega = k\cdot z,\non\\
& &\alpha \not={0} ;\non\\
&14)& \theta_{0} = \theta_{2} = \theta_{3} = 0,
\quad\theta_{1} = (c\cdot z)/2,
\quad\omega = k\cdot z ;\label{2.6.3z}\\
&15)& \theta_{0} = \theta_{2} = \theta_{3} = 0,
\quad\theta_{1} = - (k\cdot z)/2, \quad\omega = 2b\cdot z 
+ (k\cdot z)^{2} ;\non\\
&16)& \theta_{0} = \theta_{2} = \theta_{3} = 0,
\quad\theta_{1} = - (k\cdot z)/2,
\quad\omega = 2(c\cdot z - \alpha b\cdot z) - \alpha(k\cdot z)^{2}
;\non\\ 
&17)& \theta_{0} = \alpha^{-1} \arctan(b\cdot z/c\cdot z),
\quad\theta_{1} = \theta_{2} =0,\non\\
& & \theta_{3} = - \arctan(b\cdot z/c\cdot z),
\quad\omega = (b\cdot z)^{2} + (c\cdot z)^{2}, \quad\alpha \not={0}
;\non\\ 
&18)& \theta_{0} = - \ln(k\cdot z),
\quad\theta_{1} = \theta_{2} = 0, \quad\theta_{3} = \alpha\ln(k\cdot
z),\non\\ 
& &\omega = (a\cdot z)^{2} - (d\cdot z)^2 ;\non\\
&19)& \theta_{0} = - \ln(k\cdot z),
\quad\theta_{1} = \theta_{2} = 0,\quad \theta_{3} =
-\arctan(b\cdot z/c\cdot z),\non \\
& &\omega = (b\cdot z)^{2} + (c\cdot z)^{2} ;\non\\
&20)& \theta_{0} = \theta_{3} = 0,
\quad\theta_{1} = b\cdot z/2k\cdot z,
\quad\theta_{2} = c\cdot z/2k\cdot z, \quad\omega = k\cdot z ;\non\\
&21)& \theta_{0} = \theta_{3} = 0,
\quad\theta_{1} = (1/2)[(k\cdot z + \beta)b\cdot z - \alpha c\cdot z]
[k\cdot z(k\cdot z + \beta) \non\\
& & - \alpha]^{-1},\quad \theta_{2} = (1/2)(k\cdot z c\cdot z 
- b\cdot z)[k\cdot z(k\cdot z + \beta) - \alpha]^{-1},\non\\
& &\omega = k\cdot z ;\non\\
&22)& \theta_{0} = \theta_{3} = 0,
\quad\theta_{1} = (1/2k\cdot z)(b\cdot z -
c\cdot z(k\cdot z + \beta)^{-1}), \non\\
& &\theta_{2} = (1/2) c\cdot z(k\cdot z + \beta)^{-1}, 
\quad\omega = k\cdot z;\non\\
&23)& \theta_{0} = \theta_{3} = 0,
\quad\theta_{1} = b\cdot z/2k\cdot z,
\quad\theta_{2} =(1/2) c\cdot z(k\cdot z + 1)^{-1},\non\\
& &\omega = k\cdot z ;\non\\
&24)& \theta_{0} = -\ln(k\cdot z),
\quad\theta_{1} = b\cdot z/2k\cdot z,\non\\
& & \theta_{2} =
\theta_{3} = 0, \quad\omega = (a\cdot z)^{2} 
- (b\cdot z)^{2} - (d\cdot z)^{2} ;\non\\
&25)& \theta_{0} = -\ln(k\cdot z),
\quad\theta_{1} = [b\cdot z - \alpha\ln(k\cdot z)]/2k\cdot z, \non\\
& & \theta_{2} = \theta_{3} = 0, \quad\omega 
= c\cdot z - \beta\ln(k\cdot z) ;\non\\
&26)& \theta_{0} = 0,
\quad\theta_{1} = b\cdot z/2k\cdot z, \quad\theta_{2} 
= c\cdot z/2k\cdot z, \non\\
& & \theta_{3} = -z\cdot z/4k\cdot z, \quad\omega = k\cdot z ;\non\\
&27)& \theta_{0} = -\ln(k\cdot z),
\quad\theta_{1} = b\cdot z/2k\cdot z,
\quad\theta_{2} =
c\cdot z/2k\cdot z,\non\\
& & \theta_{3} = \alpha\ln(k\cdot z), \quad\omega = z\cdot z.\non
\end{eqnarray}
where $k_\mu=a_\mu+b_\mu,\ z_\mu=x_\mu + \tau_\mu,\ 
\tau_\mu =\mbox{\rm const},\ 
\mu={0,\ldots,3}$.

Let us consider an example of construction of an Ansatz for the
vector field $A_{\mu}(x)$ by taking as $\psi(x)$ the Ansatz 
\begin{displaymath}
\psi(x)=\exp \{(1/2) \g_0 \g_3 \ln (x_0 + x_3)\}
        \vp\, (x_0^2 -   x_3^2)
\end{displaymath}
invariant under the three-dimensional algebra $\langle J_{03},\ P_1,\
P_2 \rangle\in AP(1,3)$.

It is not difficult to check that a $P(1,3)$-ungenerable
Ansatz for the spinor field is obtained by making the following
change: 
\begin{eqnarray*}
& &\ga_0\to \ga\cdot a,
\quad
\ga_1\to \ga\cdot b,
\quad
\ga_2\to \ga\cdot c,
\quad
\ga_3\to \ga\cdot d,\\
& &x_0\to a\cdot z,
\quad
x_1\to b\cdot z,
\quad
x_2\to c\cdot z,
\quad
x_3\to d\cdot z
\end{eqnarray*}
in the above Ansatz. 

As a result, we have
\begin{eqnarray*}
\psi(x)&=&\exp \{(1/2)\ga\cdot a\ga \cdot d\ln 
(k\cdot z)\}\varp\\
& &=\{\cosh [(1/2)
\ln (k\cdot z)]+
\ga\cdot a\ga\cdot d\sinh[(1/2)\ln 
(k\cdot z)] \}\varp,\\
\bar{\psi}(x)&=&\bar{\varphi}\exp \{-(1/2)\ga\cdot a\ga \cdot d\ln 
(k\cdot z)\}\\
& &=\{\cosh [(1/2)\ln (k\cdot z)] - \ga\cdot a
\ga\cdot d\sinh[(1/2)\ln (k\cdot z)] \},
\end{eqnarray*}
where $\varp$ is an arbitrary complex-valued four-component
function of $(a\cdot z)^2-(d\cdot z)^2$. 
Substitution of the formulae obtained into (\ref{2.6.1b}) yields 
\begin{eqnarray*}
A_{\mu}(x)&=&\bar{\varp}\{\cosh \theta-\ga\cdot a\ga\cdot d
\sinh\theta\}\ga_{\mu}\{\cosh \theta + \ga\cdot a
\ga\cdot d\sinh\theta\}\varp\\
&=&\bar{\varp}\ga_{\mu}\varp-(\sinh\theta)\bar{\varp}[\ga\cdot a
\ga\cdot d,\, \ga_{\mu}](\cosh\theta+\ga\cdot a\ga\cdot d
\sinh\theta)\varp\\
&=&\bar{\varp}\ga_{\mu}\varp+2a_{\mu}\bar{\varp}
(\ga\cdot d\cosh \theta+\ga\cdot a\sinh\theta)
\varp\sinh\theta -2d_{\mu}\\
& &\times\bar{\varp}(\ga\cdot a\cosh \theta+\ga\cdot
d\sinh\theta)\varp\sinh\theta =\{(a_\mu a_\nu-d_\mu d_\nu)\\
& &\times\cosh 2\theta
+(a_\mu d_\nu-d_\mu a_\nu)\sinh 2\theta
-b_\mu b_\nu - c_\mu c_\nu\}\bar{\varphi}\gamma^\nu\varphi,
\end{eqnarray*}
where $\theta=(1/2)\ln (k\cdot z)$. Designating the real-valued 
functions $\bar{\varp}\ga_{\mu}\varp$ by $B_{\mu},\ \mu={0,\ldots,3}$  
we arrive at the Ansatz 4 from (\ref{2.6.3}).

To obtain from (\ref{2.6.3}) Ans\"atze for the vector field invariant
under the three-dimensional subalgebras of the Poincar\`e algebra
(\ref{2.2.7}) we put 
\begin{displaymath}
\tau_{\mu}=0,\quad a_{\mu}=\delta_{\mu 0},\quad b_{\mu}= -\delta_{\mu
  1},\quad c_{\mu}=-\delta_{\mu 2},\quad d_{\mu}= -\delta_{\mu 3}.
\end{displaymath}

Let us emphasize that the above procedure of construction of Ans\"atze 
for the vector, scalar and tensor fields is based on
transformational properties of the spinor field with respect to the
Poincar\`e group only and the explicit form of the function $\psi(x)$
is not used.  There arises a natural question: what equations are
satisfied by functions $u(x)$,\ $A_{\mu}(x)$,\ $F_{\mu\nu}(x)$
defined by formulae (\ref{2.6.1a})--(\ref{2.6.1c}) provided
$\psi=\psi(x)$ is a solution of the nonlinear Dirac equation? In other
words, is it possible to construct exact solutions of equations
describing fields with spins $s = 0, 1, 3/2,\ldots$ with the help of
exact solutions of a nonlinear PDE for the field with the spin
$s=1/2$?

It occurs that for some classes of fields the answer to this
question is positive \cite{103}.

We look for solutions of the complex nonlinear d'Alembert 
equation\index{d'Alembert equation} 
\begin{equation}
\pa_{\mu}\pa^{\mu}u=\la_1|u|^{k_1}u,
\label{2.6.4}
\end{equation}
where $\la_1,\ k_1$ are constants, in the form
\begin{equation}
u(x)=\bar{\psi}\psi\, e^{i\theta(x)}.
\label{2.6.5}
\end{equation}
Here $\psi=\psi(x)$ is a solution of nonlinear spinor equation
(\ref{2.4.1}) and $\theta(x)\in C^2(\R^2,\R^1)$ is a phase of the
field $u(x)$. With the use of exact solutions of the nonlinear Dirac
equation listed in Section 2.4 we have obtained a number of exact
solutions of the nonlinear d'Alembert equation which are adduced in
the Table 2.6.1.

Thus, spinors $\psi=\psi(x)$ satisfying nonlinear PDE (\ref{2.4.1})
give rise to complex scalar fields $u=u(x)$\index{Scalar field} which
are described by the nonlinear d'Alembert equation (\ref{2.6.4}). It
is interesting to note that the inverse procedure is also possible.
Namely, starting from a special subclass of exact solutions of the
nonlinear d'Alembert equation we can obtain exact solutions of the
nonlinear Dirac equation (see Section 2.1).  
\index{Exact solutions!of the nonlinear d'Alembert equation}

As straightforward computation shows, the vector field
constructed with the help of formula (\ref{2.6.1b}), where $\psi(x)$
is a solution of the nonlinear spinor equation (\ref{2.4.1}),
satisfies the following system of PDEs:
\begin{equation}
\begin{array}{l}
\Bigl (\pa_{\mu}\pa^{\mu}+M^2(x)\Bigr )A_{\mu}
(x)=j_{\nu}(x),\\[2mm]
\pa_{\mu}A_{\mu}(x)=0,
\end{array}
\label{2.6.6}
\end{equation}
functions $M(x),\ j_{\mu}(x)$ depending on the choice of
$\psi(x)$. 

For example, if we take $\psi=\psi_1(x)$, then
\begin{eqnarray*}
& &M(x)=\la(\bar{\chi}\chi)^{1/2k}= \mbox{\rm const},\\
& &j_{\mu}(x)=\la\bar{\chi}(\cos\la x_0-i\ga_0
\sin\la x_0)\ga_{\mu}(\cos\la x_0+i\ga_0
\sin\la x_0)\chi.
\end{eqnarray*}

Consequently, the  nonlinear spinor field gives rise to a field
$A_{\mu}(x)$ which can be interpreted as the vector 
field with a variable mass $M(x)$.
\vspace{2mm}

\noindent
{\em Table 2.6.1.}\ {\bf Exact solutions of the nonlinear}\\
\phantom{{\em Table 2.6.1.}\ }{\bf d'Alembert equation}
\vspace{1.5mm}

\noindent
\begin{tabular}{|c|c|c|}
\hline
  & & \\
$N$ &  $u(x)$ & $k_1$ \\ [2.5mm]
\hline
  & &  \\[2mm]
1--10 & $C\exp\{i\alpha \cdot x\}$ & $k_1\in \R^1$  \\[2mm]
11 & $C(x_1^2+x_2^2)^{-1/2}\exp\{iw_0\}$ & 2  \\[2mm]
12 & $C[(x_1+w_1)^2+(x_2+w_2)^2]^{-1/2}
\exp\{iw_0\}$ & 2  \\[2mm]
13 & $C(x_1^2+x_2^2)^{-1/2}\exp\{iw_0\}$ & 2  \\[2mm]
14 & $C(x_0^2-x_1^2-x_3^2)^{-1}$ & 1  \\[2mm]
15 & $C(x\cdot x)^{-3/2}$ & 2/3  \\[2mm]
16 & $C(x_0^2-x_3^2)^{-1/2}$ & 2  \\[2mm]
17 & $C(x_1^2+x_2^2)^{-1/2}\exp\{iw_0\}$ & 2  \\[2mm]
18 & $C[(x_1+w_1)^2+(x_2+w_2)^2]^{-1/2}
\exp\{iw_0\}$ & 2  \\[2mm]
19 & $Cw_0^{-2}\exp\bigl \{i(x_1+w_1)\bigr\}$ & 0  \\[2mm]
21 & $C(x_0^2-x_1^2-x_2^2)^{-2}$ & 1/2  \\[2mm]
22 & $C(x_0^2-x_1^2-x_2^2)^{-2}$ & 1/2  \\[2mm]
24 & $C(x_1^2+x_2^2+x_3^2)^{-2}$ & 1/2  \\[2mm]
26 & $C(x_0^2-x_1^2-x_3^2)^{-1}$ & 1  \\[2mm]
27 & $C(x\cdot x)^{-3/2}$ & 2/3  \\[2mm]
28 & $C\{[x_2+\beta(x_0+x_3)]^2+[x_1 + (1/2)(x_0+x_3)^2]^2\}^{-k}$ &
$1/k,\, k<0$  \\[2mm] 
29 & $C\{[x_2+\beta(x_0+x_3)]^2+[x_1
+(1/2)(x_0+x_3)^2]^2\}^{-1}$ & 1  \\[2mm]
\hline
\end{tabular}
\vspace{1.5mm}

\noindent
Here $N$ denotes the number of the corresponding solution of the
nonlinear Dirac equation, $w_0,\ w_1,\ w_2$ are arbitrary smooth
functions of $x_0+x_3;\ C,\ \alpha_{\mu},\ \beta$ are constants.
\vspace{2mm}

Let us adduce an example of a tensor field\index{Tensor field} with
the spin $s = 1$ constructed with the use of an exact solution of
nonlinear PDE (\ref{2.4.1}). Substituting the four-component function
$\psi(x)=\exp\{-i\la\ga_0 x_0\}$ into the formulae
\begin{displaymath}
E_a=i\bar{\psi}\ga_0\ga_a\psi,\quad
H_a= (i/2)\varepsilon_{abc}\bar{\psi}\ga_b\ga_c\psi
\end{displaymath}
we get the exact solution of the Maxwell 
equations\index{Maxwell equations} with the current
\begin{displaymath}
j_0=0,\quad
j_a=-2i\la(\chi^{\dagger}\ga_a\chi)\sin 2\la x_0-
2\la(\bar{\chi}\ga_a\chi)\cos 2\la x_0.
\end{displaymath}

In conclusion of this section we give the formula for 
construction of Poinca\-r\`e-in\-variant
Ans\"atze for the field with the spin 
$s = 3/2$\index{Ansatz!for spinor-vector field}
\begin{equation}
\Lambda_{\mu}(x)=(\bar{\psi}\ga_{\mu}\psi)\psi,
\ \
\mu={0,\ldots,3}.
\label{2.6.7}
\end{equation}

Substitution of the $P(1,3)$-invariant Ans\"atze (\ref{2.2.8}) into
(\ref{2.6.7}) gives rise to Ans\"atze for the field $\Lambda_{\mu}(x)$
with the spin $s=3/2$ reducing the corresponding Poincar\'e-invariant
equation to systems of ODEs.  
\vspace{10mm}

\noindent
{\large\bf 2.7. Exact solutions of the Dirac-d'Alembert
  equation\label{s2.7}} 

\markboth{Chapter 2. EXACT SOLUTIONS}
{2.7. Exact solutions of the Dirac-d'Alembert equation}
\def\theequation{2.\arabic{section}.\arabic{equation}}
\setcounter {section} {7}
\setcounter {equation}{0}
\vspace{7mm}

\noindent
Few works containing exact solutions of systems of nonlinear
PDEs of the form (\ref{1.4.8}) \cite{19,20,217} use essentially the
Ansatz for the spinor field\index{Heisenberg Ansatz} 
\begin{equation}
        \psi(x)=\{\g \cdot xf(x \cdot x)+ig(x\cdot x)\}\chi
\label{2.7.1}
\end{equation}
suggested by Heisenberg \cite{116}. In (\ref{2.7.1}) $\{f, g\} \subset
C^1(\R^1,\R^1)$ are arbitrary real-valued functions.

The scalar field $u=u(x)$ is looked for in the form
\begin{equation}
        u(x)=\vp(x\cdot x), \quad \vp\in C^2(\R^1,\C^1).
\label{2.7.2}
\end{equation}

Substitution of (\ref{2.7.1}), (\ref{2.7.2}) into (\ref{1.4.8}) under
\begin{equation}
        F=R_2(uu^*,\, \bar\psi\psi)\psi, \quad 
        H=R_1(uu^*,\, \bar\psi\psi)u, \quad
        R_i\in C(\R^2,\R^1)
\label{2.7.3}
\end{equation}
gives rise to a system of three ordinary differential equations for
functions $f,\ g,\ \vp$. Consequently, a reduction of PDE
(\ref{1.4.8}) both by the number of independent variables and by the
number of dependent variables takes place. We recall that Ans\"atze
constructed in Section 2.2 reduce  Poincar\'e-invariant spinor
equations by the number of independent variables only.

In \cite{96} we have suggested a generalization of the Heisenberg
Ansatz which made it possible to obtain broad classes of the exact
solutions of the multi-dimensional Dirac and Dirac-d'Alembert
equations\index{Dirac-d'Alembert equations}.

Following \cite{99} we look for a solution of system of PDEs 
(\ref{1.4.8}), (\ref{2.7.3}) of the 
form\index{Ansatz!for spinor field}\index{Ansatz!for scalar field}
\begin{equation}
        \psi(x)=\{f(\om)\g_\mu\p_\mu\om+ig(\om)\}\chi, \quad
        u(x)=\vp(\om),
\label{2.7.4}
\end{equation}
where $\{f, g\} \subset C^1(\R^1,\R^1)$,\ $\vp \in C^2(\R^1,\R^1)$,\
$\om=\om(x)$ is an arbitrary smooth real-valued function.

Since
\begin{eqnarray*}
& &        i\g_\mu\p_\mu\psi(x)=\{-\dot g\g_\mu\p_\mu\om+if\p_\mu
        \p^\mu\om+i\dot f(\p_\mu\om)(\p^\mu\om)\}\chi,\\
& &        \p_\mu\p^\mu u(x)=\ddot \vp(\p_\mu\om)(\p^\mu\om)+
        \dot \vp\p_\mu \p^\mu \om,\\
& &        \bar\psi\psi=\bar \chi \chi \Bigl(g^2+f^2(\p_\mu \om)
        (\p^\mu \om)\Bigr),
\end{eqnarray*}
substitution of formulae (\ref{2.7.4}) into (\ref{1.4.8}) yields the
system of PDEs for $f,\ g,\ \vp,\ \om$:
\begin{eqnarray}
& &     \ddot \vp(\p_\mu\om)(\p^\mu\om)+\dot \vp\p_\mu
        \p^\mu\om=R_1\vp,\non\\
& &     \dot f(\p_\mu\om)(\p^\mu\om)+f\p_\mu\p^\mu
\om=R_2g,\label{2.7.5} \\ 
& &     \dot g=-R_2f,\non
\end{eqnarray}
where $R_i=R_i(\vp\vp^*,\, g^2+f^2(\p_\mu\om)(\p^\mu \om)), \
i={1,2}$.

If we resolve two last equations with respect to
$(\p_\mu\om)(\p^\mu \om)$ and $\p_\mu\p^\mu \om$, then the following 
necessary compatibility conditions arise
\begin{displaymath}
    \p_\mu\p^\mu\om=F_1(\om), \quad (\p_\mu\om)(\p^\mu\om)=F_2(\om).
\end{displaymath}

In other words, the scalar function $\om=\om(x)$ has to satisfy the
d'Alembert-Hamilton system (\ref{2.1.22}) and besides the functions
$F_1,\ F_2$ do not vanish simultaneously. Since functions $f(\om),\
g(\om)$ are arbitrary, we can choose them in such a way that
$\om=\om(x)$ satisfies system of PDEs (\ref{2.1.27}), equations
(\ref{2.7.5}) taking the form 
\begin{eqnarray}
& &        \ve \ddot \vp+F(\om)\dot \vp=R_1(\vp\vp^*,\, g^2+\ve
f^2)\vp, \non\\
& &        \ve \dot f+F(\om)f=R_2(\vp\vp^*,\, g^2+\ve f^2)g,
\label{2.7.6}\\
& &        \dot g=-R_2(\vp\vp^*,\, g^2+\ve f^2)f,\non
\end{eqnarray}
where $F(\om)=\ve N\om^{-1}, \ N={0,\ldots,3}, \ \ve=\pm 1$.

Thus, the problem of constructing particular solutions of the
multi-dimen\-sional system of five PDEs (\ref{1.4.8}) is reduced to
integration of a system of three ODEs. If we succeed in integrating
system (\ref{2.7.6}), then substitution of the obtained results into
Ansatz (\ref{2.7.4}), where $\om=\om(x)$ is the solution of the
d'Alembert-Hamilton system\index{d'Alembert-Hamilton system}
(\ref{2.1.27}), gives rise to exact solutions of the initial system of
PDEs (\ref{1.4.8}), (\ref{2.7.3}).

Let us note that Ansatz (\ref{2.7.4}) can be interpreted as the
formula for construction of the nonlinear spinor field $\psi(x)$
satisfying the Dirac-d'Alembert system with the help of the nonlinear
scalar field $\om(x)$ satisfying the nonlinear d'Alembert-Hamilton
system.

It is clear that the $P(1,3)$-invariant Ans\"atze obtained in 
Section 2.2 can also be applied to reduce the Poincar\'e-invariant
equation (\ref{1.4.8}) but the resulting systems of ODEs prove to be
much more complicated than system (\ref{2.7.6}).

We will construct exact solutions of system of PDEs (\ref{1.4.8}),
(\ref{2.7.3}) having the following nonlinearities: 
\begin{equation} 
\begin{array}{rcl}
R_1&=&-\{\mu_1|u|^{k_1}+\mu_2(\bar\psi\psi)^{k_2}\}^2, \\[2mm]
R_2&=&\lbd_1|u|^{k_1}+\lbd_2(\bar\psi\psi)^{k_2}, \end{array}
\label{2.7.8}
\end{equation}
where $|u|^2=uu^*;\ \lbd_1,\ \lbd_2,\ \mu_1,\ \mu_2,\ k_1,\ k_2$ are
constants. 

Substitution of Ansatz (\ref{2.7.4}) into system of PDEs
(\ref{1.4.8}), (\ref{2.7.8}) yields the following equations for
unknown functions $f,\ g,\ \vp$: 
\begin{eqnarray}
& &     \lbd \ddot \vp+F(\om)\dot \vp=-\{\mu_1|\vp|^{k_1}+
        \tilde \mu_2(g_2+\lbd f^2)^{k_2}\}^2\vp,
\non\\
& &
        \lbd\dot f+F(\om)f=\{\lbd_1|\vp|^{k_1}+
        \tilde \lbd_2(g^2+\lbd f^2)^{k_2}\}g,
\label{2.7.9}\\
& &
        \dot g=-\{\lbd_1|\vp|^{k_1}+
        \tilde \lbd_2(g^2+\lbd f^2)^{k_2}\}f.\non
\end{eqnarray}

Here $\tilde \mu_2=\mu_2(\bar\chi\chi)^{k_2}$,\ 
$\tilde \lbd_2=\lbd_2 (\bar\chi \chi)^{k_2}$;\
$F(\om)=N\lbd\om^{-1}$,\ $N=0, 1, 2, 3$,\ $\lbd=\ve=\pm 1$.

We have succeeded in constructing the general solution of 
system (\ref{2.7.9}) provided $N=0$. Under $N\ne 0$ some particular  
solutions are obtained.
\vspace{1.5mm}

\noindent
1) \ $N=0, \ \lbd=1$

Multiplying the second equation of system (\ref{2.7.9}) by $f$, the
third by $g$ and summing we have $f\dot f + g\dot g=0\ \Rightarrow \ 
f^2+g^2=C_1^2=\mbox{\rm const}$. Due to this fact equations for
$g,\ f$ are easily integrated
\begin{equation}
\begin{array}{l}
        f(\om)=C_1\sin\Biggl\{\lbd_1{\displaystyle\mathop{\int}
        \limits^{\edi\om}}|\vp(z)|        
        ^{k_1}dz+\tilde \lbd_2 C_1^{2k_2}\om+C_2\Biggr\},\\[2mm]
        g(\om)=C_1\cos\Biggl\{\lbd_1{\displaystyle\mathop{\int}
        \limits^{\edi\om}}|\vp(z)|^{k_1}dz+\tilde
      \lbd_2C_1^{2k_2}\om+C_2\Biggr\}, 
\end{array}
\label{2.7.10}
\end{equation}
where $C_2=\mbox{\rm const}$.

Substituting (\ref{2.7.10}) into the first equation of system 
(\ref{2.7.9}) we come to the following ODE for $\vp(\om)$:
\begin{displaymath}
        \ddot \vp=-\{\mu_1|\vp|^{k_1}+\tilde \mu C_1^{2k_2}\}^2\vp.
\end{displaymath}

On representing the complex-valued function $\vp$ in the form
\begin{equation}
        \vp(\om)=\rho(\om)e^{i\theta(\om)},
\label{2.7.11}
\end{equation}
where $\rho(\om)$,\ $\theta(\om)$ are real-valued functions, we
rewrite this ODE as follows
\begin{equation}
 \ddot \rho-\rho \dot \theta^2=-\{\mu_1\rho^{k_1}
 +\tilde \mu_2C_1^{2k_2}\}^2\rho,\quad
        2\dot \theta \dot \rho+\ddot \theta\rho=0.
\label{2.7.12}
\end{equation}

The second equation of the above system implies that
$\dot \theta=C_3 \rho^{-1/2}, \ C_3= \mbox{\rm const}$. Substitution of 
this result into the first equation of system (\ref{2.7.12}) yields  
the ODE for $\rho=\rho(\om)$
\begin{displaymath}
        \ddot \rho=C_3^2-\mu_1^2\rho^{2k_1+1}-2\mu_1
        \tilde \mu_2C_1^{2k_2}\rho^{k_1+1}-
        \tilde \mu_2^2C_1^{4k_2}\rho\equiv a_+(\rho),
\end{displaymath}
whose general solution is given by the implicit formula
\begin{equation}
        \mathop{\int}\limits^{\edi\rho(\om)}
        \Biggl(2\int a_+(z)dz+C_4\Biggr)^{-1/2}dz=\om.
\label{2.7.13}
\end{equation}

Thus, the general solution of system of ODEs (\ref{2.7.9}) 
under $N=0, \ \lbd=1$ has the form
\begin{eqnarray*}
        f(\om)&=&C_1\sin\Biggl\{\lbd_1 \mathop{\int}\limits^
        {\edi\om}\rho^{k_1}(z)dz+ 
        \tilde \lbd_2C_1^{2k_2}\om+C_2\Biggr\},\\
        g(\om)&=&C_1 \cos \Biggl\{\lbd_1 \mathop{\int}\limits^
        {\edi\om}\rho^{k_1}(z)dz+ 
        \tilde \lbd_2C_1^{2k_2}\om+C_2\Biggr\},\\
        \vp(\om)&=&\rho(\om)\exp\Biggl\{iC_3\mathop{\int}
        \limits^{\edi\om}  
        \rho^{-1/2}(z)dz\Biggr\},
\end{eqnarray*}
where $C_1,\ C_2,\ C_3$ are constants, $\rho(\om)$ is defined by
(\ref{2.7.13}). 
\vspace{1.5mm}

\noindent
2) \ $N=2,3, \ \lbd=1$.

We look for particular solutions of system (\ref{2.7.9}) in the form
of power functions
\begin{displaymath}
   f(\om)=C\om^{\al_1}, \quad g(\om)=D\om^{\al_2}, \quad
   \vp(\om)=E\om^{\al_3}. 
\end{displaymath}

Substituting these expressions into (\ref{2.7.9}) and equating
exponents of $\om$ yield
\begin{displaymath}
        \al_1=\al_2, \quad \al_1-1=\al_2(1+2k_2), \quad 
        \al_3k_1=2\al_2k_2,
\end{displaymath}
whence $\al_1=\al_2=-1/2k_2, \ \al_3=1/k_1$.

Consequently,
\begin{equation}
        f(\om)=C\om^{-1/2k_2}, \quad g(\om)=D\om^{-1/2k_2}, \quad
        \vp(\om)=E\om^{-1/k_1},
\label{2.7.14}
\end{equation}
parameters $C,\ D,\ E$ satisfying the system of nonlinear algebraic
equations 
\begin{eqnarray}
& &   k_1^{-2}(Nk_1-k_1-1)=\{\mu_1|E|^{k_1}+\tilde
\mu_2(C^2+D^2)^{k_2}\}^2, 
\non\\
& &
        (2k_2)^{-1}D=\{\lbd_1|E|^{k_1}+\tilde
        \lbd_2(C^2+D^2)^{k_2}\}C, 
\label{2.7.15}\\
& &
        (2k_2)^{-1}(2Nk_2-1)C=\{\lbd_1|E|^{k_1}+
        \tilde \lbd_2(C^2+D^2)^{k_2}\}D.\non
\end{eqnarray}

From the second and the third equations we get the equality
\begin{equation}
        D^2C^{-2}=1-2Nk_2.
\label{2.7.16}
\end{equation}

The first equation of system (\ref{2.7.15}) and equality
(\ref{2.7.16}) yield the following restrictions on the choice of
parameters $k_1,\ k_2$:\ \ $ k_1>(N-1)^{-1}, \quad k_2>(2N)^{-1}$.

Therefore, relations (\ref{2.7.15}) can be rewritten in the form
\begin{eqnarray}
& &        D=\ve(2Nk_2-1)^{1/2}C,
\non\\
& &
        \{\mu_1|E|^{k_1}+\tilde \mu_2(2Nk_2C^2)^{k_2}\}^2=
        (1+k_1-Nk_1)k_1^{-2},
\label{2.7.17}\\
& &
        \{\lbd_1|E|^{k_1}+\tilde \lbd_2(2Nk_2C^2)^{k_2}\}=
        \ve(2Nk_2-1)^{1/2}(2k_2)^{-1},\non
\end{eqnarray}
where $\ve=\pm 1, \ k_1>(N-1), \ k_2>(2N)^{-1}$.

Under $k_1=2(N-1)^{-1}, \ k_2=N^{-1}$ system (\ref{2.7.9}) possesses
the following class of particular solutions:
\begin{eqnarray*}
        f(\om)&=&\theta\om(1+\theta^2\om^2)^{-(N+1)/2},
\\
        g(\om)&=&(1+\theta^2\om^2)^{-(N+1)/2},
\\
        \vp(\om)&=&E(1+\theta^2\om^2)^{(1-N)/2},
\end{eqnarray*}
parameters $\theta,\ E$ being determined by the system of nonlinear
algebraic equations
\begin{equation}
\begin{array}{l}
        \theta^2(N^2-1)=\{\mu_1|E|^{2/(N-1)}+\tilde \mu_2\}^2\\[2mm]
        (N+1)\theta=\{\lbd_1|E|^{2/(N-1)}+\tilde \lbd_2\}.
\end{array}
\label{2.7.18}
\end{equation}
3) \ $N=0, \ \lbd=-1$.

Multiplying the second equation of system (\ref{2.7.9}) by $f$, the
third by $g$ and summing we have $ f^2-g^2=-C_1^2=\mbox{\rm const}$.
Due to this fact equations for $g,\ f$ are easily integrated 
\begin{eqnarray*}
        f(\om)&=&C_1\sinh \Biggl\{-\lbd_1\displaystyle\mathop 
        {\int}\limits^{\edi\om}
        |\vp(z)|^{k_1}dz-\tilde \lbd_2C_1^{2k_2}\om+C_2\Biggr\},\\
        g(\om)&=&C_1\cosh  \Biggl\{-\lbd_1\displaystyle\mathop
        {\int}\limits^{\edi\om} 
        |\vp(z)|^{k_1}dz-\tilde \lbd_2C_1^{2k_2}\om+C_2\Biggr\},
\end{eqnarray*}
where $C_2=\mbox{\rm const}$.

Substitution of the above formulae into the first equation of system
(\ref{2.7.9}) gives rise to the ODE for $\vp(\om)$
\begin{displaymath}
        \ddot \vp=\{\mu_1|\vp|^{k_1}+\tilde \mu_2C_1^{2k_2}\}^2\vp.
\end{displaymath}

Representing $\vp(\om)$ in the form (\ref{2.7.11}) we come to the
following system of ODEs for $\rho(\om), \ \theta(\om)$:
\begin{displaymath}
    \ddot \rho-\rho{\dot\theta}^2=\{\mu_1\rho^{k_1}
    +\tilde \mu_2C_1^{2k_2}\}^2\rho,
\quad
        \ddot \theta\rho+2\dot \theta\dot \rho=0.
\end{displaymath}

The general solution of the above system is given implicitly
\begin{equation}
         \theta=C_3 \mathop {\int}\limits^{\edi\om} \rho^{-1/2}(z)dz,
\quad
        \mathop {\int}\limits^{\edi\rho(\om)} \Biggl(2\int a_-(z)dz
        +C_4\Biggl)^{-1/2}dz=\om,
\label{2.7.19}
\end{equation}
where $a_-(z)=\mu_1^2z^{2k_1+1}+2\mu_1 \tilde \mu_2
C_1^{2k_2}z^{k_1+1}+ \tilde \mu_2^2 C_1^{4k_1}z+C_3^2$,\ $C_3$,\ $C_4$
are cons\-tants. 

Consequently, the general solution of system of ODEs 
(\ref{2.7.9}) under $N=0$,\ $\lbd=1$ has the form
\begin{eqnarray*}
        f(\om)&=&C_1 \sinh \Biggl\{-\lbd_1\mathop {\int}\limits^
        {\edi\om}\rho^{k_1}(z)dz - \tilde \lbd_2C_1^{2k_2}\om +
        C_2\Biggr\}, \\
        g(\om)&=&C_1 \cosh  \Biggl\{-\lbd_1\mathop {\int}\limits
        ^{\edi\om}\rho^{k_1}(z)dz - \tilde \lbd_2C_1^{2k_2}\om +
        C_2\Biggr\}, \\
        \vp(\om)&=&\rho(\om)\exp\Biggl\{iC_3\mathop
        {\int}\limits^{\edi\om}\rho^{-/2}(z)dz\Biggr\},
\end{eqnarray*}
function $\rho(\om)$ being determined by (\ref{2.7.19}).
\vspace{1.5mm}

\noindent
4) \ $N=1,2,3,\ \lbd=-1$.

Solutions of equations (\ref{2.7.9}) are looked for in the form
(\ref{2.7.14}), whence we get the following system of nonlinear
algebraic equations for $C,\ D,\ E$:
\begin{eqnarray}
& &     k_1^{-2}(k_1N-k_1-1)=-\{\mu_1 |E|^{k_1}+
        \tilde \mu_2(D^2-C^2)^{k_2}\}^2,
\non\\
& &
        (2k_2)^{-1}(1-2Nk_2)=\{\lbd_1|E|^{k_1}+
        \tilde \lbd_2(D^2-C^2)^{k_2}\}DC^{-1},
\label{2.7.20}\\
& &
        (2k_2)^{-1}=\{\lbd_1|E|^{k_1}+
        \tilde \lbd_2(D^2-C^2)^{k_2}\}CD^{-1}.\non
\end{eqnarray}

Analysis of the above equations yields the restriction on the choice
of $C,\ D$:\ \ $D^2C^{-2}=1-2Nk_2$. Due to this fact equations
(\ref{2.7.20}) are rewritten in the form 
\begin{eqnarray}
& &     D=\ve C(1-2Nk_2)^{1/2},
\non\\
& &
        (1+k_1-Nk_1)k_1^{-2}=\{\mu_1|E|{k_1}+
        \tilde \mu_2(-2Nk_2C^2)^{k_2}\}^2,
\label{2.7.21}\\
& &
        \ve(1-2Nk_2)^{1/2}(2k_2)^{-1}=\{\lbd_1|E|^{k_1}+
        \tilde \lbd_2(-2Nk_2C^2)^{k_2}\},\non
\end{eqnarray}
where $\ve=\pm 1, \ k_1<(N-1)^{-1}, \ k_2<(2N)^{-1}$.

Substitution of the results obtained into Ansatz (\ref{2.7.4}) gives
the following classes of exact solutions of the nonlinear
Dirac-d'Alembert equations (\ref{1.4.8}), 
(\ref{2.7.8})
\index{Exact solutions!of the Dirac-d'Alembert equations}:
\vspace{1.5mm}

\noindent
\underline{the case of arbitrary $k_1\in \R^1, \ k_2\in \R^1$}
\begin{eqnarray*}
        \psi_1(x)&=&\Biggl\{i\cos \Biggl(\lbd_1 \mathop
        {\int}\limits^{\edi x_0}
        \rho^{k_1}(z)dz+\lbd_2(\bar \chi\chi)^{k_2}x_0+C_2\Biggr)
\\
& &
        + \g_0\sin \Biggl(\lbd_1\mathop{\int}\limits^{\edi x_0}
        \rho^{k_1} 
        (z)dz+\lbd_2(\bar\chi \chi)^{k_2}x_0+C_2\Biggr)\Biggr\}\chi,
\\
        u_1(x)&=&\rho(x_0)\exp\Biggl\{iC_3\mathop {\int}\limits^{\edi
          x_0} 
        \Bigl(\rho(z)\Bigr)^{-1/2}dz\Biggr\},
\end{eqnarray*}
where $C_2,\ C_3$ are arbitrary real constants, function
$\rho=\rho(\om)$ is determined by formula (\ref{2.7.13}) under
$C_1=1$: 
\begin{eqnarray*}
        \psi_2(x)&=&\Biggl\{i\cosh \Biggl(-\lbd_1\mathop
        {\int}\limits^{\edi\om} 
        \rho^{k_1}(z)dz-\lbd_2(\bar \chi\chi)^k_2\om+C_2\Biggr)
\\
& &
        + (\g_\mu\p_\mu\om)\sinh \Biggl(-\lbd_1
        \mathop {\int}\limits^{\edi\om}\rho^{k_1}(z)dz-\lbd_2(\bar
        \chi\chi)^{k_2} \om+C_2\Biggr)\Biggr\}\chi,
\\
        u_2(x)&=&\rho(\om)\exp\Biggl\{iC_3 \mathop
        {\int}\limits^{\edi\om} 
        \Bigl(\rho(z)\Bigr)^{-1/2}dz\Biggr\},
\end{eqnarray*}
where $C_2,\ C_3$ are arbitrary real constants, function
$\rho=\rho(\om)$ is determined by formula (\ref{2.7.19}) under
$C_1=1$,\ $\om=\om(x)$ is given by one of the formulae listed in
(\ref{2.1.33}); 
\vspace{1.5mm}

\noindent
\underline
{the case $k_1>1/2, \ k_2>1/6$}
\begin{eqnarray*}
        \psi_3(x)&=&\om^{-1/4k_2}\{\ve i(6k_2-1)^{1/2}+
        \g_\mu\p_\mu\om\}\chi,
\\
        u_3(x)&=&E\om^{-1/2k_1},
\end{eqnarray*}
where $E,\ \chi$ are defined by (\ref{2.7.17}) under $C=1, \ N=3, \
\om=\om(x)$ is given by (\ref{2.1.32});
\vspace{1.5mm}

\noindent
\underline
{the case $k_1>1, \ k_2>1/4$}
\begin{eqnarray*}
   \psi_4(x)&=&\om^{-1/4k_2}\{\ve
   i(4k_2-1)^{1/2}+\g_\mu\p_\mu\om\}\chi, 
\\
        u_4(x)&=&\om^{-1/2k_1},
\end{eqnarray*}
where $E,\ \chi$ are defined by (\ref{2.7.17}) under $C=1, \ N=2, \
\om=\om(x)$ is given by (\ref{2.1.31});
\vspace{1.5mm}

\noindent
\underline
{the case $k_1<1/2, \ \ k_2<1/6$}
\begin{eqnarray*}
 \psi_5(x)&=&\om^{-1/4k_2}\{\ve i(1-6k_2)^{1/2}+\g_\mu\p_\mu\om\}\chi,
\\
        u_5(x)&=&E\om^{-1/2k_1},
\end{eqnarray*}
where $E,\ \chi$ are defined by (\ref{2.7.21}) under $C=1, \ N=3, \
\om=\om(x)$ is given by (\ref{2.1.36});
\vspace{1.5mm}

\noindent
\underline
{the case $k_1<1, \ k_2<1/4$}
\begin{eqnarray*}
 \psi_6(x)&=&\om^{-1/4k_2}\{\ve i(1-4k_2)^{1/2}+\g_\mu\p_\mu\om\}\chi,
\\
        u_6(x)&=&E\om^{-1/2k_1},
\end{eqnarray*}
where $E,\ \chi$ are defined by (\ref{2.7.21}) under $C=1, \ N=2, \
\om=\om(x)$ is given by (\ref{2.1.35});
\vspace{1.5mm}

\noindent
\underline
{the case $k_1\in \R^1, \ \ k_2<1/2$}
\begin{eqnarray*}
 \psi_7(x)&=&\om^{-1/4k_2}\{\ve i(1-2k_2)^{1/2}+\g_\mu\p_\mu\om\}\chi,
\\
        u_7(x)&=&E\om^{-1/2k_1},
\end{eqnarray*}
where $E,\ \chi$ are defined by (\ref{2.7.21}) under $C=1, \ N=1, \
\om=\om(x)$ is given by (\ref{2.1.34});
\vspace{1.5mm}

\noindent
\underline
{the case $k_1=2, \ k_2=1/2$}
\begin{eqnarray*}
 \psi_8(x)&=&(1+\theta^2\om^2)^{-3/2}\{i+\theta\om\g_\mu\p_\mu\om\}\chi,
\\
        u_8(x)&=&E(1+\theta^2\om^2)^{-1/2},
\end{eqnarray*}
where $\theta,\ E,\ \chi$ are defined by (\ref{2.7.18}) under 
$N=2, \ \om=\om(x)$ is given by (\ref{2.1.31});
\vspace{1.5mm}

\noindent
\underline
{the case $k_1=1, \ k_2=1/3$}
\begin{eqnarray*}
 \psi_9(x)&=&(1+\theta^2\om^2)^{-2}\{i+\theta\om\g_\mu\p_\mu\om\}\chi,
\\
        u_9(x)&=&E(1+\theta^2\om^2)^{-1},
\end{eqnarray*}
where $\theta,\ E,\ \chi$ are defined by (\ref{2.7.18}) 
under $N=3, \ \om=\om(x)$ is given by (\ref{2.1.32}).

According to Theorem 1.4.1 system of PDEs (\ref{1.4.8}),
(\ref{2.7.8}) under $k_1=1, \ k_2=1/3$ admits the conformal group
$C(1,3)$.  Therefore we can apply to solutions $\{\psi_1(x),\,
u_1(x)\}$,\ $\{\psi_2(x), \, u_2(x)\}$,\ $\{\psi_3(x), \, u_3(x)\}$,\ 
$\{\psi_7(x), \, u_7(x)\}$,\ $\{\psi_9(x), \, u_9(x)\}$ with $k_1=1$,\ 
$k_2=1/3$ the procedure of generating solutions by means of the
four-parameter group of special conformal 
transformations\index{Conformal!transformations}

\begin{eqnarray}
& &     \psi_{II}(x)=\s^{-2}(x)(1-\g \cdot \theta \g \cdot
x)\psi_I(x'), \non\\
& &
        u_{II}(x)=\s^{-1}(x)u_I(x'),
\label{2.7.22}\\
& &
        x'_\mu=(x_\mu-\theta_\mu x\cdot x)\s^{-1}(x),\non
\end{eqnarray}
where $\s(x)=1-2\theta \cdot x + \theta \cdot \theta x \cdot x$,\
$\theta_\mu$ are constants.

The above formulae are obtained from (\ref{1.4.14}) with the help of
Theorem 2.4.1.

The nonlinear functions $\lbd_1|u|^{k_1}+\lbd_2(\bar
\psi\psi)^{k_2}$,\ $\mu_1|u|^{k_1}+\mu_2(\bar \psi\psi)^{k_2}$ can be
interpreted as "masses" of the spinor $(M(\psi))$ and scalar
$(M(u))$ particles created because of interaction of these particles.
As straightforward computation shows, the following relations hold
\begin{displaymath}
        \Bigl(M(\psi)/M(u)\Bigr)^2=(1/4)k_1^2(2Nk_2-1)
         k_2^{-2}(Nk_1-k_1-1)^{-1},
\end{displaymath}
where the cases $N=2, \ N=3$ correspond to the solutions
$\{\psi_4,\, u_4\}$,\ $\{\psi_3,\, u_3\}$;
\begin{displaymath}
        \Bigl(M(\psi)/M(u)\Bigr)^2=(1/4)k_1^2(2Nk_2-1)
        k_2^{-2}(Nk_1-k_1-1)^{-1},
\end{displaymath}
where the cases $N=1,\ N=2,\ N=3$ correspond to the solutions 
$\{\psi_7,\, u_7\}$,\ $\{\psi_6,\, u_6\}$,\ $\{\psi_5,\, u_5\}$;
\begin{displaymath}
        \Bigl(M(\psi)/M(u)\Bigr)^{-2}=(N+1)(N-1)^{-1},
\end{displaymath}
where the cases $N=2,\ N=3$ correspond to the solutions $\{\psi_8,\,
u_8\}$,\ $\{\psi_9,\, u_9\}$.

Consequently, in spite of the fact that "masses" of the spinor and
scalar particles described by equations (\ref{1.4.8}),
(\ref{2.7.8}) are variable their ratio is the constant determined by
the exponents $k_1,\ k_2$ and by some discrete parameter $N$. Thus,
the above relations can be interpreted as the formulae for the mass
spectrum of the spinor and scalar fields.  It is worth noting
that the discrete parameter $N$ arises because of the fact that the
nonlinear differential operator $\pm \om^2\Box$ has the discrete
spectrum $N=0,1,2,3$ on the set of solutions of the equation
$(\p_\mu\om)(\p^\mu\om)=\pm 1$ (see Section 2.1).

The solutions $\{\psi_3(x),\, u_3(x)\}-\{\psi_9(x),\, u_9(x)\}$ vanish
at the infinity under positive $k_1,\ k_2$ and besides they have a
non-integrable singularity \cite{99}.

Using the fact that system (\ref{1.4.8}), (\ref{2.7.8}) under
$\lbd_1=\mu_2=0$ splits into the nonlinear Dirac and d'Alembert
equations we can get from $\{\psi_1(x),\, u_1(x)\}-\{\psi_9(x),\, 
u_9(x)\}$ their exact solutions by putting $\lbd_1=0, \mu_2=0$. In
particular, the solutions $\{\psi_2(x),\, u_2(x)\}$, $\{\psi_5(x),\, 
u_5(x)\}$, $\{\psi_6(x),\ u_6(x)\}$ give rise to new solutions of
the nonlinear Dirac equation which differ from those constructed in
Section 2.4.

In conclusion we will say a few words about exact solutions of the
conforma\-lly-invariant system of PDEs
\begin{equation}
\begin{array}{l}
        \p_\mu\p^\mu u=\lbd_3u^3-\lbd_1\bar \psi\psi,\\
        i\g_\mu\p_\mu\psi=\{\lbd_1u+\lbd_2(\bar \psi\psi)^{1/3}\}\psi,
\end{array}
\label{2.7.27}
\end{equation}
where $\lbd_1,\ \lbd_2,\ \lbd_3$ are constants, obtained in \cite{19,20}
with the help of Heisenberg Ansatz (\ref{2.7.1}),
(\ref{2.7.2}). Since Ansatz (\ref{2.7.1}) is a particular case of
Ansatz (\ref{2.7.4}) (under $\om(x)=x\cdot x$), the above mentioned
solutions can be constructed within the framework of our approach. In
particular, functions $\{\psi_3(x),\, u_3(x)\}$, $\{\psi_9(x),\,
u_9(x)\}$ with $k_1=1,\ k_2=1/3$ satisfy system of PDEs (\ref{2.7.27})
provided the constants $E,\ \chi^\mu,\ \theta$ satisfy the algebraic
relations  
\begin{displaymath}
        \lbd_1E+\lbd_2 2^{1/3}(\bar \chi\chi)^{1/3}=3\ve/2,\quad
        -2\lbd_1(\bar \chi\chi)+\lbd_3E^3=3
\end{displaymath}
and
\begin{displaymath}
        \lbd_1E+\lbd_2(\bar \chi\chi)^{1/3}=4\theta,\quad
        \lbd_1(\bar \chi\chi)-\lbd_3E^3=8\theta^2,
\end{displaymath}
correspondingly.
\vspace{10mm}

\noindent
{\large\bf 2.8. Exact solutions of the nonlinear electrodynamics
equations\label{s2.8} } 

\markboth{Chapter 2. EXACT SOLUTIONS}
{2.8. Exact solutions of the nonlinear electrodynamics equations }
\def\theequation{2.\arabic{section}.\arabic{equation}}
\setcounter {section} {8}
\setcounter {equation}{0}
\vspace{7mm}

\noindent
Let us carry out reduction of Poincar\'e-invariant equations
(\ref{1.4.7}) for the spinor and vector fields using Ans\"atze
constructed in Section 2.6. Substitution of $P(1,3)$-ungenerable
Ans\"atze for the spinor field (we recall that these are obtained by
making the change 
\begin{equation}
\begin{array}{l}
\ga_0\to \ga\cdot a,
\quad
\ga_1\to \ga\cdot b,
\quad
\ga_2\to \ga\cdot c,
\quad
\ga_3\to \ga\cdot d,\\[2mm]
x_0\to a\cdot z,
\quad
x_1\to b\cdot z,
\quad
x_2\to c\cdot z,
\quad
x_3\to d\cdot z
\end{array}
\label{2.8.0z}
\end{equation} in $P(1,3)$-invariant Ans\"atze (\ref{2.2.8})) and
$P(1,3)$-ungenerable Ans\"atze for the vector field (\ref{2.6.3}),
(\ref{2.6.3z}) into system of PDEs (\ref{1.4.7}), (\ref{1.4.19})
yields after rather cumbersome computations 27 systems of ODEs for
functions $\vp(\om)$,\ $B_{\mu}(\om)$. Systems of ODEs for $\vp(\om)$
are obtained from (\ref{2.3.5}) if we replace $\g_\mu,\ x_\mu$ by
the expressions given in (\ref{2.8.0z}) and put

\begin{displaymath}
R=\{\g\cdot B(f_1 + f_2\g_4) + f_3 + f_4\g_4\}\vp,
\end{displaymath}
where$f_1,\ldots,f_4$ are arbitrary smooth functions of 
\begin{equation}
\bar{\vp}\vp,\quad  \bar{\vp}\g_4\vp,
\quad \bar{\vp}\g\cdot B\vp,\quad 
\bar{\vp}\g_4\g\cdot B\vp,\quad   
\vp^T\g_0\g_2 \g\cdot B\vp,\quad 
B\cdot B.
\label{2.8.1z}
\end{equation}

Reduced systems of ODEs for the vector field are written in the
following unified form:
\begin{equation}
\begin{array}{rcl}
k_{\mu\gamma}{\ddot B}^{\gamma} + l_{\mu\gamma}{\dot B}^{\gamma} +
m_{\mu\gamma}B^{\gamma}
&=& g_1B_{\mu}+ g_2\bar\vp\g_{\mu}\vp 
+ g_3\bar\vp\g_4\g_{\mu}\vp\\[2mm] 
& &+ g_4\vp^T\g_0\g_2\g_{\mu}\vp, \ \ \mu={0,\ldots,3},
\end{array}
\label{2.8.1}
\end{equation}
where $g_1,\ldots,g_4$ are arbitrary smooth functions of the
variables (\ref{2.8.1z}) and $k_{\mu\g},\ l_{\mu\g}$,\ $m_{\mu\g}$ are
functions of $\om$ listed below
\begin{eqnarray*}
& 1)&  k_{\mu\gamma} = - g_{\mu\gamma} - d_{\mu}d_{\gamma},
\quad l_{\mu\gamma} =
m_{\mu\gamma} = 0;\\
& 2)&  k_{\mu\gamma} = g_{\mu\gamma} - a_{\mu}a_{\gamma},\quad
l_{\mu\gamma} = 
m_{\mu\gamma} = 0;\\
& 3)&  k_{\mu\gamma} = k_{\mu}k_{\gamma},
\quad l_{\mu\gamma} = m_{\mu\gamma} = 0;\\
& 4)&  k_{\mu\gamma} = 4g_{\mu\gamma}\om - a_{\mu}a_{\gamma}(\om +
1)^{2} - (a_{\mu}d_{\gamma} + d_{\mu}a_{\gamma})(\om^{2} - 1) - 
d_{\mu}d_{\gamma}(\om - 1)^{2},\\
& &l_{\mu\gamma} = 4[g_{\mu\gamma} + (a_{\mu}d_{\gamma} -
 a_{\gamma}d_{\mu})] - 2[a_{\mu}(\om + 1) 
+ d_{\mu}(\om - 1)]k_{\gamma},\\
& & m_{\mu\gamma} = 0;\\
& 5)&   k_{\mu\gamma} = - g_{\mu\gamma} - b_{\mu}b_{\gamma},\quad
l_{\mu\gamma} = - b_{\mu}k_{\gamma},\quad m_{\mu\gamma} = 0;\\
& 6)&  k_{\mu\gamma} = - g_{\mu\gamma} - b_{\mu}b_{\gamma},\quad
l_{\mu\gamma} = 0,\quad
m_{\mu\gamma} = - (a_{\mu}a_{\gamma} 
- d_{\mu}d_{\gamma})/\alpha^{2};\\
& 7)&  k_{\mu\gamma} = - g_{\mu\gamma} 
- (\alpha k_{\gamma} e^{-\om/\alpha} - c_{\gamma})
(\alpha k_{\mu} e^{-\om/\alpha} - c_{\mu}),\quad
l_{\mu\gamma} = (2/\alpha)(a_{\mu}d_{\gamma} \\
& &- a_{\gamma}d_{\mu}) + (\alpha k_{\mu} e^{-\om/\alpha} 
- c_{\mu}) k_{\gamma} e^{-\om/\alpha},\quad
m_{\mu\gamma} = -(a_{\mu}a_{\gamma} - d_{\mu}d_{\gamma})/\alpha^{2};\\ 
& 8)&  k_{\mu\gamma} = - 4\om(g_{\mu\gamma} + c_{\mu}c_{\gamma}),\quad
l_{\mu\gamma} = - 4(g_{\mu\gamma} + c_{\mu}c_{\gamma}),
\quad m_{\mu\gamma} = -(b_{\mu}b_{\gamma})/\om;\\
& 9)&  k_{\mu\gamma} = - g_{\mu\gamma} - d_{\mu}d_{\gamma},
\quad l_{\mu\gamma} = 0,\quad m_{\mu\gamma} = (b_{\mu}b_{\gamma} +
c_{\mu}c_{\gamma})/\alpha^{2};\\
& 10)&  k_{\mu\gamma} = g_{\mu\gamma} - a_{\mu}a_{\gamma},
\quad l_{\mu\gamma} = 0,\quad m_{\mu\gamma} = - (b_{\mu}b_{\gamma} 
+ c_{\mu}c_{\gamma})/\alpha^2;\\
& 11)&  k_{\mu\gamma} = - k_{\mu}k_{\gamma},
\quad l_{\mu\gamma} = 2(c_{\mu}b_{\gamma} -
 b_{\mu}c_{\gamma}),\quad m_{\mu\gamma} = 0;\\
& 12)&  k_{\mu\gamma} = - k_{\mu}k_{\gamma},
\quad l_{\mu\gamma} = - (k_{\mu}k_{\gamma})/\om,\quad
m_{\mu\gamma} = 0;\\
& 13)&  k_{\mu\gamma} = - k_{\mu}k_{\gamma},
\quad l_{\mu\gamma} = - (k_{\mu}k_{\gamma})/\om,\quad
m_{\mu\gamma} = - (k_{\mu}k_{\gamma})/(\alpha^{2}\om^{2});\\
& 14)&  k_{\mu\gamma} = - k_{\mu}k_{\gamma},\quad
l_{\mu\gamma} = 0,\quad m_{\mu\gamma} = - k_{\mu}k_{\gamma};\\
& 15)&  k_{\mu\gamma} = - 4g_{\mu\gamma} - 4b_{\mu}b_{\gamma},\quad
m_{\mu\gamma} = l_{\mu\gamma} = 0;\\
& 16)&  k_{\mu\gamma} = - 4(1 + \alpha^{2})g_{\mu\gamma} 
- 4(c_{\mu} -\alpha b_{\mu})
(c_{\gamma} - \alpha b_{\gamma}),\quad m_{\mu\gamma} = l_{\mu\gamma} =
0;\\ 
& 17)&  k_{\mu\gamma} = - 4\om(g_{\mu\gamma} 
+ c_{\mu}c_{\gamma}),\quad l_{\mu\gamma} 
= - 4(g_{\mu\gamma} + c_{\mu}c_{\gamma}),\\
& &   m_{\mu\gamma} =
- (1/\om) \lbrack(a_{\mu}a_{\gamma} - d_{\mu}d_{\gamma})\alpha^{-2} +
b_{\mu}b_{\gamma}\rbrack ;\\
& 18)&  k_{\mu\gamma} = 4[g_{\mu\gamma}\om - (a_{\mu} 
- d_{\mu})(a_{\gamma} - d_{\gamma})],\quad
l_{\mu\gamma} = 4[g_{\mu\gamma} + (a_{\mu}d_{\gamma} 
- a_{\gamma}d_{\mu})\\ 
& &+ \alpha(b_{\mu}c_{\gamma} - c_{\mu}b_{\gamma})] 
- 2(a_{\mu} - d_{\mu})k_{\gamma},\quad  m_{\mu\gamma} = 0;\\
& 19)&  k_{\mu\gamma} = - 4\om(g_{\mu\gamma} +
c_{\mu}c_{\gamma}),\quad 
l_{\mu\gamma} = - 4g_{\mu\gamma} - 4c_{\mu}c_{\gamma} -
2c_{\mu}k_{\gamma}\om^{1/2};\\
& & 
m_{\mu\gamma} = -(b_{\mu}b_{\gamma})\om^{-1};\\
& 20)&  k_{\mu\gamma} = - k_{\mu}k_{\gamma},
\quad l_{\mu\gamma} = - (2k_{\mu}k_{\gamma})/\om,\quad
m_{\mu\gamma} = 0;\\
& 21)&  k_{\mu\gamma} = - k_{\mu}k_{\gamma},
\quad l_{\mu\gamma} = -k_{\mu}k_{\gamma}(2\om + \beta)
[\om(\om + \beta) - \alpha]^{-1},\\
& &  m_{\mu\gamma} = - k_{\mu}k_{\gamma}(\alpha -
1)^{2}[\om(\om + \beta) - \alpha]^{-2};\\
& 22)&  k_{\mu\gamma} = - k_{\mu}k_{\gamma},
\quad l_{\mu\gamma} = - k_{\mu}k_{\gamma}(2\om + \beta)
[\om(\om + \beta)]^{-1},\\
& &m_{\mu\gamma} = - k_{\mu}k_{\gamma}[\om(\om + \beta)]^{-2};\\
& 23)&  k_{\mu\gamma} = - k_{\mu}k_{\gamma},\quad l_{\mu\gamma} =
-k_{\mu}k_{\gamma}(2\om + 1)[\om(\om + \beta)]^{-1},\quad
m_{\mu\gamma} = 0;\\
& 24)&  k_{\mu\gamma} = 4\om g_{\mu\gamma} 
- (k_{\mu}\om + a_{\mu} - d_{\mu})
(k_{\gamma}\om + a_{\gamma} - d_{\gamma}),
\quad l_{\mu\gamma} = 6g_{\mu\gamma}  \\
& &+ 4(a_{\mu}d_{\gamma} - a_{\gamma}d_{\mu})-3(k_{\mu}\om + a_{\mu} -  
d_{\mu})k_{\gamma},\quad m_{\mu\gamma} = - k_{\mu}k_{\gamma};\\
& 25)&  k_{\mu\gamma} = - g_{\mu\gamma} - (c_{\mu} - \beta k_{\mu})
(c_{\gamma} - \beta k_{\gamma}),
\quad l_{\mu\gamma} = 2(\beta k_{\mu} -c_{\mu})k_{\gamma},\\
& &m_{\mu\gamma} = - k_{\mu}k_{\gamma};\\
& 26)&   k_{\mu\gamma} = - k_{\mu}k_{\gamma},
\quad l_{\mu\gamma} = (c_{\mu}b_{\gamma} -
b_{\mu}c_{\gamma} + 2 k_{\mu} k_{\gamma})/\om,\\ 
& &m_{\mu\gamma} = (c_{\mu}b_{\gamma} - b_{\mu}c_{\gamma})/\om ;\\
& 27)&  k_{\mu\gamma} = 4\om g_{\mu\gamma} - (a_{\mu} - d_{\mu} +
k_{\mu}\om)(a_{\gamma} - d_{\gamma} + k_{\gamma}\om),
\quad l_{\mu\gamma} = 4[2g_{\mu\gamma} \\
& &+ \alpha (b_{\mu}c_{\gamma} - c_{\mu}b_{\gamma}) 
- k_{\mu}k_{\gamma}\om - (a_{\mu}a_{\gamma} -
d_{\mu}d_{\gamma})], \quad m_{\mu\gamma} =
- 2k_{\mu}k_{\gamma}.
\end{eqnarray*}

Integration of the above systems of ODEs even under specific $F,\ 
R_{\mu}$ is an extremely hard problem. So it is not surprising that up
to now there is practically no papers devoted to construction of exact
solutions of the Maxwell-Dirac equations
(\ref{1.4.1}).\index{Maxwell-Dirac equations}

The fact that ODEs obtained are integrable with some specific $ F,\ 
R_{\mu}$ is a consequence of their nontrivial symmetry. Using 
Theorem 2.3.1 we can prove that these systems admit invariance
algebras which are isomorphic to algebras (\ref{2.3.12}).

In the present section we will construct multi-parameter families of
exact solutions of classical electrodynamics equations 
(\ref{1.4.1}) and of the system of nonlinear 
PDEs\index{Nonlinear!electrodynamics equations}
\begin{equation}
\begin{array}{l}
(i\g_{\mu}\p_{\mu}- e\g_{\mu}A^{\mu})\psi(x) =0,\\[2mm]
\p_{\nu}\p^{\nu}A_{\mu} - \p^{\mu}\p_{\nu}A_{\nu} =
    -e\bar \psi \g_{\mu}\psi + \lbd A_{\mu}A_{\nu}A^{\nu},
\end{array}
\label{2.8.2}
\end{equation}
where $e,\ \lbd $ are constants.
\vspace{2mm}

\noindent
{\bf 1. Exact solutions of the classical electrodynamics equations. }
\ We have made an observation that integrable cases of the systems of
ODEs obtained by means of reduction of (\ref{2.4.1}) with the help of
$P(1,3)$-ungenerable Ans\"atze for the spinor and vector fields
give rise to the exact solutions of system of nonlinear
PDEs (\ref{1.4.1}) of the form
\begin{equation}
\begin{array}{l}
\psi(x) = (\g\cdot a + \g \cdot d)\vp(\om_1,\,  \om_2,\,
\om_3),\\[2mm] 
A_{\mu}(x) = (a_{\mu} + d_{\mu})u(\om_1,\,  \om_2,\,  \om_3),
\end{array}
\label{2.8.4}
\end{equation}
where $\vp(\vec \om)$ is a four-component complex-valued function,
$u(\vec\om)$
is a scalar real-valued function; $\om_1=b\cdot x,\ \om_2 = c\cdot x,\
\om_3 = a\cdot x + d \cdot x$.

Formulae (\ref{2.8.4}) imply the following method of constructing
particular solutions of equation (\ref{1.4.1}): not to fix {\em a
  priori} the functions $\vp,\ u$ in (\ref{2.8.4}) but to consider
them as the new dependent variables.  Such an approach proved to be
very efficient because it enabled us to obtain exact solutions of the
classical electrodynamics equations containing arbitrary functions
\cite{103}. Substituting Ansatz (\ref{2.8.4}) into (\ref{1.4.1}) and
taking into account the identities
\begin{eqnarray*}
& &(\g\cdot a + \g\cdot d)^2 = a\cdot a + d \cdot d = 0,\\
& &\bar\vp(\g\cdot a + \g\cdot d)\g_{\mu}(\g\cdot a
   + \g\cdot d)\vp =2(a_{\mu} + d_{\mu}) 
   \bar\vp(\g\cdot a + \g\cdot d)\vp
\end{eqnarray*}
we come to the system of two-dimensional PDEs for $\vp(\vec \om),\
u(\vec\om)$ 
\begin{eqnarray}
& &   \g\cdot b\, \vp_{\om_1} + \g \cdot c\, \vp_{\om_2} - im\vp = 0,
\label{2.8.5a}\\
& &   u_{\om_1\om_1} + u_{\om_2\om_2} =
   2e\bar \vp(\g\cdot a + \g \cdot d)\vp,\label{2.8.5b}
\end{eqnarray}
where $\vp_{\om_i} = \p \vp /\p\om_i,\ u_{\om_i\om_i}=
   \p^2u/\p\om^2_i,\ i=1,2 $.

Let us emphasize that in the above equations there is no
differentiation with respect to the variable $\om_3$.
Consequently, functions $\vp,\ u$ contain $\om_3$ as a parameter.

The general solution of PDE (\ref{2.8.5b}) is given by the d'Alembert 
formula for the two-dimensional Poisson equation \cite{41}
\begin{equation}
\begin{array}{rcl}
 u(\vec\om)&=& w(z,\om_3) + w(z^*, \om_3)\\[2mm]
& &      -ie{\edi\int\limits^{\edi\om_2}\ \ 
         \int\limits_{\edi \om_1 - i(\om_2-\tau)}
         \limits^{\edi \om_1 +
           i(\om_2-\tau)}}\bar\vp(\xi,\eta)(\g\cdot a + 
         \g \cdot d)\vp(\xi,\eta) d\xi  d\eta,
\end{array}
\label{2.8.6}
\end{equation}
where $w$ is an arbitrary analytical with respect 
to the variable $z=\om_1 + i\om_2$ function.

Consequently, the problem of construction of exact solutions of system
of nonlinear PDEs (\ref{1.4.1}) is reduced via Ansatz (\ref{2.8.4}) to
integration of the two-dimensional linear Dirac equation
(\ref{2.8.5a}). Using the Fourier transform we can obtain its general
solution in the form of the Fourier integral \cite{26,41} but we
restrict ourselves to the cases when it is possible to construct exact
solutions in explicit form.

Choosing the eigenfunction of the Hermitian operator $-i\p_{\om_1}$ as
a particular solution of equation (\ref{2.8.5a}) yields
\begin{equation}
   \vp(\vec\om) = \exp\{i\lbd\om_1 + i\g\cdot c(\lbd\g\cdot b 
                    -  m)\om_2\}\vp_0\, (\om_3),
\label{2.8.7}
\end{equation}
where $\vp_0 \in C^1(\R^1, \C^4)$. Imposing on solution 
(\ref{2.8.7}) the condition of $2\pi$-periodicity with respect to 
$\om_1$ we get
\begin{equation}
   \lbd=\lbd_n= 2\pi n,\ \ n \in \Z.
\label{2.8.8}
\end{equation}

Substituting formulae (\ref{2.8.7}), (\ref{2.8.8}) into (\ref{2.8.6})
and  computing the integral we  arrive at the explicit expression 
for $u(\vec \om)$
\begin{equation}
\begin{array}{rcl}
u(\vec\om) &=& (1/2)(m^2 + \lbd^2_n)^{-1}
\{\tau_1 \cosh[2(m^2 + \lbd^2_n)^{1/2}\om_2]\\[2mm] 
& &+ \tau_2\sinh[2(m^2 + \lbd^2_n)^{1/2}\om_2]\} 
+w(z,\, \om_3) + w(z^*,\, \om_3).
\end{array}
\label{2.8.9}
\end{equation}

Here $ z=\om_1 + i\om_2,\ \tau_1 = e\bar\vp_0(\g\cdot a +\g\cdot
d)\vp_0$,\ $ \tau_2 = ie(m^2 + \lbd_n^2)^{-1/2}$ $\bar\vp_0$ $\times
(\g\cdot a +\g\cdot d)$ $(\lbd_n\g\cdot b - m)\vp_0$.

Substitution of formulae (\ref{2.8.7}), (\ref{2.8.9}) 
into Ansatz (\ref{2.8.4}) gives a multi-parameter family 
of the exact solutions of the classical electrodynamics 
equations (\ref{1.4.1}) containing three arbitrary functions:
\index{Exact solutions!of the classical electrodynamics equations}
\begin{eqnarray}
 \psi(x)&=& (\g\cdot k) \exp \{i\lbd_n b\cdot x 
    + i\g\cdot c(\lbd_n \g \cdot b - m) c\cdot x\} \vp_0(k\cdot
    x)\non\\
    & &\equiv\psi^{(n)}(x),\non\\
   A_{\mu}(x)&=&k_{\mu}\Bigl\{w(z,\, k\cdot x) 
   + w(z^*,\, k\cdot x)\label{2.8.10}\\
   & &+ (1/2)(m^2 + \lbd_n^2)^{-1} \{\tau_1 \cosh[2(m^2 
   + \lbd_n^2)^{1/2}c\cdot x]\non\\
   & &+ \tau_2\sinh[2(m^2 + \lbd_n^2 )^{1/2}c\cdot x]\}\Bigr\}
   \equiv A_{\mu}^{(n)}(x),\non
\end{eqnarray}
where $ k_\mu = a_\mu + d_\mu $.

Similarly, if we choose a particular solution of equation 
(\ref{2.8.5a}) in the form
\begin{eqnarray*}
   \vp(\vec\om)&=&(\om_1^2 + \om_2^2)^{-1/4}
      \exp\{-(1/2)(\g\cdot b)(\g\cdot c)\arctan(\om_1/\om_2)\}\\
      & &\times\exp\{-im(\g\cdot c)(\om_1^2 + \om_2^2)^{1/2}\}\vp_0\,
      (\om_3), 
\end{eqnarray*}
where $\vp_0 \in C^1(\R^1, \C^4)$, then formulae (\ref{2.8.4}),
(\ref{2.8.6}) give rise to the following family of exact solutions:
\index{Exact solutions!of the classical electrodynamics equations}
\begin{equation}
\begin{array}{rcl}
   \psi(x)&=&|z|^{-1/2}(\g\cdot k)\exp\{-(1/2)
      (\g\cdot b)(\g\cdot c)\arctan[(b\cdot x)\\[2mm]
      & &\times(c\cdot x)^{-1}]\}\exp\{-im\g\cdot c\, |z|\}\vp_0\,
      (k\cdot x),\\[2mm] 
   A_{\mu}(x)&=&k_{\mu}\Biggl\{w(z,\, k\cdot x) + w(z^*,\, k\cdot
   x)\\[2mm] 
      & &+ \edi\int\limits^{\edi |z|}[\tau_1\sinh(2my) 
      + \tau_2\cosh(2my)]y^{-1}dy \Biggr\}.
\end{array}
\label{2.8.11}
\end{equation}

In the above formulae $w$ is an arbitrary analytic with respect to 
 $z=b\cdot x + i c\cdot x$ function, $ |z|^2=zz^* $ and
\begin{displaymath}
   \tau_1 = -2e\bar\vp_0(\g\cdot k)\vp_0,\quad
   \tau_2 = 2ie\bar\vp_0(\g\cdot k)(\g\cdot c)\vp_0.
\end{displaymath}

We will consider the solution (\ref{2.8.10}) in more detail putting
$w=0,\ \vp_0=\exp\{-\alpha^2(k\cdot x)^2\}\chi $,
where $\chi$ is an arbitrary constant four-component column, $\alpha = 
\mbox{\rm const} $. A direct check shows that the identities 
\begin{equation}
\begin{array}{l}
-\p_{\mu}\p^{\mu}A_{\nu}^{(n)} =
 4(m^2 + 4\pi^2n^2)A_{\nu}^{(n)},\quad \p_{\mu}A_{\mu}^{(n)} =
 0,\\[2mm] 
 -\p_{\mu}\p^{\mu}\psi^{(n)} =m^2\psi^{(n)},
\end{array}
\label{2.8.12}
\end{equation}
where $n \in \Z,\ \p_{\mu}= \p/\p x_{\mu},\ \mu={0,\ldots,3}$, hold.
The operator $-\p_{\mu}\p^{\mu}$ is one of the Casimir
operators\index{Casimir operator} of the Poincar\'e algebra (see the
Appendix 1). Its eigenvalues are interpreted as masses of 
particles described by the corresponding motion equations. If such an
interpretation is extended to a nonlinear case, then relations
(\ref{2.8.12}) can be treated as follows: interaction of the spinor
and vector fields according to nonlinear equations (\ref{1.4.1}) gives
rise to the massive vector field\index{Vector field}
$A_{\mu}^{(n)}(x)$ with the mass $M_n = 2(m^2 + 4\pi^2n^2)^{1/2},\ n
\in \Z$ (in other words, the nonlinear interaction of the fields
$\psi(x),\ A_{\mu}(x)$ generates the mass spectrum). Let us emphasize
that the effect described is nonlinear because in the case of
the linear Maxwell equations the Casimir operator $\p_{\mu}\p^{\mu}$
has the zero eigenvalue (this is seen from (\ref{2.8.10}), where $
A_{\mu}^{(n)} = 0 $ under $ e=0 $). 

Since solutions (\ref{2.8.10}), (\ref{2.8.11}) depend analytically on
$m$, solutions of the massless classical electrodynamics equations are
obtained from (\ref{2.8.10}), (\ref{2.8.11}) by putting $m=0$.

This case deserves a special consideration because the invariance
group of equations (\ref{1.4.1}) under $m=0$ is the 15-parameter
conformal group (Theorem 1.4.2). The general solution of the
two-dimensional Dirac equation under $m=0$ has been constructed in
\cite{103} 
\begin{equation}
\vp(\vec \om)= (\g\cdot b + i \g\cdot c)\vp_1(z,\om_3) 
  + (\g\cdot b - i \g\cdot c)\vp_2(z^*,\om_3),
\label{2.8.13}
\end{equation}
where $\vp_1,\ \vp_2$ are arbitrary four-component functions whose
components are analytical functions of $z= b\cdot x + ic\cdot x $ and
$z^* = b \cdot x - i c \cdot x $, respectively.

Substitution of (\ref{2.8.13}) into (\ref{2.8.6}) yields the following 
expression for $u(\vec\om)$:
\begin{eqnarray*}
        u(\vec\om)&=&w(z,\om_3) + w(z^*,\om_3) \\
        & &+e\Biggl\{z^*\edi\int g_1(z,\om_3)dz 
        + z\int g_2(z^*,\om_3) dz^*\Biggr\},
\end{eqnarray*}
where $g_1={\bar\vp}_1(\g\cdot k)(1-i\g\cdot b\g\cdot c)\vp_2,\
 g_2={\bar\vp}_2(\g\cdot k)(1+i\g\cdot b\g\cdot c)\vp_1$.

Substituting the above formulae into the Ansatz (\ref{2.8.4}) we come
to the multi-parameter family of the exact solutions which contains
five arbitrary complex-valued functions
\index{Exact solutions!of the classical electrodynamics equations}
\begin{eqnarray}
   \psi(x)&=&(\g\cdot k)\{(\g\cdot b + i\g \cdot c )\vp_1
             (z,\, k\cdot x) + (\g\cdot b - i \g\cdot c)\non\\
             & &\times\vp_2 (z^*,\, k\cdot x)\},\non\\
   A_{\mu}(x)&=&k_{\mu}\Biggl\{w(z,\, k\cdot x) 
             + w(z^*,\, k\cdot x)\label{2.8.14}\\
             & &+ e\Biggl(z^*\edi\int g_1(z,\, k\cdot x)dz 
             +z\edi\int g_2(z^*,\, k\cdot x) dz^*\Biggr)\Biggr\}.\non
\end{eqnarray}

To obtain $C(1, 3)$-ungenerable family of solutions of system of
PDEs (\ref{1.4.1}) with $m=0$ we employ the solution 
generation procedure. The formulae for generating solutions of the
classical electrodynamics equations (\ref{1.4.1}) by the
four-parameter special conformal transformation group have been
obtained in \cite{86}
\begin{eqnarray}
   \psi_{II}(x)&=&\s^{-2}(x)(1-\g\cdot x\g\cdot
   \theta)\psi_I(x'),\non\\ 
   A_{II\mu}(x)&=&\s^{-2}(x)\{g_{\mu\nu}\s(x) + 2(x_{\nu}\theta_{\mu} 
   -x_{\mu}\theta_{\nu})\label{2.8.16}\\
   & &+ 2\theta\cdot x x_{\mu}\theta_{\nu} - x\cdot x 
   \theta_{\mu}\theta_{\nu} - \theta\cdot 
   \theta x_{\mu}x_{\nu})\} A_{I}^{\nu}(x'),
\end{eqnarray}
where $x^{\prime}_\mu=(x_\mu-\theta_\mu x\cdot x)\sigma^{-1}(x)$,\
$\sigma(x)=$ $1-2\theta\cdot x$ $+\theta\cdot\theta$ $x\cdot x$,\
$\theta_\mu$ are arbitrary real constants.
 
Substitution of expressions (\ref{2.8.14}) into (\ref{2.8.16}) gives
rise to the $C(1, 3)$ - ungenerable family of exact solutions of
system (\ref{1.4.1}) with $m=0$. We omit the corresponding formulae
because of their awkwardness.
\vspace{2mm}

\noindent
{\bf 2. Exact solutions of system of nonlinear PDEs (\ref{2.8.2})}.
\ To obtain exact solutions of equations (\ref{2.8.2}) we apply the
Ansatz  
\begin{equation}
\begin{array}{l}
   \psi(x)=(\g\cdot a - \g\cdot d)\exp\{if(k\cdot x)\}\chi,\\[2mm]
   A_{\mu}(x) = (a_{\mu} - d_{\mu})g_1 (k\cdot x) 
    + k_{\mu}g_2(k\cdot x).
\end{array}
\label{2.8.17}
\end{equation}

Here $\{f, g_1, g_2\} \subset C^1(\R^1, \R^1),\ \chi$ is an arbitrary
four-component constant column.

The Ansatz (\ref{2.8.17}) reduces equations (\ref{2.8.2}) to the
system of three ODEs for $f(\om),\ g_1(\om),\ g_2(\om)$
\begin{equation}
   \dot f = -eg_2,\quad 
   \ddot g_1 = - 2\lbd g_1g_2^2,\quad
   g_1^2g_2 = (e/2\lbd)\bar\chi(\g\cdot a - \g \cdot d)\chi.
\label{2.8.18}
\end{equation}

On eliminating the function $g_2$ from the second equation we get the
second-order ODE for $g_1$ 
\begin{equation} 
\ddot g_1 = - (\tau^2/\lbd)g_1^{-3},
\label{2.8.19}
\end{equation}
where $\tau = 2^{-1/2}e\bar\chi(\g\cdot a - \g\cdot d)\chi$.

The above equation is integrated in elementary functions, its general 
solution having the form \cite{132}
\begin{equation}
   g_1(\om)=\ve C_1^{-1/2}\Bigl((C_1\om + C_2) -
   \tau^2/\lbd\Bigr)^{1/2}. 
\label{2.8.20a}
\end{equation}

In addition, ODE (\ref{2.8.19}) with $\lbd>0$ possesses the one-parameter
family of singular solutions
\begin{equation}
   g_1(\om) = \ve(2|\tau||\lbd|^{-1/2}\om + C_2)^{1/2}.
\label{2.8.20b}
\end{equation}

In (\ref{2.8.20a}), (\ref{2.8.20b}) $C_1,\ C_2 $ are real 
constants, $\ve = \pm 1$.

Substituting formulae (\ref{2.8.20a}), (\ref{2.8.20b}) into the 
second equation of system (\ref{2.8.18}) yields
\begin{eqnarray*}
& &   g_2(\om) = C_1 \tau \lbd^{-1}\Bigl((C_1\om + C_2)^2 
            - \tau^2/\lbd\Bigr)^{-1},\\
& &   g_2(\om) = -\tau|\lbd|^{-1}(2|\tau| |\lbd|^{-1/2}\om +
C_2)^{-1}. 
\end{eqnarray*}

Integrating the first equation of system (\ref{2.8.18}) we get the
explicit form of the function $f(\om)$
\begin{eqnarray*}
& &  f(\om) = e(-\lbd)^{1/2}\arctan\Bigl(\tau^{-1}(-\lbd)^{1/2}
  (C_1\om + C_2)\Bigr),\\
& &  f(\om) = -e|\lbd|^{-1/2}\ln(2\tau|\lbd|^{-1/2}\om + C_2).
\end{eqnarray*}

Substitution of the above formulae into the Ansatz (\ref{2.8.17})
gives rise to the multi-parameter families of the exact solutions of
system (\ref{2.8.2}) 
\index{Exact solutions!of the nonlinear electrodynamics equations}
\vspace{1.5mm}

\noindent
\underline{the case $\lbd \in \R^1,\ \lbd\ne 0$}
\begin{eqnarray*}
\psi(x)&=&(\g\cdot a - \g \cdot d)\exp\Bigl\{-ie(-\lbd)^{1/2}
\arctan\Bigl(\tau^{-1}(-\lbd)^{-1/2}\quad\quad\, \\
& &\times[C_1(k\cdot x) + C_2]\Bigr)\}\chi,\\
A_{\mu}(x)&=&\ve (a_{\mu}- d_{\mu})C_1^{-1/2}
\Bigl\{(C_1k\cdot x+ C_2)^2 - \tau^2\lbd^{-1}\Bigr\}^{-1/2} \\
& & + C_1 \tau \lbd^{-1}k_{\mu} 
 \Bigl\{(C_1k\cdot x + C_2)^2 - \tau^2\lbd^{-1}\Bigr\}^{-1};\\
\end{eqnarray*}
\underline{the case $\lbd < 0$}
\begin{eqnarray*}
\psi(x)&=&(\g\cdot a - \g \cdot
d)\exp\Bigl\{-ie|\lbd|^{-1/2}\ln\Bigl(2\tau 
|\lbd|^{-1/2}k\cdot x + C_2\Bigr)\Bigr\}\chi,\\
A_{\mu}(x)&=&\ve (a_{\mu}- d_{\mu})\Bigl\{2|\tau||\lbd|^{-1/2}k\cdot x  
  + C_2\Bigr\}^{1/2}- \tau|\lbd|^{-1}k_\mu\\
& &\times \Bigl\{2|\tau||\lbd|^{-1/2}k\cdot x +C_2\Bigr\}^{-1},
\end{eqnarray*}
where $\tau = 2^{-1/2}e \bar\chi(\g\cdot a - \g \cdot d) \chi,\ C_1,\ 
C_2 $ are real constants.

Let us note that the solutions obtained are singular with respect to
the coupling constant $\lbd$. That is why they cannot be obtained in
the framework of the perturbation theory by expanding with respect to
a small parameter $\lbd$.  
\vspace{2mm}

\noindent
{\bf 5. Exact solutions of the Maxwell-Born-Infeld equations.}\
By the Maxwell-Born-Infeld
equations\index{Maxwell-Born-Infeld equations} we mean the Maxwell
equations\index{Maxwell equations} 
\begin{equation}
\begin{array}{ll}
\p_t\vec D = {\rm rot\,} \vec H,& {\rm div\,}  \vec D =0,\\
\p_t\vec B = - {\rm rot\,}  \vec E,& {\rm div\,}  \vec B =0
\end{array}
\label{2.8.22}
\end{equation}
supplemented by the constitutive equations suggested by Born and
Infeld (see, e.g. \cite{94}) 
\begin{equation}
   \vec D = \tau\vec E + \tau_1 \vec B, 
   \quad \vec H = \tau\vec B -\tau_1\vec E.
\label{2.8.23}
\end{equation}

Here $\vec E$,\ $\vec H$ are field intensities, $\vec B$,\ 
$\vec D$ are inductions,
\begin{eqnarray*}
   \tau &=& \{1+\vec B^2 - \vec E^2 - (\vec B\vec E)^2\}^{-1/2},\\
   \tau_1 &=& (\vec B\vec E)\tau.
\end{eqnarray*}

Till now, up to our knowledge, there are no papers containing exact
solutions of system (\ref{2.8.22}), (\ref{2.8.23}) in explicit form.
We will construct multi-parameter families of exact solutions of
system of nonlinear PDEs (\ref{2.8.22}), (\ref{2.8.23}) using the
following simple assertion.  
\vspace{2mm}

\noindent
{\bf Lemma 2.8.1.}\ {\em The general solution of system
of PDEs (\ref{2.8.22}) is given by the formulae
\begin{equation}
\begin{array}{l}
    \vec B = {\rm rot\,}  \vec u,\quad \vec D = {\rm rot\,}  \vec
    w,\\[2mm]  
    \vec H = \p_t\vec w,\quad \vec E = -\p_t\vec u,
\end{array}
\label{2.8.24}
\end{equation}
where $\vec u = (u_1, u_2, u_3),\ \vec w=(w_1, w_2, w_3)$ are
arbitrary smooth vector-functions.}

To prove the lemma we make use of the well-known fact that the general 
solutions of equations
\begin{displaymath}
     {\rm div\,} \vec r =0,\quad {\rm rot\,} \vec\rho = \vec 0
\end{displaymath}
are given by the formulae
\begin{displaymath}
  \vec r = {\rm rot\,}  \vec R,\quad \vec\rho = {\rm grad} R_0,
\end{displaymath}
where $R_a,\ R_0$ are arbitrary twice differentiable functions. $\rhd$ 

According to Lemma 2.8.1, the Maxwell-Born-Infeld equations are
represented in the form (\ref{2.8.24}), where $\vec u,\ \vec w$ are
smooth vector-functions satisfying the first-order system of
nonlinear PDEs 
\begin{equation}
\begin{array}{l}
  {\rm rot\,} \vec w = -\tau\{\p_t\vec u + 
  [(\p_t\vec u)({\rm rot\,} \vec u)]{\rm rot\,} \vec u\},\\[2mm]
  \p_t\vec w=
  \tau\{{\rm rot\,}  \vec u - [(\p_t\vec u)({\rm rot\,}  \vec
  u)]\p_t\vec u\}, 
\end{array}
\label{2.8.25}
\end{equation}
with $ \tau = \{ 1+ ({\rm rot\,}  \vec u)^2 - (\p_t\vec u)^2 -
       [(\p_t\vec u)({\rm rot\,}  \vec u)]^2\}^{-1/2}$.

Thus, the over-determined system of fourteen equations (\ref{2.8.22}),
(\ref{2.8.23}) for twelve functions $E_a,\ H_a,\ D_a,\ B_a$ is reduced
to the system of six nonlinear PDEs for six functions $u_a,\ w_a$.

To construct exact solutions of (\ref{2.8.25}) we apply the Ansatz
\cite{108.5}\index{Ansatz!for tensor field}
\begin{equation}
   \vec u = \vec a\vp(t,\, \vec b\,\vec x,\, \vec c\,\vec x) 
   \equiv \vec a\vp(\om_0, \om_1, \om_2).
\label{2.8.26}
\end{equation}

Here $\vp\in C^2(\R^3, \R^1);\ \vec a,\ \vec b,\ \vec c$ are 
arbitrary constant vectors satisfying the conditions
\begin{eqnarray*}
  &&\vec a^{\, 2} = \vec b^{\, 2} =\vec c^{\, 2} = 1,\\
  &&\vec a\,\vec b = \vec b\,\vec c = \vec c\,\vec a =0.
\end{eqnarray*}

Since ${\rm rot\,}  \vec u = -\vec c\vp_{\om_1} + \vec b\vp_{\om_2}$, 
the equality $(\p_t\vec u)({\rm rot\,}  \vec u ) = 0$ holds. 
Consequently, system (\ref{2.8.25}) takes the form
\begin{equation}
\begin{array}{l}
        {\rm rot\,} \vec w = -\tau \vec a\vp_{\om_0},\\[2mm]
        \p_t\vec w =\tau(-\vec c\vp_{\om_1} + \vec b\vp_{\om_2}),
        \end{array}
\label{2.8.27}
\end{equation}
where
\begin{displaymath}
\tau = ( \vp^2_{\om_1}+ \vp^2_{\om_2} -\vp^2_{\om_0}+ 1)^{-1/2}.
\end{displaymath} 

The  compatibility condition $\p_t({\rm rot\,}  \vec w) 
= {\rm rot\,} (\p_t\vec w)$ when  applied to (\ref{2.8.27}) yields
\begin{displaymath}
   \p_t(-\tau\vec a\vp_{\om_0}) = {\rm rot\,}  [\tau(-\vec c\vp_{\om_1}+
     \vec b \vp_{\om_2})]
\end{displaymath}
or
\begin{displaymath}
   \vec a(1-\vp_{\om_{\mu}}\vp_{\om^{\mu}} )^{-3/2}\{(1-\vp_{\om_{\mu}}
    \vp_{\om^{\mu}})\Box \vp 
    + \vp_{\om_{\mu}\om_{\nu}}\vp_{\om^{\mu}}\vp_{\om^{\nu}}\} = \vec 0.
\end{displaymath}

Here summation over the repeated indices in the pseudo-Euclidean
space $R(1,2)$ is used.

Consequently, provided $\vp(\om)$ satisfies the nonlinear scalar PDE
\begin{equation}
 (1-\vp_{\om_{\mu}}\vp_{\om^{\mu}})\Box\vp + \vp_{\om_{\mu}\om_{\nu}}
 \vp_{\om^{\mu}}\vp_{\om^{\nu}} = 0
\label{2.8.28}
\end{equation}
with $ 1-\vp_{\om_{\mu}}\vp_{\om^{\mu}} \ne 0 $, formulae
(\ref{2.8.24}), (\ref{2.8.26}), (\ref{2.8.27}) give a particular
solution of system of nonlinear PDEs (2.2.22), (\ref{2.8.23}).

Wide classes of exact solutions of nonlinear equation (\ref{2.8.28})
were constructed in \cite{89}. Inserting these into formulae
(\ref{2.2.24}) and (\ref{2.2.26}) we get the following multi-parameter
families of exact solutions of the Maxwell-Born-Infeld equations:
\index{Exact solutions!of the Maxwell-Born-Infeld equations}
\begin{eqnarray*}
\vec E &=& -\vec a(\dot h_1\vec c\, \vec x + \dot h_2),
\\
\vec H &=&(1+h^2_1)^{-1/2}[h_1\vec b -(\dot h_1\vec c\, 
\vec x + \dot h_2)\vec c\,],
\\
\vec B &=& h_1\vec b - \vec c(\dot h_1\vec c\, \vec x + \dot h_2),
\\
\vec D &=&-\vec a(1+h^2_1)^{-1/2}(\dot h_1\vec c\, \vec x + \dot h_2),
\\[4mm]
\vec E &=& -(C_1 t/\om)\vec a (1+C_2\om^4)^{-1/2},
\\
\vec H &=& (C_1/\om)[-\vec b(\vec c\, \vec x) 
+ \vec c(\vec b\, \vec x)] (1+C_2\om^4 - C_1^2)^{-1/2};
\\
\vec B &=& (C_1/\om)[-\vec b(\vec c\, \vec x) 
+ \vec c(\vec b\, \vec x)] (1+C_2\om^4)^{-1/2},
\\
\vec D &=& -(C_1 t/\om) \vec a (1+C_2\om^4 - C_1^2)^{-1/2},
\\[4mm]
\vec E &=& \mp (1/4)\vec a\{C_1^{-1}(t-\vec b\, \vec x)^{-1}\coth
 [C_1(t + \vec b\, \vec x) + C_2]\}^{1/2}
\\
& &\times \{2C_1(t - \vec b\, \vec x) + \sinh 2
 [C_1(t + \vec b\, \vec x) + C_2]\}
\\
& &\times\cosh^{-2} [C_1(t + \vec b\, \vec x) + C_2],
\\
\vec H &=& \mp 2^{-3/2}\vec c\{2C_1(t-\vec b\, \vec x) - \sinh 2
 [C_1(t + \vec b\, \vec x) + C_2]\}
\\
& &\times C_1^{-1/2}(t - \vec b\, \vec x)^{-1/2}\{ \sinh 2
 [C_1(t + \vec b\, \vec x) + C_2]\}^{-1/2},
\\
\vec B &=& \mp (1/4)\vec c\{C_1^{-1}(t-\vec b\, \vec x)^{-1}\coth
 [C_1(t + \vec b\, \vec x) + C_2]\}^{1/2}
\\
& &\times \{2C_1(t - \vec b\, \vec x) - \sinh 2
 [C_1(t + \vec b\, \vec x) + C_2]\} 
\\
& &\times\cosh^{-2} [C_1(t + \vec b\, \vec x) + C_2],
\\
\vec D &=& \mp 2^{-3/2}\vec a\{2C_1(t-\vec b\, \vec x) + \sinh 2
 [C_1(t + \vec b\, \vec x) + C_2]\}
\\
& &\times C_1^{-1/2}(t - \vec b\, \vec x)^{-1/2}\{ \sinh 2
 [C_1(t + \vec b\, \vec x) + C_2]\}^{-1/2},
\\[4mm]
\vec E &=& \mp (1/2)\vec a\Bigl\{2C_3^{-1} +C_2C_3
\exp\{C_3(t-\vec b\, \vec x)\}\Bigr\}
\\
& &\times \Bigl\{C_2 \exp\{C_3(t-\vec b\, \vec x)\} +
 2C_3^{-1}(t + \vec b\, \vec x)\Bigr\}^{-1/2},
\\
\vec H &=& \mp (1/2)\vec c\Bigl\{2C_3^{-1} - C_2C_3
\exp\{C_3(t-\vec b\, \vec x)\}\Bigr\}
\\
& &\times \Bigl\{C_2 \exp\{C_3(t-\vec b\, \vec x)\} +
 2C_3^{-1}(t + \vec b\, \vec x)\Bigr\}^{-1/2},
\\
\vec B &=& \mp (1/2)\vec c\Bigl\{2C_3^{-1} - C_2C_3
\exp\{C_3(t-\vec b\, \vec x)\}\Bigr\}
\\
& &\times \Bigl\{C_2 \exp\{C_3(t-\vec b\, \vec x)\} +
 2C_3^{-1}(t + \vec b\, \vec x)\Bigr\}^{-1/2},
\\
\vec D &=& \mp (1/2)\vec a\Bigl\{2C_3^{-1} + C_2C_3
\exp\{C_3(t-\vec b\, \vec x)\}\Bigr\}
\\
& &\times \Bigl\{-C_2 \exp\{C_3(t-\vec b\, \vec x)\} +
 2C_3^{-1}(t + \vec b\, \vec x)\Bigr\}^{-1/2},
\end{eqnarray*}
where $h_i=h_i(t+\vec b\, \vec x) \in C^2(\R^1, \R^1)$ are arbitrary
functions; $C_1,\ C_2,\ C_3 $ are arbitrary real constants; $ \om^2 =
\om_0^2 - \om_1^2 - \om_2^2 \equiv t^2-(\vec b\, \vec x)^2 - (\vec c\,
\vec x)^2$.

Other classes of exact solutions of system of PDEs 
(\ref{2.8.22}), (\ref{2.8.23}) are obtained by putting 
\begin{equation}
   {\rm rot}\vec u = \vec 0,\quad  \vec u_{tt}=\vec 0
\label{2.8.29}
\end{equation}
in (\ref{2.8.25}).

Resulting from (\ref{2.8.29}) equations (\ref{2.8.25}) take the form
\begin{equation}
\begin{array}{l}
{\rm rot}\vec w = -\{1-({\rm grad }\vp)^2\}^{-1/2}{\rm grad}\vp,\\[2mm]
\vec u = {\rm grad }(t\vp +v),
\end{array}
\label{2.8.30}
\end{equation}
where $\{\vp(\vec x), v(\vec x)\} \subset C^2(\R^3, \R^1)$ are 
arbitrary functions.

Since ${\rm div\,} ({\rm rot} \vec w)=0$, from (\ref{2.8.30}) it
follows that 
\begin{displaymath}
   {\rm div\,} \{[1-({\rm grad}\,\vp)^2]^{-1/2}{\rm grad}\vp \} = 0,
\end{displaymath}
whence
\begin{displaymath}
   [1-({\rm grad}\,\vp)^2]^{-3/2}
   \{[1-({\rm grad}\,\vp)^2]\Delta \vp +
   \vp_{x_ax_b}\vp_{x_a}\vp_{x_b}\}=0. 
\end{displaymath}

The above equation with $ ({\rm grad}\vp)^2 \ne 1 $ is equivalent
to the elliptic analogue of PDE (\ref{2.8.28})
\begin{equation}
   (1-({\rm grad}\vp)^2)\Delta \vp + \vp_{x_ax_b}\vp_{x_a}\vp_{x_b}=0.
\label{2.8.31}
\end{equation}

In \cite{89} the following two classes of exact solutions of nonlinear
PDE (\ref{2.8.31}) were constructed
\begin{eqnarray*}
    \vp(\vec x) &=& C_1\ln\Bigl\{\Bigl((\vec a\, \vec x + C_2)^2 +
    (\vec b\, \vec x + C_3)^2\Bigr)^{1/2} \\
   & &+ \Bigl((\vec a\, \vec x + C_2)^2 + (\vec b\, \vec x + C_3)^2 
   + C_1^2\Bigr)^{1/2}\Bigr\},\\[2mm]
    \vp(\vec x) &=&\edi\int\limits_{\edi\om_0}\limits^{\edi(\vec
      x^2)^{1/2}} (1+ C_1^2\tau^4)^{-1/2}d\tau,
\end{eqnarray*}
where $C_a,\ a={1,2,3},\ \om_0$ are arbitrary real constants.

Inserting the above formulae into (\ref{2.8.24}), (\ref{2.8.30}) we
get two multi-parameter families of exact solutions of the 
Maxwell-Born-Infeld equations
\index{Exact solutions!of the Maxwell-Born-Infeld equations}
\begin{eqnarray*}
  \vec B&=&0,\quad \vec H\ \,=\ \,\vec 0,\\
  \vec D &=& C_1 \{\vec a(\vec a\, \vec x + C_2) +
            \vec b (\vec b \, \vec x + C_3)\} 
  [(\vec a\, \vec x + C_2)^2 +(\vec b \, \vec x + C_3)^2]^{-1},\\
  \vec E &=& C_1 \{\vec a(\vec a\, \vec x + C_2) +
            \vec b (\vec b \, \vec x + C_3)\}[(\vec a\, \vec x +
            C_2)^2 \\ 
   & &+ (\vec b \, \vec x + C_3)^2]^{-1/2}[(\vec a\, \vec x + C_2)^2 
   + (\vec b \, \vec x + C_3)^2 + C_1^2]^{-1/2};\\[4mm]
  \vec B&=&\vec 0,\quad \vec H\ \, =\ \, \vec 0,\\
   \vec D &=& -(1/C_1)\vec x(\vec x^{\, 2})^{-3/2},\\
   \vec E &=& -\vec x(\vec x^{\, 2})^{-1/2}\{1 + C_1^2(\vec
   x^{\, 2})^2\}^{-1/2}, 
\end{eqnarray*}
where $C_1,\ C_2,\ C_3$ are arbitrary real constants, 
$C_1\ne 0$.
\newpage
\phantom{.}

\newpage
\pagestyle{myheadings}                
\thispagestyle{empty}
\noindent
{\sl
C H A P T E R \ \  3\label{ch3}}
\vspace{2mm}

\hrule
\vspace{35mm}

\rightline
{\large\bf
TWO-DIMENSIONAL}
\vspace{2mm}

\rightline
{\large\bf
SPINOR MODELS}
\vspace{7mm}

\noindent
In this chapter nonlinear spinor PDEs with two independent variables
$x_0$, \ $x_1$ invariant under infinite-parameter groups are
considered. Such a broad symmetry makes it possible to obtain changes of
variables which linearize equa\-ti\-ons considered and to construct
their general solutions. Partial linearization of the nonlinear
Thirring system of PDEs is carried out.
\vspace{10mm}

\noindent
{\large \bf 3.1. Two-dimensional spinor equations invariant 
\vspace{1.5mm}
 
\noindent
\phantom{\large \bf 3.1. }under infinite-parameter groups\label{s3.1}}

\markboth{Chapter 3. TWO-DIMENSIONAL SPINOR MODELS}
{3.1. Two-dimensional spinor equations} 
\def\theequation{3.\arabic{section}.\arabic{equation}}
\setcounter {section} {1}
\setcounter {equation}{0}
\vspace{7mm}

\noindent
Invariance of PDEs under study with respect to some
infinite-parameter Lie groups makes it possible to construct their
exact solutions containing arbitrary functions. Of special interest
are two-dimensional equations possessing such a property since many of
them can be integrated in closed form. Methods used to construct the
general solutions of the two-dimensional d'Alembert \cite{78},
Liouville \cite{92}, Born-Infeld \cite{92}, Monge-Amp\`ere \cite{92,93},
gas dynamics \cite{82}, massless Thirring \cite{175,195} equations
are, in fact, based on the unique symmetry of the equations
enumerated.  

It is worth noting that most of the equations which are integrable by
the inverse scattering method also possess broad symmetry. They are
invariant under infinite-parameter Lie-B\"acklund groups
\index{Lie-B\"acklund group} 
\cite{6,127,161}. 

We will show that the list of integrable two-dimensional PDEs can be
supplemented by the following equations:
\begin{eqnarray}
& &\Bigl ( i\g_{\mu}\p_{\mu}-\lambda\g_{\mu} (\bar\psi\g^{\mu}\psi)
\Bigr )\psi=0;\label{3.1.1}\\[3mm]
& &\left \{\begin{array}{l} 
( i\g_{\mu}\p_{\mu}-e\g_{\mu}A^{\mu} )\psi
  =0,\\[2mm]
\p_{\nu}\p^{\nu}A_{\mu}-\p^{\mu}\p_{\nu}A_{\nu}=
-e\bar\psi\g_{\mu}\psi;\end{array}\right.\label{3.1.2}\\[3mm]
& &\Bigl (i(\g_0+\g_4)\p_0+i\g_1\p_1-\lambda(\psi^{\dagger}
\psi+\bar\psi\g_4\psi)^{1/2k}\Bigr )\psi=0.\label{3.1.3}
\end{eqnarray}

In (\ref{3.1.1})--(\ref{3.1.3}) $\psi=\psi(x_0, x_1)$ is a
four-component complex-valued func\-ti\-on-column; \ $A_0(x_0, x_1), \ 
A_1(x_0,x_1)$ \ are real-valued functions; \ $\mu,\nu=0,1$; \
$\lambda, \ e$ are constants. 

Dirac matrices\index{Dirac!matrices} are chosen in the form
$$
\g_0=\pmatrix{
       0&i\sigma_2\cr\cr
       -i\sigma_2&0\cr},\qquad
\g_1=\pmatrix{
       0&\sigma_1\cr\cr
       -\sigma_1&0\cr}.
$$

Let us note that PDE (\ref{3.1.1}) is a two-dimensional analogue of
the Dirac-Heisenberg equation\index{Dirac-Heisenberg equation}
\cite{116,121}, system (\ref{3.1.2}) is a two-dimensio\-nal system of
massless classical electrodynamics equations, PDE (\ref{3.1.3}) is a
two-dimensional Galilei-invariant equation for a massless particle
with the spin $s = 1/2$ (see also the Section 4.1).

Symmetry properties of equations (\ref{3.1.1})--(\ref{3.1.3}) 
are described by the following assertions.
\vspace{1.5mm}

\noindent
{\bf Theorem 3.1.1}\cite{93,206}.\ {\em System of PDEs (\ref{3.1.1})
is invariant under the infinite-parameter transformation group
$$
G_{\infty}=(O_{\xi}\otimes O_{\eta})\, \mbox{$\stimes$} \widetilde G,
$$
where  $O_{\xi}$ is the group of linear transformations of the
space $(\psi^0, \psi^{0*},\psi^2,\psi^{2*})$ preserving
the quadratic form
$|\psi^0|^2+|\psi^2|^2,$
its parameters being arbitrary smooth functions of
$\xi=x_0+x_1, \ |\psi^0|, \ |\psi^2|$; \ $O_{\eta}$
is the group of linear transformations in the space
$(\psi^1,\psi^{1*},\psi^3,\psi^{3*})$
preserving the quadratic form
$|\psi^1|^2+|\psi^3|^2,$
its parameters being arbitrary smooth functions of
$\eta=x_0-x_1, \ |\psi^1|, \ |\psi^3| $;
the group $\widetilde G$ is given by the formulae
\begin{equation}
\begin{array}{l}
       x_0^\prime=\edi{1\over 2}\Biggl (\edi\mathop\int\limits^{\edi
         x_0+x_1} 
       f_0^{-2}(z)dz+ \edi\mathop\int\limits^{\edi
         x_0-x_1}f_1^{-2}(z)dz 
       \Biggr ), \\[2mm]
       x_1^\prime=\edi{1\over 2}\Biggl (\edi\mathop\int\limits^{\edi 
         x_0+x_1} 
       f_0^{-2}(z)dz- \edi\mathop\int\limits^{\edi
         x_0-x_1}f_1^{-2}(z)dz 
       \Biggr ), \\[5mm]
       \psi^{\prime 0}=f_0(x_0+x_1)\psi^0, \
       \psi^{\prime 1}=f_1(x_0-x_1)\psi^1, \\[2mm]
       \psi^{\prime 2}=f_0(x_0+x_1)\psi^2, \
       \psi^{\prime 3}=f_1(x_0-x_1)\psi^3,  
\label{3.1.4}
\end{array}
\end{equation}

In (\ref{3.1.4}) $ f_0, \ f_1$ are arbitrary smooth real-valued
functions.} 
\vspace{1.5mm}

\noindent
{\em Proof.}$\quad$ After writing component-wise we represent 
(\ref{3.1.1}) in the form
\begin{eqnarray*}
i(\partial_0-\partial_1)\psi^0&=&2\lambda(|\psi^1|^2
+|\psi^3|^2)\psi^0, \\ 
i(\partial_0+\partial_1)\psi^1
&=&2\lambda(|\psi^0|^2+|\psi^2|^2)\psi^1, \\
i(\partial_0-\partial_1)\psi^2&=&2\lambda(|\psi^1|^2
+|\psi^3|^2)\psi^2, \\ 
i(\partial_0+\partial_1)\psi^3
&=&2\lambda(|\psi^0|^2+|\psi^2|^2)\psi^3.
\end{eqnarray*}

In the cone variables $\xi=x_0+x_1, \ \eta=x_0-x_1$ the above 
system reads
\begin{equation}
\begin{array}{rcl}
i\p_\eta\psi^0&=&\lambda(|\psi^1|^2+|\psi^3|^2)\psi^0,
 \\ [2mm]
i\p_\xi\psi^1&=&\lambda(|\psi^0|^2+|\psi^2|^2)\psi^1,
 \\ [2mm]
i\p_\eta\psi^2&=&\lambda(|\psi^1|^2+|\psi^3|^2)\psi^2,
 \\ [2mm]
i\p_\xi\psi^3&=&\lambda(|\psi^0|^2+|\psi^2|^2)\psi^3,
\end{array}
\label{3.1.5}
\end{equation}
the group $\widetilde G$ being given by the formulae
\begin{equation}
\begin{array}{l}
       \xi^\prime=\edi\mathop\int\limits^{\edi\xi}f_0^{-2}(z)dz, \
       \eta^\prime=\edi\mathop\int\limits^{\edi\eta}f_1^{-2}(z)dz,
       \\[3mm] 
       \psi^{\prime 0}=f_0(\xi)\psi^0, \
       \psi^{\prime 1}=f_1(\eta)\psi^1, \\[2mm]
       \psi^{\prime 2}=f_0(\xi)\psi^2, \
       \psi^{\prime 3}=f_1(\eta)\psi^3.  
\label{3.1.6}
\end{array}
\end{equation}

Applying to both parts of the first equation of system (\ref{3.1.5})
the operation of complex conjugation we have
\begin{eqnarray*}
-i\p_\eta\psi^{*0}&=&\lambda(|\psi^1|^2
+|\psi^3|^2)\psi^{*0},
\end{eqnarray*}
whence
\begin{displaymath}
\psi^0\p_\eta\psi^{*0}+\psi^{*0}\p_\eta\psi^0=0
\end{displaymath}
or
\begin{displaymath}
\p_\eta|\psi^0|=0.
\end{displaymath}

Similarly,
\begin{displaymath}
\p_\xi|\psi^1|=0,\quad \p_\eta|\psi^2|=0,\quad 
\p_\xi|\psi^3|=0.
\end{displaymath}
From the equalities obtained it follows that system of PDEs
(\ref{3.1.5}) is invariant under the group $O_\xi\otimes O_\eta$.

Let us prove that system (\ref{3.1.5}) admits transformation 
group (\ref{3.1.6}). To this end we make in (\ref{3.1.5}) the 
change of variables according to formulae (\ref{3.1.6}) thus
obtaining the following equations:
\begin{eqnarray*}
i\p_{\eta^\prime}\psi^{\prime 0}-\lambda(|\psi^{\prime 1}
|^2 +|\psi^{\prime 3}|^2)\psi^{\prime 0}&=&
f_0f_1^2\Bigl (i\p_\eta\psi^0-\lambda(|\psi^1|^2
+|\psi^3|^2)\psi^0\Bigr ), \\
i\p_{\xi^\prime}\psi^{\prime 1}-\lambda(|\psi^{\prime 0}
|^2 +|\psi^{\prime 2}|^2)\psi^{\prime 1}&=&
f_1f_0^2\Bigl (i\p_\xi\psi^1-\lambda(|\psi^0|^2
+|\psi^2|^2)\psi^1\Bigr ), \\
i\p_{\eta^\prime}\psi^{\prime 2}-\lambda(|\psi^{\prime 1}
|^2 +|\psi^{\prime 3}|^2)\psi^{\prime 2}&=&
f_0f_1^2\Bigl (i\p_\eta\psi^2-\lambda(|\psi^1|^2
+|\psi^3|^2)\psi^2\Bigr ), \\
i\p_{\xi^\prime}\psi^{\prime 3}-\lambda(|\psi^{\prime 0}
|^2 +|\psi^{\prime 2}|^2)\psi^{\prime 3}&=&
f_1f_0^2\Bigl (i\p_\xi\psi^3-\lambda(|\psi^0|^2
+|\psi^2|^2)\psi^2\Bigr ), 
\end{eqnarray*}
whence the validity of the theorem follows. $\rhd$
\vspace{1.5mm}

\noindent
{\bf Theorem 3.1.2.} {\em System of PDEs (\ref{3.1.2}) is invariant
  under the infinite-parameter transformation group of the form
\begin{eqnarray*}
  G_\infty=(O_\xi\otimes O_\eta)\mbox{$\stimes$} \widetilde
  P(1,1)\mbox{$\stimes$} U(1),
\end{eqnarray*}
where $ \widetilde P(1,1)$ is the extended Poincar\'e group, U(1) 
is the group of gauge transformations\index{Gauge transformation group}
\begin{eqnarray*}
\psi^{\prime \alpha}&=&\psi^\alpha \exp  \{-ief\}, \\
A_{\mu}^\prime&=& A_\mu +\p^\mu f 
\end{eqnarray*}
with} $f=f(x_0,x_1)\in C^3(\R^2,\R^1)$.
\vspace{1.5mm}

\noindent
{\bf Theorem 3.1.3.} {\em System of PDEs (\ref{3.1.3}) is invariant
  under the infinite-parameter transformation group  having the 
following generators:
\vspace{1.5mm}

\noindent 
\underline{under  $k\ne 1/2 $}
\begin{eqnarray*}
& &P_0=\p_0, \quad P_1=\p_1, \\
& &D_1=x_0\p_0+x_1\p_1+k,\\
& &D_2=2x_0\p_0+x_1\p_1+k
+(1/2)(1-\g_0\g_4),\\
& &G=w_1(x_0)\p_1-(1/2)\dot w_1(x_0)(\g_0+\g_4)\g_1,\\ 
& &Q=(\g_0+\g_4)\Bigl (\g_2w_2(x_0)+\g_3w_3(x_0)\Bigr );
\end{eqnarray*}
\underline{under  $k=1/2$}
\begin{eqnarray*}
& &\widetilde A=w_0(x_0)\p_0+\dot w_0(x_0)x_1\p_1
+(1/2)\dot w_0(x_0)\quad\\
& &\quad-(1/2)\ddot w_0(x_0)x_1(\g_0+\g_4)\g_1,\\ 
& &G=w_1(x_0)\p_1-(1/2)\dot w_1(x_0)(\g_0+\g_4)\g_1,\\ 
& &Q=(\g_0+\g_4)\Bigl (\g_2w_2(x_0)+\g_3w_3(x_0)\Bigr ),\\ 
& &D=2x_0\p_0+x_1\p_1 + 1 - (1/2)\g_0\g_4.
\end{eqnarray*}

Here $w_0,\ldots,w_3$ are arbitrary smooth real-valued functions}.

Theorem 3.1.2 is proved in the same way as Theorem 3.1.1. To
prove Theorem 3.1.3 it is necessary to apply the Lie method.

Let us note that symmetry properties of equations
(\ref{3.1.1})--(\ref{3.1.3}) are not exhausted by the invariance under
the local symmetry groups described above. As shown in \cite{93}
system of PDEs (\ref{3.1.1}) is invariant under the group of nonlocal
(integral) transformations\index{Nonlocal!transformation group}
\begin{eqnarray*}
\psi^{\prime 0}&=&\theta_0\psi^0\exp  \Biggl \{ -i\lambda 
\edi\mathop\int\limits^{\edi x_0-x_1}\Bigl ((|\theta_1
|^2-1)|\psi^1|^2+(|\theta_3
|^2-1)|\psi^3|^2\Bigr )d\eta\Biggr \}, \\
\psi^{\prime 1}&=&\theta_1\psi^1\exp  \Biggl \{ -i\lambda 
\edi\mathop\int\limits^{\edi x_0+x_1}\Bigl ((|\theta_0
|^2-1)|\psi^0|^2+(|\theta_2
|^2-1)|\psi^2|^2\Bigr )d\xi\Biggr \}, \\
\psi^{\prime 2}&=&\theta_2\psi^2\exp  \Biggl \{ -i\lambda 
\edi\mathop\int\limits^{\edi x_0-x_1}\Bigl ((|\theta_1
|^2-1)|\psi^1|^2+(|\theta_3
|^2-1)|\psi^3|^2\Bigr )d\eta\Biggr \}, \\
\psi^{\prime 3}&=&\theta_3\psi^3\exp  \Biggl \{ -i\lambda 
\edi\mathop\int\limits^{\edi x_0+x_1}\Bigl ((|\theta_0
|^2-1)|\psi^0|^2+(|\theta_2
|^2-1)|\psi^2|^2\Bigr )d\xi\Biggr \}, 
\end{eqnarray*}
where $\{\theta_0,\ldots,\theta_3\} \subset \C^1$.
\vspace{10mm}

\noindent
{\large\bf 3.2. Nonlinear two-dimensional Dirac-Heisenberg
  equations\label{s3.2}}  

\markboth{Chapter 3. TWO-DIMENSIONAL SPINOR MODELS}
{3.2. Nonlinear two-dimensional Dirac-Heisenberg equations} 
\def\theequation{3.\arabic{section}.\arabic{equation}}
\setcounter {section} {2}
\setcounter {equation}{0}
\vspace{7mm}

\noindent
In this section we will construct the general solution of system
(\ref{3.1.1}) with the help of the nonlocal linearization 
method\index{Nonlocal!linearization method}
\cite{92,93}. In other words, a nonlocal change of variables reducing
(\ref{3.1.1}) to a system  of linear PDEs will be suggested.

The form of the change of variables is implied by the structure of the 
group of integral transformations given at the end of the previous
section. We introduce new dependent variables $\varphi^0(\xi,\eta),
\ldots,\varphi^3(\xi,\eta)$ in the following way:
\begin{eqnarray}
\psi^0&=&\varphi^0\exp  \Biggl \{ -i\lambda 
\edi\mathop\int\nolimits\Bigl (|\varphi^1|^2+|\varphi^3|^2\Bigr )
d\eta\Biggr \},\non \\
\psi^1&=&\varphi^1\exp  \Biggl \{ -i\lambda 
\edi\mathop\int\nolimits\Bigl (|\varphi^0|^2+|\varphi^2|^2\Bigr )
d\xi\Biggr \},\non\\
\psi^2&=&\varphi^2\exp  \Biggl \{ -i\lambda 
\edi\mathop\int\nolimits\Bigl (|\varphi^1|^2+|\varphi^3|^2\Bigr )
d\eta\Biggr \},\label{3.2.1}\\
\psi^3&=&\varphi^3\exp  \Biggl \{ -i\lambda 
\edi\mathop\int\nolimits\Bigl (|\varphi^0|^2+|\varphi^2|^2\Bigr )
d\xi\Biggr \}.\non
\end{eqnarray}

Substituting (\ref{3.2.1}) into (\ref{3.1.5}) we get a system of
linear equations for $\varphi^0,\ldots,\varphi^3$
\begin{equation}
\begin{array}{l}
\p_\eta\varphi^0=0, \quad \p_\xi\varphi^1=0, \\[2mm]
\p_\eta\varphi^2=0,\quad \p_\xi\varphi^3=0.
\end{array}
\label{3.2.2} 
\end{equation} 

Integration of the above equations yields the following expressions
for $\varphi^\mu, \ \mu ={0,\ldots,3}$:
\begin{equation}
\begin{array}{l}
\varphi^0=U^0(\xi), \quad\varphi^1=U^1(\eta), \\[2mm]
\varphi^2=U^2(\xi), \quad \varphi^3=U^3(\eta),
\end{array}
\label{3.2.3}
\end{equation}
where $U^\mu\in C^1(\R^1,\C^1)$ are arbitrary functions.

Substitution of formulae (\ref{3.2.3}) into (\ref{3.2.1}) with
subsequent change of independent variables $\xi\to x_0+x_1, \ \eta\to 
x_0-x_1$ gives the general solution of the nonlinear Dirac-Heisenberg
equation (\ref{3.1.1})
\index{General solution!of the two-dimensional Dirac-Heisenberg equation}
\begin{equation}
\begin{array}{l}
\psi^0(x)=U^0(x_0+x_1)\exp  \Biggl \{ -i\lambda 
\edi\mathop\int\limits^{\edi x_0-x_1}\Bigl (| U^1|^2+| U^3|^2\Bigr )
d\tau\Biggr \}, \\[1mm]
\psi^1(x)=U^1(x_0-x_1)\exp  \Biggl \{ -i\lambda 
\edi\mathop\int\limits^{\edi x_0+x_1}\Bigl (| U^0|^2+| U^2|^2\Bigr )
d\tau\Biggr \}, \\[1mm]
\psi^2(x)=U^2(x_0+x_1)\exp  \Biggl \{ -i\lambda 
\edi\mathop\int\limits^{\edi x_0-x_1}\Bigl (| U^1|^2+| U^3|^2\Bigr )
d\tau\Biggr \}, \\[1mm]
\psi^3(x)=U^3(x_0-x_1)\exp  \Biggl \{ -i\lambda 
\edi\mathop\int\limits^{\edi x_0+x_1}\Bigl (| U^0|^2+| U^2|^2\Bigr )
d\tau\Biggr \}. 
\end{array}
\label{3.2.4}
\end{equation}

The result obtained enables us to construct in explicit form solution
of the classical Cauchy problem\index{Cauchy problem} for system of 
PDEs (\ref{3.1.1})
\begin{equation}
\begin{array}{l}
\Bigl ( i\g_{\mu}\p_{\mu}-\lambda\g_{\mu} (\bar\psi\g^{\mu}\psi)
\Bigr )\psi=0, \\[2mm]
\psi^\alpha(0,x_1)=f^\alpha(x_1), \ x_1\in\R^1,  
\end{array}
\label{3.2.5}
\end{equation}
where $f^\alpha\in C^1(\R^1,\C^1)$, \ $\alpha ={0,\ldots,3}$.

Imposing on the solution (\ref{3.2.4}) the initial conditions of the
Cauchy problem (\ref{3.2.5}) we get functional relations for
determination of $U^\mu, \ \mu={0,\ldots,3}$
\begin{eqnarray*}
f^0(z)&=&U^0(z)\exp  \Biggl \{ -i\lambda 
\edi\mathop\int\limits^{\edi -z}\Bigl (| U^1|^2+| U^3|^2\Bigr )
d\tau\Biggr \}, \\
f^1(z)&=&U^1(-z)\exp  \Biggl \{ -i\lambda 
\edi\mathop\int\limits^{\edi z}\Bigl (| U^0|^2+| U^2|^2\Bigr )
d\tau\Biggr \}, \\
f^2(z)&=&U^2(z)\exp  \Biggl \{ -i\lambda 
\edi\mathop\int\limits^{\edi -z}\Bigl (| U^1|^2+| U^3|^2\Bigr )
d\tau\Biggr \}, \\
f^3(z)&=&U^3(-z)\exp  \Biggl \{ -i\lambda 
\edi\mathop\int\limits^{\edi z}\Bigl (| U^0|^2+| U^2|^2\Bigr )
d\tau\Biggr \}, 
\end{eqnarray*}
whence it follows   
\begin{eqnarray*}
U^0(z)&=&f^0(z)\exp  \Biggl \{ i\lambda 
\edi\mathop\int\limits^{\edi -z}\Bigl (| f^1(-\tau)|^2
+| f^3(-\tau)|^2\Bigr )d\tau \Biggr \}, \\
U^1(-z)&=&f^1(z)\exp  \Biggl \{ i\lambda 
\edi\mathop\int\limits^{\edi z}\Bigl (| f^0(\tau)|^2
+| f^2(\tau)|^2\Bigr )d\tau \Biggr \}, \\
U^2(z)&=&f^2(z)\exp  \Biggl \{ i\lambda 
\edi\mathop\int\limits^{\edi -z}\Bigl (| f^1(-\tau)|^2
+| f^3(-\tau)|^2\Bigr )d\tau\Biggr \}, \\
U^3(-z)&=&f^3(z)\exp  \Biggl \{ i\lambda 
\edi\mathop\int\limits^{\edi z}\Bigl (| f^0(\tau)|^2
+| f^2(\tau)|^2\Bigr )d\tau\Biggr \}. 
\end{eqnarray*}

Substitution of the above equalities into (\ref{3.2.4}) gives the
solution of the Cauchy problem (\ref{3.2.5})
\begin{eqnarray*}
\psi^0(x)&=&f^0(x_1+x_0)\exp  \Biggl \{ i\lambda 
\edi\mathop\int\limits^{\edi x_1-x_0}_{\edi x_1+x_0}
\Bigl (| f^1|^2+| f^3|^2\Bigr )
d\tau\Biggr \}, \\
\psi^1(x)&=&f^1(x_1-x_0)\exp  \Biggl \{ -i\lambda 
\edi\mathop\int\limits^{\edi x_1+x_0}_{\edi x_1-x_0}
\Bigl (| f^0|^2+| f^2|^2\Bigr )
d\tau\Biggr \}, \\
\psi^2(x)&=&f^2(x_1+x_0)\exp  \Biggl \{ i\lambda 
\edi\mathop\int\limits^{\edi x_1-x_0}_{\edi x_1+x_0}
\Bigl (| f^1|^2+| f^3|^2\Bigr )
d\tau\Biggr \}, \\
\psi^3(x)&=&f^3(x_1-x_0)\exp  \Biggl \{ -i\lambda 
\edi\mathop\int\limits^{\edi x_1+x_0}_{\edi x_1-x_0}
\Bigl (| f^0|^2+| f^2|^2\Bigr )
d\tau\Biggr \}. 
\end{eqnarray*}

Thus, the Cauchy problem (\ref{3.2.5}) with $f^\alpha \in C^1(\R^1,
\C^1)$, \ $\alpha={0,\ldots,3}$ has the unique solution.

In the case involved we have succeeded in integrating a nonlinear
system of PDEs due to the fact that it is equivalent to the linear
system (\ref{3.2.2}). In some cases the nonlocal linearization method
makes it possible to construct wide classes of exact solutions of
essentially nonlinear PDEs. This is achieved by imposing such
additional constraints on the equation under study that the system
obtained is linearizable. A peculiar example is the generalized
Thirring model\index{Thirring model}
\begin{equation}
\begin{array}{l}
iu_y=mv+\lambda_1|v|^2u, \\[2mm]
iv_x=mu+\lambda_2|u|^2v,
\end{array}
\label{3.2.6}
\end{equation}
where $u=u(x,y), \ v=v(x,y)$ are complex-valued functions, $m, \
\lambda_1, \ \lambda_2$ are real constants.

Provided $\lambda_1=\lambda_2=\lambda$, system of PDEs (\ref{3.2.6})
coincides with the classical Thirring model that is integrable by
means of the inverse scattering method \cite{175,195}. As established 
by David \cite{45.1} the generalized Thirring model (\ref{3.2.6}) 
is also integrable by the mentioned method and, therefore, has soliton 
solutions.

Here we restrict ourselves to the case
\begin{displaymath}
\lambda_1=\lambda,\quad \lambda_2=-\lambda
\end{displaymath}
and consider the following system:
\begin{equation}
\begin{array}{l}
iu_y=mv+\lambda|v|^2u, \\[2mm]
iv_x=mu-\lambda|u|^2v.
\end{array}
\label{3.2.7}
\end{equation}

We will show that there exists a map of the set of solutions of the
linear Klein-Gordon equation\index{Klein-Gordon equation}
\begin{equation}
w_{xy}+m^2w=0
\label{3.2.8}
\end{equation}
into the set of solutions of system of PDEs (\ref{3.2.7}).

To this end we apply the following Ansatz \cite{209.3}:
\begin{equation}
\begin{array}{l}
u=F_1\exp  \{iG+(i\pi/4)\}, \\[2mm]
v=F_2\exp  \{iG-(i\pi/4)\}
\end{array}
\label{3.2.9}
\end{equation}
where $F_1, \ F_2, \ G$ are some real-valued functions.

Substitution of (\ref{3.2.9}) into (\ref{3.2.7}) yields an
over-determined system of four nonlinear PDEs for $F_1, \ F_2, \ G$
\begin{eqnarray*}
&&F_{1 y}=-mF_2, \quad F_{2 x}=mF_1, \\
&&G_x=\lambda F_1^2, \quad G_y=-\lambda F_2^2.
\end{eqnarray*}

Since
\begin{eqnarray*}
(G_x)_y=2\lambda F_1F_{1 y}=-2\lambda mF_1F_2=-2\lambda F_2F_{2
x}=(G_y)_x,
\end{eqnarray*}
the above system is compatible and its general solution can be
represented in the form
\begin{eqnarray*}
& &F_1=w(x,y), \quad F_2=-m^{-1}w_y(x,y), \\
& &G=\lambda\edi\mathop\int\limits^{\edi x}_{\edi A}w^2(\tau,y)d\tau 
-\lambda m^{-2}\mathop\int\limits^{\edi y}_{\edi B}w_y^2(A,\tau)d\tau, 
\end{eqnarray*}
where $A, \ B$ are some real constants and $w(x,y)$ is an arbitrary 
solution of (\ref{3.2.8}).

Thus, each solution of the linear Klein-Gordon equation (\ref{3.2.8})
gives rise to the exact solution of the nonlinear system (\ref{3.2.7}) 
of the form
\begin{eqnarray*}
u&=&w\exp  \Biggl \{(i\pi/4)+i\lambda\edi\mathop\int
\limits^{\edi x}_{\edi A}w^2(\tau,y)d\tau 
-i\lambda m^{-2}\edi\mathop\int
\limits^{\edi y}_{\edi B}w_y^2(A,\tau)d\tau\Biggr \}, \\
v&=&-m^{-1}w_y\exp  \Biggl \{(-i\pi/4)+i\lambda\edi\mathop\int
\limits^{\edi x}_{\edi A}w^2(\tau,y)d\tau 
-i\lambda m^{-2}\edi\mathop\int
\limits^{\edi y}_{\edi B}w_y^2(A,\tau)d\tau\Biggr \}. \\
\end{eqnarray*}

Due to invariance of system (\ref{3.2.7}) under the one-parameter
group of gauge transformations
\begin{eqnarray*}
u^\prime=u\exp  \{i\theta\}, \ v^\prime=v\exp  \{i\theta\}, \
\theta\in \R^1
\end{eqnarray*}
the solution obtained can be rewritten in the following equivalent
form:  
\index{Exact solutions!of the Thirring model}
\begin{equation}
\begin{array}{rcl}
u&=&w\exp  \Biggl \{i\lambda\edi\mathop\int
\limits^{\edi x}_{\edi A}w^2(\tau,y)d\tau 
-i\lambda m^{-2}\edi\mathop\int
\limits^{\edi y}_{\edi B}w_y^2(A,\tau)d\tau\Biggr \}, \\[2mm]
v&=&im^{-1}w_y\exp  \Biggl \{i\lambda\edi\mathop\int
\limits^{\edi x}_{\edi A}w^2(\tau,y)d\tau 
-i\lambda m^{-2}\edi\mathop\int
\limits^{\edi y}_{\edi B}w_y^2(A,\tau)d\tau\Biggr \}. 
\end{array}
\label{3.2.10}
\end{equation}

The above formulae can be interpreted as a linearizing nonlocal
transformation, since functions (\ref{3.2.10}) satisfy system of PDEs
(\ref{3.2.7}) iff the function $w(x,y)$ satisfies the linear Klein-Gordon
equation (\ref{3.2.8}). However in this way only a part of solutions of
system under study is obtained. Therefore, we can speak about partial
linearization of the generalized Thirring model (another example of
partial linearization\index{Partial linearization} is considered in
Section 2.8).  
\vspace{10mm}

\noindent
{\large\bf 3.3. Two-dimensional classical electrodynamics
  equations\label{s3.3}} 

\markboth{Chapter 3. TWO-DIMENSIONAL SPINOR MODELS}
{3.3. Two-dimensional classical electrodynamics equations}
\def\theequation{3.\arabic{section}.\arabic{equation}}
\setcounter {section} {3}
\setcounter {equation}{0}
\vspace{7mm}

\noindent
The change of variables (\ref{3.2.1}) proves to be efficient when
constructing the general solution of the system of nonlinear PDEs
(\ref{3.1.2}).

Writing the first equation (\ref{3.1.2}) component-wise and passing to 
the cone variables $\xi, \ \eta$ we come to the following system of
PDEs for the functions $\psi^0(\xi,\eta),\ldots,\psi^3(\xi,\eta)$:
\begin{equation}
\begin{array}{rcl}
i\p_\eta\psi^0&=&(e/2)(\widetilde A_0+\widetilde A_1)\psi^0, \\[2mm] 
i\p_\xi\psi^1&=&(e/2)(\widetilde A_0-\widetilde A_1)\psi^1, \\[2mm]
i\p_\eta\psi^2&=&(e/2)(\widetilde A_0+\widetilde A_1)\psi^2, \\[2mm]
i\p_\xi\psi^3&=&(e/2)(\widetilde A_0-\widetilde A_1)\psi^3, 
\end{array} 
\label{3.3.1}
\end{equation}
where
\begin{displaymath}
\widetilde A_\mu=A_\mu\Bigl ((1/2)(\xi+\eta),\, (1/2)(\xi-\eta)\Bigr ).
\end{displaymath}

On making the change of variables
\begin{eqnarray}
\psi^0&=&\varphi^0(\xi,\eta)\exp  \Biggl \{ -(ie/2)
\edi\mathop\int\nolimits(\widetilde A_0+\widetilde A_1)d\eta\Biggr
\},\non \\ 
\psi^1&=&\varphi^1(\xi,\eta)\exp  \Biggl \{ -(ie/2)
\edi\mathop\int\nolimits(\widetilde A_0-\widetilde A_1)d\xi\Biggr
\},\non\\
\psi^2&=&\varphi^2(\xi,\eta)\exp  \Biggl \{ -(ie/2)
\edi\mathop\int\nolimits(\widetilde A_0+\widetilde A_1)d\eta\Biggr \},
\label{3.3.2}\\
\psi^3&=&\varphi^3(\xi,\eta)\exp  \Biggl \{ -(ie/2)
\edi\mathop\int\nolimits(\widetilde A_0-\widetilde A_1)d\xi\Biggr
\},\non  
\end{eqnarray}
we rewrite (\ref{3.3.1}) in the following way:
\begin{eqnarray*}
&&\p_\eta\varphi^0=0, \quad \p_\xi\varphi^1=0, \\
&&\p_\eta\varphi^2=0,\quad \p_\xi\varphi^3=0.
\end{eqnarray*}

The general solution of the above system is given by formulae
(\ref{3.2.3}). Consequently, the general solution of equations
(\ref{3.3.1}) is of the form
\begin{equation}
\begin{array}{rcl}
\psi^0&=&U^0(\xi)\exp  \Biggl \{ -(ie/2)
\edi\mathop\int\nolimits(\widetilde A_0+\widetilde A_1)d\eta\Biggr \},
\\[4mm] 
\psi^1&=&U^1(\eta)\exp  \Biggl \{ -(ie/2)
\edi\mathop\int\nolimits(\widetilde A_0-\widetilde A_1)d\xi\Biggr \},
\\[4mm] 
\psi^2&=&U^2(\xi)\exp  \Biggl \{ -(ie/2)
\edi\mathop\int\nolimits(\widetilde A_0+\widetilde A_1)d\eta\Biggr \},
\\[4mm] 
\psi^3&=&U^3(\eta)\exp  \Biggl \{ -(ie/2)
\edi\mathop\int\nolimits(\widetilde A_0-\widetilde A_1)d\xi\Biggr \}, 
\end{array}
\label{3.3.3}
\end{equation}
where $U^\mu\in C^1(\R^1,\C^1), \ \mu={0,\ldots,3}$ are
arbitrary functions.

Substituting expressions (\ref{3.3.3}) into the remaining equations of
system (\ref{3.1.2}) we get an over-determined system of PDEs for
$A_0, \ A_1$
\begin{equation}
\begin{array}{l}
\p_1(\p_1A_0+\p_0A_1)=e\Bigl (|U^0|^2+|U^1|^2+|U^2|^2
+|U^3|^2\Bigr), \\[2mm]
\p_0(\p_1A_0+\p_0A_1)=e\Bigl (|U^0|^2-|U^1|^2+|U^2|^2
-|U^3|^2\Bigr). 
\label{3.3.4}
\end{array}
\end{equation}

Introducing the new dependent variable
\begin{displaymath}
w=\p_1A_0+\p_0A_1
\end{displaymath}
we rewrite equations (\ref{3.3.4}) as follows
\begin{displaymath}
\p_\xi w=e\Bigl (|U^0(\xi)|^2+|U^2(\xi)|^2\Bigr ), \quad
\p_\eta w=e\Bigl (|U^1(\eta)|^2+|U^3(\eta)|^2\Bigr ),
\end{displaymath}
whence
\begin{displaymath}
w=e\edi\mathop\int\limits^{\edi\xi}\Bigl (|U^0(z)|^2
+|U^2(z)|^2\Bigr)dz 
+e\edi\mathop\int\limits^{\edi\eta}\Bigl(|U^1(z)|^2
+|U^3(z)|^2\Bigr )dz.
\end{displaymath}

Consequently, to determine $A_0(x_0,x_1), \ A_1(x_0,x_1)$ it is
necessary to integrate PDE
\begin{eqnarray*}
A_{0 x_1}+A_{1 x_0}
&=&e\edi\mathop\int\limits^{\edi x_0+x_1}\Bigl (|U^0(z)|^2
+|U^2(z)|^2\Bigr)dz \\
& &+e\edi\mathop\int\limits^{\edi x_0-x_1}\Bigl (|U^1(z)|^2
+|U^3(z)|^2\Bigr )dz,
\end{eqnarray*}
whose general solution is of the form
\begin{equation}
\begin{array}{rcl}
A_0&=&e\edi\mathop\int\limits^{\edi x_0+x_1}\edi\mathop\int\limits^
{\edi z}\Bigl (|U^0(\xi)|^2+|U^2(\xi)|^2\Bigr )d\xi dz+\p_0f, \\[2mm]
A_1&=&e\edi\mathop\int\limits^{\edi
  x_0-x_1}\edi\mathop\int\limits^{\edi z}
\Bigl (|U^1(\eta)|^2+|U^3(\eta)|^2\Bigr )d\eta dz-\p_1f
\end{array}
\label{3.3.5}
\end{equation}
with an arbitrary function $f=f(x_0,x_1)\in C^3(\R^2,\R^1)$.
\index{General solution!of the two-dimensional classical
electrodynamics equations}

Substitution of (\ref{3.3.5}) into (\ref{3.3.3}) gives rise to the
final expressions for the functions $\psi^0(x),\ldots,\psi^3(x)$
\begin{eqnarray}
\left\{\matrix{\psi^0(x) \cr \psi^2(x)}\right\}&=&
\left\{\matrix{U^0(x_0+x_1) \cr U^2(x_0+x_1)}\right\}
\exp  \Biggl \{-ief+(ie^2/2)\nonumber\\[1mm]
& &\times\Biggl ((x_1-x_0) \edi\mathop\int\limits^{\edi x_0+x_1}
\edi\mathop\int\limits^{\edi z}\Bigl (|U^0(\xi)|^2
+|U^2(\xi)|^2\Bigr )d\xi dz \nonumber\\[1mm] 
& &-\edi\mathop\int\limits^{\edi x_0-x_1}\edi\mathop\int\limits^{\edi
  z_2} 
\edi\mathop\int\limits^{\edi z_1}\Bigl (|U^1(\eta)|^2
+|U^3(\eta)|^2\Bigr )d\eta dz_1 dz_2\Biggr )\Biggr \}, \nonumber\\ 
\left\{\matrix{\psi^1(x) \cr \psi^3(x)}\right\}&=&
\left\{\matrix{U^1(x_0-x_1) \cr U^3(x_0-x_1)}\right\}
\exp  \Biggl \{-ief+(ie^2/2)\label{3.3.6}
\\[1mm]
& &\times\Biggl ((x_1+x_0) \edi\mathop\int\limits^{\edi x_0-x_1}
\edi\mathop\int\limits^{\edi z}\Bigl (|U^1(\eta)|^2
+|U^3(\eta)|^2\Bigr )d\eta dz \nonumber\\[1mm] 
& &-\edi\mathop\int\limits^{\edi x_0+x_1}\edi\mathop\int\limits^{\edi
  z_2} 
\edi\mathop\int\limits^{\edi z_1}\Bigl (|U^0(\xi)|^2
+|U^2(\xi)|^2\Bigr )d\xi dz_1 dz_2\Biggr )\Biggr \}.\nonumber
\end{eqnarray}

Choosing arbitrary functions $U^\mu, \ \mu={0,\ldots,3}$ in proper way
we can obtain special classes of solutions possessing some additional
properties.

If, for example, we choose in formulae (\ref{3.3.3}), (\ref{3.3.6})
\begin{equation}
\begin{array}{l}
U^\mu(z)=C^\mu\Biggl (\edi{d\over dz}\exp  \{-z^2\}\Biggr )^{1/2}, \
\mu=0,2, \\[1mm] 
U^1(z)=U^3(z)=0, \ f=0,
\end{array}
\label{3.3.7}
\end{equation}
where $C^0, \ C^2$ are complex constants, then the corresponding wave 
function $\psi(x)$ is localized in the neighborhood of the point
$x_0=x_1$. Consequently, solution (\ref{3.3.6}), (\ref{3.3.7}) is
a solitary wave propagating with the velocity $v=1$.

Substitution of expressions (\ref{3.3.7}) into (\ref{3.3.5}) yields
the following formulae for $A_0(x), \ A_1(x)$:
\begin{eqnarray*}
A_0(x)&=&e\Bigl (|C^0|^2+|C^2|^2\Bigr )\edi\mathop\int
\limits^{\edi x_0+x_1}\exp \{-\tau^2\}d\tau,\\
A_1(x)&=&0,
\end{eqnarray*}
whence it follows that the electro-magnetic field
\begin{displaymath}
F_{\mu \nu} =\p_\mu A^\nu-\p_\nu A^\mu
\end{displaymath}
is localized in the neighborhood of the point $x_1=x_0$ and vanishes
rapidly as $|x_1|\to +\infty$.

Thus, we can interpret (\ref{3.3.6}), (\ref{3.3.7}) as a wave function 
of a particle moving in the electro-magnetic field $F_{\mu \nu}$,
which is localized in the neighborhood of the line $x_1=x_0$.

In conclusion, we note that the method described above has been used
in \cite{206} to construct the general solution of the following
two-dimensional system of nonlinear PDEs:
\begin{eqnarray*}
& &\Bigl ( i\g_{\mu}\p_{\mu}-e\g_{\mu}A^{\mu}
  -\lambda\g_{\mu} (\bar\psi\g^{\mu}\psi)\Bigr )\psi=0, \\
& &\p_{\nu}\p^{\nu}A_{\mu}-\p^{\mu}\p_{\nu}A_{\nu}=
-e\bar\psi\g_{\mu}\psi, \quad \mu,\nu=0,1,
\end{eqnarray*}
which can be represented in the form
\begin{eqnarray*}
A_\mu(x)&=&\widetilde A_\mu(x), \\
\left\{\matrix{\psi^0(x) \cr \psi^2(x)}\right\}&=&
\left\{\matrix{\tilde\psi^0(x) \cr \tilde\psi^2(x)}\right\}
\exp  \Biggl \{-i\lambda\edi\mathop\int\limits^{\edi x_0-x_1}
\Bigl (|U^1(\eta)|^2+|U^3(\eta)|^2\Bigr )d\eta\Biggr \},\\
\left\{\matrix{\psi^1(x) \cr \psi^3(x)}\right\}&=&
\left\{\matrix{\tilde\psi^1(x) \cr \tilde\psi^3(x)}\right\}
\exp  \Biggl \{-i\lambda\edi\mathop\int\limits^{\edi x_0+x_1}
\Bigl (|U^0(\xi)|^2+|U^2(\xi)|^2\Bigr )d\xi\Biggr \},
\end{eqnarray*}
functions $\widetilde A_0(x), \ \widetilde A_1(x), \
\tilde\psi^0(x),\ldots,\tilde\psi^3(x)$ being given by formulae
(\ref{3.3.5}), (\ref{3.3.6}) correspondingly.
\vspace{10mm}

\noindent
{\large\bf 3.4. General solutions of Galilei-invariant spinor
equations\label{s3.4}} 

\markboth{Chapter 3. TWO-DIMENSIONAL SPINOR MODELS}
{3.4. General solutions of Galilei-invariant spinor equations}
\def\theequation{3.\arabic{section}.\arabic{equation}}
\setcounter {section} {4}
\setcounter {equation}{0}
\vspace{7mm}

\noindent
Let us rewrite system (\ref{3.1.3}) in the equivalent form by
introducing new functions $F(x), \ f(x)$
\begin{eqnarray}
\psi_{x_0}&=&-F, \nonumber\\
\psi_{x_1}&=&i\lambda f_{x_1}\g_1\psi-\g_1(\g_0+\g_4)F,
\label{3.4.1}\\ 
f_{x_1}&=&(\psi^\dagger\psi+\bar\psi\g_4\psi)^{1/2k}.\nonumber
\end{eqnarray}

Consider now the second equation of system (\ref{3.4.1}) as a system
of ODEs with respect to $x_1$. Since this system is linear, it is
possible to apply the standard method of variation of an arbitrary
constant \cite{132}. The general solution of the homogeneous part of
the system in question is given by the formula
\begin{eqnarray*}
\psi(x)=\exp  \{i\lambda \g_1f(x)\}\varphi(x_0),
\end{eqnarray*}
where $\varphi(x_0)$ is an arbitrary four-component
function. Consequently, the general solution of the second equation 
from system (\ref{3.4.1}) has the form
\begin{equation}
\begin{array}{rcl}
\psi(x)&=&\exp  \{i\lambda \g_1f(x)\}\Biggl (\varphi(x_0)+\g_1
\edi\mathop\int\limits^{\edi x_1}\exp  \{-i\lambda\g_1f(x_0,z)\}
\\[2mm] 
& &\times(\g_0+\g_4)F(x_0,z)dz\Biggr )
\end{array}
\label{3.4.2}
\end{equation}

Multiplying both parts of the first equation from (\ref{3.4.1}) by
the matrix $\g_0+\g_4$ we have
\begin{displaymath}
(\g_0+\g_4)F=-(\g_0+\g_4)\psi_{x_0},
\end{displaymath}
whence
\begin{displaymath}
(\g_0+\g_4)F=-\exp  \{-i\lambda\g_1f\}(\g_0+\g_4)
(\dot\varphi+i\lambda\g_1f_{x_0}\varphi).
\end{displaymath}

Consequently, formula (\ref{3.4.2}) takes the form
\begin{equation}
\begin{array}{rcl}
\psi(x)&=&\exp  \{i\lambda \g_1f(x)\}\Biggl (\varphi(x_0)+\g_1
(\g_0+\g_4)\edi\mathop\int\limits^{\edi x_1}\exp  \{2i\lambda\g_1
\\[2mm] 
& &\times f(x_0,z)\}\Bigl (\dot\varphi(x_0)+i\lambda\g_1f_{x_0}(x_0,z)
\varphi(x_0)\Bigr )dz\Biggr ).
\end{array}
\label{3.4.3}
\end{equation}

Substitution of the above expression into the third equation of system
(\ref{3.4.1}) yields the nonlinear ODE for a function $f(x)$
\begin{equation}
f_{x_1}=(A_1\cosh 2\lambda f+A_2\sinh 2\lambda f)^{1/2k},
\label{3.4.4}
\end{equation}
where
\begin{displaymath}
A_1=\bar\varphi(\g_0+\g_4)\varphi, \quad 
A_2=i\bar\varphi(\g_0+\g_4)\g_1\varphi.
\end{displaymath}

The general solution of (\ref{3.4.4}) is given by the quadrature
\begin{equation}
\edi\mathop\int\limits^{\edi f(x_0,x_1)}(A_1\cosh 2\lambda z
+A_2\sinh 2\lambda z)^{-1/2k}dz=x_1+C(x_0),
\label{3.4.5}
\end{equation}
$C(x_0)$ being an arbitrary smooth real-valued function.

Thus, the general solution of nonlinear system of PDEs (\ref{3.1.3})
has the form (\ref{3.4.3}), function $f=f(x_0,x_1)$ being determined
by implicit formula (\ref{3.4.5}).

Using formulae (\ref{3.4.3}), (\ref{3.4.5}) it is not difficult to 
obtain the general solution of the linear equation
\begin{eqnarray*}
\Bigl (i(\g_0+\g_4)\p_0+i\g_1\p_1-\lambda\Bigr )\psi(x)=0,
\end{eqnarray*}
which is of the form
\vspace{1.5mm}

\noindent 
1) under $\lambda\ne 0 $
\begin{displaymath}
\psi(x)=\exp  \{i\lambda \g_1x_1\}\Bigl (\varphi(x_0)+(i/2\lambda)
(\g_0+\g_4)\exp  \{2i\lambda \g_1x_1\}\dot\varphi(x_0)\Bigr ),
\end{displaymath}
2) under $\lambda=0$
\begin{displaymath}
\psi(x)=\varphi(x_0)+x_1\g_1(\g_0+\g_4)\dot\varphi(x_0).
\end{displaymath}

Reduction of the nonlinear Dirac equation (\ref{2.4.1}) by means of
the Ansatz 
\begin{equation}
\begin{array}{l}
\psi(x)=\varphi(\xi,\omega), \\[2mm]
\xi=x_0+x_3, \ \omega =x_2
\end{array}
\label{3.4.6}
\end{equation}
invariant under the two-parameter group with generators
$\p_0-\p_3, \ \p_1$ yields the two-dimensional system of PDEs
\begin{equation}
\Bigl (i(\g_0+\g_3)\p_\xi+i\g_2\p_2
-\lambda(\bar\psi\psi)^r\Bigr )\psi=0, \ r=1/2k,
\label{3.4.7}
\end{equation}
which can also be integrated by means of the above described trick 
\cite{100,211.1}.

Rewriting system of PDEs (\ref{3.4.7}) in the equivalent form
(\ref{3.4.1}) we have
\begin{eqnarray}
\varphi_{\xi}&=&F, \label{3.4.8}\\
\varphi_{\omega}&=&if_{\omega}\g_2\varphi+\g_2(\g_0
+\g_3)F, \label{3.4.9}\\
f_{\omega}&=&\lambda (\bar\psi\psi)^r.\label{3.4.10}
\end{eqnarray}

Integration of system (\ref{3.4.9}) by the method of variation of an
arbitrary constant with respect to $\omega$ yields the following
expression for $\varphi$:

\begin{eqnarray*}
\varphi(\xi,\omega)&=&\exp  \{i\g_2 f\}\Biggl (\Theta(\xi)+\g_2
(\g_0+\g_3)\edi\mathop\int\limits^{\edi \omega}_0\exp  \{i\g_2
f(\xi,z)\} \\ 
& &\times F(\xi,z)dz\Biggr),
\end{eqnarray*}
where $\Theta(\xi)$ is an arbitrary four-component function-column.

As due to (\ref{3.4.8}) the equation
\begin{displaymath}
(\g_0+\g_3)\varphi_\xi =(\g_0+\g_3)F
\end{displaymath}
holds, we can exclude from the above equality the function $F$
\begin{equation}
\begin{array}{rcl}
\varphi(\xi,\omega)&=&\exp \{i\g_2 f\}\Biggl (\Theta+\g_2
(\g_0+\g_3)\edi\mathop\int\limits^{\edi \omega}_0\exp \{2i\g_2
f(\xi,z)\} \\[2mm] 
& &\times\Bigl (\dot\Theta +if_\xi(\xi,z)\g_2\Theta\Bigr )dz\Biggr ).
\label{3.4.11}
\end{array}
\end{equation}

The only thing left is to substitute (\ref{3.4.11}) into
(\ref{3.4.10}). As a result, we get an integro-differential equation
for $f=f(\xi,\omega)$
\begin{equation}
f_\omega=\lambda\Biggl (A+B\edi\mathop\int\limits^{\edi \omega}_0\cosh
2fdz  
+C\edi\mathop\int\limits^{\edi \omega}_0\sinh   2fdz\Biggr )^r,
\label{3.4.12}
\end{equation}
where
\begin{eqnarray*}
& &A=\bar\Theta\Theta,\\
& &B=\bar\Theta\g_2(\g_0+\g_3)\dot\Theta -
\dot{\bar\Theta}\g_2(\g_0+\g_3)\Theta, \\
& &C=i\Bigl (\bar\Theta(\g_0+\g_3)\dot\Theta 
-\dot{\bar\Theta}(\g_0+\g_3)\Theta \Bigr ).
\end{eqnarray*}

The general solution of equation (\ref{3.4.12}) has been constructed
in \cite{211.1}. Since its explicit form depends on relations between  
$B$ and $C$, we have to consider four inequivalent cases.
\vspace{1.5mm}

\noindent
{\bf Case 1.} \ $B=\pm C, \ B\ne 0$
\vspace{1.5mm}

\noindent
a) \ $r\ne -1$
\begin{eqnarray*}
& &f=\pm (1/2)\ln  \Bigl (\varepsilon \pm 2\lambda
B^{-1}(r+1)^{-1}(A+Bg)^{r+1}\Bigr ), \\
& &\edi\mathop\int\limits^{\edi g(\xi,\omega)}_0\biggl [\varepsilon
\pm 2\lambda
B^{-1}(r+1)^{-1}(A+B\tau)^{r+1}\biggr ]^{-1}d\tau=\omega;
\end{eqnarray*}
b) \ $r=1$
\begin{eqnarray*}
& &f=\pm (1/2)\ln  \Bigl (\varepsilon \pm 2\lambda
B^{-1}\ln  (A+Bg)\Bigr ), \\
& &\edi\mathop\int\limits^{\edi g(\xi,\omega)}_0\biggl [\varepsilon
\pm 2\lambda
B^{-1}\ln  (A+B\tau)\biggr ]^{-1}d\tau=\omega.
\end{eqnarray*}
{\bf Case 2.} $B^2>C^2 \Leftrightarrow B=\alpha(\xi)\cosh
2\beta(\xi), 
\ B=\alpha(\xi)\sinh  2\beta(\xi)$
\vspace{1.5mm}

\noindent
a) \ $r\ne -1$
\begin{eqnarray*}
& &\cosh  2(f+\beta)=\Bigl [1+ \Bigl (\varepsilon +2\lambda
\alpha^{-1}(r+1)^{-1}(A+\alpha g)^{r+1}\Bigr )^2\Bigr ]^{1/2}, \\
& &\edi\mathop\int\limits^{\edi g(\xi,\omega)}_0\biggl [1 + \Bigl(
\varepsilon +2\lambda \alpha^{-1}(r+1)^{-1}(A+\alpha\tau)^{r+1}\Bigr
)^2 \biggr ]^{-1/2}d\tau=\omega;
\end{eqnarray*}
b) \ $r=-1$
\begin{eqnarray*}
& &\cosh  2(f+\beta)=\Bigl [1+ \Bigl (\varepsilon +2\lambda
\alpha^{-1}\ln  (A+\alpha g)\Bigr )^2\Bigr ]^{1/2}, \\
& &\edi\mathop\int\limits^{\edi g(\xi,\omega)}_0\biggl [1 + \Bigl(
\varepsilon +2\lambda \alpha^{-1}\ln (A+\alpha\tau)\Bigr )^2
\biggr ]^{-1/2}d\tau=\omega.
\end{eqnarray*}

{\bf Case 3.} $B^2<C^2 \Leftrightarrow B=\alpha(\xi)\sinh
2\beta(\xi), 
\ B=\alpha(\xi)\cosh   2\beta(\xi)$

a) \ $r\ne -1$
\begin{eqnarray*}
& &\sinh  2(f+\beta)=\Bigl [-1+ \Bigl (\varepsilon +2\lambda
\alpha^{-1}(r+1)^{-1}(A+\alpha g)^{r+1}\Bigr )^2\Bigr ]^{1/2}, \\
& &\edi\mathop\int\limits^{\edi g(\xi,\omega)}_0\biggl [-1 + \Bigl(
\varepsilon +2\lambda \alpha^{-1}(r+1)^{-1}(A+\alpha\tau)^{r+1}\Bigr
)^2 \biggr ]^{-1/2}d\tau=\omega.
\end{eqnarray*}

b) \ $r=-1$
\begin{eqnarray*}
& &\sinh 2(f+\beta)=\Bigl [-1+ \Bigl (\varepsilon +2\lambda
\alpha^{-1}\ln  (A+\alpha g)\Bigr )^2\Bigr ]^{1/2}, \\
& &\edi\mathop\int\limits^{\edi g(\xi,\omega)}_0\biggl [-1 + \Bigl(
\varepsilon +2\lambda \alpha^{-1}\ln (A+\alpha\tau)\Bigr )^2
\biggr ]^{-1/2}d\tau=\omega.
\end{eqnarray*}
{\bf Case 4.} $B=C=0$
\begin{eqnarray*}
& &f=\lambda A^r\omega.
\end{eqnarray*}

In the above formulae parameter $\varepsilon$ takes the 
values \ $-1, \ 0, \ 1$. 

Thus, we have constructed the general solution of system
(\ref{3.4.7}). Substitution of the obtained expression for the
four-component function $\varphi=\varphi(\xi,\omega)$ into Ansatz
(\ref{3.4.6}) with $r=1/2k$ yields a class of exact solutions of the
nonlinear Dirac equation (\ref{2.4.1}).  And what is more, this class
contains four arbitrary complex functions of $\xi=x_0+x_3$ (components
of the function $\Theta(\xi)$). Such arbitrariness enables us to solve
a wide class of Cauchy problems for the system of nonlinear PDEs
(\ref{2.4.1}).

\newpage
\thispagestyle{empty}

\noindent
{\sl
C H A P T E R \ \  4\label{ch4}}
\vspace{2mm}

\hrule
\vspace{35mm}

\rightline
{\large\bf
NONLINEAR}
\vspace{2mm}

\rightline
{\large\bf
GALILEI-INVARIANT}
\vspace{2mm}

\rightline
{\large\bf
SPINOR EQUATIONS}
\vspace{7mm}

In the present chapter we investigate linear and nonlinear systems of
PDEs for the spinor field admitting the Galilei group $G(1,3)$.
Wide classes of nonlinear first-order spinor PDEs invariant under
the group $G(1,3)$ and its extensions, groups $G_1(1,3)$ and
$G_2(1,3)$, are described. All Ans\"atze for the spinor field $\psi
(t,\, \vec x)$ invariant under the $G(1,3)$ non-conjugate
three-parameter subgroups of the Galilei group are obtained. With the
use of these Ans\"atze the multi-parameter families of exact solutions
of a nonlinear Galilei-invariant spinor equation are constructed.
In addition, we briefly consider the second-order spinor PDEs invariant
under the group $G(1,3)$.  
\vspace{10mm}

\noindent
{\large\bf 4.1. Nonlinear equations for the spinor field invariant
\vspace{1.5mm}

\noindent
\phantom{\large\bf 4.1. }under the group {\boldmath $G$}(1,3) and its
extensions\label{s4.1}} 
\markboth{Chapter 4. NONLINEAR GALILEI-INVARIANT SPINOR EQUATIONS }
   {4.1. Nonlinear equations for the spinor field }
\def\theequation{4.\arabic{section}.\arabic{equation}}
\setcounter {section} {1}
\setcounter {equation}{0}
\vspace{7mm}

\noindent
In spite of the fact that the Galilei relativity principle is known
for more than 300 years, the concept of the Galilei
group\index{Galilei!group} has arisen only recently (1950--1970).  It
is even more surprising, if we take into account that Sophus Lie has
discovered this group as early as in 1889. It was Lie who established
that the one-dimensional linear heat-transfer equation (which up to
the constant factor coincided with the Schr\"odinger equation) was
invariant with respect to the translation group, Galilei
transformation\index{Galilei!transformations}, scale and projective
transformations.  Simultaneously, he discovered a projective
representation of the Galilei group $G(1,1)$.

Bargmann and Wigner \cite{17,128} have rediscovered projective
representations of the Galilei group and showed the fundamental role
played by these in the quantum theory. Since Bargmann's and Wigner's
works the Galilei group is intensively used by specialists dealing
with mathematical physics problems.

A Galilei-invariant equation for a particle with the spin $s=1/2$ was
suggested in \cite{109,143}. Systematic study of the first-order
equations invariant under the group $G(1,3)$ was begun by
L\'evy-Leblond \cite{143,144} and Hagen, Hurley \cite{114,123}. The
algebraic-theoretical derivation and detailed investigation of the new
classes of linear Galilei-invariant equations for particles with
arbitrary spins were carried out in \cite{73}--\cite{75},
\cite{77,77.0,83}. Some nonlinear Galilei-invariant systems of PDEs
were considered in \cite{83,181,209}.

A Galilei-invariant equation for a particle with the spin $s=1/2$ can
be represented in the form \cite{209}
\begin{equation}
\{-i (\gamma_0 +\gamma_4 )\p_t  +i\gamma_a \p_a
+ m(\gamma_0 -\gamma_4) \} \psi (t,\, \vec x )=0,
\label{4.1.1}
\end{equation}
where $\p_t =\p / \p t,\ \p_a =\p / \p_{x_a},\ a={1,2,3}$,\
$m= \mbox{\rm const}$,\ $\psi =\psi (t,\, \vec x )$ is a
four-component complex-valued function (spinor), $\vec x \in \R^3$,\
$t\in \R^1$. 

In the process of derivation of equation (\ref{4.1.1}) the Dirac's
heuristic trick was used. Name\-ly, one looked for a  first-order
system of PDEs with constant matrix coeffic\-ients for a spinor $\psi
(t,\, \vec x )$ whose components satisfied the Schr\"odinger 
equation\index{Schr\"odinger!equation}
\begin{equation}
(4im \p_t -\p_a \p_a ) \psi^{\alpha} (t,\, \vec x) =0,\ \
                             \alpha={0,\ldots,3},
\label{4.1.2}
\end{equation}
whence it immediately followed that up to equivalence the equation
required had the form (\ref{4.1.1}) (to obtain (\ref{4.1.2}) one has
to act with the operator $-i (\gamma_0 +\gamma_4) \p_t +i \gamma_a 
\p_a +m(\gamma_0 -\gamma_4)$) on system (\ref{4.1.1}). Let us
note that the more traditional notation of the equation for a 
Galilean particle with the spin $s=1/2$
\begin{displaymath}
\left\{
i(1+\gamma_0) \p_t +i \gamma_a \p_a +m (1-\gamma_0) \right\}
\psi (t,\, \vec x) =0
\end{displaymath}
is obtained if we multiply (\ref{4.1.1}) by the matrix $\gamma_4$ and
change the dependent variable $\psi \to \psi ' =2^{-1/2}
(1+\gamma_4)\psi $. 
\vspace{2mm}

\noindent
{\bf 1. Local symmetry of system of PDEs (\ref{4.1.1}).}\ The Lie
symmetry of PDE (\ref{4.1.1}) for $m \ne 0$ is well-known. In
particular, in \cite{77.0,83,181} it was established that
(\ref{4.1.1}) is invariant under the 13-parameter generalized Galilei
group\index{Generalized Galilei group} $G_2(1,3)$ (it is also called
the Schr\"odinger group\index{Schr\"odinger!group} and denoted
$Sch(1,3)$). We will prove the assertions describing the maximal (in
Lie sense) invariance group admitted by equation (\ref{4.1.1}) for
both cases $m\ne 0$ and $m=0$.  \vspace{1.5mm}

\noindent
{\bf Theorem 4.1.1.} \ {\em The maximal local invariance group of
  equation (\ref{4.1.1}) with $m \ne 0$ is the $14$-parameter group
  $G^{(1)} =G_2(1,3) \otimes I(1)$,\label{page213}\footnote{Since
    equation (\ref{4.1.1}) is linear, it admits an infinite-parameter
    group\index{Infinite-parameter Lie group}
    $\psi'=\psi+\theta\Psi(t,\, \vec x)$, where $\theta$ is a group
    parameter and $\Psi$ is an arbitrary solution of the system of PDEs
    (\ref{4.1.1}). Such a symmetry gives no essential information
    about the structure of the solutions of the equation under
    consideration and therefore is neglected.} where $G_2(1,3)$ is
  the $13$-parameter generalized Galilei
  group\index{Generalized Galilei group} having
  the generators
  \index{Maximal symmetry!of the Galilei-invariant spinor equation}
\begin{eqnarray}
P_0 &=& \p_t,\quad P_a\ \, =\ \, \p_a,\quad M\ \, =\ \, 2im,\non\\
J_{ab} &=& x_a \p_b-x_b \p_a -(1/2 )\gamma_a \gamma_b,\non\\
G_a &=& t\p_a +2im x_a +(1/2) (\gamma_0 +\gamma_4)
\gamma_a,\label{4.1.3}\\ 
D &=& 2t\p_t +x_a \p_a +2 -(1/2) \gamma_0 \gamma_4,\non\\
A &=& tD -t^2 \p_t +imx_a x_a +(1/2) (\gamma_0 
+\gamma_4) \gamma_a x_a \non
\end{eqnarray}
and $I(1)$ is the following one-parameter group  
\begin{displaymath} 
x_{\mu}' =x_{\mu},\quad \psi' (t^{\prime},\, \vec x^{\prime} )
=e^{\theta } \psi(t,\, \vec x). 
\end{displaymath}

In the above formulae $a, b={1,2,3},\ a\ne b$,\ $\theta=\mbox{\rm
  const}$ is a group parameter.}

The proof is carried out by means of the Lie method. According to 
\cite{164} the operator
\begin{equation}
\begin{array}{rcl}
 Q &=& \xi_0 (x, \psi^*, \psi )\p_t 
+\xi_a (x, \psi^*,\psi )\p_a\\[2mm]
& & + \eta^{\alpha} (x,\psi^*, \psi )\p_{\psi^{\alpha}} +
\eta^{*\alpha} (x, \psi^*, \psi) \p_{\psi^{*\alpha}}
\end{array}
\label{4.1.4}
\end{equation}
generates an invariance group of PDE (\ref{4.1.1}) iff the
following relations hold 
\begin{equation}
\begin{array}{l}
\left.\matrix{\widetilde Q \{-i (\gamma_0 
+\gamma_4) \psi_t +i \gamma_a \psi_{x_a} +
m(\gamma_0 -\gamma_4) \psi \}\cr\cr}\right.
\left|\matrix{&=0,\cr[L]&\cr}\right.\\[4mm]
\left.\matrix{\widetilde Q \{i (\gamma_0^* 
+\gamma_4^*) \psi_t^* -i \gamma_a^* \psi_{x_a}^* +
m(\gamma_0^* -\gamma_4^*) \psi^* \}\cr\cr}\right.
\left|\matrix{&=0,\cr[L]&\cr}\right.
\end{array}
\label{4.1.5}
\end{equation}
where $\widetilde Q$ is the first prolongation of the  operator $Q$. 
By the symbol $[L]$ we designate the set of solutions of equation
(\ref{4.1.1}). 

Relations  (\ref{4.1.5}) yield the following determining  equations for
the coeffici\-ents of the operator $Q$:
\begin{eqnarray}
& &\eta = -(\tilde a + b_{\mu} \gamma^{\mu} + c_{\mu\nu}
\gamma^{\mu} \gamma^{\nu} +d_{\mu} \gamma^{\mu} \gamma_4 + e)\psi
+\Omega \psi^*+\Psi,\non\\
& &\eta^*  = -(\tilde a ^* +b^*_{\mu} \gamma^{\mu *} +c ^*_{\mu \nu}
\gamma^{\mu *} \gamma^{\nu *} + d^*_{\mu } \gamma^{\mu *} \gamma_4 ^* +
e^* )\psi^* +\Omega^* \psi+\Psi^*,\non\\
& &\p_t \xi_0 + 2d_0=\p_1 \xi_1 = \p_2 \xi_2  =\p_3 \xi_3,
\quad \p_a \xi_0 = 0,\non\\
& &\p_t \xi_a =2d_a =-4c_{0a},\quad \p_b \xi_a = -\p_a \xi_b 
=4 c_{ab},\non\\
& &\p_a e =\p_t \p_c \xi_b,\ \ (a, b, c)=\ {\rm cycle}\ (1, 2, 3),
\quad b_a =0,\label{4.1.6}\\
& &m\, e = 0,\quad m(\p_t\xi_0+4d_0) =0,\quad 
\p_t \Bigl(\tilde a -d_0 -(1/2) \p_a \xi_a \Bigr) =0, \non\\
& &\p_b \tilde a - (1/2) \p_a \p_a \xi_b  -
2im \p_t \xi_b =0,\quad \p_a d_0 =0,\non\\
& &\p_t \Omega =\p _a \Omega =0,\quad 
(\g_0+\g_4)\Omega=-\Theta(\g_0-\g_4),\non\\
& &\g_a\Omega=-\Omega\g_a^*,\quad
m(\g_0-\g_4)\Omega=m\Theta(\g_0+\g_4),\non
\end{eqnarray}
where $\eta$ is the four-component function $\{\eta^0,\eta^1,
\eta^2,\eta^3\}^T$,\ $\Psi$ is an arbitrary solution of the system of PDEs 
(\ref{4.1.1}), $\Omega,\ \Theta$ are complex $(4\times 4)$-matrices,
indices $a,\ b,\ c$ take the values 1,\ 2,\ 3 and what is more $a\ne b$.

Since $m\ne 0$, from (\ref{4.1.6}) it follows that $e=0$,\ $f=-2d_0$
and besides $d_0 =d_0 (t)$. Due to this fact the equations for
functions $\xi_0,\ \xi_1,\ \xi_2,\ \xi_3$ are rewritten in the form
\begin{eqnarray*}
& &\p_a \xi_0 =0, \quad \p_t \xi_0 =-4d_0 (t),\quad
\p_b \xi_a =-\p_a \xi_b,\ \ a\ne b,\\
& &\p_1\xi_1 =\p_2 \xi_2=\p_3\xi_3=-2d_0(t),\quad
\p_t \p_a \xi_b =0,\ \ a\ne b,
\end{eqnarray*}
whence it follows that $\p_b \p_a \xi_a =0$ (no summation over $a$)
and what is more the equalities
\begin{displaymath}
\p_a \p_b \xi_c =-\p_a \p_c\xi_b=-\p_c\p_a\xi_b=\p_c \p_b \xi_a=
\p_b \p_c \xi_a =-\p_b \p_a \xi_c,
\end{displaymath}
where $(a,b,c)={\rm cycle} (1, 2, 3)$, hold. Consequently,
$\p_a\p_b\xi_c =-\p_a \p_b \xi_c =0$ which implies that the functions
$\xi_{\mu}$ are linear in the variables $x_1$,\ $x_2,\ x_3$. Due 
to this fact it is not difficult to integrate the
system of PDEs (\ref{4.1.6}). Its general solution under $m\ne 0$ has
the form 
\begin{eqnarray*}
& &\xi_0 = A_1 t^2 +2A_2 t +A_3,\\
& &\xi_a =B_{ab} x_b +(A_1 t +A_2)x_a +C_at +D_a,\\
& &\tilde a = im A_1x_ax_a +2im C_ax_a +2A_1t+2A_2+A_4,\\
& &c_{ab} =(1/4)B_{ab},\quad d_a=(1/2) (A_1x_a +C_a),\\
& &c_{0a}=-(1/4) (A_1x_a +C_a),\quad
d_0=-(1/2) (A_1t+A_2),\\
& &b_0 = b_a = e = 0,\quad
\Omega=0,\quad \Theta=0,
\end{eqnarray*}
where $A_1,\ldots,A_4,\ B_{ab},\ C_a,\ D_a$ are arbitrary real 
constants, $B_{ab} =-B_{ba}$,\ $a, b={1,2,3}$.

Substitution of the above results into (\ref{4.1.4}) shows that the
most general infinitesimal operator $Q$ admitted by the equation
under study is a linear combination of operators (\ref{4.1.3}),
$I=\psi^{\alpha} \p_{\psi^{\alpha}} +\psi^{\alpha *} \p_{\psi^{\alpha
 *}}$ and $Q^\prime=\Psi^\alpha(t,\, \vec x)\p_{\psi^\alpha}
+\Psi^{*\alpha}(t,\, \vec x)\p_{\psi^{*\alpha}}$. Consequently,
operators (\ref{4.1.3}) together with the operators $I,\ Q^\prime$ form
the basis of the maximal invariance algebra admitted by the system of PDEs
(\ref{4.1.1}). The theorem is proved. $\rhd$ 
\vspace{1.5mm}

\noindent
{\bf Note 4.1.1.}\ Operators $P_0,\ P_a,\ M,\  J_{ab},\ G_a$ form a
basis of the Lie algebra of the Galilei group\index{Galilei!group} 
which is called the Galilei algebra $AG(1,3)$.\index{Galilei!algebra}
\vspace{1.5mm}

\noindent
{\bf Note 4.1.2.}\ Operators $P_0,\ P_a,\ M,\  J_{ab},\ G_a,\ D$ form
a basis of the Lie algebra of the extended Galilei 
group\index{Extended!Galilei group} $G_1(1,3)$
which is called the extended Galilei algebra\index{Extended!Galilei algebra}
$A\widetilde G_1(1,3)$.
\vspace{1.5mm}

\noindent
{\bf Theorem 4.1.2.}\ {\em The maximal local invariance group 
admitted  by (\ref{4.1.1}) with $m=0$ is the infinite-parameter Lie  
group having the generators\footnote{See the footnote on the page
  \pageref{page213}.} 
\begin{eqnarray}
A_{\infty} &=&\vp_0 (t)\p_t +\dot\vp_0(t) x_a\p_a +
  (3/2)\dot \vp_0(t)\non\\
  & &+(1/2) \ddot\vp_0 (t)(\gamma_0 +\gamma_4) \gamma_a x_a,\non\\
G_{\infty} &=&\vp_a(t) \p_a +(1/2) (\gamma_0 +\gamma_4)
  \gamma_a \dot \vp_a(t),\non\\
D_{\infty}&=& \vp_4(t) \p_t +(1/2) \dot \vp_4(t) 
  (1-\gamma_0 \gamma_4),\non\\
T_{\infty}&=&(\gamma_0 +\gamma_4)  \vp_5(t),\label{4.1.7}\\
J_{\infty}&=& \ve_{abc} \vp_{5+a} (t) \Bigl(x_c \p_b +
 (1/4) \gamma_b \gamma_c\Bigr)\non\\
 & &+(1/2) (\gamma_0 +\gamma_4)\gamma_a \dot \vp_{5+a} (t) 
 \gamma_b x_b,\non\\
M_1&=&\{ C_1 \psi \}^{\alpha} \p_{\psi^{\alpha}} +
    \{ C_1^* \psi^* \}^{\alpha} \p_{\psi^{*\alpha}},\non\\
M_2 &=&\{C_2 \gamma_2 \gamma_4 \psi^*\}^{\alpha}  \p_{\psi^{\alpha}}
     +\{ C_2^* \gamma_2^* \gamma_4^*\psi\}^{\alpha}
     \p_{\psi^{*\alpha}},\non\\ 
M_3&=&\{C_3(\g_2+\g_3\g_1)\psi^*\}\p_{\psi^\alpha}
+\{C_3^*(\g_2^*+\g_3^*\g_1^*)\psi\}\p_{\psi^{*\alpha}},\non
\end{eqnarray}
where $\vp_0(t),\ \vp_1(t),\ldots,\vp_8(t)$ are arbitrary smooth
functions, $\dot \vp_s = d\vp_s / dt$,\ $s={0,\ldots,8}$, the symbol
$\{ \psi\} ^{\alpha}$ denotes the $\alpha$-${\rm th}$ component of
$\psi$,\ $C_1,\ C_2,\ C_3$ are arbitrary complex constants and}
\begin{displaymath}
\ve_{abc}=\cases{1,\ \ (a, b, c) = {\rm cycle} (1, 2, 3),\cr
-1,\ \ (a, b, c) = {\rm cycle} (2, 1, 3),\cr
0,\ \ \mbox{\it in the remaining cases.} \cr } 
\end{displaymath}

\noindent
{\em Proof.}$\quad$ The determining equations for coefficients of the 
infinitesimal ope\-ra\-tor of the invariance group of equation
(\ref{4.1.1})  are of the form (\ref{4.1.6}) under $m=0$. The general
solution of system of PDEs (\ref{4.1.6}) with $m=0$ is given by the
following formulae: 
\begin{eqnarray*}
& &\xi_0 =\vp_0(t) -2\vp_4 (t),\quad \xi_a =\ve_{abc} x_b\vp_{5+c}(t)
+ \dot \vp (t) x_a +\vp_a (t),\\
& &\tilde a =(3/2) \dot \vp_0(t) +C_1,\quad
b_0 =e =\vp_5 (t)-(1/2) x_a \vp_{5+a}(t),\\
& &b_a=0,\quad c_{ab} =(1/4) \ve_{abc} \vp_{5+c}(t),\\
& &d_a =-2c_{0a} =-(1/2) \Bigl(\ddot \vp _0(t) x_a +\dot \vp_a(t)
      + \ve_{abc} \dot \vp_{5+c} (t) x_b \Bigr),\\
& &\Omega =C_2 \gamma_2 \gamma_4 + C_3(\g_2+\g_3\g_1),
\end{eqnarray*}
where $\{C_1, C_2, C_3\} \subset \C^1; \ \vp_0(t),\ldots,\vp_8(t)$ are
arbitrary smooth functions. Substituting the above result into
(\ref{4.1.4}) we come to the conclusion that the most general
infinitesimal operator admitted by equation (\ref{4.1.1}) under
$m=0$ is a linear combination of operators (\ref{4.1.7}) and
$Q^\prime=\Psi^\alpha(t,\vec x)\p_{\psi^\alpha}+\Psi^{*\alpha}(t,\,
\vec x)\p_{\psi^{*\alpha}}$.  Consequently, the operators listed in
(\ref{4.1.7}) together with the operator $Q^\prime$ form the basis of
the maximal (in Lie sense) invariance algebra of (\ref{4.1.1}). The
theorem is proved.  $\rhd$ 
\vspace{1.5mm}

\noindent
{\bf Note 4.1.3.}\ The algebra (\ref{4.1.7}) contains as a subalgebra
the infinite-dimen\-si\-on\-al centerless Virasoro 
algebra\index{Infinite-dimensional Lie algebra}\index{Virasoro algebra} 
with the following basis operators: 
\begin{eqnarray*}
q_n &\equiv&A_{\infty} (t^n) =t^n\p_t +
           nt^{n-1} x_a\p_a +(3n/2) t^{n-1}\\ 
           & & +(1/2) n(n-1) t^{n-2} (\gamma_0 +\gamma_4) \gamma_a x_a,
\end{eqnarray*}
which satisfy the commutation relations
\begin{displaymath} 
[q_n,\, q_m]=(m-n) q_{n+m-1}, \ \ n, m \in \Z.  
\end{displaymath}

The Virasoro algebra is a Kac-Moody-type algebra which plays an
important role in the theory of two-dimensional dynamical systems
(see, for example, \cite{46,194,202}).
\vspace{1.5mm}

\noindent
{\bf Note 4.1.4.} \ On the set of solutions of system of PDEs
(\ref{4.1.1}) with $m=0$ two inequivalent representations of the Lie
algebra of the generalized Galilei group are realized

\begin{eqnarray*}
&1)&
P_0 =\p_t,\quad P_a=\p_a,\\
& &G_a =t\p_a +(1/2) (\gamma_0 +\gamma_4 )\gamma_a,\\
& &J_{ab} = x_a \p_b -x_b \p_a  
-(1/2) \gamma_a \gamma_b, \ \ a\ne b,\\     
& & D= 2t\p_t +x_a \p_a +2 -(1/2) \gamma_0 \gamma_4,\\
& &A= tD -t^2 \p_t +(1/2) (\gamma_0 +\gamma_4) \gamma_a x_a;\\[2mm] 
&2)&
P_0 =\p_t,\quad P_a =\p_a,\\
& &G_a =t \p_a  + (1/2) (\gamma_0 +\gamma_4) \gamma_a,\\
& &J_{ab} =x_a \p_b -x_b\p_a -(1/2) \gamma_a \gamma_b,\ \ a\ne b,\\
& &D= t\p_t +x_a \p_a +3/2,\\
& &A= 2tD -t^2 \p_t +(\gamma_0 +\gamma_4 )\gamma_a x_a.
\end{eqnarray*}

Further, we adduce transformation groups generated by the
operators (\ref{4.1.3}). To obtain a one-parameter transformation
group generated by operator $Q$ (\ref{4.1.4}) it is necessary to solve
the following Cauchy problem\index{Cauchy problem} (the Lie equations): 
\begin{eqnarray} 
& &
{dt'\over d\tau}=\xi_0 (t', \vec x', \psi^{\prime *}, \psi' ),
\quad {dx'_a\over d\tau}=\xi_a (t', \vec x', 
\psi^{\prime *}, \psi' ),\non\\ 
& &{d\psi^{\prime\alpha}\over d\tau} = \eta^{\alpha} (t', \vec x ',
\psi^{\prime *},\psi'),\quad 
{d \psi^{\prime *\alpha}\over d\tau}=\eta^{*\alpha}(t', \vec x ', 
\psi^{\prime *},\psi' ),\label{4.1.8}\\ 
& &t'(0) =t,\quad x_a'(0) =x_a,\quad 
\psi^{\prime\alpha}(0)=\psi^{\alpha},\quad 
\psi^{\prime *\alpha}(0) =\psi^{*\alpha}.\non
\end{eqnarray} 
Substituting into (\ref{4.1.8}) functions $\xi_{\mu},\ \eta^{\alpha}, \ 
\eta^{*\alpha}$ corresponding to the operators (\ref{4.1.3}) and
integrating the equations obtained we arrive at the following 
transformation groups:
\begin{eqnarray}
P&:&\cases{t'=t+\theta_0,\cr
 x'_a=x_a +\theta_a,\cr
   \psi'(t^{\prime},\, \vec x^{\prime})=\psi(t,\, \vec x);\cr}
\label{4.1.9}\\
J&:&\cases{t'=t,\cr
 x_a'=\Bigl(\delta_{ab}\cos \theta +\ve_{abc} \theta_c 
 \theta^{-1} \sin \theta \cr
 \quad +\theta_a \theta_b \theta^{-2} (1-\cos \theta)\Bigr) x_b,\cr
 \psi'(t^{\prime},\, \vec x^{\prime})= \exp \{ -(1/4) 
 \ve_{abc} \theta_a \gamma_b\gamma_c \}
 \psi(t,\, \vec x);\cr}
\label{4.1.10}\\
G&:&\cases{t'=t,\cr
 x_a'= x_a +\theta_a t,\cr
 \psi'(t^{\prime},\, \vec x^{\prime})= \exp \Bigl\{ -2im
 \Bigl(\theta_a x_a  
 +(t/2)\theta_a \theta_a \Bigr) \cr
 \quad -(1/2)(\gamma_0 +\gamma_4) \gamma_a \theta_a \Bigr\} 
 \psi(t,\, \vec x);\cr}
\label{4.1.11}\\
D&:& \cases{t'=t e^{2\theta _0},\cr
 x'_a=x_a e^{\theta_0},\cr
 \psi' (t^{\prime},\, \vec x^{\prime}) = \exp \{ -2\theta_0 
 +(1/2) \theta_0 \gamma_0 
 \gamma_4 \} \psi(t,\, \vec x);\cr}
\label{4.1.12}\\
A&:& \cases{t'= t(1-\theta_0 t)^{-1},\cr
 x'_a =x_a (1-\theta_0 t)^{-1},\cr
 \psi' (t^{\prime},\, \vec x^{\prime}) =(1-\theta_0 t)^2 \exp \Bigl\{
 -im\theta_0 (1-\theta_0 t)^{-1} x_a x_a \cr
 \quad- (1/2t)\ln (1-\theta_0 t) \Bigl(t\gamma_0\gamma_4 
 +(\gamma_0 +\gamma_4) \gamma_a x_a \Bigr)\Bigr\}\psi(t,\, \vec
 x);\cr} 
\label{4.1.13}\\
M&:& \cases{t'=t,\cr
x_a'= x_a,\cr
\psi' (t^{\prime},\, \vec x^{\prime}) =e^{-2 im \theta_0}
\psi(t,\, \vec x);\cr}\label{4.1.14} 
\end{eqnarray}
where $P=\theta_0 P_0+\theta_a P_a,\ J=(1/2)\ve_{abc}\theta_a J_{bc},\
G=\theta_aG_a$,\ $\theta_0,\ \theta_a$ are group parameters,
$\theta=(\theta_a \theta_a)^{1/2}$.

One can check by a direct computation that equation (\ref{4.1.1}) 
is invariant under groups (\ref{4.1.9})--(\ref{4.1.14}).
\vspace{1.5mm}

\noindent
{\bf Note 4.1.5.}\ Transformation groups corresponding to the
operators (\ref{4.1.7}) are given in \cite{106.2}.
\vspace{2mm}

\noindent
{\bf 2. Non-Lie symmetry of system of PDEs {\bf(\ref{4.1.1}).}}
As earlier (see Section 1.1) we designate 
by ${\cal M}_1$ the class of the first-order differential 
operators with complex matrix coefficients
\begin{displaymath}
X_0 =A_0 (t,\, \vec x)\p_t +A_b (t,\vec x)\p_b +B(t,\vec x)
\end{displaymath}
acting on the space of four-component complex-valued functions
$\psi=\psi(t,\, \vec x)$. Below we adduce the assertions describing
the symmetry of equation (\ref{4.1.1}) in the class ${\cal M}_1$.
\vspace{1.5mm}

\noindent
{\bf Theorem 4.1.3.}\ {\em System of PDEs (\ref{4.1.1}) with $m\ne 0$ 
has 34 linearly-inde\-pen\-dent symmetry operators belonging to the 
class ${\cal M}_1$. The list of these operators is exhausted 
by the generators of the generalized Galilei group (\ref{4.1.3}) and by
the following 21 operators}\/:
\index{Nonlocal (non-Lie) symmetry!of the Galilei-invariant spinor equation}
\begin{eqnarray*}
M_1&=&I,\quad M_2\ \, =\ \, iI, \\
W_0 &=&(1/2) (\gamma_0 +\gamma_4)\p_t -(im/2)(\gamma_0 -\gamma_4),\\
W_a &=&(1/2) \ve_{abc} \Bigl((1/2)(\gamma_0 +\gamma_4 )
   (\gamma_b \p_c -\gamma_c \p_b) + im \gamma_b\gamma_c \Bigr),\\
S_a &=&\gamma_0\gamma_4 \p_a +(\gamma_0 +\gamma_4 )\gamma_a \p_t -
   im(\gamma_0 -\gamma_4) \gamma_a,\\
T_a &=&(1/2) \ve_{abc} \Bigl( (1/2) (\gamma_0 -\gamma_4)
   (\gamma_b \p_c -\gamma_c \p_b ) +\gamma_b \gamma_c \p_t \Bigr),\\
R_0 &=&t W_0 +x_a W_a +(3/4) (\gamma_0 +\gamma_4),\\
R_a &=&2t T_a +2x_a W_0 +\ve_{abc}\Bigl(x_b S_c 
   +(1/2) \gamma_b \gamma_c\Bigr) +(3/2)\gamma_a,\\
N_0 &=& x_a S_a +\g_0\g_4,\\
N_a &=&tS_a +2 \ve_{abc} x_b W_c +(\gamma_0 +\gamma_4) \gamma_a,\\
K_a &=&2x_a R_0 -(x_b x_b ) W_a +\ve_{abc}
   \Bigl(tx_b S_c +(1/2) t\gamma_b \gamma_c
   +(\gamma_0 +\gamma_4) x_b \gamma_c\Bigr)\\
   & &+t^2 T_a +(3/2) t \gamma_a, 
\end{eqnarray*}

\noindent
where $I$ is the unit $(4\times 4)$-matrix.
\vspace{1.5mm}

\noindent
{\bf Theorem 4.1.4.}\ {\em Basis of the infinite-dimensional 
vector space of symmetry operators of system (\ref{4.1.1}) with $m=0$ 
belonging to the class ${\cal M}_1$ can be chosen as follows }
\begin{eqnarray*}
I_1&=&I,\quad I_2\ \, =\ \, iI,\quad
  A_{\infty},\quad G_{\infty},\quad D_{\infty},
  \quad T_{\infty},\quad J_{\infty},\\
W_{\infty}&=&(\gamma_0 +\gamma_4)(\vp_0^6 \p_t +\vp_a^6 \p_a ),\\
S_{\infty}&=&\vp_a^7 \Bigl((\gamma_0 +\gamma_4)\gamma_a \p_t +
  \gamma_0 \gamma_4\p_a \Bigr) +(1/2)(\gamma_0+\gamma_4)
  \gamma_a \dot \vp_a^7,\\
P_{\infty}&=&\vp_a^8 \Bigl( 2\gamma_a \p_t -(\gamma_0 -\gamma_4)\p_a
\Bigr) 
  +(1/4) (2 \gamma_a +\ve_{abc} \gamma_b \gamma_c )\dot \vp _a^8,\\
Q_{\infty}&=&\vp^9_0(\g_0+\g_4)x_a\p_a,\\
R_{\infty}&=&\vp_a^{10}\Bigl\{ \ve_{abc}\Bigl((\gamma_0
+\gamma_4)x_b\gamma_c 
  + \gamma_0 \gamma_4 x_b \p_c\Bigr) +(1/2) \gamma_a \Bigr\} \\
  & &-(1/2)(\gamma_0+\gamma_4) \gamma_a \dot \vp_a ^{10} \gamma_b x_b,
  \\   
N_{\infty}&=& \vp_0^{11}\Bigl((\gamma_0+\gamma_4) \gamma_a x_a \p_t +
  \gamma_0  \gamma_4 x_a \p_a \\
  & &+(1/2) (1+\gamma_0\gamma_4)\Bigr)+(1/2)\dot\vp_0^{11}
  (\gamma_0 +\gamma_4) \gamma_a x_a,\\
K_{\infty}&=&\vp_a^{12}\Bigl(-(\gamma_0 +\gamma_4)(x_b x_b)\p_a +
  2x_a (\gamma_0 +\gamma_4) x_b \p_b \\
  & &+2x_a (\gamma_0+\gamma_4) +\ve_{abc} x_b \gamma_c\Bigr),\\
L_{\infty}&=&\ve_{abc}\vp_a^{13}(\gamma_0+\gamma_4)\Bigl(x_b \p_c  
  +(1/4)\gamma_b\gamma_c\Bigr).
\end{eqnarray*}

Here $A_{\infty},\ldots,J_{\infty}$ are operators listed in 
(\ref{4.1.7}), $\vp_{\mu}^N,\ \mu={0,\ldots,3},\ N={6,\ldots,13}$ are
arbitrary smooth functions of $t$, an overdot means differentiation 
with respect to $t$.
\vspace{1.5mm}

\noindent
{\em Proof.}$\quad$ We give the main idea of the  proof omitting 
very cumbersome intermediate calculations. According to the
definition of a symmetry operator, to describe all
linearly independent symmetry operators of equation (\ref{4.1.1})
belonging to the class ${\cal M}_1$ it is necessary to construct 
a general solution of the operator equation 
\begin{displaymath}
[L,\, X]=(R_0 \p_t +R_a \p_a +R) L,
\end{displaymath}
where $L=-i (\gamma_0 +\gamma_4) \p_t +i \gamma_a \p_a +m (\gamma_0
-\gamma_4)$;\ $R_0,\ R_a,\ R$ are variable $(4\times 4)$-matrices. 

Computing the commutator and equating coefficients of
linearly-inde\-pen\-dent operators $\p^2_t$,\ $\p_t\p_a$,\ $\p_a
\p_b$,\ $\p_t,\ \p_a$ yield a system of matrix PDEs for $A_0,\ A_b$,\ 
$B,\ R_0$,\ $R_a,\ R$. Eliminating matrix functions $R_0,\ R_a,\ R$ we
arrive at the over-determined system of PDEs for 80 functions $A^{\mu
  \nu}_0$,\ $A_b^{\mu\nu}$,\ $B^{\mu\nu}$ (by $A^{\mu\nu}$ we
designate the entries of the matrix $A$) $\mu, \nu ={0,\ldots,3}$
which general solution gives rise to a complete set of symmetry
operators of equation (\ref{4.1.1}). $\rhd$

The complete set of symmetry operators of equation (\ref{4.1.1})
belonging to the class ${\cal M}_1$ does not form a Lie algebra.  But
it contains some subsets which have very interesting algebraic
properties. In particular, the basis generators of the Galilei group
$P_0,\ P_a,\ J_{ab},\ G_a,\ M$ are even basis elements and the
operators $W_0,\ W_a$ are odd basis elements of a superalgebra. This
superalgebra can be considered as a superextension of the Galilei
algebra $AG(1,3)$ \cite{211.2}.

A detailed account of symmetry properties of system of linear PDEs
(\ref{4.1.1}) in the class of differential operators of the order
higher than 1 and in the class of integro-differential operators can  
be found in \cite{77.0}.
\vspace{2mm}

\noindent
{\bf 3. Nonlinear spinor equations invariant under the group
{\boldmath $G$}(1,3)}.
In this subsection we will obtain a complete description of
Galilei-invariant systems of PDEs 
\begin{equation}
\{ -i (\gamma_0 +\gamma_4) \p_t +i\gamma_a\p_a +m( \gamma_0 -\gamma_4) 
\} \psi = F(\psi^*,\, \psi),
\label{4.1.21}
\end{equation}
where $F(\psi^*,\, \psi )$ is a complex-valued four-component
function. In addition, all the functions $F(\psi^*,\psi )$ such that
equation (\ref{4.1.21}) admits wider symmetry groups (in particular,
the generalized Galilei group $G_2(1,3)$) will be constructed.
\vspace{1.5mm}

\noindent
{\bf Theorem 4.1.5.}\ {\em The system of nonlinear PDEs (\ref{4.1.21}) is
  invariant under the Galilei group iff
\begin{equation}
F(\psi^*,\, \psi )=\Bigl(f_1 +(\gamma_0 +\gamma_4) f_2 \Bigr)\psi,
\label{4.1.22}
\end{equation}
where $f_1,\ f_2$ are arbitrary smooth functions of}\
$w_1 =\bar{\psi} \psi $,\
$w_2=\psi^\dagger \psi +\bar \psi \gamma_4 \psi $.
\vspace{1.5mm}

\noindent
{\em Proof.}$\quad$ It is convenient to represent a four-component
function $F(\psi^*,\, \psi)$ in the form $F=H(\psi^*,\, \psi)\psi $,
where $H$ is a variable $(4 \times 4)$-matrix.

At first, we select from the class of equations (\ref{4.1.21})
those which are invariant under the rotation group $O(3) \subset
G(1,3)$. Acting by the first prolongation of the generator of the
group $O(3)$ 
\begin{displaymath}  
Q=\alpha_{ab} x_a \p_b -(1/4) \{ \alpha_{ab} \gamma_a 
 \gamma_b \psi \}^{\alpha}\p_{\psi^{\alpha}} -(1/4) 
 \{\alpha_{ab} \gamma_a^* \gamma^*_b \psi^*\}^{\alpha}
 \p_{\psi^{* \alpha }},
\end{displaymath}
where $\alpha_{ab} =-\alpha_{ba}$ are real constants,
on (\ref{4.1.21}) we arrive at the following relation for
$H=H(\psi^*,\, \psi)$: 
\begin{equation}
\left.\matrix{\widetilde Q \{-i (\gamma_0 
+\gamma_4) \psi_t +i \gamma_a \psi_{x_a} +
m(\gamma_0 -\gamma_4) \psi - H\psi\}\cr\cr}\right.
\left|\matrix{&=0.\cr[L]&\cr}\right.
\label{4.1.23}
\end{equation}

Here $[L]$ is the set of solutions of PDE (\ref{4.1.21}). 

Designating
\begin{displaymath}
Q_{ab} =-(1/2) \{ \gamma_a \gamma_b \psi \}^{\alpha}
\p_{\psi^{\alpha}} 
-(1/2)\{ \gamma^*_a \gamma^*_b \psi^* \}^{\alpha} \p_{\psi^{\alpha
    *}}, 
\end{displaymath}
we rewrite equation (\ref{4.1.23}) as follows
\begin{equation}
Q_{ab} H+(1/2) [\gamma_a \gamma_b,\, H]=0.
\label{4.1.24}
\end{equation}

Expanding the matrix $H$ in the complete system of the Dirac matrices
\begin{equation}
\begin{array}{rcl}
H&=&\tilde a(\psi^*,\,  \psi )I +b_{\mu} (\psi^*,\, \psi) \gamma^{\mu}
+ c_{\mu \nu} (\psi^*,\,  \psi ) \gamma^{\mu} \gamma^{\nu} \\[2mm]
& &+d_{\mu} (\psi^*,\, \psi)
\gamma_4 \gamma^{\mu} +e (\psi^*,\,  \psi ) \gamma_4
\end{array}
\label{4.1.25}
\end{equation}
and substituting it into (\ref{4.1.24}) we get
\begin{eqnarray*}
& &Q_{ab} ( \tilde a I +b_{\mu} \gamma^{\mu} +c_{\mu\nu} \gamma^{\mu}
\gamma^{\nu}+ 
d_{\mu} \gamma_4 \gamma^{\mu} +e \gamma_4 )\\
& &\quad=b^{\mu} (g_{\mu a } \gamma_b -g_{\mu b} \gamma_a) 
  + d^{\mu} \gamma_4(g_{\mu a } \gamma_b - g_{\mu b} \gamma_a) \\
& &\quad -c^{\mu\nu} (g_{a\nu } \gamma_b \gamma_{\nu} 
  +g_{b\mu }\gamma_a \gamma_{\nu} -
  g_{a\mu} \gamma_b \gamma_{\nu} -g_{b \nu} \gamma_a \gamma_{\mu} ).
\end{eqnarray*}

Equating the coefficients of linearly independent matrices we
arrive at the following system of PDEs:
\begin{eqnarray}
& &Q_{ab}\tilde a = Q_{ab} e =Q_{ab} b_0 =Q_{ab} d_0 =0,\non\\
& &Q_{ab}b_k = b_c (g_{ca } \delta_{kb} -g_{cb} \delta_{ka} ),\non\\
& &Q_{ab}d_k = d_c (g_{ca } \delta_{kb} 
  -g_{cb} \delta_{ka} ),\label{4.1.26}\\
& &Q_{ab}c_{0k} = c_{0c} (g_{ca } \delta_{kb} 
  -g_{cb} \delta_{ka} ),\non\\
& &Q_{ab}c_{kc} = c_{mn} (g_{an } \delta_{bm}^{kc} 
  +g_{bm} \delta_{an}^{kc} - g_{am} \delta_{bn}^{kc} 
  -g_{bn} \delta_{am}^{kc}).\non
\end{eqnarray}

In (\ref{4.1.26}) $\delta_{mn}^{kc} =\delta_{m k} \delta_{n c} 
-\delta_{n k} \delta_{m c}$,\ $a, b, c$, $k, m, n={1,2,3}$.

Integration of system (\ref{4.1.26}) is carried out in the same way as
integration of (\ref{1.2.8a})--(\ref{1.2.8d}). That is why we omit
intermediate computations and give the final result
\begin{eqnarray}
\tilde a &=&E_1 (\vec y),\quad b_0\ \, =\ \, E_2 (\vec y),\quad
  d_0\ \, =\ \, E_3 (\vec y), \quad e\ \, =\ \, E_4(\vec y),\non\\
b_k &=&\psi^\dagger \gamma_k \psi B_1 (\vec y) +\psi^\dagger
  \gamma_4 \gamma_k \psi B_2 (\vec y) + \psi^T \gamma_0 
  \gamma_2 \gamma_k \psi B_3 (\vec y ),\non\\
c_{0k} &=&\psi^\dagger \gamma_k \psi C_1 (\vec y) +\psi^\dagger
  \gamma_4 \gamma_k \psi C_2 (\vec y) +
  \psi^T \gamma_0 \gamma_2 \gamma_k \psi C_3 (\vec y
  ),\label{4.1.27}\\ 
d_{k} &=&\psi^\dagger \gamma_k \psi D_1 (\vec y) +\psi^\dagger
  \gamma_4 \gamma_k \psi D_2 (\vec y) +
  \psi^T \gamma_0 \gamma_2 \gamma_k \psi D_3 (\vec y ),\non\\
c_{ab} &=&\psi^\dagger \gamma_a \gamma_b \psi C_4 (\vec y) +\psi^\dagger
  \gamma_4 \gamma_a \gamma_b \psi C_4 (\vec y) +
  \psi^T \gamma_0 \gamma_2 \gamma_a \gamma_b \psi C_6 (\vec y ),\non
\end{eqnarray}
where $B_1,\ B_2,\ldots, E_4 $ are arbitrary smooth complex-valued
functions; $\vec y$ is a complete set of functionally-independent
invariants of the group $O(3)$ which can be chosen in the form $\vec y
=( \bar {\psi} \psi,\, \psi ^\dagger\psi,\, \psi ^\dagger\gamma_4
\psi,\, \bar\psi \gamma_4 \psi,\, \psi^T \gamma_2\psi )$.

Substitution of (\ref{4.1.25}), (\ref{4.1.27}) into (\ref{4.1.21})
gives rise to the following class of $O(3)$-invariant spinor
equations: 
\begin{eqnarray}
& &\{ -i (\gamma_0 +\gamma_4 )\p_t +i \gamma_a \p_a 
  +m (\gamma_0 -\gamma_4) \} \psi \non\\
& &\quad=\Bigl\{E_1 +\gamma_0 E_2 +\gamma_0\gamma_4 E_3 
  +\gamma_4 E_4 +\gamma_a (\psi^\dagger \gamma_a \psi B_1\non\\
& &\quad +\psi^\dagger \gamma_4 \gamma_a \psi B_2 +\psi^T \gamma_0
\gamma_2 
  \gamma_a \psi B_3 )+\gamma_4 \gamma_a  
  (\psi^\dagger \gamma_a \psi D_1 \label{4.1.28}\\
& &\quad +\psi^\dagger \gamma_4 \gamma_a \psi D_2 
  +\psi^T \gamma_0 \gamma_2 \gamma_a \psi D_3 )
  +\gamma_0 \gamma_a (\psi^\dagger \gamma_a \psi C_1\non\\
& &\quad + \psi^\dagger \gamma_4 \gamma_a \psi C_2 +
  + \psi^T \gamma_0 \gamma_2 \gamma_a \psi C_3 ) + \gamma_a \gamma_b
  (\psi^\dagger \gamma_a \gamma_b \psi C_4 \non\\
& &\quad +\psi^\dagger \gamma_4 \gamma_a \gamma_b \psi C_5 +
  \psi^T \gamma_0 \gamma_2 \gamma_a \gamma_b \psi C_6 \Bigr\}\psi.\non
\end{eqnarray}

Formulae (\ref{4.1.28}) are substantially simplified if we use 
identity (1.2.18) rewritten in the form
\begin{displaymath}
(\bar \psi_1 \gamma_a \psi_2 ) \gamma_a \psi_2 =
(\psi^\dagger_1 \psi_2 ) \gamma_0 \psi_2 - (\bar \psi_1 \psi_2)\psi_2
- (\bar \psi _1 \gamma_4 \psi_2 )\gamma_4 \psi_2.
\end{displaymath}

Here $\psi_1$,\ $\psi_2$ are arbitrary four-component functions.

Due to the above identity equation (\ref{4.1.28}) takes a more 
compact form 
\begin{equation}  
\begin{array}{l}
\{ -i (\gamma_0 +\gamma_4) \p_t +i\gamma_a \p_a +
  m(\gamma_0 -\gamma_4)\}\psi\\[2mm]
\quad =(h_1 +h_2 \gamma_0 +h_3 \gamma_0 \gamma_4  
  +h_4 \gamma_4 )\psi,
\end{array}
\label{4.1.29}
\end{equation}
where $h_i=h_i (y_1,\,  y_2,\,  y_3,\,  y_4,\,  y_5)$,\ 
$i={1,\ldots,4}$ are arbitrary smooth complex-valued functions.

Next, acting with the first prolongation of the generator of 
group (\ref{4.1.11}) on (\ref{4.1.29}) and using the Lie invariance
criterion we get the following equations for $H=h_1 +h_2 \gamma_0 +h_3
\gamma_0 \gamma_4 +h_4 \gamma_4$: 
\begin{equation}
Q_a H + (1/2) [ H,\, (\gamma_0 +\gamma_4) \gamma_a ]=0,\ \
               a={1,2,3},
\label{4.1.30}
\end{equation}
where $Q_a =-(1/2) \{ (\gamma_0 +\gamma_4) \gamma_a 
\psi \}^{\alpha} \p_{\psi^{\alpha}}
-(1/2) \{ (\gamma_0^* +\gamma_4^*) \gamma_a^* \psi^* \}^{\alpha} 
\p_{\psi^{* \alpha}}$.

Computing commutators in the left-hand sides of system (\ref{4.1.30})
and equating to zero the coefficients of linearly independent
$\gamma$-matrices we come to the system of PDEs for $h_1,\ h_2,\ h_3,\
h_4$ 
\begin{equation}
Q_a h_1 =0,\quad Q_a h_2 =0,\quad h_2 =h_4,\quad h_3 =0.
\label{4.1.31}
\end{equation}

Integration of the above equations yields
\begin{displaymath}
h_1 =f_1 (w_1,\, w_2),\quad h_2 =h_4 =f_2 (w_1,\, w_2),\quad h_3 =0,
\end{displaymath}
where $w_1 =\bar\psi \psi,\ \ w_2= \psi^\dagger \psi +\bar\psi
\gamma_4 \psi$. 

Generally speaking, the group $G(1,3)$ is not the maximal invariance
group of equation (\ref{4.1.21}) with $F$ of the form (\ref{4.1.22}).
Below we give without proof the assertions describing functions
$F=F(\psi^*,\, \psi )$ such that the system of PDEs (\ref{4.1.21}) admits
wider groups.  
\vspace{1.5mm}

\noindent
{\bf Theorem 4.1.6.}\ {\em Equation (\ref{4.1.21}) is invariant under
  the group $G_1(1,3) = G(1,3)\mbox{$\stimes$} D(1)$, where $D(1)$ is
  the one-parameter group of scale transformations, only in the
  following cases:}
\begin{eqnarray}
&1)& f_1 =(\psi^\dagger \psi +\bar\psi \gamma_4 \psi )
  ^{1/(2k-1)} \tilde f_1\Bigl((\bar\psi \psi )^{1-2k}(\psi^\dagger \psi 
  +\bar \psi \gamma_4 \psi )^{2k}\Bigr),\non\\
& & f_2 =(\psi^\dagger\psi +\bar\psi \gamma_4 \psi )^
  {2/(2k-1)} \tilde f_2\Bigl((\bar\psi \psi )^{1-2k}(\psi^\dagger \psi 
  +\bar \psi \gamma_4 \psi )^{2k}\Bigr),\label{4.1.32}\\
& & D(1)\ \mbox{\em being\ of\ the\ form}\non\\
& & t'=te^{2\theta_0},\quad x'_a =x_a e^{\theta_0},\non\\
& & \psi' (t^{\prime},\, \vec x^{\prime})= \exp \Bigl\{\theta_0
  \Bigl(-k +(1/2)\gamma_0 \gamma_4 \Bigr)\Bigr\}
  \psi(t,\, \vec x),\label{4.1.33}\\
& & \mbox{\em under}\ k\ne 1/2;\non\\
&2)& f_1=\bar\psi\psi \tilde f_1 (\psi^\dagger \psi
  +\bar \psi \gamma_4 \psi),\quad 
  f_2=(\bar\psi \psi )^2 \tilde f_2 (\psi^\dagger \psi 
  +\bar \psi \gamma_4 \psi),\non\\
& & D(1)\ \mbox{\em being\ of\ the\ form}\ (\ref{4.1.33})\ 
\mbox{\em under}\ k=1/2;\non\\ 
&3)& m=0,\quad f_1 =(\psi^\dagger \psi +\bar\psi \gamma_4 \psi )^{1/2 k}
  \tilde f_1 \Bigl((\bar\psi \psi)(\psi^\dagger \psi 
  +\bar\psi \gamma_4 \psi)^{-1}\Bigr),\non\\
& & f_2 =(\psi^\dagger \psi +\bar\psi \gamma_4 \psi )^{1/2 k} 
  \tilde f_2 \Bigl((\bar\psi \psi)(\psi^\dagger \psi 
  +\bar\psi \gamma_4 \psi)^{-1}\Bigr),\label{4.1.34}\\
& & D(1)\ \mbox{\em being\ of\ the\ form}\non\\
& &t'=te^{\theta_0},\quad x_a' =x_a e^{\theta_0}, \quad
\psi '(t^{\prime},\, \vec x^{\prime})=e^{-k \theta_0 }
\psi(t,\, \vec x).\label{4.1.35}
\end{eqnarray}
{\bf Theorem 4.1.7.}\ {\em Equation (\ref{4.1.21}) is invariant under
  the generalized Galilei group $G_2(1,3)$ only in the following
  cases: 

1) $f_1,\ f_2$ are of the form (\ref{4.1.32}) under $k=3/2$, the
groups of scale and projective transformations are given by formulae
(\ref{4.1.33}) (with $k=3/2$) 
and (\ref{4.1.13});

2) $m=0,\ f_1,\ f_2$ are of the form (\ref{4.1.34}) under $k=3/2$,
the group of scale transformations is of the form (\ref{4.1.35}) with
$k=3/2$ and the group of the projective
transformations\index{Projective transformation group} has the form}
\begin{eqnarray*}
& &t^{\prime}=t(1-\theta_0 t)^{-1},\quad
x_a^{\prime}\ \, =\ \, x_a(1-\theta_0 t)^{-2},\\
& &\psi^\prime (t^{\prime},\, \vec x^{\prime})=(1-\theta_0 t)^3
\exp \{\theta_0(1-\theta_0 t)^{-1}(\g_0+\g_4)\g_a x_a\}
\psi(t,\, \vec x).
\end{eqnarray*}
\vspace{10mm}

\noindent
{\large\bf 4.2. Exact solutions of Galilei-invariant
\vspace{1.5mm}

\noindent
\phantom{\large\bf 4.2. }spinor equations\label{s4.2} }
\markboth{Chapter 4. NONLINEAR GALILEI-INVARIANT SPINOR EQUATIONS }
{4.2. Exact solutions of Galilei-invariant spinor equations}
\def\theequation{4.\arabic{section}.\arabic{equation}}
\setcounter {section} {2}
\setcounter {equation}{0}
\vspace{7mm}

The present section is devoted to reduction and construction of the
multi-parameter families of exact solutions of the nonlinear
Galilei-invariant systems of PDEs
\begin{equation}
\{ -i (\gamma_0 +\gamma_4 )\p_t +i \gamma_a \p_a 
  +m(\gamma_0 -\gamma_4) -f_1 -f_2 
  (\gamma_0 +\gamma_4) \bigr\}\psi =0,
\label{4.2.1}
\end{equation}
where $f_i =f_i (\bar \psi \psi,\, \psi^\dagger \psi +\bar \psi \psi
)$. 
\vspace{2mm}

\noindent
{\bf 1. Ans\"atze for the spinor field}.\
Since a linear representation of the Galilei algebra is realized on
the set of solutions of the system of PDEs (\ref{4.2.1}), we can look for 
Ans\"atze reducing (\ref{4.2.1}) to systems of ODEs in the form
\begin{equation}
\psi (t,\, \vec x) =A(t, \vec x) \vp (\omega ),
\label{4.2.2}
\end{equation}
where $\vp =\vp (\omega)$ is a complex-valued four-component
function. A variable
$(4 \times 4)$-matrix $A(t,\, \vec x )$ and a real-valued function
$\omega =\omega (t,\, \vec x )$ are determined by equations
(\ref{1.5.15}), (\ref{1.5.16}), where operators $Q_1,\ Q_2,\ Q_3$ are
the basis elements of some three-dimensional subalgebra of the Galilei  
algebra $AG(1,3)$.

A classification of the $G(1,3)$ non-conjugate subalgebras of the
algebra $AG(1,3)$ has been carried out in \cite{186} (we use a more
convenient classification given in \cite{66}). Each three-dimensional
subalgebra $\langle Q_1,\, Q_2,\, Q_3 \rangle $ satisfying condition
(\ref{1.5.2}) gives rise to an Ansatz of the form (\ref{4.2.2}) which
reduces the $G(1,3)$-invariant system of PDEs (\ref{4.2.1}) to a system of
ODEs for $\vp\, (\omega )$ (Theorem 1.5.1).

It should be noted that the subalgebraic structure of the algebra
$AG(1,3)$ in the case $m\ne 0$ differs essentially from the one in the 
case $m=0$. That is why  the cases $m\ne 0$ and $m=0$ lead to
principally different sets of Ans\"atze for the spinor field. 

Since the system of nonlinear PDEs (\ref{4.2.1}) with $m=0$ admits the
infinite-parameter Lie group with generators $G_\infty,\ J_\infty$
from (\ref{4.1.7}) \cite{205}, which contains the group $G(1,3)$ 
as a subgroup, it makes no sense reducing it by means of subgroups 
of the Galilei groups. That is why we restrict ourselves to the 
case $m\ne 0$ (Galilei-invariant Ans\"atze for the case $m=0$ 
are constructed in \cite{106.2}).

At first, we will write down the complete list of inequivalent
Ans\"atze for the spinor field invariant under the $G(1,3)$
non-conjugate three-dimensional subalgebras of the algebra $AG(1,3)$ 
and then consider an example of integration of the over-determined 
system of equations (\ref{1.5.15}), 
(\ref{1.5.16}).\index{Ansatz!$G(1,3)$-invariant}
\index{Subalgebras!of the Galilei algebra}
\begin{eqnarray}
&1)&\langle P_0,\, P_1,\, P_2\rangle,\non\\
& &\psi(t,\, \vec x)=\vp\, (x_3);\non\\
&2)&\langle J_{12} +\al P_0,\, P_1,\, P_2\rangle,\non\\
& &\psi(t,\, \vec x)=\exp \{ (t/2\al)\gamma_1 \gamma_2\}\vp\,
(x_3);\non\\ 
&3)& \langle P_{0} +i\al m,\, P_1,\, P_2\rangle,\non\\
& &\psi(t,\, \vec x)=\exp\{-i\al mt\}\vp\, (x_3);\non\\
&4)& \langle J_{12},\, P_0,\, P_3\rangle,\non\\
& &\psi(t,\, \vec x)=\exp \{ -(1/2) \gamma_1 \gamma_2
\arctan(x_1/x_2)\} \vp\, (x_1^2 +x_2^2);\non\\
&5)& \langle J_{12}+\al P_0 +\beta G_3,\, P_1,\, P_2\rangle,\non\\ 
& &\psi(t,\, \vec x)=\exp \{(2im/3)\beta\al^{-2}t(\beta t^2 -3\al x_3) 
-(\beta t/\al)\eta_3 +(t/2\al )\gamma_1 \gamma_2\}\non\\
& &\times\vp\, (\beta t^2 -2\al x_3);\non\\
&6)&\langle J_{12} +\al G_3,\, P_1,\, P_2\rangle,\non\\
& &\psi(t,\, \vec x)=\exp \{ (1/2\al t)x_3 (\gamma_1 \gamma_2 -2\al
\eta_3  
-2im\al x_3 )\}\vp\, (t);\non\\
&7)& \langle J_{12} +\al G_3,\, G_1,\, G_2\rangle,\non\\
& &\psi(t,\, \vec x)=\exp \{ -(im/t)(x_1^2 +x_2^2 ) - (1/t)(\eta_1 x_1
+\eta_2 x_2)\}\non\\ 
& &\times\exp \{ (1/2\al t)x_3 (2i\al m x_3 +\al \eta_3 -
\gamma_1 \gamma_2 )\}\vp\, (t);\non\\
&8)& \langle J_{12} +\al P_3,\, P_1,\, P_2\rangle,\non\\
& &\psi(t,\, \vec x)=\exp \{ -(1/2\al )x_3 \gamma_1 \gamma_2 \}\vp\,
(t);\non\\ 
&9)& \langle J_{12} +\al P_3,\, G_1,\, G_2\rangle,\label{4.2.2z}\\
& &\psi(t,\, \vec x)=\exp \{ -(im/t)(x_1^2 +x_2^2 ) 
- (1/t)(\eta_1 x_1 +\eta_2 x_2 )\}\non\\
& &\times\exp \{ -(1/2\al ) x_3 \gamma_1 \gamma_2\}\vp\, (t);\non\\
&10)& \langle P_1,\, P_2,\, P_3\rangle,\non\\
& &\psi(t,\, \vec x)=\vp\, (t);\non\\
&11)& \langle G_1,\, P_2,\, P_3\rangle,\non\\
& &\psi(t,\, \vec x)=\exp \{ -(im/t)x_1^2 -(1/t)x_1\eta_1\}\vp\,
(t);\non\\ 
&12)& \langle G_{1} +\al P_1,\, G_2,\, P_3\rangle,\non\\
& &\psi(t,\, \vec x)=\exp \{ -im [t^{-1}x_2^2  +(t-\al )^{-1} x_1^2]
-t^{-1}x_2 \eta_2\non\\
& & +(\al -t)^{-1} x_1\eta_1\}\vp\, (t);\non\\
&13)& \langle G_{1} +\al P_1,\, G_2 +\beta P_2,\, G_3\rangle,\non\\
& &\psi(t,\, \vec x)=\exp \{ im [(\al -t)^{-1} x_1^2 +
(\beta -t )^{-1} x_2^2  -t^{-1}x_3^2]\non\\
& &+(\al -t)^{-1}x_1\eta_1 +(\beta -t)^{-1}x_2\eta_2 -
t^{-1}x_3\eta_3\}\vp\, (t);\non\\
&14)& \langle G_1 +\al P_0,\, P_2,\, P_3\rangle,\non\\
& &\psi(t,\, \vec x)=\exp \{ (2im/3)\al^{-2}t(t^2 -3\al
x_1)-(t/\al)\eta_1\}\vp\, (t);\non\\ 
&15)& \langle J_{12}+i\al m,\, P_0,\, P_3\rangle,\non\\
& &\psi(t,\, \vec x)=\exp \{ [i\al m -(1/2)\gamma_1 \gamma_2]
\arctan (x_1/x_2)\}\vp\, (x_1^2 +x_2^2);\non\\
&16)& \langle J_{12}+i\al m,\, P_0+i\beta m,\, P_3\rangle,\non\\
& &\psi(t,\, \vec x)=\exp \{ i\beta mt +[i\al m -(1/2) \gamma_1
\gamma_2] 
\arctan (x_1/x_2)\}\vp\, (x_1^2 +x_2^2);\non\\
&17)& \langle G_1 +\al P_2,\, 
G_2 +\al P_1 +\beta P_2 +\tau P_3,\, G_3 -\rho G_1 
-\delta G_2 -\al\delta P_1\rangle,\non\\
& &\psi(t,\, \vec x)=\exp \{-(im/t)x_1^2 -(1/t)x_1\eta_1\}
\exp\Bigl\{-(im/t)(\al x_1 +t x_2)^2\non\\
& &\times[t(t-\beta )-\al^2]^{-1}
+(1/\tau t)(\al \eta_1 +t\eta_2)x_3\Bigr\}\exp \Bigl\{im w^2\Bigl(f(t)[t
(t-\beta )\non\\ 
& & -\al^2 ]\Bigr)^{-1}-[f(t)]^{-1}w\Bigl((\delta t^{-1} - \tau^{-1})
(\al\eta_1 +t\eta_2)-\eta_3\Bigr)\Bigr\}\vp\, (t).\non
\end{eqnarray}

In the above formulae $\al,\ \beta,\ \rho,\ \delta $ are arbitrary
real parameters; $\eta_a =(1/2)(\gamma_0 +\gamma_4)\gamma_a$,\ 
$a={1,2,3}$;\ $\vp\, (\omega)$ is an arbitrary four-component
function; 
\begin{eqnarray*}
& &w=\tau (\al x_1 +tx_2 ) +\Bigl(t(t-\beta ) -\al^2\Bigr)x_3,\quad
\tau  = \al \rho +\beta \delta,\quad\\
& &f(t)=\tau \Bigl(\al (\rho t -\al \delta )+\delta 
t^2 \Bigr) - t\Bigl(t(t - \beta )-\al^2 \Bigr).
\end{eqnarray*}

As an example, we construct the Ansatz N $11$ from (\ref{4.2.2z}).
Substitution of $Q_1 =t\p_{x_1} +2im x_1 +(1/2)(\gamma_0
+\gamma_4)\gamma_1$,\ $Q_2 =\p_{x_2},\ Q_3=\p_{x_3}$ into the system of
PDEs (\ref{1.5.15}), (\ref{1.5.16}) gives the following equations for
$A(t,\, \vec x )$,\ $\omega (t,\, \vec x)$:
\begin{eqnarray}
& &\begin{array}{l}
  \p_{x_2} A=\p_{x_3} A=0,\\[2mm]
  t\p_{x_1} A +\Bigl(2 im x_1 +(1/2)(\gamma_0 
  +\gamma_4)\gamma_1\Bigr) A =0,
  \end{array}
  \label{4.2.3}\\[2mm]
& &\begin{array}{l}
  \p_{x_2} \omega =0,\quad \p_{x_3} \omega =0,
  \quad t\p_{x_1} \omega =0.
  \end{array}
  \label{4.2.4}
\end{eqnarray}

The first integral of system (\ref{4.2.4}) has the form $\omega =t$.
Next, from (\ref{4.2.3}) it follows that $A=A(t,\, x_1)$ and
in addition
\begin{displaymath}
{\p A\over \p x_1} =-(1/t)\Bigl(2im x_1 
+(1/2)(\gamma_0 +\gamma_4 )\gamma_1\Bigr)A.
\end{displaymath}
Integrating the above equation we get the expression for $A$.
\vspace{2mm}

\noindent
{\bf 2. Reduction of nonlinear equation (\ref{4.2.1})}.\
We will carry out reduction of PDE (\ref{4.2.1}) to systems of 
ODEs provided $m\ne 0$. Substitution of Ans\"atze (\ref{4.2.2}) into
(\ref{4.2.1}) gives rise to equations of the form
\begin{equation} 
A(t,\, \vec x) L\Biggl(\omega,\vp^*, \vp, {d\vp\over
  d\omega}\Biggr)=0. 
\label{4.2.5}
\end{equation}

Since ${\rm det}A(t,\, \vec x)\ne 0$, the above equation is rewritten
as follows 
\begin{displaymath}
L\Biggl(\omega,\, \vp^*,\, \vp,\, {d\vp\over d\omega}\Biggr)=0.
\end{displaymath}

Below we give explicit forms of systems of PDEs for $\vp =\vp\,
(\omega )$ corresponding to the Galilei-invariant Ans\"atze for the
spinor field $\psi(t,\, \vec x)$ listed in (\ref{4.2.2z})
\begin{eqnarray}
&1)& i\gamma_3 \dot\vp + m(\gamma_0 -\gamma_4 )\vp=\widetilde F,
  \non\\
&
2)&  i\gamma_3 \dot\vp +\Bigl((i/2\alpha)(\gamma_0 +\gamma_4 )\gamma_3
  +m(\gamma_0 -\gamma_4 )\Bigr)\vp =\widetilde F,
  \non\\
&
3)& i\gamma_3\dot\vp + \Bigl( \alpha m (\gamma_0 +\gamma_4 )
  +m(\gamma_0 -\gamma_4 )\Bigr)\vp =\widetilde F,
  \non\\
&
4)& 2i \omega^{1/2} \gamma_2 \dot\vp +\Bigl((i/2)\omega^{-1/2}
\gamma_2 + m(\gamma_0 -\gamma_4)\Bigr)\vp =\widetilde F,
  \non\\
&
5)& -2i \alpha \gamma_3 \dot\vp +\Bigl(m(\gamma_0 -\gamma_4 )
  +\alpha^{-2} \beta \omega (\gamma_0 +\gamma_4 ) 
  \non\\
&
& +(i/2\alpha )(\gamma_0 +\gamma_4 )\gamma_3 \Bigr)\vp =\widetilde F,
  \non\\
&
6)& -i(\gamma_0 +\gamma_4)\dot\vp +\Bigl(m(\gamma_0 -\gamma_4 )
+(i/2\alpha \omega)[\gamma_0\gamma_4 
  \non\\
&
& -\alpha (\gamma_0 +\gamma_4)]\Bigr)\vp =\widetilde F,
  \non\\
&
7)& -i(\gamma_0 +\gamma_4 )\dot\vp +\Bigl((i/2\alpha \omega)
  \gamma_0 \gamma_4 -(i/\omega)(\gamma_0 +\gamma_4) 
  +m(\gamma_0 -\gamma_4 )\Bigr)\vp\non\\
& & =\widetilde F,
  \non\\
&
8)& -i(\gamma_0 +\gamma_4)\dot\vp +\Bigl(
  m(\gamma_0 -\gamma_4) -(i/2\alpha)\gamma_0\gamma_4\Bigr)\vp
  =\widetilde F,
\non\\
&
9)& -i(\gamma_0 +\gamma_4)\dot\vp +\Bigl(
  m(\gamma_0 -\gamma_4) -i \omega^{-1} (\gamma_0 +\gamma_4) 
  -i(2\alpha)^{-1} \gamma_0 \gamma_4  \Bigr) \vp\non\\
& &=\widetilde F,
\non\\
&
10)& -i(\gamma_0 +\gamma_4 )\dot\vp +m(\gamma_0 -\gamma_4)\vp
=\widetilde F, 
\label{4.2.6}\\
&
11)& -i(\gamma_0 +\gamma_4 )\dot\vp +
  \Bigl(m(\gamma_0 -\gamma_4) -(i/2\omega)(\gamma_0 
  +\gamma_4 )\Bigr)\vp =\widetilde F,
\non\\
&
12)&  -i(\gamma_0 +\gamma_4 )\dot\vp +
\Bigl(m(\gamma_0 -\gamma_4 ) -i(2\omega -\alpha )
[2\omega(\omega -\alpha)]^{-1}\non\\
&
&\times (\gamma_0 +\gamma_4)\Bigr)\vp=\widetilde F,
\non\\
&
13)& -i(\gamma_0 +\gamma_4 )\dot\vp + \Bigl(m(\gamma_0 -\gamma_4) -
i[3\omega^2 -2( \alpha +\beta )\omega +\alpha \beta]
\non\\
&
& \times [2\omega (\omega -\alpha ) (\omega -\beta )]^{-1}
  (\gamma_0 +\gamma_4 )\Bigr\}\vp =\widetilde F,
  \non\\
&
14)& -2i\alpha \gamma_1\dot\vp +\Bigl(
  m(\gamma_0 -\gamma_4 ) +m\alpha ^{-2}\omega(\gamma_0 
  +\gamma_4 )\Bigr)\vp = \widetilde F,
  \non\\
&
15)& 2i\omega ^{1/2} \gamma_2 \dot \vp + \Bigl(i\omega^{-1/2}
  [i\alpha m \gamma_1 +(1/2) \gamma_2] + m(\gamma_0
  -\gamma_4)\Bigr)\vp 
  =\widetilde F,
  \non\\
&
16)& 2i\omega ^{1/2}\gamma_2\dot\vp + \Bigl( i\omega^{-1/2}
  [i\alpha m \gamma_1 +(1/2) \gamma_2] + m(\gamma_0 -\gamma_4 )
\non\\
&
& + m\beta(\gamma_0 +\gamma_4)\Bigr)\vp =\widetilde F,
\non\\
&
17)& -i(\gamma_0 +\gamma_4)\dot\vp +\Bigl(i[2\omega f(\omega )]^{-1}
  [\omega^3 + \alpha (\alpha +\rho \tau )\omega - 2\tau \alpha^2
  \delta] 
  \non\\
&
&\times (\gamma_0 +\gamma_4 )-(i/\omega)(\gamma_0 +\gamma_4)
  +m (\gamma_0 -\gamma_4) \Bigr)\vp =\widetilde F.\non
\end{eqnarray}

Here a dot over $\vp$ means differentiation with respect to $\omega$,
\begin{eqnarray*}
& &\widetilde F=\{f_1\,
(\bar\vp\vp,\, \vp^\dagger\vp +\bar\vp\gamma_4\vp) 
+(\gamma_0 +\gamma_4)f_2\, (\bar\vp\vp,\, 
\vp^\dagger\vp+\bar\vp\gamma_4\vp)\}\vp,\\
& &\tau =\alpha\rho +\delta\beta,\quad f(\omega)
=\tau[\alpha(\rho\omega -\alpha\delta) +
\delta\omega^2]-[\omega(\omega-\beta)-\alpha^2]\omega.
\end{eqnarray*}
{\bf 3. Exact solutions of nonlinear equation (\ref{4.2.1}).}\
We will construct the multi-parameter families of exact solutions of
nonlinear Galilei-invariant system of PDEs of the form (\ref{4.2.1})
with the power nonlinearity 
\begin{equation}
\Bigl\{ -i (\gamma_0 +\gamma_4 )\p_t +i \gamma_a \p_a
+m (\gamma_0 -\gamma_4 )-\lambda (\psi^\dagger \psi + \bar \psi
\gamma_4 \psi  )^{1/2 k} \Bigr\} \psi =0, 
\label{4.2.7}
\end{equation}
where $\lambda,\ k$ are constants, by means of the Ans\"atze 
for the spinor field $\psi(t,\, \vec x)$ invariant under the
$G(1,3)$ non-conjugate three-dimensional subalgebras of the Galilei
algebra $AG(1,3)$. 

According to the results obtained in the previous subsection
substitution of Ans\"atze (\ref{4.2.2z}) into (\ref{4.2.7}) gives rise
to systems of ODEs (\ref{4.2.6}) with $\widetilde F = \lambda
(\vp^\dagger \vp +\bar \vp\gamma_4 \vp )^{1/2 k} \vp $.

ODEs 6--13 prove to be integrable due to the following assertion.
\vspace{1.5mm}

\noindent
{\bf Lemma 4.2.1.}\ {\em The quantities
\begin{eqnarray*}
& &I_6 =\bar \vp (\gamma_0 +\gamma_4 )\vp\, \omega,\quad
 I_7 = \bar \vp (\gamma_0 +\gamma_4 )\vp\, \omega^2,\\
& &I_8=\bar \vp (\gamma_0 +\gamma_4 )\vp,\quad
  I_9=\bar \vp (\gamma_0 +\gamma_4 )\vp\, \omega^2,\\
& &I_{10} =\bar \vp (\gamma_0 +\gamma_4 )\vp,\quad
  I_{11} =\bar \vp (\gamma_0 +\gamma_4 )\vp\, \omega,\\
& &I_{12}= \bar \vp (\gamma_0 +\gamma_4 )\vp\, (\omega^2 -\alpha
\omega ),\\ 
& &I_{13}= \bar \vp (\gamma_0 +\gamma_4 )\vp\, 
\omega(\omega -\alpha )(\omega -\beta)
\end{eqnarray*}
are the first integrals of the systems of PDEs 6,7,\ldots,13,
respectively.}
\vspace{1.5mm}

Proof is carried out by a direct check. $\rhd$

According to the above lemma, the systems of nonlinear ODEs 6--13 
are linearized. And what is more, the linear systems obtained 
are integrable in quadratures. This fact enables us to construct 
the general solutions of the reduced ODEs 6--13 from (\ref{4.2.6}). 
These solutions are represented in the following unified form:
\begin{equation}
\vp_N (\omega ) =(1/2) \{f_N (\omega)
(\gamma_0 +\gamma_4) + g_N(\omega)
(1+ \gamma_0 \gamma_4 )\}\chi,
\label{4.2.8}
\end{equation}
where $N={6,\ldots,13}$,\ $\chi $ is an arbitrary constant 
four-component column,
\begin{eqnarray}
f_6 (\omega ) &=&(1/2m)[\tilde \lambda\, \omega^{-1/2k}-
(i/2\alpha \omega )]g_6,
\non\\
g_6 (\om) &=&\om ^{-1/2 }\exp\{
-(i/16)(\alpha^2 m\om)^{-1} +i W_1 (k,\om )\};
\non\\
f_7 (\om) &=&(1/2m)[\tilde \lbd\, \om^{-1/ k} -(i/2\alpha \om )]g_7,
\non\\
g_7 (\om)&=& \om^{-1} \exp\{-(i/16)(\alpha^2 m\om)^{-1}
+i W_2 (k,\om )\};
\non\\
f_8 (\om ) &=&(1/2m)[\tilde\lbd +(i/2\alpha)]g_8,
\non\\
g_8 (\om )&=&\exp \{
i(1+4\alpha ^2 \tilde \lbd ^2 ) (16 \alpha^2 m )^{-1} \om \};
\non\\
f_9 (\om )&=& (1/2m)[\tilde \lbd \om ^{-1/k }+(i/2\alpha)]g_9,
\non\\
g_9 (\om)&=&(1/\om)\exp \{ i\om (16\alpha^2 m )^{-1} +
iW_2 (k,\om )\};
\non\\
f_{10} (\om) &=&(\tilde\lbd/2m)\exp \{ (i/4m)
\tilde\lbd^2\om\},
\non\\
g_{10} (\om ) &=& \exp \{ (i/4m)\tilde\lbd ^2 \om\};
\label{4.2.9}\\
f_{11} (\om )&=& (\tilde\lbd/2m)\om^{1/2 k}g_{11},
\non\\
g_{11} (\om ) &=&\om ^{1/2 } \exp \{ iW_1 (k,\om )\};
\non\\
f_{12} (\om )&=& (\tilde\lbd/2m)(\om^2 -\alpha \om )^{-1/2k}g_{12},
\non\\
g_{12} (\om )&=& (\om^2 -\alpha \om )^{-1/2} \exp \Biggl\{
i\tilde \lbd ^2 (4m)^{-1}\edi\int\limits^{\edi\om }
(z^2 -\alpha z )^{-1/k} dz\Biggr\};
\non\\
f_{13} (\om )&=& (\tilde\lbd/2m)
[\om (\om -\alpha )(\om -\beta )]^{-1/2k }g_{13},
\non\\
g_{13} (\om )&=& [\om (\om -\alpha )(\om -\beta )] ^{-1/2}\non\\
& &\times\exp \Biggl\{(i\tilde\lbd^2/4m)
\edi\int\limits^{\edi\om }[z(z-\alpha)
(z-\beta )]^{-1/k} dz\Biggr\}.\non
\end{eqnarray}
\noindent
In (\ref{4.2.9}) $\tilde\lbd =\lbd (\chi^\dagger \chi +\bar \chi
\gamma_4 
\chi )^{1/2k }$, 
\begin{displaymath}
W_n (k,\om ) =(\tilde\lbd ^2/4m)\left\{
\begin{array}{ll}k(k-n)^{-1} \om ^{(k-n)/k },& {\rm under}\ k\ne n,\\[2mm] 
\ln \om,& {\rm under}\ k=n. \end{array}\right.
\end{displaymath}

Substitution of the above results into the corresponding Ans\"atze for
the spinor field $\psi (t,\,\vec x)$ yields the following classes of
exact solutions of nonlinear equation (\ref{4.2.7}):
\index{Exact solutions!of nonlinear Galilei-invariant spi\-nor equations}
\begin{eqnarray}
\psi (t,\, \vec x)&=&\exp \{x_3 (2\alpha t )^{-1}
  (\gamma_1\gamma_2 -2\alpha\eta_3 -2i\alpha m x_3)\}\vp_6\, (t),
\non\\
\psi (t,\, \vec x)&=&\exp \{
  -imt^{-1}x_a x_a-t^{-1}\eta_a x_a \}
  \exp\{ (2\alpha t)^{-1}x_3\gamma_1\gamma_2\}\vp_7\, (t),
\non\\
\psi (t,\, \vec x)&=&\exp \{
-(2\alpha)^{-1}x_3\gamma_1\gamma_2\}\vp_8 \, (t),
\non\\
\psi (t,\, \vec x)&=&\exp \{
-imt^{-1} (x_1^2+x_2^2)-t^{-1}(x_1\eta_1 +x_2\eta_2 )\}
\non\\
& &\times\exp \{-(2\alpha)^{-1}x_3\gamma_1\gamma_2\} \vp_9 \, (t),
\non\\
\psi (t,\, \vec x)&=&\vp_{10} \, (t),
\label{4.2.10}\\
\psi (t,\, \vec x)&=&\exp \{
-imt^{-1}x_1^2 -t^{-1}x_1\eta_1\}\vp_{11}\, (t),
\non\\
\psi (t,\, \vec x)&=&\exp \{
-imt^{-1}x_2^2 -t^{-1}x_2\eta_2\}\exp \{im(\alpha -t )^{-1}x_1^2
\non\\
& &+(\alpha -t)^{-1}x_1\eta_1\}\vp_{12}\, (t),
\non\\
\psi (t,\, \vec x)&=&\exp \{
-im[t^{-1}x_3^2 +(t-\alpha)^{-1}x_1^2 +(t-\beta)^{-1}x_2^2]
\non\\ 
& &-t^{-1}x_3\eta_3 +(\alpha -t)^{-1}x_1\eta_1 +
(\beta -t)^{-1}x_2\eta_2\}\vp_{13}\, (t).\non
\end{eqnarray}

To obtain $G(1,3)$-ungenerable families of exact solutions of system
of nonlinear PDEs (\ref{4.2.7}) it is necessary to apply the procedure
of generating solutions by transformations from the symmetry group of
system of PDEs (\ref{4.2.7}).

Using Theorem 2.4.1 it is not difficult to obtain the formulae of
generating solutions by transformation groups
(\ref{4.1.9})--(\ref{4.1.14})  
\begin{eqnarray*}
\index{Solution generation!with translations}
P&:&\cases{
 \psi_{II}(t,\, \vec x)  =\psi_{I}(t^{\prime},\, \vec x^{\prime}),\cr 
 t'=t+\theta_0,\cr
 x'_a=x_a +\theta_a;
\cr}\\
\index{Solution generation!with rotation group}
J&:&\cases{
 \psi_{II}(t,\, \vec x)= \exp \{(1/4) 
 \ve_{abc} \theta_a \gamma_b\gamma_c \}
 \psi_{I}(t^{\prime},\, \vec x^{\prime}),\cr
 t'=t,\cr
 x_a'=\Bigl(\delta_{ab}\cos \theta +\ve_{abc} \theta_c 
 \theta^{-1} \sin \theta \cr
 \quad +\theta_a \theta_b \theta^{-2} (1-\cos \theta)\Bigr) x_b;
\cr }\\
\index{Solution generation!with Galilei transformations}
G&:&\cases{
 \psi_{II}(t,\, \vec x)= \exp \Bigl\{2im \Bigl(\theta_a x_a 
 +(t/2)\theta_a \theta_a \Bigr) \cr
 \quad +(1/2)(\gamma_0 +\gamma_4) \gamma_a \theta_a \Bigr\} 
 \psi_{I}(t^{\prime},\, \vec x^{\prime}),\cr  
 t'=t,\cr
 x_a'= x_a +\theta_a t;
\cr }\\
\index{Solution generation!with scale transformations}
D&:& \cases{
 \psi_{II}(t,\, \vec x) = \exp \{2\theta_0 
 -(1/2) \theta_0 \gamma_0 
 \gamma_4 \} \psi_{I}(t^{\prime},\, \vec x^{\prime}),\cr  
 t'=t e^{2\theta _0},\cr
 x'_a=x_a e^{\theta_0};
\cr }\\
\index{Solution generation!with projective transformations}
A&:& \cases{
 \psi_{II}(t,\, \vec x) =(1-\theta_0 t)^{-2} \exp \Bigl\{
 im\theta_0 (1-\theta_0 t)^{-1} x_a x_a\cr
 \quad +(1/2t)\ln (1-\theta_0 t) \Bigl(t\gamma_0\gamma_4 
 +(\gamma_0 +\gamma_4) \gamma_a x_a
 \Bigr)\Bigr\}\psi_{I}(t^{\prime},\,  
 \vec x^{\prime}),\cr
 t'= t(1-\theta_0 t)^{-1},\cr
 x'_a =x_a (1-\theta_0 t)^{-1};
\cr  }\\
M&:& \cases{
\psi_{II}(t,\, \vec x) =e^{2 im \theta_0}
\psi_{I}(t^{\prime},\, \vec x^{\prime}),\cr 
 t'=t,\cr
 x_a'= x_a,
\cr } 
\end{eqnarray*}
where $\theta_0,\ \theta_a$ are arbitrary parameters,\
$\theta=(\theta_a\theta_a)^{1/2}$.

Applying the solution generation formulae to (\ref{4.2.10}) and making 
some rather cumbersome computations yield the $G(1,3)$-ungenerable
families of exact solutions of nonlinear equation (\ref{4.2.7}). Below 
we present one of them
\begin{eqnarray*}
\psi(t,\, \vec x)&=&(1/2)\exp \{im(2\theta_a x_a +t\theta_a\theta_a)\}
  \exp \Bigl\{ -(1/2T)a_b z_b \Bigl(2im a_b z_b \\
  & &+(\gamma_0 +\gamma_4)\gamma_b a_b\Bigr)\Bigr\}
  \Bigl\{(\gamma_0 +\gamma_4)\Bigl(f_{11}(T)
  +g_{11}(T)\gamma_a \theta_a\Bigr)\\
  & &+g_{11}(T)(1+ \gamma_0\gamma_4)\Bigr\} \chi,
\end{eqnarray*}
where $z_a =x_a +t\theta_a +\tau _a$,\ $T=t+\tau_0$;\
$\{\theta_a,\ \tau_{\mu}\} \subset \R^1$ are arbitrary parameters; 
$\vec a$ is an arbitrary constant unit vector; the functions 
$f_{11}(T)$,\ $g_{11}(T)$ are given in (\ref{4.2.9}).

It is worth noting that all solutions (\ref{4.2.10}) have a singularity
at the point $m=0$. Consequently, it is impossible to obtain
solutions of massless equation (\ref{4.1.1}) putting in (\ref{4.2.10})
$m=0$.
\vspace{10mm}

\noindent
{\large\bf 4.3. Galilei-invariant second-order spinor
  equations\label{s4.3} } 
\markboth{Chapter 4. NONLINEAR GALILEI-INVARIANT SPINOR EQUATIONS }
   {4.3. Galilei-invariant second-order spinor equations}
\def\theequation{4.\arabic{section}.\arabic{equation}}
\setcounter {section} {3}
\setcounter {equation}{0}
\vspace{7mm}

\noindent
As noted in Section 4.1 the substitution
\begin{equation}
\psi (t,\, \vec x) =\{-i(\gamma_0 +\gamma_4)\p _t +i\gamma_a \p_a +
m(\gamma_0 -\gamma_4)\}\Psi(t,\, \vec x)
\label{4.3.1}
\end{equation}
reduces equation (\ref{4.1.1}) to a system of splitting Schr\"odinger
equations 
\begin{equation}
(4im\p_t -\p_a\p_a )\Psi^{\alpha}(t,\, \vec x) =0,
\label{4.3.2}
\end{equation}
where $\Psi^{\alpha} $ are components of the spinor $\Psi$.

Thus, the problem of integration of system of linear PDEs
(\ref{4.1.1}) is reduced to the integration of the scalar
Schr\"odinger equation. That is why system of the second-order PDEs
(\ref{4.3.2}) can also be used to describe a Galilean particle with
spin $s=1/2$. However equations (\ref{4.3.2}) describe particles with
different spins because they are invariant under the Galilei group
having the generators
\begin{eqnarray}
& &P_0=\p_t,\quad P_a =\p_a,\non\\
& &J_{ab} =-x_a \p_b +x_b \p_a +S_{ab},\ \ a\ne b,\label{4.3.3}\\
& &G_a =t\p_a +2im x_a +\eta_a,\non
\end{eqnarray}
where $S_{ab},\ \eta_a $ are arbitrary constant matrices of 
the corresponding dimensions which satisfy the commutation relations
\begin{equation}
\begin{array}{l}
[S_{ab},\, S_{cd} ] = -\delta_{ad}S_{bc} -\delta_{bc}S_{ad} +
\delta_{ac}S_{bd} +\delta_{bd}S_{ac},\\[2mm]
[\eta_a,\, S_{bc} ]=\delta_{ac} \eta_b -\delta_{ab} \eta_c,\quad
[\eta_a,\, \eta_b] =0.
\end{array}
\label{4.3.4}
\end{equation}

To obtain from (\ref{4.3.2}) a system of PDEs describing a particle
with spin $s=1/2$ one should impose an additional constraint on
the set of solutions of (\ref{4.3.2}). For example, if equations
(\ref{4.3.2}) are considered together with (\ref{4.1.1}) 
\begin{equation}
\begin{array}{l}
(4im\p_t-\p_a\p_a) \psi (t,\, \vec x ) =0,\\[2mm]
\{-i (\gamma_0 +\gamma_4 )\p_t +i \gamma_a \p_a 
+m(\gamma_0 -\gamma_4 )\}\psi (t,\, \vec x ) =0,
\end{array}
\label{4.3.5}
\end{equation}
then the maximal (in Lie sense) invariance group of the system
obtained is the generalized Galilei group having the generators
(\ref{4.1.3}). This assertion follows from the fact that the set of
solutions of system (\ref{4.3.5}) coincides with the set of solutions of
equation (\ref{4.1.1}).

Imposing on solutions of system (\ref{4.3.2}) the weaker nonlinear
constraint 
\begin{equation}
\p_t \Bigl(\bar \psi (\gamma_0 +\gamma_4 ) \psi\Bigr) =\p_a (\bar \psi 
\gamma_a \psi) 
\label{4.3.6}
\end{equation}
we get another example of a Galilei-invariant system of PDEs 
for a particle with spin $s=1/2$. Let us note that additional
constraint (\ref{4.3.6}) is an algebraic consequence of equation
(\ref{4.1.1}). Therefore, the set of solutions of system
(\ref{4.3.2}), (\ref{4.3.6}) contains all solutions of equation
(\ref{4.1.1}).  

By a direct check we can become convinced of the fact that the system of
PDEs (\ref{4.3.2}) is not invariant under the generalized Galilei
group with generators (\ref{4.1.3}). The same assertion holds for
nonlinear equations of the 
form\index{Galilei-invariant second-order spinor equation} 
\begin{equation}
(4im \p_t -\p_a \p_a )\psi + F(\psi^*, \psi ) =0,
\label{4.3.7}
\end{equation}
where $F$ is a complex-valued four-component function.
\vspace{1.5mm}

\noindent
{\bf Theorem 4.3.1.}\ {\em The system of PDEs (\ref{4.3.7}) is invariant
  under the Galilei group with the generators $P_0,\ P_a,\ J_{a b},\
G_a,\ M$ from (\ref{4.1.3}) iff
\begin{equation}
F=\{f_1 + (\gamma_0 +\gamma_4)f_2\} \psi,
\label{4.3.8}
\end{equation}
where $f_1,\ f_2 $ are arbitrary smooth functions of $w_1 =\bar \psi
\psi $,\ $w_2 =\psi^\dagger \psi +\bar \psi \gamma_4 \psi
$. Furthermore, the class of PDEs (\ref{4.3.7}) contains no equations
admitting the group $G_2(1,3)$ with generators (\ref{4.1.3}).}
\vspace{1.5mm}

\noindent
{\em Proof.}$\quad$ Invariance of system (\ref{4.3.7}) with respect to
the group of translations (\ref{4.1.9}) is evident. Consequently, to
prove the theorem we have to study the restrictions imposed on the
four-component function $F(\psi^*,\, \psi )$ by the requirement that
(\ref{4.3.7}) admits the Lie groups with the generators $J_{ab},\ 
G_a$. Acting by the first prolongation of the operators $J_{ab},\ 
G_a$ on (\ref{4.3.7}) and applying the Lie invariance criterion we
get an over-determined system of linear PDEs for $F(\psi^*,\, \psi )$.
If we rewrite the function $F$ in the equivalent form
$H(\psi^*,\, \psi )\psi$, where $H(\psi^*,\, \psi )$ is a $4\times
4$-matrix, then the system of PDEs in question takes the form
\begin{equation}
\begin{array}{l}
\Bigl(\{ \gamma_a \gamma_b \psi \}^{\alpha} \p_{\psi^{\alpha}} +
\{ \gamma_a^* \gamma_b^* \psi^* \}^{\alpha } \p_{\psi^{\alpha *
    }}\Bigr)H= [\gamma_a \gamma_b,\, H ],\\
\Bigl(\{(\gamma_0 +\gamma_4)\gamma_a\psi\}^{\alpha }\p_{\psi^{\alpha}}
+ \{(\gamma_0^* +\gamma_4^*)\gamma_a^* \psi^* \}^{\alpha }
\p_{\psi^{*\alpha}} \Bigr)H\\
\quad=[H,\, (\gamma_0 +\gamma_4)\gamma_a].
\end{array}
\label{4.3.9}
\end{equation}

Here $\{\psi \}^{\alpha} $ is the $\alpha$-th component of $\psi $,\
$[\quad,\quad ]$ is the commutator.

Equations (\ref{4.3.9}) coincide with (\ref{4.1.24}), (\ref{4.1.30}),
whose general solution after being substituted into the equality
$F(\psi^*,\, \psi )=H(\psi^*,\, \psi )\psi $ gives rise to formula
(\ref{4.3.8}).

On applying the Lie method we come to the conclusion that the
necessary and sufficient conditions for system (\ref{4.3.7}),
(\ref{4.3.8}) to be invariant under the group of projective
transformations (\ref{4.1.13}) are as follows 
\begin{eqnarray*}
& &(w_1 \p_{w_1} +w_2 \p_{w_2} -2/3)f_i =0,\ \ i=1,2,\\
& &(\gamma_0 +\gamma_4)\Bigl(i\gamma_a\p_a +
m(\gamma_0-\gamma_4)\Bigr)\psi=0.
\end{eqnarray*}

Since the last equation is not a consequence of system (\ref{4.3.7}),
(\ref{4.3.8}), equation (\ref{4.3.7}) is not invariant with respect to
the generalized Galilei group having generators (\ref{4.3.1}). The
theorem is proved. $\rhd$

According to the above theorem, to obtain a $G_2(1,3)$-invariant
nonlinear generalization of system (\ref{4.3.2}) we have to study the
wider class of PDEs 
\begin{equation}
(4im \p_t -\p_a \p_a )\psi -F (\psi^*, \psi,
{\mathop{\psi}\limits_{\scriptscriptstyle 1}}^*,  
\mathop{\psi}\limits_{\scriptscriptstyle 1}) = 0,
\label{4.3.10}
\end{equation}
where the notation $\mathop{\psi}\limits_{\scriptscriptstyle 1} =\{
\p_t \psi, \p_a \psi \}$ is used.

Here we adduce only one example of the equation of the form
(\ref{4.3.10}) invariant under the group $G_2(1,3)$ with generators
(\ref{4.1.3})
\begin{eqnarray*}
& &(4im\p_t -\p_a \p_a )\psi +(1/3)(\bar\psi \psi )^{-1} 
\Bigl\{\Bigl(i(\gamma_0 +\gamma_4 )\p_t 
-i\gamma_a \p_a\Bigr)\bar\psi\psi\Bigr\}\\
& &\quad\times\Bigl\{-i(\gamma_0 +\gamma_4 ) \p_t +i\gamma_a \p_a 
+m(\gamma_0 -\gamma_4 )\Bigr\}\psi\\
& &\quad +(\bar\psi
\psi)^{2/3}\Bigl(f_1+f_2(\gamma_0+\gamma_4)\Bigr)\psi=0, 
\end{eqnarray*}
where $f_i =f_i[(\psi^\dagger \psi + \bar \gamma_4 \psi )^3 (\bar \psi
\psi)^{-2}], \ i=1,2$ are arbitrary smooth functions.

There exist second-order PDEs invariant under the Galilei group which
are principally different from (\ref{4.3.2}). For example, in
\cite{73,77.0} the following $G(1,3)$-invariant system of 
PDEs\index{Galilei-invariant second-order spinor equation}
\begin{displaymath}
(i\gamma_{\mu } \p_{\mu } -m ) \psi(t,\, \vec x) =(1/2m)(1-\gamma_0
-i\gamma_4) \p_a\p_a\psi(t,\, \vec x)
\end{displaymath}
was obtained. It is invariant under the Galilei group with the
generators 
\begin{eqnarray*}
& &P_0=\p_t,\quad P_a =\p_a,\\
& &J_{ab} =-x_a \p_b +x_b \p_a +(1/2) \gamma_a \gamma_b,\ \ a\ne b,\\
& &G_a =t\p_a -imx_a +(1/2)(1+i\gamma_4)\gamma_a.
\end{eqnarray*}

\newpage
\thispagestyle{empty}
\noindent
{\sl C H A P T E R \ \  5\label{ch5}}
\vspace{2mm}

\hrule
\vspace{35mm}

\rightline
{\large\bf
SEPARATION}
\vspace{2mm}

\rightline
{\large\bf
OF VARIABLES}
\vspace{7mm}

\noindent
In this Chapter we present the basis of the symmetry approach
to the separation of variables in systems of linear PDEs. A generalization 
of the St\"ackel method of separation of variables \cite{51,187}
for the case of systems of differential equations is suggested.
Separation of variables in some Galilei-invariant PDEs is performed. 
\vspace{10mm}

\noindent
{\large\bf 5.1. Separation of variables and symmetry of systems
\vspace{1.5mm}

\noindent
\phantom{\large\bf 5.1. }of partial differential
equations\label{s5.1}} 

\markboth{Chapter 5. SEPARATION OF VARIABLES }
   {5.1. Separation of variables and symmetry}
\def\theequation{5.\arabic{section}.\arabic{equation}}
\setcounter {section} {1}
\setcounter {equation}{0}
\vspace{7mm}

\noindent
The Dirac equation (\ref{1.1.1}) is called separable in Cartesian
coordinates if it has exact solutions of the form
\begin{equation}
    \psi(x)=V_0(x_0)V_1(x_1)V_2(x_2)V(x_3)\chi,
\label{5.1.1}
\end{equation}
where $V_{\mu}$ are nonsingular ($4\times 4$)-matrices, $\chi$ is a
constant four-component column.

It is well-known that there exists a deep relation between variable
separation and symmetry properties of PDEs \cite{97,136,155}. This
relation can be characterized in the following way: a solution with
separated variables is a common eigenfunction of some set of commuting
symmetry operators of the equation considered. To demonstrate the main
steps of applying the method of separation of variables within
the framework of the symmetry approach we will consider an example.
A particular solution of (\ref{1.1.1}) is looked for as a solution of
the following over-determined system of PDEs:
\begin{equation}
\begin{array}{l}
    (i\g_{\mu}\p_{\mu}-m)\psi(x)=0,\quad
    (\p_0-\lbd_1)\psi(x)=0,\\[2mm]
    (\p_1-\lbd_2)\psi(x)=0,\quad
    (\p_2-\lbd_3)\psi(x)=0.
\end{array}
\label{5.1.2}
\end{equation}

Integration of the last three equations of system (\ref{5.1.2}) yields 
\begin{displaymath}
    \psi(x)=\exp\{\lbd_1x_0 + \lbd_2x_1 + \lbd_3x_2\}\vp\,(x_3).
\end{displaymath}

Substitution of the above expression into the first equation from
(\ref{5.1.2}) gives rise to the system of ODEs for the four-component
function $\vp\,(\om)$
\begin{displaymath}
   i\g_3\dot\vp + (-m+i\lbd_1\g_0 + i\lbd_2\g_1 + i\lbd_3\g_2)\vp=0,
\end{displaymath}
whose general solution reads
\begin{displaymath}
    \vp\,(\om)=\exp\{i\g_3(-m + i\lbd_1\g_0  + i\lbd_2\g_1 +
   i\lbd_3\g_2)\om\}\chi.
\end{displaymath}

Hence we conclude that the general solution of (\ref{5.1.2}) is of the
form (\ref{5.1.1}) 
\begin{equation}
\begin{array}{rcl}
    \psi(x)&=&\exp\{\lbd_1x_0\}\exp\{\lbd_2x_1\}\exp\{\lbd_3x_2\}\\[2mm]  
    &&\times \exp\{i\g_3(-m+i\lbd_1\g_0 + i\lbd_2\g_1 
    + i\lbd_3\g_2)x_3\}\chi.
\end{array}
\label{5.1.3}
\end{equation}

Comparing (\ref{5.1.1}) with (\ref{5.1.3}) we come to a conclusion
that the solution with separated variables (in Cartesian coordinates)
is the eigenfunction of operators $\p_0,\ \p_1,\ \p_2$ which form a
commutative subalgebra of the invariance algebra of the Dirac
equation.  Consequently, classification of commuting symmetry
operators is a part of the method of separation of variables.

From the above example it is seen that the solution with separated
variables contains arbitrary parameters $\lbd_1$,\ $\lbd_2$,\ $\lbd_3$  
which are called separation constants.

Now we turn to the problem of variable separation in arbitrary systems 
of linear first-order PDEs
\begin{equation}
      \{L_{\mu}(x)\p_{\mu}+M(x)\}u(x)=0,
\label{5.1.4}
\end{equation}
where $u=\Bigl(u^0(x),\, u^1(x),\ldots, u^{m-1}(x)\Bigr)^T;$\
$x=(x_0,\, x_1,\ldots, x_{n-1}); \ \{n, m\} \subset \N $;\ $L_\mu,\ M$
are $(m\times m)$-matrices ($M$ is supposed to be nonsingular).

In what follows, a block $(n_1N_1\times n_2N_2)$-matrix $B$, whose
entries are $(N_1\times N_2)$-matrices $B_{\mu\nu},\ 
\mu={1,\ldots,n_1},\ \nu={1,\ldots,n_2}$, is designated for brevity as
$B=\|B_{\mu\nu}\|_{\mu=1\nu=1}^{n_1\; \; \, n_2}$. Such a notation
is very convenient and simplifies considerably all manipulations with
block matrices. For example, a product of two block $(n_1N_1\times
n_2N_2)$- and $(n_2N_2\times n_3N_3)$-matrices
\begin{displaymath}
B=\|B_{\mu\nu}\|_{\mu=1\nu=1}^{n_1\; \; \,  n_2},\quad
C=\|C_{\mu\nu}\|_{\mu=1\nu=1}^{n_2\; \; \,  n_3}
\end{displaymath}
is a block $(n_1N_1\times n_3N_3)$-matrix 
\begin{displaymath}
BC=\|B_{\mu\alpha}C_{\alpha\nu}\|_{\mu=1\nu=1}^{n_1\; \; \,  n_3},
\end{displaymath}
where summation over the repeated indices from $1$ to $n_2$ is
understood. More details about operations with block matrices can be
found in \cite{110}. 

In the theory of variable separation in linear PDEs with one dependent
variable a very important role is played by the St\"ackel matrices
$C=\|c_{\mu\nu}\|,$\ ${\rm det}\, C\ne 0$, where $c_{\mu\nu}$ are smooth
functions depending on the variable $x_\mu$ only. Separable PDEs admit
rather natural and simple description in terms of the St\"ackel
matrices \cite{136}. It is believed that the above matrices when
properly generalized should be of importance for variable separation
in multi-component systems of linear PDEs as well \cite{156}.

Below we present an approach to variable separation in systems of
linear PDEs (\ref{5.1.4}) which uses essentially a generalized
block St\"ackel matrix introduced below.\index{St\"ackel matrix}
\vspace{1.5mm}

\noindent
{\bf Definition 5.1.1.}\ Block $(nm\times nm)$-matrix
$C=\|C_{\mu\nu}(x_\mu)\|_{\mu,\nu=0}^{n-1}$, where
$C_{\mu\nu}(x_\mu)$ are square $(m\times m)$ matrices, is called
the St\"ackel matrix if the following conditions are fulfilled:
\begin{eqnarray*}
&1)& {\rm det}\, C\ne 0,\\
&2)& [C_{\mu\nu},\, C_{\alpha\beta}]+[C_{\mu\beta},\,
C_{\alpha\nu}]=0.
\end{eqnarray*}

Evidently, provided $m=1$, the above definition coincides with the
usual definition of the St\"ackel matrix (see e.g. \cite{136,156}).
\vspace{1.5mm}

\noindent
{\bf Definition 5.1.2.}\ A set of smooth real-valued functions
$z_\mu=z_\mu(x),\ \mu={0,\ldots,n-1}$ is called a coordinate system if the
condition ${\rm det}\, \|\p_{x_\mu}z_\nu(x)\|_{\mu,\nu=0}^{n-1}$ $\ne 0$ is
satisfied. 

Now we are ready to give a precise definition of separation of
variables in systems of linear PDEs which has been suggested for the
first time in \cite{108.4}. 
\vspace{1.5mm}

\noindent
{\bf Definition 5.1.3.}\ Let $(m\times m)$-matrix functions
$V_{\mu}(z_\mu),\ \mu={0,\ldots,n-1}$ satisfy the system of $n$ matrix 
ODEs\index{Separation of variables}
\begin{equation}
\begin{array}{l}
   {\edi dV_{\mu}\over \edi dz_{\mu}} =\Bigl(C_{\mu 0}(z_{\mu})
   +C_{\mu a}(z_\mu)\lbd_a\Bigr)V_{\mu},\ \ \mu={0,\ldots,n-1},\\[2mm]
\left.\matrix{V_\mu \cr\cr\cr}\right.
\left|\matrix{&=I,\cr z_\mu=\theta_\mu &\cr\lbd_a=0 &\cr}\right.
\end{array}
\label{5.1.5}
\end{equation}
where $C=\|C_{\mu\nu}(z_\mu)\|_{\mu,\nu=0}^{n-1}$ is a St\"ackel
matrix, $\lbd_1,\ldots,\lbd_{n-1}$ are arbitrary parameters taking
values in some open domain $\Lambda \subset \R^{n-1}$,\ $I$ is the unit
$(m\times m)$-matrix,\ $\theta_\mu$ are arbitrary fixed real
constants. We say that the system of linear PDEs (\ref{5.1.4}) is
separable in the coordinate system $z_0(x),\ z_1(x),\ldots,
z_{n-1}(x)$ if there exist such a $(m\times m)$-matrix $A(x)$ and such
a St\"ackel matrix $C$ that substitution of the Ansatz\index{Ansatz}
\begin{equation}
   u(x)=A(x)\prod_{\mu=0}^{n-1}V_{\mu}(z_{\mu},\vec\lbd)\chi,
\label{5.1.6}
\end{equation}
where $ V_\mu(z_\mu),\ \mu={0,\ldots,n-1} $ are solutions of system of
ODEs (\ref{5.1.5}) and $\chi$ is an arbitrary $m$-component constant
column, into (\ref{5.1.4}) yields an identity with respect to $\vec
\lbd \in \Lambda $.  

Our aim is to solve the following mutually related problems:
\begin{itemize}
\item{to describe separable systems of PDEs (\ref{5.1.4}) in
    terms of the correspon\-ding St\"ackel matrices,}
\item{to establish a correspondence between separability of
    systems of PDEs and their symmetry properties.} 
\end{itemize}

Solution of the first problem is necessary for general
understanding of the mechanism of variable separation in systems of
linear PDEs and for classification of separable systems. Solving the
second problem we obtain a practical tool for finding coordinate
systems providing variable separation in a given system of linear
PDEs.

Before adducing the principal assertions we make an important
remark. It is readily seen that if a system of linear ODEs
(\ref{5.1.4}) is separable in a coordinate system $z_\mu=z_\mu(x)$,
then the equation 
\begin{displaymath}
      \{L_{\mu}^\prime (z)\p_{z_{\mu}}+M^\prime(x)\}w(z)=0,
\end{displaymath}
obtained from (\ref{5.1.4}) by means of the change of variables
\begin{displaymath}
   z_{\mu}=z_{\mu}(x),\quad w(z) = A^{-1}(x)u(x),
\end{displaymath}
is separable in the coordinate system  $z_\mu^\prime=z_\mu$ and what is
more, the solution with separated variables (\ref{5.1.6}) reads
\begin{displaymath}
w(z)=\prod_{\mu=0}^{n-1}V_{\mu}(z_{\mu},\vec\lbd)\chi.
\end{displaymath}

Consequently, when classifying separable systems we can consider 
separation in Cartesian coordinates $z_\mu=x_\mu$ only and 
choose $A(x)=I$. With this remark the solution with separated
variables (\ref{5.1.6}) takes the form
\begin{equation}
u(x)=\prod_{\mu=0}^{n-1}V_{\mu}(x_{\mu},\vec\lbd)\chi.
\label{5.1.7}
\end{equation}
{\bf Theorem 5.1.1.}\ {\em Equation (\ref{5.1.4}) is separable iff
there exists a St\"ackel matrix $C$ satisfying the condition}
\begin{equation}
L_{\mu}(x) C_{\mu\nu}(x_\mu)=-\delta_{\nu 0}M(x).
\label{5.1.8}
\end{equation}

\noindent
{\em Proof.}$\quad$ The necessity. Let system of PDEs (\ref{5.1.4}) be
separable. Then, according to Definition 5.1.3 there is such a
St\"ackel matrix $C$ that solutions $V_{\mu}(x_{\mu})$ of the matrix
system of ODEs
\begin{equation}
\begin{array}{l}
   {\edi dV_{\mu}\over \edi dx_{\mu}} =\Bigl(C_{\mu 0}(x_{\mu})
+C_{\mu a}(x_\mu)\lbd_a\Bigr)V_{\mu},\ \ \mu={0,\ldots,n-1},\\[2mm]
\left.\matrix{V_\mu \cr\cr\cr}\right.
\left|\matrix{&=I,\cr x_\mu=\theta_\mu &\cr\lbd_a=0 &\cr}\right.
\end{array}
\label{5.1.9}
\end{equation}
after being substituted into (\ref{5.1.7}) give rise to an exact
solution of the initial system of PDEs (\ref{5.1.4}) with an arbitrary
$\vec\lbd \in \Lambda$.

Inserting (\ref{5.1.7}) into (\ref{5.1.4}) with account of
(\ref{5.1.9}) we get
\begin{eqnarray}
&&L_0(C_{0 0}+C_{0 a}\lbd_a\Bigr)V_1V_2\cdot\cdots\cdot V_{n-1}
\chi\non\\ 
&&\quad +L_1V_0(C_{1 0}+C_{1 a}\lbd_a\Bigr)V_2V_3\cdot\cdots\cdot
V_{n-1}\chi 
+\ldots\label{5.1.10}\\
&&\quad +L_{n-1}V_0V_1\cdot\cdots\cdot V_{n-2}(C_{n-1 0}+C_{n-1
  a}\lbd_a\Bigr)\chi+M\chi =0.  \non
\end{eqnarray}

Using properties of the St\"ackel matrix $C$ it is not difficult to
prove that the matrices $A_\mu(x_\mu)=C_{\mu 0}(x_{\mu})
+C_{\mu a}(x_\mu)\lbd_a,\ \mu={0,\ldots,n-1}$ are mutually
commuting. 

Indeed,
\begin{eqnarray*}
[A_\mu,\, A_\nu]&=&[C_{\mu 0},\, C_{\nu 0}]
+\lbd_a\Bigl([C_{\mu 0},\, C_{\nu a}]+[C_{\mu a},\, C_{\nu 0}]\Bigr)\\ 
&&+ \lbd_a\lbd_b\Bigl([C_{\mu a},\, C_{\nu b}]+[C_{\mu b},\, C_{\nu
  a}]\Bigr)=0.\\ 
\end{eqnarray*}

Since $V_\mu(x_\mu)$ are solutions of the Cauchy problem
(\ref{5.1.9}), they can be represented by the following converging
series \cite{132}:
\begin{displaymath}
V_\mu=I+\int\limits_{\edi\theta_\mu}^{\edi x_\mu}A_\mu(\tau)d\tau+
\int\limits_{\edi\theta_\mu}^{\edi x_\mu}A_\mu(\tau)
\int\limits_{\edi\theta_\mu}^{\edi\tau}A_\mu(\tau_1)d\tau_1d\tau
+\ldots, \ \ \mu={0,\ldots,n-1}.
\end{displaymath}

Hence, it follows that $[A_\mu,\, V_\nu]=0$ under $\mu\ne\nu$. With
this fact relation (\ref{5.1.10}) is rewritten in the form
\begin{equation}
\Bigl(L_\mu(C_{\mu 0}+C_{\mu
  a}\lbd_a)+M\Bigr)\prod_{\mu=0}^{n-1}V_\mu\chi=0. 
\label{5.1.11}
\end{equation}

Since $\chi$ is an arbitrary $m$-component constant column and
matrices $V_\mu$ are invertible, the above equality is equivalent to
the following one:
\begin{displaymath}
L_\mu(C_{\mu 0}+C_{\mu a}\lbd_a)+M=0.
\end{displaymath}
Splitting the equality obtained with respect to $\lbd_a$ we arrive at
the conditions (\ref{5.1.8}). 

The sufficiency. Let $V_\mu$ be solutions of (\ref{5.1.9}) with
a St\"ackel matrix $C$ satisfying (\ref{5.1.8}). Inserting
the Ansatz (\ref{5.1.7}) into (\ref{5.1.4}) and taking into account
the relations $[A_\mu,\, V_\nu]=0,\ \mu\ne\nu$ we get the equality
(\ref{5.1.11}). Hence it follows that the function (\ref{5.1.7})
satisfies the initial system of PDEs (\ref{5.1.4}) identically with
respect to $\vec\lbd \in \Lambda$. The theorem is proved. $\rhd$

Let $B=\|B_{\mu\nu}(x) \|^{n-1}_{\mu,\nu=0} $ be the inverse of the
St\"ackel matrix $C=
\linebreak \|C_{\mu\nu}(x_{\mu})\|^{n-1}_{\mu,\nu=0} $, i.e.,\ 
\begin{displaymath}
  B_{\mu\alpha} C_{\alpha\nu}=C_{\mu\al}B_{\al\nu}=\delta_{\mu\nu}I.
\end{displaymath}
Then, multiplying (\ref{5.1.8}) by $B_{\nu\mu}$ on the right and
summing over $\nu$ we arrive at the following representation of the
matrices  $L_{\mu}$:
\begin{equation}
   L_{\mu}= -M B_{0\mu},\ \ \mu={0,\ldots,n-1}.
\label{5.1.12}
\end{equation}

Consequently, Theorem 5.1.1 admits an equivalent formulation:\
{\em the system of linear PDEs (\ref{5.1.4}) is separable iff the matrix
coefficients $L_{\mu},\ M$ are given in the St\"ackel form
(\ref{5.1.12})}. Thus, we have proved an analogue of the well-known
theorem about variable separation in PDEs with one dependent variable
\cite{136,156}.

Theorem 5.1.1 provides a description of separable systems of PDEs
via the corresponding St\"ackel matrices but it gives no method for
construction of solutions with separated variables for specific
equations. As stated above, the most effective method for separating
variables in systems of linear PDEs is utilization of their symmetry
properties. We will show that our definition of variable separation
in a system of PDEs is consistent with its symmetry properties.
Furthermore, we will obtain a simple description of a solution with
separated variables in terms of the first-order symmetry operators
of the system under consideration

First,we will prove an auxiliary lemma.
\vspace{1.5mm}

\noindent
{\bf Lemma 5.1.2}.\ {\em Let $\|B_{\mu\nu}(x)\|^{n-1}_{\mu,\nu=0}$ be 
a block nonsingular $(nm\times nm)$-matrix. The inverse of it is
designated as $\|H_{\mu\nu}(x)\|^{n-1}_{\mu,\nu=0}$. Then matrix
functions $ B_{\mu\nu}(x)$, $ B_{\mu}(x)$ satisfy the system of PDEs

\begin{eqnarray}
&1)& [B_{\mu\al},\, B_{\nu\beta}] + [B_{\mu\beta},\, B_{\nu\al}] =
0,\non\\ 
&2)& [B_{\mu\al},\, B_{\nu}] - [B_{\nu\al},\, B_{\mu}] 
+B_{\mu\beta}\p_{\beta}B_{\nu\al} -
B_{\nu\beta}\p_{\beta}B_{\mu\al} =0,\label{5.1.13}\\
&3)& B_{\mu\al}\p_{\al}B_{\nu} -
B_{\nu\al}\p_{\al}B_{\mu} + [B_{\mu},\, B_{\nu}]=0,\non
\end{eqnarray}
iff matrix functions $H_{\mu\nu}(x),\ H_{\mu}(x) = -
H_{\mu\nu}B_{\nu}$ satisfy the system of PDEs
\begin{eqnarray}
&1)& [H_{\mu\al},\, H_{\nu\beta}] + [H_{\mu\beta},\, H_{\nu\al}] =
0,\non\\ 
&2)& \p_{\nu}H_{\mu\al}- \p_{\mu} H_{\nu\al} + [H_{\mu\al},\,
   H_{\nu}] - [H_{\nu\al},\, H_{\mu}]=0,\label{5.1.14}\\
&3)& \p_{\nu}H_{\mu} - \p_{\mu}H_{\nu} + [H_{\mu},\, H_{\nu}]=0.\non
\end{eqnarray}

In (\ref{5.1.13}), (\ref{5.1.14}) subscripts $\mu,\ \nu,\ \al,\ \beta
$ take the values $0, 1, 2, \ldots, n-1 $. }
\vspace{1.5mm}

\noindent
{\em Proof.}$\quad$ Consider an over-determined system of PDEs
\begin{equation}
   (B_{\mu\nu}\p_{\nu} + B_{\mu})u =
     \lbd_{\mu}u,\ \ \mu={0,\ldots,n-1}.
\label{5.1.15}
\end{equation}

According to Theorem 1.5.3 the above system is compatible iff
conditions 
\begin{equation}
   [B_{\mu\al}\p_{\al} + B_{\mu},\,
    B_{\nu\beta}\p_{\beta} + B_{\nu}] = 0
\label{5.1.16}
\end{equation}
hold true. Computing commutators in the left-hand sides of
(\ref{5.1.16}) and equating to zero coefficients of the
linearly independent operators $\p_{\mu}\p_{\nu},\ \p_{\mu},\ I$ we
get equations (\ref{5.1.13}). Consequently, the system of PDEs
(\ref{5.1.13}) provides the necessary and sufficient compatibility
conditions for system (\ref{5.1.15}).

Next, multiplying both parts of (\ref{5.1.15}) by $H_{\al\mu}$ 
on the left and summing over $\mu$ we have
\begin{equation}
   \p_{\mu}u = H_{\mu\nu}(\lbd_{\nu}-B_{\nu})u.
\label{5.1.17}
\end{equation}

The compatibility criterion\
$\p_{\mu}(\p_{\nu}u)=\p_{\nu}(\p_{\mu}u)$ for the system
(\ref{5.1.17}) yields the identities
\begin{displaymath}
 \p_{\mu}\Bigl(H_{\nu\al}(\lbd_{\al}-B_{\al})u\Bigr)=
  \p_{\nu}\Bigl(H_{\mu\al}(\lbd_{\al}-B_{\al})u\Bigr),
\end{displaymath}
whence it follows that $(m\times m)$-matrices $ H_{\mu\nu}(x),\
H_{\mu}(x)=-H_{\mu\nu}B_{\nu}$ satisfy the system of PDEs
(\ref{5.1.14}). The lemma is proved. $\rhd$
\vspace{1.5mm}

\noindent
{\bf Theorem 5.1.2.}\ {\em Let the system of PDEs (\ref{5.1.4}) be
  separable. Then, a solution with separated variables $u(x)$ is a
  common eigenfunction of commuting first-order differential
  operators   $Q_1,\ Q_2, \ldots, Q_{n-1}$ which are symmetry
  operators of system   (\ref{5.1.4}).} 
\vspace{1.5mm}

\noindent
{\em Proof.}$\quad$  As earlier, we designate by the symbol
$B=\|B_{\mu\nu}(x)\|^{n-1}_{\mu,\nu=0}$ the inverse of the St\"ackel
matrix  $C=\|C_{\mu\nu}(x_\mu)\|^{n-1}_{\mu,\nu=0}$. Due to
the properties of the St\"ackel matrix $C$, the matrix functions
$H_{\mu\nu}=C_{\mu\nu}(x_{\mu}),\ H_{\mu}=0$ satisfy system
(\ref{5.1.14}). Hence it follows (Lemma 5.1.2) that the matrix
functions $B_{\mu\nu}(x),\ B_{\mu}(x) =0$ satisfy equations
(\ref{5.1.13}). Consequently, the operators
$Q_{\mu}=B_{\mu\nu}\p_{\nu}$ commute. 

By definition the solution with separated variables
$u(x)=\prod_{\mu=0}^{n-1}V_{\mu}(x_{\mu},\vec\lbd)\chi $ satisfies 
the system of PDEs 
\begin{equation}
   \p_{\mu}u = \Bigl(C_{0\mu}(x_{\mu}) + C_{\mu
     a}(x_{\mu})\lbd_a\Bigr)u. 
\label{5.1.18}
\end{equation}
Multiplying both parts of (\ref{5.1.18}) by $B_{\al\mu}(x) $ on the
left and summing over $\mu$ we obtain
\begin{equation}
B_{\al\mu}\p_{\mu} u = (\delta_{\al 0}I +
  \delta_{\al a}\lbd_a)u.
\label{5.1.19}
\end{equation}

Putting in (\ref{5.1.19}) $\alpha = 0, 1, 2, \ldots, n-1$ we arrive
at the relations 
\begin{equation}
\begin{array}{rcl}
  B_{0\mu}\p_{\mu} u&=&u,\\[2mm]
  B_{a\mu}\p_{\mu} u&=&\lbd_a u, \ \ a={1,\ldots,n-1}.
\end{array}
\label{5.1.20}
\end{equation}

But according to (\ref{5.1.12}) $B_{0\mu}=-M^{-1}L_{\mu},\ 
\mu={0,\ldots,n-1} $, whence 
\begin{displaymath}
   [B_{a\mu}\p_{\mu},\, M^{-1}L_{\mu}\p_{\mu} + I]=
   -[B_{a\mu}\p_{\mu},\, B_{0\mu}\p_{\mu}] =0.
\end{displaymath}

Now we will show that $Q_a$ are symmetry operators, which will
complete the proof. Indeed,
\begin{eqnarray*}
&&[Q_a,\, L_\mu \p_\mu +M]\equiv [Q_a,\, M(M^{-1}L_{\mu}\p_{\mu} +
I)]\\ 
&&\quad =M[Q_a,\, M^{-1}L_{\mu}\p_\mu + I]+[Q_a,\,
M](M^{-1}L_{\mu}\p_\mu + I)\\ 
&&\quad =(B_{a\mu}\p_\mu M+B_a M)M^{-1}(L_\mu \p_\mu +M)\equiv
R_a(x)(L_\mu \p_\mu +M),
\end{eqnarray*}
the same which is required. The theorem is proved. $\rhd$
\vspace{1.5mm}

\noindent 
{\bf Note 5.1.1.}\ A class of solutions with separated variables of a
given system of linear PDEs can be considerably extended if we
define these by formula (\ref{5.1.6}) without imposing
additional constraints on the matrix functions $V_{\mu}(z_\mu, \vec
\lbd)$. A peculiar example is the four-component complex-valued
function: 
\begin{equation}
\psi(\vec x)=\exp\{-i\lbd_1(\g_0+\g_4)x_1\}
\exp\Bigl\{-\Bigl(\lbd_2+(1/2)\g_0\g_4\Bigr)\ln x_2\Bigr\}\vp\,
(x_2/x_3), 
\label{5.1.21}
\end{equation}
which is a solution with separated variables in the coordinate
system $z_0=x_1,\ z_1=\ln x_2,\ z_2=x_2/x_3$ of the spinor equation:
\begin{equation}
\Bigl(\ve_{abc}\g_a\g_b\p_c+m/x_2+f(x_2/x_3)(\g_0+\g_4)\Bigr)\psi(\vec
x)=0, 
\label{5.1.22}
\end{equation} 
where $m=\mbox{\rm const}$,\ $f$ is an arbitrary real-valued function.

The function (\ref{5.1.21}) is a ``generalized'' eigenfunction of the
symmetry operators $Q_1=\p_1$,\ $Q_2=x_1\p_1+x_2\p_2+(1/2)\g_0\g_4$ of
the system of PDEs (\ref{5.1.22}) in a sense that it satisfies the
following equalities:
\begin{displaymath}
Q_1\psi=\lbd_1(\g_0+\g_4)\psi,\quad Q_2\psi=\lbd_2\psi,
\end{displaymath}
and what is more, the operators $Q_1$ and $Q_2$ do not commute.

However such a class of solutions with separated variables is too large to
be described by means of the classical symmetry of the equation under
study. To give a symmetry interpretation of these solutions it is
necessary to study {\em conditional
  symmetry}\index{Conditional!symmetry} of systems of linear PDEs
\cite{97,100}. Unlike the classical case, the determining
equations\index{Determining equations} for conditional symmetry
operators are nonlinear. By this reason, a systematic description of
solutions with separated variables (\ref{5.1.6}) without imposing
additional constraints on the form of functions $V_{\mu}(z_\mu, \vec
\lbd)$ seems to be impossible.

According to Theorem 5.1.2, a solution with separated variables in
the sense of Definition 5.1.3 has to be looked for as an
eigenfunction of some commuting symmetry operators of the equation
under study. Consequently, we can formulate the following symmetry
approach to the problem of variable separation in systems of linear
PDEs of the form (\ref{5.1.4}):
\begin{itemize}
\item{at the first step, the symmetry properties of (\ref{5.1.4})
    in the class ${\cal M}_1$ of the first-order differential
    operators with matrix coefficients are investigated;} 
\item{at the second step, the $(n-1)$-dimensional commutative
    subalgebras of the symmetry algebra are classified;}
\item{at the third step, a compatible over-determined system of PDEs}
\begin{equation}
\begin{array}{l}
   (L_{\mu}\p_{\mu} + M)u = 0,\\[2mm]
   Q_au = (B_{a\mu}(x)\p_{\mu} + B_a(x))u = \lbd_au,\ \
   a={1,\ldots,n-1}, 
\end{array}
\label{5.1.23}
\end{equation}
\item[{}]{where $Q_1,\ Q_2,\ldots,Q_{n-1}$ are commuting symmetry
    operators (Lie or non-Lie ones) of equation (\ref{5.1.4}), is
    transformed to a separated form}
\begin{equation}
   \p_{z_\mu}w =\Bigl(C_{\mu 0}(z_\mu)+C_{\mu a}(z_\mu)\lbd_a\Bigr)w,
\label{5.1.24}
\end{equation}
\item[{}]{by a proper change of variables}
\begin{equation}
   z_{\mu}=z_{\mu}(x),\quad w(z)=A^{-1}(x)u(x).
\label{5.1.25}
\end{equation}
\end{itemize}

If it is possible to implement the above three steps, then due
to Theorem 5.1.2 the initial system of PDEs (\ref{5.1.4}) is 
separable in coordinates $z_{\mu}=z_{\mu}(x),\ \mu={0,\ldots,n-1}$ and 
solution with separated variables has the form (\ref{5.1.6}), where
$V_{\mu}(z_{\mu}, \vec\lbd)$ are $(m\times m)$-matrices satisfying
systems of ODEs
\begin{displaymath}
 {\edi d V_{\mu}\over \edi d z_{\mu}}=\Bigl(C_{\mu 0}(z_\mu)
+C_{\mu a}(z_\mu)\lbd_a\Bigr)V_{\mu},\ \ \mu={0,\ldots,n-1}
\end{displaymath}
(no summation over $\mu$).

The most difficult problem to be solved in the framework of the above
approach is a choice of an appropriate change of variables
(\ref{5.1.25}). A regular method for finding such a change is known
only for the case, when operators $Q_a$ are Lie symmetry operators.
Otherwise, we have to solve a {\em nonlinear} \/problem in order to get
an explicit form of the ``new'' variables $z_\mu=z_\mu(x)$ and the
matrix function $A(x)$.
 
In the next two sections we will apply the approach suggested to some 
Galilei-invariant PDEs.
\vspace{10mm}

\noindent
{\large\bf 5.2. Separation of variables in the Galilei-invariant
\vspace{1.5mm}

\noindent
\phantom{\large\bf 5.2. }spinor equation\label{s5.2} }
\markboth{Chapter 5. SEPARATION OF VARIABLES}
   {5.2. Separation of variables in the Galilei-invariant spinor
     equation} 
\def\theequation{5.\arabic{section}.\arabic{equation}}
\setcounter {section} {2}
\setcounter {equation}{0}
\vspace{7mm}

\noindent
The problem of variable separation in the Dirac 
equation\index{Dirac!equation} (\ref{1.1.1}) was studied intensively 
by many researchers \cite{11,12,35,40,131,156,179}, a number of 
important results were obtained. Nevertheless, they
did not succeed in creating the complete theory (as it was the case
for the Hamilton-Jacobi equation) of variable separation in equation
(\ref{1.1.1}).

Analyzing the methods applied we come to the conclusion that the most
effec\-tive ones are those based on symmetry properties of the Dirac
equation.  V.N. Shapovalov and G.G. Ekle in \cite{179} described
solutions of the system of PDEs (\ref{1.1.1}) with separated variables
which were eigenfunctions of triplets of mutually commuting
first-order symmetry operators (a complete description of such
operators is given by Theorem 1.1.3). They have obtained 29
inequivalent ($P$(1,3) non-conjugate) triplets of mutually commuting
first-order symmetry operators, each one giving rise to a solution of
the Dirac equation with separated variables.

In addition, we can construct a solution with separated variables by using 
symmetry operators of the order higher than one. In particular, a
number of papers (see \cite{11,12} and references therein) are
devoted to the application of the second-order symmetry operators 
to variable separation in the Dirac equation.

At the same time, the problem of variable separation in spinor PDEs
invariant under the Galilei group has not been studied yet. In the
present section we will carry out separation of variables in the
system of linear PDEs for the spinor field (\ref{4.1.1}) by using its
Lie and non-Lie symmetry\index{Non-Lie!symmetry} described by Theorems
4.1.1, 4.1.3.

To apply the approach developed in the previous section we have,
first of all, to describe inequivalent triplets of mutually
commuting symmetry operators of equation (\ref{4.1.1}). To this
end, we need the following assertion.
\vspace{1.5mm}

\noindent
{\bf Theorem 5.2.1.}\ {\em Let $Q_1=Q_1^{(\ell)}+Q_1^{(n)} $,\
$Q_2=Q_2^{(\ell)} + Q_2^{(n)} $ be linear combinations of the
first-order symmetry operators of equation (\ref{4.1.1}) with real
coefficients and besides $ Q_1^{(\ell)},\ Q_2^{(\ell)}$ be linear
combinations of Lie symmetry operators and $Q_1^{(n)},\ Q_2^{(n)}$
be linear combinations of non-Lie ones. Let the operators $ Q_1,\ Q_2$
commute, then $C_1Q_1^{(n)} + C_2Q_2^{(n)} = 0$ with some non-vanishing
simultaneously real constants $C_1,\ C_2$.} 
\vspace{1.5mm}

{\em Proof.}$\quad$ The proof of the assertion demands very involved
computations, therefore only a general scheme of it will be given.

We declare the operators $t,\ x_a$ to be of the degree $+ 1$, the
operators $\p_t,\ \p_a,\ im$ to be of the degree $- 1$, the operators
$I$,\ $\g_{\mu}$ to be of the degree $0$. In addition, we assume that
the zero operator 0 has an arbitrary degree. With such assumptions the
set of the symmetry operators of equation (\ref{4.1.1}) separates into
the three classes 
\vspace{1.5mm}

\noindent
1)\ operators of the degree\ $-1$
\begin{displaymath}
P_0,\quad  P_a,\quad  W_0,\quad  W_a,\quad  S_a,\quad  T_a;
\end{displaymath}
2)\ operators of the degree\ $0$
\begin{displaymath}
J_{ab},\quad  G_a,\quad  D,\quad  M_1,\quad  M_2,\quad  R_0,\quad
R_a,\quad  N_0,\quad  N_a; 
\end{displaymath}
3)\ operators of the degree\ $+1$
\begin{displaymath}
A,\quad  K_a. 
\end{displaymath}

It is easy to see that the relation
\begin{equation}
   [Q(n),\, Q(\ell)]= Q(n+\ell),
\label{5.2.1}
\end{equation}
where $Q(k)$ is a symmetry operator of the degree $k$,
holds. Representing the operators  $Q_1,\ Q_2$ in the form
\begin{displaymath}
   Q_i=Q_i(-1)+Q_i(0)+Q_i(+1),\ \ i=1,2
\end{displaymath}
and using (\ref{5.2.1}) we get
\begin{eqnarray*}
[Q_1,\, Q_2]&=&[Q_1(-1),\, Q_2(-1)] + [Q_1(-1),\, Q_2(0)] 
  + [Q_1(0),\, Q_2(-1)]\\ 
& & + [Q_1(+1),\, Q_2(-1)] + [Q_1(0),\, Q_2(+1)] + [Q_1(+1),\,
Q_2(0)]\\  
& & + [Q_1(+1),\, Q_2(+1)] + [Q_1(0),\, Q_2(0)] + [Q_1(-1),\,
Q_2(+1)]\\ 
& & = Q(-2) +Q(-1) + Q(0)+ Q(+1) + Q(+2) = 0.
\end{eqnarray*}

From the above equalities we obtain the following relations:
\begin{displaymath}
Q(-2) =Q(-1) = Q(0)= Q(+1) = Q(+2) = 0.
\end{displaymath}

Consequently, $[Q_1(-1),\, Q_2(-1)]= 0$ or
\begin{eqnarray*}
& &[a_0^{(1)}W_0 + a_a^{(1)}W_a + b_a^{(1)}S_a + c_a^{(1)}T_a +
   d_0^{(1)}P_0 + d_a^{(1)}P_a,\\
& &\quad a_0^{(2)}W_0 + a_a^{(2)}W_a + b_a^{(2)}S_a + c_a^{(2)}T_a +
d_0^{(2)}P_0 + d_a^{(2)}P_a ] = 0.
\end{eqnarray*}

Computing the commutator in the left-hand side of the above equality 
and equating coefficients of linearly independent operators we
arrive at the conclusion that there exist such real constants $C_1,\
C_2$ that 
\begin{equation}
\begin{array}{l}
   C_1a_0^{(1)} + C_2a_0^{(2)} = 0,\quad
   C_1a_a^{(1)} + C_2a_a^{(2)} = 0,\\
   C_1b_a^{(1)} + C_2b_a^{(2)} = 0,\quad
   C_1c_a^{(1)} + C_2c_a^{(2)} = 0,
\end{array}
\label{5.2.2}
\end{equation}
where $a={1,2,3}$, and what is more $C_1^2 + C_2^2 \ne 0$ (without
loss of generality we may choose $C_2\ne 0$).

Due to (\ref{5.2.2}) the equality
\begin{displaymath}
   0 = Q(-1) = [Q_1(-1),\, Q_2(0) ] + [ Q_1(0),\, Q_2(-1)]
\end{displaymath}
takes the form
\begin{displaymath}
   [ Q_1^{(n)}(-1),\, C_1Q_1(0) + C_2Q_2(0) ] +
   [\al_0 P_0 + \al_a P_a,\, Q_1(0)]=0
\end{displaymath}
with some real constants $\al_0,\ \al_1,\ \al_2,\ \al_3 $.

Computing the commutator and equating coefficients of the
linearly-inde\-pen\-dent operators we arrive at the condition
\begin{displaymath}
   C_1Q_1^{(n)}(0) + C_2Q_2^{(n)}(0) = 0.
\end{displaymath}

Similarly,
\begin{displaymath}
C_1Q_1^{(n)}(+1)+C_2Q_2^{(n)}(+1) = 0.
\end{displaymath}

Thus, we have established that there exist such non-vanishing
simultaneously real numbers $C_1,\ C_2 $ that
\begin{eqnarray*}
  C_1Q_1^{(n)} + C_2Q_2^{(n)} &=&
  C_1\Bigl(Q_1^{(n)}(-1) + Q_1^{(n)}(0) + Q_1^{(n)}(+1)\Bigr)\\ 
  & &+ C_2\Bigl(Q_2^{(n)}(-1) +
  Q_2^{(n)}(0) + Q_2^{(n)}(+1)\Bigr) = 0.
\end{eqnarray*}

The theorem is proved. $\rhd$
\vspace{1.5mm}

\noindent
{\bf Note 5.2.1} \ As established in \cite{133,179} the above
assertion holds true for the first-order symmetry operators of the
Dirac equation.  

Theorem 5.2.1 simplifies substantially the problem of
classification of inequivalent triplets of the mutually commuting
symmetry operators of equation (\ref{4.1.1}). Since we look for a
solution with separated variables as a solution of over-determined
system of PDEs (\ref{5.1.23}), triplets of the symmetry operators $
\langle Q_1,\, Q_2,\, Q_3 \rangle $ and $ \langle Q_1,\, Q_2,\, C_1Q_1
+ C_2Q_2 + C_3Q_3 \rangle $ with $C_3\ne 0$ are equivalent. Hence, by
using Theorem 5.2.1, it follows that triplets of the mutually
commuting symmetry operators belong to one of the following classes:
\begin{equation}
\begin{array}{l}
{\rm I}.\ \langle Q_1^{(\ell)} + Q_1^{(n)},\, Q_2^{(\ell)},\,
 Q_3^{(\ell)} \rangle,\quad\\
{\rm II}.\ \langle Q_1^{(n)},\, Q_2^{(\ell)},\, Q_3^{(\ell)} \rangle,
\end{array}
\label{5.2.3}
\end{equation}
where we designate by the symbol $Q_a^{(\ell)}$ a linear combination
of the Lie symmetry operators and by the symbol $Q_1^{(n)}$ a linear
combination of the non-Lie symmetry operators.

By the arguments used while proving Theorem 5.2.1 we establish
that the operators $Q_1^{(\ell)} + Q_1^{(n)},\ Q_2^{(\ell)} $ and
$Q_3^{(\ell)}$ commute iff
\begin{eqnarray}
& & [Q_1^{(\ell)},\, Q_2^{(\ell)}]=[Q_1^{(n)},\, Q_2^{(\ell)}] =
0,\quad\non\\ 
& & [Q_1^{(\ell)},\, Q_3^{(\ell)}]=[Q_1^{(n)}, Q_3^{(\ell)}] =
0,\label{5.2.4}\\ 
& & [Q_2^{(\ell)},\, Q_3^{(\ell)}] = 0.\non
\end{eqnarray}

Consequently, to classify $G_2(1,3)$ inequivalent triplets of
commuting symmetry operators of equation (\ref{4.1.1}) we can make
use of the results of subalgebraic analysis of the Lie algebra of the
generalized Galilei group $G_2(1,3)$ which has been carried out in
\cite{15,66}.

According to \cite{15,66} there are 5 three-dimensional and 14
two-dimensi\-on\-al $G_2(1,3)$ non-conjugate commutative subalgebras of
the algebra $AG_2(1,3)$.  Solving for each of them equations
(\ref{5.2.4}) we get the following assertion.  
\vspace{1.5mm}

\noindent
{\bf Theorem 5.2.2.}\ {\em The list of $G_2(1,3)$ non-conjugate
  triplets of commuting first-order symmetry operators of
  equation (\ref{4.1.1}) is exhausted by the following ones:

\begin{eqnarray}
&
 1)&\langle G_1+\alpha P_0,\, P_2,\, P_3\rangle,
\non\\
&
 2)&\langle G_1+\alpha P_1,\, G_2,\, P_3\rangle,
\non\\
&
 3)&\langle G_1+\alpha P_1,\, G_2+\beta P_2,\, P_3\rangle,
\non\\
&
 4)& \langle J_{12},\,  P_0,\,  \al W_0 + \al_1 N_0 + \al_2 W_3 +
    \al_3 T_3 + \delta S_3 \rangle,
\non\\
&
 5)& \langle J_{12},\,  A+P_0,\,  \al W_3 + \beta N_0 +
    \delta(T_3 + K_3)\rangle,
\non\\
&
 6)& \langle J_{12},\,  D,\,  \al W_3 + \beta R_3 +
    \delta N_0\rangle,
\non\\
& 7)& \langle J_{12},\,  G_3+P_0,\,  \al W_3 + \beta (R_0 +
    T_3) + \delta S_3 \rangle,
\non\\
&
 8)& \langle J_{12},\,  P_3,\,  \al W_0 + \al_1 W_3 +\al_2T_3 +\al_3
 S_3 
 +\delta N_3 \rangle,
\non\\
&
 9)& \langle G_1,\,  P_2,\,  \al_a W_a + \beta N_2 + \delta S_1\rangle,
\non\\
&
 10)& \langle G_1+ P_2,\,  P_3,\,  \al_a W_a + \beta(2W_0 - N_3)
 +\delta S_1 \rangle,
\non\\
&
 11)& \langle P_0,\,  P_1,\,  \al W_0 + \al_a W_a +\beta_a T_a
 +\delta_a S_a \rangle,
\non\\
&
 12)& \langle P_1,\,  P_2,\,  \al W_0 + \al_a W_a +\beta_a T_a
 +\delta_a S_a \rangle,
\non\\
&
 13)& \langle J_{12} + P_3,\,  P_0,\,  \al W_0 + \al_1 W_3 +\al_2T_3
+\delta S_3 \rangle,
\non\\
&
 14)& \langle G_3 + P_0,\,  P_2,\,  \al_a W_a + \beta(T_1-N_2)
 +\delta S_3 \rangle,
\non\\
&
 15)& \langle G_1 + P_2,\,  J_{12}+A+P_0,\,  \al W_3 + \beta(T_3 + N_0 
 + K_3)   \rangle,
\non\\
&
 16)& \langle J_{12} + P_0,\,  P_3,\,  \al W_0 + \al_1 W_3 +\al_2T_3
 +\al_3 S_3 \rangle,
\non\\
&
 17)& \langle G_1 + P_1 +\al P_3,\,  G_2,\,  \al_a W_a +
   \beta[(1+\al^2)S_2 +N_2
\non\\
& & - \al R_0] + \delta(N_1+\al N_3)\rangle,
\label{5.2.5}\\
&
 18)& \langle G_1 + P_1 +\al P_3,\,  G_2+\beta P_3,\,  \al_3W_a 
+\delta(N_1
\non\\
& & - 2\beta W_0 - \beta^2 S_1 - \beta R_0 + \al\beta S_2
+\al N_3)  
\non\\
& &+\rho[N_2 - \beta S_3 -\al\beta S_1 -\al R_0+ (1+\al^2)S_2 - \beta
N_3] \rangle, 
\non\\
&
 19)& \langle P_0 + \al W_0 + \al_a W_a +\beta_aT_a +\delta_a S_a,\,
 P_1,\,  P_2 \rangle,
\non\\
&
 20)& \langle P_0,\,  P_1+\al W_0 + \al_a W_a +\beta_aT_a 
+\delta_a S_a,\,  P_2   \rangle,
\non\\
&
 21)& \langle P_0 + \al W_0 + \al_a W_a +\beta_aT_a +\delta_a S_a,\,
P_2,\,  P_3   \rangle,
\non\\
&
 22)& \langle G_1 + \al W_0 + \al_a W_a +\beta_aT_a +\delta_a S_a,\,
 P_2,\,  P_3   \rangle,
\non\\
&
 23)& \langle G_1,\,  P_2 + \al_a W_a + \beta N_3 
+\delta S_1,\,  P_3 \rangle,
\non\\
&
 24)& \langle G_1 + P_0 +\al W_0 + \al_a W_a +\beta_aT_a +\delta_a
 S_a,\,  
P_2,\,  P_3 \rangle,
\non\\
&
 25)& \langle G_1+P_0,\,  P_2 + \al_a W_a + \beta(T_2 - N_3) +\delta
 S_1,\,  
P_3 \rangle,
\non\\
&
 26)& \langle G_1+P_1 + \al_a W_a + \beta N_3 +\delta S_2,\,  G_2,\, 
P_3 \rangle,
\non\\
&
 27)& \langle G_1+P_1,\,  G_2 + \al_a W_a + \beta S_1,\, 
P_3 \rangle,
\non\\
&
 28)& \langle G_1+P_1,\,  G_2,\,  P_3 + \al_a W_a + \beta N_1 \rangle,
\non\\
&
 29)& \langle J_{12} + \al W_0 + \al_a W_a + \beta_a T_a +
 \delta_aS_a,\, P_0,\,  P_3 \rangle,
\non\\
&
 30)& \langle J_{12},\,  P_0 + \al W_0 + \al_1 W_3 + \al_2 T_3 +
\al_3 S_3 +\beta N_3,\,  P_3 \rangle,
\non\\
&
 31)& \langle J_{12},\,  P_0,\,  P_3 + \al W_0 + \al_1 N_0 + \al_2 W_3 +
\al_3 T_3 +\beta S_3 \rangle,
\non
\end{eqnarray}
where $\al,\ \al_a,\ \beta,\ \beta_a,\ \delta,\ \delta_a,\ \rho $ are
arbitrary real constants.}

We have not succeeded yet in relating each triplet from the list
(\ref{5.2.5}) to some coordinate system providing variable separation
in system of PDEs (\ref{4.1.1}) (so far it is not clear whether such a
relation exists). Another problem is that there exist different
triplets yielding the same coordinate system. For example, triplets 8
and 9 from (\ref{5.2.5}) give rise to solutions of (\ref{4.1.1}) with
separated variables in Cartesian coordinates $t,\ x_a$. In such a case
we adduce the most simple triplet of symmetry operators corresponding
to a given coordinate system.

We have obtained 16 coordinate systems providing variable separation
in equation (\ref{4.1.1}). As an example, we will consider a procedure
of variable separation in the case when all elements of the triplet
\begin{equation}
 Q_a = \xi_{a0}(t,\vec x)\p_t + \xi_{ab}(t,\vec x)\p_b + \eta_a(t,
 \vec x) 
\label{5.2.6}
\end{equation}
belong to the Lie algebra admitted by the equation under study.

Since the above operators commute, there exists such a
change of variables \cite{106.1}
\begin{equation}
\begin{array}{l}
   z_{\mu}=z_{\mu}(t, \vec x),\ \ \mu={0,\ldots,3},\\[2mm]
   \tilde\psi(z) = A(t,\vec x)\psi(t, \vec x),
\end{array}
\label{5.2.7}
\end{equation}
where $A(t,\, \vec x)$ is some invertible $(4\times 4)$-matrix, that
operators (\ref{5.2.6}) take the form $ Q_a = \p_{z_a}$. And what is
more due to Theorem 1.5.1 the initial equation (\ref{4.1.1})
on the set of solutions of the system of PDEs
\begin{equation}
   Q_a\tilde\psi = \lbd_a\tilde\psi
\label{5.2.8}
\end{equation}
is rewritten as follows
\begin{displaymath}
R_0(z_0)\tilde\psi_{z_0} + R_1(z_0;\, \lbd_1, \lbd_2,
\lbd_3)\tilde\psi = 0 
\end{displaymath}
with some matrices $R_0,\ R_1 $.

Thus, the system of PDEs (\ref{5.1.19}) rewritten in the new variables
$z_{\mu},\ \tilde\psi(z)$ takes the form
\begin{eqnarray*}
& &  R_0(z_0)\tilde\psi_{z_0} + R_1(z_0; 
  \lbd_1, \lbd_2, \lbd_3)\tilde\psi = 0,\\
& &  \tilde\psi_{z_a}=\lbd_a\tilde\psi,\ \ a={1,2,3}
\end{eqnarray*}
i.e.,\  the variables $z_{\mu}$ separate.

On integrating the above systems of ODEs and substituting the
result into (\ref{5.2.7}) we get the solution of equation
(\ref{4.1.1})  with separated variables.

Provided one element of the triplet of symmetry operators is a
non-Lie one, there is no general approach to the problem of
transforming system (\ref{5.1.23}) to the "separated" form
\begin{equation}
R_{1\mu}(z_{\mu})\tilde\psi_{z_{\mu}} + R_{2\mu}(z_{\mu};\,
\lbd_1, \lbd_2, \lbd_3)\tilde\psi = 0,\ \ \mu={0,\ldots,3},
\label{5.2.9}
\end{equation}
(no summation over $\mu$ is carried out), where $R_{1\mu},\ R_{2\mu}$
are some $(4\times 4)$-matrices. Each triplet containing non-Lie
symmetry operator demands specific and very involved computations. 

In the case considered, the problem is a little bit simplified
since two elements of the triplet are Lie symmetry operators.
Transforming these to the form $Q_a=\p_{z_a},\ a=1,2$ we get two new
variables $z_1(t,\, \vec x),\ z_2(t,\, \vec x)$. The third new
variable is always $z_0=t$. So it is necessary to guess 
the fourth variable $z_3=z_3(t,\, \vec x)$ and the 
$(4\times 4)$-matrix $A(t,\, \vec x)$
transforming the system of PDEs (\ref{4.1.1}) to a separated form
(\ref{5.2.9}).  Omitting details of derivation of the corresponding
formulae we present the final result: triplets of commuting symmetry
operators, coordinate systems providing variable separation and
corresponding systems of separated ODEs of the form (\ref{5.2.9}).
\index{Separation of variables!in the Galilei-invariant spinor equation}
\begin{eqnarray*}
& 
 1)& \langle P_0,\, P_1,\, P_2 \rangle, \\
& & A(t,\, \vec x)=I,\quad z_0=x_3,\quad z_1=t,\quad
z_2 = x_1,\quad z_3 = x_2, \\
& & \tilde \psi_{z_0} +\{\lbd_1\g_3(\g_0 + \g_4) - \lbd_2\g_3\g_1 -
\lbd_3\g_3\g_2 + im\g_3(\g_0 - \g_4)\}\tilde\psi=0, \\
& & \tilde \psi_{z_a} = \lbd_a\tilde \psi,\ \ a={1,2,3}; \\[3mm]
&
 2)& \langle J_{12},\, P_0,\, P_3 \rangle,
\\
& & A(t,\, \vec x)=\exp\{-(1/2)z_2\g_1\g_2\},\\
& & z_0=(x_1^2 + x_2^2)^{1/2}, \quad
z_1 =t,\quad z_2 = \arctan (x_2/x_1),\quad z_3 = x_3, \\
& & \tilde \psi_{z_0} +\{\lbd_1\g_1(\g_0 + \g_4) -
  \lbd_2z_0^{-1}\g_1\g_2 -\lbd_3\g_1\g_3 \\
& &\quad + im\g_1(\g_0 - \g_4) + (1/2)z_0^{-1}\}\tilde\psi=0, \quad
\tilde \psi_{z_a} = \lbd_a\tilde \psi,\ \ a={1,2,3}; \\[3mm]
&
 3)& \langle G_{1}+\alpha P_0,\, P_2,\, P_3 \rangle,
\\
& & A(t,\, \vec x)=\exp\{2imz_0z_1 + (i/3)\al m z_1^3 
+ (1/2)z_1\g_1(\g_0 + \g_4) \},\\
& & z_0= x_1 - t^2/2\al,\quad z_1 =t/\al,\quad z_2 =x_2,\quad z_3=
x_3, \\ 
& & \tilde \psi_{z_0} +\{\al^{-1}(\lbd_1-2imz_0)\g_1(\g_0 + \g_4)
-\lbd_2\g_1\g_2 - \lbd_3\g_1\g_3 \\ 
& &\quad + im\g_1(\g_0 + \g_4)\}\psi=0, \quad
\tilde \psi_{z_a} = \lbd_a\tilde \psi,\ \ a={1,2,3}; \\[3mm]
&
 4)& \langle G_1 + \al P_1,\, G_2,\, P_3 \rangle,
\\
& & A(t,\, \vec x)=\exp\{imz_0z_2^2 + (1/2)z_2(\g_0 + \g_4)\g_2 +
(1/2)z_1(\g_0 + \g_4)\g_1 \\ 
& &\quad  + im(z_0 + \al)z_1^2 \},\\
& &  z_0= t,\quad z_1 =x_1/(t+\al),\quad z_2 =x_2/t,\quad z_3= x_3, \\
& & -i(\g_0 + \g_4)\tilde\psi_{z_0} +\{-(1/2)
 (z_0 + \al)^{-1}(\g_0 + \g_4)+(i/2)z_0^{-1}(\g_0 + \g_4) \\
& &\quad +i\lbd_1(z_0+\al)^{-1}\g_1 + i\lbd_2z_0^{-1}\g_2 
+ i \lbd_3\g_3 + m(\g_0 - \g_4)\}\tilde\psi=0, \\
& & \tilde\psi_{z_a} = \lbd_a\tilde\psi,\ \ a={1,2,3};\\[3mm]
&
 5)& \langle G_{1}+\al P_1,\, G_2 + \beta P_2,\, G_3 \rangle,\\
& & A(t,\, \vec x)=\exp\{im[(z_0 + \al)z_1^2 + (z_0 + \beta) z_2^2 +
z_0z_3^2]\\ 
& &\quad +(1/2)(\g_0 + \g_4)\g_az_a\},\\
& &z_0=t,\quad z_1 =x_1/(t+\al),\quad z_2 =x_2/(t+\beta),\quad
z_3= x_3/t,\\ 
& & - i(\g_0 + \g_4)\tilde\psi_{z_0} +
\Bigl\{i\lbd_1(z_0+\al)^{-1}\g_1 + 
i\lbd_2(z_0+\beta)^{-1}\g_2 \\
& &\quad +i\lbd_3z_0^{-1}\g_3 +  (i/2)\Bigl((z_0 + \al)^{-1} + (z_0 + 
\beta)^{-1}+ z_0^{-1}\Bigr)(\g_0 + \g_4)\\
& &\quad + m(\g_0 - \g_4)\Bigr\}\tilde\psi =0, \quad
\tilde \psi_{z_a} = \lbd_a\tilde \psi,\ \  a={1,2,3}; \\[3mm]
&
 6)& \langle N_0 + \al W_0,\, J_{12},\, P_0 \rangle, \\
& & A(t,\, \vec x)=\exp\{-(1/2)\g_1\g_3 z_3 \}
\exp\{-(1/2)\g_1\g_2 z_2\}, \\
& & z_0=t,\quad z_1 =(x_1^2 + x_2^2 + x_3^2)^{1/2},\quad
z_2 = \arctan (x_2/x_1), \\
& & z_3 = \arctan [x_3(x_1^2 + x_2^2)^{-1/2}], \\
& &\tilde\psi_{z_0}=\lbd_1\tilde\psi,\quad \tilde\psi_{z_2}=
-\lbd_2\tilde\psi, \\
& &\tilde\psi_{z_1}= \Bigl\{\Bigl((\al/2) z^{-1}_1 -
\g_1\Bigr)\Bigl(\lbd_1(\g_0+\g_4) + im(\g_0-\g_4)\Bigr) \\
& &\quad +\lbd_3z_1^{-1}\g_0\g_4 - z_1^{-1}\Bigr\}\tilde\psi, \\
& &\tilde\psi_{z_3}= \Bigl\{(1/2)\tan z_3 + \lbd_2(\cos z_3)^{-1}
\g_2\g_3  -(\al/2)\Bigl(\lbd_1(\g_0+\g_4)\\ 
& &\quad - im(\g_0-\g_4)\Bigr)\g_2 - \lbd_3\g_2\Bigr\}\tilde\psi;
\\[3mm] 
&
 7)& \langle N_3 + \al W_3,\, J_{12},\, P_3 \rangle, \\
& & A(t,\, \vec x)=\exp\{imz_0z_1^2 + (1/2)(\g_0+\g_4)\g_1z_1 \}
\exp\{-(1/2)\g_1\g_2 z_2\}, \\
& & z_0=t,\quad z_1 =(x_1^2 + x_2^2)^{1/2}/t,\quad
z_2 = \arctan (x_2/x_1),\quad z_3=x_3,\\
& &z_0(\g_0+\g_4)\g_3\tilde\psi_{z_0} + \Bigl\{(\g_0+\g_4)\g_3  +
\lbd_3 z_0\g_0\g_4 + imz_0\g_3(\g_0-\g_4) \\
& &\quad +(\al/2)\Bigl(\lbd_3(\g_0 + \g_4) -2im\g_3\Bigr) -
\lbd_1\Bigr\}\tilde\psi=0, \\
& &\g_2\tilde\psi_{z_1} + \{(1/2)z_1^{-1}\g_1\g_2 +
\lbd_3z_1^{-1}\g_1\}\tilde\psi = 0,\\
& &\tilde\psi_{z_2}= -\lbd_2\tilde\psi,\quad
\tilde\psi_{z_3}= \lbd_3\tilde\psi; \\[3mm]
&
 8)& \langle G_1,\, P_2,\, N_2 + \ve S_1 \rangle,\ \ \ve=\pm1,
\\
& & A(t,\vec x)=\exp\{-\ve (\g_0+\g_4)z_3(1+z_0^2)^{-1/2} \}
\exp\{imz_0z_3^2+ (1/2)z_0z_3\\
& &\quad\times (1+z_0^2)^{-1/2}(\g_0+\g_4)\g_3\}
\exp\{-(\ve/2)\g_1\g_2\arctan z_0\}\\
& &\quad\times \exp\{imz_0z_1^2 + (1/2)z_1(\g_0 + \g_4)\g_1\},\\
& & z_0 =t,\quad z_1 =x_1/t,\quad
z_2 = x_2,\quad z_3=x_3(1+t^2)^{-1/2}, \\
& & (1+z_0^2)^{1/2}(\g_0 + \g_4)\g_1\tilde\psi_{z_0} +
 \{(1+z_0^2)^{1/2}(2z_0)^{-1}(\g_0 + \g_4)\g_1 \\
& &\quad + (\ve\lbd_1z_0^{-1} + \lbd_2z_0) \g_0\g_4 -
  im(1+z_0^2)^{1/2}(\g_0-\g_4)\g_1 -\lbd_3\}\tilde\psi = 0,\\
& &\tilde\psi_{z_1}= -\lbd_1\tilde\psi,\quad
\tilde\psi_{z_2}= \lbd_2\tilde\psi, \\
& &\tilde\psi_{z_3}+\{2imz_3\g_3 - (\lbd_1-\ve\lbd_2)\g_2\g_3
 - \lbd_3\g_2\}\tilde\psi = 0;\\[3mm]
&
 9)& \langle D,\, J_{12},\, N_0\rangle, \\
& & A(t,\, \vec x)=\exp\{(1/4)(\g_0+\g_4)\g_1\}
   \exp\{-(1/2)\g_0\g_4 z_0 +2\ln z_1 - 2z_0\}\\
& &\quad\times \exp\{-(1/2)\g_1\g_3 z_3\}
   \exp\{-(1/2)\g_1\g_2 z_2 \}, \\
& & z_0 =(1/2)\ln t,\quad z_1 =(x_1^2 + x_2^2 +
x_3^2)^{1/2}t^{-1/2},\\ 
& &  z_2 = \arctan (x_2/x_1),\quad
z_3 =\arctan [x_3(x_1^2 +x_2^2)^{-1/2}], \\
& &\tilde\psi_{z_0} =\lbd_1\tilde\psi,\quad
\tilde\psi_{z_2} = -\lbd_2\tilde\psi, \\
& &\tilde\psi_{z_1} = \{-(1/4)(\g_0 + \g_4)\g_1(1+2\lbd_1) -
im(\g_0 -\g_4)\g_1 +im\\
& &\quad  + (im/4)\g_1(\g_0 + \g_4) +
\lbd_3z_1^{-1}\g_4\g_0\}\tilde\psi,\\ 
& &\tilde\psi_{z_3} = \{\lbd_3\g_2 + (1/2)\tan z_3 +
\lbd_2(\cos z_3)\g_2\g_3\}\tilde\psi;\\[3mm]
& 10)&\langle D,\, J_{12},\, W_3 \rangle, \\
& & A(t,\, \vec x)= \exp\{-(1/2)\g_0\g_4 z_0\}
 \exp\{-(1/2)\g_1\g_2 z_2 \}, \\
& & z_0 =(1/2)\ln t,\quad z_1 =(x_1^2 + x_2^2)^{1/2}t^{-1/2},\quad
z_2 = \arctan (x_2/x_1),\\
& & z_3 = x_3 t^{-1/2},\\ 
& &\tilde\psi_{z_0} =\lbd_1\tilde\psi, \quad
\tilde\psi_{z_2} = -\lbd_2\tilde\psi, \\
& & \g_1(\g_0+ \g_4)\tilde\psi_{z_1} +\{(1/2)z_1^{-1}\g_1(\g_0 + \g_4)
- 
\lbd_2z_1^{-1}\g_2(\g_0 + \g_4) \\
& &\quad -2\lbd_3\g_3 - 2im\g_0 \g_4 \}\tilde\psi = 0, \\
& & (\g_0 + \g_4)\tilde\psi_{z_3} - \{2im\g_3 + 2\lbd_3\}\tilde\psi =
0;\\[3mm] 
&
 11)& \langle A+P_0,\, J_{12},\, N_0 \rangle, \\
& & A(t,\, \vec x)= \exp\{imz_1^2\tan z_0 - 2\ln(\cos z_0)
\}\Bigl\{\exp\{(1/2)\g_0\g_4 \\ 
& &\quad \times \ln(\cos z_0) \} + (1/2)z_1\sin z_0(\cos z_0)^{-1/2}
(\g_0 + \g_4)\g_1\Bigr\}\\
& &\quad\times\exp\{-(1/2)\g_1\g_3 z_3\}\exp\{-(1/2)\g_1\g_2 z_2 \},
\\ 
& & z_0 = \arctan  t,\quad z_1 = (x_1^2 + x_2^2 +
x_3^2)^{1/2}(1+t^2)^{-1/2},\\ 
& & z_2 = \arctan (x_2/x_1)\quad, z_3 = \arctan [x_3(x_1^2 +
x_2^2)^{-1/2}], \\ 
& &\tilde\psi_{z_0} =\lbd_1\tilde\psi, \quad
\tilde\psi_{z_2} = -\lbd_2\tilde\psi, \\
& & \tilde\psi_{z_1} =\Bigl\{-z_1^{-1} + \lbd_1(\g_0 + \g_4)\g_1 
- im\g_1(\g_0 - \g_4) + \lbd_1 z_1^{-1}\g_0\g_4\Bigr\}\tilde\psi,  \\
& & \tilde\psi_{z_3} = \{(1/2)\tan z_3 - \lbd_2(\cos z_3)^{-1}\g_2\g_3
+ \lbd_3\g_2\}\tilde\psi; \\[3mm]
&
 12)& \langle A+P_0,\, J_{12},\, W_3 \rangle, \\
& & A(t,\, \vec x)= \exp\{im(z_1^2 + z_3^2)\tan z_0 - 2\ln(\cos z_0)
\}\Bigl\{\exp\{(1/2)\g_0\g_4\\ 
& &\quad \times \ln(\cos z_0) \} + (1/2)(\g_1z_1 + \g_3z_3) 
\sin z_0(\cos z_0)^{-1/2}\Bigr\}\\
& &\quad\times\exp\{-(1/2)\g_1\g_2z_2 \},\\
& & z_0 =\arctan  t,\quad z_1 =(x_1^2 + x_2^2)^{1/2}(1+t^2)^{-1/2}, \\
& & z_2 = \arctan (x_2/x_1),\quad z_3 = x_3, \\
& &\tilde\psi_{z_0} =\lbd_1\tilde\psi, \quad
\tilde\psi_{z_2} = -\lbd_2\tilde\psi, \\
& &(\g_0+ \g_4)\g_2\tilde\psi_{z_1}  + \{(2z_1)^{-1}(\g_0 + \g_4)\g_2 +
\lbd_2z_1^{-1}(\g_0 + \g_4)\g_1   \\
& & \quad + 2(\lbd_3 - im\g_1 \g_2) \}\tilde\psi = 0, \\
& & (1/2)(\g_0 + \g_4)\tilde\psi_{z_3} -\{\lbd_3 + im\g_3\}\tilde\psi
= 0 ;\\[3mm] 
&
 13)& \langle G_3+P_0,\, J_{12},\, S_3 \rangle, \\
& & A(t,\, \vec x)= \exp\{2im[z_0z_3 +(1/6) z_0^3]
+ (1/2)z_0(\g_0 + \g_4)\g_3\\
& &\quad - (1/2)\g_1\g_2z_2\}, \\
& & z_0 = t,\quad z_1 =(x_1^2 + x_2^2)^{1/2},\quad z_2 = \arctan
(x_2/x_1),\\ 
& & z_3 = x_3-(1/2)t^2, \\
& &\tilde\psi_{z_0} =\lbd_1\tilde\psi, \quad
\tilde\psi_{z_2} = -\lbd_2\tilde\psi, \\
& & \tilde\psi_{z_1} = \{(2z_1)^{-1} -
   \lbd_2z_1^{-1}\g_1\g_2+ \lbd_3\g_2\}\tilde\psi, \\
& &\tilde\psi_{z_3} = \{\lbd_1(\g_0 + \g_4)\g_3 -
 im\g_3(\g_0 - \g_4)- 2im z_3(\g_0 + \g_4)\g_3 \\
& &\quad + \lbd_3\g_0 \g_4\})\tilde\psi;\\[3mm]
&
 14)& \langle J_{12}+P_3,\, P_0,\, S_3 \rangle, \\
& & A(t,\, \vec x)= \exp\{(1/2)\g_1\g_2(z_3 - z_2)\},\\
& &  z_0=t,\quad z_1 =(x_1^2 + x_2^2)^{1/2},\quad
z_2 = \arctan (x_2/x_1) + x_3,\quad z_3 = x_3, \\
& &\tilde\psi_{z_0} =\lbd_1\tilde\psi, \quad
\tilde\psi_{z_3} = \lbd_3\tilde\psi, \\
& & \tilde\psi_{z_1} = \{(2z_1)^{-1} - \lbd_2z_1^{-1}\g_1\g_2+
\lbd_3\g_2\}\tilde\psi, \\
& &\tilde\psi_{z_2} = \{-\lbd_2 +\lbd_3\g_0 \g_4 -
\lbd_1(\g_0+\g_4)\g_3- 
im\g_3(\g_0 - \g_4)\}\tilde\psi; \\[3mm]
&
 15)& \langle  J_{12}+P_0,\, P_3,\, W_0 \rangle, \\
& & A(t,\, \vec x)= \exp\{-(1/2)\g_1\g_2 z_2\},  \\
& &z_0=t,\quad z_1 =(x_1^2 + x_2^2)^{1/2},\quad
z_2 = t + \arctan (x_2/x_1),\quad z_3 = x_3, \\
& &\tilde\psi_{z_0} = \lbd_2\tilde\psi, \quad
\tilde\psi_{z_3} = \lbd_3\tilde\psi, \\
& &(\g_0+\g_4)\g_1\tilde\psi_{z_1} + \Bigl\{z_1^{-1}\g_2
\Bigl(-2\lbd_1 - \lbd_2
(\g_0+\g_4)+im(\g_0-\g_4)\Bigr) \\
& & \quad  + \Bigl(-(2z_1)^{-1}\g_1 + 2\lbd_1 -
\lbd_3\g_3\Bigr)(\g_0 + \g_4)\Bigr\}\tilde\psi = 0, \\
& & (\g_0 + \g_4)\tilde\psi_{z_2} + \{2\lbd_1 +
\lbd_2(\g_0 + \g_4) - im(\g_0 - \g_4)\}\tilde\psi = 0; \\[3mm]
&
 16)& \langle  G_1 + P_2,\, P_3,\, S_1 \rangle, \\
& & A(t,\, \vec x)= \exp\{imz_1^2z_0 + (z_1/2)\g_1(\g_0 + \g_4)\},\\
& & z_0=t,\quad z_1 =x_1/t,\quad z_2 = x_2 - x_1/t,\quad z_3 = x_3, \\ 
& & -i(\g_0+\g_4)\g_1\tilde\psi_{z_0} + \{-i(2z_0)^{-1}(\g_0 + \g_4)
 + i\lbd_1z_0^{-1}\g_1 + i\lbd_2 \g_2\g_3 \\
& & + m(\g_0 - \g_4)\}\tilde\psi = 0, \quad
\tilde\psi_{z_1} = \lbd_1\tilde\psi, \quad
\tilde\psi_{z_3} = \lbd_3\tilde\psi,\\ 
& & \tilde\psi_{z_2} = \{\lbd_2\g_3 + \lbd_3\g_2 \g_3\}\tilde\psi.
\end{eqnarray*}
In the above formulae $\alpha,\ \beta$ are arbitrary real parameters,
$\lbd_1,\ \lbd_2,\ \lbd_3$ are separation constants.

Note that coordinate systems given in the formulae 1--5 correspond
to the Lie symmetry of the system of PDEs (\ref{4.1.1}) and the ones given in
the formulae 6--16 correspond to its non-Lie symmetry.
\vspace{10mm}

\noindent
{\large\bf 5.3. Separation of variables in the Schr\"odinger
equation\label{s5.3}} 
\markboth{Chapter 5. SEPARATION OF VARIABLES}
   {5.3.  Separation of variables in the Schr\"odinger equation}
\def\theequation{5.\arabic{section}.\arabic{equation}}
\setcounter {section} {3}
\setcounter {equation}{0}
\vspace{7mm}

\noindent
The problem of separation of variables in the two-dimensional
Schr\"odinger equation\index{Schr\"odinger!equation}
\begin{equation}
i u_t+u_{x_1x_1}+u_{x_2x_2}=V(x_1,x_2)u
\label{5.3.1}
\end{equation}
as well as a majority of classical problems of mathematical physics
can be formulated in a very simple way (but this simplicity does not,
of course, imply existence of an easy way to its solution). To 
separate variables in equation (\ref{5.3.1}) we have to construct such
functions $R(t,\vec x), \ \omega _1(t,\vec x), \ \omega _2(t,\vec x)$
that the Schr\" odinger equation (\ref{5.3.1}) after being rewritten
in the new variables 
\begin{equation}
\begin{array}{l}
z_0=t, \ z_1=\omega _1(t,\vec x), \ z_2=\omega _2(t,\vec x),\\[1mm]
v(z_0,\vec z)=R(t,\vec x)u(t,\vec x)\\
\end{array}
\label{5.3.2}
\end{equation}
separates into three ordinary differential equations (ODEs) by means
of the substitution $v=\vp_0(z_0)\vp_1(z_1)\vp_2(z_2)$. From this
point of view the problem of separation of variables in equation
(\ref{5.3.1}) is studied in \cite{27.1,27.2,115,179.1}. 

But of no less importance is the problem of describing the
potentials $V(x_1,x_2)$ for which the Schr\" odinger equation admits
variable separation. Thus by a separation of
variables in equation (\ref{5.3.1}) we imply two mutually connected
problems. The first one is to describe all such functions $V(x_1,x_2)$
that the corresponding Schr\" odinger equation (\ref{5.3.1}) can be
separated into three ODEs in some coordinate system of the form
(\ref{5.3.2}) (classification problem). The second problem is to
construct for each function $V(x_1,x_2)$ obtained in this way all
coordinate systems (\ref{5.3.2}) enabling us to carry out separation
of variables in equation (\ref{5.3.1}).

As far as we know, the second problem has been solved provided $V=0$
\cite{27.2} and $V=\alpha x_1^{-2}+\beta x_2^{-2}$\cite{27.1}. The
first one was considered in a restricted sense in \cite{179.1}. Using
the symmetry approach to classification problem the authors obtained
some potentials providing separability of equation (\ref{5.3.1}) and
carried out separation of variables in the corresponding Schr\"
odinger equations.  But their results are far from being complete and
systematic. The necessary and sufficient conditions imposed on the
potential $V(x_1,x_2)$ by the requirement that the Schr\" odinger
equation admits symmetry operators of an arbitrary order are obtained
in \cite{159.1}.  But so far there is no systematic and exhaustive
description of potentials $V(x_1,x_2)$ providing separation of
variables in equation (\ref{5.3.1}).

To have a right to claim a description of {\em all} \/potentials and
{\em all} \/coordinate systems, which make it possible to separate the
 Schr\"odinger equation, it is necessary to have a definition of separation of
variables. It is natural to utilize Definition 5.1.3 adapted to the case of a
second-order PDE with one dependent variable. Consider the following system
of ODEs:
\begin{eqnarray}
i \edi {d\varphi _0\over dt}&=&U_0(t,\varphi _0;\lambda _1,
\lambda _2),\nonumber\\  
\edi {d^2\varphi _1\over d\omega _1^2}&=&U_1\Biggl(\omega _1,
\varphi _1, \edi {d\varphi _1\over d\omega _1};\lambda _1,\lambda
_2\Biggr),\label{5.3.3a}\\ 
\edi {d^2\varphi _2\over d\omega _2^2}&=&U_2\Biggl(\omega _2,
\varphi _2, \edi {d\varphi _2\over d\omega _2};
\lambda _1,\lambda _2\Biggr),\nonumber
\end{eqnarray}
where $U_0, \ U_1, \ U_2$ are some smooth functions of the
corresponding arguments, $\{\lambda _1, \lambda _2\} \subset \R^1$ are
arbitrary parameters (separation constants) and what is
more\index{Separation of variables}
\begin{equation}
{\rm rank}\, \|\partial U_\mu/\partial \lambda _a\|
_{\mu =0\;a=1}^{2\;\;\;\;\;\;2}=2
\label{5.3.3b}
\end{equation}
(the last condition ensures essential dependence of the corresponding
solution with separated variables on $\lambda _1, \ \lambda _2$, see
\cite{136}).
\vspace{1.5mm}

\noindent
{\bf Definition 5.3.1.}\ We say that equation (\ref{5.3.1}) admits a
separation of variables in the coordinate system $t, \ \omega
_1(t,\vec x),$ $ \ \omega _2(t,\vec x)$ if there exists such a function
$Q(t,\vec x)$ that substitution of the Ansatz
\begin{equation}
u=Q(t,\vec x)\varphi _0(t)\varphi _1\Bigl (\omega _1(t,\vec x)
\Bigr)\varphi_2\Bigl (\omega _2(t,\vec x)\Bigr )
\label{5.3.4}
\end{equation}
into (\ref{5.3.1}) with subsequent elimination of the derivatives
$\dot\vp_0, \ \ddot\vp_1, \ \ddot\vp_2$ according to equations
(\ref{5.3.3a}) yields an identity with 
respect to $\varphi _0$,\ $\varphi _1$,\ $\varphi _2$,\ $\dot\vp_1$,\ 
$\dot\vp_2$, $\lambda _1$,\ $\lambda _2$.
Thus, according to the above definition to separate variables in
equation (\ref{5.3.1}) we have
\begin{itemize}
\item{to substitute the expression (\ref{5.3.4}) into (\ref{5.3.1}),}
\item{to eliminate the derivatives $\dot\vp_0, \ \ddot\vp_1, \ 
    \ddot\vp_2$ with the help of equations (\ref{5.3.3a}),}
\item{to split the equality obtained with respect to the variables
    $\varphi _0$,\ $\varphi _1$,\ $\varphi _2$,\ $\dot\vp_1$,\ 
    $\dot\vp_2$, \ $\lambda _1$,\ $\lambda _2$ considered as
    independent.}
\end{itemize}

As a result, we get some over-determined system of PDEs for the
functions $Q(t,\vec x),$ \ $\omega _1(t,\vec x),$ $\ \omega _2(t,\vec
x)$.  On solving it we obtain a complete description of all coordinate
systems and potentials providing separation of variables in the Schr\"
odinger equation.  Naturally, the words {\em complete description}
\/make sense only within the framework of our definition. So if one
uses a more general definition it may be possible to construct new
coordinate systems and potentials providing separability of equation
(\ref{5.3.1}). But all solutions of the Schr\" odinger equation with
separated variables known to us fit into the scheme suggested by us
and can be obtained in the above described way.  
\vspace{2mm}

\noindent
{\bf 1. Classification of potentials {\boldmath $V(x _1,x _2)$}.}\
We do not adduce in full detail computations needed because
they are very cumbersome. We will restrict ourselves to 
pointing out main steps of the realization of the above 
suggested algorithm.

First of all we make a remark, which makes life a little
bit easier. It is readily seen that a substitution of the form
\begin{eqnarray} 
Q&\to &Q^\prime =Q\Psi _1(\omega _1)\Psi _2(\omega _2),\nonumber\\
\omega _a& \to &\omega _a^\prime =\Omega _a(\omega _a),\ \
a=1,2,\label{5.3.5}\\ 
\lambda _a&\to &\lambda _a^\prime =\Lambda _a(\lambda _1,
\lambda _2),\ \ a=1,2,\nonumber
\end{eqnarray} 
does not alter the structure of relations (\ref{5.3.3a}),
(\ref{5.3.3b}), (\ref{5.3.4}). That is why  we can introduce the
following equivalence relation: 
\begin{displaymath}
(\omega _1, \; \omega _2, \; Q) \ \sim \ (\omega ^\prime _1, \;
\omega ^\prime _2, \; Q^\prime ),
\end{displaymath}
provided (\ref{5.3.5}) holds with some $\Psi _a, \ \Omega _a, \
\Lambda _a$. 

Substituting (\ref{5.3.4}) into (\ref{5.3.1}) and excluding the
derivatives $\dot\vp_0, \ \ddot\vp_1,\ \ddot\vp_2$ with the use of
equations (\ref{5.3.3a}) we get
\begin{eqnarray*}
& &i (Q_t\varphi _0\varphi _1\varphi _2 
+QU _0\varphi _1\varphi _2
+Q\omega _{1t}\varphi _0\dot \varphi _1\varphi _2
+Q\omega _{2t}\varphi _0\varphi _1\dot \varphi _2)\\
& &\quad +(\triangle Q)\varphi _0\varphi _1\varphi _2
+2Q _{x_a}\omega _{1x_a}\varphi _0\dot \varphi _1\varphi _2
+2Q _{x_a}\omega _{2x_a}\varphi _0\varphi _1\dot \varphi _2\\
& &\quad +Q\Bigl ((\triangle \omega _1)\varphi _0\dot \varphi
_1\varphi _2 
+(\triangle \omega _2)\varphi _0\varphi _1\dot \varphi _2
+\omega _{1x_a}\omega_ {1x_a}\varphi _0U_1\varphi _2\\
& &\quad +\omega _{2x_a}\omega_ {2x_a}\varphi _0\varphi _1U_2
+2\omega _{1x_a}\omega_ {2x_a}\varphi _0\dot \varphi _1\dot 
\varphi _2\Bigr )
=VQ\varphi _0\varphi _1\varphi _2,
\end{eqnarray*}
where the summation over the repeated index $a$ from 1 to 2 is
understood, $\triangle =\partial _{x_1}^2+\partial _{x_2}^2$.

Splitting the equality obtained with respect to the independent variables
$\varphi _1$,\ $\varphi _2$,\ $\dot\vp_1$,\ $\dot\vp_2$,\ $\ 
\lambda _1, \ \lambda _2$ we conclude that ODEs (\ref{5.3.3a}) are
linear and up to the equivalence relation (\ref{5.3.5}) can be written
in the form
\begin{eqnarray*}
i \edi {d\varphi _0\over dt}&=&\Bigl (\lambda _1R_ 1(t)+
\lambda _2R_ 2(t) +R_ 0(t)\Bigr )\varphi _0,\\  
\edi {d^2\varphi _1\over d\omega _1^2}&=&\Bigl (\lambda
_1B_{11}(\omega _1) 
+\lambda _2B_{12}(\omega _1)+B_{01}(\omega _1)\Bigr )\varphi _1,\\
\edi {d^2\varphi _2\over d\omega _2^2}&=&\Bigl (\lambda
_1B_{21}(\omega _2) 
+\lambda _2B_{22}(\omega _2)+B_{02}(\omega _2)\Bigr )\varphi _2
\end{eqnarray*}
and what is more, functions $\omega _1, \ \omega _2, \ Q$ satisfy an 
over-determined system of nonlinear PDEs
\begin{eqnarray}
&1)&\omega _{1x_b}\omega _{2x_b}=0,\nonumber\\[2mm] 
&2)&B_{1a}(\omega _1)\omega _{1x_b}\omega _{1x_b}+
B_{2a}(\omega _2)\omega _{2x_b}\omega _{2x_b}+R_a(t)=0,
\nonumber\\[2mm] 
&3)&2\omega _{ax_b}Q_{x_b}+Q(i \omega _{at}+\triangle \omega _a)=0, 
\label{5.3.6}\\[2mm] 
&4)&\Bigl (B_{01}(\omega _1)\omega _{1x_b}\omega _{1x_b}+
B_{02}(\omega _1)\omega _{2x_b}\omega _{2x_b}\Bigr )Q+
i Q_t+\triangle Q\nonumber\\[2mm]
& &+R_0(t)Q-V(x _1,x _2)Q=0,\nonumber\
\end{eqnarray}
where $a,b=1,2$.

Thus, to solve the problem of separation of variables for the linear
Schr\" odin\-ger equation it is necessary to construct a general
solution of the system of nonlinear PDEs (\ref{5.3.6}). Roughly speaking,
to solve a linear equation we have to solve a system of {\em nonlinear
  equations}\/! This is the reason why so far there is no complete
description of all coordinate systems providing separability of the
four-dimensional d'Alembert equation \cite{155}.

However in the present case we have succeeded in integrating the system of
nonlinear PDEs (\ref{5.3.6}). Our approach to its integration is based 
on the following change of variables (hodograph transformation)
\begin{eqnarray*}
& &z_0=t, \ z_1=Z_1(t,\omega _1,\omega _2), \
z_2=Z_2(t,\omega _1,\omega _2),\\
& &v_1=x_1, \ v_2=x_2,
\end{eqnarray*}
where $z_0, \ z_1, \ z_2$ are new independent and $v_1, \ v_2$ are new
dependent variables, correspondingly.

Using the hodograph transformation determined above we have 
con\-struc\-ted the general solution of equations 1--3 from
(\ref{5.3.6}). It is given up to the equivalence relation
(\ref{5.3.5}) by one of the following 
formulae:
\begin{eqnarray}
&1)&\omega _1=A(t)x_1+W_1(t), \quad \omega
_2=B(t)x_2+W_2(t),\nonumber\\ 
& &Q(t,\vec x)=\exp \Bigl \{-(i/4)\Bigl ((\dot A/A)x_1^2+
(\dot B/B)x_2^2\Bigr )-(i/2)\Bigl 
((\dot W_1/A)x_1\non\\
& &\phantom{Q(t,\vec x)=} +(\dot W_2/B)x_2\Bigr )\Bigr \};\nonumber\\ 
&2)&x_1=W(t){\rm e}^{\omega_1} \sin \omega_2 + W_1(t), \quad
x_2=W(t){\rm e}^{\omega_1} \cos \omega_2+W_2(t),\non\\
& &Q(t,\vec x)=\exp \Bigl \{(i \dot W/4W)\Bigl ((x_1-W_1)^2
+(x_2-W_2)^2\Bigr )\non\\
& &\phantom{Q(t,\vec x)=}+(i/2)(\dot W_1x_1+\dot W_2x_2)\Bigr
\};\nonumber\\  
&3)&x_1=(1/2)W(t)(\omega _1^2-\omega _2^2)+W_1(t), \quad
x_2=W(t)\omega _1\omega _2+W_2(t),\label{5.3.7}\\
& &Q(t,\vec x)=\exp \Bigl \{(i \dot W/4W)\Bigl ((x_1-W_1)^2
+(x_2-W_2)^2\Bigr )\non\\
& &\phantom{Q(t,\vec x)=}+(i/2)(\dot W_1x_1+\dot W_2x_2)\Bigr
\};\nonumber\\ 
&4)&x_1=W(t)\cosh \omega _1\cos \omega _2+W_1(t), \quad
x_2=W(t)\sinh \omega _1\sin \omega _2+W_2(t),\nonumber\\
& &Q(t,\vec x)=\exp \Bigl \{(i \dot W/4W)\Bigl ((x_1-W_1)^2
+(x_2-W_2)^2\Bigr )\non\\
& &\phantom{Q(t,\vec x)=} +(i/2)(\dot W_1x_1+\dot W_2x_2)\Bigr
\}.\nonumber 
\end{eqnarray}

Here $A, \ B, \ W, \ W_1, \ W_2$ are arbitrary smooth functions of
$t$.  

Substituting the obtained expressions for the functions $Q, \ \omega
_1, \ \omega _2$ into the last equation from the system (\ref{5.3.6})
and splitting with respect to variables $x_1, \ x_2$ we get explicit
forms of potentials $V(x_1, x_2)$ and systems of nonlinear ODEs for
unknown functions $A(t), \ B(t), \ W(t), \ W_1(t), \ W_2(t)$. We have
succeeded in integrating these and in constructing all coordinate
systems providing the separation of variables in the initial equation
(\ref{5.3.1}) \cite{213.5}. Integration has been carried out up to the
equivalence relation which is introduced below in Notes 5.3.1--5.3.3.
\vspace{1.5mm}

\noindent
{\bf Note 5.3.1.}\ The Schr\"odinger equation with the potential
\begin{equation}
V(x_1, x_2)=k_1x_1+k_2x_2 + k_3+V_1(k_2x_1-k_1x_2),
\label{5.3.8z}
\end{equation}
where $k_1, \ k_2, \ k_3$ are constants, is transformed to 
the Schr\"odinger equation with the potential

\begin{equation}
V^\prime(x^{\prime}_1, x^{\prime}_2)=V_1 
(k_2x^{\prime}_1-k_1x^{\prime}_2)
\label{5.3.9z}
\end{equation}
by means of the following change of variables:
\begin{equation}
\begin{array}{l}
t^\prime =t, \quad \vec x^\prime =\vec x+t^2\vec k, \\[2mm]
u^\prime =u\exp \{(i/3)(k_1^2+k_2^2)t^3
+i t(k_1x_1+k_2x_2)+i k_3t \}.
\end{array}
\end{equation}

It is readily seen that the class of Ans\"atze (\ref{5.3.4}) is
transformed into itself by the above change of variables. That is why
potentials (\ref{5.3.8z}) and (\ref{5.3.9z}) are considered as
equivalent.  
\vspace{1.5mm}

\noindent
{\bf Note 5.3.2.}\ The Schr\"odinger equation with the potential  
\begin{equation}
V(x_1, x_2)=k(x_1^2+x_2^2) + V_1(x_1/x_2)(x_1^2+x_2^2)^{-1}
\label{5.3.8}
\end{equation}
with $k=\mbox{\rm const}$ is reduced to the Schr\" odinger equation
with the potential
\begin{equation}
V^\prime (x_1, x_2)=V _1(x^\prime _1/x^\prime _2)
(x_1^{\prime 2}+x_1^{\prime 2})^{-1}
\label{5.3.9}
\end{equation}
by means of the change of variables
\begin{displaymath}
t^\prime =\alpha (t), \quad \vec x^\prime =\beta (t)\vec x, \quad
u^\prime =u\exp \Bigl \{i \gamma (t)(x_1^2+x_2^2)+\delta (t)\Bigr \},
\end{displaymath}
where $\Bigl (\alpha (t), \; \beta (t), \; \gamma (t), \; \delta
(t)\Bigr )$ is an arbitrary solution of the system of ODEs 
\begin{eqnarray*}
& &\dot \gamma -4\gamma ^2=k, \quad \dot \beta -4\gamma \beta=0, 
\quad \dot \alpha -\beta ^2=0, \quad \dot \delta +4\gamma =0 
\end{eqnarray*}
such that $\beta \ne 0$.

Since the above change of variables does not alter the structure of
the Ansatz (\ref{5.3.4}), when classifying potentials $V(x_1, x_2)$
providing separability of the corresponding Schr\"odinger equation we
consider potentials (\ref{5.3.8}), (\ref{5.3.9}) as equivalent.
\vspace{1.5mm}

\noindent
{\bf Note 5.3.3.}\ It is well-known (see e.g. \cite{113.1,159.2}) that 
the general form of the invariance group admitted by equation 
(\ref{5.3.1}) is as follows:
\begin{eqnarray*}
& &t^\prime =F(t,\vec \theta ), \quad x_a^\prime =g_a(t, \vec x, \vec
\theta ), \ \ a=1,2,\\ 
& &u^\prime =h(t, \vec x, \vec \theta )u
+U(t,\vec x),
\end{eqnarray*}
where $\vec \theta =(\theta _1, \theta _2, \ldots,\theta _n)$ are
group parameters and $U(t,\vec x$) is an arbitrary solution of
equation (\ref{5.3.1}). 

The above transformations also do not alter the structure of the
Ansatz (\ref{5.3.4}). That is why  systems of coordinates $t ^\prime,
\ x_1^\prime, \ x_2^\prime $ and $t, \ x_1, \ x_2$ are considered as
equivalent.

Below we give without derivation a list of potentials $V(x_1, x_2)$
providing separability of the Schr\"odinger equation (\ref{5.3.1})
(some details can be found in \cite{213.5}).
\begin{enumerate}
\item{$V(x_1, x_2)=V_1(x_1)+V_2(x_2)$;}
\begin{enumerate}
\item{$V(x_1, x_2)=k_1x_1^2+k_2x_1^{-2}+V_2(x_2), \quad k_2\ne
    0$;} 
\begin{enumerate}
\item{$V(x_1,x_2)=k_1x_1^2+k_2x_2^2+k_3x_1^{-2}+k_4x_2^{-2}, \quad
    k_3k_4\ne 0$,}
\item[{}]{$k_1^2+k_2^2\ne 0, \ k_1\ne k_2$;} 
\item{$V(x_1, x_2)=k_1x_1^2+k_2x_1^{-2}, \quad k_1k_2\ne
    0$;} 
\item{$V(x_1, x_2)=k_1x_1^{-2}+k_2x_2^{-2}$;}
\end{enumerate}
\end{enumerate}
\begin{enumerate}
\item[{(b)}]{$V(x_1, x_2)=k_1x_1^2+V_2(x_2)$;}
\begin{enumerate}
\item{$V(x_1, x_2)=k_1x_1^2+k_2x_2^2+k_3x_2^{-2}, \quad
    k_1k_3\ne 0, \ k_1\ne k_2$;}
\item{$V(x_1, x_2)=k_1x_1^2+k_2x_2^2, \quad k_1k_2\ne 0, \
    k_1\ne k_2$;} 
\item{$V(x_1, x_2)=k_1x_1^2+k_2x_2^{-2}, \quad k_1\ne 0$;}
\end{enumerate}
\end{enumerate}
\item{$V(x_1, x_2)=V_1(x_1^2+x_2^2)+V_2\biggl (\edi {x_1\over
      x_2}\biggr ) (x_1^2+x_2^2)^{-1}$;}
\begin{enumerate}
\item{$V(x_1, x_2)=V_2\biggl (\edi {x_1\over x_2}\biggr
    )(x_1^2+x_2^2)^{-1}$;} 
\item{$V(x_1, x_2)=(x_1^2+x_2^2)^{-1/2}(k_1+k_2x_1x_2^{-2})
+k_3x_2^{-2}, \quad k_1^2+k_3^2\ne 0$;}
\end{enumerate}
\item{$V(x_1,x_2)=\Bigl (V_1(\omega _1)+V_2(\omega _2)\Bigr )
(\omega _1^2+\omega _2^2)^{-1}$,} 
\item[{}]{${\rm where} \quad \omega _1^2-\omega _2^2=2x_1, \
\omega _1\omega _2=x_2$;}
\item{$V(x_1, x_2)=\Bigl (V_1(\omega _1)+V_2(\omega _2)\Bigr )
(\sinh ^2\omega _1+\sin ^2\omega _2)^{-1}$,}
\item[{}]{${\rm where} \quad \cosh \omega _1\cos \omega _2=x_1, \
\sinh \omega _1\sin \omega _2=x_2$;}
\item{$V(x_1, x_2)=0$.}
\end{enumerate}

In the above formulae $V_1, \ V_2$ are arbitrary smooth functions,
$k_1, \ k_2, \ k_3, \ k_4$ are real arbitrary constants.

It should be emphasized that the above potentials are not inequivalent 
in a usual sense. These potentials differ from each other
by the fact that the coordinate systems providing separability of the
corresponding Schr\" odinger equations are different.  Moreover, in
some cases the form of coordinate systems depends essentially on the
signs of the parameters $k_j, \ j ={1,\ldots,4}$.

Next, we consider in detail separation of variables in the
Schr\"odinger equation with the anisotropic harmonic oscillator
potential $V(x_1, x_2)= k_1x_1^2+k_2x_2^2$ and the Coulomb potential
$V(x_1, x_2)= k_1(x_1^2+x_2^2)^{-1/2}$.  
\vspace{1.5mm}

\noindent
{\bf 2. Separation of variables in the Schr\"odinger equation with
the anisotropic harmonic oscillator and the Coulomb potentials.}\
Here we will obtain all coordinate systems providing separability 
of the Schr\"odinger equation  with the potential $V(x_1, x_2)=
k_1x_1^2+k_2x_2^2$
\begin{equation}
i u_t+u_{x_1x_1}+u_{x_2x_2}=(k_1x_1^2+k_2x_2^2)u.
\label{5.3.11}
\end{equation}

In the following, we consider the case $k_1\ne k_2$, because otherwise 
equation (\ref{5.3.1}) is reduced to the free Schr\" odinger equation
(see Note 5.3.2) which has been studied in detail in \cite{155}.

Explicit forms of the coordinate systems to be found depend
essentially on the signs of the parameters $k_1, \ k_2$. We consider
in some detail the case, when $k_1<0, \ k_2>0$ (the cases $k_1>0, \ 
k_2>0$ and $k_1<0, \ k_2<0$ are handled in an analogous way). This
means that equation (\ref{5.3.11}) can be written in the form
\begin{equation}
i u_t+u_{x_1x_1}+u_{x_2x_2}+(1/4)(a^2x_1^2-b^2x_2^2)u=0.
\label{5.3.12}
\end{equation}
where $a,\ b$ are arbitrary non-zero real constants (the factor 
$1/4$ is introduced for further convenience).
 
As stated above to describe all coordinate systems $t,\ \omega
_1(t,\vec x),\ \omega _2(t,\vec x)$ providing separability of equation
(\ref{5.3.11}) it is necessary to construct the general solution of
system (\ref{5.3.7}) with $V(x_1,x_2)= -(1/4)(a^2x_1^2-b^2x_2^2)$. The
general solution of equations 1--3 from (\ref{5.3.6}) splits into four
inequivalent classes listed in (\ref{5.3.7}). Analysis shows that only
solutions belonging to the first class can satisfy the fourth equation
of (\ref{5.3.6}).

Substituting the expressions for $\omega _1, \ \omega _2, \ Q$ given
by the formulae 1 from (\ref{5.3.7}) into the equation 4 from
(\ref{5.3.6}) with $V(x_1,x_2)=-(1/4)(a^2x_1^2-b^2x_2^2)$ and
splitting with respect to $x_1, \ x_2$ yield
\begin{eqnarray}
& &B_{01}(\omega _1)=\alpha _1\omega _1^2+\alpha _2\omega _1, \quad
B_{02}(\omega _2)=\beta _1\omega _2^2+\beta _2\omega _2,
\nonumber\\[2mm] 
& &(\dot A/A)^{^{\edi.}} -(\dot A/A)^2
-4\alpha _1A^4+a^2=0,\label{5.3.13a} \\
& &(\dot B/B)^{^{\edi.}}-(\dot B/B)^2
-4\beta _1B^4-b^2=0,\label{5.3.13b} \\
& &\ddot \theta _1-2\dot \theta _1(\dot A/
A)-2(2\alpha_1\theta _1+\alpha _2)A^4=0,\label{5.3.13c} \\
& &\ddot \theta _2-2\dot \theta _2(\dot B/
B)-2(2\beta_1\theta _2+\beta _2)B^4=0.\label{5.3.13d} 
\end{eqnarray}

Here $\alpha _1, \ \alpha _2, \ \beta _1, \ \beta _2$ are arbitrary
real constants.

Evidently, equations (\ref{5.3.13a})--(\ref{5.3.13d}) can be rewritten  
in the following unified form:
\begin{equation}
(\dot y/y)^{^{\edi.}}-(\dot y/y)^2-4\alpha y^4=k,\quad
\ddot z-2\dot z(\dot y/y)-2(2\alpha z+\beta )y^4=0.
\label{A.1}
\end{equation}

Provided $k=-a^2<0$, system (\ref{A.1}) coincides with equations
(\ref{5.3.13a}), (\ref{5.3.13c}) and under $k=b^2>0$ with equations
(\ref{5.3.13b}), (\ref{5.3.13d}).

First of all, we note that the function $z=z(t)$ is determined up
to addition of an arbitrary constant. Indeed, the coordinate
functions $\omega _a$ have the following structure:
\begin{displaymath}
\omega _a=yx_a+z,\ \ a=1,2.
\end{displaymath}
But the coordinate system $t, \ \omega _1, \ \omega _2$ is equivalent 
to the coordinate system $t, \ \omega _1+C_1, \ \omega _2+C_2, \
C_a\in \R^1$. Hence it follows that the function $z(t)$ is equivalent
to the function $z(t)+C$ with arbitrary real constant
$C$. Consequently, provided $\alpha \ne 0$, we can choose in
(\ref{A.1}) $\beta =0$. 
\vspace{1.5mm}

\noindent
{\bf Case 1.}\ $\alpha =0$

On making in (\ref{A.1}) the change of variables
\begin{equation}
w=\dot y/y, \quad v=z/y
\label{A.2}
\end{equation}
we get
\begin{equation}
\dot w=w^2+k, \quad \ddot v+kv=2\beta y^3.
\label{A.3}
\end{equation}

First, we consider the case $k=-a^2<0$. Then, the general solution
of the first equation from (\ref{A.3}) is given by one of the
formulae 
$$w=-a\coth a(t+C_1), \quad  w=-a\tanh a(t+C_1), \quad w=\pm a, \ \
C_1\in \R^1,$$ 
whence
\begin{equation}
\begin{array}{l}
y(t)=C_2\sinh ^{-1} a(t+C_1), \quad y(t)=C_2\cosh ^{-1}
a(t+C_1),\\[2mm]  
y(t)=C _2\exp (\pm at), \ \ C_2\in \R^1.
\end{array}
\label{A.4}
\end{equation}

The second equation of system (\ref{A.3}) is a linear inhomogeneous
ODE. We substitute its general solution into (\ref{A.2})
and get the following expressions for $z(t)$:
\begin{eqnarray}
z(t)&=&(C_3\cosh at+C_4\sinh at)\sinh ^{-1}a(t+C_1)\non\\
&& +(\beta C_2^4/a^2)\sinh ^{-2} a(t+C_1),\non \\
z(t)&=&(C_3\cosh at+C_4\sinh at)\cosh ^{-1}a(t+C_1)\label{A.5}\\
&& +(\beta C_2^4/a^2)\cosh ^{-2} a(t+C_1),\non \\
z(t)&=&(C_3\cosh at+C_4\sinh at)\exp (\pm at)\non\\
&& +(\beta C _2^4/4a^2)\exp (\pm 4at),\non 
\end{eqnarray}
where $\{C_3, C_4\} \subset \R^1$.

The case $k=b^2>0$ is treated in a similar way, the general
solution of (\ref{A.1}) being given by the formulae
\begin{eqnarray}
y(t)&=&D_2\sin ^{-1}b(t+D_1),\non\\
z(t)&=&(D_3\cos bt+D_4\sin bt)\sin ^{-1}b(t+D_1)\label{A.6}\\
& &+(\beta D_2^4/b^2)\sin ^{-2} b(t+D_1),\non
\end{eqnarray}
where $D_1, \ D_2, \ D_3, \ D_4$ are arbitrary real constants.
\vspace{1.5mm}

\noindent
{\bf Case 2.}\ $\alpha \ne 0, \ \beta =0$

On making in (\ref{A.1}) the change of variables
\begin{displaymath}
y=\exp w, \quad v=z/y
\end{displaymath}
we have

\begin{equation}
\ddot w-\dot w^2=k+\alpha \exp 4w, \quad  \ddot v +kv=0.
\label{A.7}
\end{equation}

The first ODE from (\ref{A.7}) is reduced to the first-order linear
ODE 
\begin{displaymath}
(1/2){dp(w)\over dw}-p(w)=k+\alpha \exp 4w
\end{displaymath}
by the substitution $\dot w=[p(w)]^{1/2}$, whence
\begin{displaymath}
p(w)=\alpha \exp 4w +\gamma \exp 2w -k, \ \ \gamma \in \R^1.
\end{displaymath}

The equation $\dot w=[p(w)]^{1/2}$ has a singular solution
$w=C=\mbox{\rm const}$ such that $p(C)=0$. If $\dot w \ne 0$, then
integrating the equation $\dot w=p(w)$ and returning to the initial
variable $y$ we get
\begin{displaymath}
\int\limits^{\edi y(t)}\tau^{-1}(\alpha \tau ^4+\gamma \tau ^2-k)
^{-1/2} d\tau = t +C_1.
\end{displaymath}

Taking the integral in the left-hand side of the above equality we 
obtain the general solution of the first ODE from (\ref{A.1}). It is
given by the following formulae:
\vspace{1.5mm}

\noindent
\underline {under $k=-a^2<0$}
\begin{equation}
\begin {array} {rcl} 
y(t)&=&C_2\Bigl (\alpha +\sinh 2a(t+C_1)\Bigr )^{-1/2},\\[2mm]
y(t)&=&C_2\Bigl (\alpha +\cosh 2a(t+C_1)\Bigr )^{-1/2},\\[2mm]
y(t)&=&C_2\Bigl (\alpha +\exp (\pm 2at)\Bigr )^{-1/2},
\end{array}
\label{A.8}
\end{equation}
\underline {under $k=b^2>0$}
\begin{eqnarray}
y(t)&=&D_2\Bigl (\alpha +\sin 2b(t+D_1)\Bigr )^{-1/2}.
\label{A.9}
\end{eqnarray}

Here $C_1, \ C_2, \ D_1, \ D_2$ are arbitrary real constants.

Integrating the second ODE from (\ref{A.7}) and returning to the
initial variable $z$ we have
\vspace{1.5mm}

\noindent
\underline {under $k=-a^2<0$}
\begin{eqnarray}
z(t)&=&y(t)(C_3\cosh at+C_4\sinh at),
\label{A.10}
\end{eqnarray}
\underline {under $k=b^2>0$}
\begin{eqnarray}
z(t)&=&y(t)(D_3\cos bt+D_4\sin bt),
\label{A.11}
\end{eqnarray}
where $C_3, \ C_4, \ D_3, \ D_4$ are arbitrary real constants.

Thus, we have constructed the general solution of the system
of nonlinear ODEs (\ref{A.1}) which is given by the formulae
(\ref{A.5})--(\ref{A.11}). 

Substitution of the formulae (\ref{A.2}), (\ref{A.4})--(\ref{A.6}),
(\ref{A.8})--(\ref{A.11}) into the corresponding expressions 1 from
(\ref{5.3.7}) yields a complete list of coordinate systems providing
separability of the Schr\" odinger equation (\ref{5.3.12}). These
systems can be transformed to canonical form if we use Note 5.3.3.

The invariance group of equation (\ref{5.3.12}) is generated by the
following basis operators \cite{213.4}:
\begin{eqnarray}
& &P_0=\partial _t, \quad I=u\partial _u, \quad M=i u\partial _u, 
\quad Q_{\infty}=U(t,\vec x)\partial _u, \nonumber\\
& &P_1=\cosh at \, \partial _{x_1}+(i a/2)
(x_1\sinh at)u\partial _u,\nonumber\\
& &P_2=\cos bt \, \partial _{x_2}-(i b/2)
(x_2\sin bt)u\partial _u,\label{5.3.14} \\
& &G_1=\sinh at \, \partial _{x_1}+(i a/2)
(x_1\cosh at)u\partial _u,\nonumber\\
& &G_2=\sin bt \, \partial _{x_2}+(i b/2)
(x_2\cos bt)u\partial _u,\nonumber
\end{eqnarray}
where $U(t, \vec x)$ is an arbitrary solution of equation
(\ref{5.3.12}). 

Making use of the finite transformations generated by the infinitesimal
operators (\ref{5.3.14}) and Note 5.3.3 we can choose in the
formulae (\ref{A.4})--(\ref{A.6}), (\ref{A.8}), (\ref{A.10}),
(\ref{A.11}) $C_3=C_4=D_1=0, \ D_3=D_4=0, \ C_2=D_2=1$. As a result,
we come to the following assertion.
\index{Separation of variables!in the Schr\"odinger equation}
\vspace{1.5mm}

\noindent
{\bf Theorem 5.3.1.}\ {\em The Schr\"odinger equation (\ref{5.3.12})
  admits separation of variables in 21 inequivalent coordinate systems
  of the form 
\begin{equation}
\omega _0=t, \quad \omega _1=\omega _1(t, \vec x), \quad
\omega _2=\omega _2(t, \vec x),
\label{5.3.15}
\end{equation}
where $\omega _1$ is given by one of the formulae from the first and $ 
\omega_2$ by one of the formulae from the second column of the
Table 5.3.1.}

There is no necessity to consider specially the case when in
(\ref{5.3.11}) $k_1>0, \ k_2<0$, since such an equation by the change
of independent variables $u(t,x_1,x_2)\to u(t,x_2,x_1)$ is reduced to
equation (\ref{5.3.12}).

Below we adduce without proof the assertions describing coordinate
systems providing separation of variables in equation (\ref{5.3.11})
with $k_1<0, \ k_2<0$ and $k_1>0, \ k_2>0$ and in the Schr\"odinger
equation with the Coulomb potential $k_1(x_1^2+x_2^2)^{-1/2}$.  
\vspace{2mm}

\noindent
{\em Table 5.3.1.}\ {\bf Coordinate systems providing
separability}\\
\phantom{{\em Table 5.3.1.}\ }{\bf of the Schr\"odinger
equation (\ref{5.3.12})} 
\vskip 1.5mm 

\noindent
\begin{tabular}{|c||c|}\hline
 & \\ 
$\omega_1(t,\vec x)$ & $\omega_2(t,\vec x)$\\ 
 & \\ \hline 
 & \\ 
$x_1\Bigl (\sinh a(t+C)\Bigl )^{-1}+
\alpha \Bigl (\sinh a(t+C)\Bigr )^{-2}$ & $x_2(\sin bt)^{-1}+\beta
(\sin bt)^{-2}$\\ 
$x_1\Bigl (\cosh a(t+C)\Bigl )^{-1}+
\alpha \Bigl (\cosh a(t+C)\Bigr )^{-2}$ & $x_2(\beta +\sin
2bt)^{-1/2}$\\ 
$x_1\exp (\pm at)+ \alpha \exp (\pm 4at)$ & $x_2$ \\ 
$x_1\Bigl (\alpha +\sinh 2a(t+C)\Bigr )^{-1/2}$ & \\ 
$x_1\Bigl (\alpha +\cosh 2a(t+C)\Bigr )^{-1/2}$ & \\ 
$x_1\Bigl (\alpha +\exp (\pm 2at)\Bigr )^{-1/2}$ & \\ 
$x_1$ & \\ 
 & \\ \hline
\end{tabular}
\vspace{1.5mm}

\noindent
Here $C, \ \alpha, \ \beta $ are arbitrary real
constants. 
\vspace{2mm}

\noindent
{\bf Theorem 5.3.2.}\ {\em The Schr\" odinger equation  
\begin{equation}
i u_t+u_{x_1x_1}+u_{x_2x_2}+ (1/4)(a^2x_1^2+b^2x_2^2)u = 0
\label{5.3.18}
\end{equation}
with $a^2\ne 4b^2$ admits separation of variables in 49 inequivalent
coordinate systems of the form (\ref{5.3.15}), where $\omega_1$ is
given by one of the formulae from the first and $\omega_2$ by one of
the formulae from the second column of the Table 5.3.2. Provided
$a^2=4b^2$, one more coordinate system should be included into the
above list, namely,}
\begin{equation}
\omega_0=t, \quad \omega_1^2-\omega_2^2 =2x_1, \quad \omega
_1\omega_2=x_2.
\label{5.3.19}
\end{equation}
{\bf Theorem 5.3.3.}\ {\em The Schr\" odinger equation
\begin{equation}
i u_t+u_{x_1x_1}+u_{x_2x_2}-(1/4)(a^2x_1^2+b^2x_2^2)u=0
\label{5.3.20}
\end{equation}
with $a^2\ne 4b^2$ admits separation of variables in 9 inequivalent
coordinate systems of the form (\ref{5.3.15}), where $\omega _1$ is
given by one of the formulae from the first and $\omega_2$ by one of
the formulae from the second column of the Table 5.3.3. Provided
$a^2=4b^2$, the above list should be supplemented by the coordinate
system (\ref{5.3.19}).} 
\vspace{2mm}

\noindent
{\em Table 5.3.2.}\ {\bf Coordinate systems providing
separability}\\
\phantom{{\em Table 5.3.2.}\ }{\bf of the Schr\"odinger
equation (\ref{5.3.18})} 
\vspace{1.5mm} 

\noindent
\begin{tabular}{|c||c|}\hline
 & \\
$\omega_1(t,\vec x)$ & $\omega_2(t,\vec x)$\\
 & \\ \hline
 & \\
$x_1\Bigl (\sinh a(t+C)\Bigl )^{-1}+
\alpha \Bigl (\sinh a(t+C)\Bigr )^{-2}$ & $x_2(\sinh bt)^{-1}+\beta
(\sinh bt)^{-2} $\\ 
$x_1\Bigl (\cosh a(t+C)\Bigl )^{-1}+
\alpha \Bigl (\cosh a(t+C)\Bigr )^{-2}$ & $x_2(\cosh bt)^{-1}+\beta
(\cosh bt)^{-2}$\\ 
$x_1\exp (\pm at)+
\alpha \exp (\pm 4at)$ & $x_2\exp (\pm bt)+\beta \exp (\pm  4bt)$ \\ 
$x_1\Bigl (\alpha +\sinh 2a(t+C)\Bigr )^{-1/2}$ &$x_2(\beta +\sinh
2bt)^{-1/2}$ \\  
$x_1\Bigl (\alpha +\cosh 2a(t+C)\Bigr )^{-1/2}$ &$x_2(\beta +\cosh
2bt)^{-1/2}$ \\  
$x_1\Bigl (\alpha +\exp (\pm 2at)\Bigr )^{-1/2}$ &$x_2\Bigl (\beta
+\exp (\pm 2bt)\Bigr )^{-1/2}$ \\  
$x_1$ & $x_2$\\
 & \\ \hline
\end{tabular}
\vspace{1.5mm}

\noindent
Here $C,\ \alpha,\ \beta$ are arbitrary constants.
\vspace{2mm}

\noindent
{\bf Theorem 5.3.4.}\ {\em The Schr\" odinger equation with the
  Coulomb potential  
$$
i u_t + u_{x_1x_1}+u_{x_2x_2} - k_1(x_1^2+x_2^2)^{-1/2}u = 0
$$
admits separation of variables in two coordinate systems. One of them
is the polar coordinate system
\begin{displaymath}
t=\omega_0,\quad x_1 = {\rm e}^{\omega_1}\sin \omega_2, \quad
x_2 = {\rm e}^{\omega_1}\cos \omega _2
\end{displaymath}
and another is the parabolic coordinate system (\ref{5.3.19}).}

It is important to note that explicit forms of coordinate systems
providing separability of equations (\ref{5.3.12}), (\ref{5.3.18}),
(\ref{5.3.20}) depend essentially on the parameters $a, \ b$ contained
in the potential $V(x_1,x_2)$.  It means that the free Schr\"odinger
equation ($V=0$) does not admit separation of variables in such
coordinate systems. Consequently, they are essentially new.
\vspace{2mm}

\noindent
{\bf 3. Conclusion.}\
In the present section we have studied the case when the
Schr\"odinger equation (\ref{5.3.1}) separates into one first-order
and two second-order ODEs. It is not difficult to prove
that there are no functions $Q(t,\vec x), 
\linebreak \omega _{\mu}(t,\vec x), \ \mu ={0,\ldots,2}$ such that
the Ansatz 
\begin{displaymath}
u=Q(t,\vec x)\varphi _0\Bigl(\omega _0(t,\vec x)\Bigr)
\varphi _1\Bigl (\omega _1(t,\vec x)
\Bigr)\varphi_2\Bigl (\omega _2(t,\vec x)\Bigr )
\end{displaymath}
separates equation (\ref{5.3.1}) into three second-order ODEs (see
\cite{213.4}). Nevertheless, there exists a possibility for equation
(\ref{5.3.1}) to be separated into two first-order and one
second-order ODEs or into three first-order ODEs. This is a probable
source of new potentials and new coordinate systems providing
separability of the Schr\"odinger equation. It should be mentioned that
separation of the two-dimensional d'Alembert equation
\begin{displaymath}
u_{tt}-u_{xx}=V(x)u
\end{displaymath}
into one first-order and one second-order ODEs gives no new 
potentials as compared with separation of it into two second-order
ODEs. But for some already known potentials new coordinate systems
providing separability of the above equation are obtained 
\cite{213.1,213.3}.
\vspace{2mm}

\noindent
{\em Table 5.3.3.}\ {\bf Coordinate systems providing
separability}\\
\phantom{{\em Table 5.3.3.}\ }{\bf of the Schr\"odinger
equation (\ref{5.3.20})} 
\vspace{1.5mm} 

\noindent
\begin{tabular}{|c||c|}\hline
 & \\
$\omega_1(t,\vec x)$ & $\omega_2(t,\vec x)$\\
 & \\ \hline
 & \\
$x_1\Bigl (\sin a(t+C)\Bigl )^{-1}+
\alpha \Bigl (\sin a(t+C)\Bigr )^{-2}$ & $x_2(\sin bt)^{-1}+\beta
(\sin bt)^{-2}$\\ 
$x_1\Bigl (\beta +\sin 2a(t+C)\Bigl )^{-1/2}$ & $x_2(\beta +\sin
2bt)^{-1/2}$\\ 
$x_1$ & $x_2$ \\
 & \\ \hline 
\end{tabular}
\vspace{1.5mm}

\noindent
Here $C,\ \alpha,\ \beta$ are arbitrary constants.
\vspace{2mm}

Let us briefly analyze the connection between separability and symmetry
properties of equation (\ref{5.3.1}). It is well-known that each
solution of the free Schr\" odinger equation with separated variables
is a common eigenfunction of its two mutually commuting second-order
symmetry operators \cite{155}. And what is more, separation constants
$\lambda _1, \ \lambda _2$ are eigenvalues of these symmetry
operators.

We will establish that the same assertion holds for the 
Schr\" odinger equation (\ref{5.3.1}). Let us make in equation
(\ref{5.3.1}) the following change of variables:
\begin{equation}
u=Q(t,\vec x)U\Bigl (t,\omega _1(t,\vec x),
\omega _2(t,\vec x)\Bigr ),
\label{5.3.21}
\end{equation}
where $(Q, \; \omega _1, \; \omega _2)$ is an arbitrary solution 
of the system of PDEs (\ref{5.3.6}). 

Substituting the expression (\ref{5.3.21}) into (\ref{5.3.1}) and
taking into account equations (\ref{5.3.6}) we get
\begin{equation}
\begin{array}{l}
Q\Bigl (i U_t+[U_{\omega _1\omega _1}-
B_{01}(\omega _1)U]\omega _{1x_a}\omega _{1x_a}+
[U_{\omega _2\omega _2}-B_{02}(\omega _2)U]\\[2mm]
\quad\times\omega _{2x_a}\omega _{2x_a}\Bigr )=0.
\end{array}
\label{5.3.22}
\end{equation}

Resolving equations 2 from the system (\ref{5.3.6}) with respect to
$\omega_{1x_a}\omega_{1x_a}$ and $\omega_{2x_a}\omega_{2x_a}$    
we have
\begin{eqnarray*}
\omega_{1x_a}\omega_{1x_a}&=&(1/\delta)\Bigl (R_2(t)
B_{21}(\omega _2)-R_1(t)B_{22}(\omega _2)\Bigr ), \\
\omega_{2x_a}\omega_{2x_a}&=&(1/\delta)\Bigl (R_1(t)
B_{12}(\omega _1)-R_2(t)B_{11}(\omega _1)\Bigr ),
\end{eqnarray*}
where $\delta =B_{11}(\omega _1)B_{22}(\omega _2)-
B_{12}(\omega _1)B_{21}(\omega _2)$ ($\delta \ne 0$ resulting from
the condition (\ref{5.3.3b})).

Substitution of the above equalities into equation (\ref{5.3.22}) with
subsequent division by $Q\ne 0$ yields the following PDE:
\begin{eqnarray}
&&i U_t+(1/\delta)R_1(t)\Bigl (B_{12}(\omega _1)
[U_{\omega _2\omega _2} - B_{02}(\omega _2)U]-
B_{22}(\omega _2)\non\\
&&\quad\times[U_{\omega _1\omega _1}-B_{01}(\omega _1)U]\Bigr ) 
+(1/\delta)R_2(t)\Bigl (B_{21}(\omega _2)
[U_{\omega _1\omega _1}\label{5.3.23}\\
&&\quad -B_{01}(\omega _1)U]-B_{11}(\omega _1)[U_{\omega _2\omega _2}- 
B_{02}(\omega _2)U]\Bigr )=0.\non
\end{eqnarray}

Thus, in the new coordinates $t, \ \omega _1, \ \omega _2, \ 
U(t,\omega _1, \omega_2)$ equation (\ref{5.3.1}) takes the form
(\ref{5.3.23}).

By direct (and very cumbersome) computation one can check 
that the following second-order differential operators
\begin{eqnarray*}
X_1&=&
(1/\delta)B_{22}(\omega _2)\Bigl (\partial _{\omega _1}^2-
B_{01}(\omega _1)\Bigr )-(1/\delta)B_{12}(\omega _1)
\Bigl (\partial _{\omega _2}^2 - B_{02}(\omega _2)\Bigr ),\\
X_2&=&
-(1/\delta)B_{21}(\omega _2)\Bigl (\partial _{\omega _1}^2-
B_{01}(\omega _1)\Bigr )+(1/\delta)B_{11}(\omega _1)
\Bigl (\partial _{\omega _2}^2 - B_{02}(\omega _2)\Bigr )
\end{eqnarray*}
commute under arbitrary $B_{0a}, \ B_{ab}, \ a,b=1,2$, i.e.,\ 
\begin{equation}
[X_1,\, X_2]\equiv X_1X_2-X_2X_1=0.
\label{5.3.24}
\end{equation}
 
After being rewritten in terms of the operators $X_1$,\ $X_2$ 
equation (\ref{5.3.23}) reads

\begin{displaymath}
\Bigl (i \partial _t-R_1(t)X_1-R_2(t)X_2\Bigr )U=0.
\end{displaymath}

Since the relations
\begin{equation}
[i \partial_t-R_1(t)X_1-R_2(t)X_2,\, X_a]=0, \ \ a=1,2
\label{5.3.25}
\end{equation}
hold, operators $X_1$, \ $X_2$ are mutually commuting symmetry
operators of equation (\ref{5.3.23}). Furthermore, the solution of
equation (\ref{5.3.23}) with separated variables $U=\varphi
_0(t)\varphi _1(\omega _1) \varphi _2(\omega _2)$ satisfies the
identities 
\begin{equation}
X_aU=\lambda _aU,\ \ a=1,2.
\label{5.3.26}
\end{equation}

Consequently, if we designate by $X_1^{\prime }$, \ $X_2^{\prime }$ 
the operators $X_1$, \ $X_2$ written in the initial variables 
$t, \ \vec x, \ u$, then we get from (\ref{5.3.24})--(\ref{5.3.26})
the following equalities:
\begin{eqnarray*}
& &[i \partial _t+\triangle -V(x_1,x_2),\,
X_a^{\prime }]=0, \ \ a=1,2, \\[2mm]
& &[X_1^{\prime },\, X_2^{\prime }]=0, \quad
X_a^{\prime }u=\lambda _au, \ \ a=1,2,
\end{eqnarray*}
where $u=Q(t,\vec x)\varphi _0(t)\varphi _1(\omega _1)
\varphi _2(\omega _2)$.

This means that each solution with separated variables is a common
eigenfunction of two mutually commuting symmetry operators
$X_1^{\prime }, \ X_2^{\prime }$ of the Schr\" o\-din\-ger equation
(\ref{5.3.1}), separation constants $\lambda _1, \ \lambda _2$ being
their eigenvalues.

So, we have exposed two possible approaches to variable separation in
linear PDEs which are based on their symmetry properties. The first
one is to start with a set of commuting symmetry operators of the
equation under study and to finish with the Ansatz (\ref{5.1.6})
\cite{11,155,178}.  Another approach suggested for the first time
in \cite{108.4} is closer to the original understanding of the
separation of variables in PDEs. A desired form (\ref{5.3.4}) of the
Ansatz for a solution with separated variables is postulated and then
it turns out that the solution obtained can be related to a set of
mutually-commuting symmetry operators of the equation under
consideration.

Both approaches have their merits and drawbacks. We think that the
utilization of the first approach is the only way to separate
variables in multi-component systems of PDEs. But to separate
variables in PDEs with one dependent variable it is preferable to
apply the second approach, since a computation of symmetry operators
is an extra step which is not, in fact, necessary for obtaining
solutions with separated variables. Another benefit of the approach in
question is its simplicity, only some basics of the standard
university course of mathematical physics are required for
understanding and implementing it.

One more merit is that the second approach in contrast to the first 
one can be easily generalized in order to separate variables in {\em 
  nonlinear} PDEs \cite{213.3}. Using such a generalization we have 
  classified in the paper \cite{209.4} all nonlinear d'Alembert 
equations
\begin{displaymath}
u_{x_0 x_0}-u_{x_1 x_1}=F(u),
\end{displaymath}
which separate into two first-order ODEs
\begin{displaymath}
\dot\vp_1=R_1(\vp_1),\quad \dot\vp_2=R_1(\vp_2)
\end{displaymath}
by means of the Ansatz
\begin{displaymath}
u(x_0,x_1)=f\Bigl(\vp_1(x_0)+\vp_2(x_1)\Bigr).
\end{displaymath}

It turned out that nonlinear d'Alembert equations admitting variable
se\-paration in the above sense are equivalent to one of the following 
PDEs: 
\index{Separation of variables!in the nonlinear d'Alembert equation}
\begin{eqnarray*}
\Box u&=&\lbd_1\Bigl(\cosh u + (\sinh 2u)\arctan {\rm
  e}^u\Bigr)+\lbd_2\sinh 2u,\\
\Box u&=&\lbd_1{\rm e}^u+\lbd_2{\rm e}^{-2u},\\
\Box u&=&\lbd_1\Bigl(\sinh u - (\sinh 2u)\,{\rm arctanh}\, {\rm
  e}^u\Bigr)+\lbd_2\sinh 2u,\\
\Box u&=&\lbd_1\Bigl(2\sin u + (\sin 2u)\ln\tan (u/2)\Bigr)+\lbd_2\sin
2u,\\ 
\Box u&=&\lbd_1 u +\lbd_2 u\ln u,
\end{eqnarray*}
where $\Box u=u_{x_0 x_0}-u_{x_1 x_1}$,\ $\lbd_1,\ \lbd_2$ are arbitrary
real constants.

This fact enabled us to construct exact solutions of the above
nonlinear PDEs which could not be found by the symmetry reduction
procedure. 

Let us also mention {\em anti-reduction} \/of
\index{Anti-reduction} PDEs \cite{106.2a,106.3}
which is also a ge\-neralization of a traditional notion of separation
of variables specially designed to handle nonlinear PDEs.

\newpage
\thispagestyle{empty}
\noindent
{\sl
C H A P T E R \ \  6\label{ch6}}
\vspace{2mm}

\hrule
\vspace{35mm}

\rightline
{\large\bf
CONDITIONAL SYMMETRY}
\vspace{2mm}

\rightline
{\large\bf
AND REDUCTION}
\vspace{2mm}

\rightline
{\large\bf
OF SPINOR EQUATIONS}
\vspace{7mm}

In this chapter a non-Lie method of reduction of nonlinear Poincar\'e-
and Galilei-invariant systems of PDEs to differential equations of
lower dimension is suggested. With the use of this method we construct
the wide classes of conditionally-invariant Ans\"atze reducing
nonlinear $P(1,3)$- and $G(1,3)$-invariant spinor equations to systems
of ODEs.  
\vspace{10mm}

\noindent
{\large \bf 6.1. Non-Lie reduction of Poincar\'e-invariant spinor
  equations\label{s6.1}} 

\markboth{Chapter 6. CONDITIONAL SYMMETRY AND REDUCTION}
   {6.1. Non-Lie reduction of Poincar\'e-invariant spinor equations}
\def\theequation{6.\arabic{section}.\arabic{equation}}
\setcounter {section} {1}
\setcounter {equation}{0}
\vspace{7mm}

\noindent
In Section 2.3 we have constructed a number of Ans\"atze for
the spinor field $\psi (x)$ reducing $P(1,3)$-invariant 
equation\index{Nonlinear!Dirac equation}
\begin{equation}
       \Bigl (i\g _\mu \p _\mu -\tilde f_1(\bar \psi \psi,
        \bar \psi \g _4\psi )-\tilde f_2(\bar \psi \psi,
        \bar \psi \g _4\psi )\g _4\Bigr )\psi =0 
\label{6.1.1}
\end{equation}
to systems of ODEs which cannot be obtained within the framework of the
classical Lie approach. Existence of such Ans\"atze is a consequence of
{\em conditional symmetry}\index{Conditional!symmetry} of equation 
(\ref{6.1.1}).
\vspace{1.5mm}

\noindent
{\bf Definition 6.1.1.}\ Equation (\ref{6.1.1}) is
conditionally-invariant under the involutive set\index{Involutive set} 
of operators
\begin{eqnarray*}
        Q_a=\xi_{a\mu }(x)\p_\mu +\eta_a(x),\ \ a={1,\ldots,N},
\end{eqnarray*}
if the system of PDEs
\begin{equation}
       (i\g_\mu \p_\mu -\tilde f_1-\tilde f_2\g_4)\psi =0, \quad
       Q_a\psi =0, \ \ a={1,\ldots,N}
\label{6.1.2}
\end{equation}
is invariant in Lie sense with respect to the one-parameter groups 
generated by the operators $Q_a$.

Due to Theorem 1.5.1 conditional invariance of PDE (\ref{6.1.1}) 
under the involutive set of operators $Q_a$ ensures its reducibility 
and, consequently, can be used to construct exact solutions of the 
(\ref{6.1.1}).

A usual approach to investigation of conditional symmetry of a given
PDE is application of the infinitesimal Lie method. But the
problem is that the determining equations for functions
$\xi_{a\mu}(x), \ \eta_a(x)$ prove to be nonlinear ones. That is why 
there is a little hope to describe all conditional symmetries of
multi-dimensional system of PDEs (\ref{6.1.1}).  It should be said
that more or less systematic results on conditional symmetry of PDEs
are obtained for two-dimensional equations only \cite{89}.

In the present section we suggest a method making it possible to get
both invariant and conditionally-invariant Ans\"atze constructed in
Sections 2.2, 2.4. Moreover, applying this method we obtain some
essentially new Ans\"atze for spinor field $\psi=\psi(x)$ reducing
system of PDEs (\ref{6.1.1}) to systems of ODEs.  \vspace{2mm}

\noindent
{\bf 1. Reduction of the nonlinear Dirac equation (\ref{6.1.1}).}\
Analysis of Ans\"atze for the spinor field invariant under the one-
and three-parameter subgroups of the Poincar\'e group shows that all
of them have the following structure:
\begin{equation}
\begin{array}{rcl}
        \psi (x)&=&\exp \{\theta_{\ssl A}\g_{\ssl A}(\g_0+\g_3)\} 
                \exp \{(1/2)\theta_0\g_0\g_3
                +(1/2)\theta_3\g_1\g_2\} \\[2mm]
        & &\times \cases {\vp(\om_1, \om_2, \om_3),\cr \vp(\om_1),\cr}
\end{array}
\label{6.1.3a}
\end{equation}
where $\vp$ is an arbitrary four-component function-column;
$\theta_\mu, 
\ \om_a$ are some real-valued scalar functions, the constraint holding
\begin{equation}
        \theta_{\ssl A}=\theta_{\ssl A}(x_0+x_3, x_1,
        x_2).\label{6.1.3b} 
\end{equation}

Hereafter the subscripts denoted by Latin alphabet letters $A$, $B$
take the values 1, 2 and summation over the repeated indices is
understood. 

The key idea of the approach suggested can be formulated in a rather
simple and natural way: we impose no {\em a priori} \/constraints on
the functions $\theta_\mu$, $\om_a$, they are obtained from the
requirement that substitution of expression (\ref{6.1.3a}) into
(\ref{6.1.1}) yields a system of PDEs for the function $\vp(\vec \om)$
(or a system of ODEs for the function $\vp(\om_1)$) with coefficients
depending on the new variables $\om_1,\ \om_2,\ \om_3$ only.
\index{Non-Lie reduction!of Poincar\'e-invariant spinor equations}

In the following we describe all Ans\"atze of the form (\ref{6.1.3a}),
(\ref{6.1.3b}) reducing the system of nonlinear four-dimensional PDEs 
(\ref{6.1.1}) to a system of ODEs.

Substituting the 
Ansatz\index{Ansatz!for spinor field}\index{Ansatz!$P(1,3)$-invariant}
\begin{equation}
    \psi(x)=\exp\{\theta_{\ssl A}\g_{\ssl
      A}(\g_0+\g_3)\}\exp\{(1/2)\theta_0\g_0\g_3+    
        (1/2)\theta_3\g_1\g_2\}\vp(\om)
\label{6.1.4}
\end{equation}
into equation (\ref{6.1.1}) and multiplying the expression obtained by 
the matrix
\begin{eqnarray*}
        \exp\{-(1/2)\theta_0\g_0\g_3-(1/2)\theta_3\g_1\g_2\}
        \exp\{-\theta_{\ssl A}\g_{\ssl A}(\g_0+\g_3)\}
\end{eqnarray*}
on the left yield
\begin{equation}
        iR_1\vp+iR_2\dot \vp
        =\Bigl (\tilde f_1(\bar \vp\vp,\bar \vp\g_4\vp)
        +\tilde f_2(\bar \vp\vp,\bar \vp\g_4\vp)\g_4\Bigr )\vp,
\label{6.1.5}
\end{equation}
where $R_1=R_1(x), \ R_2=R_2(x)$ are $(4\times 4)$-matrices determined 
by the following equalities:
\begin{eqnarray*}
 R_1&=&2e^{\theta_0}\Bigl(-\p_{\ssl A}\theta_{\ssl
   A}+\g_1\g_2(\p_1\theta_2 -\p_2\theta_1)\Bigr )(\g_0+\g_3)+\Bigl
 [(\g_0\p_0 \theta_0+\g_3\p_3\theta_0)\\
 & &\times(\cosh \theta_0+\g_0\g_3\sinh \theta_0)+\g_{\ssl A}\Bigl
        (\p_{\ssl A}\theta_0+2\theta_{\ssl
          A}(\p_3\theta_0-\p_0\theta_0)\Bigr ) (\cos \theta_3\\
 & &+\g_1\g_2\sin \theta_3)-2e^{\theta_0}\theta_{\ssl A}(\p_{\ssl
   A}\theta_0) (\g_0+\g_3)\Bigr ]\g_0\g_3+2e^{\theta_0}\theta_{\ssl
   A}\theta_{\ssl A} (\p_0\theta_0\\
 & &-\p_3\theta_0)(\g_0+\g_3)+\Bigl [(\g_0\p_0\theta_3+
        \g_3\p_3\theta_3)(\cosh\theta_0+ \g_0\g_3\sinh\theta_0)\\
 & &+\g_{\ssl A}\Bigl (\p_{\ssl A}\theta_3+2\theta_{\ssl
   A}(\p_3\theta_3 - \p_0\theta_3)\Bigr)   
        (\cos \theta_3+\g_1\g_2 \sin
        \theta_3)-2e^{\theta_0}\theta_{\ssl A}\\ 
 & &\times(\p_{\ssl A}\theta_3)(\g_0+\g_3)\Bigr ]\g_1\g_2+
        2e^{\theta_0}\theta_{\ssl A}\theta_{\ssl
          A}(\p_0\theta_3-\p_3\theta_3) (\g_0+\g_3)\g_1\g_2,\\   
 R_2&=&(\g_0\p_0\om+\g_3\p_3\om)(\cosh\theta_0+\g_0
       \g_3\sinh\theta_0)+\g_{\ssl A}\Bigl (\p_{\ssl
         A}\om+2\theta_{\ssl A}(\p_3\om\\ 
    & &-\p_0\om)\Bigr )(\cos \theta_3 + \g_1\g_2\sin
    \theta_3)-2e^{\theta_0}\theta_{\ssl A}(\p_{\ssl A}\om)(\g_0 +
    \g_3)\\  
    & &+2e^{\theta_0}\theta_{\ssl A}\theta_{\ssl A}(\p_0\om-
    \p_3\om)(\g_0+\g_3).
\end{eqnarray*}

Consequently, Ansatz (\ref{6.1.4}) reduces equation (\ref{6.1.1}) to a
system of ODEs iff there exist such $(4\times 4)$-matrices
$Q_1(\om)$, $Q_2(\om)$ that 
\begin{equation}
        R_1(x)=Q_1(\om), \quad R_2(x)=Q_2(\om).
\label{6.1.6}
\end{equation}

Expanding matrices $Q_1(\om), \ Q_2(\om)$ in the complete system of
the Dirac matrices and equating coefficients of the matrices $I, \ 
\g_\mu, \ S_{\mu\nu}, \ \g_4\g_\mu, \ \g_4$ we obtain from
(\ref{6.1.6}) the over-determined system of nonlinear PDEs for
functions $\theta_\mu, \ \om$
\begin{eqnarray}
        &1)& (\p_0\theta_0)\sinh \theta_0+(\p_3\theta_0)\cosh
          \theta_0-2e^{\theta_0} 
            \p_{\ssl A}\theta_{\ssl A}-2e^{\theta_0}\theta_{\ssl
              A}\p_{\ssl A}\theta_0 \non \\ 
          & &+2e^{\theta_0}\theta_{\ssl A}\theta_{\ssl
            A}(\p_0\theta_0-\p_3\theta_0) = f_1(\om), \non\\          
        &2)& (\p_0\theta_0)\cosh \theta_0+(\p_3\theta_0)\sinh
            \theta_0-2e^{\theta_0}
          \p_{\ssl A}\theta_{\ssl A}-2e^{\theta_0}\theta_{\ssl
            A}\p_{\ssl A}\theta_0\non \\   
          & &+2e^{\theta_0}\theta_{\ssl A}\theta_{\ssl
            A}(\p_0\theta_0-\p_3\theta_0) = f_2(\om),\non \\
        &3)& \Bigl (\p_2\theta_3+2\theta_2(\p_3\theta_3-\p_0\theta_3)
        \Bigr ) 
            \cos  \theta_3-\Bigl
            (\p_1\theta_3+2\theta_1(\p_3\theta_3\non \\ 
          & &-\p_0\theta_3)\Bigr )\sin  \theta_3=f_3(\om),\non \\
        &4)& \Bigl (\p_1\theta_3+2\theta_1(\p_3\theta_3-\p_0\theta_3) 
            \Bigr )\cos  \theta_3
            +\Bigl (\p_2\theta_3+2\theta_2(\p_3\theta_3 \non \\
            & & -\p_0\theta_3)\Bigr )\sin  \theta_3=f_4(\om),\non \\
        &5)&
        2e^{\theta_0}(\p_1\theta_2-\p_2\theta_1)+(\p_0\theta_3)
        \cosh\theta_0  
            +(\p_3\theta_3)\sinh\theta_0\non \\
            & &+2e^{\theta_0}\theta_{\ssl A}\theta_{\ssl
              A}(\p_0\theta_3-\p_3 \theta_3) 
            -2e^{\theta_0}\theta_{\ssl A}\p_{\ssl
              A}\theta_3=f_5(\om),\non \\ 
        &6)&
        2e^{\theta_0}(\p_1\theta_2-\p_2\theta_1)+(\p_0\theta_3)
        \sinh\theta_0 + (\p_3\theta_3)\cosh \theta_0\non \\
            & & +2e^{\theta_0}\theta_{\ssl A}\theta_{\ssl
              A}(\p_0\theta_3-\p_3 \theta_3)- 
            2e^{\theta_0}\theta_{\ssl A}\p_{\ssl
              A}\theta_3=f_6(\om),\non \\ 
        &7)& \Bigl (\p_1\theta_0+2\theta_1(\p_3\theta_0-\p_0\theta_0)
            \Bigr )\cos  \theta_3
            +\Bigl (\p_2\theta_0+2\theta_2(\p_3\theta_0\non \\
            & & -\p_0\theta_0)\Bigr )\sin  \theta_3=f_7(\om),
            \label{6.1.7}\\ 
        &8)& \Bigl (\p_2\theta_0+2\theta_2(\p_3\theta_0-\p_0\theta_0)
             \Bigr )\cos  \theta_3
            -\Bigl (\p_1\theta_0+2\theta_1(\p_3\theta_0\non \\
            & & -\p_0\theta_0)\Bigr )\sin  \theta_3=f_8(\om),\non \\
        &9)& (\p_0\om)\cosh\theta_0+(\p_3\om)\sinh\theta_0-
            2e^{\theta_0}\theta_{\ssl A}\p_{\ssl A}\om
            +2e^{\theta_0}\non \\ 
        & &  \times \theta_{\ssl A}\theta_{\ssl
          A}(\p_0\om-\p_3\om)=f_9(\om),\non \\ 
        &10)& (\p_3\om)\cosh\theta_0+(\p_0\om)
            \sinh \theta_0-2e^{\theta_0}\om_{\ssl A}\p_{\ssl A}\om
            +2e^{\theta_0} \non \\ 
        & &  \times \theta_{\ssl A}\theta_{\ssl
          A}(\p_0\om-\p_3\om)=f_{10}(\om), \non \\ 
        &11)& \Bigl (\p_1\om+2\theta_1(\p_3\om-\p_0\om)
              \Bigr )\cos  \theta_3
              +\Bigl (\p_2\om+2\theta_2(\p_3\om\non \\
        & &  -\p_0\om)\Bigr )\sin  \theta_3=f_{11}(\om),\non \\
        &12)& \Bigl (\p_2\om+2\theta_2(\p_3\om-\p_0\om)
              \Bigr )\cos  \theta_3
              -\Bigl (\p_1\om+2\theta_1(\p_3\om\non \\
        & &  -\p_0\om)\Bigr )\sin  \theta_3=f_{12}(\om),\non
\end{eqnarray}
where $f_1(\om),\ldots,f_{12}(\om)$ are arbitrary smooth real-valued
functions. 

Thus, the problem of construction of Ans\"atze (\ref{6.1.4}) reducing
the nonlinear Dirac equation (\ref{6.1.1}) to systems of ODEs is
equivalent to the one of integration of the over-determined system of
PDEs (\ref{6.1.7}). Let us emphasize that the above system is
compatible because Poincar\'e-invariant Ans\"atze obtained in Section
2.2 are contained in class (\ref{6.1.4}).

Integration of the system of nonlinear PDEs (\ref{6.1.7}) is
substantially simplified if we utilize an equivalence relation
which is introduced below.

First of all, we note that the class of Ans\"atze (\ref{6.1.4}) is
transformed into itself if we generate the spinor field
(\ref{6.1.4}) by the 8-parameter transformation group $G_8 \subset
P(1,3)$ with the generators $P_{\mu },\ J_{12},\ J_{03},\ 
J_{01}-J_{13}$, \ $J_{02}$ $-$ $J_{23}$.

The above assertion is checked by a direct verification. Take, as an
example, the one-parameter transformation group having the generator
$J_{03}$. Applying formula (\ref{2.4.37}) with $a=3$ to (\ref{6.1.4}) 
we get
\begin{eqnarray*}
\psi(x)&=&\exp\{\theta_{\ssl A}^{\prime}(x^{\prime})\g_{\ssl
  A}(\g_0+\g_3)\} \exp\{(1/2)\theta_0^{\prime}(x^{\prime})\g_0\g_3\\
           & &+(1/2)\theta_3^{\prime}(x^{\prime})\g_1\g_2\}
          \vp\Bigl (\om^{\prime}(x^{\prime})\Bigr ),
\end{eqnarray*}
where
\begin{eqnarray*}
& &x_0^{\prime}=x_0\cosh \tau +x_3\sinh \tau,\quad x_1^{\prime}=x_1,
\\ 
& &x_2^{\prime}=x_2,\quad x_3^{\prime}=x_3\cosh \tau +x_0\sinh \tau,
\\ 
& &\theta _0^{\prime}=\theta _0+\tau, \quad \theta _1^{\prime}=\theta
_1e^{-\tau }, \quad  
\theta _2^{\prime}=\theta _2e^{-\tau },\quad \\
& & \theta _3^{\prime}=\theta_3, \quad \om ^{\prime}=\om.
\end{eqnarray*}

Consequently, the group $G_8$ induces in the space of variables $x$, \
$\theta _{\mu}(x)$, \ $\om (x)$ some transformation group $\widetilde
G_8$. It is not difficult to establish that $\widetilde G_8$
is the invariance group of system of PDEs (\ref{6.1.7}).

Another transformation leaving the class of Ans\"atze (\ref{6.1.4})
invariant is the following one:

\begin{eqnarray}
     & &   \theta_0\to \theta_0+g_0(\om), \quad \theta_3\to
     \theta_3+g_3(\om),\quad \om\to g(\om),\quad\non \\   
     & &   \theta_1\to \theta_1+e^{-\theta_0}\Bigl (g_1(\om)\cos
     \theta_3 
                    -g_2(\om)\sin  \theta_3\Bigr ),\label{6.1.8}\\
     & &   \theta_2\to \theta_2+e^{-\theta_0}\Bigl (g_2(\om)\cos
     \theta_3 
                    +g_1(\om)\sin  \theta_3\Bigr ).\non
\end{eqnarray}

That is why it is natural to introduce the following equivalence
relation $E$. We say that solutions of system (\ref{6.1.4}) $\theta
_{\mu}(x), \ \om (x)$ and $\theta _{\mu}^{\prime}(x), \ \om
^{\prime}(x)$ are equivalent if they can be transformed one into
another by

1) a suitable transformation from the group $\widetilde G_8$, or

2) a suitable transformation of the form (\ref{6.1.8}).

An easy check shows that $E$ is indeed an equivalence
relation. It divides the set of solutions of the system of PDEs under
study into inequivalent classes which are described by the following
assertion.
\vspace{1.5mm}

\noindent
{\bf Theorem 6.1.1.}\ {\em The general solution of system of PDEs
  (\ref{6.1.7}) determined up to the equivalence relation $E$ is given
  by one of the following formulae:
\begin{eqnarray}
        &1)& \theta_1=\theta_2=0,\quad\theta_0=\ln     (x_0+x_3),
        \quad 
        \theta_3=C\ln    (x_0+x_3), \quad\om=x_0^2-x_3^2; \non \\
        &2)& \theta_{\ssl A}=-x_{\ssl A}\Bigl (2(x_0+x_3)\Bigr )^{-1},
        \quad \theta_0=\ln (x_0+x_3), \quad
        \theta_3=C \ln     (x_0+x_3), \non \\ 
        & &\om=x_0^2-x_1^2-x_2^2-x_3^2; \non \\
        &3)& \theta_1=0, \quad\theta_2=-x_2\Bigl (2(x_0+x_3)\Bigr
        )^{-1}, \quad \theta_0=\ln (x_0+x_3),\non \\ & &\theta_3=C \ln     
        (x_0+x_3), \quad 
        \om=x_0^2-x_2^2-x_3^2; \non \\
        &4)& \theta_1=\theta_2=0, \quad\theta_0=0, \quad
        \theta _3=C_1(x_0+x_3), \quad\om=x_0-x_3\non\\
        & &+C_2(x_0+x_3); \non \\
        &5)& \theta_1=\theta_2=0, \quad\theta_0=Cx_1, \quad\theta_3=0, 
        \quad 
        \om=Cx_1+\ln (x_0-x_3); \non \\
        &6)& \theta_{\ssl A}=\p_{\ssl A}W, \quad\theta_0=0,
        \quad\theta_3=(C/2)(x_0-x_3+4W), \quad\om=x_0+x_3,\non \\ 
        & & W=\tau_1z^2+\tau_2z+\tau_1^*z^{*2}
        +\tau_2^*z^*+\tau_3zz^*, \non \\
        & &{ where} \quad z=x_1+ix_2 \ { and \ the \ functions} \
        \tau _j(x) \ { are \ determined \ by } \non \\
        & &{  one \ of \ the \ formulae \ a \ - \
          c \ given \ below } \non \\
        &a)& \tau_1=C_2\Bigl (64C_2^2(x_0+x_3)^2-1\Bigr
        )^{-1}e^{iC_1},\non \\ 
        & &\tau_2=C_3\exp \Biggl
        \{16C_2\mathop{\int}\limits^{\edi (1/2)(x_0+x_3)}
        (256C_2^2\xi^2-1)^{-1} 
        \Bigl [-16C_2\xi \non \\ 
        & &\quad +\cos  \Bigl (2R_1(\xi)-C_1\Bigr )\Bigr ]d\xi
        +iR_1\Bigl ((1/2)(x_0+x_3)\Bigr )\Biggr \}, \non \\
        & &\tau_3=16C_2^2(x_0+x_3)\Bigl (1-64C_2^2(x_0+x_3)^2\Bigr
        )^{-1}, \non \\ 
        & &\dot R_1(\xi)=16C_2(1-256 C_2^2\xi^2)^{-1}\Bigl [16C_2\xi 
           +\sin  \Bigl (2R_1(\xi)-C_1\Bigr )\Bigr ]; \non \\
        &b)& \tau_1=\Bigl (16(x_0+x_3)\Bigr )^{-1}e^{iC_1},
        \label{6.1.9}\\ 
        & &\tau_2=C_2\exp \Biggl \{(1/2)
        \mathop{\int}\limits^{\edi (1/2)(x_0+x_3)} \Bigl [\cos  \Bigl  
        (2R_2(\xi)-C_1\Bigr ) -1\Bigr ] \xi^{-1}d\xi \non \\
        & &\quad +iR_2\Bigl
        ((1/2)(x_0+x_3)\Bigr )\Biggr \}, \quad
        \tau_3=-\Bigl (8(x_0+x_3)\Bigr )^{-1}, \non \\
        & & 2\xi\dot R_2(\xi)+\sin  \Bigl (2R_2(\xi)-C_1\Bigr )+1=0;
        \non \\ 
        &c)& \tau_1=0, \quad\tau_2=(C_1+iC_2)(x_0+x_3)^{-1}, \quad
        \tau_3=\Bigl (4(x_0+x_3)\Bigr )^{-1}; \non \\
        &7)& \theta_{\ssl A}=(1/2)\dot w_{\ssl A}+C_2\arctan  (\tilde
        x_1/ \tilde x_2)(\tilde x_1^2+\tilde x_2^2)^{1/2} \non \\
        & &\quad \times \exp \{-C_1\arctan  (\tilde x_1/
        \tilde x_2)\}\p_{\ssl A}\Bigl(\arctan   
        (\tilde x_1/\tilde x_2)\Bigr), \non \\
        & &\theta_0=C_1 \arctan   (\tilde x_1/\tilde
        x_2),  \quad
        \theta_3=-\arctan   (\tilde x_1/\tilde x_2),\quad
        \om=\tilde x_1^2+\tilde x_2^2; \non \\
        &8)& \theta_{\ssl A}=(1/2)\dot w_{\ssl A}+(\tilde x_1^2+
        \tilde x_2^2)^{1/2}\Bigl(C_1(x_0+x_3)^{-1} \non \\
        & &\quad \times \arctan  (\tilde x_1/\tilde x_2)
        +w_3\Bigr)\p_{\ssl A}\Bigl(\arctan  (\tilde x_1/
        \tilde x_2)\Bigr), \non \\
        & &\theta_0=\ln     (x_0+x_3), \quad\theta_3=-\arctan   
        (\tilde x_1/\tilde x_2), \quad
        \om=\tilde x_1^2+\tilde x_2^2; \non \\
        &9)& \theta_{\ssl A}=x_1w_{\ssl A}+\p_{\ssl A}\Bigl
        (U(z,x_0+x_3)+U(z^*,x_0+x_3)\Bigr ), \quad z=x_1+ix_2, \non \\  
        & &\theta_0=\theta_3=0, \quad\om=x_0+x_3; \non \\
        &10)& \theta_1=(x_1\sin  w_2-x_2\cos  w_2)\Bigl [
        \Bigl ((1/2)\dot w_1+Ce^{-w_1}\Bigr )\sin  w_2 \non \\
        & &\quad -(1/2) \dot w_2 \cos  w_2\Bigr ]+w_4\sin  w_2 
        +(1/2)\dot w_3\cos  w_2, \non \\
        & & \theta_2=(x_1\sin  w_2-x_2\cos  w_2)\Bigl [-\Bigl
        ((1/2)\dot w_1+Ce^{-w_1}\Bigr )\cos  w_2 \non \\
        & &\quad -(1/2)\dot w_2\sin  w_2\Bigr ] -w_4\cos  w_2
        +(1/2)\dot w_3\sin  w_2, \non \\ 
        & & \theta_0=w_1, \quad\theta_3=w_2, \quad\om=x_1\cos 
        w_2+x_2\sin  w_2+w_3; \non \\  
        &11)& \theta_{\ssl A}=(1/2) \dot w_{\ssl A},
        \quad\theta_0=C(x_2+w_2), \quad \theta_3=0,
        \quad\om=x_1+w_1. \non 
\end{eqnarray}

In the above formulae $\tilde x_{\ssl A}=x_{\ssl A}+w_{\ssl A}$; \ 
$A=1,2$; \ $w_1, \ w_2, \ w_3, \ w_4$ are arbitrary smooth real-valued
functions of $x_0+x_3$; \ $U$ is an arbitrary analytic function of
$z$; \ $C, \ C_1, \ C_2, \ C_3$ are arbitrary real constants.}
\vspace{1.5mm}

\noindent
{\em Proof.}$\quad$ On introducing new independent variables
$\xi=(1/2)(x_0+x_3), \ \eta=(1/2)(x_0-x_3)$ we rewrite system
(\ref{6.1.7}) in the form 
\begin{eqnarray}
        &1)& \p_\eta \theta_0=f_1(\om)e^{\theta_0}, \non \\
        &2)& \p_\xi\theta_0-4\p_{\ssl A}\theta_{\ssl A}-4\theta_{\ssl
          A}\p_{\ssl A}\theta_0+4f_1(\om) 
        e^{\theta_0}\theta_{\ssl A}\theta_{\ssl
          A}=f_2(\om)e^{-\theta_0}, \non \\ 
        &3)& \p_1\theta_3=2\theta_1e^{\theta_0}f_1(\om)+f_4(\om)\cos
         \theta_3- 
        f_3(\om)\sin  \theta_3, \non \\
        &4)& \p_2\theta_3=2\theta_2e^{\theta_0}f_1(\om)+f_3(\om)\cos
         \theta_3+ 
        f_4(\om)\sin  \theta_3, \non \\
        &5)& \p_\eta \theta_3=f_5(\om)e^{\theta_0}, \non \\
        &6)&
        \p_{\xi}\theta_3+4(\p_1\theta_2-\p_2\theta_1)+4f_5(\om)
        e^{\theta_0}   
        \theta_{\ssl A}\theta_{\ssl A}-4\theta_{\ssl A}\p_{\ssl
          A}\theta_3 \non \\ 
        & &=f_6(\om)e^{-\theta_0},  \label{6.1.10}\\
        &7)& \p_1\theta_0=2\theta_1f_1(\om)e^{\theta_0}+f_7(\om)\cos
        \theta_3- 
        f_8(\om) \sin  \theta_3, \non \\
        &8)& \p_2\theta_0=2\theta_2f_1(\om)e^{\theta_0}+f_8(\om)\cos
         \theta_3+ 
        f_7(\om)\sin  \theta_3, \non \\
        &9)& \p_\eta \om=f_9(\om)e^{\theta_0}, \non \\
        &10)& \p_\xi
        \om-4\theta_{\ssl A}\p_{\ssl
          A}\om+4f_9(\om)e^{\theta_0}\theta_{\ssl A}\theta_{\ssl A}=  
        f_{10}(\om)e^{-\theta_0}, \non \\
        &11)& \p_1\om=2\theta_1f_9(\om)e^{\theta_0}+f_{11}(\om)
        \cos  \theta_3-f_{12}(\om)\sin  \theta_3, \non \\
        &12)& \p_2\om=2\theta_2f_9(\om)e^{\theta_0}+f_{12}(\om)\cos
        \theta_3 
        +f_{11}(\om)\sin  \theta_3.\non
\end{eqnarray}

Now we see that the above system contains a subsystem of PDEs
1, 5, 9 

$$
\p_\eta \theta_0=f_1(\om)e^{\theta_0}, \quad \p_\eta \theta_3
=f_5(\om)e^{\theta_0}, \quad \p_\eta \om=f_9(\om)e^{\theta_0},
$$   
which can be considered as a system of ODEs with respect to the 
variable $\eta $. Transforming $\theta _0, \ \theta _3, \ \om$
according to (\ref{6.1.8}) we can put $f_1f_9=0$. With this remark the 
above system is easily integrated. Its general solution determined up
to the equivalence relation $E$ is given by one of the following
formulae: 
\begin{eqnarray*}
       &{\rm I}.& {\rm under} \ f_1=f_5=f_9=0, \\ 
        & & \theta_0=F_1, \quad \theta_3=F_2, \quad \om=F_3; \\
       &{\rm II}.& {\rm under} \ f_1=f_5=0, \ f_9\ne 0, \\
       & & \theta_0=\ln F_1, \ \om=\eta F_1+F_2, \ \theta_3=F_3; \\
       &{\rm III}.& {\rm under} \ f_9=0, \ f_1\ne 0,\\ 
        & & \theta_0=-\ln  (\eta+F_2), \quad \om=F_1, \quad
        \theta_3=f_5(F_1)\ln (\eta+F_2)+F_3; \\
       &{\rm IV}.& {\rm under} \ f_1=f_9=0, \ f_5\ne 0, \\
        & & \theta_0=-\ln F_2, \quad
        \theta_3=F_2^{-1}f_5(F_1)\eta+F_3, \quad \om=F_1, 
\end{eqnarray*}
where $F_1, \ F_2, \ F_3$ are arbitrary smooth real-valued functions
of $\xi, \ x_1, \ x_2$.

Thus, to prove the theorem we have to consider four inequivalent cases
{\rm I}--{\rm IV}. We will integrate system of PDEs (\ref{6.1.10}) in
the case $f_1=f_5=f_9=0$, the remaining cases are handled in an
analogous way.

When proving the theorem, we will use essentially the following
assertion.
\vspace{1.5mm}

\noindent
{\bf Lemma 6.1.1.}\ {\em General solution of system of PDEs
\begin{eqnarray*}
        \p_1u&=&A_1(u)\cos  v-A_2(u)\sin  v, \\
        \p_2u&=&A_2(u)\cos  v+A_1(u)\sin  v, \\
        \p_1v&=&B_1(u)\cos  v-B_2(u)\sin  v, \\
        \p_2v&=&B_2(u)\cos  v+B_1(u)\sin  v, 
\end{eqnarray*}
determined up to the equivalence relation
\begin{eqnarray*}
        u\to h_1(u), \quad v\to v+h_2(u), \quad h_i\in C^1(\R^1,\R^1)
\end{eqnarray*}
is given by one of the formulae
\begin{equation}
\begin{array}{rl}
        1)& u=(x_1+w_1)^2+(x_2+w_2)^2, \\[2mm] 
          & v=\arctan   \Bigl ((x_1+w_1)(x_2+w_2)^{-1}\Bigr ); \\[2mm]
        2)& u=x_1\cos w_2+x_2\sin  w_2+w_1, \ v=w_2; \\[2mm]
        3)& u=w_1, \ v=w_2.
\end{array}
\label{6.1.11}
\end{equation}
Here $w_1, \ w_2$ are arbitrary smooth real-valued functions of
$\xi$.}

Proof of the above assertion is carried out with the help of rather
simple but very cumbersome computations, therefore it is omitted.

Substituting $\theta_0=F_1(\xi,x_1,x_2),\ \theta_3 = F_2(\xi,x_1,x_2), 
\ \om = F_3(\xi,x_1,x_2)$ into system (\ref{6.1.10}) we have

\begin{eqnarray}
         &1)& \p_\xi F_1-4\p_{\ssl A}\theta_{\ssl A}-4\theta_{\ssl
           A}\p_{\ssl A}F_1=f_2e^{-F_1}, 
         \non \\ 
         &2)& \p_1 F_2=f_4\cos  F_2-f_3\sin  F_2, \non \\
         &3)& \p_2 F_2=f_3\cos  F_2+f_4\sin  F_2, \non \\
         &4)& \p_\xi F_2+4(\p_1\theta_2-\p_2\theta_1)-4\theta_{\ssl
           A}\p_{\ssl A} F_2 = f_6e^{-F_1}, \non \\ 
         &5)& \p_1 F_1=f_7\cos  F_2-f_8\sin  F_2, \label{6.1.12}\\ 
         &6)& \p_2 F_1=f_8\cos  F_2+f_7\sin  F_2, \non \\
         &7)& \p_\xi F_3-4\theta_{\ssl A}\p_{\ssl
           A}F_3=f_{10}e^{-F_1}, \non \\ 
         &8)& \p_1 F_3=f_{11}\cos  F_2-f_{12}\sin  F_2, \non \\
         &9)& \p_2 F_3=f_{12}\cos  F_2+f_{11}\sin  F_2,\non
\end{eqnarray} 
where $f_2,\ldots,f_{12}$ are arbitrary smooth real-valued functions
of $F_3$. 

According to Lemma 6.1.1 a subsystem of equations 2, 3, 8, 9 has
three inequivalent classes of solutions given by formulae
(\ref{6.1.11}).

\noindent
{\bf Case 1.} $F_2=-\arctan   \Bigl ((x_1+w_1)(x_2+w_2)^{-1}\Bigr
), \quad F_3=(x_1+w_1)^2 \\ \phantom{{\bf Case 1.}} +(x_2+w_2)^2$.

Substitution of the above expressions into the fifth and sixth
equations of system (\ref{6.1.12}) yields the following system of PDEs
for the function $F_1$: 
\begin{equation}
\begin{array}{rcl}
 \p_1F_1&=&\tilde x_2\tilde f_7(\tilde x_1^2+\tilde x_2^2)
           +\tilde x_1\tilde f_8(\tilde x_1^2+\tilde x_2^2), \\[2mm]
 \p_2F_1&=&\tilde x_2\tilde f_8(\tilde x_1^2+\tilde x_2^2)
           -\tilde x_1\tilde f_8(\tilde x_1^2+\tilde x_2^2),
\end{array}
\label{6.1.13}
\end{equation}
where $\tilde x_{\ssl A}=x_{\ssl A}+w_{\ssl A}, \ A=1,2$.

Taking into account the compatibility condition $\p_1(\p_2F_1)=
\p_2(\p_1F_1)$ we have $\dot{\tilde f_7}=0$ or $\tilde f_7=C_1=
\mbox{\rm const}$. Hence it follows that up to the equivalence
relation $E$ the general solution of system (\ref{6.1.13}) can be
represented in the form
\begin{eqnarray*}
        F_1=C_1\arctan   (\tilde x_1/\tilde x_2)
        +w_3(\xi), \ \ w_3\in C^1(\R^1,\R^1).
\end{eqnarray*}

From the seventh equation of system (\ref{6.1.12}) it follows that
functions $\theta_1, \ \theta_2$ satisfy the equality
\begin{eqnarray*}
  \tilde x_{\ssl A}(\dot w_{\ssl A}-4\theta_{\ssl A})=f_{10}(\tilde
  x_1^2+\tilde x_2^2) \exp \{-C_1 \arctan   (\tilde x_1/ 
   \tilde x_2)-w_3\},
\end{eqnarray*}
whose general solution reads
\begin{eqnarray*}
        \theta_{\ssl A}=(1/4)\dot w_{\ssl A}+W(\xi,x_1,x_2)\p_{\ssl
          A}\Bigl (\arctan (\tilde x_1/\tilde x_2)\Bigl ), \ \ A=1,2.
\end{eqnarray*}

Here $W$ is an arbitrary smooth real-valued function.

Substituting the above results into the first and fourth equations
of system (\ref{6.1.12}) we arrive at the following system of two PDEs
for $W$: 
\begin{eqnarray}
        & &\tilde x_{\ssl A}\p_{\ssl A}W=W+\alpha_1\exp \{-C_1\arctan   
        (\tilde x_1/\tilde x_2)-w_3\}, \non \\
        & &(\tilde x_2\p_1-\tilde x_1\p_2)W=-C_1W+(1/4)\dot w_3
        (\tilde x_1^2+\tilde x_2^2) \label{6.1.14}\\
        & &\quad +\alpha_2\exp\{-C_1\arctan   (\tilde x_1/
        \tilde x_2)-w_3\},\non
\end{eqnarray}
where $\alpha_{\ssl A}=\alpha_{\ssl A}(\tilde x_1^2+\tilde x_2^2)$.
Integration of system of linear PDEs (\ref{6.1.14}) yields two
inequivalent classes of solutions
\vspace{1.5mm}

\noindent
\underline {under $C_1\ne 0$}
\begin{eqnarray*}
        W&=&\Bigl (C_3+C_2\arctan   (\tilde x_1/
            \tilde x_2)\Bigr )
         (\tilde x_1^2+\tilde x_2^2)^{1/2} \\
         & &\times\exp \{-C_1\arctan   (\tilde x_1/
           \tilde x_2)\}, \quad w_3=0;
\end{eqnarray*}
\underline{under $C_1= 0$}
\begin{eqnarray*}
   W&=&\Bigl (w_0(\xi)+C\xi^{-1}\arctan   (\tilde x_1/
   \tilde x_2)\Bigr )(\tilde x_1^2+\tilde x_2^2)^{1/2},
      \quad  w_3=\xi.
\end{eqnarray*}

Here $C, \ C_1, \ C_2, \ C_3$ are real constants, $w_0\in
C^1(\R^1,\R^1)$ is an arbitrary function.

Substituting the results obtained into the corresponding expressions
for $\theta_\mu $, \ $\om$ and returning to the initial independent
variables $x_\mu$ we get up to the equivalence relation $E$ the
formulae 8, 9 from (\ref{6.1.9}).  
\vspace{1.5mm}

\noindent
{\bf Case 2.} $F_2=w_2(\xi), \quad F_3=w_3(\xi)$.

Up to the equivalence relation $E$ we can choose $F_3=\xi, \ F_2=0$.
Substitution of these expressions into (\ref{6.1.12}) gives rise to the
following system of PDEs for $F_1, \ \theta_1, \ \theta_2$:
\begin{eqnarray}
        &1)& \p_\xi
        F_1-4\theta_{\ssl A}\p_{\ssl A}F_1-4\p_{\ssl A}\theta_{\ssl
          A}=f_2(\xi)e^{-F_1}, \non \\ 
        &2)& \p_2\theta_1-\p_1\theta_2=f_6(\xi)e^{-F_1}, \non \\
        &3)& \p_1F_1=f_7(\xi),  \label{6.1.15} \\
        &4)& \p_2F_1=f_8(\xi), \non \\
        &5)& 1=f_{10}(\xi)e^{-F_1}.\non
\end{eqnarray}

From the last three equations we conclude that within the equivalence
relation $F_1=0$. Integrating the remaining equations and returning to
the initial independent variables we obtain within the equivalence
relation $E$ formulae 9 from (\ref{6.1.9}).  
\vspace{1.5mm}

\noindent
{\bf Case 3.} $F_2=w_2(\xi), \quad F_3=x_1\cos w_2(\xi)+x_2\sin 
w_2(\xi)+ w_3(\xi)$.

Substitution of the above expressions into equations 1, 4--7 from
(\ref{6.1.12}) gives rise to the over-determined system of PDEs for
functions $F_1,$\ $ \theta_1,$ \ $\theta_2$
\begin{eqnarray}
        &1)& 4\p_{\ssl A}\theta_{\ssl A}=\p_\xi F_1-4\theta_{\ssl
          A}\p_{\ssl A} F_1+f_2e^{-F_1}, \non \\ 
        &2)& 4(\p_2\theta_1-\p_1\theta_2)=\dot w_2+f_6e^{-F_1}, \non
        \\ 
        &3)& \p_1F_1=f_7\cos  w_2-f_8\sin  w_2, \label{6.1.16}\\
        &4)& \p_2F_1=f_8\cos  w_2+f_7\sin  w_2, \non \\
        &5)& \dot w_2(x_2\cos  w_2-x_1\sin  w_2)+\dot w_3 \non \\
          & &-4(\theta_1\cos  w_2+\theta_2\sin
          w_2)=f_{10}e^{-F_1},\non 
\end{eqnarray}
where $f_2,\ f_6, \ f_7, \ f_8, \ f_{10}$ are arbitrary smooth 
functions of $x_1\cos w_2$ $+x_2\sin w_2$ $+w_3$.

The necessary and sufficient compatibility condition of a subsystem of
equations 3, 4 reads $\p_1(\p_2 F_1)=\p_2(\p_1 F_1)$, whence it
follows that $f_8=C_1=\mbox{\rm const}$. Substituting $f_8=C_1$ into
equations 3, 4 from (\ref{6.1.16}) and integrating the equations
obtained we have
\begin{eqnarray*}
   F_1=C_1(x_2\cos  w_2-x_1\sin  w_2)+w_1(\xi), \ \ w_1\in
   C^1(\R^1,\R^1). 
\end{eqnarray*}

With account of the above formula system (\ref{6.1.16}) is rewritten
in the following way:
\begin{eqnarray}
        &1)& \p_{\ssl A}\theta_{\ssl A}=-(1/4)C_1\dot
        w_2(x_1\cos  w_2+x_2\sin  w_2)+(1/4)\dot w_1 \non \\
        & &\quad +C_1(\theta_1\sin  w_2-\theta_2\cos w_2)
        +f_2\exp\{-C_1(x_2\cos  w_2 \non \\
        & &\quad -x_1\sin  w_2)-w_1\}, \non \\
        &2)& \p_2\theta_1-\p_1\theta_2=(1/4)\dot
        w_2+f_6\exp\{-C_1(x_2\cos  w_2 \label{6.1.17}\\
        & &\quad -x_1\sin w_2)-w_1\}, \non \\
        &3)& \theta_1\cos w_2+\theta_2\sin  w_2=(1/4)\dot w_2(x_2\cos
         w_2-x_1\sin  w_2) \non \\
        & &\quad +(1/4)\dot w_3+f_{10}\exp \{-C_1(x_2\cos  w_2-
        x_1\sin  w_2) -w_1\}.\non
\end{eqnarray}

Integrating equations (\ref{6.1.17}) we get up to the equivalence
relation $E$ the formulae 10, 11 from (\ref{6.1.9}) under $C_1=0$ and 
$C_1\ne 0$, respectively. The theorem is proved. $\rhd$

Choosing in an appropriate way parameters and arbitrary functions we
can obtain from (\ref{6.1.4}) and (\ref{6.1.9}) Ans\"atze invariant
under the $P(1,3)$ non-conjugate three-dimensional subalgebras of the
algebra $AP(1,3)$ constructed in Section 2.2. Hence it follows, in
particular, that the classical Lie approach gives no complete
description of Ans\"atze reducing nonlinear PDE (\ref{6.1.1}) to ODEs.
Additional possibilities of reduction of equation (\ref{6.1.1}) are
the consequence of its conditional symmetry. To become convinced of
this fact we will construct involutive sets of the first-order
differential operators
\begin{eqnarray*}
        Q_a=\xi_{a\mu}(x)\p_\mu+\eta_a(x),\ \ a={1,2,3},
\end{eqnarray*}
where $\xi_{a\mu}(x)$ are real-valued scalar functions, $\eta_a(x)$
are $(4\times 4)$-matrices, such that Ans\"atze (\ref{6.1.4}),
(\ref{6.1.9}) are invariant with respect to these operators. Then, we
will show that the nonlinear Dirac equation (\ref{6.1.1}) is
conditionally-inva\-ri\-ant with respect to so obtained involutive
sets of differential operators.

According to Definition 1.5.2 Ansatz (\ref{6.1.4}) is invariant
with respect to the involutive set of operators $Q_1, \ Q_2, \ Q_3$ if
the conditions
\begin{equation}
      Q_a\psi(x)\equiv Q_a\Bigl (A(x)\vp (\om )\Bigr )=0, \
      a={1,2,3},
\label{6.1.18}
\end{equation}
where $A(x)=\exp \{\g _{\ssl A}\theta _{\ssl A}(\g _0+\g
_3)\}\exp\{(1/2)\theta _0\g _0\g _3+(1/2)\theta_3\g _1\g _2\}$, hold
with an arbitrary four-component function $\vp (\om )$.

Equating coefficients of $\vp (\om )$ and $\dot \vp (\om )$ in the
left-hand side of (\ref{6.1.18}) to zero we get
\begin{eqnarray}
        & &\xi_{a\mu}(x)\p_\mu \om(x)=0, \
        a={1,2,3},\label{6.1.19} \\
        & &\eta_a(x)=-\Bigl (\xi_{a\mu}(x)\p_\mu A\Bigr )A^{-1}, \ 
        a={1,2,3}. \label{6.1.20} 
\end{eqnarray}
Thus, to obtain the involutive set of operators $O_a$ such that the
Ansatz (\ref{6.1.4}) is invariant with respect to it we have

\begin{itemize}
\item{to solve equations (\ref{6.1.19}) which should be considered as
    a system of linear algebraic equations with respect to $\xi _{a\mu
      }(x), \ a={1,2,3}, \ \mu={0,\ldots,3}$;}
\item{to get explicit expressions for $\eta _a, \ a={1,2,3}$ from
    (\ref{6.1.20}).}
\end{itemize}

On solving equations (\ref{6.1.19}), (\ref{6.1.20}) for each class of
functions $\theta_{\mu }(x)$, \ $\om(x)$ from (\ref{6.1.9}) we obtain
the following sets of operators 
$Q_a$:\index{Conditional symmetry!of the Dirac equation}
\begin{eqnarray}
        &1)& Q_1=\p_1, \ Q_2=\p_2, \
             Q_3=x_0\p_3+x_3\p_0-(1/2)\g_0\g_3-C\Bigl (x_2\p_1 \non \\
             & &\quad -x_1\p_2+(1/2)\g_1\g_2\Bigr ); \non \\
        &2)& Q_1=(x_0+x_3)\p_1+x_1(\p_0-\p_3)+
             (1/2)\g_1(\g_0+\g_3), \non \\
             & &Q_2=(x_0+x_3)\p_2+x_2(\p_0-\p_3)+
             (1/2)\g_2(\g_0+\g_3), \non \\
             & &Q_3=x_0\p_3+x_3\p_0-(1/2)\g_0\g_3-
             C\Bigl (x_2\p_1-x_1\p_2+(1/2)\g_1\g_2\Bigr ); \non \\
        &3)& Q_1=\p_1, \ Q_2=(x_0+x_3)\p_2+x_2(\p_0-\p_3)+
             (1/2)\g_2(\g_0+\g_3), \non \\
             & &Q_3=x_0\p_3+x_3\p_0-(1/2)\g_0\g_3-
             C\Bigl (x_2\p_1-x_1\p_2+(1/2)\g_1\g_2\Bigr ); \non \\
        &4)& Q_1=\p_1, \ Q_2=\p_2, \
             Q_3=(1-C_2)\p_0+(1+C_2)\p_3 \non \\ 
             & &\quad -2C_1\Bigl
             (x_2\p_1-x_1\p_2+(1/2)\g_1\g_2\Bigr );\non \\ 
        &5)& Q_1=\p_0+\p_3, \ Q_2=\p_2, \non \\
              & &Q_3=\p_1+C\Bigl
             (x_0\p_3+x_3\p_0-(1/2)\g_0\g_3\Bigr ); \non \\
        &6)& Q_{\ssl A}=\p_{\ssl A}-\g_{\ssl B}(\p_{\ssl B}\p_{\ssl
          A}W)(\g_0+\g_3) 
             -2C(\p_{\ssl A}W)\Bigl (\g_1\g_2 \non \\ 
              & &\quad +2(\g_1\p_2W-\g_2\p_1W)(\g_0+\g_3)\Bigr ), \
             A= 1,2, \non \\
              & &Q_3=\p_0-\p_3-C\g_1\g_2-2C(\g_1\p_2W-
             \g_2\p_1W)(\g_0+\g_3); \non \\
        &7)& Q_1=\p_0-\p_3, \ Q_2=\tilde x_1\p_2-\tilde x_2\p_1
             -(1/2)(\g_1\g_2-C_1\g_0\g_3) \non \\
              & &\quad +C_1(\tilde x_1^2+\tilde
             x_2^2)^{-1/2}\exp\ \{-C_1 \arctan  (\tilde x_1/
             \tilde x_2)\} \non \\ 
          & &\quad \times (\g_2\tilde x_1-\g_1\tilde
          x_2)(\g_0+\g_3)+(1/2)(\g_1\dot w_2-\g_2\dot w_1)
          \label{6.1.21}\\  
          & &\quad \times(\g_0+\g_3) 
          -(C_1/2)\g_{\ssl A}\dot w_{\ssl A}(\g_0+\g_3), \non \\
          & &Q_3=\p_0+\p_3-2\dot w_{\ssl A}\p_{\ssl A}-\g_{\ssl
            A}\ddot w_{\ssl A}(\g_0+\g_3); \non \\
        &8)&Q_1=\p_0-\p_3, \ Q_2=\tilde x_1\p_2-\tilde
          x_2\p_1-(1/2)\g_1\g_2-C_1(\tilde x_1^2 \non \\
          & &\quad +\tilde x_2^2)^{-1/2}(x_0+x_1)^{-1}(\tilde x_1
          \g_2-\tilde x_2\g_1)(\g_0+\g_3) \non \\
          & &\quad -(1/2)(\g_1\dot w_2-\g_2\dot w_1)(\g_0+\g_3), \non
          \\ 
          & & Q_3=x_0\p_3+x_3\p_0-(x_0+x_3)\dot w_{\ssl A}
          \p_{\ssl A}-(1/2)\g_0\g_3 \non \\
          & &\quad -(1/2)\g_{\ssl A}\Bigl (\dot w_{\ssl
            A}+(x_0+x_3)\ddot w_{\ssl A}\Bigr )(\g_0 
          +\g_3)-(\tilde x_1^2+\tilde x_2^2)^{-1/2} \non \\
          & &\quad \times \Bigl (w_3+(x_0+x_3)\dot
          w_3\Bigr )(\g_1\tilde x_2 
          -\g_2\tilde x_1)(\g_0+\g_3); \non \\
        &9)& Q_{\ssl A}=\p_{\ssl A}-\g_{\ssl B}\Bigl [ \p_{\ssl
          B}\p_{\ssl A}\Bigl (U(x_1+ix_2,\, x_0+x_3) \non \\ 
          & &\quad +U(x_1-ix_2,\,
          x_0+x_3)\Bigr )\Bigr ](\g_0+\g_3)+\delta_{A1}\g_{\ssl
            B}W_{\ssl B} \non \\ 
          & &\quad \times (\g_0+\g_3), \ A= 1,2, 
          \ Q_3=\p_0-\p_3;\non \\
          &10)& Q_1=\p_0-\p_3, \
          Q_2=(\sin  w_2)\p_1-(\cos w_2)\p_2 \non \\
          & &\quad -\biggl [\g_1\Bigl [\Bigl ((1/2)\dot w_1+Ce^{-w_1}
          \Bigr ) 
          \sin w_2-(1/2)\dot w_2 \non \\
          & &\quad \times \cos  w_2\Bigr ]-\g_2\Bigl 
          [\Bigl ((1/2)\dot w_1+Ce^{-w_1}\Bigr )\cos  w_2 \non \\
          & &\quad +(1/2)\dot w_2\sin  w_2\Bigr ]\biggr ](\g_0+\g_3), \ 
          Q_3=\dot w_2(x_1\p_2-x_2\p_1) \non \\
          & &\quad +\dot w_3\Bigl ((\cos  w_2)\p_1+(\sin  w_2)
          \p_2\Bigr )+(1/2)(\p_0+\p_3) \non \\
          & &\quad -(x_1\sin  w_2-x_2\cos  w_2)\Bigl ((1/2)(\ddot
          w_1+\dot w_1^2) 
          (\g_1\sin w_2 \non \\
          & &\quad -\g_2\cos  w_2)-(1/2)(\ddot w_2+\dot w_1\dot w_2)
          (\g_1\cos  w_2 +\g_2\sin  w_2)\Bigr )\non \\ 
          & &\quad \times (\g_0+\g_3)-
          (\dot w_4+\dot w_1w_4)(\g_1\sin  w_2 -\g_2\cos 
          w_2)(\g_0+\g_3)\non \\ 
          & &\quad -(1/2)(\ddot w_3+\dot w_1\dot w_3)(\g_1\cos  w_2 
          +\g_2\sin w_2)(\g_0+\g_3)\non \\
          & &\quad -(1/2)\dot w_2\g_1\g_2-(1/2)\dot w_1\g_0\g_3, \non
          \\ 
          &11)& Q_1=\p_0-\p_3, \ Q_2=\p_2-(C/2)\g_0\g_3-
          (C/2)\g_{\ssl A}\dot w_{\ssl A}(\g_0+\g_3),\non \\
          & &Q_3=-\dot w_1\p_1+(1/2)(\p_0+
          \p_3)-(1/2)\g_{\ssl A}\ddot w_{\ssl A}(\g_0+\g_3) \non \\
          & & \quad -(C/2)\dot w_2 
          \g_0\g_3-(C/2)\dot w_2\g_{\ssl A}\dot w_{\ssl
            A}(\g_0+\g_3).\non  
\end{eqnarray} 

Analyzing the above formulae we come to a conclusion that only the
operators 1--5 from (\ref{6.1.21}) are linear combinations of the
generators of the Poincar\'e group $P(1,3)$ (\ref{1.1.20}).  The
remaining triplets of operators cannot be represented as linear
combinations of operators (\ref{1.1.20}).  Consequently, Ans\"atze
6--11 from (\ref{6.1.9}) are not invariant with respect to
three-parameter subgroups of the group $P(1,3)$ and cannot, in
principle, be constructed within the framework of the Lie approach.
They correspond to conditional symmetry of the nonlinear Dirac
equation (\ref{6.1.1}).

Let us consider as an example the eighth triplet of operators
$Q_1$, \ $Q_2$, \ $Q_3$. Rather tiresome computations yield the
following relations:
\begin{eqnarray}
        & & \tilde Q_1L=0, \non \\ 
        & &\tilde Q_2L=-2iC_1(\tilde x_1^2+\tilde
        x_2^2)^{-1/2}(x_0+x_3)  
        \Bigl ((\g_0+\g_3) Q_2\psi \non \\ 
        & &\quad +(\g_1\tilde x_2-\g_2\tilde x_1)Q_1 
        \psi\Bigr )+i(\g_1\dot w_2-
        \g_2\dot w_1)Q_1\psi+\Bigl (C_1(\tilde x_1^2 \non \\
        & &\quad +\tilde x_2^2)^{-1/2}(x_0+x_3)^{-1}
        (\tilde x_1\g_2-\tilde x_2\g_1)(\g_0+\g_3) \non \\
        & &\quad +(1/2)\g_1\g_2+(1/2)(\g_1\dot w_2-\g_2\dot w_1)
        (\g_0+\g_3)\Bigr )L, \label{6.1.22}\\
        & &\tilde Q_3L=2i\Bigl (w_3+(x_0+x_3)\dot w_3\Bigr )
        (\tilde x_1^2+\tilde x_2^2)^{-1/2}
        \Bigl ((\g_0+\g_3)Q_2\psi \non \\
        & &\quad +(\g_1\tilde x_2-\g_2\tilde x_1)Q_1\psi\Bigr )+
        \Bigl [(1/2)\g_0\g_3+(1/2)\g_{\ssl A}\Bigl (\dot w_{\ssl A}
        \non \\ 
        & &\quad +(x_0+x_3)\ddot w_{\ssl A}\Bigr )(\g_0+\g_3)+
        (\tilde x_1^2+\tilde x_2^2)^{-1/2}\Bigl (w_3+(x_0+x_3) \non \\ 
        & &\quad \times \dot w_3\Bigr )(\g_1\tilde x_2-\g_2\tilde x_1)
        (\g_0+\g_3)\Bigr ]L.\non\\
        & &[Q_1,\, Q_2]=[Q_2,\, Q_3]=0, \quad [Q_3,\, Q_1]=Q_1, \non 
\end{eqnarray}

In (\ref{6.1.22}) we designate by the symbol $\tilde Q_a$ the first   
prolongation of operator $Q_a$,\ $L=i\g_\mu\psi_{x_\mu}-\tilde 
f_1\psi-\tilde f_2\g_4\psi $.

Thus, the nonlinear Dirac equation (\ref{6.1.1}) is
conditionally-invariant  with respect to the involutive set 
of operators $Q_1, \ Q_2, \ Q_3$. 

Substitution of the Ans\"atze (\ref{6.1.4}), 
(\ref{6.1.9}) into (\ref{6.1.1}) gives rise to the following 
systems of ODEs for the four-component function $\vp=\vp(\om)$:
\begin{eqnarray*}
        &1)& (1/2)(\g_0+\g_3)(1+C\g_1\g_2)\vp+\Bigl (\g_0-\g_3+
           \om(\g_0+\g_3)\Bigr )\dot \vp=R, \\
        &2)& (1/2)(\g_0+\g_3)(3+C\g_1\g_2)\vp+\Bigl (\g_0-\g_3+
           \om(\g_0+\g_3)\Bigr )\dot \vp=R, \\
        &3)& (1/2)(\g_0+\g_3)(2+C\g_1\g_2)\vp+\Bigl (\g_0-\g_3+
           \om(\g_0+\g_3)\Bigr )\dot \vp=R, \\
        &4)& (C_1/2)(\g_0+\g_3)\g_4\vp+\Bigl (C_2(\g_0+\g_3)+
           \g_0-\g_3\Bigr )\dot \vp=R, \\
        &5)& (1/2)\g_2\g_4\vp+\Bigl (C\g_1+e^{-\om}
           (\g_0+\g_3)\Bigl )\dot \vp=R, \\
        &6)& -\Bigl (C(\g_0-\g_3)\g_4+4\tau_3(\om)(\g_0+\g_3)+
           8C |\tau_2(\om)|^2 \\
        & &\times (\g_0+\g_3)\g_4\Bigr )\vp+(\g_0+\g_3)\dot
           {\vp}=R, \\
        &7)& \om^{-1/2}\Bigl ((1/2)\g_2-C_2(\g_0+\g_3)+
           (C_1/2)\g_2\g_4\Bigl )\vp+2\om^{1/2}\g_2\dot \vp=R, \\
        &8)& (1/2)\Bigl ((1-2C_1\om^{-1/2})(\g_0+\g_3)+\om^{-1/2}
           \g_2\Bigr )\vp+2\om^{1/2}\g_2\dot \vp=R, \\
        &9)& (\g_0+\g_3)\Bigl (w_2(\om)\g_4-w_1(\om)\Bigr )\vp+
           (\g_0+\g_3)\dot \vp=R, \\
        &10)& -C(\g_0+\g_3)\vp+\g_1\dot \vp=R, \\
        &11)& -C\g_1\g_4\vp+\g_1\dot \vp=R, \\
\end{eqnarray*}
where $R=-i\Bigl (\tilde f_1(\bar \vp \vp,\bar \vp\g_4\vp)+
\tilde f_2(\bar \vp\vp, \bar \vp\g_4\vp)\g_4\Bigr )\vp$.

It is important to emphasize a very important difference between
Poincar\'e-invariant An\-s\"atze for the spinor field and
conditionally-invariant Ans\"atze given in (\ref{6.1.9}). As it was said above, 
$P(1,3)$-invariant Ans\"atze for the spinor field reduce any 
Poincar\'e-invariant spinor equation to systems of ODEs, provided the
generators of the Poincar\'e group have the form (\ref{1.1.20}). But
for Ans\"atze (\ref{6.1.9}) it is not the case. Each specific
equation gives rise to a specific system of PDEs for functions
$\theta _{\mu }, \ \om $. This means that the approach suggested makes
it possible to take into account a structure of solutions of the
equation under study more precisely than the Lie approach does.

It is worth noting that the formula (\ref{6.1.4}) can be easily 
adapted to the case of a field with an arbitrary spin $s$. Let us
rewrite it in the following way:\index{Ansatz!$P(1,3)$-invariant}
\begin{equation}
        \psi(x)=\exp\{2\theta_{\ssl A}(S_{0{\ssl A}}-S_{{\ssl
            A}3})\}\exp\{\theta_0S_{03} + \theta_3S_{12}\}\vp(\om),
\label{6.1.23}
\end{equation}
where $S_{\mu\nu}=(1/4)(\g_\mu\g_\nu-\g_\nu\g_\mu)$. Ansatz
(\ref{6.1.23}) can be applied to reduce any Poincar\'e-invariant
equation (by means of the method described above) provided it admits
the group $P(1,3)$ with the following generators:
\begin{eqnarray*}
        P_\mu=\p^\mu, \quad J_{\mu\nu}=x_{ \mu }P_\nu-x _{\nu}
        P_\mu+S_{\mu \nu }.
\end{eqnarray*}

Here $S_{\mu\nu}$ are constant matrices of the corresponding dimension 
satisfying the commutation relations of the Lie algebra $AO(1,3)$.
\vspace{2mm}

\noindent
{\bf 2. Non-Lie reduction of spinor equations invariant under the
  extended Poincar\'e group.} We look for solutions of the nonlinear
$\tilde P(1,3)$-invariant equations
\begin{equation}
       \Bigl \{i\g_\mu\p_\mu-(\bar \psi\psi)^{1/2k}\Bigl [g_1\Bigl
       (\bar \psi\psi 
       (\bar \psi\g_4\psi)^{-1}\Bigr )+g_2\Bigl (\bar \psi \psi
        (\bar \psi\g_4\psi)^{-1}\Bigr )\g _4\Bigr ]\Bigl \}\psi=0
\label{6.1.24}
\end{equation}
in the 
form\index{Ansatz!for spinor field}\index{Ansatz!$\wid P(1,3)$-invariant}
\index{Non-Lie reduction!of Poincar\'e-invariant spinor equations}
\begin{equation}
        \psi(x)=\exp\{\theta_0+\g_{\ssl A}\theta_{\ssl
          A}(\g_0+\g_3)\}\vp(\om), 
\label{6.1.25}
\end{equation}
where $\theta_0, \ \theta_1, \ \theta_2, \ \om$ are arbitrary
smooth real-valued functions of $x_0+x_3,\  x_1$,
\linebreak $x_2$; \ $\vp$ is an
arbitrary complex-valued four-component function.

Substituting the Ansatz (\ref{6.1.25}) into (\ref{6.1.24}) and
multiplying the expression obtained by the matrix $\exp\{-\theta_0
-\theta_{\ssl A}\g_{\ssl A}(\g_0+\g_3)\}$ yield
\begin{eqnarray*}
        iR_1(x)\vp+iR_2(x)\dot \vp=(\bar
        \vp\vp)^{1/2k}(g_1+g_2\g_4)\vp, 
\end{eqnarray*}
where
\begin{eqnarray*}
        g_{\ssl A}&=&g_{\ssl A}\Bigl (\bar \vp\vp(\bar \vp
        \g_4\vp)^{-1}\Bigr ),\ \ A= {1,2}, \\
        R_1&=&(\g_0+\g_3)\p_\xi\theta_0+\g_{\ssl A}\p_{\ssl
          A}\theta_0 + \g_{\ssl A}\g_{\ssl B}\p_{\ssl A}\theta_{\ssl
          B}(\g_0+\g_3)-2\theta_{\ssl A}\p_{\ssl A} 
              \theta_0(\g_0+\g_3),\\
        R_2&=&(\g_0+\g_3)(\p_\xi\om-2\theta_{\ssl A}\p_{\ssl
          A}\om)+\g_{\ssl A}\p_{\ssl A}\om 
\end{eqnarray*}
(as earlier, the notation $\xi=x_0+x_3$ is used).

Consequently, Ansatz (\ref{6.1.25}) reduces the initial equation
(\ref{6.1.24}) to a system of ODEs if there exist such $(4\times
4)$-matrices $G_1(\om), \ G_2(\om)$ that $R_{\ssl A}(x)=G_{\ssl A}(\om), 
\ A=1,2$. Hence we get the system of nonlinear PDEs for unknown
functions $\theta_0, \ \theta_1, \ \theta_2, \ \om$
\begin{eqnarray}
        &1)& (\p_\xi-2\theta_{\ssl A}\p_{\ssl A})\theta_0-\p_{\ssl
          A}\theta_{\ssl A}=f_1(\om)\exp \{\theta_0k^{-1}\}, \non \\
        &2)& \p_1\theta_0=f_2(\om)\exp\{\theta_0k^{-1}\}, \non \\
        &3)& \p_2\theta_0=f_3(\om)\exp\{\theta_0k^{-1}\}, \non \\
        &4)& \p_2\theta_1-\p_1\theta_2=f_4(\om)\exp\{\theta_0k^{-1}\},
        \label{6.1.26}\\ 
        &5)& (\p_\xi-2\theta_{\ssl A}\p_{\ssl
          A})\om=f_5(\om)\exp\{\theta_0k^{-1}\}, \non \\ 
        &6)& \p_1\om=f_6(\om)\exp\{\theta_0k^{-1}\}, \non \\
        &7)& \p_2\om=f_7(\om)\exp\{\theta_0k^{-1}\}.\non
\end{eqnarray}

In (\ref{6.1.26}) $f_1,\ldots,f_7$ are arbitrary smooth real-valued
functions. 

Solutions of the above system of nonlinear PDEs are looked for up to
the equivalence relation $E$ which is introduced in the following way.
We say that the solutions of equations (\ref{6.1.26}) $\theta _0(x), \ 
\theta _{\ssl A}(x), \ \om (x)$ and $\theta ^{\prime}_0(x), \ \theta
^{\prime}_{\ssl A}(x), \ \om ^{\prime}(x)$ are equivalent if they are
transformed one into another by

1) a suitable transformation from the group $\widetilde G_8$, which is
induced in the space of variables $ x,\ \theta _0(x)$,\ $\theta
_{\ssl A}(x)$,\ $\om (x)$ by the action of the transformation group
$G_8 \subset \widetilde P(1,3)$ with generators $P_{\mu }$,\ 
$J_{01}-J_{13}$,\ $J_{02}-J_{23}$,\ $J_{12}$,\ $D$ on Ansatz
(\ref{6.1.25}), or

2) a suitable transformation of the form
\begin{equation}
        \om \to h(\om), \quad \theta_0\to \theta_0+h_0(\om), \quad
        \theta_{\ssl A}\to\theta_{\ssl A}+h_{\ssl A}(\om),
\label{6.1.27}
\end{equation}
where $\{h, h_0, h_1, h_2\} \subset C^1(\R^1,\R^1)$.

Due to the fact that system (\ref{6.1.26}) is over-determined we
have succeeded in constructing its general solution. Up to the 
equivalence relation $E$ it is given by one of the formulae  
\begin{eqnarray}
        &1)& \theta_0=k \ln     w_1, \quad \theta_1=(2w_1)^{-1} (\dot
          w_1x_1+\dot w_2), \non \\ 
          & &\theta_2=(2w_1)^{-1}\Bigl ((2k-1) \dot w_1x_2+w_3\Bigr ),
          \quad  
          \om=w_1x_1+w_2; \non \\ 
        &2)&  \theta_0=-k \ln     (x_0+w_1), \quad
          \theta_{\ssl A}=w_3\Bigl ((x_1+w_1)^2 \non \\
          & &\quad +(x_2+w_2)^2\Bigr )^{k-1} 
          (x_{\ssl A}+w_{\ssl A})+(1/2)\dot w_{\ssl A},\ \ A= 1,2,
          \label{6.1.28}\\  
          & &\om =(x_1+w_1)(x_2+w_2)^{-1}; \non \\
        &3)& \theta_0=0, \quad  \om=x_0+x_3, \quad
          \theta_{\ssl A}=\p_{\ssl A}\Bigl (U(x_1+ix_2,\, x_0+x_3)
          \non \\ 
          & &\quad +U(x_1-ix_2, \ x_0+x_3)\Bigr )+w_{\ssl A}x_1, \
          A=1,2.\non 
\end{eqnarray}

Here $w_1, \  w_2, \  w_3$ are arbitrary smooth real-valued functions
of $x_0+x_3; \ U$ is an arbitrary function analytic in the first
variable.  

Substitution of Ans\"atze (\ref{6.1.25}), (\ref{6.1.28}) into equation
(\ref{6.1.24}) gives rise to the following systems of ODEs:
\begin{eqnarray*}
        &1)& i\g_1\dot \vp=R, \\
        &2)& i(\g_2-\om \g _1)\dot \vp=R, \\
        &3)& i(\g_0+\g_3)\dot \vp+(\g_0+\g_3)(w_2\g_1
             \g_2-w_1)\vp=R, 
\end{eqnarray*}
where $R=(\bar \vp \vp)^{1/2k}\Bigl [g_1\Bigl (\bar \vp \vp
(\bar \vp\g_4\vp)^{-1}\Bigr )+\g_4g_2\Bigl (\bar \vp\vp
(\bar \vp \g_4\vp)^{-1}\Bigr )\Bigr ]\vp$.

Generally speaking, Ans\"atze (\ref{6.1.25}), (\ref{6.1.28}) are
not invariant with respect to the three-parameter subgroups of the
group $\tilde P(1,3)$ (description of inequivalent $\tilde
P(1,3)$-invariant Ans\"atze for the spinor field is given in 
Section 2.2). In the case involved we deal with reduction via
conditionally-invariant Ans\"atze. For example, the involutive set of
operators $Q_a$ corresponding to the Ansatz 1 from (\ref{6.1.28}) is
of the form\index{Conditional symmetry!of the Dirac equation}
\begin{eqnarray*}
       & & Q_1=(1/2)(\p_0-\p_3), \quad 
           Q_2=w_1\p_2+(1/2)(1-2k)\dot w_1\g_2(\g_0+\g_3), \\
       & & Q_3=(1/2){w_1}(\p_0+\p_3)-\dot w_1 x_{\ssl A}\p_{\ssl
         A}-\dot w_2\p_1 - k\dot w_1+(2w_1)^{-1} \\
           & &\quad \times (2\dot w_1^2-w_1\ddot
              w_1)\Bigl (\g_{\ssl A}x_{\ssl A}+2(k-1)\g_2x_2\Bigr
              )(\g_0 + \g_3)+(2w_1)^{-1} \\
           & &\quad \times \Bigl ((2\dot w_1\dot w_2-w_1
              \ddot w_2)\g_1+(w_3\dot w_1-w_1\dot w_3)\g_2\Bigr )
              (\g_0+\g_3).
\end{eqnarray*}

The above operators satisfy the following relations:
\begin{eqnarray*}
        & &[Q_1,\, Q_2]=[Q_1,\, Q_3]=0, \quad [Q_2,\, Q_3]=-2w_1Q_2,\\
        & &\tilde Q_1L=0, \quad
        \tilde Q_2L=A_1L+A_2Q_1\psi+A_3Q_2\psi,\\
        & &\tilde Q_3L=B_0L+B_aQ_a\psi,
\end{eqnarray*}
where $\tilde Q_a$ is the first prolongation of operator $Q_a; \,
L=i\g_\mu \p_\mu\psi -(\bar \psi\psi)^{1/2k}(g_1+g_2\g_4)\psi$; \ 
$A_a$, \ $B_0$, \ $B_a$ are some variable $(4\times 4)$-matrices.
Hence it follows that the nonlinear Dirac equation (\ref{6.1.24}) is
conditionally-invariant with respect to the involutive set of
operators $Q_1, \ Q_2, \ Q_3$.

In conclusion we adduce the two classes of new exact solutions of the
nonlinear spinor equation
\begin{eqnarray*}
        \Bigl (i\g_\mu\p_\mu-\lbd (\bar \psi\psi)^{1/2k}\Bigr )\psi=0
\end{eqnarray*}
constructed with the use of conditionally-invariant Ans\"atze
(\ref{6.1.9}), (\ref{6.1.28})
\index{Exact solutions!of the nonlinear Dirac equation}
\begin{eqnarray*}
      \psi(x)&=&\exp\biggl \{\g_1(\g_0+\g_3)\biggl [(x_1\sin  w_2-x_2
                \cos  w_2)\Bigl [\Bigl ((1/2)\dot w_1 \\
             & &+Ce^{-w_1}\Bigr )\sin  w_2-(1/2)\dot
             w_2\cos  w_2\Bigr ]+w_4\sin  w_2+(1/2)\dot w_3 \\
             & &\times \cos  w_2\biggr ]-\g_2(\g_0+\g_3)
             \biggl [(x_1\sin  w_2-x_2\cos  w_2)\Bigl 
             [\Bigl ((1/2)\dot w_1 \\& &+Ce^{-w_1}\Bigr )
             \cos  w_2+(1/2)\dot w_2\sin w_2\Bigr ] 
             +w_4\cos  w_2 \\
             & &-(1/2)\dot w_3\sin  w_2\biggr ]\biggr
             \}\exp\{(1/2)w_1\g_0\g_3  
             +(1/2)w_2\g_1\g_2\} \\
             & &\times \exp\Bigl \{\Bigl (i\lbd(\bar
             \chi\chi)^{1/2k}\g_1-C\g_1(\g_0+\g_3)\Bigr )
             (x_1\cos  w_2 \\
             & &+x_2\sin w_2+w_3)\Bigr \}\chi, \\
      \psi(x)&=&w_1^k\exp\Bigl \{(2w_1)^{-1}\Bigl [(\dot w_1x_1+\dot
             w_2)\g_1+ 
             \Bigl ((2k-1)\dot w_1x_1+w_3\Bigr )\g_2\Bigr ] \\
             & &\times (\g_0+\g_3)\Bigr \}\exp\{i\lbd\g_1(\bar \chi 
             \chi )^{1/2k}
             (w_1x_1+w_2)\}\chi.
\end{eqnarray*}

Here $w_1,w_2,w_3,w_4$ are arbitrary smooth real-valued functions 
of $x_0+x_3; \ \chi $ is an arbitrary four-component constant column.
\vspace{10mm}

\noindent
{\large \bf 6.2. Non-Lie reduction of Galilei-invariant spinor
  equations\label{s6.2}} 
\markboth{Chapter 6. CONDITIONAL SYMMETRY AND REDUCTION}
   {6.2. Non-Lie reduction of Galilei-invariant spinor equations}
\def\theequation{6.\arabic{section}.\arabic{equation}}
\setcounter {section} {2}
\setcounter {equation}{0}
\vspace{7mm}

\noindent
Taking into account the classical ideas and methods of symmetry
analysis of differential equations we generalize results obtained in
the previous section in the form of the following non-Lie algorithm of 
reduction of PDEs:\index{Non-Lie!reduction}
\begin{itemize}
\item{the maximal (in Lie sense) invariance group of the equation
    under study is found by the Lie method;}
\item{subgroup analysis of the invariance group is carried out, each
    subgroup giving rise to some Ansatz which reduces PDE in question
    to an equation having a smaller dimension. As a rule, Ans\"atze
    obtained in this way have a quite definite structure which is
    determined by the representation of the symmetry 
group.}
\item{the general form of the invariant Ansatz is obtained. This
    Ansatz includes several scalar functions $\theta_1, \ldots,
    \theta_N$ satisfying some compatible over-determined system of
    nonlinear PDEs (reduction conditions).} 
\item{equations for $\theta_1,\ldots,\theta_N$ are integrated.}
\end{itemize}

Let us realize the above algorithm for the following system of
nonlinear spinor PDEs:
\begin{equation}
        \{-i(\g_0+\g_4)\p_t+i\g_a\p_a+m(\g_0-\g_4)-
        F(\psi^*,\psi)\}\psi=0,
\label{6.2.1}
\end{equation}
where $F$ is a variable $(4\times 4)$-matrix.

According to Theorem 4.1.5 equation (\ref{6.2.1}) is invariant
under the Galilei group iff
\begin{equation}
        F=\tilde f_1(\bar \psi \psi,\, \psi ^\dagger\psi+\bar \psi
        \g_4\psi ) + \tilde f_2(\bar \psi \psi,\,
          \psi ^\dagger\psi+\bar \psi \g_4\psi )(\g_0+\g_4),
\label{6.2.2}  
\end{equation}
where $\{\tilde f_1, \tilde f_2\} \subset C^1(\R^2,\C^2)$ are arbitrary
functions. In Section 4.2 we have constructed
$G(1,3)$-inequivalent Ans\"atze for the spinor field $\psi(t,\vec
x)$ invariant under three-parameter subgroups of the Galilei
group. One can become convinced of the fact that these Ans\"atze have 
the 
form\index{Ansatz!for spinor field}\index{Ansatz!$G(1,3)$-invariant}
\index{Non-Lie reduction!of Galilei-invariant spinor equations}
\begin{equation}
        \psi(t,\vec x)=\exp \{i\theta_0+\g_a\theta_a(\g_0+\g_4)\}
        \exp \{\theta_4\g_1\g_2\}\vp(\om),
\label{6.2.3}
\end{equation}
where $\theta_\mu, \ \theta_4, \ \om$ are smooth real-valued functions  
on $t,\ \vec x; \ \vp=\vp(\om)$ is an arbitrary complex-valued
four-component function. 

In the following, we will describe all Ans\"atze (\ref{6.2.3}) with
$\theta_4=0$ reducing the Galilei-invariant equation (\ref{6.2.1}),
(\ref{6.2.2}) to systems of ODEs.

Substituting (\ref{6.2.3}) with $\theta_4=0$ into (\ref{6.2.1}), 
(\ref{6.2.2}) and requiring for the obtained equation be 
equivalent to a system of ordinary differential equations for 
$\vp(\om)$ we have
\begin{eqnarray}
        &1)& \p_2\theta_3-\p_3\theta_2=f_1(\om), \non \\
        &2)& \p_3\theta_1-\p_1\theta_3=f_2(\om), \non \\
        &3)& \p_1\theta_2-\p_2\theta_1=f_3(\om), \non \\
        &4)& \p_a\theta_a=f_4(\om), \label{6.2.4}\\
        &5)& (\p_t+2\theta_a\p_a)\theta_0+4m\theta_a\theta_a=f_5(\om), 
        \non \\ 
        &6)& (\p_t+2\theta_a\p_a)\om=f_6(\om), \non \\
        &7)& \p_a\om=f_{6+a}(\om), \non \\
        &8)& \p_a\theta_0+4m\theta_a=f_{9+a}(\om).\non
\end{eqnarray}

Here $f_1,\ldots,f_{12}$ are arbitrary smooth real-valued functions,
$a={1,2,3}$.

As earlier (see Section 6.1), we introduce an equivalence relation $E$
on the set of solutions of system of PDEs (\ref{6.2.4}). We say that
solutions of equations (\ref{6.2.4}) $\theta _0(t, \vec x), \ \theta
_a(t, \vec x), \ \om (t, \vec x)$ and $\theta _0^{\prime }(t, \vec x),
\ \theta _a^{\prime }(t, \vec x), \ \om ^{\prime }(t, \vec x)$ are
equivalent if they are transformed one into another by

1) a suitable transformation from the group $\widetilde G_{11}$ which
is induced in the space of the variables $t, \ \vec x, \ \theta _0(t,
\vec x), \ \theta _a(t, \vec x), \ \om (t, \vec x)$ by the action of
the Galilei group $G(1,3)$ on Ansatz (\ref{6.2.3}), or

2) a suitable transformation of the form 
\begin{eqnarray*}
        & &\theta_0\to \theta_0+h_0(\om),\\
        & &\theta_a\to \theta_a+h_a(\om),\\
        & &\om\to h(\om),
\end{eqnarray*}
where $\{h_\mu, h\} \subset C^1(\R^1,\R^1)$ are arbitrary functions.
\vspace{1.5mm}

\noindent
{\bf Theorem 6.2.1.}\ {\em General solution of system of PDEs
  (\ref{6.2.4}) determined up to the equivalence relation $E$ is given 
  by one of the following formulae:
\index{Conditional symmetry!of Galilei-invariant spinor equations}
\begin{eqnarray*}
        &{\rm I.}& m=0 \\
        &1)& \om=x_1+w_1(t), \quad 
             \theta_0=C_3\Bigl
             (x_2-2w_2(t)\Bigr)+C_4\Bigl(x_3-2w_3(t)\Bigr)+C_5t, \\ 
        & &  \theta_1=-(1/2)\dot w_1(t), \quad
             \theta_2=-\alpha(C_3x_2+C_4x_3)+\dot w_2(t)+C_1x_2, \\
        & &  \theta_3=\alpha(C_3x_3-C_4x_2)+\dot w_3(t)+C_2x_2, \quad 
             \alpha=(C_1C_3+C_2C_4) \\
        & &  \quad \times (C_3^2+C_4^2)^{-1}; \\
        &2)& \om=x_1+w_0(t), \quad \theta_0=C_3t, \quad
        \theta_1=-(1/2)\dot w_0(t), \\ 
        & &  \theta_2=U(x_2+ix_3,t)+U(x_2-ix_3,t)+C_1x_2, \\
        & &  \theta_3=iU(x_2+ix_3,t)-iU(x_2-ix_3, t)+C_2x_2; \\
        &3)& \om=t, \quad \theta_0=x_ag_a(t), \quad
        \theta_a=\varepsilon_{abc}h_b(t)x_c+ 
             \p_aW+w_0(t)x_a, \\
        & &  { function} \ W=W(t,\vec x) \  
             { being \ given} \
             { by \ one \ of \ the \ relations \ a \ - \ c} \\
        &a)& { under} \ g_1=g_2=g_3=0 \\ 
        & &  \p_a\p_aW=0; \\
        &b)& { under} \ g_2=0, \ g_3\ne 0 \\
        & &  W=g_3^{-1}\Bigl (r_1x_1x_3+r_2x_2x_3+r_4x_3+(1/2)r_3x_3^2
             -(1/2)g_3^{-1}g_1r_1x_3^2 \\
        & &  \quad +(1/2)(g_3^{-1}g_1r_1-r_3)x_2^2\Bigr) +
             U(z,t)+U(z^*,t), \\  
        & &  z=(g_1^2+g_3^2)^{-1/2}(g_1x_3-g_3x_1)+ix_2;\\
        &c)& { under} \ g_1^2+g_2^2\ne 0, \ g_3\ne 0 \\
        & &  W=(1/2)g_3^{-2}r_1(2g_3x_1x_3-g_1x_3^2)+(1/2)g_3^{-2}
             r_2(2g_3x_2x_3 \\
        & &  \quad -g_2x_3^2)+(1/2)g_3^{-1}(r_3x_3+2r_4x_3)+(1/2)
             g_3^2(g_1^2+g_2^2)^{-1}(r_1g_1 \\
        & &  \quad +r_2g_2-r_3g_3)(g_2x_1-g_1x_2)^2+U(z,t)+U(z^*,t),
        \\ 
        & &  z=\Bigl ((g_1^2+g_2^2)^{-1}(g_2^2+g_3^2)-g_1^2g_3^2
             (g_1^2+g_2^2)^{-2}\Bigr )^{1/2}(g_2x_1-g_1x_2) \\
        & &  \quad +i\Bigl (g_1g_3(g_1^2+g_2^2)^{-1}(g_2x_1-g_1x_2)
             +g_3x_2-g_2x_3\Bigr ), \\
        & &  { where} \\ 
        & &  r_a=-\Bigl (g_aw_0+\varepsilon_{abc}g_bh_c + (1/2)\dot
             g_a\Bigr ), \quad r_4=g_0. \\
        &{\rm II.}& m\ne 0 \\
        &1)& \om=x_1+(4m)^{-1}C_5t^2+C_7t, \quad
             \theta_0=(2mC_7+C_5t)\om \\
        & &  \quad +(C_3-4mC_1)x_2+(C_4-4mC_2)x_3-(12m)^{-1}C_5^2t^3
             -(1/2)C_5C_7t^2\\
        & &  \quad +C_6t, \quad  \theta_1=-(4m)^{-1}C_5t-(1/2)C_7,
        \quad  
             \theta_2=C_1, \quad \theta_3=C_2; \\
        &2)& \om=t, \quad \theta_0=-2mR_0x_ax_a+R_ax_a-4m(T_{ab}
             x_ax_b+T_ax_a), \\
        & &  \theta_a=R_0x_a+2T_{ab}x_b+T_a,
\end{eqnarray*}
where $R_0(t), \ R_b(t), \ T_{bc}(t), \ T_b(t)$ are real-valued
functions satisfying the Ric\-ca\-ti-type systems of ODEs
\index{Riccati system}
\begin{eqnarray*}
        & &(\dot R_0+2R_0^2)\delta _{ab}+2\dot T_{ab}+8T_{ac}T_{bc}+
           8R_0T_{ab}=0, \\
        & &\dot R_a-4m\dot T_a-8mR_0T_a-16 mT_{ab}T_b+
           4T_{ab}R_b+2R_0R_a=0 
\end{eqnarray*}
and besides
\begin{displaymath}
T_{ab}=T_{ba},\quad T_{11}+T_{22}+T_{33}=0.
\end{displaymath}

In the above formulae $w_0, \ w_a, \ g_0, \ g_a, \ h_a$ are arbitrary
smooth real-valued functions of $t$; \ $a, b, c={1,2,3}$; \ $U$
is an arbitrary function analytic in the variable $z$; $C_1,\ C_2,
\ldots, C_7$ are arbitrary constants.}

A detailed proof of this assumption can be found in \cite{7,211}.

Substantial extension of the class of Ans\"atze (\ref{6.2.3}) reducing
nonlinear PDE (\ref{6.2.1}), (\ref{6.2.2}) to systems of ODEs as
compared with the class of Lie Ans\"atze (see Section 4.2) is achieved
due to the conditional symmetry\index{Conditional!symmetry} of
equation (\ref{6.2.1}).

Computing involutive sets of operators $Q_a=\xi _{a\mu }(t, \vec
x)\p _{\mu } + \eta _a(t, \vec x) \ (\p _0 \equiv \p _t), \ 
a={1,2,3}$ with the use of formulae (\ref{6.1.19}),
(\ref{6.1.20}), $(4\times 4)$-matrix $A=A(t, \vec x)$ and scalar
function $\om (t, \vec x)$ being determined by the formulae
I.1--II.2, we can become convinced of that Ans\"atze I.2, I.3
correspond to conditional symmetry of system of PDEs (\ref{6.2.1}), 
(\ref{6.2.2}).

Substitution of the Ans\"atze obtained above into the initial equation
(\ref{6.2.1}), (\ref{6.2.2}) yields systems of ODEs for a
four-component function $\vp(\om)$
\begin{eqnarray*}
        {\rm I.}& 1)& i\g_1\dot\vp+i\Bigl ((C_2\g_1-C_1-iC_5)
         (\g_0+\g_4)+iC_3\g_2+iC_4\g_3\Bigr )\vp=R, \\
         &2)& i\g_1\dot \vp+i(C_2\g_1-C_1-iC_3)
         (\g_0+\g_4)\vp=R, \\
         &3)&-i(\g_0+\g_4)\dot \vp+i\Bigl ((2h_a
         \g_a-3w_0-ig_0)(\g_0+\g_4) \\
         & &+ig_a\g_a\Bigr )\vp=R, \\
        {\rm II.}& 1)& i\g_1\dot \vp+\Bigl ((C_5\om+C_6
         -4mC_1^2-4mC_2^2+mC_7^2+2C_1C_3 \\
         & &+2C_2C_4)(\g_0+\g_4)-C_3\g_2-C_4\g_3+
         m(\g_0-\g_4)\Bigr )\vp=R, \\
         &2)&-i(\g_0+\g_4)\dot \vp+\Bigl (-3iR_0(\g_0+\g_4)+
         (2R_aT_a-4mT_aT_a) \\
         & &\times (\g_0+\g_4)-R_a\g_a+m(\g_0-
         \g_4)\Bigr )\vp=R,
\end{eqnarray*}

Here $R=\Bigl (\tilde f_1(\bar \vp\vp,\vp^\dagger\vp+\bar \vp\g_4\vp)
+\tilde f_2(\bar \vp\vp, \vp^\dagger\vp+\bar \vp\g_4\vp)(\g_0
+\g_4)\Bigr )\vp;$ \ $w_0$, \ $w_a$, \ $h_a$, \ $g_0$, \ $g_a$, \ 
$T_a$, \ $R_a$ are functions of $\om$ determined in Theorem 6.2.1; \ 
$C_1,\ldots,$ $C_7$ are constants.

A particular or general solution $\vp=\vp(\om)$ of one of the
above equations after being substituted into corresponding Ansatz 
(\ref{6.2.3}) gives rise to a class of exact solutions of the 
initial nonlinear PDE.

As an example, we adduce the class of solutions of system of 
nonlinear PDEs (\ref{6.2.1}), (\ref{6.2.2}) with $m=0, \ \tilde f_1=0, 
\ \tilde f_2=\lbd(\psi^\dagger\psi+\bar \psi\g_4\psi)^k,$
$\lbd=\mbox{\rm const}, \ k=\mbox{\rm const}$ constructed with use 
of the Ansatz I.3 
\index{Exact solutions!of nonlinear Galilei-invariant spi\-nor equations}
\begin{eqnarray*}
        \psi(x)=\exp\{i\lbd(\chi^\dagger\chi+\bar \chi\g_4\chi)^k t+
        (\g_0+\g_4)\g_a\p_aW\}\chi,
\end{eqnarray*}
where $\chi$ is an arbitrary constant four-component column,
$W=W(t,\vec x)$ is an arbitrary solution of the three-dimensional
Laplace equation\index{Laplace equation} 
\begin{eqnarray*}
        \Delta_3W=\p_a\p_aW=0.
\end{eqnarray*}

The approach to the problem of reduction of $G(1,3)$-invariant
equations for the spinor field of the form (\ref{6.2.1}) suggested
above can be generalized for the case of an arbitrary
Galilei-invariant system of PDEs admitting the group $G(1,3)$ with
generators
\begin{eqnarray*}
      & &  P_0=\p_t,\quad P_a=\p_a, \\
      & &  J_{ab}=x_b\p_a-x_a\p_b+S_{ab},\quad \\
      & &  G_a=t\p_a+i\lbd x_a+\eta_a, 
\end{eqnarray*}
where $\lbd=\mbox{\rm const}; \ a,b={1,2,3}; \ S_{ab},\ \eta_a$ 
are arbitrary constant matrices satisfying the commutation relations 
of the Lie algebra $AE(3)$. Exact solutions of such a system are 
looked for in the form
\begin{eqnarray*}
\psi(t,\vec
x)=\exp\{\theta_0+\theta_a\eta_a\}\exp\{\theta_4S_{12}\}\vp(\om), 
\end{eqnarray*}
where $\{\theta_0, \theta_1, \theta_2, \theta_3, \theta_4, \om\} 
\subset C^1(\R^4,\R^1)$.

\newpage
\thispagestyle{empty}

\noindent
{\sl
C H A P T E R \ \  7\label{ch7}}
\vspace{2mm}

\hrule
\vspace{35mm}

\rightline
{\large\bf
REDUCTION AND EXACT SOLUTIONS}
\vspace{2mm}

\rightline
{\large\bf
OF {\boldmath $SU$}(2) YANG-MILLS EQUATIONS}
\vspace{7mm}

\noindent
In the present chapter a detailed account of symmetry properties of
$SU(2)$ Yang-Mills equations is given. Using a subgroup structure of
the Poincar\'e and conformal groups we have constructed all
$C(1,3)$-inequivalent Ans\"atze for the Yang-Mills field which are
invariant under three-parameter subgroups of the Poincar\'e
group.  With the aid of these Ans\"atze reduction of Yang-Mills
equations to systems of ordinary differential equations is carried out
and wide families of their exact solutions are obtained. A number of
generalizations of the Lie Ans\"atze are suggested making it possible
to construct broad families of exact solutions of the Yang-Mills
equations containing arbitrary functions. It is shown that a
possibility of such generalizations is provided by nontrivial
conditional symmetry of the Yang-Mills equations.
\vspace{10mm}

\noindent
{\large\bf 7.1. Symmetry reduction and exact solutions of the 
\vspace{1.5mm}

\noindent
\phantom{\large\bf 7.1. }Yang-Mills equations\label{s7.1}} 

\markboth {Chapter 7. REDUCTION AND EXACT SOLUTIONS}
{7.1. Symmetry reduction and exact solutions}
\def\theequation{7.\arabic{section}.\arabic{equation}}
\setcounter {section} {1}
\setcounter {equation}{0}
\vspace{7mm}

\noindent
{\bf 1. Introduction.}\ A majority of papers devoted to construction
of the explicit form of exact solutions of the SU(2) Yang-Mills
equations (YMEs)\index{Yang-Mills!equations}
\begin{equation}
\begin{array}{l}
\partial_{\nu}\partial^{\nu}{\vec  A }_{\mu} - \partial^{\mu} 
\partial_{\nu}{\vec A }_{\nu} + e \Bigl(( \partial_{\nu}
{\vec A }_{\nu})\times{\vec A }_{\mu} -
2(\partial_{\nu}{\vec A}_{\mu})\times{\vec A}_{\nu} 
\\[2mm]
\quad + (\partial^{\mu}{\vec A}_{\nu})\times{\vec A}^{\nu}\Bigr)
+ e^{2}{\vec A}_{\nu}\times({\vec A}^{\nu}\times{\vec A}_{\mu}) =
\vec{0}.
\end{array}
\label{1.1}
\end{equation}
are based on the Ans\"atze for the three-component vector-potential of
the Yang-Mills field $\vec{A}_{\mu}(x_{0},\, x_{1},\, x_{2},\, x_{3})$
(called, for brevity, the Yang-Mills field)\index{Yang-Mills!field}
suggested by Wu and Yang, Rosen, 't Hooft, Corrigan and Fairlie,
Wilczek, Witten (see \cite{1.1} and references therein). And what is
more, Ans\"atze mentioned are obtained in a non-algorithmic way, i.e.,\ 
there is no regular and systematic method for constructing these
Ans\"atze.

Since there are only a few distinct exact solutions of YMEs, it is
difficult to give their reliable and self-consistent physical
interpretation. That is why the problem of prime importance is the 
development of an effective regular approach for constructing new
exact solutions of system of nonlinear PDEs (\ref{1.1}).

A natural approach to construction of particular solutions of YMEs
(\ref{1.1}) is to utilize their symmetry properties. Apparatus of the
theory of Lie transformation groups makes it possible to reduce system
of PDEs (\ref{1.1}) to systems of ODEs by using invariant Ans\"atze.
If we succeed in constructing its general or particular solutions,
then substituting the results obtained into the corresponding
Ans\"atze we obtain exact solutions of YMEs.  Let us note that
symmetry reductions of the Euclidean self-dual YMEs (which form the
first-order system of PDEs) by means of the subgroups of the Euclid
group\index{Euclid!group} $E(4)$ have been performed in the paper
\cite{138.1}. It is interesting to note that many integrable
two-dimensional PDEs are obtained as symmetry reductions of the
self-dual YMEs (see \cite{34.1} and references therein).

Another possibility of construction of exact solutions of YMEs is to
use their conditional symmetry. To this end, we apply the same approach
which enables us to obtain broad families of conditionally-invariant
Ans\"atze for the nonlinear Dirac equation (see Chapter 6).

In the present chapter we exploit both possibilities mentioned
above. In the first section symmetry reduction of system of PDEs
(\ref{1.1}) by means of three-parameter subgroups of the
Poincar\'e group is carried out and a number of its non-Abelian exact
solutions are constructed. The second section is devoted to
investigation of conditional symmetry of YMEs.
\vspace{2mm}

\noindent
{\bf 2. Symmetry and solution generation for the Yang-Mills
  equations.}\ It was known long ago that YMEs are invariant with
respect to the group $C(1,3)\otimes{SU(2)}$, where $C(1,3)$ is the
15-parameter conformal group\index{Conformal!group} having the  
following generators:\index{Maximal symmetry!of the Yang-Mills equations}
\begin{equation}
\begin{array}{l}
P_{\mu}= \partial_{\mu},\\[2mm]
J_{{\alpha}{\beta}}= x^{\alpha}\partial_{\beta} -
x^{\beta}\partial_{\alpha} +
A^{a{\alpha}}\partial_{\edi A_{\beta}^{a}} - A^{a{\beta}}
\partial_{\edi A_{\alpha}^{a}},\\[2mm]
D=x_{\mu}\partial_{\mu} - A_{\mu}^{a}\partial_{\edi
  A_{\mu}^{a}},\\[2mm]
K_{\mu}=2x^{\mu}D - x_{\nu}x^{\nu}\partial_{\mu} +
2A^{a\mu}x_{\nu}\partial_ {\edi A_{\nu}^{a}} -
2A_{\nu}^{a}x^{\nu}\partial_{\edi A^{a\mu}},
\end{array}
\label{2.1}
\end{equation}
and $SU(2)$ is the infinite-parameter special unitary group with the 
following basis generator:
\begin{equation}
Q = \Bigl(\varepsilon_{abc}A_{\mu}^{b}w^{c}(x) +
e^{-1}\partial_{\mu}w^{a}(x)\Bigr) \partial_{A_{\mu}^{a}}.
\label{2.2}
\end{equation}

In (\ref{2.1}), (\ref{2.2})
$\partial_{A_{\mu}^{a}}$=$\partial/\partial A_{\mu}^{a}$, \ $w^{c}(x)$
are arbitrary smooth functions, $\varepsilon_{abc}$ is the third-order
anti-symmetric tensor with $\varepsilon_{123} = 1$.

But the fact that the group with generators (\ref{2.1}), (\ref{2.2})
is a maximal (in Lie sense) invariance group admitted by YMEs was
established only recently \cite{176.1} with the use of a symbolic
computation technique. The only explanation for this situation is a
very cumbersome structure of the system of PDEs (\ref{1.1}). As a
consequence, realization of the Lie algorithm of finding the maximal
invariance group admitted by YMEs demands a huge amount of
computations.  This difficulty has been overcome with the aid of
computer facilities.

One of the remarkable consequences of the fact that the equation
under study admits a nontrivial symmetry group is a possibility of
getting new solutions from the known ones by the solution generation
technique (see Theorem 2.4.1).

To make use of Theorem 2.4.1 we need formulae for finite
transformations generated by the infinitesimal operators (\ref{2.1}),
(\ref{2.2}). We adduce them following \cite{1.1,89}.
\vspace{1.5mm}

\noindent
1)\ The group of translations\index{Translation group} 
(generator $ X = \tau_{\mu} P_{\mu})$

\begin{displaymath}
x_{\mu}'=x_{\mu} + \tau_{\mu},\quad A_{\mu}^{d\prime}=A_{\mu}^{d}.
\end{displaymath}
2)\ The Lorentz group $O(1,3)$\index{Lorentz!group}
\vspace{1.5mm}

\noindent
a)\ the group of rotations\index{Rotation group} 
(generator $X = \tau J_{ab}$)
\begin{eqnarray*}
&&x_{0}'=0,\quad x_{c}' = x_{c},\ \ c \not= {a},\ \ c\not={b},\\
&&x_{a}'=x_{a} \cos{\tau} + x_{b}\sin{\tau},\\
&&x_{b}'=x_{b}\cos{\tau} - x_{a} \sin{\tau},\\
&&A_{0}^{d\prime}=A_{0}^{d},\quad A_{c}^{d\prime}=A_{c}^{d},\ \
c\not={a},\ \ 
c\not={b},\\ 
&&A_{a}^{d\prime}=A_{a}^{d}\cos{\tau} + A_{b}^{d}\sin{\tau},\\
&&A_{b}^{d\prime}=A_{b}^{d}\cos{\tau} - A_{a}^{d}\sin{\tau};
\end{eqnarray*}
b)\ the group of Lorentz transformations (generator $ X = \tau 
J_{0a}$)\index{Lorentz!transformations}
\begin{eqnarray*}
&&x_{0}'=x_{0} \cosh{\tau} + x_{a}\sinh{\tau},\\
&&x_{a}'=x_{a}\cosh{\tau} + x_{0}\sinh{\tau},
\quad x_{b}'=x_{b},\ \ b\not={a},\\
&&A_{0}^{d\prime}=A_{0}^{d}\cosh{\tau} + A_{a}^{d}\sinh{\tau},\\
&&A_{a}^{d\prime}=A_{a}^{d}\cosh{\tau} + A_{0}^{d}\sinh{\tau},\quad
A_{b}^{d\prime}=A_{b}^{d},\ \ b\not={a}.
\end{eqnarray*}
3)\ The group of scale transformations (generator $X = 
\tau D$)\index{Scale transformation group}
\begin{displaymath}
x_{\mu}' = x_{\mu} e^{\edi\tau},\quad A_{\mu}^{d\prime}=A_{\mu}^{d}
e^{-\edi\tau}. 
\end{displaymath}
4)\ The group of special conformal 
transformations\index{Conformal!transformations} (generator $X =
\tau_{\mu} K^{\mu}$) 
\begin{eqnarray*}
&&x_{\mu}'=(x_{\mu} - \tau_{\mu}x_{\nu}x^{\nu})\sigma^{-1}(x),\\ 
&&A_{\mu}^{d\prime}=\Bigl(g_{\mu\nu}\sigma(x) + 2 (x_{\mu}\tau_{\nu}
  -x_{\nu}\tau_{\mu} + 2\tau_{\alpha}x^{\alpha}\tau_{\mu}x_{\nu}
\\
&&\quad - x_{\alpha}x^{\alpha}\tau_{\mu}\tau_{\nu} -
 \tau_{\alpha}\tau^{\alpha}x_{\mu}x_{\nu}\Bigr)A^{d\nu}.
\end{eqnarray*}
5)\ The group of gauge 
transformations\index{Gauge transformation group}   
(generator $X = Q$)
\begin{eqnarray*}
&&x_{\mu}'=x_{\mu},\\
&&A_{\mu}^{d\prime}= A_{\mu}^{d}\cos{w} +
\varepsilon_{dbc}A_{\mu}^{b}n^{c} 
\sin{w} + 2 n^{d}n^{b}A_{\mu}^{b}\sin^{2}(w/2)\\
&&\quad + e^{-1}\Bigl((1/2)n^{d}\partial_{\mu}w +
(1/2)(\partial_{\mu}n^{d})\sin{w} +
\varepsilon_{dbc}(\partial_{\mu}n^{b})n^{c} \Bigr).
\end{eqnarray*}

In the above formulae $\sigma(x) = 1 - \tau_{\alpha}x^{\alpha} +
(\tau_{\alpha}\tau^{\alpha})(x_{\beta}x^{\beta})$,\ $n^{a} = n^{a}(x)$ 
is the unit vector determined by the equality $w^{a}(x) = w(x)
n^{a}(x),\ a = {1,2,3}$.

Using Theorem 2.4.1 it is not difficult to obtain formulae for
generating solutions of YMEs by the above transformation groups. We
adduce these omitting the derivation (see also \cite{86.1}).
\vspace{1.5mm}

\noindent
1)\ The group of translations
\index{Solution generation!with translations}
\begin{displaymath}
A_{\mu}^{a}(x) = u_{\mu}^{a}(x + \tau).
\end{displaymath}
2)\ The Lorentz group
\index{Solution generation!with rotation group}
\index{Solution generation!with Lorentz transformations}
\begin{eqnarray*}
&&A_{\mu}^{d}(x)=a_{\mu} u_{0}^{d}(a\cdot x,\, b\cdot x,\, c\cdot x,\, 
d\cdot x) + b_{\mu}u_{1}^{d}(a\cdot x,\, b\cdot x,\, c\cdot x,\,
d\cdot x) \\  
&&\quad + c_{\mu}u_{2}^{d}(a\cdot x,\, b\cdot x,\, c\cdot x,\, d\cdot
x) + 
d_{\mu}u_{3}^{d}(a\cdot x,\, b\cdot x,\, c\cdot x,\, d\cdot x).
\end{eqnarray*}
3)\ The group of scale transformations
\index{Solution generation!with scale transformations}
\begin{displaymath}
A_{\mu}^{d}(x) = e^{\edi\tau}u_{\mu}^{d}(xe^{\edi\tau}).
\end{displaymath}
4)\ The group of special conformal transformations
\index{Solution generation!with special conformal transformations}
\begin{eqnarray*}
&&A_{\mu}^{d}(x)=\Bigl(g_{\mu \nu}\sigma^{-1}(x) +
2\sigma^{-2}(x)(x_{\nu} \tau_{\mu} - x_{\mu}\tau_{\nu} +
2\tau_{\alpha}x^{\alpha}x_{\mu}\tau_{\nu} \\ 
&&\quad - x_{\alpha}x^{\alpha}\tau_{\mu}\tau_{\nu} -
\tau_{\alpha} \tau^{\alpha}x_{\mu}x_{\nu})\Bigr) u^{d\nu}\Bigl([x-
\tau(x_{\alpha}x^{\alpha})] \sigma^{-1}(x)\Bigr). 
\end{eqnarray*}
5)\ The group of gauge transformations
\begin{eqnarray*}
&&A_{\mu}^{d}(x)=u_{\mu}^{d}\cos{w} + \varepsilon_{dbc}u_{\mu}^{b}
n^{c}  
\sin{w} + 2n^{d}n^{b}u_{\mu}^{b}\sin^{2}(w/2)\\
&&\quad + e^{-1} \Bigl( (1/2) n^{d}\partial_{\mu}w +
(1/2)(\partial_{\mu}n^{d})\sin{w} +
\varepsilon_{dbc}(\partial_{\mu}n^{b})n^{c}\Bigr). 
\end{eqnarray*}

Here $u_{\mu}^{d}(x)$ is a given solution of YMEs;
$A_{\mu}^{d}(x)$ is a new solution of YMEs; $\tau,\ \tau_{\mu}$ are
arbitrary parameters; \ $a_{\mu},\ b_{\mu},\ c_{\mu},\ d_ {\mu} $ are
arbitrary parameters satisfying the equalities
\begin{eqnarray*}
&&a_{\mu}a^{\mu} = - b_{\mu}b^{\mu} = - c_{\mu}c^{\mu} =
-d_{\mu}d^{\mu} = 1,\\
&&a_{\mu}b^{\mu} = a_{\mu}c^{\mu} = a_{\mu}d^{\mu} = b_{\mu}c^{\mu} =
b_{\mu}d^{\mu} = c_{\mu}d^{\mu} = 0.
\end{eqnarray*}

In addition, we use the following notations:
$x + \tau = \{x_{\mu} + \tau_{\mu}, \mu = {0,\ldots,3} \}$,\
$a\cdot x = a_{\mu}x^{\mu}$.

Thus, each particular solution of YMEs gives rise to a multi-parameter
family of exact solutions by virtue of the above solution generation
formulae.\index{Solution generation} 
\vspace{2mm}

\noindent
{\bf 3. Ans\"atze for the Yang-Mills field.}\ Let us recall that the
key idea of the symmetry approach to the problem of reduction of PDEs
is a special choice of the form of a solution. This choice is dictated
by a structure of the symmetry group admitted by the equation under
study.

In the case involved, to reduce YMEs by $N$ variables we have to
construct Ans\"atze for the Yang-Mills field $A_{\mu}^{a}(x)$
invariant under $(4-N)$-dimensional subalgebras of the algebra with
the basis elements (\ref{2.1}), (\ref{2.2}).  Since we are looking
for Poincar\'e-invariant Ans\"atze reducing YMEs to systems of ODEs,
$N$ is equal to 3. Due to invariance of YMEs under the conformal group
$C(1,3)$ it is enough to consider only subalgebras which cannot be
transformed one into another by a group transformation from $C(1,3)$,
i.e.,\   $C(1,3)$-inequivalent subalgebras. Complete description of
$C(1,3)$-inequivalent subalgebras of the Poincar\'e algebra was
obtained in \cite{66}.

According to Theorem 1.5.1 to construct an Ansatz 
invariant under the invariance algebra having the basis elements
\begin{equation}
X_{a} = \xi_{a\mu}(x,A)\partial_{\mu} + \eta_{a\mu}^{b}
(x,A)\partial_{A_{\mu}^{b}},\ \ a = {1,2,3},
\label{3.1}
\end{equation}
where $A = \{A_{\mu}^{a},\ a = {1,2,3},\
\mu = {0,\ldots,3}\}$, we have
\begin{itemize}
\item{to construct a complete system of functionally-independent
    invariants of the operators (\ref{3.1}) \ $\Omega =
    \{\om_{j}(x,A),\ j = {1,\ldots,13}\}$;}
\item{to resolve the relations
\begin{equation}
F_{j}\Bigl(\om_{1}(x,A),\ldots,\om_{13}(x,A)\Bigr) = 0, \ \ j =
{1,\ldots,12} 
\label{3.2}
\end{equation}
    with respect to the functions $A_{\mu}^{a}$.}
\end{itemize}

As a result, we get an Ansatz for the field $A_{\mu}^{a}(x)$ which
reduces YMEs to the system of twelve nonlinear ODEs.
\vspace{2mm}

\noindent
{\bf Remark 7.1.1.}\ Equalities (\ref{3.2}) can be resolved with
respect to $A_{\mu}^{a}, \ a = {1,2,3},\ \mu = {0,\ldots,3}$ provided
the condition
\begin{equation}
{\rm rank}\, {\|{\xi_{a\mu}(x,A)}\|}_{a=1\mu=0}^{3 \ \ \ 3} = 3
\label{3.3}
\end{equation}
holds. If (\ref{3.3}) does not hold, the above procedure leads to
partially-invariant solutions\index{Partially-invariant solution} 
\cite{163}, which are not
considered here.

In Section 1.5 we have established that a procedure of
construction of invariant Ans\"atze could be substantially simplified
if coefficients of operators $X_{a}$ have the structure:
\begin{equation}
\xi_{a\mu} = \xi_{a\mu}(x), \quad \eta_{a\mu}^{b} =
\rho_{a\mu\nu}^{bc}(x)A_{\nu}^{c}
\label{3.4}
\end{equation}
(i.e.,\  basis elements of the invariance algebra realize a linear
representation). In this case, the invariant Ansatz for the field
$A_{\mu}^{a}(x)$ is searched for in the form
\begin{equation}
A_{\mu}^{a}(x) = Q_{\mu\nu}^{ab}(x)B^{b\nu}\Bigl(\om(x)\Bigr).
\label{3.5}
\end{equation}

Here $B_{\nu}^{b}(\om)$ are arbitrary smooth functions and $\om(x),
\ Q_{\mu\nu}^{ab}(x)$ are particular solutions of the system of PDEs
\begin{equation}
\xi_{a\mu}\om_{x_{\mu}} = 0, \quad 
(\xi_{a\nu}\partial_{\nu} - \rho_{a\mu\alpha}^{bc})
Q_{\alpha\beta}^{cd} = 0, 
\label{3.6}
\end{equation}
where $\mu = {0,\ldots,3},\ a,b,d = {1,2,3}$.

The basis elements of the Poincar\'e algebra  $P_{\mu},\
J_{\alpha\beta}$ from (\ref{2.1}) evidently satisfy conditions
(\ref{3.4}) and besides the equalities 
\begin{equation}
\eta_{a\mu}^{b} = \rho_{a\mu\nu}(x)A_{\nu}^{b},
\label{3.7}
\end{equation}
hold.

This fact allows further simplification of formulae (\ref{3.5}),
(\ref{3.6}). Namely, the Ansatz for the Yang-Mills field invariant
under a 3-dimensional subalgebra of the Poincar\'e algebra with basis
elements belonging to the class (\ref{3.1}), (\ref{3.7}) should be
looked for in the form\index{Ansatz!for Yang-Mills field} 
\begin{equation}
A_{\mu}^{a}(x) = Q_{\mu\nu}(x)B^{a\nu}\Bigl(\om(x)\Bigr),
\label{3.8}
\end{equation}
where $B_{\nu}^{a}(\om)$ are arbitrary smooth functions and $\om(x)$,\
$Q_{\mu\nu}(x)$ are particular solutions of the following system of
PDEs: 
\begin{eqnarray}
&&\xi_{a\mu} \om_{x_{\mu}} = 0,\ \ a = {1,2,3},
\label{3.9}\\
&&\xi_{a\alpha}\partial_{\alpha}Q_{\mu\nu} -
\rho_{a\mu\alpha}Q_{\alpha\nu} = 0,\ \
a = {1,2,3},\ \ \mu,\nu = {0,\ldots,3}.
\label{3.10}
\end{eqnarray}

Thus, to obtain the complete description of $C(1,3)$-inequivalent
Ans\"atze for the field $A_{\mu}^{a}(x)$ invariant under 3-dimensional
subalgebras of the Poincar\'e algebra, it is necessary to integrate 
the over-determined system of PDEs (\ref{3.9}), (\ref{3.10}) for each
$C(1,3)$-inequivalent subalgebra. Let us note that compatibility of
(\ref{3.9}), (\ref{3.10}) is guaranteed by the fact that operators
$X_{1},\ X_{2},\ X_{3}$ form a Lie algebra.

Consider, as an example, a procedure of constructing Ansatz
(\ref{3.8}) invariant under the subalgebra $\langle{P_{1},\, P_{2},\,
  J_{03}}\rangle$. In this case system (\ref{3.9}) reads
\begin{displaymath}
\om_{x_{1}} = 0,\quad \om_{x_{2}} = 0,\quad x_{0}\om_{x_{3}} +
x_{3}\om_{x_{0}} = 0,
\end{displaymath}
whence $\om = x_{0}^{2} - x_{3}^{2}$.

Next, we note that the coefficients $\rho_{1\mu\nu},\
\rho_{2\mu\nu}$ \ of the operators $P_{1},\ P_{2}$ are equal to zero,
while coefficients $\rho_{3\mu\nu}$ form the following
$(4\times{4})$ matrix
\begin{displaymath}
{\|\rho_{3\mu\nu}\|}_{\mu,\nu = 0}^{3}=
\pmatrix{0&0&0&1\cr
       0&0&0&0\cr
       0&0&0&0\cr
       1&0&0&0\cr}
\end{displaymath}
(we designate this constant matrix by the symbol $S$).

With account of the above fact, equations (\ref{3.10}) take the form
\begin{equation}
Q_{x_{1}} = 0,\quad Q_{x_{2}} = 0,\quad x_{0}Q_{x_{3}} +
x_{3}Q_{x_{0}} - SQ = 0,
\label{3.11}
\end{equation}
where $Q = {\|Q_{\mu\nu}(x)\|}_{\mu,\nu = 0}^{3}$ is a
$(4\times{4})$-matrix. 

From the first two equations of system (\ref{3.11}) it follows that
$Q = Q(x_{0}$,\ $x_{3})$.  Since $S$ is a constant matrix, a solution
of the third equation can be looked for in the form (see Section
2.2) 
\begin{displaymath}
Q = \exp\{f(x_{0},\, x_{3})S\}.
\end{displaymath}

Substituting this expression into (\ref{3.11}) we get
\begin{displaymath}
( x_{0}f_{x_{3}} + x_{3}f_{x_{0}} - 1 )\exp\{fS\} = 0
\end{displaymath}
or, equivalently,
\begin{displaymath}
x_{0}f_{x_{3}} + x_{3}f_{x_{0}} = 1,
\end{displaymath}
whence $f = \ln (x_{0} + x_{3})$.

Consequently, a particular solution of equations (\ref{3.11}) reads
\begin{displaymath}
Q = \exp\{\ln(x_{0} + x_{3})S\}.
\end{displaymath}

Using an evident identity $S = S^{3}$ we get the equalities:
\begin{eqnarray*}
Q &=& \sum_{n=0}^{\infty}{S^{n}\over n!}\Bigl(\ln(x_{0} +
x_{3})\Bigr)^{n} = I + S\biggl(\ln(x_{0} + x_{3}) + {1\over
  3!}[\ln(x_{0} + x_{3})]^{3}\\ 
&&+ \ldots\biggr) + S^{2}\biggl({1\over 2!}[\ln(x_{0} + x_{3})]^{2} +
{1\over 4!}[\ln(x_{0} + x_{3})]^{4} + \ldots \biggr)\\
&& = I + S\sinh[\ln(x_{0} + x_{3})] + S^{2}\Bigl(\cosh[\ln(x_{0} +
x_{3})] - 1\Bigr), 
\end{eqnarray*}
where $I$ is the unit $(4\times{4})$-matrix.

Substitution of the obtained expressions for functions
$\om(x),\ Q_{\mu\nu}(x)$ into (\ref{3.8}) yields the Ansatz for the
Yang-Mills field $A_{\mu}^{a}(x)$ invariant
under the algebra $\langle{P_{1},\, P_{2},\, J_{03}}\rangle$
\begin{eqnarray}
A_{0}^{a} &=& B_{0}^{a}(x_{0}^{2} - x_{3}^{2})\cosh\ln(x_{0} + x_{3}) +
B_{3}^{a}(x_{0}^{2} - x_{3}^{2})\sinh\ln(x_{0} + x_{3}),\non\\
A_{1}^{a} &=& B_{1}^{a}(x_{0}^{2} - x_{3}^{2}), \quad 
A_{2}^{a} = B_{2}^{a}(x_{0}^{2} - x_{3}^{2}),\label{3.12}\\
A_{3}^{a} &=& B_{3}^{a}(x_{0}^{2} - x_{3}^{2})\cosh\ln(x_{0} + x_{3}) +
B_{0}^{a}(x_{0}^{2} - x_{3}^{2})\sinh\ln(x_{0} + x_{3}).\non
\end{eqnarray}

Substituting (\ref{3.12}) into YMEs we get a system of ODEs for
functions $B_{\mu}^{a}$.  If we succeed in constructing its
general or particular solution, then substituting it into formulae
(\ref{3.12}) we get an exact solution of YMEs. But such a solution
will have an unpleasant feature: independent variables $x_{\mu}$ will
be included into it in an asymmetric way. At the same time, in the
initial equation (\ref{1.1}) all independent variables are on equal
rights. To remove this drawback we have to apply the solution
generation procedure by transformations from the Lorentz group. As a
result, we will obtain the Ansatz for the Yang-Mills field in the
manifestly-covariant form with symmetric dependence on $x_{\mu}$.

In the same way, we construct the rest of Ans\"atze invariant under
three-dimensional subalgebras of the Poincar\'e algebra. They are
represented in the unified form (\ref{3.8}), 
where\index{Ansatz!$P(1,3)$-invariant}
\begin{eqnarray}
Q_{\mu\nu}(x) &=& (a_{\mu}a_{\nu} -
d_{\mu}d_{\nu})\cosh\theta_{0} + 
(d_{\mu}a_{\nu} - d_{\nu}a_{\mu})\sinh\theta_0 \non\\
& &+2(a_{\mu} + d_{\mu})[(
\theta_{1}\cos\theta_{3} + \theta_{2}\sin\theta_{3})b_{\nu} +
( \theta_{2}\cos\theta_{3} \non\\
& &-\theta_{1}\sin\theta_{3})c_{\nu}+(\theta_{1}^{2} 
+ \theta_{2}^{2})e^{-\theta_{0}}(a_{\nu} +
d_{\nu})] + (b_{\mu}c_{\nu} \label{3.13}\\
& &- b_{\nu}c_{\mu})\sin\theta_{3} -
(c_{\mu}c_{\nu} + b_{\mu}b_{\nu}) 
\cos\theta_{3} - 2 e^{-\theta_{0}}\non\\
& &\times(\theta_{1}b_{\mu} + \theta_{2}c_{\mu})(a_{\nu}
+ d_{\nu})\non
\end{eqnarray}
and $ \theta_{\mu}(x),\ \om(x)  $ are some functions whose
explicit form is determined by the choice of a subalgebra of the
Poincar\'e algebra $AP(1,3)$.

Below, we adduce a complete list of 3-dimensional
$C(1,3)$-inequivalent subalgebras of the Poincar\'e algebra following
\cite{66} 
\begin{eqnarray}
&& L_{1} = \langle P_{0},\, P_{1},\, P_{2} \rangle;\quad
 L_{2} = \langle P_{1},\, P_{2},\, P_{3} \rangle;\non\\
&& L_{3} = \langle P_{0} + P_{3},\, P_{1},\, P_{2} \rangle;\quad
 L_{4} = \langle J_{03} + \alpha J_{12},\, P_{1},\, P_{2} \rangle;\non\\
&& L_{5} = \langle J_{03},\, P_{0} + P_{3},\, P_{1} \rangle;\quad
 L_{6} = \langle J_{03} + P_{1},\, P_{0},\, P_{3} \rangle;\non\\
&& L_{7} = \langle J_{03} + P_{1},\, P_{0} + P_{3},\, P_{2} \rangle;\quad
 L_{8} = \langle J_{12} + \alpha J_{03},\, P_{0},\, P_{3} \rangle;\non\\
&& L_{9} = \langle J_{12} + P_{0},\, P_{1},\, P_{2} \rangle;\quad
 L_{10} = \langle J_{12} + P_{3},\, P_{1},\, P_{2} \rangle;\label{3.14}\\
&& L_{11} = \langle J_{12} + P_{0} - P_{3},\, P_{1},\, P_{2} \rangle;\quad
 L_{12} = \langle G_{1},\, P_{0} + P_{3},\, P_{2} + \alpha P_{1}
\rangle;\non\\
&& L_{13} = \langle G_{1} + P_{2},\, P_{0} + P_{3},\, P_{1}  \rangle;\quad
 L_{14} = \langle G_{1} + P_{0} - P_{3},\, P_{0} + P_{3},\, P_{2}
 \rangle;\non\\
&& L_{15} = \langle G_{1} + P_{0} - P_{3},\, P_{0} + P_{3},\, P_{1} + \alpha
 P_{2} \rangle;\quad
 L_{16} = \langle J_{12},\, J_{03},\, P_{0} + P_{3} \rangle;\non\\
&& L_{17} = \langle G_{1} + P_{2},\, G_{2} - P_{1} + \alpha P_{2},\, P_{0} +
 P_{3} \rangle;\quad
 L_{18} = \langle J_{03},\, G_{1},\, P_{2} \rangle;\non\\
&& L_{19} = \langle G_{1},\, J_{03},\, P_{0} + P_{3} \rangle;\quad
 L_{20} = \langle G_{1},\, J_{03} + P_{2},\, P_{0} + P_{3} \rangle;\non\\
&& L_{21} = \langle G_{1},\, J_{03} + P_{1} + \alpha P_{2},\, P_{0} +
P_{3} \rangle;\quad
 L_{22} = \langle G_{1},\, G_{2},\, J_{03} + \alpha J_{12} \rangle;\non\\
&& L_{23} = \langle G_{1},\, P_{0} + P_{3},\, P_{1} \rangle;\quad
 L_{24} = \langle J_{12},\, P_{1},\, P_{2} \rangle;\non\\
&& L_{25} = \langle J_{03},\, P_{0},\, P_{3} \rangle;\quad
 L_{26} = \langle J_{12},\, J_{13},\, J_{23} \rangle;\quad
 L_{27} = \langle J_{01},\, J_{02},\, J_{12} \rangle.\non
\end{eqnarray} 

Here $G_{i} = J_{0i} - J_{i3},\ i = 1,2,\ \alpha \in \R^1$.

$P(1,3)$-invariant Ans\"atze for the Yang-Mills field
$A_{\mu}^{a}(x)$ are of the form (\ref{3.8}), (\ref{3.13}), functions 
$\theta_{\mu}(x),\ \om (x)$ being determined by one of the following
formulae:  
\begin{eqnarray}
L_{1} &:& \theta_{\mu} = 0,\quad \om  = d\cdot x ;\non\\
L_{2} &:& \theta_{\mu} = 0,\quad \om  = a\cdot x ;\non\\
L_{3} &:& \theta_{\mu} = 0,\quad \om  = k\cdot x ;\non\\
L_{4} &:& \theta_{0} = -\ln|k\cdot x|,\quad
 \theta_{1} = \theta_{2} = 0, \quad  \theta_{3} = \alpha \ln|k\cdot
 x|,\non\\
 &&\om  = (a\cdot x)^{2} - (d\cdot x)^{2} ;\non\\
L_{5} &:& \theta_{0} = -\ln|k\cdot x|,
 \quad \theta_{1} = \theta_{2} = \theta_{3} = 0,
 \quad \om  = c\cdot x ;\non\\
L_{6} &:& \theta_{0} = - b\cdot x,
 \quad \theta_{1} = \theta_{2} = \theta_{3} = 0,
 \quad \om  = c\cdot x ;\non\\
L_{7} &:& \theta_{0} = - b\cdot x,
 \quad \theta_{1} = \theta_{2} = \theta_{3} = 0,
 \quad \om  = b\cdot x - \ln|k\cdot x| ;\non\\
L_{8}&:& \theta_{0} = \alpha \arctan(b\cdot x/c\cdot x),
 \quad \theta_{1} = \theta_{2} =0,
 \quad \theta_{3} = - \arctan(b\cdot x/c\cdot x),\non\\
 &&\om  = (b\cdot x)^{2} + (c\cdot x)^{2} ;\non\\
L_{9} &:& \theta_{0} = \theta_{1} = \theta_{2} = 0,
 \quad \theta_{3} = - a\cdot x, \quad \om  = d\cdot x ;\non\\
L_{10}&:& \theta_{0} = \theta_{1} = \theta_{2} = 0,
 \quad \theta_{3} = d\cdot x, \quad \om  = a\cdot x ;\non\\
L_{11}&:& \theta_{0} = \theta_{1} = \theta_{2} = 0,
 \quad \theta_{3} = -(1/2) k\cdot x, \quad \om  = a\cdot x - d\cdot x
 ;\non\\  
L_{12}&:& \theta_{0} = 0,
 \quad \theta_{1} = (1/2)(b\cdot x - \alpha c\cdot x )(k\cdot x)^{-1},
 \quad \theta_{2} = \theta_{3} = 0,\non\\
 &&\om  = k\cdot x ;\label{3.15}\\
L_{13}&:& \theta_{0} = \theta_{2} = \theta_{3} = 0,
 \quad \theta_{1} = (1/2) c\cdot x, \quad \om  = k\cdot x ;\non\\
L_{14}&:& \theta_{0} = \theta_{2} = \theta_{3} = 0,
 \quad \theta_{1} = -(1/4) k\cdot x,
 \quad \om  = 4 b\cdot x + (k\cdot x)^{2};\non\\
L_{15}&:& \theta_{0} = \theta_{2} = \theta_{3} = 0,
 \quad \theta_{1} = -  (1/4) k\cdot x,
 \quad \om  = 4( \alpha b\cdot x - c\cdot x) \non\\
 &&+ \alpha (k\cdot x)^{2};\non\\ 
L_{16}&:& \theta_{0} = - \ln|k\cdot x|, \quad
 \theta_{1} = \theta_{2} = 0, \quad \theta_{3} = - \arctan (b\cdot
 x/c\cdot x),\non\\
 && \om  = (b\cdot x)^{2} + (c\cdot x)^{2} ;\non\\ 
L_{17}&:& \theta_{0} = \theta_{3} = 0,
 \quad \theta_{1} = (1/2)\Bigl(c\cdot x + (\alpha + k\cdot x)b\cdot
 x\Bigr) \Bigl(1 + k\cdot x \non\\
 &&\times (\alpha + k\cdot x)\Bigr)^{-1},\quad
 \theta_{2} = - (1/2)(b\cdot x - c\cdot x k\cdot x)\Bigl(1 + k\cdot x
 \non\\ 
 &&\times (\alpha + k\cdot x)\Bigr)^{-1}, \quad \om  = k\cdot x
 ;\non\\ 
L_{18}&:& \theta_{0} = - \ln|k\cdot x|,
 \quad \theta_{1} =  (1/2) b\cdot x(k\cdot x)^{-1},
 \quad \theta_{2} = \theta_{3} = 0,\non\\
 && \om  = (a\cdot x)^{2} - (b\cdot x)^{2} - (d\cdot x)^{2} ;\non\\
L_{19}&:& \theta_{0} = - \ln|k\cdot x|,
 \quad \theta_{1} =  (1/2) b\cdot x(k\cdot x)^{-1},
 \quad \theta_{2} = \theta_{3} = 0,
 \quad \om  = c\cdot x ;\non\\
L_{20}&:& \theta_{0} = - \ln|k\cdot x|, \quad \theta_{1} =
(1/2) b\cdot x(k\cdot x)^{-1},  \quad \theta_{2} = \theta_{3} =
0,\non\\  
 && \om  = \ln |k\cdot x| - c\cdot x;\non\\
L_{21}&:& \theta_{0} = - \ln |k\cdot x|,
 \quad \theta_{1} = \frac{1}{2} (b\cdot x - \ln |k\cdot x|) (k\cdot
 x)^{-1}, \quad \theta_{2} = \theta_{3} = 0,\non\\
 && \om  = \alpha \ln |k\cdot x| - c\cdot x;\non\\
L_{22}&:& \theta_{0} = -\ln|k\cdot x|,
 \quad \theta_{1} = (1/2) b\cdot x(k\cdot x)^{-1},
 \quad \theta_{2} = (1/2) c\cdot x(k\cdot x)^{-1},\non\\
 && \theta_{3} = \alpha\ln|k\cdot x|, \quad \om  = (a\cdot x)^{2} -
 (b\cdot x)^{2} - (c\cdot x)^{2} - (d\cdot x)^{2}.\non 
\end{eqnarray}

Here $k_\mu=a_\mu + d_\mu,\ \mu = {0,\ldots,3}$.
\vspace{1.5mm}

\noindent
{\bf Note 7.1.2.}\ Basis elements of subalgebras $L_{23}-L_{27}$ do
not satisfy (\ref{3.3}). That is why Ans\"atze invariant under these
subalgebras lead to partially-invariant solutions and are not
considered here.  \vspace{2mm}

\noindent
{\bf 4. Reduction of the Yang-Mills equations.}\
In order to reduce YMEs to ODE it is necessary to substitute Ansatz 
(\ref{3.8}), (\ref{3.13}) into (\ref{1.1}) and convolute the
expression obtained with $Q_{\alpha}^{\mu}(x)$. As a result, we get a
system of twelve nonlinear ODEs for functions $B_{\nu}^{a}(\om )$ of
the form 
\begin{equation}
\begin{array}{l}
  k_{\mu\gamma}\ddot{\vec{B}^{\gamma}} +
  l_{\mu\gamma}\dot{\vec{B}^{\gamma}} + m_{\mu\gamma}\vec{B}^{\gamma}
  + e g_{\mu\nu\gamma}\dot{\vec{B}^{\nu}} \times{\vec{B}^{\gamma}}+ e
  h_{\mu\nu\gamma} \vec{B}^{\nu}\times{\vec{B}^{\gamma}}\\ \quad +
  e^{2}\vec{B}_{\gamma}\times (\vec{B}^{\gamma}\times{\vec{B}_{\mu}})
  = \vec{0}.
\end{array}
\label{4.1}
\end{equation}

Coefficients of the reduced ODE are given by the following formulae:
\begin{eqnarray}
k_{\mu\gamma} &=& g_{\mu\gamma} F_{1} - G_{\mu}G_{\gamma},\non\\
l_{\mu\gamma} &=& g_{\mu\gamma} F_{2} + 2 S_{\mu\gamma} -
G_{\mu}H_{\gamma} 
- G_{\mu}\dot{G_{\gamma}},\non\\
m_{\mu\gamma} &=& R_{\mu\gamma} -
G_{\mu}\dot{H_{\gamma}},\label{4.2}\\ 
g_{\mu\nu\gamma} &=& g_{\mu\gamma} G_{\nu} +
g_{\nu\gamma} G_{\mu} - 2 g_{\mu\nu} G_{\gamma},\non\\
h_{\mu\nu\gamma} &=& (1/2)(g_{\mu\gamma} H_{\nu} -
g_{\mu\nu} H_{\gamma}) - T_{\mu\nu\gamma},\non
\end{eqnarray}
where $g_{\mu\nu}$ is a metric tensor of the Minkowski space $R(1,3)$
and $F_{1},\ F_{2},\ G_{\mu}$, $\ldots$, $T_{\mu\nu\gamma}$ are
functions of $\om $ determined by the relations
\begin{eqnarray}
&&F_{1} = \om _{x_{\mu}} \om _{x^{\mu}},
\quad  F_{2} = \Box \om,\quad
G_{\mu} = Q_{\alpha\mu} \om _{x_{\alpha}}, \quad H_{\mu} = Q_{\alpha
\mu x_{\alpha}},\non\\
&&S_{\mu\nu} = Q_{\mu}^{\alpha} Q_{\alpha\nu x_{\beta}}\om
_{x^{\beta}}, 
\quad R_{\mu\nu} = Q_{\mu}^{\alpha} \Box Q_{\alpha\nu},\label{4.3}\\
&&T_{\mu\nu\gamma} = Q_{\mu}^{\alpha} Q_{\alpha\nu x_{\beta}}
Q_{\beta\gamma} + Q_{\nu}^{\alpha} Q_{\alpha\gamma x_{\beta}}
Q_{\beta\mu} + Q_{\gamma}^{\alpha} Q_{\alpha\mu x_{\beta}}
Q_{\beta\nu}.\non 
\end{eqnarray}

Substituting functions $Q_{\mu\nu}(x)$ from (\ref{3.13}), where
$\theta_{\mu}(x),\ \om (x)$ are determined by one of the formulae
(\ref{3.15}), into (\ref{4.2}), (\ref{4.3}) we obtain coefficients of
the corresponding systems of ODEs (\ref{4.1})
\begin{eqnarray}
L_{1} &:& k_{\mu\gamma} = - g_{\mu\gamma} - d_{\mu}d_{\gamma},\quad
 l_{\mu\gamma} = m_{\mu\gamma} = 0, \non\\
 &&g_{\mu\nu\gamma} = g_{\mu\gamma}d_{\nu} + g_{\nu\gamma}d_{\mu} -
 2g_{\mu\nu}d_{\gamma},\quad h_{\mu\nu\gamma} = 0 ;\non\\
 L_{2} &:& k_{\mu\gamma} = g_{\mu\gamma} - a_{\mu}a_{\gamma},
 \quad l_{\mu\gamma} = m_{\mu\gamma} = 0,\non\\
 && g_{\mu\nu\gamma} = g_{\mu\gamma}a_{\nu} + g_{\nu\gamma}a_{\mu} -
 2 g_{\mu\nu}a_{\gamma},\quad h_{\mu\nu\gamma} = 0 ;\non\\
L_{3} &:& k_{\mu\gamma} = - k_{\mu}k_{\gamma},
 \quad l_{\mu\gamma} = m_{\mu\gamma} = 0,\non\\
 &&g_{\mu\nu\gamma} = g_{\mu\gamma}k_{\nu} + g_{\nu\gamma}k_{\mu} -
 2g_{\mu\nu}k_{\gamma},\quad h_{\mu\nu\gamma} = 0 ;\non\\
L_{4} &:& k_{\mu\gamma} = 4 g_{\mu\gamma}\om  - a_{\mu} a_{\gamma}(\om 
 + 1)^{2} - d_{\mu} d_{\gamma}(\om  - 1)^{2} - (a_{\mu} d_{\gamma} +
 a_{\gamma} d_{\mu})\non\\
 &&\times(\om ^{2} - 1),\quad l_{\mu\gamma} = 4\Bigl(g_{\mu\gamma} +
 \alpha (b_{\mu} c_{\gamma} - c_{\mu} b_{\gamma})\Bigr) - 2
 k_{\mu}(a_{\gamma} - d_{\gamma} + k_{\gamma}\om ),\non\\
 &&m_{\mu\gamma} = 0,\quad
 g_{\mu\nu\gamma} = \epsilon \Bigl(g_{\mu\gamma}(a_{\nu} - d_{\nu} +
 k_{\nu}\om ) + g_{\nu\gamma}(a_{\mu} - d_{\mu} + k_{\mu}\om )\non\\
 &&- 2 g_{\mu\nu}(a_{\gamma} - d_{\gamma} + k_{\gamma}\om )\Bigr),
 \quad h_{\mu\nu\gamma} = (\epsilon/2) (g_{\mu\gamma} k_{\nu}
 - g_{\mu\nu} k_{\gamma}) + \alpha \epsilon \Bigl((b_{\mu}
 c_{\nu}\non\\ 
 && - c_{\mu} b_{\nu})k_{\gamma} + (b_{\nu} c_{\gamma} - c_{\nu}
 b_{\gamma}) k_{\mu} + (b_{\gamma} c_{\mu} - c_{\gamma} b_{\mu})
 k_{\nu} \Bigr) ;\non\\ 
L_{5} &:&  k_{\mu\gamma} = - g_{\mu\gamma} - c_{\mu}c_{\gamma},\quad 
 l_{\mu\gamma} = - \epsilon c_{\mu}k_{\gamma},\quad m_{\mu\gamma} =
 0,\non\\   
 &&g_{\mu\nu\gamma} = g_{\mu\gamma}c_{\nu} + g_{\nu\gamma}c_{\mu} -
 2g_{\mu\nu}c_{\gamma},\quad
 h_{\mu\nu\gamma} =(\epsilon/2) (g_{\mu\gamma}k_{\nu} - g_{\mu\nu}
 k_{\gamma}) ;\non\\
L_{6} &:& k_{\mu\gamma} = - g_{\mu\gamma} - c_{\mu}c_{\gamma},\quad
 l_{\mu\gamma} = 0, \quad m_{\mu\gamma} = - (a_{\mu}a_{\gamma} -
 d_{\mu}d_{\gamma}),\non\\
 &&g_{\mu\nu\gamma} = g_{\mu\gamma}c_{\nu} + g_{\nu\gamma}c_{\mu} -
 2g_{\mu\nu}c_{\gamma},
 \quad h_{\mu\nu\gamma} = - \Bigl((a_{\mu}d_{\nu} - a_{\nu}d_{\mu})
 b_{\gamma} \non\\
 &&+ (a_{\nu}d_{\gamma} - a_{\gamma}d_{\nu})b_{\mu} +
 (a_{\gamma}d_{\mu} - a_{\mu}d_{\gamma})b_{\nu}\Bigr) ;\non\\
L_{7} &:& k_{\mu\gamma} = - g_{\mu\gamma} - (b_{\mu} - \epsilon
 k_{\mu}{\rm e}^\omega)(b_{\gamma} - \epsilon k_{\gamma}{\rm e}^\omega),\quad
 l_{\mu\gamma} = - 2(a_{\mu}d_{\gamma} - a_{\gamma}d_{\mu})\non\\
 &&+\epsilon{\rm e}^\omega k_\gamma(b_\mu-\epsilon k_\mu{\rm e}^\omega),\quad  
 m_{\mu\gamma} = -(a_{\mu}a_{\gamma} - d_{\mu}d_{\gamma}),\non\\
 &&g_{\mu\nu\gamma} = g_{\mu\gamma}(b_{\nu} - \epsilon k_{\nu}{\rm e}^\omega) +
 g_{\nu\gamma}(b_{\mu} - \epsilon k_{\mu}{\rm e}^\omega)
 - 2 g_{\mu\nu}(b_{\gamma} - \epsilon k_{\gamma}{\rm e}^\omega),\non\\
 &&h_{\mu\nu\gamma} = -  \Bigl((a_{\mu}d_{\nu} -
 a_{\nu}d_{\mu})b_{\gamma} +  (a_{\nu}d_{\gamma} -
 a_{\gamma}d_{\nu})b_{\mu} + (a_{\gamma}d_{\mu} -
a_{\mu}d_{\gamma})b_{\nu}\Bigr) ;\non\\ 
L_{8} &:& k_{\mu\gamma} = - 4\om (g_{\mu\gamma} +
 c_{\mu}c_{\gamma}),\quad  l_{\mu\gamma} = - 4(g_{\mu\gamma} +
 c_{\mu}c_{\gamma}),\non\\ 
 &&m_{\mu\gamma} = -{\om }^{-1}\Bigl(\alpha^{2}(a_{\mu}a_{\gamma} -
 d_{\mu} d_{\gamma}) + b_{\mu} b_{\gamma}\Bigr), \quad
 g_{\mu\nu\gamma} =  2 {\om}^{1/2}(g_{\mu\gamma}c_{\nu}\non\\
 &&+ g_{\nu\gamma}c_{\mu} - 2 g_{\mu\nu}c_{\gamma}),\quad
 h_{\mu\nu\gamma} = (1/2){\om}^{-1/2} (g_{\mu\gamma}c_{\nu} -
 g_{\mu\nu}c_{\gamma}) + \alpha{\om}^{-1/2}\non\\
 &&\times\Bigl((a_{\mu}d_{\nu} -
 a_{\nu}d_{\mu}) b_{\gamma}
 + (a_{\nu}d_{\gamma} - d_{\nu} a_{\gamma})
 b_{\mu} + (a_{\gamma}d_{\mu} - a_{\mu} d_{\gamma})
 b_{\nu}\Bigr);\non\\  
L_{9} &:& k_{\mu\gamma} = - g_{\mu\gamma} - d_{\mu}d_{\gamma},
 \quad l_{\mu\gamma} = 0,  \quad m_{\mu\gamma} = b_{\mu}b_{\gamma} +
 c_{\mu}c_{\gamma},\non\\
 &&g_{\mu\nu\gamma} = g_{\mu\gamma}d_{\nu} + g_{\nu\gamma}d_{\mu} -
 2g_{\mu\nu}d_{\gamma},\quad  h_{\mu\nu\gamma} =
 a_{\gamma}(b_{\mu}c_{\nu} - c_{\mu}b_{\nu})\non\\ 
 &&+ a_{\mu}(b_{\nu}c_{\gamma} - c_{\nu}b_{\gamma}) + a_{\nu}
 (b_{\gamma}c_{\mu} - c_{\gamma}b_{\mu}) ;\non\\
L_{10} &:& k_{\mu\gamma} = g_{\mu\gamma} - a_{\mu}a_{\gamma},
 \quad l_{\mu\gamma} = 0,
 \quad m_{\mu\gamma} = - (b_{\mu}b_{\gamma} + c_{\mu}c_{\gamma}),\non\\
 &&g_{\mu\nu\gamma} = g_{\mu\gamma}a_{\nu} + g_{\nu\gamma}a_{\mu} - 2
 g_{\mu\nu} a_{\gamma},\quad h_{\mu\nu\gamma} = -
 \Bigl(d_{\gamma}(b_{\mu}c_{\nu} -  c_{\mu}b_{\nu})\non\\
 &&+ d_{\mu}(b_{\nu}c_{\gamma} - c_{\nu}b_{\gamma}) +
 d_{\nu}(b_{\gamma}c_{\mu} - c_{\gamma}b_{\mu})\Bigr) ;\non\\
L_{11} &:& k_{\mu\gamma} = - (a_{\mu} - d_{\mu})(a_{\gamma} -
d_{\gamma}), 
 \quad l_{\mu\gamma} = - 2(b_{\mu}c_{\gamma} -
 c_{\mu}b_{\gamma}),\quad m_{\mu\gamma} = 0,\non\\
 && g_{\mu\nu\gamma} = g_{\mu\gamma}(a_{\nu} - d_{\nu}) +
 g_{\nu\gamma}(a_{\mu} - d_{\mu}) - 2g_{\mu\nu}(a_{\gamma} -
 d_{\gamma}), \quad h_{\mu\nu\gamma} = \non\\ 
 &&= (1/2)\Bigl( k_{\gamma}(b_{\mu}c_{\nu} -
 c_{\mu}b_{\nu}) + k_{\mu}(b_{\nu}c_{\gamma} - c_{\nu}b_{\gamma}) +  
 k_{\nu}(b_{\gamma}c_{\mu} - c_{\gamma}b_{\mu})\Bigr) ;\non\\
L_{12} &:& k_{\mu\gamma} = -  k_{\mu}k_{\gamma},\quad
 l_{\mu\gamma} = - {\om}^{-1}k_{\mu}k_{\gamma},\quad
 m_{\mu\gamma} = - \alpha^{2}\om ^{-2} k_{\mu} k_{\gamma},\non\\
 &&g_{\mu\nu\gamma} = g_{\mu\gamma}k_{\nu} + g_{\nu\gamma}k_{\mu} -
 2g_{\mu\nu}k_{\gamma},\quad
 h_{\mu\nu\gamma} =(1/2)\om^{-1}(g_{\mu\gamma}k_{\nu} -
 g_{\mu\nu}k_{\gamma})\non\\
 &&+ \alpha\om^{-1}\Bigl(( k_{\mu} b_{\nu} - k_{\nu}
 b_{\mu})c_{\gamma} + 
 (k_{\nu} b_{\gamma} - k_{\gamma} b_{\nu})c_{\mu} + (k_{\gamma}b_{\mu}
 - 
 k_{\mu} b_{\gamma})c_{\nu}\Bigr);\non\\
L_{13} &:& k_{\mu\gamma} = - k_{\mu}k_{\gamma},\quad l_{\mu\gamma} =
 0,\quad m_{\mu\gamma} = - k_{\mu}k_{\gamma},\non\\
 &&g_{\mu\nu\gamma} = g_{\mu\gamma}k_{\nu} +
 g_{\nu\gamma}k_{\mu} - 2g_{\mu\nu}k_{\gamma},
 \quad h_{\mu\nu\gamma} = - \Bigl((k_{\mu}b_{\nu} -
 k_{\nu}b_{\mu})c_{\gamma}\non\\ 
 &&+ (k_{\nu} b_{\gamma} - k_{\gamma} b_{\nu})c_{\mu}
 + (k_{\gamma}b_{\mu} - k_{\mu}b_{\gamma})c_{\nu}\Bigr) ;\non\\
L_{14} &:& k_{\mu\gamma} = - 16 (g_{\mu\gamma} +
b_{\mu}b_{\gamma}),\quad 
 l_{\mu\gamma} = m_{\mu\gamma} = h_{\mu\nu\gamma} = 0,\label{4.4}\\
 &&g_{\mu\nu\gamma} = 4(g_{\mu\gamma}b_{\nu} +
 g_{\nu\gamma}b_{\mu} - 2g_{\mu\nu}b_{\gamma}) ;\non\\
L_{15} &:& k_{\mu\gamma} = - 16\Bigl ((1 + \alpha^{2})g_{\mu\gamma} +
 (c_{\mu} -\alpha b_{\mu})(c_{\gamma} - \alpha
 b_{\gamma})\Bigr),\non\\ 
 &&l_{\mu\gamma} = m_{\mu\gamma} = h_{\mu\nu\gamma} = 0,\quad
 g_{\mu\nu\gamma} = - 4\Bigl(g_{\mu\gamma}(c_{\nu} - \alpha
 b_{\nu})\non\\  
 &&+ g_{\nu\gamma}(c_{\mu} - \alpha b_{\mu}) -
 2g_{\mu\nu}(c_{\gamma} - \alpha b_{\gamma})\Bigr);\non\\
L_{16} &:& k_{\mu\gamma} = - 4\om (g_{\mu\gamma} + c_{\mu}c_{\gamma}),  
 \quad l_{\mu\gamma} = - 4(g_{\mu\gamma} + c_{\mu}c_{\gamma}) -
 2 \epsilon \om^{1/2}k_{\gamma} c_{\mu},\non\\
 &&m_{\mu\gamma} = - {\om}^{-1} b_{\mu}b_{\gamma},
 \quad  g_{\mu\nu\gamma} = 2{\om}^{1/2}(g_{\mu\gamma}c_{\nu} +
 g_{\nu\gamma}c_{\mu} - 2g_{\mu\nu}c_{\gamma}),\non\\
 &&h_{\mu\nu\gamma} = (1/2) \Bigl(\epsilon  (g_{\mu\gamma}k_{\nu} - 
 g_{\mu\nu}k_{\gamma}) + \om^{-1/2}(g_{\mu\gamma}c_{\nu} -
 g_{\mu\nu}c_{\gamma})\Bigr) ;\non\\
L_{17} &:& k_{\mu\gamma} = - k_{\mu}k_{\gamma},\quad
 l_{\mu\gamma} = -(2\om  + \alpha)\Bigl(\om (\om  + \alpha) + 1\Bigr)
 k_{\mu} k_{\gamma},\non\\ 
 &&m_{\mu\gamma} = - 4 k_{\mu}k_{\gamma}\Bigl(1 + \om (\om + \alpha )
 \Bigr)^{-2},\quad  g_{\mu\nu\gamma} = g_{\mu\gamma}k_{\nu} + 
 g_{\nu\gamma}k_{\mu} - 2g_{\mu\nu}k_{\gamma},\non\\
 &&h_{\mu\nu\gamma} = (1/2)(2\om + \alpha)\Bigl(1 + \om (\om + \alpha 
 )\Bigr)^{-1}(g_{\mu\gamma}k_{\nu} - g_{\mu\nu}k_{\gamma})\non\\
 &&- 2\Bigl(1 + \om (\om  + \alpha)\Bigr)^{-1}
 \Bigl((k_{\mu}b_{\nu}-k_{\nu}b_{\mu})c_{\gamma} +
 (k_{\nu}b_{\gamma} - k_{\gamma}b_{\nu})c_{\mu} \non\\
 &&+ (k_{\gamma}b_{\mu} - k_{\mu}b_{\gamma})c_{\nu}\Bigr);\non\\
L_{18} &:& k_{\mu\gamma} = 4\om g_{\mu\gamma} - (k_{\mu}\om  + a_{\mu}
 - d_{\mu}) (k_{\gamma}\om  + a_{\gamma} - d_{\gamma}),\quad
 l_{\mu\gamma} = 6g_{\mu\gamma}\non\\
 && + 4(a_{\mu}d_{\gamma} - a_{\gamma}d_{\mu}) - 3k_{\gamma}
 (k_{\mu}\om  + a_{\mu} - d_{\mu}),\quad m_{\mu\gamma} = -
 k_{\mu}k_{\gamma},\non\\ 
 &&g_{\mu\nu\gamma} = \epsilon \Bigl(g_{\mu\gamma}(k_{\nu}\om  +
 a_{\nu} - 
 d_{\nu}) +  g_{\nu\gamma}(k_{\mu}\om  + a_{\mu} - d_{\mu})\non\\
 && - 2g_{\mu\nu}(k_{\gamma}\om  + a_{\gamma} - d_{\gamma})\Bigr),\quad
 h_{\mu\nu\gamma} = \epsilon (g_{\mu\gamma}k_{\nu} -
 g_{\mu\nu}k_{\gamma}) ; \non\\
L_{19} &:& k_{\mu\gamma} = - g_{\mu\gamma} - c_{\mu}c_{\gamma},
 \quad l_{\mu\gamma} = 2 \epsilon k_{\gamma}c_{\mu},\quad
 m_{\mu\gamma} = - k_{\mu}k_{\gamma},\non\\
 &&g_{\mu\nu\gamma} = g_{\mu\gamma}c_{\nu} + g_{\nu\gamma}c_{\mu} -
 2 g_{\mu\nu}c_{\gamma},\quad
 h_{\mu\nu\gamma} = \epsilon (g_{\mu\gamma} k_{\nu} - g_{\mu\nu}
 k_{\gamma}) ;\non\\
L_{20} &:& k_{\mu\gamma} = - g_{\mu\gamma} - (c_{\mu} - \epsilon
k_{\mu}) 
 (c_{\gamma} - \epsilon k_{\gamma}),\quad l_{\mu\gamma} =
 2 \epsilon k_{\gamma}c_{\mu} - 2 k_{\mu} k_{\gamma},\non\\
 &&m_{\mu\gamma} = - k_{\mu}k_{\gamma}, \quad g_{\mu\nu\gamma} =
 g_{\mu\gamma}(\epsilon k_{\nu} - c_{\nu})+ g_{\nu\gamma}(\epsilon
 k_{\mu} - c_{\mu})\non\\ 
 &&- 2g_{\mu\nu}(\epsilon k_{\gamma} - c_{\gamma}), 
 \quad h_{\mu\nu\gamma} = \epsilon (g_{\mu\gamma} k_{\nu} -
 g_{\mu\nu} k_{\gamma}) ;\non\\
L_{21} &:& k_{\mu\gamma} = - g_{\mu\gamma} - (c_{\mu} - \alpha
 \epsilon k_{\mu})(c_{\gamma} - \alpha \epsilon k_{\gamma}),\quad
 l_{\mu\gamma} = 2 (\epsilon k_{\gamma}c_{\mu} - \alpha k_{\mu}
 k_{\gamma}),\non\\ 
 &&m_{\mu\gamma} = - k_{\mu}k_{\gamma},
 \quad g_{\mu\nu\gamma} = - g_{\mu\gamma}(c_{\nu} - \alpha \epsilon
 k_{\nu}) - g_{\nu\gamma}(c_{\mu} - \alpha \epsilon k_{\mu})\non\\
 &&+ 2g_{\mu\nu}(c_{\gamma} - \alpha \epsilon k_{\gamma}),
 \quad h_{\mu\nu\gamma} = \epsilon (g_{\mu\gamma} k_{\nu} -
 g_{\mu\nu} k_{\gamma}) ;\non\\
L_{22} &:& k_{\mu\gamma} = 4\om  g_{\mu\gamma} - (a_{\mu} - d_{\mu} + 
 k_{\mu}\om )(a_{\gamma} - d_{\gamma} + k_{\gamma}\om ),\non\\
 &&l_{\mu\gamma} = 4\Bigl(2g_{\mu\gamma} + \alpha
 (b_{\mu}c_{\gamma} - c_{\mu}b_{\gamma}) - a_{\mu} a_{\gamma} +
 d_{\mu} d_{\gamma} - \om  k_{\mu}k_{\gamma} \Bigr),\non\\
 && m_{\mu\gamma} = - 2 k_{\mu}k_{\gamma},
 \quad g_{\mu\nu\gamma} = \epsilon \Bigl(g_{\mu\gamma}(a_{\nu} -
 d_{\nu} 
 + k_{\nu}\om ) + g_{\nu\gamma}(a_{\mu} - d_{\mu}\non\\ 
 &&+ k_{\mu}\om ) - 2g_{\mu\nu}(a_{\gamma} - d_{\gamma} + k_{\gamma}\om
 )\Bigr), \quad h_{\mu\nu\gamma} =(3\epsilon/2)(g_{\mu\gamma}k_{\nu} -
 g_{\mu\nu}k_{\gamma})\non\\
 &&-\epsilon \alpha\Bigl(k_{\gamma} (b_{\mu}c_{\nu} - c_{\mu}b_{\nu})
 +  k_{\mu}(b_{\nu}c_{\gamma} - c_{\nu}b_{\gamma}) +
 k_{\nu}(b_{\gamma}c_{\mu} - c_{\gamma}b_{\mu})\Bigr);\non 
\end{eqnarray}
where $\epsilon = 1$ for $a\cdot x + d\cdot x > 0$ and $ \epsilon =
-1$ for $ a\cdot x + d\cdot x < 0$. 
\vspace{2mm}

\noindent
{\bf 5. Exact solutions of the Yang-Mills equations.}\
When applying the symmetry reduction procedure to the nonlinear Dirac  
equation, we succeeded in constructing general solutions for most of
the reduced  systems of ODEs. In the case considered we are not so
lucky. Nevertheless, we obtain some particular solutions of equations
(\ref{4.1}), (\ref{4.2}), (\ref{4.4}). 

The principal idea of our approach to integration of systems of ODEs
(\ref{4.1}), (\ref{4.2}), (\ref{4.4}) is rather simple and quite
natural. It is reduction of these systems by the number of
components with the aid of {\em ad hoc} \/substitutions. Using this
trick we have constructed particular solutions of equations 1, 2, 5,
8, 14, 15, 16, 18, 19, 20, 21, 22 $(\alpha = 0)$. Below we adduce
substitutions for ${\vec B}_{\mu}(\om)$ and corresponding equations.
\begin{eqnarray}
&(1)& \vec{B}_{\mu} = a_{\mu}\vec{e}_{1}f(\omega) +
 b_{\mu}\vec{e}_{2}g(\om) +  c_{\mu}\vec{e}_{3}h(\om ),\non\\ 
&&\ddot{f} - e^{2}(g^{2} + h^{2})f = 0,\quad
 \ddot{g} + e^{2}(f^{2} - h^{2})g = 0,\non\\
&&\ddot{h} + e^{2}(f^{2} - g^{2})h = 0.\non\\
&(2)& \vec{B}_{\mu} = b_{\mu}\vec{e}_{1}f(\om ) +
 c_{\mu}\vec{e}_{2}g(\om ) +  d_{\mu}\vec{e}_{3}h(\om ),\non\\
&&\ddot{f} + e^{2}(g^{2} + h^{2})f = 0,
 \quad \ddot{g} + e^{2}(f^{2} + h^{2})g = 0,\non\\
&&\ddot{h} + e^{2}(f^{2} + g^{2})h = 0.\non\\
&(5)& \vec{B}_{\mu} = k_{\mu}\vec{e}_{1}f(\om ) +
 b_{\mu}\vec{e}_{2}g(\om ),\non\\ 
&& \ddot{f} - e^{2}g^{2}f = 0,\quad \ddot{g} = 0. \non\\
&(8.1)&({\rm under}\ \alpha = 0) \ \vec{B}_{\mu} =
 k_{\mu}\vec{e}_{1}f(\om ) + b_{\mu}\vec{e}_{2}g(\om ),\non\\
&& 4\om \ddot{f} + 4\dot{f} - e^{2}g^{2}f = 0,
 \quad 4\om \ddot{g} + 4\dot{g} - \om ^{-1} g = 0. \non\\
&(8.2)& \vec{B}_{\mu} = a_{\mu}\vec{e}_{1}f(\om ) +
 d_{\mu}\vec{e}_{2}g(\om ) + b_{\mu}\vec{e}_{3}h(\om ),\non\\
&&4\om \ddot{f} + 4\dot{f} - \alpha^{2}\om^{-1} f - 2 \alpha e
 \om^{-1/2} g h + e^{2}(h^{2} + g^{2})f = 0,\non\\
&&4\om \ddot{g} + 4\dot{g} + \alpha^{2}\om^{-1} g + 2
  \alpha e\om^{-1/2} f h + e^{2}(f^{2} - h^{2})g = 0,\non\\
&&4\om \ddot{h} + 4\dot{h} - \om ^{-1}h + 2 \alpha e\om^{-1/2}
 fg + e^{2}(f^{2} - g^{2})h = 0.\non\\
&(14.1)& \vec{B}_{\mu} = a_{\mu}\vec{e}_{1}f(\om ) +
d_{\mu}\vec{e}_{2}g(\om ) + c_{\mu}\vec{e}_{3}h(\om ),\non\\
&&16\ddot{f} - e^{2}(h^{2} + g^{2})f =0,
 \quad 16\ddot{g} + e^{2}(f^{2} - h^{2})g = 0,\non\\
&& 16\ddot{h} + e^{2}(f^{2} - g^{2})h = 0.\non\\
&(14.2)& \vec{B}_{\mu} = k_{\mu}\vec{e}_{1}f(\om ) +
 c_{\mu}\vec{e}_{2}g(\om ),\label{5.1}\\ 
&& 16\ddot{f} - e^{2}g^{2}f = 0,\quad \ddot{g} = 0.\non\\
&(15.1)& \vec{B}_{\mu} = a_{\mu}\vec{e}_{1}f(\om ) +
 d_{\mu}\vec{e}_{2}g(\om ) + (1 + \alpha^{2})^{-1/2}( \alpha c_{\mu}
 + b_{\mu})\vec{e}_{3}h(\om ),\non\\ 
&& 16(1 + \alpha^{2})\ddot{f} - e^{2}(h^{2} + g^{2})f = 0,\non\\
&& 16(1 + \alpha^{2})\ddot{g} + e^{2}(f^{2} - h^{2})g = 0,\non\\
&& 16(1 + \alpha^{2})\ddot{h} + e^{2}(f^{2} - g^{2})h = 0.\non\\
&(15.2)& \vec{B}_{\mu} = k_{\mu}\vec{e}_{1}f(\om ) + (1 +
 \alpha^{2})^{-1/2} (\alpha c_{\mu} + b_{\mu})
 \vec{e}_{2}g(\om ),\non\\
&& 16(1 + \alpha^{2})\ddot{f} - e^{2}fg^{2} = 0,
 \quad \ddot{g} = 0.\non\\
&(16)& \vec{B}_{\mu} = k_{\mu}\vec{e}_{1}f(\om ) +
 b_{\mu}\vec{e}_{2}g(\om ),\non\\ 
&& 4\om \ddot{f} + 4\dot{f} - e^{2}g^{2}f = 0,
 \quad 4\om \ddot{g} + 4\dot{g} - \om ^{-1}g = 0.\non\\
&(18)& \vec{B}_{\mu} = b_{\mu}\vec{e}_{1}f(\om ) +
 c_{\mu}\vec{e}_{2}g(\om ),\non\\ 
& & 4\om \ddot{f} + 6\dot{f} + e^{2}g^{2}f = 0,
 \quad 4\om \ddot{g} + 6\dot{g} + e^{2}f^{2}g = 0.\non\\
&(19)& \vec{B}_{\mu} = k_{\mu}\vec{e}_{1}f(\om ) +
 b_{\mu}\vec{e}_{2}g(\om ),\non\\ 
&& \ddot{f} - e^{2}g^{2}f = 0, \quad \ddot{g} = 0.\non\\
&(20)& \vec{B}_{\mu} = k_{\mu}\vec{e}_{1}f(\om ) +
 b_{\mu}\vec{e}_{2}g(\om ),\non\\ 
&& \ddot{f} - e^{2}g^{2}f = 0, \quad \ddot{g} = 0.\non\\
&(21)& \vec{B}_{\mu} = k_{\mu}\vec{e}_{1}f(\om ) +
 b_{\mu}\vec{e}_{2}g(\om ),\non\\  
&& \ddot{f} - e^{2}g^{2}f = 0, \quad \ddot{g} = 0.\non\\
&(22)& ({\rm under}\ \alpha = 0)\ \vec{B}_{\mu} =
 b_{\mu}\vec{e}_{1}f(\om ) + c_{\mu}\vec{e}_{2}g(\om ),\non\\
&&4\om \ddot{f} + 8\dot{f} + e^{2}g^{2}f = 0,
 \quad 4\om \ddot{g} + 8\dot{g} + e^{2}f^{2}g = 0.\non
\end{eqnarray}

In the above formulae we use the notations $\vec{e}_{1} = (1,0,0)$,
\ $\vec{e}_{2} = (0,1,0)$,\ $\vec{e}_{3} = (0,0,1)$.

Thus, combining symmetry reduction by the number of independent
variables and reduction by the number of dependent variables we
reduce YMEs to rather simple ODEs. 

Next, we will briefly consider a procedure of integration of systems
of nonlinear ODEs (\ref{5.1}).

Substitution $f = 0,\ g = h = u(\om )$ reduces the system of ODEs 1
from (\ref{5.1}) to the equation

\begin{equation}
 \ddot{u} = e^{2}u^{3},
\label{5.2}
\end{equation}
which is integrated in elliptic functions\index{Elliptic function}
\cite{20.3,132}. In addition, ODE (\ref{5.2}) has a solution which is
expressed in terms of elementary functions $ u = \sqrt{2}(e\om -
C)^{-1},\ C \in \R^{1}$.

ODE 2 with $ f = g = h = u(\om )$ reduces to the form
$ \ddot{u} + 2e^{2}u^{3} = 0$. This equation is also integrated
in elliptic functions \cite{20.3,132}. 

Integrating the second equation of system of ODEs 5 we get $ g =
C_{1}\om  + C_{2},\ C_{i}\in{\R^{1}}$. If $C_{1} \not={0}$, then the
constant  $C_{2}$ can be neglected, and we may put $C_{2} =
0$. Provided $C_{1} \not={0}$, the first equation from system 5 reads

\begin{equation}
\ddot{f} - e^{2}C_{1}^{2}\om ^{2}f = 0.
\label{5.3}
\end{equation}

The general solution of ODE (\ref{5.3}) is given by the formula

$$
f (\om) = \om ^{1/2}Z_{1/4}\Bigl((ieC_1/2)\om ^{2}\Bigr).
$$

Hereafter, we use the notation
$Z_{\nu}(\om ) = C_{3} J_{\nu}(\om ) + C_{4}Y_{\nu}(\om )$,
where $J_{\nu},\ Y_{\nu}$ are Bessel functions,\index{Bessel!function} 
$C_{3},\ C_{4}$ are arbitrary real constants.

In the case $C_{1} = 0,\ C_{2} \not={0}$ the general solution of the
first equation from system 5 reads $f = C_{3} \cosh{C_{2}e\om} + C_{4}
\sinh{C_{2}e\om}$, where $C_{3},\ C_{4}$ are arbitrary real
constants. 

At last, provided $C_{1} = C_{2} = 0$, the general solution of the
first equation from system 5 has the form $ f = C_{3}\om + C_{4},\ 
\{C_{3},C_{4}\} \subset {\R^{1}}$.

The general solution of the second ODE from system 8.1 is of the form 
$g = C_{1} {\om }^{1/2} + C_{2}{\om }^{-1/2}$, where $C_{1},\ C_{2}$
are arbitrary real constants. 

Substituting the expression obtained into the first equation we get

\begin{equation}
4\om ^{2} \ddot{f} + 4\om \dot{f} - e^{2}(C_{1}\om  + C_{2})^{2}f = 0. 
\label{5.4}
\end{equation}

We cannot solve ODE (\ref{5.4}) with $C_{1} C_{2} \not={0}$. In the
remaining cases its general solution reads 
\vspace{1.5mm}

\noindent
a)\ $C_{1} \not={0},\ C_{2} = 0$
\begin{displaymath}
f = Z_{0}\Bigl((ieC_1/2)\om\Bigr),
\end{displaymath}
b)\ $C_{1} = 0,\ C_{2}\not={0}$
\begin{displaymath}
f = C_{3} \om ^{eC_{2}/2} + C_{4}\om ^{-eC_{2}/2},
\end{displaymath}
c)\ $C_{1} = 0,\ C_{2} = 0$
\begin{displaymath}
f = C_{3}\ln{\om } + C_{4}.
\end{displaymath}

Here $C_{3},\ C_{4}$ are arbitrary real constants.

We did not succeed in obtaining particular solutions of the system
8.2.  Equations 14.1 coincide with equations 1, if we replace $e$ by
$e/4$. Similarly, equations 14.2 coincide with equations 5, if we
change $e$ by $e/4$. Next, equations 15.1 coincide with equations 1
and equations 15.2 with equations 5, if we replace $e$ by $(e/4)(1 +
\alpha^{2})^{-1/2}$.

System of ODEs 16 coincides with the system 8.1 and systems 19, 20, 21 
with the system 5. We did not succeed in integrating equations 18.

At last, the system 22 $({\rm under}\ \alpha = 0)$ with the
substitution $f = g = u(\om )$ reduces to the form
\begin{equation}
\om \ddot{u} + 2 \dot{u} + (e^{2}/4)u^{3} = 0.
\label{5.5}
\end{equation}

ODE (\ref{5.5}) is the Emden-Fowler equation which is integrated in
terms of elliptic functions (see, e.g. \cite{132}). It has two classes
of particular solutions which are expressed in terms of elementary
functions
\begin{displaymath}
u = e^{-1}\om ^{-1/2},\quad u= 2\sqrt{2}C_1 e^{-1}(\omega +
C_1)^{-1},\ \ C_1\in \R^1. 
\end{displaymath}

Substituting the results obtained into the corresponding formulae from
(\ref{5.1}) and then into the Ansatz (\ref{3.13}), we get exact
solutions of the nonlinear YMEs (\ref{1.1}). Let us note that
solutions of the systems of ODEs 5, 8.1, 14.2, 15.2, 16, 19, 20, 21
satisfying the condition $g = 0$ give rise to Abelian solutions of
YMEs. We do not adduce these and present non-Abelian solutions of YMEs
only.
\index{Exact solutions!of the Yang-Mills equations}
\begin{eqnarray}
&1)& \vec{A}_{\mu} = (\vec{e}_{2}b_{\mu} + \vec{e}_{3}c_{\mu})\sqrt{2} 
(ed\cdot x - \lambda)^{-1} ;\non\\
&2)& \vec{A}_{\mu} = (\vec{e}_{2}b_{\mu} + \vec{e}_{3}c_{\mu})
\lambda\, {\rm sn}\, [(e\sqrt{2}/2) \lambda d\cdot x]\,
{\rm dn}\, [(e\sqrt{2}/2) \lambda d\cdot x]\non\\
&&\times \Bigl( {\rm cn}\, [(e\sqrt{2}/2) \lambda d\cdot x]\Bigr)^{-1}
;\non\\ 
&3)& \vec{A}_{\mu} = (\vec{e}_{2}b_{\mu} + \vec{e}_{3}c_{\mu})\lambda
\Bigl( {\rm cn}\, (e \lambda d\cdot x)\Bigr)^{-1} ;\non\\
&4)& \vec{A}_{\mu} = (\vec{e}_{1}b_{\mu} + \vec{e}_{2}c_{\mu} +
\vec{e}_{3}d_{\mu}) \lambda\, {\rm cn}\, (e\sqrt{2}\lambda a\cdot x)
;\non\\ 
&5)& \vec{A}_{\mu} = \vec{e}_{1}k_{\mu}|k\cdot x|^{-1}(c\cdot
  x)^{1/2}Z_{1/4}\Bigl((ie\lambda/2)(c\cdot x)^{2} \Bigr)
+ \vec{e}_{2}b_{\mu}\lambda c\cdot x ;\non\\
&6)& \vec{A}_{\mu} = \vec{e}_{1}k_{\mu}|k\cdot x|^{-1} \Bigl( \lambda_{1}
\cosh (e\lambda c\cdot x) + \lambda_{2} \sinh (e\lambda c\cdot x)
\Bigr) + 
\vec{e}_{2}b_{\mu}\lambda ;\non\\
&7)& \vec{A}_{\mu} = \vec{e}_{1}k_{\mu}Z_{0}\Bigl((ie\lambda/2)
[(b\cdot x)^{2} + (c\cdot x)^{2}]\Bigr) + \vec{e}_{2}(b_{\mu}c\cdot x
- c_{\mu}b\cdot x)\lambda ;\non\\ 
&8)& \vec{A}_{\mu} = \vec{e}_{1}k_{\mu}\Bigl(\lambda_{1}
[(b\cdot x)^{2} + (c\cdot x)^{2}]^{e\lambda/2} + \lambda_{2}[(b\cdot
x)^{2} + (c\cdot x)^{2}]^{-e\lambda/2}\Bigr) \non\\ 
&& + \vec{e}_{2}(b_{\mu}c\cdot x - c_{\mu}b\cdot x)\lambda[(b\cdot
x)^{2} + (c\cdot x)^{2}]^{-1}  ;\non\\
&9)& \vec{A}_{\mu} = \Bigl\{\vec{e}_{2}\Bigl((1/8) [d_{\mu} - k_{\mu}
(k\cdot x)^{2}] + (1/2) b_{\mu}k\cdot x\Bigr) +
\vec{e}_{3}c_{\mu}\Bigr\} \lambda\non\\ 
&&\times{\rm sn}\,\Bigl((e\lambda\sqrt{2}/8)
[4b\cdot x + (k\cdot x)^{2}]\Bigr)\,
{\rm dn}\, \Bigl((e\lambda\sqrt{2}/8)[4b\cdot x + (k\cdot
x)^{2}]\Bigr)\non\\
&& \times \Bigl\{ {\rm cn}\, \Bigl((e\lambda\sqrt{2}/8)[4b\cdot
x + (k\cdot x)^{2}]\Bigr)\Bigr\}^{-1} ;\non\\ 
&10)& \vec{A}_{\mu} = \Bigl\{\vec{e}_{2}\Bigl((1/8)[d_{\mu} -
k_{\mu}(k\cdot x)^{2}] + (1/2)b_{\mu}k\cdot x\Bigr) +
\vec{e}_{3}c_{\mu}\Bigr\} \lambda \non\\
&&\times\Bigl\{ {\rm cn}\, \Bigl((e\lambda/4)[4b\cdot x + (k\cdot
x)^{2}]\Bigr) \Bigr\}^{-1} ;\non\\
&11)& \vec{A}_{\mu} = \Bigl\{\vec{e}_{2}\Bigl((1/8)[d_{\mu} - k_{\mu} 
(k\cdot x)^{2}] + (1/2) b_{\mu}k\cdot x\Bigr) + \vec{e}_{3}
c_{\mu}\Bigr\} 4 \sqrt{2}\non\\
&&\times \Bigl(e[4b\cdot x + (k\cdot x)^{2}] - \lambda
\Bigr)^{-1}; \non\\ 
&12)& \vec{A}_{\mu} = \vec{e}_{1} k_{\mu}[4b\cdot x + (k\cdot
x)^{2}]^{1/2} Z_{1/4} \Bigl((ie \lambda/8)[4b\cdot x + (k\cdot
x)^{2}]^{2}\Bigr)\non\\
&& + \vec{e}_{2} c_{\mu} \lambda [4b\cdot x +(k\cdot
x)^{2}];\non\\ 
&13)& \vec{A}_{\mu} = \vec{e}_{1}k_{\mu}\Bigl\{\lambda_{1}
\cosh\Bigl((e\lambda/4)[4b\cdot x +(k\cdot x)^{2}]\Bigl) + \lambda_{2}
\sinh \Bigl((e\lambda/4)[4b\cdot x \non\\
&& + (k\cdot x)^{2}]\Bigr)\Bigr\} + \vec{e}_{2} c_{\mu} \lambda;\non\\  
&14)& \vec{A}_{\mu} = \Bigl(\vec{e}_{2}[d_{\mu} - (1/8)
k_{\mu}(k\cdot x)^{2} - (1/2)b_{\mu} k\cdot x] + \vec{e}_{3}[\alpha
c_{\mu} + b_{\mu} \non\\
&&+ (1/2) k_{\mu} k\cdot x](1 + \alpha^{2})^{-1/2}\Bigr\} 
\lambda\, {\rm sn}\,  \Bigl((e \lambda \sqrt{2}/8)[4(\alpha b\cdot x
- c\cdot x)\non\\
&&+ \alpha(k\cdot x)^{2}] (1 + \alpha^{2})^{-1/2}\Bigr)\,
{\rm dn}\,  \Bigl((e \lambda \sqrt{2}/8)[4(\alpha b\cdot x
- c\cdot x)+ \alpha(k\cdot x)^{2}]\non\\
&&\times (1 + \alpha^{2})^{-1/2}\Bigr) \Bigl\{ {\rm cn}\, \Bigl((e
\lambda \sqrt{2}/8)[4(\alpha b\cdot x - c\cdot x)+ \alpha(k\cdot
x)^{2}]\non\\  
&&\times (1 + \alpha^{2})^{-1/2} \Bigr)
\Bigr\}^{-1};\non\\   
&15)& \vec{A}_{\mu} = \Bigl(\vec{e}_{2}[d_{\mu} - (1/8)
k_{\mu}(k\cdot x)^{2} - (1/2)b_{\mu} k\cdot x] + \vec{e}_{3}[\alpha
c_{\mu} + b_{\mu} \non\\
&&+ (1/2) k_{\mu} k\cdot x](1 + \alpha^{2})^{-1/2}\Bigr)
\Bigl\{{\rm cn}\, \Bigl((e \lambda/4)[4(\alpha b\cdot x
- c\cdot x) + \alpha(k\cdot x)^{2}]\non\\
&&\times(1 + \alpha^{2})^{-1/2}\Bigr)\Bigr\}^{-1};\non\\ 
&16)& \vec{A}_{\mu} = \Bigl(\vec{e}_{2}[d_{\mu} - (1/8)
k_{\mu}(k\cdot x)^{2} - (1/2)b_{\mu} k\cdot x] + \vec{e}_{3}[\alpha
c_{\mu} + b_{\mu} \non\\
&&+ (1/2) k_{\mu} k\cdot x](1 + \alpha^{2})^{-1/2}\Bigr)4
\sqrt{2}(1+\alpha^{2})^{1/2}\Bigl( e[4(\alpha b\cdot x - c\cdot
x)\non\\ 
&&+\alpha (k\cdot x)^{2}]\Bigr)^{-1};\label{5.6}\\ 
&17)& \vec{A}_{\mu} = \vec{e}_{1} k_{\mu}\Bigl\{[4 ( \alpha
  b\cdot x - c\cdot x) + \alpha(k\cdot x)^{2}]^{1/2}
Z_{1/4}\Bigl((ie\lambda/8)[4 ( \alpha b\cdot x \non\\
&&- c\cdot x) + \alpha(k\cdot x)^{2}]^{2}(1 +
\alpha^{2})^{-1/2}\Bigr)\Bigr\}  
+\vec{e}_{2}[\alpha c_{\mu} + b_{\mu} + (1/2) k_{\mu} k\cdot x]
\lambda \non\\
&&\times[4(\alpha b\cdot x - c\cdot x) +\alpha (k\cdot x)^{2}](1 +
\alpha^{2})^{-1/2};\non\\  
&18)& \vec{A}_{\mu} = \vec{e}_{1} k_{\mu}\Bigl\{\lambda_{1}
\cosh \Bigl((e \lambda/4)(1+\alpha^{2})^{-1/2}
[4 (\alpha b\cdot x - c\cdot x) + \alpha(k\cdot x)^{2}]\Bigr) \non\\
&& + \lambda_{2} \sinh \Bigl((e\lambda/4)
(1+\alpha^{2})^{-1/2}[4(\alpha b\cdot x - c\cdot x) +
\alpha (k\cdot x)^{2}] \Bigr) \Bigr\}+ \vec{e}_{2}[\alpha c_{\mu}
\non\\ 
&& + b_{\mu} + (1/2) k_{\mu} k\cdot
x]\lambda(1+\alpha^{2})^{-1/2};\non\\ 
&19)& \vec{A}_{\mu} = \vec{e}_{1} k_{\mu}|k\cdot x|^{-1} Z_{0} \Bigl( 
(ie\lambda/2)[(b\cdot x)^{2} + (c\cdot x)^{2}]\Bigr) +
\vec{e}_{2}(b_{\mu} c\cdot x\non\\
&&- c_{\mu} b\cdot x)\lambda;\non\\
&20)& \vec{A}_{\mu} = \vec{e}_{1} k_{\mu} |k\cdot x|^{-1} \Bigl(
\lambda_{1} [(b\cdot x)^{2} + (c\cdot x)^{2}]^{e \lambda/2}+
\lambda_{2} [(b\cdot x)^{2}  \non\\ 
&&+(c\cdot x)^{2}]^{-e \lambda/2}\Bigr) + \vec{e}_{2}( b_{\mu} c\cdot 
x - c_{\mu} b\cdot x)\lambda [(b\cdot x)^{2} + (c\cdot
x)^{2}]^{-1};\non\\ 
&21)& \vec{A}_{\mu} = \vec{e}_{1} k_{\mu}|k\cdot x|^{-1}(c\cdot
x)^{1/2} Z_{1/4} [(ie\lambda/2) (c\cdot x)^{2}] +
\vec{e}_{2}[b_{\mu}\non\\
&&- k_{\mu} b\cdot x(k\cdot x)^{-1}]\lambda c\cdot x ;\non\\ 
&22)& \vec{A}_{\mu} = \vec{e}_{1} k_{\mu}|k\cdot x|^{-1} \Bigl(
\lambda_{1}\cosh (\lambda e c\cdot x )+\lambda_{2} \sinh(\lambda e
c\cdot x)\Bigr)\non\\
&&+ \vec{e}_{2}[b_{\mu}- k_{\mu} b\cdot x(k\cdot x)^{-1}]\lambda
;\non\\  
&23)& \vec{A}_{\mu} = \vec{e}_{1} k_{\mu}|k\cdot x|^{-1} (\ln|k\cdot
x| -c\cdot x)^{1/2} Z_{1/4} [(ie\lambda/2) (\ln|k\cdot x| -c\cdot
x)^{2}] \non\\  
&& + \vec{e}_{2}[b_{\mu}- k_{\mu} b\cdot x(k\cdot x)^{-1}]\lambda
(\ln|k\cdot x|- c\cdot x) ;\non\\ 
&24)& \vec{A}_{\mu} = \vec{e}_{1} k_{\mu}|k\cdot x|^{-1} \Bigl(
\lambda_{1}\cosh [\lambda e (\ln|k\cdot x| - c\cdot x )]+\lambda_{2}
\sinh[\lambda e (\ln|k\cdot x|\non\\
&& - c\cdot x)] \Bigr)+ \vec{e}_{2}[b_{\mu}- k_{\mu} b\cdot x(k\cdot
 x)^{-1}]\lambda ;\non\\ 
&25)& \vec{A}_{\mu} = \vec{e}_{1} k_{\mu}|k\cdot x|^{-1} (\alpha
\ln|k\cdot x| -c\cdot x)^{1/2} Z_{1/4} [(ie\lambda/2) ( \alpha
\ln|k\cdot x| \non\\ 
&&-c\cdot x)^{2}] + \vec{e}_{2}[b_{\mu}- k_{\mu}( b\cdot x- \ln
|k\cdot x|)(k\cdot x)^{-1}]\lambda ( \alpha \ln|k\cdot x|- c\cdot x)
;\non\\ 
&26)& \vec{A}_{\mu} = \vec{e}_{1} k_{\mu}|k\cdot x|^{-1} \Bigl(
\lambda_{1}\cosh [\lambda e ( \alpha \ln|k\cdot x| - c\cdot x
)]+\lambda_{2} \sinh[\lambda e \non\\
&&\times(\alpha \ln|k\cdot x| - c\cdot x))\Bigr)+ 
\vec{e}_{2}[b_{\mu}- k_{\mu}(b\cdot x - \ln |k\cdot x|)(k\cdot
x)^{-1}]\lambda ;\non\\ 
&27)& \vec{A}_{\mu} = \Bigl(\vec{e}_{1}[b_{\mu} - k_{\mu}b\cdot
x(k\cdot x)^{-1}] + \vec{e}_{2}[c_{\mu} - k_{\mu}c\cdot x(k\cdot
x)^{-1}]\Bigr)e^{-1}\non\\
&&\times\cases{(x\cdot x)^{-1/2},\cr
2\sqrt{2}\lambda(x\cdot x + \lambda)^{-1};}\non\\
&28)& \vec{A}_{\mu} = \Bigl(\vec{e}_{1}[b_{\mu} - k_{\mu}b\cdot
x(k\cdot x)^{-1}] + \vec{e}_{2}[c_{\mu} - k_{\mu}c\cdot x(k\cdot
x)^{-1}]\Bigr)f(x\cdot x),\non\\
&&\om \ddot{f} + 2\dot{f} +(e^{2}f^{3}/4) = 0.\non
\end{eqnarray}

In the above formulae  $Z_{\alpha}(\om)$ is the Bessel function; $ 
{\rm sn},\ {\rm dn},\ {\rm cn}$  are Jacobi elliptic 
functions\index{Elliptic function}
having the modulus $\sqrt{2}/2$;\ $\lambda,\ \lambda_{1},\ \lambda_{2}
= \mbox{\rm const}$.  

Let us note that the solutions N $27$ are nothing more but the
meron\index{Meron solution} and the 
instanton\index{Instanton solution} solutions of YMEs \cite{1.1}.  
In the Euclidean space the
meron and instanton solutions were obtained by Alfaro, Fubini,
Furlan \cite{46.1} and Belavin, Polyakov, Schwartz, Tyupkin
\cite{21.1} with the use of the Ansatz suggested by 't Hooft
\cite{195.1}, Corrigan and Fairlie \cite{40.1} and Wilczek
\cite{201.1}.

Another important point is that we can obtain new exact solutions of
YMEs by applying to solutions (\ref{5.6}) the solution generation
technique. We do not adduce the corresponding formulae because of
their awkwardness. 
\vspace{2mm}

\noindent
{\bf 6. Some generalizations.}\ It was noticed in \cite{105,106} that
group-invariant solutions of nonlinear PDEs could provide us with
rather general information about the structure of solutions of the
equation under study. Using this fact, we constructed in
\cite{105,106,106.2} a number of new exact solutions of the nonlinear
Dirac equation which could not be obtained by the symmetry reduction
procedure (see also Sections 6.1 and 7.2). We will demonstrate that
the same idea proves to be efficient for constructing new solutions of
YMEs.

Solutions of YMEs numbered by 7, 8, 19, 20 can be represented in the
following unified form:

\begin{equation} 
\vec{A}_{\mu} = k_{\mu}\vec{B}(k\cdot x,\, c\cdot x) +
b_{\mu}\vec{C}(k\cdot x,\, c\cdot x). 
\label{6.1}
\end{equation}

Substituting the Ansatz (\ref{6.1}) into YMEs and splitting the
equality obtained with respect to linearly independent four-vectors
with components $k_{\mu}$,\ $b_{\mu}$,\ $c_{\mu}$, we get
\begin{eqnarray}
&1)& \vec{C}_{\om _{1}\om _{1}} = \vec{0},\non\\
&2)& \vec{C}\times\vec{C}_{\om _{1}} = \vec{0},\label{6.2}\\
&3)& \vec{B}_{\om _{1}\om _{1}} + e\vec{C}_{\om _{0}}\times\vec{C} +
  e^{2}\vec{C}\times(\vec{C}\times\vec{B}) = \vec{0}.\non
\end{eqnarray}

Here we use the notations $\om _{0} = k\cdot x,\ \om _{1} = c\cdot
x$. 

The general solution of the first two equations from (\ref{6.2}) is
given by one of the formulae
\vspace{1.5mm}

\noindent
{\rm I.}\ $\vec{C} = \vec{f}(\om _{0})$,
\vspace{1.5mm}

\noindent
{\rm II.}\ $\vec{C} = \Bigl(\om _{1} + v_{0}(\om
_{0})\Bigr)\vec{f}(\om _{0})$, 
\vspace{1.5mm}

\noindent
where $v_{0},\ \vec{f}$ are arbitrary smooth functions.

Consider the case $\vec{C} = \vec{f}(\om _{0})$. Substituting this
expression into the third equation from (\ref{6.2}) we have

\begin{equation}
\vec{B}_{\om _{1}\om _{1}} + e\vec{f}_{\om _{0}}\times\vec{f} +
e^{2}\vec{f} (\vec{f}\vec{B}) - e^{2}\vec{f}^{\, 2}\vec{B} = \vec{0}.  
\label{6.3}
\end{equation}

Since equations (\ref{6.3}) do not contain derivatives of $\vec{B}$
with respect to $\om _{0}$, they can be considered as a system of ODEs
with respect to the variable $\om _{1}$. Multiplying (\ref{6.3}) by
$\vec{f}$ we arrive at the relation $(\vec{B}\vec{f}\,)_{\om _{1}\om
_{1}} = 0$, whence

\begin{equation}
\vec{B}\vec{f} = v_{1}(\om _{0})\om _{1} + v_{2}(\om _{0}).
\label{6.4}
\end{equation}

In (\ref{6.4}) $ v_{1},\ v_{2}$ are arbitrary sufficiently smooth
functions.

With account of (\ref{6.4}) system (\ref{6.3}) reads
\begin{displaymath}
\vec{B}_{\om _{1}\om _{1}} - e^{2}{\vec f}^{\, 2}\vec{B} =
e\vec{f}\times\vec{f}_{\om _{0}} - 
e^{2}(v_{1}\om _{1} + v_{2})\vec{f}.
\end{displaymath}

The above linear system of ODEs is easily integrated. Its general
solution is given by the formula
\begin{equation}
\begin{array}{l}
\vec{B} = \vec{g}(\om _{0})\cosh e|\vec{f}|\om _{1} + \vec{h}(\om
 _{0}) \sinh e|\vec{f}|\om _{1} + e^{-1}|\vec{f}|^{-2}{\vec{f}}_{\om
 _{0}}\times\vec{f}\\[2mm]  
\quad\times |\vec{f}|^{-2} (v_{1}\om _{1} + v_{2})\vec{f},
\end{array}
\label{6.5}
\end{equation}
where $\vec{g},\ \vec{h}$ are arbitrary smooth functions.

Substituting (\ref{6.5}) into (\ref{6.4}) we get the following
restrictions on the choice  of the functions $\vec{g},\ \vec{h}$: 

\begin{equation}
\vec{f}\,\vec{g} = 0, \quad \vec{f}\,\vec{h} = 0.
\label{6.6}
\end{equation}

Thus, provided $\vec{C}_{\om _{1}} = 0$, the general solution of the 
system of ODEs (\ref{6.3}) is given by formulae (\ref{6.5}),
(\ref{6.6}). Substituting (\ref{6.5}) into the initial Ansatz
(\ref{6.1}) we obtain the following family of exact solutions of YMEs:   
\begin{eqnarray*}
\vec{A}_{\mu} &=& k_{\mu}\Bigl\{\vec{g}\cosh
e|\vec{f}|c\cdot x + \vec{h} \sinh e|\vec{f}|c\cdot x +
e^{-1}|\vec{f}|^{-2}\dot{\vec{f}}\times\vec{f}\\ 
&&+ (v_{1}c\cdot x + v_{2})\vec{f}\Bigr\} + b_{\mu}\vec{f},
\end{eqnarray*}
where $\vec{f},\ \vec{g},\ \vec{h},\ v_{1},\ v_{2}$ are arbitrary
smooth functions of $k\cdot x$ satisfying (\ref{6.6}), an overdot
denotes differentiation with respect to $\om_0=k\cdot x$.

The case $\vec{C} = [\om_{1} + v_{0}(\om_{0})]\vec{f}(\om_{0})$
is treated in a similar way. As a result, we obtain the following
family of exact solutions of YMEs:
\index{Exact solutions!of the Yang-Mills equations}
\begin{eqnarray*}
\vec{A}_{\mu} &=& k_{\mu}\Bigl\{(c\cdot x + v_{0})^{1/2}\Bigl(
\vec{g}J_{1/4} [(ie/2)|\vec{f}|(c\cdot x + v_{0})^{2}]\\ 
&& + \vec{h}Y_{1/4}[(ie/2)|\vec{f}|(c\cdot x + v_{0})^{2}]\Bigr) +  
(v_{1}c\cdot x + v_{2})\vec{f}\\
&& + e^{-1}|\vec{f}|^{-2}\dot{\vec{f}}\times\vec{f}\Bigr\} + 
b_{\mu}(c\cdot x + v_{0})\vec{f},
\end{eqnarray*}
where $\vec{f},\ \vec{g},\ \vec{h},\ v_{0},\ v_{1},\ v_{2}$
are arbitrary smooth functions of $k\cdot x$ satisfying (\ref{6.6}),
$J_{1/4}(\om),\ Y_{1/4}(\om)$ are the Bessel functions.

Another effective Ansatz for the Yang-Mills field is obtained if
we replace $c\cdot x$ in (\ref{6.1}) by $b\cdot x$
\begin{equation}
\vec{A}_{\mu} = k_{\mu}\vec{B}(k\cdot x,\, b\cdot x) +
b_{\mu}\vec{C}(k\cdot x,\, b\cdot x). 
\label{6.7}
\end{equation}

Substitution of (\ref{6.7}) into YMEs yields the following system of
PDEs for $\vec{B},\ \vec{C}$:

\begin{equation}
\vec{B}_{\om_{1}\om_{1}} - \vec{C}_{\om_{0}\om_{1}} -
e (\vec{B}\times\vec{C}_{\om_{1}} +
2\vec{B}_{\om_{1}}\times\vec{C} + \vec{C}\times\vec{C}_{\om_{0}}) +
e^{2}\vec{C}\times(\vec{C}\times\vec{B}) = \vec{0}.
\label{6.8}
\end{equation}

We have succeeded in integrating system (\ref{6.8}), provided $\vec{C}
= \vec{f}(\om_{0})$. Substituting the result obtained into
(\ref{6.7}), we come to the following family of exact solutions of
YMEs: 
\index{Exact solutions!of the Yang-Mills equations}
\begin{eqnarray*}
\vec{A}_{\mu} &=& k_{\mu}\Bigl\{(\vec{g} + b\cdot x |\vec{f}|^{-1}
\vec{g} \times\vec{f}\,) \cos ( e|\vec{f}|b\cdot x) +
(\vec{h} + b\cdot x|\vec{f}|^{-1}\vec{h}\times\vec{f}\,)\\
&&\times\sin(e|\vec{f}|b\cdot x) +
e^{-1}|\vec{f}|^{-2}\dot{\vec{f}}\times 
\vec{f} + (v_{1}b\cdot x + v_{2})\vec{f}\Bigr\} + b_{\mu}\vec{f},
\end{eqnarray*}
where $\vec{f},\ \vec{g},\ \vec{h},\ v_{1},\ v_{2}$ are arbitrary
smooth functions of $k\cdot x$.

In addition, we have constructed the following class of exact solutions
of YMEs: 
\index{Exact solutions!of the Yang-Mills equations}
\begin{eqnarray*}
\vec{A}_{\mu} &=& k_{\mu}\vec{e}_{1}vu^{2}(b\cdot x) +
b_{\mu}\vec{e}_{2}u(b\cdot x),
\end{eqnarray*}
where $\vec{e}_{1} = (1,0,0), \ \vec{e}_{2} = (0,1,0);
\ v$ is an arbitrary smooth function of $k\cdot x$; \ $ u(b\cdot
x)$ is a solution of the nonlinear ODE $\ddot{u} = e^{2}u^{5}$,
which is integrated in elliptic functions. 

In conclusion of this section we will obtain a generalization of the
plane-wave Coleman solution \cite{38.1}
\begin{equation}
\vec{A}_{\mu} = k_{\mu}\Bigl(\vec{f}(k\cdot x)b\cdot x +
\vec{g}(k\cdot x)c\cdot x\Bigr). 
\label{6.9}
\end{equation}

It is not difficult to verify that (\ref{6.9}) satisfy YMEs with
arbitrary  $\vec{f},\ \vec{g}$.

Evidently, solution (\ref{6.9}) is a particular case of the Ansatz
\begin{equation}
\vec{A}_{\mu} = k_{\mu}\vec{B}(k\cdot x,\, b\cdot x,\, c\cdot x).
\label{6.10}
\end{equation}

Substituting (\ref{6.10}) into YMEs we get

\begin{equation}
\vec{B}_{\om_{1}{\om_1}} + \vec{B}_{\om_{2}\om_{2}} = \vec{0},
\label{6.11}
\end{equation}
where $\om_{1} = b\cdot x,\ \om_{2} = c\cdot x$.

Integrating the Laplace equations\index{Laplace equation} (\ref{6.11})
and substituting the result obtained into (\ref{6.10}) we have

\begin{displaymath}
\vec{A}_{\mu} = k_{\mu} \Bigl(\vec{U}(k\cdot x,\, b\cdot x + i c\cdot
x) + \vec{U} (k\cdot x,\, b\cdot x - i c\cdot x)\Bigr).
\end{displaymath}

Here $\vec{U}(k\cdot x,\, z)$ is an arbitrary analytical with respect  
to $z$ function. Choo\-sing $\vec{U} = (1/2)[\vec{f}(k\cdot x) -
i\vec{g}(k\cdot x)]z$ we get the Coleman's solution (\ref{6.9}).
\vspace{10mm}

\noindent
{\large\bf 7.2. Non-Lie reduction of the Yang-Mills
  equations\label{s7.2}} 
\markboth{Chapter 7. REDUCTION AND EXACT SOLUTIONS}
   {7.2. Non-Lie reduction of the Yang-Mills equations}
\def\theequation{7.\arabic{section}.\arabic{equation}}
\setcounter {section} {2}
\setcounter {equation}{0}
\vspace{7mm}

\noindent
In the present section we will obtain conditionally-invariant
Ans\"atze for the Yang-Mills field $\vec A_\mu(x)$ utilizing the
idea which enables us to construct non-Lie\index{Non-Lie!Ansatz} 
(conditionally-invariant)
Ans\"atze for the spinor field $\psi(x)$. This idea proves to be 
fruitful for obtaining new reductions and constructing new exact
solutions of the $SU(2)$ Yang-Mills equations (\ref{1.1}) as
compared with those found by means of the symmetry reduction of YMEs.
\vspace{2mm}

\noindent
{\bf 1. Reduction of YMEs.}\
We are looking for a solution of YMEs of the form (\ref{3.8}),
(\ref{3.13}) without imposing {\em a priori} \/conditions on the
functions $\omega(x),\ \theta_{\mu}(x)$. They should be determined
from the requirement that substitution of the Ansatz (\ref{3.8})
into system of PDEs (\ref{1.1}) yields a system of ordinary
diffe\-ren\-ti\-al equations for a vector function $\vec
B_{\mu}(\omega)$. 

By direct check one can become convinced of that the following
assertion holds true.
\vspace{1.5mm}

\noindent
{\bf Lemma 7.2.1.}\ {\em Ansatz (\ref{3.8}), (\ref{3.13}) reduces YMEs 
  (\ref{1.1}) to a system of ODEs iff the functions $\omega(x)$,\ 
  $\theta_{\mu}(x)$ satisfy the system of PDEs }
\begin{eqnarray}
&1)&\omega_{x_{\mu}}\omega_{x^{\mu}}=F_1(\omega),\non\\
&2)&\Box\omega=F_2(\omega),\non\\
&3)&Q_{\alpha\mu}\omega_{x_{\alpha}}=G_{\mu}(\omega),\non\\
&4)&Q_{\alpha\mu x_{\alpha}}=H_{\mu}(\omega),\label{7.2.4}\\
&5)&Q^{\alpha}_{\mu}Q_{\alpha\nu x_{\beta}}
\omega_{x^\beta}=R_{\mu\nu}(\omega), \non\\
&6)&Q^{\alpha}_{\mu}\Box Q_{\alpha\nu}=S_{\mu\nu}(\omega),\non\\
&7)&Q^{\alpha}_{\mu} Q_{\alpha\nu x_{\beta}}Q_{\beta\gamma}
+Q_{\nu}^{\alpha} Q_{\alpha\gamma x_{\beta}}Q_{\beta\mu}+
Q_{\gamma}^{\alpha}Q_{\alpha\mu x_{\beta}}Q_{\beta\nu}
=T_{\mu\nu\gamma}(\omega),\non
\end{eqnarray}
{\em where $F_1,\ F_2,\ G_{\mu},\ldots,T_{\mu\nu\gamma}$
are some smooth functions},\ $\mu, \nu, \gamma={0,\ldots,3}$.
\index{Non-Lie reduction!of the Yang-Mills equations}

{\em And what is more, a reduced equation has the form
\begin{equation}
\begin{array}{l}
k_{\mu\gamma}\ddot{\vec B^{\gamma}}+
l_{\mu\gamma}\dot{\vec B^{\gamma}}+m_{\mu\gamma}\vec B^{\gamma}+
eq_{\mu\nu\gamma}\dot{\vec B^{\nu}}\times\vec B^{\gamma}+
eh_{\mu\nu\gamma}\vec B^{\nu}\times\vec B^{\gamma}\\[2mm]
\quad +e^2\vec B_{\gamma}
\times (\vec B^{\gamma}\times\vec B_{\mu})=\vec 0,
\end{array}
\label{7.2.5}
\end{equation}
where}
\begin{eqnarray}
k_{\mu\gamma}&=&g_{\mu\gamma}F_1-G_{\mu}G_{\gamma},\non\\
l_{\mu\gamma}&=&g_{\mu\gamma}F_2+2R_{\mu\gamma}-
G_{\mu}H_{\gamma}-G_{\mu}\dot G_{\gamma},\non\\
m_{\mu\gamma}&=&S_{\mu\gamma}-
G_{\mu}\dot H_{\gamma},\label{7.1.6}\\
q_{\mu\nu\gamma}&=&g_{\mu\gamma}G_{\nu}+g_{\nu\gamma}
G_{\mu}-2g_{\mu\nu}G_{\gamma},\non\\
h_{\mu\nu\gamma}&=&(1/2)(g_{\mu\gamma}H_{\nu}-
g_{\mu\nu}H_{\gamma})-T_{\mu\nu\gamma}.\non
\end{eqnarray}

Thus, to describe all Ans\"atze of the form (\ref{3.8}) reducing YMEs
to a system of ODEs we have to construct the general solution of the
over-determined system of PDEs (\ref{3.13}), (\ref{7.2.4}).  Let us
emphasize that system (\ref{3.13}), (\ref{7.2.4}) is compatible since
Ans\"atze for the Yang-Mills field $\vec Y_\mu(x)$ invariant under
the $P(1,3)$ non-conjugate subgroups of the Poincar\'e group satisfy
equations (\ref{3.13}), (\ref{7.2.4}) with some specific choice of
the functions $F_1,\ F_2,\ldots, T_{\mu\nu\gamma}$.

Computations needed to integrate system of nonlinear PDEs
(\ref{3.13}), (\ref{7.2.4}) are rather involved. In addition, they have
much in common with those performed to obtain conditionally-invariant
Ans\"atze for the spinor field (Theorem 6.1.1). That is why we
present here only a principal idea of our approach to solving the
system (\ref{3.13}), (\ref{7.2.4}). When integrating it we use
essentially the fact that the general solution of system of equations
1, 2 from (\ref{7.2.4})\index{d'Alembert-Hamilton system} is known 
(see Section 2.1). With already
known $\omega (x)$ we proceed to integration of linear PDEs 3, 4 from
(\ref{7.2.4}).  Next, we substitute the results obtained into the
remaining equations and thus get the final form of the functions
$\omega(x)$, \ $\theta_{\mu}(x)$.

Before adducing the results of integration of system of PDEs
(\ref{3.13}), (\ref{7.2.4}) we make a remark. As a direct check shows, 
the structure of the Ansatz (\ref{3.8}), (\ref{3.13}) is not
altered by the change of variables 
\begin{equation}
\begin{array}{l}
\omega\to\omega'=T(\omega),\quad
\theta_0\to\theta_0'=\theta_0+T_0(\omega),\\[2mm]
\theta_1\to\theta_1'=\theta_1+e^{\theta_0}
\Bigl(T_1(\omega)\cos\theta_3+T_2(\omega)\sin\theta_3\Bigr),\\[2mm]
\theta_2\to\theta_2'=\theta_2+e^{\theta_0}
\Bigl(T_2(\omega)\cos\theta_3-T_1(\omega)\sin\theta_3\Bigr),\\[2mm]
\theta_3\to\theta_3'=\theta_3+T_3(\omega),
\end{array}
\label{7.2.7}
\end{equation}
where $T(\omega),\ T_{\mu}(\omega)$ are arbitrary smooth
functions. That is why  solutions of system (\ref{3.13}), (\ref{7.2.4})
connected by the relations (\ref{7.2.7}) are considered as equivalent.

It occurs that the new (non-Lie) Ans\"atze\index{Non-Lie!Ansatz} 
are obtained only when the
functions $\omega(x),\ \theta_{\mu}(x)$ up to the equivalence relations 
(\ref{7.2.7}) have the form
\begin{equation}
\begin{array}{l}
\theta_{\mu}=\theta_{\mu} (\xi,\, b\cdot x,\, c\cdot x),\\[2mm]
\omega = \omega (\xi,\, b\cdot x,\, c\cdot x),
\end{array}
\label{7.2.8}
\end{equation}
where
$\xi=(1/2)k\cdot x,\ k_{\nu}=a_{\nu}+d_{\nu},\ \mu,\nu ={0,\ldots,3}$.

A list of inequivalent solutions of system of PDEs (\ref{3.13}),
(\ref{7.2.4}) belonging to the class (\ref{7.2.8}) is exhausted by the 
following solutions: 
\begin{eqnarray}
&1)&\theta_0=\theta_3=0,\quad
\omega=(1/2)k\cdot x,\quad
\theta_1=w_0(\xi)b\cdot x+w_1(\xi)c\cdot x,\non\\
& &\theta_2=w_2(\xi)b\cdot x+w_3(\xi)c\cdot x;\non\\[3mm]
&2)&\omega=b\cdot x+w_1(\xi),
\quad\theta_0=\alpha\Bigl (c\cdot x+w_2(\xi)\Bigr ),\non\\
& &\theta_a=-(1/4)\dot w_a(\xi),\ \
a=1,2,\quad \theta_3=0,\label{7.2.9}\\[3mm]
&3)&\theta_0=T(\xi), \quad \theta_3=w_1(\xi),\quad
\omega=b\cdot x\cos w_1+c\cdot x\sin w_1+
w_2(\xi),\non\\
& &\theta_1=\Bigl ((1/4)(\varepsilon e^T+\dot T)
(b\cdot x\sin w_1-c\cdot x\cos w_1)+w_3(\xi)\Bigr )
\sin w_1\non\\
& &\quad +(1/4)\Bigl ( \dot w_1(b\cdot x
\sin w_1-c\cdot x\cos w_1)-\dot w_2\Bigr )\cos w_1,\non\\
& &\theta_2=-\Bigl ((1/4)(\varepsilon e^T+\dot T)
(b\cdot x\sin w_1-c\cdot x\cos w_1)+w_3(\xi)\Bigr )
\cos w_1\non\\
& &\quad +(1/4)\Bigl ( \dot w_1(b\cdot x
\sin w_1-c\cdot x\cos w_1)-\dot w_2\Bigr )\sin w_1;\non\\[3mm]
&4)&\theta_0=0,\quad \theta_3= \arctan
\ \Bigl ([c\cdot x+w_2 (\xi)][b\cdot x+
w_1(\xi)]^{-1}\Bigr ),\non\\
& &\theta_a=-(1/4)\dot w_a (\xi),\ \ a=1,2,\non\\
& &\omega = \Bigl ([b\cdot x+
w_1(\xi)]^2+[c\cdot x+w_2(\xi)]^2\Bigr )^{1/2}.\non
\end{eqnarray}

Here $\alpha\ne 0$ is an arbitrary constant, $\varepsilon=\pm 1$,\
$w_0,\ w_1,\ w_2,\ w_3$ are arbitrary smooth functions of
$\xi= (1/2) k\cdot x,$\ $T=T(\xi)$ is a solution of 
the nonlinear ODE
\begin{equation}
(\dot T+\varepsilon e^T)^2+\dot w_1^2=\vark e^{2T},
\ \ \vark \in \R^1.
\label{7.2.10}
\end{equation}

Substitution of the Ansatz (\ref{3.8}), where
$Q_{\mu\nu}(x)$ are given by formulae (\ref{3.13}), (\ref{7.2.9}),
into YMEs yields systems of nonlinear ODEs of the form (\ref{7.2.5}),
where 
\begin{eqnarray}
&1)& k_{\mu\gamma}=-(1/4)k_{\mu}k_{\gamma},\quad
l_{\mu\gamma}=-(w_0+w_3)k_{\mu}k_{\gamma},\non\\
& & m_{\mu\gamma} = -4\ (w_0^2+w_1^2+w_2^2+w_3^2)
k_{\mu}k_{\gamma}-(\dot w_0+\dot w_3)k_{\mu}k_{\gamma},\non\\
& & q_{\mu\nu\gamma}=
(1/2)(g_{\mu\gamma}k_{\nu}+g_{\nu\gamma}k_{\mu}-
2g_{\mu\nu}k_{\gamma}),\non\\
& & h_{\mu\nu\gamma}=
(w_0+w_3)(g_{\mu\gamma} k_{\nu}-g_{\mu\nu} k_{\gamma})+
2(w_1-w_2)\Bigl( (k_{\mu}b_{\nu}-k_{\nu}b_{\mu})\ c_{\gamma}\non\\
& &\quad +(b_{\mu}c_{\nu}-b_{\nu}c_{\mu})k_{\gamma}+(c_{\mu}k_{\nu}-
c_{\nu}k_{\mu})b_{\gamma}\Bigr);\non\\[3mm]
&2)& k_{\mu\gamma}=-g_{\mu\gamma}-b_{\mu}b_{\gamma},
\quad l_{\mu\gamma}=0,\quad
m_{\mu\gamma}=-\alpha^2 (a_{\mu}
a_{\gamma}-d_{\mu}d_{\gamma}),\non\\
& &q_{\mu\nu\gamma}=g_{\mu\gamma}b_{\nu}+
g_{\nu\gamma}b_{\mu}-2g_{\mu\nu}b_{\gamma},\non\\
& &h_{\mu\nu\gamma}=\alpha\Bigl ((
a_{\mu}d_{\nu}-a_{\nu}d_{\mu})c_{\gamma}+(d_{\mu}c_{\nu}-
d_{\nu}c_{\mu}) a_{\gamma}+(c_{\mu}a_{\nu}-c_{\nu}a_{\mu})d_{\gamma}
\Bigr );\non\\[3mm]
&3)& k_{\mu\gamma}=-g_{\mu\gamma}-b_{\mu}b_{\gamma},
\quad l_{\mu\gamma}=-(\varepsilon/2)b_{\mu}
k_{\gamma},\label{7.2.11}\\ 
& &m_{\mu\gamma}=-(\vark/4)k_{\mu}k_{\gamma},
\quad q_{\mu\nu\gamma}=g_{\mu\gamma} b_{\nu}+g_{\nu\gamma}b_{\mu}-
2g_{\mu\nu}b_{\gamma},\non\\
& &h_{\mu\nu\gamma}=(\ve/4)(g_{\mu\gamma}k_{\nu}
-g_{\mu\nu}k_{\gamma});\non\\[3mm]
&4)& k_{\mu\gamma}=-g_{\mu\gamma}-b_{\mu}b_{\gamma},
\quad l_{\mu\gamma}=-\omega^{-1}(g_{\mu\gamma}+
b_{\mu} b_{\gamma}),\non\\
& &m_{\mu\gamma}=-\omega^{-2}c_{\mu}c_{\gamma},\quad
q_{\mu\nu\gamma}=g_{\mu\gamma} b_{\nu}+g_{\nu\gamma} b_{\mu}-
2g_{\mu\nu} b_{\gamma},\non\\
& &h_{\mu\nu\gamma}=(1/2)\omega^{-1}(g_{\mu\gamma}b_{\nu}
-g_{\mu\nu} b_{\gamma}).\non
\end{eqnarray}
{\bf 2. Exact solutions of the Yang-Mills equations.}\
Systems (\ref{7.2.5}), (\ref{7.2.11}) are systems of twelve nonlinear
second-order ODEs with variable coefficients. That is why  there
is a little hope to construct their general solutions. But it is
possible to obtain particular solutions of system (\ref{7.2.5}) with
coefficients given by formulae 2--4 from (\ref{7.2.11}).

Consider, as an example, system of ODEs (\ref{7.2.5}) with
coefficients given by the formulae 2 from (\ref{7.2.11}). We look
for its solutions in the form
\begin{equation}
\vec B_{\mu}= k_{\mu} \vec e_1 f(\omega)+b_{\mu} \vec e_2 g(\omega),
\ \ fg\ne 0,
\label{7.2.12}
\end{equation}
where $\vec e_1=(1, 0, 0),\ \vec e_2=(0, 1, 0)$.

Substituting the expression (\ref{7.2.12}) into the above mentioned
system we get 
\begin{equation}
\ddot f +(\alpha^2-e^2g^2)f=0,\quad
f\dot g+2\dot fg=0.
\label{7.2.13}
\end{equation}

The second ODE from (\ref{7.2.13}) is easily integrated
\begin{equation}
g=\lambda f^{-2},\ \ \lambda \in\R^1,
\ \ \lambda\ne 0.
\label{7.2.14}
\end{equation}

Substitution of the result obtained into the first ODE from
(\ref{7.2.13}) yields the Ermakov-type equation for $f(\omega)$
\begin{displaymath}
\ddot f +\alpha^2 f-e^2\lambda^2f^{-3}=0,
\end{displaymath}
which is integrated in elementary functions \cite{132}
\begin{equation}
f=\Bigl (\alpha^{-2}C^2+\alpha^{-2}
(C^4-\alpha^2e^2\lambda^2)^{1/2}
\sin 2|\alpha|\omega\Bigr )^{1/2}.
\label{7.2.15}
\end{equation}

Here $C\ne 0$ is an arbitrary constant.

Substituting (\ref{7.2.12}), (\ref{7.2.14}), (\ref{7.2.15}) into the
corresponding Ansatz for $\vec A_{\mu}(x)$ we get the following class
of exact solutions of YMEs (\ref{1.1}):
\begin{eqnarray*}
\vec A_{\mu}&=&\vec e_1 k_{\mu}\exp\ (-\alpha c\cdot x-\alpha
w_2 )\Bigl (\alpha^{-2}C^2+\alpha^{-2}(C^4-
\alpha^2 e^2\lambda^2)^{1/2}\\
&&\times \sin 2 \vert\alpha\vert
(b\cdot x+w_1)\Bigr )^{1/2}+
\vec e_2\lambda \Bigl (\alpha^{-2}C^2+\alpha^{-2}
(C^4-\alpha^2 e^2 \lambda^2)^{1/2}\\
&&\times \sin 2 \vert\alpha\vert (b\cdot x+w_1)\Bigr )^{-1} (b_{\mu}+
(1/2) k_{\mu} \dot w_1\Bigr ).
\end{eqnarray*}

In a similar way we have obtained five other classes of the exact
solutions of the Yang-Mills equations 
\index{Exact solutions!of the Yang-Mills equations}
\begin{eqnarray*}
  \vec A_{\mu}&=&\vec e_1 k_{\mu}e^{-T} (b\cdot x \cos w_1+
c\cdot x \sin w_1+w_2)^{1/2}Z_{1/4}\Bigl ((ie\lambda/2)
(b\cdot x\cos w_1\\
&&+ c\cdot x \sin w_1+w_2)^2\Bigr )+ \vec e_2\lambda\ (b\cdot x
\cos w_1+c\cdot x\sin w_1+w_2)\\
&&\times \Bigl (c_{\mu}\cos w_1-b_{\mu}\sin w_1+2 k_{\mu}
[(1/4)(\ve e^T+\dot T)(b\cdot x\sin w_1\\
&&-c\cdot x\cos w_1)+w_3]\Bigr );\\[3mm]
  \vec A_{\mu}&=&\vec e_1 k_{\mu}
e^{-T} \Bigl (C_1\cosh[e\lambda(b\cdot x\cos w_1+
c\cdot x\sin w_1+w_2)]+C_2\sinh[e\lambda\\
&&\times(b\cdot x\cos w_1+ c\cdot x \sin w_1+w_2)]\Bigr )+ \vec
e_2\lambda \Bigl ( c_{\mu}\cos w_1-b_{\mu}\sin w_1\\
&&+ 2k_{\mu}[(1/4)(\ve e^T+ \dot T)(b\cdot x\sin w_1-c\cdot x\cos
w_1)+w_3]\Bigr );\\[3mm] 
  \vec A_{\mu}&=&\vec e_1 k_{\mu}
e^{-T}\Bigl (C^2(b\cdot x\cos w_1+
c\cdot x\sin w_1+w_2)^2+\lambda^2 e^2C^{-2}\Bigr )^{1/2}\\
&&+ \vec e_2\lambda \Bigl (C^2(b\cdot x\cos w_1+ c\cdot x\sin w_1+w_2)^2+
\lambda^2 e^2 C^{-2}\Bigr )^{-1}\\
&&\times \Bigl (b_{\mu}\cos w_1+c_{\mu}\sin w_1-
(1/2)k_{\mu}[\dot w_1 (b\cdot x\sin w_1\\
&&-c\cdot x\cos w_1)-\dot w_2]\Bigr );\\[3mm]
  \vec A_{\mu}&=&\vec e_1 k_{\mu} Z_0
\Bigl ((ie\lambda/2)[(b\cdot x+w_1)^2+(c\cdot x+
w_2)^2]\Bigr )+\vec e_2\lambda \Bigl ( c_{\mu}(b\cdot x+w_1)\\
&&-b_{\mu}(c\cdot x+w_2)- (1/2)k_{\mu}[\dot w_1(c\cdot x+
w_2)-\dot w_2(b\cdot x+w_1)]\Bigr );\\[3mm]
  \vec A_{\mu}&=&\vec e_1 k_{\mu}\Bigl (C_1[(b\cdot x+w_1)^2+
(c\cdot x+w_2)^2]^{e\lambda/2} + C_2[(b\cdot x+w_1)^2\\
&&+(c\cdot x+w_2)^2]^{-e\lambda/2}\Bigr) +\vec e_2\lambda[(b\cdot x+
w_1)^2+(c\cdot x+w_2)^2]^{-1}\\
&&\times \Bigl (c_{\mu}(b\cdot x+w_1)-b_{\mu}(c\cdot x+w_2)-
(1/2)k_{\mu}[\dot w_1(c\cdot x+w_2)\\
&&-\dot w_2(b\cdot x+w_1)]\Bigr ).
\end{eqnarray*}

Here\ $C_1,\ C_2,\ C\ne 0,\ \lambda$ are arbitrary parameters; $w_1,\ 
w_2,\ w_3$ are arbitrary smooth functions of $\xi=(1/2)k\cdot x, \ 
T=T(\xi)$ is a solution of ODE (\ref{7.2.10}) and
\begin{eqnarray*}
& &Z_s(\omega)= C_1J_s (\omega)+C_2Y_s (\omega),\\
& &\vec e_1 = (1,0,0),\quad \vec e_2= (0,1,0),
\end{eqnarray*}
where $J_s,\ Y_s$ are Bessel functions.

Thus, we have obtained broad families of exact non-Abelian
solutions of YMEs (\ref{1.1}).

In conclusion of the section we will say a few words about the
symmetry interpretation of the Ans\"atze (\ref{3.8}),
(\ref{3.13}), (\ref{7.2.9}).  Let us consider as an example the
Ansatz determined by the formulae 1 from (\ref{7.2.9}). As a direct
computation shows, the generators of a three-parameter Lie group
leaving it invariant are of the form
\begin{displaymath}
\begin{array}{l}
Q_1= k_{\alpha}\partial_{\alpha},\\
Q_2=b_{\alpha}\partial_{\alpha}-
2\Biggl \{[ w_0(k_{\mu}b_{\nu}-k_{\nu}b_{\mu})+
 w_2(k_{\mu}c_{\nu}-k_{\nu}c_{\mu})]
\, {\displaystyle\mathop\sum\limits_{a=1}^3} A^{a\nu}\Biggr \}
\partial_{\displaystyle{A^{a\mu}}},\\
Q_3= c_{\alpha}\partial_{\alpha}-
2\Biggl \{[ w_1(k_{\mu}b_{\nu}-k_{\nu}b_{\mu})+
w_3(k_{\mu}c_{\nu}-k_{\nu}c_{\mu})]
\, {\displaystyle\mathop\sum\limits_{a=1}^3} A^{a\nu}\Biggr \}
\partial_{\displaystyle{A^{a\mu}}}.
\end{array}
\end{displaymath}

Evidently, system of PDEs (\ref{1.1}) is invariant under the
one-parameter group having the generator $Q_1$. But it is not
invariant under the groups having the generators $Q_2,\ Q_3$. At the
same time, the system of 
PDEs\index{Conditional symmetry!of the Yang-Mills equations} 
\begin{eqnarray*}
& &\p_{\nu}\p^{\nu}\vec A_{\mu}-\p^{\mu}\p_{\nu}
\vec A_{\nu}+e\Bigl ((\p_{\nu}\vec A_{\nu})\times
\vec A_{\mu}-2(\p_{\nu}\vec A_{\mu})\times \vec A_{\nu}\\
&&\quad+(\p^{\mu}\vec A_{\nu})\times \vec A^{\nu}\Bigr )+
e^2\vec A_{\nu} \times (\vec A^{\nu}\times \vec A_{\mu})
=\vec 0,\\
&&Q_0\vec A_{\mu}\equiv k_{\alpha}\p_{\alpha}\vec A_{\mu}
=\vec 0,\\
&&Q_1 \vec A_{\mu}\equiv b_{\alpha}\p_{\alpha}\vec A_{\mu}
+2\Bigl( w_0(k_{\mu}b_{\nu}-k_{\nu}b_{\mu})+
w_2(k_{\mu}c_{\nu}-k_{\nu}c_{\mu})\Bigr )\vec A^{\nu}=\vec 0,\\
&&Q_2\vec A_{\mu}\equiv c_{\alpha}\p_{\alpha}\vec A_{\mu}
+2\Bigl ( w_1(k_{\mu}b_{\nu}-k_{\nu}b_{\mu})+
w_3(k_{\mu}c_{\nu}-k_{\nu}c_{\mu})\Bigr )\vec A^{\nu}=\vec 0
\end{eqnarray*}
is invariant under the above mentioned group. Consequently, YMEs
(\ref{1.1}) are conditi\-on\-al\-ly-invariant under the Lie algebra
$\langle Q_1,\, Q_2,\, Q_3\rangle$. It means that solutions of YMEs
obtained with the help of Ansatz invariant under the group with
generators $Q_1,\ Q_2,\ Q_3$ cannot be found by means of the classical 
symmetry reduction procedure.

As rather tedious computations show, the Ans\"atze determined by the
formulae 2--4 from (\ref{7.2.9}) also correspond to conditional 
symmetry of YMEs. Hence it follows, 
in particular, that YMEs should be included into the long list of 
mathematical and theoretical physics equations
possessing nontrivial conditional symmetry \cite{65.3}.
\newpage
\phantom{.}

\newpage
\markboth {APPENDIX \ 1}
{The Poincar\'e group and its representations}
\thispagestyle{empty}
\noindent
{\sl
A P P E N D I X\ \ {\Large \sl 1}\label{ap1}}
\vspace{2mm}

\hrule
\vspace{35mm}

\rightline
{\large\bf
THE POINCAR\'E GROUP}
\vspace{2mm}

\rightline
{\large\bf
AND ITS REPRESENTATIONS}
\def\theequation{A.\arabic{section}.\arabic{equation}}
\setcounter {section} {1}
\setcounter {equation}{0}
\vspace{7mm}

The Poincar\'e group $P(1,3)$\index{Poincar\'e!group} is a group of
linear transformations of the Minkowski space $R(1,3)$ preserving the
quadratic form $s(x)=x_0^2-x_1^2- x_2^2-x_3^2$. We say that there is a
representation of the group 
$P(1,3)$\index{Representation!of the Poincar\'e group} in some 
linear space $H$ if the homomorphism of
this group $g\to T_g$ into the set of linear operators on $H$ is
determined, i.e.,\ the product of the group corresponds to the product
of operators $T_{g_1g_2}=T_{g_1}T_{g_2}$ and the unit element of the
group $P(1,3)$ corresponds to the identical transformation of the
space $H$. If the representation space $H$ is infinite-dimensional,
then it is assumed that the domain of definition of operators $T_g, \ 
\forall\,g\in P(1,3)$ is dense in $H$.

A representation is called irreducible if $H$ contains no subspace
invariant with respect to operators $T_g, \ \forall\,g\in P(1,3)$.

Irreducible representations of the Poincar\'e group were described by
Wigner as early as 1939. It is known that the problem of
description of representations of the Lie group $G$ can be reduced to
description of representations of its Lie algebra $AG$. An abstract
definition of the algebra $AP(1,3)$ is given by commutation
relations for the basis elements $P_\mu, \ J_{\alpha\beta}$ 
\begin{equation}
\begin{array}{l}
[P_\mu,\, P_\nu]=0,\quad [P_\mu,\, J_{\alpha\beta}]=i(g_{\mu\alpha}
P_\nu - g_{\mu\beta} P_\mu),\\[2mm]
[J_{\mu\nu},\, J_{\alpha\beta}]=i(g_{\mu\beta}J_{\nu\alpha}+
g_{\nu\alpha}J_{\mu\beta} - g_{\nu\beta}J_{\mu\alpha}-
g_{\mu\alpha}J_{\nu\beta}).
\end{array}
\label{A.1.0z}  
\end{equation}

A homomorphism $x\to T(x)$ of the algebra $AP(1,3)$ into
the set of linear operators determined in some linear space $H$
\begin{eqnarray}
        & &ax+by\to aT(x)+bT(y),\non\\
        & & [x,y]\to [T(x),T(y)]=T(x)T(y)-T(y)T(x), \quad
        \label{A.1.1}\\  
        & &\{x, y\} \subset AP(1,3), \quad \{a, b\} \subset \C^1\non
\end{eqnarray}
is called a representation of the Poincar\'e
algebra\index{Poincar\'e!algebra} $AP(1,3)$. 

Wigner's results were supplemented by Shirokov \cite{180} who was the
first to construct an explicit form of the basis elements of the
algebra $AP(1,3)$ for all classes of irreducible representations. In
many successive papers representations of this algebra in various
bases were found (see, for example, \cite{21}).

We adduce the formulae giving a complete description of irreducible
representations of the Poincar\'e algebra in the class of Hermitian
operators following \cite{74,75,77}. 

According to the Schur's lemma classification of irreducible
representations of the Lie algebra $L$ is reduced to construction of
the complete set of operators commuting with all basis elements
$x \in L$ (such operators are called the Casimir 
operators\index{Casimir operator} of the
algebra $L$) and to computation of the spectrum of their eigenvalues.
Furthermore, each set of eigenvalues of all Casimir operators
corresponds to the one and only one irreducible representation
\cite{18}. 
\vspace{1.5mm} 

\noindent
{\bf Theorem A.1.1}\cite{77}.\ {\em An arbitrary Hermitian
  representation of the Poincar\'e algebra $AP(1,3)$ can be realized
  by the following 
  operators:\index{Representation!of the Poincar\'e algebra}
\begin{eqnarray}
        P_0&=&p_0,\quad P_a\ \, =\ \, p_a,\non\\
        \vec J&=&\vec x\times \vec p+\lbd_0{(\vec n+\vec
          p\,)\over (1+\vec n\,\vec p\,)},\label{A.1.2}\\ 
        \vec N&=&-p_0\vec x+{\vec \lbd\times \vec p\over p^2}-
        (\lbd_0p_0p-\vec \lbd\,\vec p\,){(\vec p\times \vec n\,)\over
         (p+\vec n\,\vec p\,)},\non
\end{eqnarray}
where $\vec J=(J_1,J_2,J_3)$,\ $\vec N=(N_1,N_2,N_3)$,\ 
$J_a=(1/2)\ve_{abc}J_{bc}$,\ $N_a=J_{0a},$\ $a={1,2,3}$,\ 
$p_0$,\ $p_a$ are real variables connected by the relation
$p_0=\ve(C_1+p_ap_a)^{1/2}$,\ $\ve=\pm 1, \ C_1$ is an arbitrary real
number, $x_a=i\p/\p p_a$,\ $a={1,2,3}$,\ $p=(p_ap_a)^{1/2}$;\ 
$\lbd_0,\ \lbd_1,\ \lbd_2,\ \lbd_3$ are matrices satisfying the
commutation relations
\begin{equation}
\begin{array}{l}
        [\lbd_0,\, \lbd_a]=i\ve_{abc}n_b\lbd_c,\\[2mm]
        [\lbd_a,\, \lbd_b]=iC_1\ve_{abc}n_c\lbd_0
\end{array}
\label{A.1.3a}
\end{equation}
and $\vec n=(n_1,n_2,n_3)$ is an arbitrary unit vector.}

The algebra (\ref{A.1.3a}) has two Casimir operators
\begin{displaymath}
        I_1=\lbd_0^2C_1+\lbd_1^2+\lbd_2^2+\lbd_3^2,\quad
        I_2=\exp\{2i\pi\lbd_0\}.
\end{displaymath}

To obtain the explicit forms of matrices $\lbd_\mu$ realizing an
irreducible representation of algebra (\ref{A.1.3a}) we choose the
basis which consists of the complete set of eigenvectors of the
commuting operators $I_1,\ I_2$ and $\lbd_0$. On designating these
vectors by the symbol $\vert C_1,C_2,\lbd \rangle$ we have
\begin{eqnarray}
     & &\lbd_0\vert C_1,C_2,\lbd \rangle=\lbd \vert C_1,C_2,\lbd
     \rangle,\non\\ 
     & &(\lbd_1\pm i\lbd_2)\vert C_1,C_2,\lbd \rangle=(1/2)(1+n_3)
     \Bigl(C_2-C_1\lbd(\lbd\pm 1)\Bigr)^{1/2}\non\\
     & &\times\vert C_1,C_2,\lbd\pm 1\rangle
     +(n_2\mp n_1)^2\Bigl(2(1+n_3)\Bigr)^{-1}\label{A.1.3b}\\
     & &\times\Bigl(C_2-C_1\lbd(\lbd\mp1)\Bigr)^{1/2}
     \vert C_1,C_2,\lbd\mp 1\rangle,\non\\
     & &\lbd_3\vert C_1,C_2,\lbd \rangle=-(\lbd_1+\lbd_2)\vert
     C_1,C_2,\lbd \rangle.\non 
\end{eqnarray}

If the representation is irreducible, then the parameters $C_1,\ C_2$
take the fixed values from the intervals enumerated below in formulae
(\ref{A.1.4})
\begin{eqnarray}
       &1)& C_1=m^2>0, \quad C_2=C_1s(s+1),\quad
         \lbd=-s,\, -s+1,\ldots,s;\non\\
       &2)& C_1=C_2=0, \quad \lbd =\tilde \lbd;\non\\
       &3)& C_1=0,\quad C_2=\eta^2>0,\non\\
       & &\lbd=0,\, \pm 1,\, \pm 2,\, \ldots \quad{\rm or}\quad 
         \lbd=\pm {1/2},\, \pm3/2,\ldots;\non\\
       &4)& C_1=-\eta^2<0,\quad C_2=-\alpha\eta^2,\quad
         -\infty<\alpha<-1/4,\label{A.1.4}\\
         & &\lbd=\pm1/2,\, \pm3/2,\ldots \quad{\rm or}\quad 
         \lbd=0,\, \pm1,\, \pm2,\ldots;\non\\
         & &C_1=-\eta^2<0, \quad 0<C_2<(1/4)\eta^2, \quad
          \lbd=0,\pm1,\pm 2,\ldots;\non\\
         & &C_1=-\eta^2<0,\quad C_2=-l(l+1)\eta^2,
         \quad \lbd=l+1,\, l+2,\ldots;\non\\
         & &C_1=-\eta^2<0,\quad C_2=-l(l+1)\eta^2,
         \quad \lbd=-l-1,\, -l-2,\ldots,\non
\end{eqnarray}
where $s>0$ and $\tilde \lbd$ are arbitrary integer or half-integer 
numbers, $l$ is a positive integer or half-integer number satisfying
the condition $-(1/2)\le l<+\infty$ whose values in the irreducible
representation are fixed.

Formulae (\ref{A.1.2})--(\ref{A.1.4}) give all possible (up to the
equivalence relation) irreducible Hermitian representations of the
commutation relations (\ref{A.1.0z}) provided not all $P_\mu$ vanish.
If $P_\mu=0, \ \mu={0,\ldots,3}$, then algebra (1.1.31) is isomorphic
to the Lie algebra of the Lorentz group $O(1,3)$. The theory of
representations of the algebra $AO(1,3)$ is expounded with exhaustive
completeness in \cite{111}.

Among all possible representations of the Poincar\'e algebra a
specific role is played by so-called covariant 
representations\index{Covariant representation} which
are characterized by the following form of the basis elements
\begin{equation}
        P_\mu=p_\mu=ig_{\mu\nu}\p_\nu, \quad
        J_{\mu\nu}=x_\mu p_\nu-x_\nu p_\mu+S_{\mu\nu},
\label{A.1.8}
\end{equation}
where $S_{\mu\nu}$ are constant matrices.

Necessary and sufficient conditions for operators (\ref{A.1.8}) to
realize a representation of the Poincar\'e algebra are as follows:
\begin{displaymath}
        [S_{\mu\nu},\, S_{\alpha\beta}]=i(g_{\mu\beta}S_{\nu\alpha}+
        g_{\nu\alpha}S_{\mu\beta}-g_{\mu\alpha}S_{\nu\beta}-
        g_{\nu\beta}S_{\mu\alpha}).
\end{displaymath}

Let us note that operators (\ref{A.1.8}) unlike those given by
(\ref{A.1.2}) realize a reducible representation of the algebra
$AP(1,3)$. In addition, this representation is non-Hermitian if the
matrices $S_{\mu\nu}$ are finite-dimensional.

In what follows we consider the case of finite-dimensional matrices
$S_{\mu\nu}$, since it is mostly used in applications.

It is straightforward to verify that the matrices
\begin{eqnarray*}
        j_a&=&(1/2)\Bigl((1/2)\ve_{abc}S_{bc}+iS_{0a}\Bigr),\\
        \tau_a&=&(1/2)\Bigl((1/2)\ve_{abc}S_{bc}-iS_{0a}\Bigr)
\end{eqnarray*}
satisfy the following commutation relations:
\begin{equation}
        [j_a,\, j_b]=i\ve_{abc}j_c,\quad
        [\tau_a,\, \tau_b]=i\ve_{abc}\tau_c, \quad
        [j_a,\, \tau_b]=0.
\label{A.1.9}
\end{equation}

As a basis of the space of a finite-dimensional irreducible
representation of algebra (\ref{A.1.9}) we take the complete set of
eigenvectors $\vert j, m; \tau,n\rangle$ of commuting operators
$j_aj_a,\ j_3,\ \tau_a\tau_a,\ \tau_3$. In this basis the action of
the operators $j_a$ and $\tau_a$ can be represented in the form
\begin{equation}
\begin{array}{l}
        j_aj_a\vert j, m; \tau,n\rangle=j(j+1)\vert j, m;
        \tau,n\rangle,\\[2mm] 
        j_3\vert j, m; \tau,n\rangle=m\vert j, m;
        \tau,n\rangle,\\[2mm] 
        (j_1\pm ij_2)\vert j, m; \tau,n\rangle=\Bigl(j(j+1)\\[2mm]
        \quad -m(m\pm 1)\Bigr)^{1/2}\vert j, m\pm 1;
        \tau,n\rangle,\\[2mm] 
        \tau_a\tau_a\vert j, m; \tau,n\rangle=\tau(\tau+1)\vert j, m;
        \tau,n\rangle,\\[2mm]
        \tau_3\vert j, m; \tau,n\rangle=n\vert j, m;
        \tau,n\rangle,\\[2mm] 
        (\tau_1\pm i\tau_2)\vert j, m;
        \tau,n\rangle=\Bigl(\tau(\tau+1)\\[2mm] 
        \quad - n(n\pm 1)\Bigr)^{1/2}\vert j, m; \tau,n \pm 1\rangle, 
\end{array}
\label{A.1.10}
\end{equation}
where $j,\ m\, (\tau,\ n)$ are (half-) integer numbers, inequalities
holding
\begin{displaymath}
        -j\le m\le j, \quad -\tau\le n\le \tau.
\end{displaymath}

Thus, irreducible finite-dimensional representations of the algebra
$AO(1,3)$ (\ref{A.1.9}) are realized by matrices of the dimension
$(2j+1)(2\tau+1)\times (2j+1)(2\tau+1)$ with matrix elements
(\ref{A.1.10}).  The above representations are denoted by the symbol
$D(j,\tau)$.

Using formulae (\ref{A.1.10}) it is easy to check that on the set of
solutions of the Dirac equation (\ref{1.1.1}) the representation 
$D(1/2, 0)\oplus D(0,1/2)$ is realized.

The Poincar\'e algebra has two principal Casimir operators
\begin{displaymath}
        I_1=P_\mu P^\mu, \quad I_2=W_\mu W^\mu,
\end{displaymath}
where $W_0=(1/2)\ve_{abc}P_aJ_{bc},$ $W_a=(1/2)P_0\ve_{abc}J_{bc}-
\ve_{abc}P_bJ_{0c}$, whose eigen\-va\-lu\-es are considered as the
mass and the spin\index{Spin} of a particle.

We say that the Poincar\'e-invariant equation describes a particle
with the spin $s$ and the mass $m$ provided its solutions satisfy
identically the relations 
\begin{displaymath}
        I_1\psi=m^2\psi, \quad I_2\psi=s(s+1)m^2\psi.
\end{displaymath}

It is established by direct computation that solutions of the Dirac
equation satisfy the equalities $I_1\psi=m^2\psi, \ I_2\psi
=(3/4)m^2\psi $, whence it follows that the Dirac equation (1.1.1)
describes a particle with the spin $s=1/2$ and the mass $m$.
\newpage
\phantom{.}

\newpage

\noindent
{\sl
A P P E N D I X\ \ {\Large \sl 2}\label{ap2}}
\vspace{2mm}

\hrule
\vspace{35mm}

\rightline
{\large\bf
THE GALILEI GROUP}
\vspace{2mm}

\rightline
{\large\bf
AND ITS REPRESENTATIONS}
\markboth {APPENDIX\ 2}
{The Galilei group and its representations}
\def\theequation{A.\arabic{section}.\arabic{equation}}
\setcounter {section} {2}
\setcounter {equation}{0}
\thispagestyle{empty}
\vspace{7mm}

The Galilei group $G(1,3)$\index{Galilei!group} is a group 
of transformations of the
four-dimen\-si\-onal space $\R^1\times \R^3$ of the form
\begin{equation}
\begin{array}{rcl}
        t'&=&t+r_0,\\[2mm]
        x'_a&=&\theta_{ab}x_b+v_at+r_a,
\end{array}
\label{A.2.1}
\end{equation}
where $\|\theta_{ab}\|_{a,b=1}^3$ is an arbitrary orthogonal matrix,
$v_a,\ r_\mu$ are real parameters.

Since elements of orthogonal $(3\times 3)$-matrix can be expressed via
three parameters (for example, via the Euler angles), the group
(\ref{A.2.1}) is a 10-parameter Lie transformation group.

It is worth noting that a condition of invariance of physical laws
with respect to coordinate transformation (\ref{A.2.1}) is nothing
else but the mathematical formulation of the Galilei relativity
principle\index{Galilei!relativity principle}.  This principle 
establishes an equivalence of inertial
reference frames.  Therefore, the corresponding motion 
equation\index{Motion!equation} has to
be invariant under the Galilei group. In other words, some
representation of the Galilei group is to be realized on the set of
solutions of the equation in question.  Consequently, to investigate
wave equations invariant under the group $G(1,3)$ we have to know its
representations.\index{Representation!of the Galilei group}

As noted in the Appendix 1 the problem of description of
representations of the Lie group reduces to the study of
representations of its Lie algebra and besides we can restrict
ourselves to irreducible representations.

An abstract definition of the Galilei algebra 
$AG(1,3)$\index{Galilei!algebra} with basis
operators $P_0,\ P_a,\, J_a,\ G_a,\ M$ is given by the following
commutation relations 
\begin{eqnarray} 
&&[P_\mu,\, P_\nu]=0,\quad [P_\mu,\, M]=0,\non\\[1mm]
&&[J_a,\, M]=0,\quad [G_a,\, M]=0,\non\\[1mm]
&&[P_0,\, G_a]=iP_a,\quad [P_0,\, J_a]=0,\label{A.2.2}\\[1mm]
&&[P_a,\, G_b]=\delta_{ab}M,\quad [P_a,\, J_b]=i\ve_{abc}P_c,\non\\[1mm]
&&[G_a,\, J_b]=i\ve_{abc}G_c,\quad [J_a,\, J_b]=i\ve_{abc}J_c\non 
\end{eqnarray}
where $\mu,\nu=0,1,2,3;\ a,b,c=1,2,3$.  
\vspace{1.5mm}

\noindent
{\bf Note A.2.1.}\ In Section 4.1 we designate the basis elements 
of the rotation group $J_a,\ a={1,2,3}$ as $J_{ab},\ a\ne b,\
a,b={1,2,3}$. These notations are related by means of the
formula
\begin{displaymath}
J_a=(1/2)\ve_{abc}J_{bc}.
\end{displaymath}

Let us note that the Lie algebra of the group (\ref{A.2.1})
satisfies relations 
(\ref{A.2.2}) under $M=0$.

The algebra (\ref{A.2.2}) has three principal Casimir operators
\begin{eqnarray}
        C_1&=&M,\non\\
        C_2&=&(M\vec J-\vec P\times \vec G)^2,\label{A.2.3}\\
        C_3&=&2MP_0-P_aP_a.\non
\end{eqnarray}

Following \cite{73,77} we give a realization of irreducible
representations of the Galilei algebra distinguished by a universal
and quite simple form of the generators of the group $G(1,3)$.
\vspace{1.5mm}

\noindent
{\bf Theorem A.2.1.} \ {\em Irreducible Hermitian representations of
  the Galilei algebra $AG(1,3)$ are numbered by numbers  $C_1,\ C_2,\
  C_3$ (eigenvalues of the Casimir operators (\ref{A.2.3})) which take
  the values}\index{Representation!of the Galilei algebra} 
\begin{equation}
\begin{array}{rl}
      1)& C_1^2=m^2>0,\quad C_2=m^2s(s+1),\quad
        -\infty <C_3<+\infty,\\[1mm]
        &s=0,\, 1/2,\, 1,\ldots;\\[2mm]
      2)& C_1=C_2=0,\quad C_3=-k^2<0;\\[2mm]
      3)& C_1=0, \quad C_2=r^2,\quad C_3=-k^2<0.
\end{array}
\label{A.2.4}
\end{equation}

The explicit form of basis operators of an irreducible representation
is determined by the formulae
\begin{eqnarray}
        P_0&=&p_0, \quad P_a\ \, =\ \, p_a, \\
        M&=&C_1 \ \, =\ \, m,\non\\ 
        J_a&=&-i\ve_{abc}p_b{\p\over \p p_c}
          +\lbd_0{(p_a+n_a)\over (1+\vec n\,\vec p\,)},\label{A.2.5}\\
        G_a&=&-ip_a{\p\over \p p_0}-im{\p\over \p p_a}
        +\ve_{abc}{\lbd_b p_c\over (\vec p\,\vec p)}\non\\
        &&-\ve_{abc}p_bn_c{(m\lbd_0-\vec \lbd\,\vec p\,)\over
        (p+\vec n\,\vec p\,)},\non
\end{eqnarray}
where $m$ is a fixed real number, $\lbd_\mu$ are matrices
(\ref{A.1.3a})--(\ref{A.1.4}) and the variables $p_0,\ p_a$ are
connected by the relation
\begin{displaymath}
2mp_0-p_ap_a=C_3,
\end{displaymath}
$C_3$ being fixed too. 

Let us give a brief characterization of the classes of irreducible
representations enumerated in (\ref{A.2.4}):

1) representations of the class I $\ (m\ne 0,\ m^2>0)$ are
characterized by three numbers $m,\ s$ and $\ve_0$, where $m$ and
$\ve_0$ are arbitrary real numbers,
$s$ is an integer or half-integer
non-negative number. Such representations are realized in the space of  
square-integrable functions $f(\vec p,\, \lbd)$, where
\begin{displaymath}
\lbd=-s,\, -s+1,\ldots,\, s,
\end{displaymath}
i.e.,\  the dimension of $f(\vec p,\, \lbd)$ with
respect to the index $\lbd$ is equal to $2s+1$. The space of
irreducible representation of the algebra $AG(1,3)$ is usually
associated with the position space of a free particle having the mass
$m$, the spin $s$ and the internal energy\index{Energy} $\ve_0/2m$;

2) representations of the class II are given by the pair of numbers
\begin{displaymath}
C_3<0\ \mbox{\rm and}\ C_4=0,\, 1/2,\, 1,\ldots.
\end{displaymath}
These representations are one-dimensional and are realized in the
space of square-integrable functions $g(p_0,\, \vec p)$.

Representations of the Galilei algebra of the class II are realized on 
the set of solutions of equations describing fields with the zero
rest mass, for example, Galilei-invariant electro-magnetic field
\cite{143,144};

3) representations of the class III are numbered by the pair of
positive numbers $r^2,\ k^2$. These representations are realized in
the space of square-integrable functions $h(p_0,\, \vec p,\, \lbd)$,
where $\lbd$ takes the infinite number of values
\begin{displaymath}
0,\, \pm 1,\, \pm 2, \ldots\ \mbox{\rm or}\ \pm 1/2,\, \pm 3/2,\ldots.
\end{displaymath}
 
So far representations of the Galilei algebra of the class III have no
applications in physics.

The above considered classes of representations of the algebra
$AG(1,3)$ exhaust all inequivalent non-Hermitian representations of
this algebra if not all $P_\mu$ are equal to zero.

Provided
\begin{displaymath}
P_\mu=0,\ \ \mu={0,\ldots,3},
\end{displaymath}
the Galilei algebra is
isomorphic to the Lie algebra of the Euclid group\index{Euclid!group}
$AE(3)$\index{Euclid!algebra} which is determined by commutation
relations (\ref{4.3.4}). The problem of complete description of
inequivalent irreducible representations of the Euclid algebra is
reduced to a purely algebraic problem which cannot be solved by
already known methods \cite{77}. By the same reason, the problem of
description of all inequivalent covariant
representations\index{Covariant representation} of the algebra
$AG(1,3)$ having the form (4.3.3) is not solved yet.

\newpage
\pagestyle{myheadings}                
\markboth {APPENDIX \ 3}
{Representations of the Poincar\'e and Galilei groups 
by Lie vector fields}
\thispagestyle{empty}
\noindent
{\sl
A P P E N D I X\ \ {\Large \sl 3}\label{ap3}}
\vspace{2mm}

\hrule
\vspace{35mm}

\rightline
{\large\bf
REPRESENTATIONS}
\vspace{2mm}

\rightline
{\large\bf
OF THE POINCAR\'E}
\vspace{2mm}

\rightline
{\large\bf
AND GALILEI GROUPS}
\vspace{2mm}

\rightline
{\large\bf
BY LIE VECTOR FIELDS}
\def\theequation{A.\arabic{section}.\arabic{equation}}
\setcounter {section} {3}
\setcounter {equation}{0}
\vspace{7mm}

Given a fixed representation of a Lie transformations group $G$, the
problem of description of differential equations invariant under the
group $G$ is reduced with the help of the infinitesimal Lie method to
integrating some over-determined linear system of PDEs (called
determining equations). But to solve the problem of constructing {\em
  all} \/differential equations admitting the transformation group $G$
whose representation is not fixed {\em a priori} \/one has
\begin{itemize}
\item{to construct all inequivalent (in some sense) representations of
    the Lie transformation group $G$,}
\item{to solve the determining equations for each representation
    obtained.}
\end{itemize}
And what is more, the first problem, in contrast to the second one,
reduces to solving {\em nonlinear} \/systems of PDEs. It has been
completely solved by Rideau and Winternitz \cite{173.2}, Zhdanov 
and Fushchych \cite{211.4} for the generalized Galilei group 
$G_2(1,1)$ acting in the space of two dependent and two independent 
variables. 

Some new representations of the Galilei group $G(1,3)$ were suggested
in \cite{67.1}--\cite{67.3},\cite{94.1}. Yehorchenko \cite{217.2} and
Fushchych, Tsyfra and Boyko \cite{94.1} have constructed new
(nonlinear) representations of the Poincar\'e groups $P(1,2)$ and
$P(1,3)$, correspondingly. A complete description of {\em covariant}
\/representations of the conformal group $C(n,m)$ in the space of
$n+m$ independent and one dependent variables was obtained by
Fushchych, Zhdanov and Lahno \cite{70.1,106.4}. It has been
established, in particular, that any covariant representation of the
Poincar\'e group $P(n,m)$ with $\mbox{\rm max} \{n,m\} \ge 3$ in the
case of one dependent variable is equivalent to the standard
representation.  But given the condition $\mbox{\rm max} \{n,m\} < 3$,
there exist essentially new representations of the corresponding
Poincar\'e groups.

In this appendix we give a brief account of our latest results on
classification of inequivalent representation of the Euclid group
$E(3)$, which is a semi-direct product of the three-parameter rotation
group $O(3)$ and of the three-parameter Abelian group of translations
$T(3)$, acting in the space of three independent $x_1, x_2, x_3$ and
$n\in{\N}$ dependent $u_1,\ldots,u_n$ variables. Furthermore, we
adduce results on classification of representations of the Poincar\'e
and Galilei groups acting in the space of four independent $x_0, x_1,
x_2, x_3$ and $n\in {\N}$ dependent $u_1,\ldots,u_n$ variables.

It is a common knowledge that investigation of representations of a Lie 
transformation group $G$ is reduced to study of representations of
its Lie algebra $AG$ whose basis elements are first-order
differential operators (Lie vector fields) of the form 
\begin{equation}
  \label{A.3.1}
Q=\xi_\alpha(x,u)\partial_{x_\alpha} + \eta_i(x,u)\partial_{u_i},      
\end{equation}
where $\xi_\alpha,\ \eta_a$ are some real-valued smooth functions
on $x = (x_0,x_1,x_2,x_3)$ $\in {\R}^4$,\ $u = (u_1,u_2,$
$\ldots,u_n)\in {\R}^n$,\ $\partial_{x_\alpha} = {\partial\over
  \partial_{x_\alpha}}$,\ $\partial_{u_i} = {\partial\over
  \partial_{u_i}}$,\ $\alpha=0,\ldots,3$,\ $i=1,2,\ldots, n$.

In the above formulae we have two kinds of variables. The variables
$x_0,\ldots,x_3$ and $u_1,u_2,\ldots,u_n$ will be referred to as
independent and dependent variables, respectively.  Difference between
these becomes essential when we take into consideration partial
differential equations invariant under the Lie algebra $AG$.

Due to the properties of the corresponding Lie transformation
group $G$ basis operators $Q_a,\ a=1,\ldots,N$ of the Lie algebra $AG$ 
satisfy commutation relations
\begin{equation}
  \label{A.3.2}
[Q_a,\ Q_b] = C^{c}_{ab}Q_c,\quad a,b=1,\ldots,N, 
\end{equation}
where $[Q_a,\ Q_b]\equiv Q_aQ_b - Q_bQ_a$ is the commutator.

In (\ref{A.3.2}) $C^{c}_{ab}\in {\R}$ are structure
constants which determine uniquely the Lie algebra $AG$. A fixed set of
the Lie vector fields $Q_a$ satisfying (\ref{A.3.2}) is
called the representation of the Lie algebra $AG$.

Thus, the problem of description of all representations of a given Lie
algebra $AG$ reduces to solving the relations (\ref{A.3.2}) with some
fixed structure constants $C^{c}_{ab}$ in the class of Lie vector fields
(\ref{A.3.1}).

It is easy to check that the relations (\ref{A.3.2}) are not altered
with an arbitrary invertible transformation of variables $x,\ u$
\begin{equation}
  \label{A.3.3}
  \begin{array}{rcll}
y_\alpha&=&f_\alpha(x,u),\quad &\alpha=0,\ldots,3,\\[2mm]
v_i     &=&g_i(x,u),\quad &i=1,\ldots,n,
  \end{array}
\end{equation}
where $f_\alpha,\ g_i$ are smooth functions. That is why, one can
introduce on a set of representations of a Lie algebra $AG$ the
following relation:\ two representations $Q_1,\ldots, Q_N$ and
$Q'_1,\ldots, Q'_N$ are called equivalent if they are transformed one
into another by means of an invertible transformation (\ref{A.3.3}).
Since invertible transformations of the form (\ref{A.3.2}) form the
group (called diffeomorphism group), the relation above is the
equivalence relation. It divides the set of all representations of the
Lie algebra $AG$ into equivalence classes $A_1,\ldots,A_r$.
Consequently, to describe all possible representations of $AG$ it
suffices to construct one representative of each equivalence class
$A_j,\ j=1,\ldots,r$.  
\vspace{1.5mm}

\noindent
{\bf Definition A.3.1.}\ Set of first-order linearly independent
differential operators $P_a, J_b$ of the form (\ref{A.3.1}) is called
the Euclid algebra $AE(3)$ if they satisfy the following 
commutation relations:
\begin{equation}
\label{A.3.4}
[P_a,\ P_b] = 0,\quad [J_a,\ P_b] = \varepsilon_{abc}P_c,\quad
[J_a,\ J_b] =  \varepsilon_{abc}J_c,
\end{equation}
where
\[
\varepsilon_{abc}=\left\{
\begin{array}{rl}
1,&(abc)=\mbox{cycle}\, (123),\\
-1,&(abc)=\mbox{cycle}\, (213),\\
0,& \mbox{in the remaining cases}.
\end{array}\right.
\]
{\bf Definition A.3.2.}\ Set of first-order linearly independent
differential operators\ $P_\mu, J_{\alpha\beta}$ of the form
(\ref{A.3.1}) is called the Poincar\'e algebra $AP(1,3)$ if they
satisfy the following commutation relations:
\begin{equation}
  \label{A.3.5}
  \begin{array}{l}
[P_\mu,\ P_\nu]=0,\quad [P_\mu,\ J_{\alpha\beta}] =
g_{\mu\alpha}P_\beta  - g_{\mu\beta}P_\alpha,\\[2mm] 
[J_{\mu\nu},\ J_{\alpha\beta}]=g_{\mu\beta}J_{\nu\alpha} +
g_{\nu\alpha}J_{\mu\beta} - g_{\mu\alpha}J_{\nu\beta} - 
g_{\nu\beta}J_{\mu\alpha}. 
\end{array}
\end{equation}     
{\bf Definition A.3.3.}\ Set of first-order linearly independent
differential operators\ $P_0, P_a, J_b, G_c, M$ of the form
(\ref{A.3.1}) is called the Galilei algebra $AG(1,3)$ if they
satisfy the commutation relations (\ref{A.3.4}) and
\begin{equation}
\label{A.3.6}
\begin{array}{l}
[P_0,\ P_a]=0,\quad [P_0,\ J_a]=0,\quad [P_0,\ G_a]=P_a,\\[2mm]
[P_0,\ M]=0,\quad [P_a,\ G_b]=\delta_{ab}M,\quad [P_a,\ M]=0,\\[2mm]
[J_a,\ G_b]=\varepsilon_{abc}G_c,\quad [G_a,\ G_b]=0,\quad [G_a,\
M]=0. 
\end{array}
\end{equation}

We say that basis elements of the Euclid algebra $AE(3)$ realize a
covariant representation if they can be reduced to the form
\begin{equation}
\label{A.3.7}
P_a=\p_{x_a},\quad J_a=-\ve_{abc}x_b\p_{x_c} + \eta_{ai}(x,u)\p_{u_i}
\end{equation}
with the help of the transformation (\ref{A.3.3}).

Note that the case when $\eta_{a i}$ are linear in $u$ corresponds to
what is called in the classical representation theory a covariant
representation of the Euclid algebra (see Appendix 2). This is the
reason why we preserve for the more general class of representations
(\ref{A.3.7}) the term \lq covariant representation\rq.

Similarly, operators $P_\mu, J_{\alpha\beta}$ realize a covariant
representation of the Poin\-ca\-r\'e algebra $AP(1,3)$ if they can be
reduced to the form
\begin{equation}
\label{A.3.8}
P_\mu=g_{\mu\nu}\p_{x_\nu},\quad
J_{\alpha\beta}=x_{\alpha}g_{\beta\nu}\p_{x_\nu} -
x_{\beta}g_{\alpha\nu}\p_{x_\nu}+ \eta_{\alpha\beta i}(x,u)\p_{u_i} 
\end{equation}
with the help of a transformation (\ref{A.3.3}).

At last, operators $P_0, P_a, J_a, G_a, M$ realize a covariant
representation of the Galilei algebra $AG(1,3)$ if they can be reduced
to the form
\begin{equation}
\label{A.3.9}
\begin{array}{l}
P_0=\p_{x_0},\quad P_a=\p_{x_a},\quad J_a=-\ve_{abc}x_b\p_{x_c} +
\eta^{1}_{ai}(x,u)\p_{u_i},\\[2mm]
G_a=x_0\p_{x_a} + \eta^{2}_{ai}(x,u)\p_{u_i},\quad
M=\eta^{3}_{i}(x,u)\p_{u_i}
\end{array} 
\end{equation}
with the help of a transformation (\ref{A.3.3}).

A specific role played by covariant representations of the algebras
$AE(3)$, $AP(1,3)$ and $AG(1,3)$ is explained by the fact that they
are widely used in physical applications. Furthermore, the
transformation groups generated by their basis elements have a natural
physical interpretation. The operators $P_0$, $P_a$ generate
translations of the time $x_0$ and space $x_a$ variables,
correspondingly. Next, the operators $J_a$ generate rotations of
the Euclid space $\vec x$ and the operators $J_{0a}$ generate the
Lorentz transformations of the Minkowski space $x_0, \vec x$
preserving the quadratic form $x_\mu x^\mu$. The operators $G_a$
generate the Galilei transformations of the space of independent
variables $x_0, x_a$ leaving the time variable $x_0$ invariant.
 
In what follows, we will restrict our considerations to the case of
covariant representations only.
\vspace{2mm}

\noindent
{\bf 1. Covariant representations of the Euclid algebra}. Direct check
shows that the operators (\ref{A.3.7}) form a basis of the Euclid
algebra iff the following relations hold:
\[
{\partial \eta_{a i}\over \partial x_b}=0,\quad 
[\eta_{a i}\partial_{u_i},\ \eta_{bj}\partial_{u_j}]=\varepsilon_{abc}
\eta_{ci}\partial_{u_i}, 
\]
where $a,b=1,2,3,\ i=1,\ldots,n$.

Consequently, functions $\eta_{a i}$ are independent of $x$ and, in
addition, the operators
\begin{equation}
\label{A.3.10}
{\cal J}_a = \eta_{a i}(u)\partial_{u_i}
\end{equation}
satisfy the commutation relations of the Lie algebra of the rotation
group 
\begin{equation}
\label{A.3.11}
[{\cal J}_a,\ {\cal J}_a]= \varepsilon_{abc} {\cal J}_c.
\end{equation}

Thus, the problem of description of all inequivalent covariant
representations reduces to describing all functions $\eta_{a i}(u)$
such that the operators ${\cal J}_a$ fulfill the commutation relations
(\ref{A.3.11}). Solution of this problem is given by the following
lemma.  
\vspace{1.5mm}

\noindent
{\bf Lemma A.3.1.} {\em Let the differential operators (\ref{A.3.10})
satisfy the commutation relations (\ref{A.3.11}). Then, there exists
a transformation
\begin{equation}
  \label{A.3.12}
v_i=F_i(u),\ \ i=1,\ldots,n
\end{equation}
reducing these operators to one of the following forms:
\begin{eqnarray}
&1.& J_1 = -\sin u_1\tan u_2\partial_{u_1} - \cos u_1\partial_{u_2}, 
\nonumber\\  
&&  J_2 = -\cos u_1\tan u_2\partial_{u_1} + \sin u_1\partial_{u_2},
\label{A.3.13}\\  
&& J_3 = \partial_{u_1};\nonumber\\[4mm]
&2.& J_1 = -\sin u_1\tan u_2\partial_{u_1} - (\cos u_1 -
\alpha \sin u_1 \sec u_2)\partial_{u_2}\nonumber\\
&&\phantom{J_1=} + \sin u_1 \sec u_2 \partial_{u_3}, \nonumber\\    
&& J_2 = -\cos u_1\tan u_2\partial_{u_1} + (\sin u_1 +
\alpha \cos u_1 \sec u_2 )\partial_{u_2}\label{A.3.14}\\
&&\phantom{J_1=} + \cos u_1 \sec u_2 \partial_{u_3},\nonumber\\   
&& J_3 = \partial_{u_1};\nonumber\\[4mm]
&3.& J_a=0,\quad a=1,2,3.\label{A.3.15}
\end{eqnarray}
Here $\alpha$ is an arbitrary smooth function of $u_3,\ldots,u_n$.}
\vspace{1.5mm}

\noindent
{\em Proof.}$\quad$ If at least one of the operators $J_a$ (say $J_3$)
is equal to zero, then by virtue of commutation relations
(\ref{A.3.11}) two other operators $J_2,\ J_3$ are also equal to zero
and we get (\ref{A.3.15}).

Let $J_3$ be a non-zero operator. Then, using a transformation
(\ref{A.3.12}) we can always reduce the operator $J_3$ to the form
$J_3=\partial_{v_1}$ (we should write $J_3'$ but to simplify the
notations we omit hereafter primes). Next, from the commutation
relations $[J_3,\ J_1] = J_2$,\ $[J_3,\ J_2] = -J_1$ it follows that
coefficients of the operators $J_1,\ J_2$ satisfy the system of ordinary
differential equations with respect to $v_1$
\[
{\partial \eta_{2i}\over \partial v_1} = \eta_{3i},\quad
{\partial \eta_{3i}\over \partial v_1} = -\eta_{2i}.\quad
i=1,\ldots,n. 
\] 
 
Solving the above system yields 
\begin{equation}
  \label{A.3.16}
\begin{array}{l}
\eta_{2i} = f_i\cos v_1 + g_i\sin v_1,\\[2mm] 
\eta_{3i} = g_i\cos v_1 - f_i\sin v_1,
\end{array}
\end{equation}
where $f_i,\ g_i$ are arbitrary smooth functions of $v_2, \ldots,
v_n$,\ $i=1,\ldots,n$.
\vspace{1.5mm}

\noindent
{\bf Case 1.} $f_j=g_j=0,\quad j\ge2$.

In this case operators $J_1,\ J_2$ read
\[
J_1=f\cos v_1\partial_{v_1},\quad 
J_2=-f\sin v_1\partial_{v_1}
\]
with an arbitrary smooth function $f=f(v_2,\ldots,v_n)$.

Inserting the above expressions into the remaining commutation
relation $[J_1,\ J_2] = J_3$ and computing the commutator on the
left-hand side we arrive at the equality $f^2=-1$ which can not be
satisfied by a real-valued function.  
\vspace{1.5mm}

\noindent
{\bf Case 2.} Not all $f_j,\ g_j,\ j\ge2$ are equal to $0$.

Making the change of variables
\[
w_1 = v_1 + V(v_2,\ldots,v_n),\quad w_j = v_j,\quad
j=2,\ldots,n 
\]
we transform operators $J_a,\ a=1,2,3$ with coefficients
(\ref{A.3.16}) as follows
\begin{eqnarray}
J_1&=&\tilde f\sin w_1\partial_{w_1} + \sum\limits_{j=2}^n (\tilde f_j
\cos w_1 + \tilde g_j \sin w_1)\partial_{w_j},\nonumber\\
J_2&=&\tilde f\cos w_1\partial_{w_1} + \sum\limits_{j=2}^n (\tilde g_j
\cos w_1 - \tilde f_j \sin w_1)\partial_{w_j},\label{A.3.17}\\
J_3&=&\partial_{w_1}.\nonumber
\end{eqnarray}
{\bf Subcase 2.1.}\ Not all $\tilde f_j$ are equal to $0$.
Making the transformation
\[
z_1=w_1,\quad z_j=W_j(w_2,\ldots,w_n),\quad j=2,\ldots,n,
\]
where $W_2$ is a particular solution of the PDE
\[
\sum\limits_{j=2}^n \tilde f_j \partial_{w_j}W_2=1,
\]
and $W_3,\ldots,W_n$ are functionally-independent first integrals of
the PDE
\[
\sum\limits_{j=2}^n \tilde f_j \partial_{w_j}W=0
\]
we reduce the operators (\ref{A.3.17}) to be
\begin{eqnarray}
  J_1&=&F\sin z_1\partial_{z_1} + \cos z_1\partial_{z_2}
  +\sum\limits_{j=2}^n G_j \sin z_1\partial_{z_j},\nonumber\\ 
  J_2&=&F\cos z_1\partial_{z_1} - \sin z_1\partial_{z_2} +
  \sum\limits_{j=2}^n G_j \cos z_1\partial_{w_j},\label{A.3.18}\\ 
  J_3&=&\partial_{z_1}.\nonumber
\end{eqnarray}

Substituting operators (\ref{A.3.18}) into the commutation relation
$[J_1,\ J_2]$ $=$ $J_3$ and equating coefficients of the
linearly independent operators $\partial_{z_1},\ldots,\partial_{z_n}$
we arrive at the following system of PDEs for the functions $F,\ 
G_2,\ldots,G_n$:
\[
F_{z_2}-F^2 = 1,\quad G_{jz_2}-FG_j=0,\quad j=2,\ldots,n.
\]

Integration of the above equations yields
\begin{eqnarray*}
F&=&\tan (z_2 + c_1),\\
G_j&=&{c_j\over \cos(z_2+c_1)},
\end{eqnarray*}
where $c_1,\ldots,c_n$ are arbitrary smooth functions of
$z_3,\ldots,z_n$, \ $j=2,\ldots,n$. 

Replacing, if necessary, $z_2$ by $z_2+c_1(z_3,\ldots,z_n)$ we may put
$c_1$ equal to zero. Next, making the transformation
\begin{eqnarray*}
y_a&=&z_a,\quad a=1,2,3,\\ 
y_k&=&Z_k(z_3,\ldots,z_n),\quad k=4,\ldots,n,
\end{eqnarray*}
where $Z_k$ are functionally-independent first integrals of the PDE
\[
\sum\limits_{j=3}^n G_j \partial_{z_j}Z = 0,
\]
we can put $G_k=0,\ k=4,\ldots,n$.

With these remarks the operators (\ref{A.3.18}) take the form
\begin{eqnarray}
J_1&=&\sin y_1\tan y_2\partial_{y_1} + \cos y_1\partial_{y_2} + 
{\sin y_1\over\cos y_2}(f\partial_{y_2} + g\partial_{y_3}),\nonumber\\
J_2&=&\cos y_1\tan y_2\partial_{y_1} - \sin y_1\partial_{y_2} +
{\cos y_1\over\cos y_2}(f\partial_{y_2} +
g\partial_{y_3}),\label{A.3.19}\\ 
J_3&=&\partial_{y_1},\nonumber
\end{eqnarray}
where $f,\ g$ are arbitrary smooth functions of $y_3,\ldots,y_n$.

If in (\ref{A.3.19}) $g\not\equiv 0$, then replacing $y_3$ by $\tilde
y_3 = \int g^{-1}d y_3$ and $y_2$ by $\tilde y_2=-y_2$ we transform
the above operators to the form (\ref{A.3.14}).

If $g\equiv 0$, then making the transformation
\[
\tilde u_1=y_1+\arctan{f\over\cos y_2},\quad \tilde u_2 = -\arctan{\sin
  y_2\over \sqrt{\cos^2y_2 + f^2}},\quad \tilde u_k = y_k,
\]
where $k=3,\ldots,n$, we reduce the operators (\ref{A.3.19}) to the
form (\ref{A.3.13}).  
\vspace{1.5mm}

\noindent
{\bf Subcase 2.2.}\ $f_j=0,\ j=2,\ldots,n$.

Substituting operators (\ref{A.3.17}) under $f_j=0$ into the
commutation relation $[J_1,\ J_2]$ $=$ $J_3$ and equating coefficients
of the linearly independent operators
$\partial_{z_1},\ldots,\partial_{z_n}$ yield the following system of
PDEs:
\[
-f^2=1,\quad fg_j=0,\quad j=2,\ldots,n.
\]

As the function $f$ is real-valued, the system obtained is
inconsistent.

Thus, we have proved that operators (\ref{A.3.12})--(\ref{A.3.15})
exhaust a set of all possible inequivalent representations of the Lie
algebra with commutation relations (\ref{A.3.11}) in the class of the
first-order differential operators (\ref{A.3.10}). 

As an immediate consequence of Lemma A.3.1 we get the following
assertion.

{\bf Theorem A.3.1.}\ {\em Any covariant representation of the Euclid
  algebra is equivalent to one of the following representations:
\begin{eqnarray}
&1.& P_a = \partial_{x_a},\quad 
 J_a= - \varepsilon_{abc}x_b\partial_{x_c};
\label{A.3.20}\\[4mm]
&2.& P_a = \partial_{x_a},\nonumber\\
&& J_1=x_3\partial_{x_2} - x_2\partial_{x_3} - \sin u_1 \tan u_2
\partial_{u_1} - \cos u_1 \partial_{u_2}, \label{A.3.21}\\ 
&& J_2= x_1\partial_{x_3}- x_3\partial_{x_1} - \cos u_1 \tan u_2
\partial_{u_1} + \sin u_1 \partial_{u_2},\nonumber\\
&& J_3= x_2\partial_{x_1} - x_1\partial_{x_2} +
\partial_{u_1};\nonumber\\[4mm] 
&3.& P_a = \partial_{x_a},\nonumber\\
&& J_1= x_3\partial_{x_2} - x_2\partial_{x_3} - \sin u_1\tan
u_2\partial_{u_1}\nonumber\\  
&&\phantom{J_1=} - ( \cos u_1 -
\alpha \sin u_1 \sec u_2 )\partial_{u_2} + \sin u_1 \sec u_2
\partial_{u_3},\label{A.3.22}\\  
&& J_2= x_1\partial_{x_3} - x_3\partial_{x_1} - \cos u_1\tan
u_2\partial_{u_1}\nonumber\\  
&&\phantom{J_2=} + ( \sin u_1 + \alpha \cos u_1 \sec u_2
)\partial_{u_2} + \cos u_1 \sec u_2\partial_{u_3},\nonumber\\  
&& J_3= x_2\partial_{x_1} - x_1\partial_{x_2} +
\partial_{u_1}.\nonumber
\end{eqnarray}
Here $\alpha$ is an arbitrary smooth function of $u_3,\ldots,u_n$.}

In two next subsections we will give without proofs the assertions
describing inequivalent covariant representations of the Poincar\'e
and Galilei algebras.
\vspace{2mm}

\noindent
{\bf 2. Covariant representations of the Poincar\'e algebra.}\ 
Inserting the operators (\ref{A.3.8}) into commutation relations
(\ref{A.3.5}) yields that the functions $\eta_{\alpha\beta i}(x,u)$
are independent of $x$ and the operators
\begin{equation}
\label{A.3.23}
{\cal J}_{\alpha\beta}=\eta_{\alpha\beta i}(u)\partial_{u_i}
\end{equation}
satisfy the commutation relations of the Lie algebra of the Lorentz
group
\[
[{\cal J}_{\mu\nu},\ {\cal J}_{\alpha\beta}] = g_{\mu\beta} 
{\cal J}_{\nu\alpha} + g_{\nu\alpha} {\cal J}_{\mu\beta} - 
g_{\mu\alpha} {\cal J}_{\nu\beta} - g_{\nu\beta} 
{\cal J}_{\mu\alpha}.
\]

Consequently, the problem of describing inequivalent covariant
representations of the Poincar\'e algebra reduces to describing
inequivalent representations of the Lorentz algebra having the basis
elements (\ref{A.3.23}).
\vspace{1.5mm}

\noindent
{\bf Theorem A.3.2.}\ {\em Any covariant representation of the
  Poincar\'e algebra is equivalent to the representation
\begin{eqnarray*}
&&P_{\mu}=g_{\mu\nu}\partial_{x_\nu},\\
&&J_{0i}=-x_0\partial_{x_i} - x_i\partial_{x_0} + \frac{1}{2}({\cal
  P}_i + {\cal K}_i),\\ 
&&J_{i3}=x_3\partial_{x_i}- x_i\partial_{x_3} + \frac{1}{2}({\cal P}_i
  - {\cal K}_i),\\
&&J_{12} = x_2\partial_{x_1} - x_1\partial_{x_2} + {\cal J}_{12},\\
&&J_{03}=-x_0\partial_{x_3} - x_3\partial_{x_0} + {\cal D},
\end{eqnarray*}
where $i=1,2$ and the operators ${\cal P}_{i}, {\cal J}_{12}, {\cal D},
{\cal K}_{i}$ are given by one of the formulae below
\begin{eqnarray*} 
&1.&{\cal P}_1=\p_{u_1},\quad
{\cal P}_2=\p_{u_2},\\
&&{\cal J}_{12}={u_2}\p_{u_1} -{u_1}\p_{u_2},\quad 
{\cal D}=  -{u_1}\p_{u_1} -{u_2}\p_{u_2},\\
&&{\cal K}_1= (-{u_1^2} + {u_2^2})\p_{u_1} -2 {u_1} {u_2}\p_{u_2},\\
&&{\cal K}_2= -2 {u_1} {u_2}\p_{u_1} +({u_1^2} - {u_2^2})\p_{u_2};\\[4mm]
&2.&{\cal P}_1=\p_{u_1},\quad
{\cal P}_2=\p_{u_2},\\
&&{\cal J}_{12}= {u_2}\p_{u_1} -{u_1}\p_{u_2},\quad
{\cal D}= -{u_1}\p_{u_1} - {u_2}\p_{u_2} + \p_{u_3},\\
&&{\cal K}_1= (-{u_1^2} + {u_2^2} + 
   \e e^{-2 {u_3}})\p_{u_1} 
  -2 {u_1} {u_2}\p_{u_2} +2 {u_1}\p_{u_3},\\
&&{\cal K}_2=  -2 {u_1} {u_2}\p_{u_1} + ({u_1^2} - {u_2^2} + 
   \e e^{-2 {u_3}})\p_{u_2} + 2 {u_2}\p_{u_3};\\[4mm]
&3.&{\cal P}_1=\p_{u_1},\quad
{\cal P}_2=\p_{u_2},\\
&&{\cal J}_{12}= {u_2}\p_{u_1} -{u_1}\p_{u_2}+\p_{u_3},\quad 
{\cal D}= -{u_1}\p_{u_1} -{u_2}\p_{u_2},\\
&&{\cal K}_1= ( -{u_1^2} + {u_2^2})\p_{u_1} -2 {u_1} {u_2}\p_{u_2} +2
{u_2}\p_{u_3},\\ 
&&{\cal K}_2= -2 {u_1} {u_2}\p_{u_1} +({u_1^2} - {u_2^2})\p_{u_2} -2
{u_1}\p_{u_3};\\[4mm]  
&4.&{\cal P}_1=\p_{u_1},\quad
{\cal P}_2=\p_{u_2},\\
&&{\cal J}_{12}= {u_2}\p_{u_1} -{u_1}\p_{u_2} +\p_{u_3},\quad 
{\cal D}= -{u_1}\p_{u_1} -{u_2}\p_{u_2}+\p_{u_4},\\
&&{\cal K}_1= (-{u_1^2} + {u_2^2} + 
  {\e} e^{-2 {u_4}}\cos 2 {u_3})\p_{u_1} - (2 {u_1} {u_2} + {\e} e^{-2
    {u_4}}\sin 2 {u_3} )\p_{u_2}\\
&&\phantom{{\cal K}_1=} + \Bigl (2 {u_2} + (q \cos {u_3} 
        + r \sin {u_3}) e^{-{u_4}}\Bigr)\p_{u_3}
  + \Bigl( 2 {u_1} - (r \cos {u_3}\\
&&\phantom{{\cal K}_1=} - q \sin {u_3} )e^{-{u_4}}\Bigr)\p_{u_4}
 + e^{-{u_4}}\sin {u_3}\p_{u_5} +  e^{-{u_4}}\cos {u_3}\p_{u_6},\\
&&{\cal K}_2= ( -2 {u_1} {u_2} - {\e} e^{-2 {u_4}}\sin 2 {u_3})\p_{u_1} +
({u_1^2} - {u_2^2} - {\e}e^{-2 {u_4}} \cos 2 {u_3} )\p_{u_2}\\
&&\phantom{{\cal K}_1=} -\Bigl (2 {u_1} + (q \sin {u_3} 
        - r \cos {u_3}) e^{-{u_4}}\Bigr)\p_{u_3}
  + \Bigl(2 {u_2} + (r \sin {u_3}\\
&&\phantom{{\cal K}_1=} + q \cos {u_3} ) e^{-{u_4}}\Bigr)\p_{u_4}
 +  e^{-{u_4}}\cos {u_3}\p_{u_5} -  e^{-{u_4}} \sin {u_3}\p_{u_6};\\[4mm]   
&5.&{\cal P}_1=\p_{u_1},\quad
{\cal P}_2=\p_{u_2},\\
&&{\cal J}_{12}= {u_2}\p_{u_1} -{u_1}\p_{u_2}+\p_{u_3},\quad 
{\cal D}= -{u_1}\p_{u_1} -{u_2}\p_{u_2}+\p_{u_4},\\
&&{\cal K}_1= (-{u_1^2} + {u_2^2} + 
   {\e} e^{-2{u_4}}\cos 2 {u_3} )\p_{u_1} 
  - (2 {u_1} {u_2} +{\e} e^{-2{u_4}} \sin 2 {u_3} )\p_{u_2}\\
&&\phantom{{\cal K}_1=} +  \Bigl(2 {u_2} + (f \cos {u_3} + 
       g \sin {u_3})e^{-{u_4}}\Bigr)\p_{u_3}
  + \Bigl(2 {u_1} - (g\cos {u_3}\\
&&\phantom{{\cal K}_1=}  - f \sin {u_3})e^{-{u_4}}\Bigr)\p_{u_4}
  + (h \cos {u_3} + \sin {u_3}) e^{-{u_4}}\p_{u_5},\\
&&{\cal K}_2= -(2 {u_1} {u_2} +{\e} e^{2 {u_4}} \sin 2 {u_3})\p_{u_1} 
 + ({u_1^2} - {u_2^2} -{\e} e^{-2{u_4}} \cos 2 {u_3})\p_{u_2} \\ 
&&\phantom{{\cal K}_1=}  + \Bigl(-2 {u_1} + (g \cos {u_3} - 
       f \sin {u_3})e^{-{u_4}}\Bigr)\p_{u_3}
  + \Bigl(2 {u_2} + (f \cos {u_3} \\
&&\phantom{{\cal K}_1=}  + g \sin {u_3}) e^{-{u_4}}\Bigr)\p_{u_4} +
  (\cos {u_3} - h \sin {u_3}) e^{-{u_4}}\p_{u_5};\\[4mm]
&6.&{\cal P}_1=\p_{u_1},\quad
{\cal P}_2=\p_{u_2},\\
&&{\cal J}_{12}= {u_2}\p_{u_1} -{u_1}\p_{u_2} +\p_{u_3},\quad 
{\cal D}= -{u_1}\p_{u_1} -{u_2}\p_{u_2} + \p_{u_4},\\
&&{\cal K}_1= (-{u_1^2} + {u_2^2})\p_{u_1} - 2 {u_1} {u_2}\p_{u_2}
  + (2 {u_2} + {\e} e^{-{u_4}} \cos {u_3})\p_{u_3}\\
&&\phantom{{\cal K}_1=} +  \Bigl(2 {u_1} + (f e^{-{\e} {u_5}} \cos
{u_3} + {\e} \sin {u_3}) e^{-{u_4}}\Bigr)\p_{u_4}\\
&&\phantom{{\cal K}_1=}  + \Bigl ((({\e} f e^{-{\e} {u_5}}
 + g)\cos {u_3}  + \sin {u_3})e^{-{u_4}}\Bigr )\p_{u_5}
  +  h e^{-{u_4}}\cos {u_3}\p_{u_6},\\
&&{\cal K}_2= -2 {u_1} {u_2}\p_{u_1} + ({u_1^2} - {u_2^2})\p_{u_2}
  - (2 {u_1} + {\e} e^{-{u_4}} \sin {u_3} )\p_{u_3} \\
&&\phantom{{\cal K}_1=} +\Bigl (2 {u_2} + ({\e} \cos {u_3} - 
       f e^{-{\e} {u_5}} \sin {u_3})e^{-{u_4}}\Bigr)\p_{u_4}\\
&&\phantom{{\cal K}_1=} + \Bigl(\cos {u_3} - ({\e} f e^{-{\e} {u_5}}
 + g) \sin {u_3})e^{-{u_4}}\Bigr)\p_{u_5} 
  - h e^{-{u_4}}\sin {u_3}\p_{u_6};\\[4mm]
&7.&{\cal P}_1=\p_{u_1},\quad
{\cal P}_2=\p_{u_2},\\
&&{\cal J}_{12}= {u_2}\p_{u_1} -{u_1}\p_{u_2} +\p_{u_3},\quad 
{\cal D}= -{u_1}\p_{u_1} -{u_2}\p_{u_2} +\p_{u_4},\\
&&{\cal K}_1= ( -{u_1^2} + {u_2^2})\p_{u_1} -2 {u_1} {u_2}\p_{u_2} 
   +2 {u_2}\p_{u_3} + ( 2 {u_1} + f e^{-{u_4}}\cos {u_3})\p_{u_4}\\
&&\phantom{{\cal K}_1=} +\Bigl(\left ( - {u_5} f + 
         g \right)\cos {u_3} + \sin {u_3}) e^{-{u_4}}\Bigr)\p_{u_5}
   + h  e^{-{u_4}} \cos {u_3}\p_{u_6},\\
&&{\cal K}_2= -2 {u_1} {u_2}\p_{u_1} + ({u_1^2} - {u_2^2})\p_{u_2}
       -2 {u_1}\p_{u_3} + (2 {u_2} - f e^{-{u_4}} \sin {u_3})\p_{u_4}\\ 
&& \phantom{{\cal K}_1=} + \Bigl( (\cos {u_3} + \left( {u_5} f - 
         g \right) \sin {u_3}) e^{-{u_4}}\Bigr)\p_{u_5} 
      - h e^{-{u_4}} \sin {u_3} \p_{u_6};\\[4mm]
&8.&{\cal P}_1=\p_{u_1},\quad
{\cal P}_2=\p_{u_2},\\
&&{\cal J}_{12}= {u_2}\p_{u_1} -{u_1}\p_{u_2} + \p_{u_3},\quad
{\cal D}= -{u_1}\p_{u_1} -{u_2}\p_{u_2} +\p_{u_4},\\
&&{\cal K}_1= (-{u_1^2} + {u_2^2})\p_{u_1} -2 {u_1} {u_2}\p_{u_2} 
  + (2 {u_2} + {\e} e^{-{u_4}} \cos {u_3})\p_{u_3} \\
&&\phantom{{\cal K}_1=} +  (2 {u_1} + {\e} e^{-{u_4}} \sin {u_3})\p_{u_4},\\
&&{\cal K}_2= -2 {u_1} {u_2}\p_{u_1} + ({u_1^2} - {u_2^2})\p_{u_2}
  - (2 {u_1} + {\e} e^{-{u_4}} \sin {u_3})\p_{u_3}\\
&&\phantom{{\cal K}_1=} + (2 {u_2} + {\e} e^{-{u_4}} \cos {u_3})\p_{u_4};\\[4mm]
&9.&{\cal P}_1=\p_{u_1},\quad
{\cal P}_2=\p_{u_2},\\
&&{\cal J}_{12}= {u_2}\p_{u_2} -{u_1}\p_{u_2} +\p_{u_3},\quad
{\cal D}= -{u_1}\p_{u_1} -{u_2}\p_{u_2} + \p_{u_4},\\
&&{\cal K}_1= (-{u_1^2} + {u_2^2})\p_{u_1} -2 {u_1} {u_2}\p_{u_2}
       + 2 {u_2}\p_{u_3} + (2 {u_1} + {\e} e^{-{u_4}}\sin
           {u_3})\p_{u_4},\\ 
&&{\cal K}_2= -2 {u_1} {u_2}\p_{u_1} + ({u_1^2} - {u_2^2})\p_{u_2}
       -2 {u_1}\p_{u_3} + (2 {u_2} + {\e} e^{-{u_4}} \cos
           {u_3})\p_{u_4};\\[4mm] 
&10.&{\cal P}_1=\p_{u_1},\quad
{\cal P}_2=\p_{u_2},\\
&&{\cal J}_{12}= {u_2}\p_{u_1} -{u_1}\p_{u_2} + \p_{u_3},\quad
{\cal D}= -{u_1}\p_{u_1} -{u_2}\p_{u_2} + \p_{u_4},\\
&&{\cal K}_1= (-{u_1}^2 + {u_2}^2 + 
   {\e} e^{-2{u_4}} \cos 2 {u_3})\p_{u_1} -
  (2 {u_1} {u_2} + {\e} e^{-2{u_4}} \sin 2 {u_3})\p_{u_2} \\
&&\phantom{{\cal K}_1=} + 2 {u_2}\p_{u_3} + 2 {u_1}\p_{u_4},\\
&&{\cal K}_2= -(2 {u_1} {u_2} + {\e}  e^{-2 {u_4}} \sin 2
   {u_3})\p_{u_1} + ({u_1}^2 - {u_2}^2 - {\e}  e^{-2 {u_4}} \cos 2
   {u_3})\p_{u_2}\\ 
&&\phantom{{\cal K}_1=} -2 {u_1}\p_{u_3} + 2 {u_2}\p_{u_4};\\[4mm]
&11.&{\cal P}_1=\p_{u_1},\quad
{\cal P}_2=\p_{u_2},\\
&&{\cal J}_{12}= {u_2}\p_{u_1} -{u_1}\p_{u_2} + \p_{u_3},\quad
{\cal D}= -{u_1}\p_{u_1} -{u_2}\p_{u_2} + Q \p_{u_3},\\
&&{\cal K}_1= (-{u_1^2} + {u_2^2})\p_{u_1} -2 {u_1} {u_2}\p_{u_2}
  +2 ({u_2} + {u_1} Q)\p_{u_3},\\
&&{\cal K}_2= -2 {u_1} {u_2}\p_{u_1} + ({u_1^2} - {u_2^2})\p_{u_2}
  -2 ({u_1} - {u_2} Q)\p_{u_3}.
\end{eqnarray*}

Here $\e = 0,1$,\ and $f,g,h$ are arbitrary smooth functions of
$u_6,\ldots,u_n$,\ and $Q$ is an arbitrary smooth function of
$u_4,\ldots,u_n$,\ and
\begin{eqnarray*}
r&=&U(u_5+iu_6,u_7,\ldots,u_n)+U(u_5-iu_6,u_7,\ldots,u_n),\\
q&=&i\Bigl(U(u_5+iu_6,u_7,\ldots,u_n) -
U(u_5-iu_6,u_7,\ldots,u_n)\Bigr)  
\end{eqnarray*}
with an arbitrary function $U$ analytic in the variable
$u_5+iu_6$.}

Note that the operators ${\cal P}_{i}, {\cal J}_{12}, {\cal D},
{\cal K}_{i}$ fulfill the commutation relations of the Lie algebra of
the conformal group $C(2)$ (which is isomorphic to the Lorentz
algebra $AO(1,3)$) 
\begin{eqnarray*}
&&[{\cal P}_i,\, {\cal D}] = -{\cal P}_i,\quad [{\cal P}_1,\, {\cal
  J}_{12}]=-{\cal P}_2,\quad [{\cal P}_2,\, {\cal J}_{12}]= {\cal
  P}_1,\\  
&&[{\cal J}_{12},\, {\cal D}] = 0,\quad [{\cal P}_1,\, {\cal K}_1] =
[{\cal P}_2,\, {\cal K}_2] = {\cal D},\\
&&[{\cal P}_1,\, {\cal K}_2] = -2 {\cal J}_{12},\quad [{\cal P}_2,\,
{\cal K}_1] = 2{\cal J}_{12},\\
&&[{\cal K}_i,\, {\cal D}] = {\cal K}_i,\quad [{\cal K}_1,\, {\cal
    J}_{12}] = - {\cal K}_2,\quad [{\cal K}_2,\, {\cal
    J}_{12}] = {\cal K}_1.
\end{eqnarray*}

The above formulae give the list of all inequivalent representations
of the algebra $AC(2)$ by Lie vector fields.
\vspace{2mm}

\noindent
{\bf 3. Covariant representations of the Galilei algebra.}\ Inserting
the formulae (\ref{A.3.9}) into (\ref{A.3.6}) and making some simple
manipulations we conclude that the basis elements of a covariant
representation of the algebra $AG(1,3)$ necessarily take the form
\begin{equation}
\label{A.3.24}
\begin{array}{l}
P_0=\partial_{x_0},\quad P_a=\partial_{x_a},\quad
J_{a}= \varepsilon_{abc}x_c\partial_{x_b} + {\cal J}_a,\\[2mm]
G_a=x_0\partial_{x_a} + x_a{\cal M} +{\cal G}_a,\quad M={\cal M},
\end{array}
\end{equation}
where ${\cal J}_a, {\cal G}_b, {\cal M}$ are Lie vector fields of the
form $\eta_i(u)\partial_{u_i}$ satisfying the commutation relations of
the Euclid algebra
\[
[{\cal G}_a,\ {\cal G}_b] = 0,\quad [{\cal J}_a,\ {\cal G}_b] =
\varepsilon_{abc}{\cal G}_c,\quad 
[{\cal J}_a,\ {\cal J}_b] =  \varepsilon_{abc}{\cal J}_c
\]
and
\[
[{\cal M},\ {\cal J}_a]=0,\quad [{\cal M},\ {\cal G}_a]=0.
\]
 
On describing all inequivalent representations of the above Lie 
algebra we arrive at the following assertion.
\vspace{2mm}

\noindent
{\bf Theorem A.3.3.}\ {\em Any covariant representation of the Galilei
  algebra $AG(1,3)$ is equivalent to the representation having the
  basis elements (\ref{A.3.24}), operators ${\cal J}_a, {\cal G}_b,
  {\cal M}$ being given by one of the formulae below
\begin{eqnarray*}
&1.& {\cal J}_1=u_3\p_{u_2}-u_2\p_{u_3},\quad 
{\cal J}_2=u_1\p_{u_3}-u_3\p_{u_1},\quad 
{\cal J}_3=u_2\p_{u_1}-u_1\p_{u_2},\\
&&{\cal G}_1=\p_{u_1},\quad {\cal G}_2=\p_{u_2},\quad 
  {\cal G}_3=\p_{u_3},\\
&&{\cal M}={\e}\p_{u_4},\\[4mm]
&2.& {\cal J}_1= -{u_2} \cos {u_3} \tan {u_4}\p_{u_1} 
  + {u_2} \sin {u_3} \tan {u_4}\p_{u_2} + \cos {u_3} \cot
  {u_4}\p_{u_3} \\
&&\phantom{{\cal J}_1=}+ \sin {u_3}\p_{u_4} + \cos {u_3} \csc
{u_4}\p_{u_5},\\ 
&& {\cal J}_2= {u_1} \cos {u_3} \tan {u_4}\p_{u_1} 
  - {u_1} \sin {u_3} \tan {u_4}\p_{u_2}
  - \cot {u_4} \sin {u_3} \p_{u_3} \\
&&\phantom{{\cal J}_1=} + \cos {u_3}\p_{u_4} - \csc {u_4} \sin {u_3}
  \p_{u_5},\\ 
&&{\cal J}_3= u_2\p_{u_1} - {u_1}\p_{u_2} + \p_{u_3},\\
&&{\cal G}_1= \p_{u_1},\quad {\cal G}_2=\p_{u_2},\quad
{\cal G}_3= \cos {u_3} \tan {u_4}\p_{u_1} - \sin {u_3} 
  \tan {u_4}\p_{u_2},\\
&&{\cal M}= f(\cos {u_3} \cos {u_5} \sec {u_4} + \sin {u_3} \sin
     {u_5})\p_{u_1}\\ 
&&\phantom{{\cal M}=} + f (- \sin {u_3} \sec {u_4} \cos {u_5} + 
   \cos {u_3} \sin {u_5})\p_{u_2} + 
   g \p_{u_5} + {\e} \p_{u_6};\\[4mm]
&3.&{\cal J}_1=  - {u_2} \cos {u_3} \tan {u_4}\p_{u_1}
  + {u_2} \sin {u_3} \tan {u_4}\p_{u_2} + \cos {u_3} \cot
  {u_4}\p_{u_3} \\
&&\phantom{{\cal J}_1=}  +\sin {u_3}\p_{u_4},\\
&&{\cal J}_2= {u_1} \cos {u_3} \tan {u_4}\p_{u_1}
  - {u_1} \sin {u_3} \tan {u_4}\p_{u_2}
  - \sin {u_3}\cot {u_4}\p_{u_3}\\
&&\phantom{{\cal J}_1=} + \cos {u_3}\p_{u_4},\\
&&{\cal J}_3= u_2\p_{u_1} - {u_1}\p_{u_2} + \p_{u_3},\\
&&{\cal G}_1= \p_{u_1},\quad {\cal G}_2=\p_{u_2},\quad
{\cal G}_3= \cos {u_3} \tan {u_4}\p_{u_1} - \sin {u_3} 
  \tan {u_4}\p_{u_2},\\[4mm]
&4.&{\cal J}_1= F{(\sec {u_3})^2}\p_{u_1}
  + \cos {u_2} \tan {u_3}\p_{u_2} - \sin {u_2}\p_{u_3},\\
&&{\cal J}_2= (F {(\sec {u_3})^2} \tan {u_2} + 
   {u_1} \sec {u_2} \tan {u_3})\p_{u_1} + \sin {u_2} \tan
   {u_3}\p_{u_2} \\  
&&\phantom{{\cal J}_2=} + \cos {u_2}\p_{u_3},\\
&&{\cal J}_3= -u_1 \tan {u_2}\p_{u_1}-\p_{u_2},\\
&&{\cal G}_1=\p_{u_1},\quad  {\cal G}_2= \tan {u_2}\p_{u_1},\quad
{\cal G}_3= \sec {u_2} \tan {u_3}\p_{u_1},\\
&&{\cal M}= Q\sec {u_2} \sec {u_3}\p_{u_1} + \p_{u_4};\\[4mm]
&5.&{\cal J}_1= Q {(\sec {u_3})^2}\p_{u_1}
  + \cos {u_2} \tan {u_3}\p_{u_2} - \sin {u_2}\p_{u_3},\\
&&{\cal J}_2= (Q{(\sec {u_3})^2} \tan {u_2} + {u_1} \sec {u_2} \tan
   {u_3})\p_{u_1} + \sin {u_2} \tan {u_3}\p_{u_2}\\
&&\phantom{{\cal J}_2=} + \cos {u_2}\p_{u_3},\\ 
&&{\cal J}_3= -u_1 \tan {u_2}\p_{u_1}- \p_{u_2},\\
&&{\cal G}_1=\p_{u_1},\quad {\cal G}_2= \tan {u_2}\p_{u_1},\quad
{\cal G}_3= \sec {u_2} \tan {u_3}\p_{u_1},\\
&&{\cal M}= Q\sec {u_2} \sec {u_3}\p_{u_1};\\[4mm]
&6.& {\cal J}_1= \cos {u_2} \tan {u_4}\p_{u_2}
  + (\cos {u_2} + {u_3} \sin {u_2} \tan {u_4})\p_{u_3} 
  - \sin {u_2}\p_{u_4},\\
&&{\cal J}_2= {u_1} \sec {u_2} \tan {u_4}\p_{u_1}
  + \sin {u_2} \tan {u_4}\p_{u_2} + (\sin {u_2} \\
&&\phantom{{\cal J}_2=} - {u_3} \cos {u_2} \tan {u_4})\p_{u_3} + \cos
  {u_2}\p_{u_4},\\ 
&&{\cal J}_3= -u_1 \tan {u_2}\p_{u_1}-\p_{u_2},\\
&&{\cal G}_1= \p_{u_1},\quad {\cal G}_2= \tan {u_2}\p_{u_1},\quad 
{\cal G}_3= \sec {u_2} \tan {u_4}\p_{u_1},\\
&&{\cal M}= F \sec {u_2} \sec {u_4}\p_{u_1}
  + G \cos {u_4} \p_{u_3} + {\e} \p_{u_5};\\[4mm]
&7.&{\cal J}_1= \sin {u_1} \tan {u_3}\p_{u_1}
  + R \sec {u_3} \sin {u_1}\p_{u_2}
  + (Q \sin {u_1}\sec {u_3} \\
&& \phantom{{\cal J}_1=}+ \cos {u_1})\p_{u_3} + {\e} \sin {u_1} \sec
  {u_3}\p_{u_4},\\ 
&&{\cal J}_2= \cos {u_1} \tan {u_3}\p_{u_1}
  + R\cos {u_1} \sec {u_3}\p_{u_2}
  + (Q\cos {u_1} \sec {u_3} \\
&&\phantom{{\cal J}_1=} - \sin {u_1})\p_{u_3} + {\e} \cos {u_1}\sec
  {u_3}\p_{u_4},\\ 
&&{\cal J}_3= \p_{u_1},\\
&&{\cal G}_1=0,\quad {\cal G}_2=0,\quad {\cal G}_3=0,\\
&&{\cal M}= \p_{u_2};\\[4mm] 
&8.& {\cal J}_1 = -\sin u_1\tan u_2\p_{u_1} - \cos u_1\p_{u_2}, \\  
&&  {\cal J}_2 = -\cos u_1\tan u_2\p_{u_1} + \sin u_1\p_{u_2}, \\  
&& {\cal J}_3 = \p_{u_1},\\ 
&&{\cal G}_1=0,\quad {\cal G}_2=0,\quad {\cal G}_3=0,\\
&&{\cal M}=0;\\[4mm]
&9.& {\cal J}_1 = -\sin u_1\tan u_2\p_{u_1} - (\cos u_1 - 
   \alpha \sin u_1 \sec u_2)\p_{u_2}\\ 
&&\phantom{{\cal J}_1=} + \sin u_1 \sec u_2 \p_{u_3},\\  
&& {\cal J}_2 = -\cos u_1\tan u_2\p_{u_1} + (\sin u_1 + 
   \alpha \cos u_1 \sec u_2)\p_{u_2}\\ 
&&\phantom{{\cal J}_1=} + \cos u_1 \sec  u_2\p_{u_3},\\   
&& {\cal J}_3 = \partial_{u_1},\\
&&{\cal G}_1=0,\quad {\cal G}_2=0,\quad {\cal G}_3=0,\\
&&{\cal M}={\e}\p_{u_1};\\[4mm] 
&10.& {\cal J}_1=0,\quad {\cal J}_2=0,\quad {\cal J}_3=0,\\
&&{\cal G}_1=0,\quad {\cal G}_2=0,\quad {\cal G}_3=0,\\
&&{\cal M}={\e}\p_{u_1}.
\end{eqnarray*}

Here $f,g$ are arbitrary smooth functions of $u_6,\ldots,u_n$,\ $F$ is
an arbitrary smooth function of $u_5$, $\ldots$, $u_n$,\ $R,Q$ are
arbitrary smooth functions of $u_4$, $\ldots$, $u_n$,\ $\alpha$ is an
arbitrary smooth function of $u_3,\ldots,u_n$ and $\e =0,1$.}



\end{document}